\documentclass[12pt]{report} 
\usepackage{epsfig} 
\usepackage{amsmath} 
\usepackage{hhline} 
\usepackage{amssymb} 
\usepackage{times} 
\usepackage{makeidx} 
\newlength{\dinwidth} 
\newlength{\dinmargin} 
\newcounter{sss}[subsection] 
\renewcommand{\thesss}{\thesubsection \, \alph{sss}) \,} 
\newcommand{\lsim}{\raisebox{-1.5mm}{$\:\stackrel{\textstyle{<}}{\textstyle{\sim}}\:$}} 
\newcommand{\gsim}{\raisebox{-0.5mm}{$\stackrel{>}{\scriptstyle{\sim}}$}}

\def \ud {{1 \over 2} } 
\newcommand{\barre}[1]{%
        \setbox1=\hbox{$#1$} \dimen2=\ht1 \dimen3=\dp1 \dimen4=\wd1 
        \setbox2=\hbox{\sl /} 
        \dimen1=\wd1 \advance\dimen1 by -\wd2 \divide\dimen1 by 2 
        \advance\dimen1 by \wd2 \advance\dimen1 by 0.4pt 
        \setbox3=\hbox to \wd1{\hss \box1 \kern -\dimen1 \box2\hss} 
        \ht3=\dimen2 \dp3=\dimen3 \wd3=\dimen4 
        \box3         } 
\def\cm{\hbox{$\;\hbox{\rm cm}$}} 
\def\eV{\hbox{$\;\hbox{\rm eV}$}} 
\def\keV{\hbox{$\;\hbox{\rm keV}$}} 
\def\GeV{\hbox{$\;\hbox{\rm GeV}$}} 
\def\GeVcc{\hbox{$\;\hbox{\rm GeV}$}} 
\def\MeV{\hbox{$\;\hbox{\rm MeV}$}} 
\def\MeVcc{\hbox{$\;\hbox{\rm MeV}$}} 
\def\TeV{\hbox{$\;\hbox{\rm TeV}$}} 
\def\TeVcc{\hbox{$\;\hbox{\rm TeV}$}}

\newcommand{\nanob}{\mbox{{\rm ~nb}~}} 
\newcommand{\picob}{\mbox{{\rm ~pb}~}} 
\newcommand{\femtob}{\mbox{{\rm ~fb}~}} 
\def \L {\Lambda } 
\def \l {\lambda } 
 
\def \t {\theta } 
\def \a {\alpha } 
 
\def \d {\delta } 
\def \D {\Delta }

\def \g {\gamma } 
\def \G {\Gamma } 
\def \O {\Omega } 
 
\def \b {\beta } 
\def \S {\Sigma } 
\def \s {\sigma } 
\def \e {\epsilon } 
\def \cddd { {\cal D } } 
\def \cala { {\cal A } }

\def\be{\begin{equation}} 
\def\ee{\end{equation}} 
\def\bea{\begin{eqnarray}} 
\def\eea{\end{eqnarray}} 
\def\ba{\begin{array}} 
\def\ea{\end{array}}  
 
\newcommand{\Ra}{$\rightarrow$\ } 
 
\def\AJ#1#2#3    {{\rm Astrophys. J.}         {\bf#1  \,}  (#2) \, #3} 
\def\ARNS#1#2#3  {{\rm Ann.~Rev.~Nucl.~Sci.}  {\bf#1  \,}  (#2) \, #3} 
\def\CPC#1#2#3   {{\rm Comp.~Phys.~Comm.}     {\bf#1  \,}  (#2) \, #3} 
\def\EPJC#1#2#3  {{\rm Eur.~Phys.~J}          {\bf #1 \,}  (#2) \, #3} 
\def\IJMPA#1#2#3 {{\rm Int. J.~Mod.~Phys.}    {\bf{A#1} \,}(#2) \, #3} 
\def\MPLA#1#2#3  {{\rm Mod.~Phys.~Lett.}      {\bf{A#1} \,}(#2) \, #3} 
\def\NIMA#1#2#3  {{\rm Nucl.~Instr.~and~Meth.}{\bf#1  \,}  (#2) \, #3} 
\def\NIMB#1#2#3  {{\rm Nucl. Instr. Meth.}    {\bf#1  \,}  (#2) \, #3} 
\def\NPB#1#2#3   {{\rm Nucl.~Phys.}           {\bf{B#1} \,}(#2) \ #3} 
\def\NPBPS#1#2#3 {{\rm Nucl.~Phys.}           {\bf{B#1} (Proc.~Suppl.)\,}(#2)  
\ #3} 
\def\NPA#1#2#3   {{\rm Nucl.~Phys.}           {\bf{A#1} \,}(#2) \, #3} 
\def\NPAPC#1#2#3 {{\rm Nucl.~Phys.}           {\bf{#1A} \,}(#2) \, #3} 
\def\NPBPC#1#2#3 {{\rm Nucl.~Phys.}           {\bf{#1B} \,}(#2) \, #3} 
\def\PLB#1#2#3   {{\rm Phys.~Lett.}           {\bf{B#1} \,}(#2) \ #3} 
\def\PR#1#2#3    {{\rm Phys.~Rep.}            {\bf#1  \,}  (#2) \, #3} 
\def\PRC#1#2#3   {{\rm Phys.~Rev.}            {\bf{C#1} \,}(#2) \, #3} 
\def\PRD#1#2#3   {{\rm Phys.~Rev.}            {\bf{D#1} \,}(#2) \ #3} 
\def\PRL#1#2#3   {{\rm Phys.~Rev.~Lett.}      {\bf#1  \,}  (#2) \ #3} 
\def\PTP#1#2#3   {{\rm Prog.~Theor.~Phys.}    {\bf#1  \,}   (#2) \, #3} 
\def\RMP#1#2#3   {{\rm Rev.~Mod.~Phys.}       {\bf#1  \,}   (#2) \, #3} 
\def\JETP#1#2#3 {{\rm Sov. Phys. JETP} {\bf{#1} \,} (#2) \ #3}
\def\JETPL#1#2#3 {{\rm Sov. Phys. JETP Lett.} {\bf{C#1} \,} (#2) \ #3}  
\def\JHEP#1#2#3  {{\rm J. Hi. Ener. Phys.}    {\bf#1  \,} (#2) \ #3} 
\def\ZETF#1#2#3 {{\rm ZhETF}       {\bf{#1} \,} (#2) \, #3}
\def\ZHETF#1#2#3 {{\rm ZhETF Pis. Red.}       {\bf{C#1} \,} (#2) \, #3} 
\def\ZPC#1#2#3   {{\rm Z.~Phys.}              {\bf{C#1} \,} (#2) \, #3} 
\def\CERNPPE#1   {{\em Preprint CERN,} CERN-PPE/#1} 
\def\CERNEP#1    {{\em Preprint CERN,} CERN-EP/#1} 
\def\etal{{\em et al.}} 
\def \SM    {Standard Model} 
\def \SUSY  {supersymmetry} 
\def \susyq {supersymmetric} 
\def \SSM   {Supersymmetric Standard Model} 
\def \MSSM  {Minimal Supersymmetric Standard Model} 
\def \GUT   {Grand Unified Theories} 
\def \SUGRA {supergravity} 
\def \fc    {flavour changing} 
\def \fcnc  {flavour changing neutral current} 
\def \cc    {coupling constant} 
\def \ccs   {coupling constants} 
\def \VEV   {VEV} 
\def \VEVs  {VEVs}

\newcommand{\LV}{\mbox{$\not \hspace{-0.13cm} L$}} 
\newcommand{\BV}{\mbox{$\not \hspace{-0.13cm} B$}} 
\newcommand{\Rp}{\mbox{$\not \hspace{-0.15cm} R_p$}} 
\newcommand{\WRP}{$W_{\hbox{\scriptsize ${\Rp}$}}$} 
 
\newcommand{\Laba}{$\lambda_{121}$} 
\newcommand{\Laca}{$\lambda_{131}$} 
\newcommand{\Labb}{$\lambda_{122}$} 
\newcommand{\Lbcb}{$\lambda_{232}$} 
\newcommand{\Lacc}{$\lambda_{133}$} 
\newcommand{\Lbcc}{$\lambda_{233}$} 
\newcommand{\Lijk}{$\lambda_{ijk}$} 
\newcommand{\LLE}      {$ {LL \bar E}$} 
\newcommand{\LQD}      {$ {LQ \bar D}$} 
\newcommand{\UDD}      {$ {\bar U \bar D \bar D}$}

\newcommand{\LiLjEk}   {$ L_i L_j \bar{E}_k $} 
\newcommand{\LiQjDk}   {$ L_i Q_j \bar{D}_k $} 
\newcommand{\UiDjDk}   {$ \bar{U}_i \bar{D}_j \bar{D}_k$} 
\newcommand{\lam} {\mbox{$\lambda_{ijk}$}} 
\newcommand{\lamp} {\mbox{$\lambda_{ijk}'$}} 
\newcommand{\lampp} {\mbox{$\lambda_{ijk}''$}} 
\newcommand{\tanb}{tan$\beta$} 
\def \tchi  {{\tilde \chi} } 
\newcommand{\XP}{$\widetilde{\chi}^+$} 
\newcommand{\XPI}{$\widetilde{\chi}_1^+$} 
 
\newcommand{\XMI}{$\widetilde{\chi}_1^-$} 
 
\newcommand{\XPM}{$\widetilde{\chi}^{\pm}$}

\newcommand{\XO}      {$\widetilde{\chi}^0$} 
\def\XOI{\ensuremath{\tilde{\chi}_1^0~}} 
\newcommand{\XOIb}    {$\widetilde{\chi}_1^0$\ } 
\newcommand{\XOII}    {$\widetilde{\chi}_2^0$} 
 
\newcommand{\XOIII}   {$\widetilde{\chi}_3^0$} 
 
\newcommand{\XOIV}    {$\widetilde{\chi}_4^0$}

\newcommand{\SQ}{$\widetilde{\rm q}$} 
\newcommand{\SQB}{$\widetilde{\bar{\rm q}}$} 
\def\ch0{{\tilde{\chi}}^0_1} 
\newcommand{\slep}     {\mbox{$ \tilde{\ell}                         $}}

\newcommand{\sel}      {\mbox{$ \tilde{{\rm e}}                      $}} 
\newcommand{\sell}     {\mbox{$ \widetilde{{\rm e}_L}                $}} 
\newcommand{\selr}     {\mbox{$ \widetilde{{\rm e}_R}                $}} 
\newcommand{\smu}      {\mbox{$ \tilde{\mu}                          $}} 
\newcommand{\smul}     {\mbox{$ \widetilde{\mu}_L                    $}} 
\newcommand{\smur}     {\mbox{$ \widetilde{\mu}_R                    $}} 
\newcommand{\stau}     {\mbox{$ \tilde{\tau}                         $}} 
\newcommand{\staul}    {\mbox{$ \widetilde{\tau}_L                   $}} 
\newcommand{\staur}    {\mbox{$ \widetilde{\tau}_R                   $}} 
\newcommand{\snu}      {\mbox{$ \tilde{\nu}                          $}} 
\newcommand{\snue}     {\mbox{$ \widetilde{\nu}_{{\rm e}}            $}} 
\newcommand{\snum}     {\mbox{$ \widetilde{\nu}_{\mu}                $}} 
\newcommand{\snut}     {\mbox{$ \widetilde{\nu}_{\tau}               $}} 
\newcommand{\asnu}     {\mbox{$ \tilde{\bar \nu}                     $}} 
\newcommand{\asnue}    {\mbox{$ \widetilde{\bar \nu}_{{\rm e}}       $}} 
\newcommand{\asnum}    {\mbox{$ \widetilde{\bar \nu}_{\mu}           $}} 
\newcommand{\asnut}    {\mbox{$ \widetilde{\bar \nu}_{\tau}          $}}

\newcommand{\sfe}      {\mbox{$ \tilde{\mathrm f}                    $}} 
\newcommand{\sfea}     {\mbox{$ \tilde{\mathrm f}_1                  $}} 
\newcommand{\sfeb}     {\mbox{$ \tilde{\mathrm f}_2                  $}} 
\newcommand{\sfeL}     {\mbox{$ \tilde{\mathrm f}_{L}                $}} 
\newcommand{\sfeR}     {\mbox{$ \tilde{\mathrm f}_{R}                $}}

\newcommand{\thmixf}   {\mbox{$\theta_{\tilde{\mathrm f}}            $}}

\newcommand{\squ}{\mbox{$\widetilde{u}$}} 
\newcommand{\sqd}{\mbox{$\widetilde{d}$}} 
\newcommand{\sqt}{\mbox{$\widetilde{t}$}} 
\newcommand{\sqb}{\mbox{$\widetilde{b}$}}

\newcommand{\achia}    {\mbox{$ \tilde{\chi}^{0}_{1}                 $}}

\newcommand{\WW}      {\mbox{$ {\mathrm W}^+{\mathrm W}^-            $}} 
\newcommand{\ZZ}      {\mbox{$ {\mathrm Z}{\mathrm Z}                $}} 
\newcommand{\GG}      {\mbox{$ {\mathrm \gamma}{\mathrm \gamma}      $}}

\def\ee{\ensuremath{e^+e^-~}} 
\def \Eslash {E \kern-.5em\slash } 
\def \pslash {p \kern-.5em\slash } 
\def \kslash {k \kern-.5em\slash } 
\newcommand{\Emiss}{\( \not \! {E} \) }

\newcommand{\Zg}      {\mbox{$ {\mathrm f \rm\bar{f}}  \gamma         $}}

\newcommand{\MW}      {\mbox{$ m_{\mathrm W}                          $}}

\def\D0{D\O} 
\def\met{\mbox{${\hbox{$E$\kern-0.6em\lower-.1ex\hbox{/}}}_T \hspace{.3cm}$}} 
\makeindex 
\newcommand{\barbier}{hepfig} 
\newcommand{\berat}{hepfig} 
 
\newcommand{\deandrea}{hepfig} 
\newcommand{\moreau}{hepfig}

\newcommand{\diagrams}{hepdiag} 
\newcommand{\figures}{hepfig} 
\begin{document} 
\begin{titlepage} 
%
 
\vspace*{1.5cm} 
 
\begin{center} 
\begin{Large} 
\boldmath 
\bf{\LARGE {\boldmath $R$}-PARITY-VIOLATING SUPERSYMMETRY \\ } 
\unboldmath 
  
\vspace{1.6cm}  
{\bf R.~Barbier$^{1}$, 
     C.~B\'erat$^{2}$, 
     M.~Besan\c{c}on$^{3}$, 
     M.~Chemtob$^{4}$, 
     A.~Deandrea$^{1}$, 
     E.~Dudas$^{5,6}$, 
     P.~Fayet$^{7}$, 
     S.~Lavignac$^{4,8}$, 
     G.~Moreau$^{9}$, 
     E.~Perez$^{3}$
     and 
     Y.~Sirois$^{10}$} 
 
\bigskip{\it {\small  
     $ ^{1}$ IPNL, Universit\'e Claude Bernard, IN2P3-CNRS, 
             69622 Villeurbanne, France \\ 
     $ ^{2}$ LPSC, Universit\'e de Grenoble 1, 
             IN2P3-CNRS, 38026 Grenoble, France \\ 
     $ ^{3}$ DAPNIA/Service de Physique des Particules,  
             CEA-Saclay, 91191 Gif-sur-Yvette, France \\ 
     $ ^{4}$ Service de Physique Th\'eorique,  
             CEA-Saclay, 91191 Gif-sur-Yvette, France \\ 
     $ ^{5}$ Laboratoire de Physique Th\'eorique, 
             Universit\'e de Paris-Sud, 91405 Orsay, France \\ 
     $ ^{6}$ Centre de Physique Th\'eorique, 
             Ecole Polytechnique, 91128 Palaiseau, France \\ 
     $ ^{7}$ Laboratoire de Physique Th\'eorique, 
             Ecole Normale Sup\'erieure, 75005 Paris, France \\  
     $ ^{8}$ CERN Theory Division, CH-1211 Gen\`eve, Suisse \\  
     $ ^{9}$ Service de Physique Th\'eorique, 
             Universit\'e Libre de Bruxelles, 1050 Brussels, Belgium \\ 
     $ ^{10}$ Laboratoire Leprince-Ringuet, Ecole Polytechnique,  
             IN2P3-CNRS, 91128 Palaiseau, France \\ 
     }} 
 
\vspace*{0.5cm} 
 
\end{Large} 
 
\vspace*{0.8cm} 
  
\end{center} 
   
\begin{abstract} 
\noindent 
  Theoretical and phenomenological implications of $R$-parity
  violation in supersymmetric theories are discussed in the context
  of particle physics and cosmology. 
  Fundamental aspects include the relation with continuous and 
  discrete symmetries and the various allowed patterns
  of $R$-parity breaking. 
  Recent developments on the generation of neutrino
  masses and mixings within different scenarios of $R$-parity
  violation are discussed.
  The possible contribution of $R$-parity-violating Yukawa
  couplings in processes involving virtual supersymmetric
  particles and the resulting constraints are reviewed.
  Finally, direct production of supersymmetric particles and their
  decays in the presence of $R$-parity-violating couplings is discussed
  together with a survey of existing constraints from collider 
  experiments.
\vspace*{1.0cm} 
\begin{center} 
To be submitted to Physics Reports 
\end{center} 
 
\end{abstract}  
 
\vfill 
 
\end{titlepage} 
 
\newpage 
\tableofcontents 
 
\cleardoublepage
\noindent{\bf INTRODUCTORY REMARKS}
 \addcontentsline{toc}{section}{INTRODUCTORY REMARKS} 

The possible appearance of $R$-parity-violating couplings, and hence 
the question of the conservation or non-conservation of baryon and lepton numbers 
($B$ and $L$) in supersymmetric theories, has been emphasized for a long time.
The rich phenomenology implied by $R$-parity violation has gained full
attention in the search for supersymmetry.
We shall discuss here the theoretical as well as phenomenological aspects 
of \,\Rp \ supersymmetry in particle and astroparticle physics.

\vspace{1.2mm}

In chapter~\ref{chap:intro} we introduce fundamental aspects of supersymmetry,
having in mind the question of the definition of conserved baryon and lepton
numbers in supersymmetric theories. 
In supersymmetric extensions of the Standard Model $R$-parity
has emerged as a discrete remnant of a group of continuous
$U(1)$ $R$-symmetry transformations acting on the supersymmetry
generator.
$R$-parity is intimately connected with baryon and lepton numbers, 
its conservation naturally allowing for conserved baryon and lepton numbers 
in supersymmetric theories.
Conversely, the violation of $R$-parity requires violations 
of $B$ and/or $L$ conservation laws. This generally
leads to important phenomenological difficulties, 
unless $R$-parity-violating interactions are sufficiently 
small. How small they have to be, and how these difficulties may be turned into 
opportunities in some specific cases, concerning for example neutrino masses
and mixings, constitute important aspects of this review.

\vspace{1.2mm}

Chapter~\ref{chap:theory} is devoted to the discussion of {\it how
$R$-parity may be broken}. The corresponding superpotential couplings (and 
resulting Lagrangian terms) and soft supersymmetry-breaking terms are recalled.
Various possible patterns of $R$-parity breaking are discussed, including 
bilinear breaking as well as spontaneous breaking. 
Further theoretical insights on the possible origin of such terms violating 
$B$ and/or $L$ as well as the $R$-parity symmetry are reviewed.
This includes more recent developments on abelian family symmetries, 
grand-unified gauge symmetries, and other discrete symmetries, and what
they could tell us about possible \,\Rp \ terms.

\vspace{1.2mm}

The high-energy convergence of the gauge couplings obtained by 
renormalization-group evolution of low-energy measurements 
gets remarkably improved once supersymmetry is 
introduced. More generally, the renormalization group equations governing the
evolutions of the coupling and mass parameters between two energy scales
provide a way to test,
at lower energies, physical assumptions postulated at a much higher
energy scale; or conversely to translate available experimental data 
into quantities at a higher energy scale. 
In chapter~\ref{chap:evolution}
we consider {\it the effects of the renormalization group equations\,}, in
the presence of $R$-parity-violating interactions. 
We focus in particular, within the supergravity framework, 
on the evolution of the constraints associated with perturbative unitarity, 
the existence of infrared fixed points and the tests of
grand-unification schemes. The additional effects of the new soft
supersymmetry-breaking terms associated with $R_p$-violations are also
discussed. 

\vspace{1.2mm}

Supersymmetric theories with conserved $R$-parity naturally provide
a (color and electrically neutral) stable lightest supersymmetric
particle (LSP), i.e.  a weakly-interacting massive particle which turns
out to be a very good Dark Matter candidate.
In contrast, one of the most striking features of supersymmetric theories 
with $R$-parity-violating interactions stems from the fact that
{\it the LSP can now decay into Standard Model particles only}. We discuss in
chapter~\ref{chap:cosmology} how such an unstable LSP might still
remain (if its lifetime is sufficiently long) a possible Dark Matter candidate. 
We also discuss the gravitino relic issue,  
and the origin of the cosmological baryon asymmetry, 
reviewing several attempts at generating this asymmetry, as well as how it
could survive in the presence of $R$-parity-violating interactions.

\vspace{1.2mm}

The most dramatic implication of $L$-violating interactions
from $R$-parity violations is the automatic generation of
{\it neutrino masses and mixings}. 
The possibility that the results of atmospheric and solar neutrino
experiments be explained by neutrino masses and mixings originating
from $R$-parity-violating interactions has motivated a large number of
studies and models. 
$R$-parity violation in the lepton sector also leads to many
new phenomena related to neutrino and sneutrino physics.
These aspects of neutrino physics related to $L$-violating 
interactions are reviewed in chapter~\ref{chap:neutrinos}. 

\vspace{1.2mm}

In chapter~\ref{chap:indirect} we discuss the possible contribution of
$R$-parity violating couplings to processes involving the {\it virtual
effects} of supersymmetric particles.
Indeed $R$-parity-violating couplings in the Supersymmetric Standard Model
introduce new interactions between ordinary and supersymmetric particles
which can contribute to a large variety of low, intermediate and 
high-energy processes, not involving the direct production of supersymmetric
particles in the final state. 
The requirement that the $R$-parity-violating contribution to a given
observable avoids conflicting with actual experimental measurements,
leads to upper bounds on the $R$-parity-violating couplings possibly involved.
These bounds are extensively discussed, the main ones
being summarized at the end of the chapter, in section~\ref{sec:indcons}. 
Their robustness as well as phenomenological implications are also
discussed at the end of this chapter.

\vspace{1.2mm}

The search for \Rp\,-supersymmetry processes has been a major analysis
activity at high-energy colliders over the past 15 years, and 
is likely to be pursued at existing and future colliders. 
Chapter~\ref{sec:collintro} 
is dedicated to the phenomenology and {\it direct searches}, at colliders,
for supersymmetric particles involving $R$-parity-violating couplings.
The essential ingredients of the corresponding phenomenology at colliders,
including discussions on the magnitude of $R$-parity-violating couplings 
and the subsequent decay of supersymmetric particles, are reviewed. 

\vspace{.2mm}
We then discuss the main and generic features of the $R$-parity-violating 
phenomenology for {\it gaugino-higgsino  pair production\,} and {\it sfermion 
pair production}, both at leptonic and hadronic colliders. 
Furthermore, a remarkable specificity of the phenomenology of \,\Rp \
at colliders comes from possibility of producing {\it a single supersymmetric
particle}. (This is also discussed in chapter~\ref{chap:colliders}
for leptonic, lepton-hadron and hadronic colliders.) 
The phenomenology of \,\Rp \
at colliders also covers virtual effects 
such as those concerning fermion pair production, contributions
to flavor-changing neutral currents and to $CP$ violation. 
These aspects are also met in 
chapter~\ref{sec:collintro}.

Altogether, many direct experimental limits 
have accumulated during the last 15 years of searches for  
\Rp \ processes at colliders.
We do not aim here at an exhaustive (and possibly tedious) catalog 
of all these searches with the corresponding limits. 
We rather choose to refer the reader interested in specific limits and 
details of experimental analyses to the relevant literature 
and emphasize  
only the description of generic features of the phenomenology
of $R$-parity-violating processes at colliders, 
illustrated by examples from the literature.

\vspace{1.2mm}

Conclusions and prospects for supersymmetry with $R_p$-violating couplings 
are given in chapter~\ref{chap:conclusions}. 
Finally,
notations and conventions are summarized in appendix A.
The Yukawa-like \Rp \ interactions associated
with the trilinear \Rp \ superpotential couplings (given in appendix A) 
are derived in appendix B. 
Useful formulae for the production and decays of sfermions, neutralinos 
and charginos are given in appendix C.

\cleardoublepage
\chapter{WHAT IS {\boldmath{$R$}}-PARITY ?}
\label{chap:intro}
In this chapter, we recall how $\,R$-parity\index{Discrete symmetries!$R$-parity}
emerged, in \susyq\ extensions of the \SM,\, as a discrete remnant of a 
continuous $\,U(1)$ $\,R$-symmetry\index{Continuous $R$--symmetry} group
acting on the supersymmetry generator, 
necessarily broken so as to allow for the gravitino and gluinos 
to acquire masses.
\hbox{$R$-parity} naturally forbids unwanted squark and slepton exchanges,
allowing for conserved baryon ($B$) and lepton ($L$) numbers 
in \susyq\ theories. 
It guarantees the stability of the ``Lightest Supersymmetric Particle'', 
which is, also, a very good candidate for the non-baryonic Dark
Matter\index{Dark matter} of the universe.
{\it A contrario\,}, \,$R$-parity violations are necessarily 
accompanied by $\,B\,$\index{Baryon number} and/or 
$\,L\,$\index{Lepton number} violations. 
This is, usually, a source of phenomenological difficulties, 
unless $\,R$-parity-violating (\Rp) interactions are sufficiently small,
as we shall discuss in this review article. $R$-parity violations, on the
other hand, could also appear as a desired feature, 
since they may provide a source of Majorana masses for neutrinos.
\,Whether $\,R$-parity turns out to be absolutely conserved, or not, 
it plays an essential r\^ole in 
the phenomenology of \susyq\ theories, 
and the experimental searches for the new sparticles.

\section{What Is {\boldmath{$R$}}-Parity, and How Was It Introduced\,?}
\label{sec:intro}

Among the problems one had to solve before thinking of applying 
\SUSY\ to the real world, was the question of the definition of
conserved quantum numbers, like baryon number $\,B\,$\index{Baryon number} and 
lepton number $\,L\,$.\index{Lepton number} ~These are carried by Dirac 
fermions\index{Dirac fermions}, the spin-$\frac{1}{2}\,$ 
quarks and leptons.
But \susyq\ theories make a systematic use of {\it \,Majorana\,}
fermions,\index{Majorana fermions} in particular the fermionic 
partners of the spin-1 gauge bosons (now called gauginos).
This makes it very difficult, even in general practically impossible,
for them to carry additive conserved quantum numbers like $\,B\,$ and $\,L$,
in a \susyq\ theory.

Still, even Majorana fermions may be arranged into (chiral or non-chiral) 
Dirac fermions so as to carry non-zero values 
of a new additive quantum number, called $\,R\,$.
In an early $\,SU(2) \times U(1)\,$\index{Group symmetries!Early SUSY model} 
\susyq\ electroweak model with two chiral doublet
Higgs superfields, now called $\,H_d\,$ and $\,H_u \,$
(or $\,H_1\,$ and $\,H_2 \,$), the
definition
of a continuous $\,R$-symmetry\index{Continuous $R$--symmetry} 
acting on the \SUSY\ generator allowed for 
an additive conserved quantum number, $\,R$, \,one unit of which is
carried 
by the \SUSY\ generator~\cite{fayet75}. 
The values of $\,R\,$ for bosons and fermions differ by $\,\pm\,1$ unit
inside 
the multiplets of \SUSY, the photon, for example, having $\,R=0\,$
while 
its spin-$\frac{1}{2}\,$ partner, constrained from the continuous
$\,R$-invariance\index{Continuous $R$--symmetry} to remain 
massless, carries $\,R\,=\,\pm\,1\,$.
Such a quantum number might tentatively have been identified as a lepton 
number, despite the Majorana\index{Majorana fermions} nature of the 
spin-$\frac{1}{2}\,$ partner of the
photon, if the latter could have been identified as one of the
neutrinos.
This, however, is not the case. The fermionic partner of the photon should
be considered as a neutrino of a new type, a ``photonic neutrino'', 
called in 1977 the {\it \,photino\,}.

This still leaves us with the question of how to define, in such theories,
Dirac spinors\index{Dirac fermions} 
carrying conserved quantum numbers like $\,B\,$\index{Baryon number} 
and $\,L\,$.\index{Lepton number}
Furthermore, these quantum numbers, presently known to be carried by 
fundamental fermions only,
not by bosons, 
seem to appear as {\it intrinsically-fermionic\,} numbers.
Such a feature cannot be maintained in a \susyq\ 
theory (in the usual framework of the ``linear
realizations'' of \SUSY), and one had to accept the (then rather heretic)
idea of attributing baryon and lepton numbers to fundamental bosons,
as well as to fermions.
These new bosons carrying $\,B\,$\index{Baryon number} or 
$\,L\,$\index{Lepton number} are the superpartners of the 
spin-$\frac{1}{2}$ quarks and leptons, namely the now-familiar, although 
still unobserved, spin-0  {\it \,squarks\,} and {\it \,sleptons\,}.
Altogether, all known particles should be associated 
with new {\it \,superpartners\,}~\cite{fayet76}.

This introduction of squarks and sleptons now makes the definition 
of baryon and lepton numbers in \susyq\ theories a quasi-triviality
\,-- these new spin-0 particles carrying $\,B\,$ and $\,L\,$,
~respectively,
almost by definition --\, to the point that this old problem
is now hardly remembered, since we are so used to its solution.
This does not mean, however, that these newly-defined $\,B$ and $\,L\,$
should {\it \,necessarily\,} be conserved,
since new interactions that could be present in \susyq\ theories 
might spoil our familiar baryon and lepton-number 
conservation laws, even without taking into account the possibility 
of Grand Unification\,!

In fact the introduction of a large number of new bosons has a price, and
carries 
with it the risk of potential difficulties.
Could these new bosons be exchanged between ordinary particles,
concurrently 
with the gauge bosons of electroweak and strong interactions\,? 
But known interactions are due to the exchanges of spin-1 gauge bosons,
not 
spin-0 particles\,!
\,Can we then construct \susyq\ theories of weak, electromagnetic
and 
strong interactions, which would be free of this potential problem posed
by 
unwanted interactions mediated by spin-0 particles\,?
Fortunately, the answer is yes. As a matter of fact the above problem,
related 
with the conservation or non-conservation of $\,B\,$ and $\,L\,$,
comes with its own natural solution, namely 
$R$-invariance\index{Continuous $R$--symmetry} or, more 
precisely, a discrete\index{Discrete symmetries} version of it, 
known as $\,R$-parity. 
This one is closely related, of course, with the definitions of 
$\,B\,$\index{Baryon number} and $\,L\,$,\index{Lepton number}
~once we have decided, and accepted, to attribute $\,B\,$ and
$\,L\,$ to the new squarks and sleptons, as well as to the ordinary 
quarks and leptons.

$\,R$-parity is associated with a 
$\,Z_2\,$\index{Group symmetries!$Z_2$ subgroup} subgroup 
of the group of continuous $\,U(1)$\index{Group symmetries!$U(1)_R$} 
$\,R$-symmetry\index{Continuous $R$--symmetry} transformations
-- often referred to as $U(1)_R$ -- acting on the gauge superfields 
and the two chiral doublet Higgs superfields 
$\,H_d\,$ and $\,H_u\,$ responsible for electroweak symmetry
breaking~\cite{fayet75},
with their definition extended to quark and lepton superfields
so that quarks and leptons carry $\,R=0\,$, 
~and squarks and sleptons, $\,R=\pm \,1\,$~\cite{fayet76}.
As we shall see later, $R$-parity appears in fact 
as the discrete remnant of this continuous $\,U(1)$
$\,R$-invariance\index{Continuous $R$--symmetry}\index{Group symmetries!$U(1)_R$}
when gravitational interactions are introduced~\cite{fayet77},
in the framework of local \SUSY\ (\SUGRA), in which the gravitino must 
at some point acquire a mass $m_{3/2}\,$ (which breaks the continuous $R$-invariance).
In addition, either the continuous $\,R$-invariance,\index{Continuous $R$--symmetry} 
or simply its discrete\index{Discrete symmetries!$R$-parity} version of 
$\,R$-parity, if imposed, 
naturally forbid the unwanted direct exchanges of the new 
squarks and sleptons 
between ordinary quarks and leptons.
It is, therefore, no surprise if the re-introduction 
of (unnecessary) \Rp\ terms in the Lagrangian density 
generally introduces again, most of the time, the problems 
that were elegantly solved by $\,R$-parity.

The precise definition of $\,R$-invariance,\index{Continuous $R$--symmetry} 
which acts chirally on the anticommuting Grassmann coordinate 
$\,\theta\,$ appearing in the definition 
of superspace and superfields, will be given later (see
Table~\ref{tab:rinv}
in subsection \ref{sec:ssm}).
$\,R$-transformations are defined so as not to act on ordinary particles, 
which all have $\,R=0\,$, \,their superpartners having, therefore, 
$\,R=\,\pm1\,$.
\,This allows one to distinguish between two separate sectors
of $\,R$-even and $\,R$-odd particles. 
$\,R$-even particles (having $\,R$-parity $\,R_p = +\,1\,$) \,include
the 
gluons, the photon, the $\,W^\pm\,$ and $\,Z$ gauge bosons, 
the quarks and leptons,
the Higgs bosons 
originating from the two Higgs doublets (required in \SUSY\ to 
trigger the electroweak breaking and to generate quark and lepton masses)
\,-- and the graviton.
$\,R$-odd particles (having $\,R$-parity $\,R_p= -\,1\,$) 
\,include their superpartners, i.e. the gluinos 
and the various neutralinos and charginos, squarks and sleptons
\,-- and the gravitino.
According to this first definition, $\,R$-parity simply corresponds 
to the parity of the additive quantum number 
$\,R\,$
associated with the above continuous $\,U(1)\,$ 
$\,R$-invariance,\index{Continuous $R$--symmetry}
as given by the expression~\cite{fayet78rp}:
\begin{equation}
\label{eq:rp01}
\hbox {\framebox [11.8cm]{\rule[-.6cm]{0cm}{1.4cm} $ \displaystyle {
\, R\hbox{-parity}\ \ \,R_p\ \ =\ \ (\,-\,1\,)^{R}\ \ =\ \ \left\{ \ 
\begin{array}{l} 
+\,1\ \ \ \ \hbox{for ordinary particles,} \vspace{2mm} \\
-\,1\ \ \ \hbox{for their superpartners.}
\end{array}  \right.
}$}}
\end{equation}

But should we limit ourselves to the discrete\index{Discrete symmetries!$R$-parity} 
$\,R$-parity symmetry, rather than considering its full continuous parent
$\,R$-invariance\,?\index{Continuous $R$--symmetry}
This {\it \,continuous\,} $\,U(1)\,$ $\,R$-invariance\index{Group symmetries!$U(1)_R$}, from which 
$\,R$-parity has emerged, is indeed a symmetry of all
four necessary basic building blocks of the \SSM~\cite{fayet76}:

\vskip -.1truecm
1) the Lagrangian density for the $\,SU(3)\times SU(2) \times U(1)\,$ 
   gauge superfields\index{Group symmetries!Superfields} 
   responsible for strong and electroweak interactions; \\
\indent
2) the $\,SU(3)\times SU(2) \times U(1)\,$ 
   gauge interactions\index{Group symmetries!Standard Model} 
   of the quark and lepton superfields;\\
\indent
3) the $\,SU(2) \times U(1)\,$\index{Group symmetries!Superfields}
   gauge interactions of the two chiral doublet Higgs superfields $\,H_d\,$
   and $\,H_u\,$ responsible for the electroweak breaking;\\
\indent
4) and the ``super-Yukawa'' interactions responsible for quark and lepton masses,
   through the trilinear superpotential couplings of quark and lepton 
   superfields with the Higgs superfields $\,H_d\,$ and $\,H_u\,$,
\begin{equation}
\label{supot}
  W\ =\ \lambda_{ij}^e\  H_d\, L_i E_j^c\ +\
  \lambda_{ij}^d\  H_d \,Q_i D_j^c\ -\ \lambda_{ij}^u\  H_u \,Q_i U_j^c\ ,
\end{equation}
in which chiral quark and lepton superfields are all taken as left-handed and
denoted by $Q_i,U_i^c, D_i^c$ and  $L_i, E_i^c$ respectively (with
$i=1,2$ or $3$ being the generation index).

\vskip .3truecm
Since all the corresponding contributions to the Lagrangian density are 
invariant under this continuous $\,R$-symmetry,\index{Continuous $R$--symmetry} 
why not simply keep it 
instead of abandoning it in favour of its discrete\index{Discrete symmetries} 
version, $R$-parity\,?
But an unbroken continuous $\,R$-invariance,\index{Continuous $R$--symmetry}
which acts chirally on {\it \,gluinos\,}, would constrain them to remain
massless, even after a spontaneous breaking of the \SUSY.
We would then expect the existence of relatively light\index{Hadrons!$R$-hadrons}
``$\,R$-hadrons''~\cite{farrar78,farrar79} made of quarks, antiquarks and 
gluinos, which have not been observed. 
Once the continuous $\,R$-invariance\index{Continuous $R$--symmetry} 
is abandoned, and \SUSY\ is 
spontaneously broken, radiative corrections do indeed allow for the 
generation of gluino masses~\cite{fayet78}, a point to which we 
shall return later.
Furthermore, the necessity of generating a mass for the 
Majorana\index{Majorana fermions} 
spin-$\frac{3}{2}\,$ {\it \,gravitino}, \,once {\it \,local\,} 
\SUSY\ is spontaneously broken, also forces us to abandon 
the continuous $\,R$-invariance \index{Continuous $R$--symmetry} 
in favour of the discrete\index{Discrete symmetries} $\,R$-parity 
symmetry, thereby automatically allowing for gravitino, gluino,
and other gaugino masses~\cite{fayet77}.

\vskip .3truecm
Once we drop the continuous $\,R$-invariance\index{Continuous $R$--symmetry} 
in favour of its discrete\index{Discrete symmetries} 
$\,R$-parity version, it is legitimate to look back and ask:
how general is this notion of $\,R$-parity, and, correlatively, are 
we {\it \,forced\,} to have this $\,R$-parity conserved\,?
As a matter of fact, there is from the beginning a close connection 
between $\,R$-parity and baryon and lepton-number conservation laws, 
which has its origin in our desire to get \susyq\ theories 
in which $\,B\,$ and $\,L\,$ could be conserved, and, at the same time, 
to avoid unwanted exchanges of spin-0 particles.

Actually the superpotential of the \susyq\ extensions of the Standard Model
discussed in Ref.~\cite{fayet76}\, was constrained from the beginning,
for that purpose, to be an {\it \,even\,} function of the quark and lepton
superfields.
In other terms, {\it\,odd\,} gauge-invariant superpotential terms 
\,($\ W\,'$, ~also denoted \WRP\,),
~which would have violated the ``matter-parity'' symmetry 
$\,(\,-1)^{(3B+L)}$, ~were then excluded from the beginning, to be able to 
recover $\,B\,$\index{Baryon number} and $\,L\,$\index{Lepton number}
conservation laws, and avoid direct Yukawa exchanges of spin-0 squarks and
sleptons between ordinary quarks and leptons.

Tolerating unnecessary superpotential terms which are {\it odd\ } 
functions of the quark and lepton superfields \,(i.e. \Rp\ terms, 
precisely those that we are going to discuss in this review), 
does indeed create, in general, immediate problems with baryon- 
and lepton-number conservation laws~\cite{weinberg82}.
Most notably, a squark-induced proton instability with a much too fast 
decay rate, if both $\,B\,$ and $\,L\,$ violations are simultaneously 
allowed; or neutrino masses (and other effects) that could be too 
large, if $\,L$ violations are allowed so that ordinary neutrinos can mix 
with neutral higgsinos and gauginos.
The aim of this review is to discuss in detail how much of these
\Rp\ contributions -- parametrized by sets of 
coefficients $\,\lambda_{ijk},\,\lambda'_{ijk},\,\lambda''_{ijk}\,$
(and possibly $\,\mu_i$, etc.) --
may be tolerated in the superpotential and in the various 
\SUSY-breaking terms.

The above intimate connection between $\,R$-parity 
and  $\,B\,$\index{Baryon number} and $\,L\,$\index{Lepton number} 
conservation laws can be made explicit
by re--expressing the $R$-parity (\ref{eq:rp01})
in terms of the spin $\,S\,$ and a matter-parity $(-1)\,^{3B+L}\,$, 
~as follows~\cite{farrar78}\index{Discrete symmetries!$R$-parity}:
\begin{equation}
\label{eq:rp02}
\hbox {\framebox [6.8cm]{\rule[-.5cm]{0cm}{1.2cm} $ \displaystyle {
R\hbox{-parity} \ \ =\ \ (-1)\,^{2S} \ (-1)\,^{3B+L}    \ \ .     
}$}}
\end{equation}
\noindent
To understand the origin of this formula we note that, 
for all ordinary particles, 
$\,(-1)\,^{2S}$ coincides with $\, \ (-1)\,^{3B+L}$,
~expressing that
among Standard  Model fundamental particles,
{\it \,leptons and quarks, and only them, are fermions,}
i.e. that $\,B\,$ and $\,L\,$ normally appear as intrinsically-fermionic 
numbers.
The quantity $\ (-1)\,^{2S}$ \ $(-1)\,^{3B+L}\ $ is always, trivially,
identical to unity for all known particles (whether fundamental or
composite) 
and for Higgs bosons as well, 
all of them previously defined as having $R$-parity $\,+1\,$.
\,(Indeed expression (\ref{eq:rp01}) of $\,R$-parity comes from the fact
that
   the (additive) quantum number $\,R\,$ was defined so as to vanish for 
   ordinary particles, which then have $\,R$-parity $\,+\,1\,$, their 
   superpartners having, therefore, $\,R$-parity $\,-\,1\,$.)\,
This immediately translates into the equivalent expression (\ref{eq:rp02}) 
of $\,R$-parity.

$R$-parity may also be rewritten as $\ (-1)^{2S} \ (-1)\,^{3\,(B-L)}\,$, 
~showing that this discrete\index{Discrete symmetries} symmetry 
(now allowing for gravitino and gluino masses) may still be conserved 
even if baryon and lepton numbers are separately violated,
as long as their difference ($\,B-L\,$) remains 
conserved, even only modulo 2.
Again, it should be emphasized that the conservation (or non-conservation) 
of $\,R$-parity is closely related with the conservation (or non-conservation) 
of baryon and lepton numbers, $\,B\,$ and $\,L\,$. 
Abandoning $\,R$-parity by tolerating 
both $\,B\,$\index{Baryon number} and $\,L\,$\index{Lepton number} 
violations, simultaneously, would allow for the proton to decay, 
with a very short lifetime\,!

The $\,R$-parity operator plays an essential r\^ole in the construction 
of \susyq\ theories of interactions, and the discussion of the experimental
signatures of the new particles.
$\,R$-invariance,\index{Continuous $R$--symmetry} or simply its 
discrete\index{Discrete symmetries} version of $\,R$-parity, 
guarantees that {\it \,the new spin-0 squarks and sleptons cannot be
directly exchanged\ } between ordinary quarks and leptons.
It ensures that the new $\,R$-odd sparticles can only be pair-produced,
and that the decay of an $\,R$-odd sparticle should always lead to 
another one (or an odd number of them).
Conserved $\,R$-parity also ensures the stability of the ``Lightest 
Supersymmetric Particle'' (or LSP), a neutralino for example (or conceivably
a sneutrino, or gravitino)\,\footnote{The possibility 
of a {\it \,charged\,} or {\it \,colored\,} LSP may also be considered, 
although it seems rather strongly disfavoured, 
as it could lead to new heavy isotopes of hydrogen and other elements, which have not been observed
(cf. subsection \ref{subsec:mc5} in chapter \ref{chap:cosmology}).},
which appears as an almost ideal candidate to constitute the non-baryonic
Dark Matter~\index{Dark matter} that seems to be present in our universe.

Expression (\ref{eq:rp02}) of $\,R$-parity in terms of 
$\,B\,$\index{Baryon number} and $\,L\,$\index{Lepton number} 
makes very apparent that imposing  $\,R$-parity is
equivalent 
to imposing a matter-parity symmetry.
Still the definition of $\,R$-parity offers the additional  advantage of 
identifying the two separate sectors of $\,R_p= +1\,$ particles and
$\,R_p=-1$ 
sparticles, making apparent the pair-production law of the new $\,R$-odd 
sparticles, and the stability of the LSP, if $\,R$-parity is conserved. 
Considering ``matter-parity'' alone would only imply directly the 
stability of the lightest ``matter-odd'' particle, not a very useful
result\,!

Obviously, in the presence of $\,R$-parity violations, the 
LSP is no longer required to be stable, superpartners being allowed to 
decay into ordinary particles.

\section{Nature Does Not Seem To Be Supersymmetric\,!}
\label{sec:back}

The algebraic structure of \SUSY\ involves a spin-$\frac{1}{2}\,$ fermionic
symmetry generator $\,Q\,$ satisfying the (anti)\,commutation relations 
in four dimensions~\cite{golfand71,volkov73,wess74}: 
\begin{equation}
   \label{alg}
   \left\{\ 
   \begin{array}{cccc}
      \{ \, Q , \bar{Q} \, \} & = & \, - \ 2 \ \gamma_{\mu} P^{\mu} &  , 
      \vspace*{0.3cm} \\
      \left[ \, Q ,  P^{\mu} \, \right] & = &  \, 0  &  .
   \end{array} 
   \right. 
\end{equation}
This spin-$\frac{1}{2}\,$ \SUSY\ generator $\,Q\,$, \,here written as a 
4-component Majorana\index{Majorana fermions} spinor,
was originally introduced as relating fermionic with bosonic fields, 
in relativistic quantum field theories.
The presence of the generator of space-time translations $\,P^\mu\,$ 
on the right-hand side of the anticommutation relations (\ref{alg})
is at the origin of the relation of \SUSY\ with general relativity 
and gravitation, since a locally \susyq\ theory 
must be invariant under local coordinate transformations~\cite{sugra}.

The \SUSY\ algebra (\ref{alg}) was introduced with quite different 
motivations: in connection with parity violation, with the hope of 
understanding parity violation in weak interactions as a consequence 
of a (misidentified) intrinsically parity-violating nature 
of the \SUSY\ algebra~\cite{golfand71};
in an attempt to explain the masslessness of the neutrino 
from a possible interpretation 
as a spin-$\frac{1}{2}\,$ Goldstone particle~\cite{volkov73}; 
or by extending to four dimensions the \SUSY\ transformations
acting on the two-dimensional string worldsheet~\cite{wess74}.
However, the mathematical existence of an algebraic structure does 
not mean that it could play a r\^ole as an invariance 
of the fundamental laws of Nature\,\footnote{Incidentally 
while \SUSY\ is commonly referred to as
``relating fermions with bosons'', \,its algebra (\ref{alg}) does not even 
require the existence of fundamental bosons\,! 
\,With non-linear 
realizations of \SUSY\ a fermionic field can be 
transformed into a {\it \,composite\,} bosonic field made of fermionic 
ones~\cite{volkov73}.}.

Indeed many obstacles seemed, long ago, to prevent \SUSY\ 
from possibly being a fundamental symmetry of Nature.
Is spontaneous \SUSY ~breaking possible at all\,?
Where is the spin-$\frac{1}{2}\,$ Goldstone fermion of \SUSY, 
if not a neutrino\,?
Can we use \SUSY\ to relate directly known bosons and fermions\,?
And, if not, why\,?
\,If known bosons and fermions cannot be directly related by \SUSY, 
do we have to accept this as the sign that \SUSY\ is {\it \,not\,}
a symmetry of the fundamental laws of Nature\,?
Can one define conserved baryon and lepton numbers in such theories,
although they systematically involve {\it \,self-conjugate\,}
Majorana fermions\index{Majorana fermions}, (so far) unknown in Nature\,?
And finally, if we have to postulate the existence of new bosons 
carrying $\,B\,$ and $\,L\,$ 
\,-- the new spin-0 squarks and sleptons --\,
can we prevent them from mediating new unwanted interactions\,? 

While bosons and fermions should have equal masses 
in a \susyq\ theory, this is certainly not the case in Nature.
Supersymmetry should clearly be broken.
But it is a special symmetry, 
since the Hamiltonian, which appears on the right-hand side 
of the anticommutation relations  (\ref{alg}), can be expressed 
proportionally to the sum of the squares of the components 
of the \SUSY\ generator, as
$\,H=\frac{1}{4}\ \sum_\alpha\,Q_\alpha^{\ 2}\,$.
~This implies that a \SUSY-preserving vacuum state must 
have vanishing energy, while a state which is not invariant 
under \SUSY\ could na\"{\i}vely be expected to have a larger, 
positive, energy. As a result, potential candidates for
\SUSY-breaking vacuum states
seemed to be necessarily unstable. This led to the question
\begin{equation}
\hbox{Q}1:\ \ \ \ \ 
\hbox {\it{Is spontaneous \SUSY ~breaking possible at all\,?}}
\end{equation}

As it turned out, and despite the above argument, 
several ways of breaking spontaneously global or local 
\SUSY\ have been found~\cite{fayet74,cremmer79}.
\,But spontaneous \SUSY\ breaking remains in general
rather difficult to obtain, at least in global \SUSY \ (and even without adressing yet the issue
of how this breaking could lead to a realistic theory),
since theories tend to prefer systematically,
for energy reasons, \susyq\ vacuum states. 
Only in very exceptional situations can the existence of 
such states be avoided\,!
\,In local \SUSY, which includes gravity,
one also has to arrange, at the price of a very severe fine-tuning,
for the energy density of the vacuum to vanish exactly, 
or almost exactly, to an extremely good accuracy,
so as not to generate an unacceptably large value of the 
cosmological constant $\,\Lambda\,$.

We still have to break \SUSY\
in an acceptable way, so as to get  \,-- if this is indeed possible --\,
a physical world which looks like the one we know\,!
\,Of course just accepting explicit \SUSY-breaking terms
without worrying too much about their possible origin
would make things much easier
\,(unfortunately also at the price of introducing a large number of arbitrary
parameters).
But such terms must have their origin 
in a spontaneous \SUSY-breaking mechanism, 
if we want \SUSY\ to play a fundamental r\^ole, 
especially if it is to be realized as a local fermionic gauge symmetry, 
as it should in the framework of \SUGRA\ theories.

But the spontaneous breaking of the global \SUSY\ must
generate a massless spin-$\frac{1}{2}\,$ Goldstone particle, leading to 
the next question,
\begin{equation}
\hbox{Q}2: \ \ 
\hbox {\it{Where is the spin-$\frac{1}{2}\,$ 
Goldstone fermion of \SUSY\ ?}}
\end{equation}
\noindent
Could it be one of the neutrinos~\cite{volkov73}\,?
A first attempt\index{Group symmetries!Early SUSY model} 
at implementing this idea within a $\,SU(2) \times U(1)\,$ 
electroweak model of ``leptons''~\cite{fayet75} quickly illustrated 
that it could not be pursued very far. 
The ``leptons'' of this model were soon to be reinterpreted as the 
``charginos'' and ``neutralinos'' of the \SSM.

If the Goldstone fermion of \SUSY\ is not one of the neutrinos, 
why hasn't it been observed\,?
Today we tend not to think at all about the question,
since: 1) the generalized use of soft terms breaking {\it \,explicitly\,} 
the \SUSY\ seems to render this question irrelevant; \ 
2) since \SUSY\ has to be realized locally anyway, 
within the framework of \SUGRA~\cite{sugra}, 
the massless \hbox{spin-$\frac{1}{2}\,$} Goldstone fermion (``goldstino'') 
should in any case be eliminated 
in favour of extra degrees of freedom for a massive 
spin-$\frac{3}{2}\,$ gravitino~\cite{fayet77,cremmer79}.

But where is the gravitino, and why has no one ever seen a fundamental 
spin-$\frac{3}{2}\,$ particle\,?
To discuss this properly we need to know which bosons and fermions 
could be associated under \SUSY.
Still, even before adressing this crucial question we might already anticipate 
that the interactions of the 
gravitino, with amplitudes proportional to the square root of the
Newton constant $\,\sqrt{G_N}\simeq 10^{-19} \GeV^{-1}$,
should in any case be absolutely negligible in particle physics experiments,
so that we don't have to worry about the fact that no gravitino
has been observed.

This simple but na\"{\i}ve answer is, however, not true 
in all circumstances\,! 
It could be that the gravitino is light, possibly
even extremely light, so that it would still interact very much like the
massless Goldstone fermion of global \SUSY, according to the
``equi\-valence theorem'' of \SUSY~\cite{fayet77}.
Its interaction amplitudes are then determined by the ratio 
$\,\sqrt{G_N}/m_{3/2}\,$ (i.e. are inversely proportional to the square
of the ``supersymmetry-breaking scale'' $\Lambda_{\hbox{\scriptsize ss}}$).
As a result a sufficiently light gravitino could have non-negligible
interactions, which might even make it observable in particle physics 
experiments, provided that the \SUSY-breaking scale parameter fixing
the value of its mass $\,m_{3/2}$\, is not too 
large~\cite{fayet77,fayet86}\,!
Because of the conservation of $\,R$-parity, at least 
to a good approximation, in the \SSM, the 
$R$-odd gravitino should normally be produced in association with another 
$\,R$-odd superpartner, provided the available energy is sufficient. 
Gravitinos could also be pair-produced, although these processes are 
normally suppressed at lower energies. But they 
would remain essentially ``invisible'' in particle physics, 
as soon as the supersymmetry-breaking scale is large enough
(compared to the electroweak scale), 
which is in fact the most plausible and widely considered situation.

In any case, much before getting to the \SSM, 
the crucial question to ask, if \SUSY\ is to be relevant in 
particle physics, is:
\begin{equation}
\hbox{Q}3: \ \ 
\hbox {\it{Which bosons and fermions 
could be related by \SUSY\ ?}}
\end{equation}
But there seems to be no answer since known bosons and fermions do not 
appear to have much in common (excepted, maybe, for the photon and the
neutrino).
In addition the number of (known) degrees of freedom is significantly 
larger for fermions than for bosons.
And these fermions and bosons have very different gauge symmetry
properties\,! Furthermore, as discussed in 
subsection \ref{sec:intro}, the question 
\begin{equation}
\hbox{Q}4: \ \ 
\begin{array}{c} \hbox {\it How could one define (conserved)} \\
\hbox{\it baryon and lepton numbers, 
in a \susyq\ theory ?}
\end{array}
\end{equation}
once appeared as a serious difficulty, 
owing in particular to the presence of {\it \,self-conjugate\,} 
Majorana fermions\index{Majorana fermions} in \susyq\ theories.
Of course nowadays we are so used to dealing with spin-0 squarks and
sleptons, 
carrying baryon and lepton numbers almost by definition, 
that we can hardly imagine this could once have appeared as a problem.
Its solution required accepting the idea 
of attributing baryon or lepton numbers to a large number of new 
fundamental bosons.
Even then, if such new spin-0 squarks and sleptons are introduced,
their direct (Yukawa) exchanges between ordinary 
 quarks and leptons, if allowed, 
could lead to an immediate disaster, preventing us from getting a theory 
of electroweak and strong interactions mediated by spin-1 
gauge bosons, and not spin-0 particles, 
with conserved $\,B\,$ and $\,L\,$ quantum numbers\,!
This may be expressed by the question
\begin{equation}
\hbox{Q}5: \ \
\begin{array}{c} \hbox {\it{How can we avoid unwanted interactions}} \\ 
\hbox{\it{mediated by spin-0 squark and slepton exchanges\,?}}
\end{array}
\end{equation}

\noindent
Fortunately, we can naturally avoid 
such unwanted interactions, thanks to $\,R$-parity, which, if present,
guarantees that squarks and sleptons can\ {\it not\,} be
directly exchanged between ordinary quarks and leptons, 
allowing for conserved baryon and lepton numbers 
in \susyq\ theories.

\section{\sloppy Continuous {\boldmath{$R$}}-Invariance, and Electroweak 
Breaking}
\label{sec:R}

The definition of the continuous $\,R$-invariance\index{Continuous $R$--symmetry} 
we are using arose
from an early attempt at relating known bosons and fermions together,
in particular the spin-1 photon with a spin-$\frac{1}{2}\,$ neutrino.
If we want to try to identify the companion of the photon as being a
``neutrino'', although it initially appears as a
self-conjugate Majorana fermion,\index{Majorana fermions} 
we need to understand how it could carry
a conserved quantum number that we might attempt to interpret
as a ``lepton'' number.
This led to the definition of {{\it a continuous $\,U(1)\,$
$\,R$-invariance}}~\cite{fayet75},\index{Continuous $R$--symmetry} 
which also guaranteed the
masslessness of this ``neutrino'' (``$\nu_L$'', ~carrying $\,+1\,$
unit of $\,R\,$), by acting chirally on the Grassmann coordinate
$\,\theta\,$ which appears in the expression of the various
gauge and chiral superfields\,\footnote{
This $R$-invariance itself originates from an analogous
$Q$-invariance used, in a two-Higgs-doublet presupersymmetry model, 
to restrict the allowed Yukawa and $\,\varphi^4$ interactions, 
in a way which prepared for the two Higgs doublets and chiral
fermion doublets to be related by supersymmetry~\cite{fayet74_bis}. 
$Q$-transformations were then modified
into $R$-symmetry transformations, which survive the electroweak breaking 
and allow for massive Dirac fermions carrying the new quantum number $R$.
\,Transformations similar to $R$-transformations
were also considered in \cite{salam75}, but acted differently 
on two disconnected sets of chiral superfields 
$\,\phi_+\,$ and $\,\phi_-\,$, \,with no mutual interactions;
i.e. they acted differently on the Grassmann coordinate 
$\,\theta$, \,depending on whether the superfields considered 
belonged to the first or second set.}.

Attempting to relate the photon with one of the neutrinos
could only be an exercise of limited validity.
The would-be ``neutrino'', in particular, while having in this model
a $\ V-A\ $ coupling to its associated ``lepton'' and the charged
$\,W^\pm$ boson, was in fact what we would now call a ``photino'',
not directly coupled to the $\,Z\,$!
~Still this first attempt, which became a part of the \SSM,
illustrated how one can break spontaneously a
$\,SU(2) \times U(1)$\index{Group symmetries!Electroweak symmetries} 
electroweak gauge symmetry in a \susyq\ theory,
using
{\it a {\bf pair} \,of  chiral doublet Higgs superfields},
now known as $\,H_d\,$ and $\,H_u\,$!
~Using only a single doublet Higgs superfield
would have left us with {\it a massless charged chiral fermion}, which is,
evidently, unacceptable.
~Our previous charged ``leptons''
were in fact what we now call two winos, or charginos,
obtained from the mixing of charged gaugino and higgsino components,
as given by the mass matrix
\begin{equation}
\label{mwino}
 M_C\ =\ 
 \hbox{\small $\displaystyle
 \left( \begin{array}{cc}
   (\,M_2\,=\,0\,) &
   \displaystyle{\frac{g\,v_u}{\sqrt 2}\,= \,M_{W}\sqrt{2}\,\sin \beta} \\
   \displaystyle{\ \frac{g\,v_d}{\sqrt 2}\,=\,M_{W}\sqrt{2}\,\cos \beta} &
   \mu\,=\,0
 \end{array} \right)
 $}
\end{equation}
\noindent
in the absence of a direct higgsino mass originating from a
$\ \mu\ H_u H_d\ $ mass term 
in the superpotential\,\footnote{This $\,\mu\,$ term initially written in 
\cite{fayet75} \,(which would have broken explicitly the continuous $\,U(1)$
$R$-invariance\index{Continuous $R$--symmetry}\index{Group symmetries!$U(1)_R$}) 
was immediately replaced by a $\ \lambda\ \,H_u H_d\,N\ $ trilinear 
coupling involving an 
{\it \,extra neutral singlet chiral superfield\,} $\,N\,$:
$\ \mu\ \,H_u H_d\ \ \longmapsto \  \lambda\  H_u H_d\,N\,$,
\,as in the so-called Next-to-Minimal Supersymmetric Standard Model
(NMSSM).}.
The whole construction showed that one could deal elegantly 
with elementary spin-0 Higgs fields
(not a very popular ingredient at the time),
in the framework of spontaneously-broken \susyq\ theories.  
Quartic Higgs couplings are no longer completely arbitrary, 
but fixed by the values of the electroweak gauge couplings
$g\,$ and $\,g'\,$ through the following ``$D$-terms''
in the scalar potential given~\footnote{With a different denomination for
the two Higgs doublets, such that $\ \varphi'' \ \mapsto \ $ Higgs doublet
$ h_d,$ \linebreak $(\varphi')^c\ \mapsto \ $ Higgs doublet $ \,h_u,$ 
$\ \,\tan \delta = v'/v''\ \mapsto\ \tan \beta = v_u/v_d\ $.}
in~\cite{fayet75}:
\begin{equation}
  \begin{array}{ccl}
    \vspace{-.4cm} \\
    V_{\hbox{\footnotesize{Higgs}}} & = & 
     \displaystyle{\frac{g^2}{8} \
                   (h_d^\dagger\ \vec{\tau} \, h_d +
                    h_u^\dagger\ \vec{\tau} \, h_u )^2 \, + \,
                   \frac{{g'}^2}{8} \     
                   ( h_d^\dagger h_d-h_u^\dagger h_u )^2 \, + \, ... } 
     \vspace{3mm} \\  & = &
     \displaystyle{\frac{g^2 + {g'}^2}{8} \
                    ( h_d^\dagger h_d-h_u^\dagger h_u )^2 \, + \,
                   \frac{g^2}{2} \
                    | h_d^\dagger h_u |^2 \, + \, ... \ \ .      }
  \end{array}
\end{equation}

\noindent
This is precisely the quartic Higgs potential of the ``minimal'' version
of the \SSM, the so-called \MSSM\ (MSSM).
Further contributions to this quartic Higgs potential also appear in the
presence of additional superfields, such as the neutral singlet chiral
superfield $\,N\,$ already introduced in the previous  model, which
plays an important r\^ole in the NMSSM, i.e. in ``next-to-minimal'' or
``non-minimal'' versions of the \SSM.
Charged Higgs bosons (called $\,H^\pm$) are always present in this
framework, as
well as several neutral ones, three of them at least.
Their mass spectrum depends on the details of the \SUSY-breaking
mechanism considered: soft-breaking terms, possibly ``derived from
\SUGRA'', presence or absence of extra $U(1)\,$ gauge 
fields\index{Group symmetries!extra $U(1)$}
and/or additional chiral superfields, r\^ole of radiative corrections,
etc..

\section{{\boldmath{$R$}}-Invariance and {\boldmath{$R$}}-Parity in the \SSM}
\label{sec:ssm}

These two Higgs doublets $\,H_d\,$ and $\,H_u\,$ are precisely those 
used to generate the masses of charged leptons and down 
quarks, and of up quarks, in \susyq\ extensions of the
\SM~\cite{fayet76}. 
Note that at the time having to introduce Higgs fields was generally 
considered as rather unpleasant.
While one Higgs doublet was taken as probably unavoidable to get to the 
\SM\ or at least simulate the effects of the spontaneous
breaking of the electroweak symmetry, having to consider two doublets, 
necessitating charged Higgs bosons as well as several neutral ones,
in addition to the ``doubling of the number of particles'', was usually 
considered as further indication of the irrelevance of \SUSY.
As a matter of fact, considerable work was devoted for a while on 
attempts to avoid fundamental \hbox{spin-0} Higgs fields (and extra 
sparticles), before returning to fundamental Higgs bosons, precisely in this 
framework of \SUSY.

In the previous $\,SU(2)\times U(1)\,$
\index{Group symmetries!Early SUSY model} 
model~\cite{fayet75}, \,it was
impossible to view seriously for very long ``gaugino'' and ``higgsino''
fields as possible building blocks for our familiar lepton fields.
This led to consider that all quarks, and leptons, ought to be associated
with new bosonic partners, the {\it \,spin-0 squarks and sleptons}.
Gauginos and higgsinos, mixed together by the spontaneous breaking of 
the electroweak symmetry, correspond to a new class of fermions, now 
known as ``charginos'' and ``neutralinos''.
In particular, the partner of the photon under \SUSY, which 
cannot be identified with any of the known neutrinos, should be viewed 
as a new ``photonic neutrino'', the {\it \,photino\,}; the fermionic 
partner of the gluon octet is an octet of self-conjugate Majorana 
fermions\index{Majorana fermions} called 
{\it \,gluinos\,} (although at the time 
{\it \,colored fermions\,} belonging to {\it \,octet\,} 
representations of the colour 
$\,SU(3)\,$\index{Group symmetries!$SU(3)_C$}
gauge group were generally believed not to exist), etc..

The two doublet Higgs superfields~\footnote{The correspondence 
  between earlier notations and modern ones is as follows:
  \\ [.2 true cm]
  \scriptsize
  \begin{tabular}{ccc}
  $S\ =\ \left( \begin{array}{cc} S^0 \vspace{.1truecm}\\ S^-
  \end{array} \right)\ \,\hbox{and}\ \ \,
  T\ = \ \left( \begin{array}{cc} T^0 \vspace{.1truecm}\\ T^-
  \end{array} \right)$   &  $\longmapsto $   &  
  $H_d\ =\ \left( \begin{array}{cc} H_d^{\,0} \vspace{.1truecm}\\
H_d^{\,-}
  \end{array} \right)\ \,\hbox{and}\ \ \,
  H_u\ = \ \left( \begin{array}{cc} H_u^{\,+} \vspace{.1truecm}\\
H_u^{\,0}
  \end{array} \right)\ .$
                           \\  [-.1 true cm] && \\
  \ \ \ \ \ \  (left-handed) \ \ \ \ \ \ \ \ \ \ \ \ \ \ \ \ \ \ \ \
(right-handed) \  &   &   (both
left-handed) 
  \\     && \\ 
\end{tabular} }
$\,H_d$ and $\,H_u\,$ generate quark and lepton masses~\cite{fayet76},
in the usual way, through the familiar trilinear superpotential of
Eq. (\ref{supot}).
The vacuum expectation values of the two corresponding spin-0 Higgs doublets 
$\,h_d\,$ 
and $\,h_u\,$ generate charged-lepton and down-quark masses, and up-quark 
masses, with mass matrices given by
$\,m^e_{ij}\,=\,\lambda^e_{ij}\,v_d/\sqrt 2\,,\
\,m^d_{ij}\,=\,\lambda^d_{ij}\,v_d /\sqrt 2\,,$ ~and 
$\,m^u_{ij}\,=\,\lambda^u_{ij}\,v_u /\sqrt 2\,$,
~respectively.
This constitutes the basic structure of the {\bf\, \SSM}, which involves,
at least, the basic ingredients 
shown in Table~\ref{tab:basic}.
Other ingredients, such as a  $\,\mu\ H_u H_d\,$ direct Higgs superfield
mass 
term in the superpotential, or an extra singlet chiral superfield $\,N\,$ 
with a trilinear superpotential coupling 
$\ \lambda \ H_u H_d\,N\,+ \, ... \ $
possibly acting as a replacement for a direct $\,\mu\ H_u H_d\,$ 
mass term~\cite{fayet75}, and/or extra $\,U(1)\,$ 
factors in the gauge\index{Group symmetries!extra $U(1)$}
group\index{Group symmetries!Superfields}, may or may not be present,
depending on the particular version of the \SSM\ considered.

\begin{table}[t]
\begin{center}
\begin{tabular}{|l|} \hline \\ 
\ \ {\small 1) \ the  three $\,SU(3)\times SU(2)\times U(1)\,$ gauge superfields;}  \
\\[.25 truecm]
\ \  {\small 2) \ chiral superfields for the three quark and lepton families;} \\ [.25 truecm]
\ \  {\small 3) \ the two doublet Higgs superfields $\,H_d\,$ and $\,H_u\,$
responsible} \ \\ 
\hskip 2truecm  {\small for the spontaneous electroweak breaking,} \\
\hskip 2truecm  {\small and the generation of quark and lepton masses;}  \\ [.25 truecm]
\ \  {\small 4) \  the trilinear superpotential of Eq.~(\ref{supot}).}
\\  \\ \hline
\end{tabular}
\end{center}
\caption{{\it The basic ingredients\index{Group symmetries!Superfields} 
              of the \SSM.}}
\label{tab:basic}
\end{table}

\vskip .3truecm

\normalsize

\begin{table}[t]
\begin{center}
\begin{tabular}{|c|c|c|} \hline 
&&\\ [-0.2true cm]
Spin 1       &Spin 1/2     &Spin 0 \\ [.1 true cm]\hline  \hline
&&\\ [-0.2true cm]
gluons ~$g$              &gluinos ~$\tilde{g}$        &\\
photon ~$\gamma$          &photino ~$\tilde{\gamma}$   &\\ 
------------------&$ - - - - - - - - - - $&--------------------------- \\
 

$\begin{array}{c}
W^\pm\\ [.1 true cm]Z \\ 
\\ \\
\end{array}$

&$\begin{array}{c}
\hbox {winos } \ \widetilde W_{1,2}^{\,\pm} \\ 
[0 true cm]
\,\hbox {zinos } \ \ \widetilde Z_{1,2} \\ 
\\ 
\hbox {higgsino } \ \tilde h^0 
\end{array}$

&$\left. \begin{array}{c}
H^\pm\\
[0 true cm] H\ \\
\\
h, \ A
\end{array}\ \right\} 
\begin{array}{c} \hbox {Higgs}\\ \hbox {bosons} \end{array}$  \\ &&\\ 
[-.1true cm]
\hline &&

\\ [-0.2cm]
&leptons ~$l$       &sleptons  ~$\tilde l$ \\
&quarks ~$q$       &squarks   ~$\tilde q$\\ [-0.3 cm]&&
\\ \hline
\end{tabular}
\end{center}
\caption{{\it Minimal particle content of the \SSM.} }
\label{tab:SSM}
\end{table}

In any case, independently of the details of the \SUSY-breaking 
mechanism ultimately considered and of the 
absence or presence of \Rp\ interactions, we obtain the 
following minimal particle content of the \SSM, 
as summarized in \hbox{Table~\ref{tab:SSM}}.
Each spin-$\frac{1}{2}\,$ quark $\,q\,$ or charged lepton $\,l^-\,$
is associated with {\it \,two\,} spin-0 partners collectively denoted by
$\,\tilde q\,$ or $\,\tilde l^-\,$, ~while a left-handed neutrino
$\,\nu_L\,$ 
is associated with a {\it \,single\,} spin-0 sneutrino $\,\tilde \nu_L$.
~We have ignored for simplicity, in this table, further mixing between
the various ``neutralinos'' described by neutral gaugino and higgsino
fields, schematically denoted by $\,\tilde\gamma,\ \tilde Z_{1,2}$, 
and $\tilde h^0$.
More precisely, all such models include four neutral Majorana 
fermions\index{Majorana fermions} at least,
mixings of the fermionic partners of the two neutral 
$\ SU(2) \times U(1)$\index{Group symmetries!Superfields} 
gauge bosons (usually denoted by $\,\tilde Z\,$ and $\,\tilde\gamma$, 
~or $\,\tilde{W_3}\,$ and $\,\tilde B\,$) ~and of the 
two neutral higgsino components 
($\,\tilde{h}_d^{\,0}\,$ and $\,\tilde{h}_u^{\,0}$). 
\,Non-minimal models also involve additional 
gauginos and/or higgsinos.

Let us return to the definition of the continuous $\,U(1)\,$ 
$\,R$-symmetry,\index{Continuous $R$--symmetry}\index{Group symmetries!$U(1)_R$}
and discrete\index{Discrete symmetries!$R$-parity} $\,R$-parity,
transformations. 
As explained earlier, the new {\it \,additive\,}  quantum number $\,R\,$
~associated with this continuous $\,U(1)$ $\,R$-symmetry
is carried by the \SUSY\ generator,
and distinguishes between bosons and fermions 
within the multiplets of \SUSY~\cite{fayet75}. 
Gauge bosons and Higgs bosons
have $\,R=0\,$ while their partners under \SUSY,
now interpreted as gauginos and higgsinos, have $\,R=\,\pm1\,$.
This definition 
is extended to the chiral quark and lepton superfields,
spin-$\frac{1}{2}\,$ quarks and leptons having $R=0$, 
~and their spin-0 superpartners, 
$\,R=+\,1\ $ (for $\,\tilde q_L,\,\tilde l_L\,$) ~or $\ R=-\,1\,$ 
(for $\,\tilde q_R,\,\tilde l_R\,$) ~\cite{fayet76}.
~The action of these 
$R$-symmetry transformations, which survive the spontaneous breaking 
of the electroweak symmetry
(see also footnote~3 in section~\ref{sec:R}),
 is given in \hbox{Table~\ref{tab:rinv}.}

\vskip .3truecm

\begin{table}[t]
\begin{center}
\vspace{4mm}
\begin{tabular}{|cccl|}
\hline &&& \\
$\ V(\,x,\,\theta,\,\bar\theta\ )$&$\rightarrow$ &  
$V (\,x, \,\theta \,e^{-i\alpha},\,\bar\theta \, e^{i\alpha}\,)$ &
{\small for the  $\,SU(3)\times SU(2)\times U(1)\,$} \\
&&&  \ \ \ \ {\small gauge superfields}   \\ [.3truecm]
$\ H_{d,u} \,(\,x,\,\theta\,)$ &$\rightarrow$  &
$H_{d,u}\, (\,x, \,\theta \,e^{-i\alpha}\,) $ & 
{\small for the left-handed doublet}\\
&&&  \ \ \ \ {\small Higgs superfields $H_d\,$ and $\,H_u\,$} \\  [.3truecm]
$\ S (\,x,\,\theta\,)$ &$\rightarrow $ & $e^{i\alpha}  \ \ 
S (\,x, \,\theta \,e^{-i\alpha}\,)$  & {\small for the left-handed} \\
&&&  \ \ \ \ {\small (anti)quark and (anti)lepton}  \\  [.1truecm]
&&&   \ \ \ \ {\small superfields $\,Q_i, \,U^c_i, \,D^c_i,\ \,
L_i, \,E^c_i $} \\  [.3truecm]
\hline
\end{tabular}
\end{center}
\caption{{\it Action of a continuous $\,U(1)\,$ 
              $\,R$-symmetry\index{Continuous $R$--symmetry} transformation 
              on the gauge and chiral superfields of the \SSM.} }
\label{tab:rinv}
\end{table}

This continuous $\,U(1)\,$
$\,R$-symmetry\index{Continuous $R$--symmetry}\index{Group symmetries!$U(1)_R$} 
($U(1)_R$) is indeed a symmetry of the four basic building blocks of the \SSM 
~(cf. Table~\ref{tab:basic}).
This includes the self-interactions of the 
$\,SU(3)\times SU(2)\times U(1)\,$
\index{Group symmetries!Superfields}
gauge superfields, and their interactions with 
the chiral  quark and lepton superfields, and the two doublet Higgs superfields
$\,H_d$ and $\,H_u\,$.
Also invariant under the continuous 
$\,U(1)$ $\,R$-symmetry\index{Continuous $R$--symmetry} 
are the super-Yukawa interactions of $\,H_d$ and $\,H_u\,$,
responsible for the generation of quark and lepton masses
through the superpotential (\ref{supot}).
\,Indeed it follows from Table~\ref{tab:rinv}
that this trilinear superpotential $W$ transforms under the continuous 
$\,R$-symmetry with ``$\,R$-weight'' $\ n_W\,=\,\sum_i\,n_i\,=\,2\,$,
~i.e. according to
\begin{equation}
W\,(\,x,\,\theta\,) \ \ \rightarrow\ \  e^{2\,i\alpha}\ \ 
W\,(\,x, \,\theta \,e^{-i\alpha}\,)\ \ .
\label{eq:R_weight}
\end{equation}

\noindent
Its auxiliary ``$\,F$-component'', obtained from the coefficient of the 
bilinear $\,\theta \,\theta \,$ term in the expansion of this
superpotential
$\,W\,$, \,is therefore $\,R$-invariant, 
generating $\,R$-inva\-riant interaction terms 
in the Lagrangian density.
Note, however, that a direct Higgs superfield mass term
$\,\mu\ H_u H_d\,$ in the superpotential, which has $\,R$-weight
$\,n=0\,$,
\,does {\it \,not\,} lead to interactions invariant 
under the continuous $\,R$-symmetry\index{Continuous $R$--symmetry}
(see also footnote~4 in section~\ref{sec:R} and~\cite{fayet75}
for a replacement of the $\mu$ term by a trilinear coupling
with an extra singlet chiral superfield, as in the NMSSM). 
But it gets in general re-allowed as soon as the continuous $\,R$-symmetry
gets reduced to its discrete\index{Discrete symmetries} version of 
$\,R$-parity.

This $\,R$-invariance\index{Continuous $R$--symmetry} 
led us to distinguish between a sector of 
$\,R$-{\it even particles\,},
which includes all those of the \SM,
with $\,R=0\,$ ~(and therefore $\,R_p = (-1)^R=+1\,$);
\,and their $\,R$-{\it odd superpartners\,}, gauginos and higgsinos, 
sleptons and squarks, with $\,R=\pm \,1\,$ 
\,(and $\,R_p =-1\,$), as indicated in 
Table~\ref{tab:Rp}. 

More precisely the necessity of generating masses for the 
(Majorana)\index{Majorana fermions}
\hbox{spin-$\frac{3}{2}\,$} gravitino~\cite{fayet77} 
and the spin-$\frac{1}{2}\,$ gluinos did not allow us to keep a
distinction between $\,R\,=\,+\,1\,$ and $\,R\,=\,-\,1\,$ particles,
forcing us to abandon the continuous $\,R$-invariance\index{Continuous $R$--symmetry} 
in favour of its discrete\index{Discrete symmetries} version, $R$-parity.
~The \,-- even or odd --\, parity character of the (additive) $\,R\,$ 
quantum number corresponds to the well-known 
$\,R${\it \,-parity},\index{Discrete symmetries!$R$-parity} 
first defined as $\,+\,1\,$ for the ordinary
particles and $\,-\,1\,$ for their superpartners, simply written 
as $\,(-1\,)^{R}\ $ in (\ref{eq:rp01})~\cite{fayet78rp},
then re-expressed as $ (-1)^{2S} (-1)^{3B+L} $     
in (\ref{eq:rp02}) as an effect of the close connection between
$\,R$-parity 
and baryon and lepton-number conservation laws.

\begin{table}[t]
\begin{center}
\small
\begin{tabular}{|c|c|} \hline 
&\\ [-0.2true cm]
Bosons       & Fermions     \\ [.2 true cm]\hline \hline 
&\\ [-0.1true cm]
$\begin{array}{ccc}
\begin{array}{c}
\hbox{gauge and Higgs bosons}  \\ [.1 true cm]
\hbox{graviton}
\end{array}
&\!&
(\,R\,=\,0\,)
\end{array}
$
& 
$\begin{array}{ccc}
\begin{array}{c}
\hbox{gauginos and higgsinos}  \\ [.1 true cm]
\hbox{gravitino}
\end{array}
&\!&
(\,R\,=\,\pm\,1\,)
\end{array}
$
\\ [-0.1 cm] & \\
$R$-parity\ \ $+$  & $R$-parity\ \ $-$ 
\\ & \\ \hline & \\[-0.1cm]
sleptons and squarks\ \ \ (\,$R\,=\,\pm\,1$\,) & 
leptons and quarks\ \ \ \ \ \ (\,$R\,=\,0$\,)  \\ [-0.1 cm] & \\
$R$-parity\ \ $-$  & $R$-parity\ \ $+$ \\ & \\ \hline
\end{tabular}
\end{center}
\caption{{\it $R$-parities\index{Discrete symmetries!$R$-parity} in the \SSM.} }
\label{tab:Rp}
\end{table}

This $\,R$-parity symmetry operator may also be viewed as a non-trivial
geometrical discrete\index{Discrete symmetries!$R$-parity} symmetry associated
with a reflection of the anticommuting fermionic Grassmann coordinate, 
$\,\theta\ \to -\,\theta\,$, in superspace~\cite{geom96}.
This $\,R$-parity operator plays an essential r\^ole in the construction 
of \susyq\ theories of interactions, 
and in the discussion of the experimental signatures of the new particles.
A conserved $\,R$-parity guarantees 
that {\it \,the new spin-0 squarks and sleptons 
cannot be directly exchanged\ } between ordinary quarks and leptons, 
as well as the absolute stability of the LSP.
But let us discuss more precisely the reasons
which led to discarding the continuous 
$\,R$-invariance\index{Continuous $R$--symmetry} 
in favour of its discrete version, 
$\,R$-parity\index{Discrete symmetries!$R$-parity}.

\section{Gravitino and Gluino Masses: From {\boldmath{$R$}}-Invariance to
{\boldmath{$R$-Parity}}}
\label{sec:grav}

There are two strong reasons, at least, to abandon the continuous 
$\,R$-invariance\index{Continuous $R$--symmetry} in favour of 
its discrete 
$\,Z_2\,$\index{Discrete symmetries!$Z_2$}\index{Group symmetries!$Z_2$ subgroup} 
subgroup generated by the $\,R$-parity transformation.
One is theoretical, the necessity \,-- once gravitation is introduced --\,
of generating a mass for the (Majorana)\index{Majorana fermions} 
\hbox{spin-$\frac{3}{2}\,$} {\it \,gravitino\,\,} in the framework of 
spontaneously-broken locally \susyq\ theories~\cite{fayet77}.
The other is phenomenological, the non-observation of massless 
(or even light) {\it \,gluinos}.
Both particles would have to stay massless in the absence of a breaking
of the continuous $\,U(1)\,$\index{Group symmetries!$U(1)_R$}
$\,R$-invariance\index{Continuous $R$--symmetry}, thereby preventing, in the case
of the gravitino, 
\SUSY\ from being spontaneously broken.
\,(A third reason could now be the non-observation at LEP
 of a charged {\it \,wino\,} \,-- also called {\it \,chargino\,}~--
lighter than the $\,W^\pm$, \,that would exist 
in the case of a continuous $\,U(1)\,$\index{Group symmetries!$U(1)_R$}
$\,R$-invariance~\cite{fayet75,fayet76},
as shown by the mass matrix $\,M_C\,$ given 
in Eq.\,(\ref{mwino}).)

All this is, therefore, also connected with the mechanism by which the
\SUSY\ should get spontaneously broken, in the \SSM.
The question has not received a definitive answer yet.
The inclusion of universal soft \SUSY-breaking terms for all
squarks and 
sleptons,
\begin{equation}
-\ \sum_{\tilde q,\,\tilde l}\ m_0^{\,2}\ \ 
(\,{\tilde q}^\dagger\,\tilde q\ +\ {\tilde l}^\dagger \,\tilde l\,)\ \ ,
\end{equation}
was already considered in 1976, for lack of a better solution.
But such terms should in fact be generated by some spontaneous 
\SUSY-breaking mechanism, if \SUSY\ is to be realized
locally.
In any case one now considers in general all soft \SUSY-breaking 
terms~\cite{girardello82} (possibly ``induced by \SUGRA''),
which essentially serve as a parametrisation of our ignorance about the 
true mechanism of \SUSY\ breaking chosen by Nature
to make superpartners heavy.

But let us return to gluino masses.
Since $\,R$-transformations act {\it \,chirally\,} 
on the Majorana\index{Majorana fermions} octet of gluinos,
\begin{equation}
\tilde g\ \ \to\ \ e^{\,\gamma_5\,\alpha}\ \tilde g\ \ 
\end{equation}
a continuous $\,R$-invariance\index{Continuous $R$--symmetry} 
would require the gluinos to remain massless,
even after a spontaneous breaking of the \SUSY\ !
As mentioned before (see section~\ref{sec:intro}), one would then
expect the existence of $\,``R$-hadrons'' which have not been 
observed~\cite{farrar78,farrar79}.
Present experimental constraints indicate that gluinos, if they exist,
must be very massive~\cite{pdg04}, requiring a significant breaking of the
continuous $\,R$-invariance,\index{Continuous $R$--symmetry} in addition
to the necessary breaking of the \SUSY.

In the framework of global \SUSY\ it is not so easy to generate
large gluino masses.
Even if global \SUSY\ is spontaneously broken, and if the
continuous $R$-symmetry\index{Continuous $R$--symmetry} is not present, it
is still in general rather difficult to obtain large masses for gluinos, 
since: \ 
i) no direct gluino mass term is present in the Lagrangian density; 
and \ 
ii) no such term may be generated spontaneously, at the tree 
approximation, via gluino couplings involving {\it coloured} spin-0 fields. 

A gluino mass may then be generated by radiative corrections
involving a new sector of quarks sensitive to the source 
of supersymmetry breaking,
that would now be called 
``messenger quarks''~\cite{fayet78},
but \ iii) this can only be through diagrams which ``know'' both about:
a) the spontaneous breaking of the global \SUSY,
through some appropriately generated \VEVs\ for auxiliary gauge or chiral 
components,
$\,<\!D\!>,\ <\!F\!>\,$ or $\,<\!G\!>\,$'s;\ \ 
b) the existence of superpotential interactions which do not preserve 
the continuous $\,U(1)\,$ 
$\,R$-symmetry.\index{Continuous $R$--symmetry}\index{Group symmetries!$U(1)_R$}
Such radiatively generated gluino masses, however, 
generally tend to be rather small, unless one introduces, in 
some often rather complicated ``hidden sector'', 
very large mass scales $\,\gg\,M_W\,$.

Fortunately gluino masses may also result directly from \SUGRA, 
as observed long ago~\cite{fayet77}. Gravitational 
interactions require, within local \SUSY, that the spin-2 
graviton be associated with a spin-$3/2\,$ partner~\cite{sugra}, the 
gravitino. Since the gravitino is the fermionic gauge particle of 
\SUSY\ it must acquire a mass $\,m_{3/2}$ as soon as the local \SUSY\ 
gets spontaneously broken\,\footnote{Depending on the notation 
this mass may be expressed as $\,m_{3/2} = \kappa\, d / \sqrt{6}\,$, \,or
$\,m_{3/2} = \kappa\, F_S / \sqrt{3}
 = \sqrt{\,{8\,\pi/3}}\ \,F_S / M_P\,$, \,where $F_S$ (or $d$) 
is the \SUSY-breaking scale parameter,
$\,\kappa^2 =8\pi\,G_N$,\, and $\,M_P\,$ is the Planck mass.
Supersymmetry is often said to be broken ``at the scale'' $\,\sqrt{F_S}\,$ 
(or $\sqrt d$) $\,=\Lambda_{\hbox{\scriptsize ss}} 
\,\approx \sqrt{m_{3/2}\,M_P}$.
}.
Since the gravitino is a self-conjugate Majorana 
fermion\index{Majorana fermions}
its mass breaks the continuous $\,R$-invariance\index{Continuous $R$--symmetry} 
which acts chirally on it, just as for the gluinos, 
\,forcing us to abandon the continuous $U(1)$\index{Group symmetries!$U(1)_R$}
$R$-invariance, in favour of its discrete 
$\,Z_2\,$\index{Discrete symmetries!$Z_2$}\index{Group symmetries!$Z_2$ subgroup} 
subgroup generated by the $R$-parity transformation.
We can no longer distinguish between the values $+1$ and $-1$ of the
 ({\it additive}) quantum number $R$; 
\,but only between \hbox{``$R$-odd''} particles 
(having $\,R=\pm1\,$) ~and ``$R$-even'' ones, ~i.e. between 
particles having $R$-parities $\,R_p=(-1)^R=-\,1,$ ~and $\,+\,1$, 
\,respectively (cf. \hbox{Table~\ref{tab:Rp}}).

In particular, when the spin-$\frac{3}{2}\,$ gravitino mass term
$\,m_{3/2}\,$,
~which corresponds to a change in $\,R\,$ $\ \Delta \,R\,=\,\pm \,2\,$,
~is introduced, the ``left-handed sfermions''
$\,\tilde f_L$, ~which carry $\,R\,=\,+\,1$, ~can mix with the 
``right-handed'' ones  $\,\tilde f_R$, 
~carrying $\,R\,=\,-\,1$, ~through mixing terms having 
$\ \Delta \,R\,=\,\pm \,2\,$, ~which may naturally \,(but not necessarily)
be of order $\ m_{3/2}\ m_f\,$
(so that the lightest of the squarks and sleptons may well turn out to 
be one of the two stop quarks $\tilde t\,$).
Supergravity theories offer a natural framework 
in which to include, in addition, direct gaugino 
Majorana\index{Majorana fermions} mass terms 
\begin{equation}
-\ \frac{1}{2}\ M_3\ \bar{\tilde g}^a {\tilde g}^a\ \
-\ \frac{1}{2}\ M_2\ \bar{\tilde W}^i {\tilde W}^i\ \ 
-\ \frac{1}{2}\ M_1\ \bar{\tilde B} {\tilde B}\ \ ,
\end{equation}
which also correspond to $\,\Delta \,R\,=\,\pm \,2\,$,
\,just as for the gravitino mass itself.
~The mass parameters $M_3,\ M_2\,$ and $\,M_1$ associated with 
the $\,SU(3) \times SU(2) \times U(1)\,$
\index{Group symmetries!Superfields} 
gauginos may then naturally \,(but not necessarily)\,
be of the same order as the gravitino mass $\,m_{3/2}\,$.

Furthermore, once the continuous $R$-invariance 
is reduced to its discrete $R$-parity subgroup, 
a direct Higgs superfield mass term $\ \mu \ H_u H_d$
~which was not allowed by the continuous $U(1)\,$
$R$-symmetry\index{Continuous $R$--symmetry} (but could be replaced by a
trilinear $\,\lambda\ H_u H_d\,N\,$
superpotential term), gets reallowed in the superpotential,
as for example in the MSSM. The size of this 
supersymmetric $\,\mu\,$ parameter might conceivably have been 
a source of difficulty, in case this parameter, present even 
if there is no supersymmetry breaking, turned out to be large.
But since the $\mu$ term breaks explicitly 
the continuous $\,R$-invariance of Table \ref{tab:rinv}
(and, also, another ``extra $\,U(1)$''\, symmetry
\index{Group symmetries!extra $U(1)$} 
acting axially on quark and lepton superfields~\cite{fayet76}),
its size may be controlled by considering one or the other of these two
symmetries. Even better, since $\mu$ got reallowed just as we abandoned 
the continuous $\,R$-invariance so as to allow for gluino and gravitino
masses, the size of $\,\mu\,$ may naturally be 
of the same order as the \SUSY-breaking gaugino mass parameters $M_i$, 
\,or the gravitino mass $\,m_{3/2}$,
\,since they all appear in violation of the continuous $\,R$-symmetry of
Table \ref{tab:rinv}~\cite{fayetminneap}.
Altogether there is here no specific hierarchy problem associated with 
the size of $\,\mu\,$.

In general, irrespective of the \SUSY-breaking mechanism
considered, still unknown
(and parametrized using a variety of possible
soft supersymmetry-breaking terms),
\,one normally expects the various superpartners not to be too heavy. 
Otherwise the corresponding new mass scale introduced in the game 
would tend to contaminate the electroweak scale, thereby
{\it creating\,} a hierarchy problem in the \SSM.
Superpartner masses are then normally expected to be naturally of the
order of $\,M_W$, ~or at most in the $\ \sim\,$ TeV range. 

Beyond that, in a more ambitious framework involving extra spacetime
dimensions, \,$R$-parity may also be responsible 
for an elegant way of implementing
\SUSY\ breaking by dimensional reduction, by demanding {\it \,discrete\,}
\,-- periodic or antiperiodic --\,
boundary conditions for ordinary $R$-even particles and their $R$-odd 
superpartners, respectively. The masses of the lowest-lying
superpartners would now get fixed 
by the compactification scale \,--\, e.g. in the simplest case  
and up to radiative correction effects, by\
\begin{equation}
m_{3/2}\ =\ M_i\ =\ \frac{\pi\,\hbar}{L\,c}\ =\ \,\frac{\hbar}{2\,R\,c}\ \,,
\end{equation}
in terms of the size $L$ of the extra dimension responsible for the \SUSY\ 
breaking\,\cite{extradim}.
This led to consider, already a long time ago, 
the possibility of relatively ``large'' extra dimensions
(as compared for example to the Planck length of $\,10^{-33}\,$ cm),
associated with a compactification scale that could then be 
as ``low'' as $\,\sim $ TeV scale.
If this is true, the discovery of superpartners in the 
$\,\mathrel{\rlap{\raise 0.511ex \hbox{$<$}}{\lower 0.511ex \hbox{$\sim$}}}\,$ 
TeV scale should be followed by the discovery of series of heavy copies 
for all particles,
corresponding to the Kaluza-Klein excitations of the extra dimensions 
\,-- quite a spectacular discovery\,!

Landing back on four dimensions,
the \SSM\ (whether ``minimal'' or not), with its $R$-parity 
symmetry (whether it turns out to be absolutely conserved or not), 
provided the basis for the
experimental 
searches for the new superpartners and Higgs bosons, starting with 
the first searches for gluinos and photinos, selectrons and smuons, 
in 1978-1980.
But how the \SUSY\ should actually be broken \,-- if indeed it is a 
symmetry of Nature\,! --\, is not known yet, and this concentrates a large 
part of the remaining uncertainties in the \SSM.
Furthermore, it is worth to discuss more precisely the question of the 
conservation, or non-conservation, of $\,R$-parity.
In other terms, how much violation of the $\,R$-parity symmetry may be 
tolerated, without running in conflict with existing experimental results on
proton decay, neutrino masses, and various accelerator or astrophysical data ?
Or, conversely, could $\,\Rp$ effects be responsible for the generation 
of small Majorana masses for neutrinos\,?
\,This is the subject of the present review article.

\cleardoublepage
\chapter{HOW CAN {\boldmath{$R$}}-PARITY BE VIOLATED?}
\label{chap:theory}

The non-conservation of baryon ($B$) and lepton ($L$) numbers is a generic 
feature of numerous extensions of the \SM. In theories involving new 
symmetries valid at some high energy scale such as \GUT\ (GUTs), such 
$B$- and $L$-violating effects are usually suppressed by powers of the high 
scale, and therefore generally small - even though some of them, like proton 
decay, may be observable.
By constrast, in \susyq\ extensions of the \SM, the scale of possible baryon 
and lepton-number violations is associated with the masses of the
superpartners (squarks and leptons) responsible for the violations, which
may lead to unacceptably large effects. 
Avoiding this, was the main interest of the introduction 
of $R$-parity\index{Discrete symmetries!$R$-parity}, 
as discussed in the previous chapter.
In view of the important phenomenological differences between \susyq\ models 
with and without $R$-parity, it is worth studying the extent to which 
$R$-parity can be broken. Furthermore, there are in principle other 
symmetries (discrete or continuous, global or local) that can forbid 
proton decay while still allowing for the presence of some $R$-parity-violating 
couplings. Their classification will allow to explore which kind 
of $R$-parity-violating terms can possibly appear.


\section{{\boldmath{$R$}}-Parity-Violating Couplings}
\label{sec:rpvgeneral}

In the \SM\ (assuming two-component, massless neutrino fields) it is impossible
to write down renormalizable, gauge-invariant interactions that violate
baryon or lepton numbers. This is no longer the case in \susyq\ extensions
of the \SM\ where, for each ordinary fermion (boson), the introduction of a 
scalar (fermionic) partner allows for new interactions
that do not preserve baryon or lepton number. As
explained in chapter~\ref{chap:intro}, these interactions can be forbidden
by introducing $R$-parity\index{Discrete symmetries!$R$-parity}.
This leads in particular to the popular ``\MSSM'' (MSSM), the \susyq\
extension of the \SM\ with gauge symmetry $SU(3)_C \times SU(2)_L \times U(1)_Y$,
\index{Group symmetries!Minimal SUSY}
with minimal particle content and for which $R$-parity invariance is generally assumed.
Throughout most of this review, the MSSM will normally be
our reference model, although the discussion of $R$-parity violation
does not in general depend much on the specific version of the \SSM\
which is considered.

\Rp\ couplings originate either from the superpotential itself, or from soft
\SUSY-breaking terms. There are various kinds of such couplings,
of dimensions 4, 3 or 2 only, with a potentially rich flavour structure.
In this section, we shall write down explicitly all possible \Rp\ terms
in the framework of the MSSM, assuming the most general breaking of
$R$-parity. We shall then consider particular 
scenarios which allow to reduce the number of independent couplings used to 
parametrize $R$-parity violation.

\subsection{Superpotential Couplings}
\label{subsec:couplings}

Assuming $R$-parity invariance, 
the \index{Superpotential}superpotential 
of the \SSM\ with minimal particle content
contains only one \susyq\ Higgs mass term, the $\mu$-term, and the
\susyq\ Yukawa interactions generating masses for the quarks and charged
leptons (see section~\ref{sec:intro}, and Appendix \ref{chap:appendixA}
for the definition of superfields),
\begin{equation}
  W_{R_p}\ \equiv\ W_{MSSM}\ =\ \mu\, H_u H_d\
  +\ \lambda^e_{ij}\, H_d L_i E^c_j\ +\ \lambda^d_{ij}\, H_d Q_i D^c_j\
  -\ \lambda^u_{ij}\, H_u Q_i U^c_j\ .
\label{eq:W_Rp_even}
\end{equation}
Other versions of the \SSM, with an extended Higgs sector and/or
additional $U(1)$\index{Group symmetries!extra $U(1)$} gauge factors, 
may have a slightly different $R$-parity conserving 
superpotential\index{Superpotential!$R_p$ conserving}, 
especially since they involve in general additional chiral superfields. 
This is for instance the case in the NMSSM where an extra neutral 
singlet is coupled to the two doublet Higgs superfields $H_d$ and $H_u$.

In the absence of $R$-parity, however, $R$-parity odd terms 
allowed by renormalizability and gauge invariance must also, in principle, be
included in the superpotential.
The ones that violate lepton-number conservation can be easily found by noting 
that the lepton superfields $L_i$ and the Higgs superfield $H_d$ have exactly 
the same gauge quantum numbers.
Thus, gauge invariance allows for bilinear and trilinear 
lepton-number-violating superpotential couplings obtained by replacing
$H_d$ by $L_i$ in Eq. (\ref{eq:W_Rp_even}).
The only other renormalizable \Rp\
superpotential term allowed by gauge invariance, $U^c_i D^c_j D^c_k\,$,
breaks baryon-number conservation.
Therefore the most general renormalizable, $R$-parity odd superpotential
\index{Superpotential!$R_p$ violating} consistent with the gauge 
symmetry and field content of the
MSSM\index{Group symmetries!Minimal SUSY} is\footnote{Other versions of the \SSM\ may
allow for additional \Rp\ superpotential terms.}~\cite{weinberg82}
(see also \cite{sakai}),
%
%
\begin{equation}
  W_{\Rp}\ =\ \mu_i\, H_u L_i\
  +\ \frac{1}{2}\, \lambda_{ijk}\, L_i L_j E^c_k\
  +\ \lambda'_{ijk}\, L_i Q_j D^c_k\
  +\ \frac{1}{2}\, \lambda''_{ijk}\, U^c_i D^c_j D^c_k \ ,
\label{eq:W_Rp_odd}
\end{equation}
where, like in Eq. (\ref{eq:W_Rp_even}), there is a summation over the 
generation indices $i,j,k = 1,2,3$, and summation over gauge indices is 
understood. One has for example $L_i L_j E^c_k \equiv
(\epsilon_{ab} L^a_i L^b_j) E^c_k = (N_i E_j - E_i N_j) E^c_k$ and
$U^c_i D^c_j D^c_k \equiv
\epsilon^{\alpha \beta \gamma} U^c_{i \alpha} D^c_{j \beta} D^c_{k \gamma}$,
where $a,b = 1,2$ are $SU(2)_L$\index{Group symmetries!$SU(2)_L$} indices, 
$\alpha, \beta, \gamma = 1,2,3$ are $SU(3)_C$\index{Group symmetries!$SU(3)_C$}
indices, and
$\epsilon_{ab}$ and $\epsilon_{\alpha \beta \gamma}$ are totally antisymmetric
tensors. Gauge invariance enforces antisymmetry of 
the $\lambda_{ijk}$ couplings with respect to their first two indices.
As a matter of fact, one has
\begin{equation}
  \lambda_{ijk}\, L_i L_j E^c_k\
  =\ \lambda_{ijk}\, \epsilon_{ab} L_i^a L_j^b E^c_k\
  =\ - \lambda_{jik}\, \epsilon_{ba} L_j^a L_i^b  E^c_k\
  =\ - \lambda_{jik}\, L_i L_j  E^c_k\ ,
\end{equation}
which leads to
\begin{equation}
\lambda_{ijk}\ =\ - \lambda_{jik}\ .
\end{equation}
Gauge invariance also enforces antisymmetry of the $\lambda''_{ijk}$ couplings 
with respect to their last two indices:
\begin{equation}
\lambda''_{ijk}\ =\ - \lambda''_{ikj}\ .
\end{equation}

At this point one should make a comment on the 
\index{Superpotential!$R_p$ violating!bilinear \Rp\ terms}bilinear \Rp\ 
superpotential terms $\mu_i H_u L_i$.
These terms can be rotated away from the superpotential upon suitably 
redefining the lepton and Higgs superfields (with
$H_d \rightarrow H'_d \propto \mu H_d + \mu_i L_i$)~\cite{hall84}.
However, this rotation will generate \Rp\ scalar mass terms
(see subsection~\ref{subsec:soft_terms})
from the ordinary, $R$-parity conserving soft \SUSY-breaking terms of
dimension 2, so that bilinear $R$-parity-violating terms
will then reappear in the scalar potential. The fact that one can make $\mu_i = 0$
in Eq. (\ref{eq:W_Rp_odd}) does not mean that the Higgs-lepton mixing 
\index{Mixing!higgs--lepton} associated with bilinear $R$-parity breaking is 
unphysical, but rather that there is no unique way of parametrizing it, as will 
be discussed insubsection~\ref{subsec:basis}.

Altogether, Eq. (\ref{eq:W_Rp_odd}) involves 48 ({\it a priori} complex)
parameters: 3 dimensionful parameters $\mu_i$ mixing 
\index{Mixing!higgs--lepton} the charged lepton
and down-type Higgs superfields, and 45 dimensionless Yukawa-like couplings
divided into 9 $\lambda_{ijk}$ and 27 $\lambda'_{ijk}$ couplings which
break lepton-number conservation, and 9 $\lambda''_{ijk}$ couplings which
break baryon-number conservation.

\subsection{\sloppy Lagrangian Terms Associated with the Superpotential
                    Couplings}
\label{subsec:Lagrangian}

\begin{figure}[t]
\begin{center}
\mbox{\epsfxsize=1.0\textwidth \epsffile{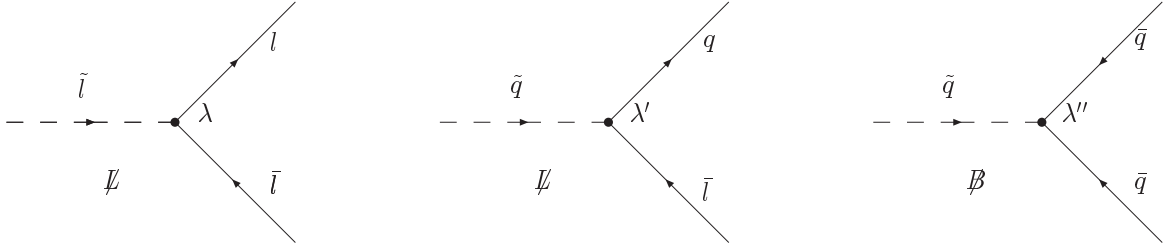}}
\caption{{\it Basic tree diagrams associated with the trilinear \Rp\
superpotential interactions involving the Yukawa couplings $\lambda$ or
$\lambda'$ (\LV), or $\lambda''$ (\BV). $q$ ($\tilde q$) and $l$ ($\tilde l$) 
denote (s)quarks and (s)leptons. The arrows on the (s)quark and (s)lepton lines
indicate the flow of the baryon (resp. lepton) number.}}
\label{fig:triplel}
\end{center}
\end{figure}

We now derive the interaction terms in the
\index{Lagrangian}Lagrangian density generated from the $R$-parity
odd superpotential of Eq. (\ref{eq:W_Rp_odd}). 

\noindent
{\bf i) {\boldmath{\Rp}} Yukawa couplings}

Let us first consider the terms
\index{Lagrangian!\Rp\ terms involving fermions}involving fermions.
They consist of fermion bilinears associated with the bilinear 
superpotential terms $\mu_i H_u L_i$, and of trilinear, Yukawa-like
interactions associated with the superpotential couplings  $\lambda$,
$\lambda'$ and $\lambda''$. In two-component notation for spinors, the fermion
bilinears read (see Appendix \ref{chap:appendixA} for the definition of
fields, and Ref. \cite{haber85} for the two-component notation):
\begin{equation}
  {\cal L}_{H_u L_i}\ =\ \mu_i \left( \tilde h_u^0 \nu_i
  - \tilde h_u^+ l_i^- \right) + \ \mbox{h.c.}\ .
\label{eq:lagmu}
\end{equation}
These terms, which mix lepton with higgsino fields,
will be discussed in section \ref{sec:bilinear}. Expanded in standard
four-component Dirac notation, the trilinear interaction terms associated
with the $\lambda$, $\lambda'$ and $\lambda''$ couplings read, respectively
(see Appendix \ref{chap:appendixA} for the definition of fields,
and Appendix \ref{chap:appendixB} for the derivation of this 
\index{Lagrangian!\Rp\ trilinear interactions}Lagrangian density):
\begin{equation}
  {\cal L}_{L_{i}L_{j}E^c_{k}}\ =\ - \frac{1}{2} \lambda_{ijk}
  \bigg ( \tilde  \nu_{iL}\bar l_{kR}l_{jL} +
  \tilde l_{jL}\bar l_{kR}\nu_{iL} + \tilde l^\star _{kR}\bar \nu^c_{iR}
  l_{jL} - (i \leftrightarrow j) \bigg ) + \ \mbox{h.c.}\ ,
\label{eq:laglambda}
\end{equation}
%
%
\begin{eqnarray}
{\cal L}_{L_i Q_j D^c_k} & = &
  - \lambda '_{ijk}  \bigg ( \tilde  \nu_{iL}\bar d_{kR}d_{jL} 
  + \tilde d_{jL}\bar d_{kR}\nu_{iL} 
+ \tilde d^\star _{kR}\bar \nu^c_{iR}d_{jL} \nonumber \\
  &  & 
  -\tilde  l_{iL}\bar d_{kR}u_{jL} 
  -\tilde u_{jL}\bar d_{kR}l_{iL} 
  - \tilde d^\star _{kR}\bar l^c_{iR} u_{jL}
                     \bigg ) + \ \mbox{h.c.}\ ,
\label{eq:laglambdap}
\end{eqnarray}
and\footnote{Due to the antisymmetry property of the $\lambda''_{ijk}$
couplings and to $SU(3)_C$\index{Group symmetries!$SU(3)_C$}
invariance, the second and third terms in
Eq. (\ref{eq:laglambdapp}) are actually identical.}
\begin{equation}
  {\cal L}_{U_i^c D_j^c D_k^c}\ =\ - \frac{1}{2} {\lambda ''}_{ijk} 
  \bigg (\tilde  u^\star _{i R}\bar d_{j R}d^c_{k L} +
  \tilde  d^\star _{k R}\bar u_{i R}d^c_{j L} +
  \tilde  d^\star _{j R}\bar u_{i R}d^c_{k L} \bigg ) 
  + \ \mbox{h.c.}\ .
\label{eq:laglambdapp}
\end{equation}
In these equations, the superscript $^c$ denotes the charge conjugate
of a spinor (for instance $\bar \nu^c_{iR} = \overline{(\nu^c_i)_R}$ 
is the adjoint of the charge conjugate of $\nu_{iL}$), the superscript $^\star$ 
the complex conjugate of a scalar field, and the $R$ and $L$ chirality indices on 
scalar fields distinguish between independent fields corresponding to the
superpartners of right- and left-handed fermion fields, respectively.  
Like in Eq. (\ref{eq:W_Rp_odd}), an implicit sum runs freely over the 
quark and lepton generations labelled by the indices $i,j,k$,
and summation over gauge indices is understood.
The trilinear interactions of Eqs. (\ref{eq:laglambda}), (\ref{eq:laglambdap})
and (\ref{eq:laglambdapp}) couple a scalar field and two fermionic fields 
(see Fig.~\ref{fig:triplel}), and are indeed forbidden by the 
$R$-parity\index{Discrete symmetries!$R$-parity} symmetry defined 
by Eq. (\ref{eq:rp02}). 

\noindent
{\bf ii) {\boldmath{\Rp}} scalar interactions}

Let us now consider the \Rp\ scalar interactions associated with the \Rp\
\index{Lagrangian!\Rp\ scalar terms} superpotential 
couplings (\ref{eq:W_Rp_odd}).
The bilinear superpotential terms induce dimension-2 and -3 terms,
\index{Lagrangian!\Rp\ scalar terms!dimension 2 and 3 terms}
\begin{equation}
 V^{\mu_i}_{\Rp}\ =\ \mu^\star\! \mu_i\, h^{\dagger}_d \tilde{L}_i\
  -\ \mu^\star_i
   \lambda^u_{jk} (\tilde{L}^{\dagger}_i \tilde{Q}_j) \tilde{u}^c_k\
  +\ \sum_l \mu^\star_l \lambda^e_{li} (h^{\dagger}_u h_d) \tilde{l}^c_i\
  +\ \mbox{h.c.}\ ,
\end{equation}
while the trilinear superpotential couplings induce dimension-4 terms,
\index{Lagrangian!\Rp\ scalar terms!dimension 4 terms}
\begin{eqnarray}
 V^{\lambda, \lambda', \lambda''}_{\Rp} & = &
  - \sum_l \lambda^{e\star}_{lj} \lambda_{lik}
  (h^{\dagger}_d \tilde{L}_i) \tilde{l}^{c\star}_j \tilde{l}^c_k\
  +\ \frac{1}{2} \sum_l \lambda^{e\star}_{il} \lambda_{jkl}
  (h_d \tilde{L}_i)^\star (\tilde{L}_j \tilde{L}_k)  \nonumber \\
  & - & \sum_l \lambda^{e\star}_{lj} \lambda'_{lik}
  (h^{\dagger}_d \tilde{Q}_i) \tilde{l}^{c\star}_j \tilde{d}^c_k\
  +\ \sum_l \lambda^{d\star}_{lj} \lambda'_{ilk}
  (h^{\dagger}_d \tilde{L}_i) \tilde{d}^{c\star}_j \tilde{d}^c_k  \nonumber \\
  & - & \sum_l \lambda^{u\star}_{lj} \lambda'_{ilk}
  (h^{\dagger}_u \tilde{L}_i) \tilde{u}^{c\star}_j \tilde{d}^c_k\
  +\ \sum_l \lambda^{d\star}_{il} \lambda'_{jkl}
  (h_d \tilde{Q}_i)^\star (\tilde{L}_j \tilde{Q}_k)  \nonumber \\
  & + & \sum_l \lambda^{d\star}_{il} \lambda''_{jkl}
  (h_d \tilde{Q}_i)^\star \tilde{u}^c_j \tilde{d}^c_k\
  -\ \frac{1}{2} \sum_l \lambda^{u\star}_{il} \lambda''_{ljk}
  (h_u \tilde{Q}_i)^\star \tilde{d}^c_j \tilde{d}^c_k\ +\ \mbox{h.c.}\ .
\end{eqnarray}

\noindent
{\bf iii) Additional {\boldmath{$R_p$}} conserving scalar couplings}

The superpotential (\ref{eq:W_Rp_odd}) also yields 
\index{Lagrangian! $R_p$ conserving scalar terms} 
scalar couplings that depend quadratically on the \Rp\ couplings.
Since $R$-parity acts as a $Z_2$ symmetry\index{Discrete symmetries!$Z_2$},
these terms correspond in fact to $R$-parity conserving interactions, 
even if they vanish in the limit of exact $R$-parity,
\begin{equation}
 V^{\mu_i}_{R_p}\ =\ \sum_i |\mu_i|^2 \left(h^{\dagger}_u h_u
  + \tilde{L}^{\dagger}_i \tilde{L}_i \right)\ ,
\end{equation}
\begin{eqnarray}
 V^{\lambda, \lambda', \lambda''}_{R_p} & = &
  \sum_m \lambda^\star_{mik} \lambda_{mjl}
  (\tilde{L}^{\dagger}_i \tilde{L}_j) \tilde{l}^{c\star}_k \tilde{l}^c_l\
  +\ \frac{1}{4} \sum_m \lambda^\star_{ijm} \lambda_{klm}
  (\tilde{L}_i \tilde{L}_j)^\star (\tilde{L}_k \tilde{L}_l)  \nonumber \\
  & + & \sum_m \left( \lambda^\star_{mik} \lambda'_{mjl}
  (\tilde{L}^{\dagger}_i \tilde{Q}_j) \tilde{l}^{c\star}_k \tilde{d}^c_l
   + \mbox{h.c.} \right)\
  +\ \sum_m \lambda^{\prime \star}_{mik} \lambda'_{mjl}
  (\tilde{Q}^{\dagger}_i \tilde{Q}_j) \tilde{d}^{c\star}_k \tilde{d}^c_l
  \nonumber \\
  & + & \sum_m \lambda^{\prime \star}_{imk} \lambda'_{jml}
  (\tilde{L}^{\dagger}_i \tilde{L}_j) \tilde{d}^{c\star}_k \tilde{d}^c_l\
  +\ \sum_m \lambda^{\prime \star}_{ijm} \lambda'_{klm}
  (\tilde{L}_i \tilde{Q}_j)^\star (\tilde{L}_k \tilde{Q}_l)  \nonumber \\
  & + & \sum_m \left( \lambda^{\prime \star}_{ijm} \lambda''_{klm}
  (\tilde{L}_i \tilde{Q}_j)^\star \tilde{u}^c_k \tilde{d}^c_l + \mbox{h.c.}
  \right)\
  +\ \frac{1}{2} \sum_m \lambda^{\prime \prime \star}_{mik} \lambda''_{mjl}
  (\tilde{d}^{c\star}_i \tilde{d}^c_j) (\tilde{d}^{c\star}_k \tilde{d}^c_l)
  \nonumber \\
  & + & \sum_m \lambda^{\prime \prime \star}_{ikm} \lambda''_{jlm}  
  \left[ (\tilde{u}^{c\star}_i \tilde{u}^c_j)
    (\tilde{d}^{c\star}_k \tilde{d}^c_l)
   - (\tilde{u}^{c\star}_i \tilde{d}^c_l) (\tilde{d}^{c\star}_k \tilde{u}^c_j)
  \right]\ ,
\end{eqnarray}
and
\begin{equation}
 V^{\mu_i, \lambda, \lambda'}_{R_p}\ =\
  - \sum_l \mu^\star_l \lambda_{lij}
   (h^{\dagger}_u \tilde{L}_i) \tilde{l}^c_j\
  -\ \sum_l \mu^\star_l \lambda'_{lij}
   (h^{\dagger}_u \tilde{Q}_i) \tilde{d}^c_j\
  +\ \mbox{h.c.}\ .
\end{equation}

\subsection{Soft Supersymmetry-Breaking Terms}
\label{subsec:soft_terms}

Since \SUSY\ is necessarily broken, a proper treatment of $R$-parity
violation must also include \Rp\ soft terms.
\index{Lagrangian!$R_p$ conserving soft terms}
In the \SSM, we parametrize
our ignorance about the mechanism responsible for \SUSY\ breaking by
introducing the most general terms that break \SUSY\ in a soft way,
i.e. without reintroducing quadratic divergences. The possible soft terms
were classified by Girardello and Grisaru \cite{girardello82}. They consist
of mass terms for the gauginos ($M \lambda \lambda$),
analytic\footnote{``Analytic'' means here that these couplings involve
only products of the scalar components of the (left-handed) chiral superfields, 
and not their complex conjugates.} couplings for the scalar fields known as ``A-terms''
($A_{ijk}\, \phi_i \phi_j \phi_k$,), analytic scalar mass terms known as
``B-terms'' ($B_{ij}\, \phi_i \phi_j$), and scalar mass terms
($m^2_{ij}\, \phi^{\dagger}_i \phi_j$).
This leads to the following soft \SUSY-breaking Lagrangian density for the
\SSM\ (see Appendix \ref{chap:appendixA}), given by:
\begin{eqnarray}   
  -\, {\cal L}^{soft}_{Rp} & = &
     (m^2_{\tilde Q})_{ij}\, {\tilde Q}^{\dagger}_i {\tilde Q}_j\
  +\ (m^2_{\tilde u^c})_{ij}\, {\tilde u}^{c \dagger}_i {\tilde u}^c_j\
  +\ (m^2_{\tilde d^c})_{ij}\, {\tilde d}^{c \dagger}_i {\tilde d}^c_j\
  +\ (m^2_{\tilde L})_{ij}\, {\tilde L}^{\dagger}_i {\tilde L}_j\
  +\ (m^2_{\tilde l^c})_{ij}\, {\tilde l}^{c \dagger}_i {\tilde l}^c_j\
  \nonumber  \\  & &
  + \left( A^e_{ij}\, h_d {\tilde L}_i {\tilde l}^c_j\
  +\ A^d_{ij}\, h_d {\tilde Q}_i {\tilde d^c}_j\
  -\ A^u_{ij}\, h_u {\tilde Q}_i {\tilde u^c}_j\
  +\ \mbox{h.c.} \right)  \nonumber  \\  & &
  +\ {\tilde m}_d^2\,  h_d^{\dagger} h_d\
  +\ {\tilde m}_u^2\,  h_u^{\dagger} h_u\
  + \left( B\, h_u h_d\ +\ \mbox{h.c.} \right)
  \nonumber  \\  & &
  +\ \frac{1}{2} M_1\, \bar{\tilde B} {\tilde B}\
  +\ \frac{1}{2} M_2\, \bar{\tilde W}^3 {\tilde W}^3\
  +\ M_2\, \bar{\tilde W}^+ {\tilde W}^+\
  +\ \frac{1}{2} M_3\, \bar{\tilde g}^a {\tilde g}^a\ .
\label{eq:V_Rp_even}
\end{eqnarray}
In the absence of $R$-parity, one must in principle also consider the most
general \index{Lagrangian!\Rp\ soft terms} soft terms associated with
$W_{\slash \!\!\!\! {R_p}}$:
\begin{eqnarray}   
  V^{soft}_{\slash \!\!\!\! {R_p}} & = &
  \frac{1}{2} A_{ijk}\, {\tilde L}_i {\tilde L}_j {\tilde l^c}_k\
  +\ A'_{ijk}\, {\tilde L}_i {\tilde Q}_j {\tilde d^c}_k 
  +\ \frac{1}{2} A''_{ijk}\, {\tilde u^c}_i {\tilde d^c}_j {\tilde d^c}_k
     \nonumber \\
  & & +\ B_i\, h_u {\tilde L}_i\
  +\ \widetilde{m}^2_{di}\, h_d^{\dagger}\, {\tilde L}_i\ +\ \mbox{h.c.}\ .
\label{eq:V_Rp_odd}
\end{eqnarray}
Again Eqs. (\ref{eq:V_Rp_even}) and (\ref{eq:V_Rp_odd}) assume the particle
content of the MSSM; more general versions of the \SSM\ may allow for
additional soft terms, both $R$-parity even and $R$-parity odd.

The soft potential (\ref{eq:V_Rp_odd}) introduces 51 new ({\it a priori}
complex) \Rp\ parameters: $9 + 27 + 9 = 45\ $ \Rp\ $A$-terms with the same
antisymmetry properties as the corresponding trilinear superpotential
couplings, 3 $B_i$ associated with the bilinear superpotential terms,
and 3 \Rp\ soft mass parameters\index{Parameters for \Rp\ models!soft masses}
$\widetilde{m}^2_{di}$ mixing the down-type Higgs boson with the slepton fields
\index{Mixing!higgs--slepton}.
In the presence of the bilinear \Rp\ soft terms, the radiative
electroweak symmetry breaking generally leads to non-vanishing sneutrino
vacuum expectation values $<{\tilde \nu}_i>\ \equiv v_i / \sqrt{2}$,
together with the familiar Higgs \VEVs\ $<h^0_d>\ \equiv v_d / \sqrt{2}$ and
$<h^0_u>\ \equiv v_u / \sqrt{2}$. Indeed, the bilinear terms in
Eq. (\ref{eq:V_Rp_odd}) yield linear terms in the ${\tilde \nu}_i$ fields
after translation of the Higgs fields,
$h^0_d \rightarrow h^0_d + v_d / \sqrt{2}$ and
$h^0_u \rightarrow h^0_u + v_u / \sqrt{2}$, which destabilizes the scalar
potential and leads to $<{\tilde \nu}_i>\ \neq 0$ -- unless particular
conditions are satisfied by the bilinear soft terms.

Since the sneutrino fields correspond to $R$-parity odd particles,
these \VEVs\ in turn induce new \Rp\ terms in the Lagrangian.
In particular new mixing terms between lepton
and chargino/neutralino fields (resp. between slepton and Higgs boson fields)
\index{Mixing!higgs--slepton}
are generated from the gauge and $R$-parity conserving
Yukawa couplings (resp. from the $R$-parity conserving $A$-terms).
It is important to keep in mind, however, that the slepton \VEVs\ are not
independent \Rp\ parameters, since they are functions of the \Rp\ couplings; 
as we shall see below it is always possible to
find a weak eigenstate basis in which $v_i = 0$.

\subsection{Choice of the Weak Interaction Basis}
\label{subsec:basis}

In the absence of $R$-parity and lepton-number conservation, there is no
{\it a priori} distinction between the $Y=-1$ Higgs ($H_d$) and the
lepton ($L_i$) superfields, which have the same gauge interactions.
One can therefore freely rotate the weak eigenstate basis by a unitary
transformation \cite{hall84}:
\begin{equation}
   \left( \begin{array}{l} H_d \\ L_i \end{array} \right)\ \, \rightarrow \, \
  \left( \begin{array}{l} H'_d \\ L'_i \end{array} \right)\ =\
  U \left( \begin{array}{l} H_d \\ L_i \end{array} \right)\ ,
\label{eq:SU4_rotation}
\end{equation}
where $U$ is an $SU(4)$\index{Group symmetries!$SU(4)$ rotations} matrix 
with entries $U_{\alpha \beta}$,
$\alpha \equiv (0,i) = (0,1,2,3)$, $\beta \equiv (0,j) = (0,1,2,3)$. Under
(\ref{eq:SU4_rotation}), \Rp\ couplings and slepton \VEVs\ transform as
follows (we use the notation $\widetilde m^2_{ij}$ instead of
$(m^2_{\tilde L})_{ij}$):
\begin{eqnarray}
 \mu'_i  & = &  U^\star_{i0}\, \mu\ +\ \sum_j U^\star_{ij}\, \mu_j\ ,  \\
 \widetilde{m}^{'2}_{di} &  = &  U_{00} U^\star_{i0}\, \widetilde m^2_d\
   +\ \sum_{l,m} U_{0l} U^\star_{im}\, \widetilde m^2_{lm}\ +\ \sum_l
   \left( U_{00} U^\star_{il}\, \widetilde{m}^2_{dl}\,
   +\, U_{0l} U^\star_{i0}\, (\widetilde{m}^2_{dl})^\star \right) ,  \\
 B'_i &  = &  U^\star_{i0}\, B\
    +\ \sum_j U^\star_{ij}\, B_j\ ,  \qquad
   \qquad  v'_i\ =\ U_{i0}\, v_d\ +\ \sum_j U_{ij}\, v_j\ ,  \\
 (\lambda_{ijk})' &  = &  \sum_l \left( U^\star_{i0} U^\star_{jl}\,
   -\, U^\star_{il} U^\star_{j0} \right) \lambda^e_{lk}\
   +\ \sum_{l,m} U^\star_{il} U^\star_{jm}\, \lambda_{lmk}\ ,  \\
 (\lambda'_{ijk})' &  = &  U^\star_{i0}\, \lambda^d_{jk}\
   +\ \sum_l U^\star_{il}\, \lambda'_{ljk}\ ,  \qquad \qquad 
   (\lambda''_{ijk})'\ =\ \lambda''_{ijk}\ .
\end{eqnarray}
Since the $A$-terms transform exactly in the same way as the trilinear
superpotential couplings, we do not write explicitly the corresponding
transformation for $A_{ijk}$, $A'_{ijk}$ and $A''_{ijk}$.

From the above equations it is clear that the values of the
lepton-number-violating couplings are 
\index{Parameters for \Rp\ models!basis--dependent couplings}basis-dependent. 
There is also a redundancy between the bilinear and trilinear \Rp\ parameters,
as can be seen from the fact that 3 among the 9 bilinear \Rp\ parameters
$\mu_i$, $B_i$ and $\widetilde{m}^2_{di}$ can be rotated away upon a suitable
superfield redefinition, as expressed by Eq. (\ref{eq:SU4_rotation}).
This leaves us with $48 + 51 - 3 = 96$ physically meaningful (in general complex)
parameters.

As already noticed, it is always possible to choose a basis in which $\mu'_i = 0$,
eliminating all bilinear $R$-parity violation from the superpotential; but then 
in general non-vanishing slepton \VEVs\ are induced by the presence of the
bilinear \Rp\ soft terms.
Alternatively, one can choose a basis in which all slepton \VEVs\
vanish\footnote{This does not mean, however, that the scalar potential
is free from any term mixing \index{Mixing!higgs--slepton}
the slepton and Higgs fields in this basis.
Instead $B_i$ and $\widetilde{m}^2_{di}$ satisfy a particular relation
that prevents the slepton fields from acquiring a \VEV\ (see the discussion
at the end of subsection \ref{subsec:H_L_basis}).}, $v'_i = 0$,
but then in general $\mu'_i \neq 0$.
It is therefore crucial, when discussing \Rp\ effects, to specify in which
basis one is working. A particular choice of basis,
in which the Higgs-lepton mixing \index{Mixing!higgs--lepton} induced by 
bilinear $R$-parity violationis parametrized in an economical and physically 
sensible way, will bepresented in section~\ref{sec:bilinear}. 
Another option (to be discussed in subsection~\ref{sec:invariants}) is to define
a complete set of basis-independent quantities parametrizing 
\Rp\ effects~\cite{davidson97,davidson98,ferrandis99}.

Before closing this subsection, let us write down explicitly the infinitesimal
$SU(4)$\index{Group symmetries!$SU(4)$ rotations} transformation that rotates 
away small bilinear \Rp\ terms from the
superpotential (\ref{eq:W_Rp_odd}), $\mu_i \ll \mu$.
Defining $\epsilon_i \equiv \frac{\mu_i}{\mu}$, we can write the corresponding
unitary matrix $U$, up to  ${\cal O} (\epsilon^2)$ terms, as\footnote{For $\mu_i \sim \mu$,
Eq. (\ref{eq:SU4_small}) is no longer a good approximation.
The explicit form of a unitary matrix $U$ that rotates away arbitrarily large
bilinear \Rp\ parameters from the superpotential can be found in Ref. \cite{dreiner04}.}
\cite{hall84}:
\begin{equation}
  U\ = \left( \begin{array}{cc}
  1 & - \epsilon_i  \\  \epsilon^\star_i & 1_{3 \times 3}
  \end{array} \right)\ .
\label{eq:SU4_small}
\end{equation}
In the new basis $\mu'_i=0$, and the trilinear \Rp\ superpotential couplings
are modified to $(\lambda_{ijk})' = \lambda_{ijk} + (\epsilon_i \lambda^e_{jk}
- \e_j \lambda^e_{ik})$
and $(\lambda'_{ijk})' = \lambda'_{ijk} + \epsilon_i \lambda^d_{jk}$, where
$\lambda^e_{jk}$ and $\lambda^d_{jk}$ are the Yukawa couplings in the initial
basis. In the \SUSY-breaking sector, the mass parameters 
transform (omitting ${\cal O} (\epsilon^2)$ terms) as: 
\begin{eqnarray}  \hskip -.5cm
  & B' = B  - \sum_i \epsilon^\star_i B_i\ ,  \quad
  B'_i = B_i + \epsilon_i B\ ,  \label{eq:B_mu_prime}  \\
  & \widetilde{m}^{'2}_d = \widetilde{m}^2_d
  - (\sum_i \epsilon^\star_i \widetilde{m}^2_{di}
  + \mbox{h.c.})\ ,  \quad
  \widetilde{m}^{'2}_{ij} = \widetilde{m}^2_{ij}
  + (\epsilon^\star_i \widetilde{m}^2_{dj}
  + \mbox{h.c.})\ ,  \\
  & \widetilde{m}^{'2}_{di} = \widetilde{m}^2_{di}
  + \epsilon_i \widetilde{m}^2_d
  - \sum_j \epsilon_j \widetilde{m}^2_{ji}\ .
  \label{eq:m2_di_prime}
\end{eqnarray}
As stressed above, such a pattern of soft
mass parameters\index{Parameters for \Rp\ models!soft masses} generally induces
non-vanishing slepton vacuum expectation values, unless some very particular
constraints, that shall be discussed in section~\ref{sec:bilinear}, 
are satisfied.

\subsection{Constraints on the Size of {\boldmath{\Rp}} Couplings}
\label{subsec:constraints}

Being renormalizable, the \Rp\ couplings of Eqs. (\ref{eq:W_Rp_odd}) and
(\ref{eq:V_Rp_odd}) are not expected to be suppressed by any large mass
scale, and may thus induce excessively large baryon and 
lepton-number-violating effects.
In particular, the combination of couplings $\lambda'_{imk} \lambda^{''\star}_{11k}$
($i = 1, 2, 3$, $m = 1, 2$) would lead to proton decay via 
tree-level down squark exchange at an unacceptable rate, unless 
$\mid\! \lambda'_{imk} \lambda^{''\star}_{11k}\! \mid$ is smaller than
about $10^{-26}$  for a typical squark mass in the
$300 \GeV$ range, smaller by more than twelve orders of magnitude than 
a typical GUT scale (see Section \ref{secxxx3d} for a discussion of this bound).
There are many other constraints, both theoretical and phenomenological,
on the superpotential and soft \SUSY-breaking \Rp\ parameters. 
Bounds on the \Rp\ superpotential couplings coming
from the non-observation of baryon and lepton-number-violating processes and
from direct searches at colliders are reviewed in chapters~\ref{chap:indirect}
and~\ref{chap:colliders} respectively. Astrophysical and cosmological bounds
are presented in chapter~\ref{chap:cosmology}, and constraints associated with
the effects of the renormalization group evolution (perturbative unitarity bounds and
unification constraints), in chapter~\ref{chap:evolution}.

\section{Patterns of {\boldmath{$R$}}-Parity Breaking}
\label{subsec:patterns}

Several patterns of $R$-parity breaking can be considered, depending on
whether the breaking is explicit or spontaneous -- and, in the case of an
explicit breaking, on which type of couplings, bilinear, trilinear, or both,
are present.
Before classifying the various patterns considered in the literature,
let us make some general comments on subtleties associated with the
lepton-number-violating couplings.

First of all, if lepton-number conservation is not associated with a symmetry 
of the theory, there is in general no preferred ($H_d$, $L_i$) basis,
as was seen in the previous section, and the statement that only a certain 
class of lepton-number-violating \Rp\ couplings are present is basis-dependent. 

Still it is not an empty statement,
especially if one thinks in terms of the number of independent parameters. 
Indeed, if there is a basis in which all $\lambda_{ijk}$ and 
$\lambda'_{ijk}$ superpotential couplings
as well as their associated soft terms vanish, 
lepton-number violation can be parametrized by 9 parameters only
(3 $\mu_i$ and 6 bilinear soft terms). In a different basis the
$\lambda_{ijk}$ and $\lambda'_{ijk}$ couplings do not vanish, but their
values are determined, through the rotation (\ref{eq:SU4_rotation}), from
the original down quark and charged lepton $R_p$-conserving
Yukawa couplings.

Secondly, in a consistent quantum field theory, all operators breaking a
symmetry up to some dimension $d$ must normally be included in the Lagrangian
density, since if some of them are absent at tree level they are in general 
expected to be generated by radiative corrections. A well-known manifestation 
of this effect is the generation of bilinear \Rp\ terms through one-loop diagrams
involving lepton-number-violating trilinear \Rp\ interactions~\cite{decarlosl,nardi}
(see also Ref.~\cite{dreiner95}). For example, the $d=4$ couplings (\ref{eq:laglambda})
induce the $d=3$ higgsino-lepton mixing \index{Mixing!higgsino-lepton}
terms $\mu_i \tilde{h}_u L_i$ as well
as the $d=2$ Higgs-slepton mixing \index{Mixing!higgs-slepton}
mass terms $B_i h_u \tilde{L}_i$ and
$\widetilde{m}^2_{di} h^{\dagger}_d \tilde{L}_i$ (where $L_i$ and
$\tilde{L}_i$ denote the fermionic and scalar components of the lepton doublet
superfields $L_i$, respectively). This provides at least three consistent
patterns of
$R$-parity violation in the lepton sector at the quantum level\footnote{We
do not assume the conservation of lepton or baryon number {\it a priori}.
Of course, if lepton-number conservation (resp. baryon-number conservation) 
were assumed, \LV\ couplings (resp. \BV\ couplings) would never be generated
by radiative corrections.}:
\begin{itemize}
  \item[{\bf (a)}] {\bf {\boldmath{$R$}}-parity violation through $d=2$, $d=3$ and $d=4$
terms}: this corresponds to the most general explicit breaking of $R$-parity,
with all superpotential and soft \Rp\ terms allowed by gauge invariance and
renormalizability present in the Lagrangian density.
In this case one has to deal with
\index{Parameters for \Rp\ models}99 new (in general complex) parameters
beyond the ($R_p$-conserving) MSSM:
\index{Parameters for \Rp\ models!bilinear $\mu$}3 bilinear ($\mu_i$) and 
\index{Parameters for \Rp\ models!trilinear $\lambda, \lambda', \lambda''$}45 
trilinear (9 $\lambda_{ijk}$, 27 $\lambda'_{ijk}$, 9 $\lambda''_{ijk}$) 
couplings in the superpotential, together with their associated 
\index{Parameters for \Rp\ models!$B$--terms}$B$-terms (3 $B_i$) and 
\index{Parameters for \Rp\ models!$A, A', A''$}$A$-terms 
(9 $A_{ijk}$, 27 $A'_{ijk}$, 9 $A''_{ijk}$), 
and\index{Parameters for \Rp\ models!soft masses} 3 additional \Rp\ 
soft masses ($\widetilde{m}^2_{di}$) in the scalar potential. (The 3 slepton
\VEVs\ $v_i$ are not independent \Rp\ parameters, since they can be expressed
in terms of the MSSM parameters and \Rp\ couplings.) Due to the
ambiguity in the choice of the ($H_d$, $L_i$) basis, however, only 6 among
the 9 bilinear \Rp\ parameters are physical, thus reducing the number of
physically meaningful \Rp\ parameters to 96.
  \item[{\bf (b)}] {\bf {\boldmath{$R$}}-parity violation through $d=2$ and $d=3$
terms}: in this case all soft \Rp\ terms are present in the scalar potential
(both the bilinear terms $B_i h_u \tilde{L}_i$ and $\widetilde{m}^2_{di}
h^{\dagger}_d \tilde{L}_i$ and the triscalar couplings $A_{ijk}$, $A'_{ijk}$
and $A''_{ijk}$), while the superpotential contains only the bilinear \Rp\
terms $\mu_i H_u L_i$, which corresponds to both $d=2$ and $d=3$ terms in
the Lagrangian density. We are thus left with 54 \Rp\ parameters (3 $\mu_i$
in the superpotential; 3 $B_i$, 3 $\widetilde{m}^2_{di}$,
9 $A_{ijk}$, 27 $A'_{ijk}$ and 9 $A''_{ijk}$ in the scalar potential),
all of them physically meaningful.
  \item[{\bf (c)}] {\bf {\boldmath{$R$}}-parity violation through $d=2$ terms}:
in this case, $R$-parity violation originates solely from the soft
terms $B_i h_u \tilde{L}_i$ and $\widetilde{m}^2_{di}
h^{\dagger}_d \tilde{L}_i$, and can therefore be parametrized by 6
parameters only.
\end{itemize}

A closer look at the renormalization group equations (see chapter
\ref{chap:evolution}) shows that there actually exist other consistent
patterns of $R$-parity violation. In particular the popular ``bilinear
$R$-parity breaking'' scenario, in which $R$-parity is explicitly broken
by bilinear (superpotential and soft) terms only, is perfectly consistent
at the quantum level since \Rp\ dimension-3 $A$-terms are not generated
from the bilinear superpotential $\mu_i$ terms, also of dimension 3.
On the other hand another popular scenario in which $R$-parity is broken
solely by trilinear terms is not consistent since, as already mentioned,
bilinear \Rp\ terms are then generated from quantum corrections. The absence
of bilinear \Rp\ terms can only be assumed, strictly speaking, at the level of
the classical Lagrangian.

With the above remarks in mind, we can now comment on the scenarios of
$R$-parity violation that have received most attention in the
literature: explicit \Rp\ by trilinear terms, explicit \Rp\ by bilinear terms
(``bilinear $R$-parity breaking'') and spontaneous $R$-parity violation.

\begin{itemize}
  \item[{\bf (i)}] {\it explicit \Rp\ by trilinear terms}: in this case
one assumes that all bilinear \Rp\ terms are absent from the tree-level 
Lagrangian. One is then left with 45 trilinear \Rp\ couplings in the
superpotential, and their associated $A$-terms. This is the situation
considered in most phenomenological studies of $R$-parity violation, and
in the major part of this review. The absence of bilinear couplings and
sneutrino \VEVs\ also has the practical advantage of removing the ambiguity
associated with the choice of the weak interaction
basis (cf. subsection~\ref{subsec:basis}).
One must keep in mind, however,
that bilinear \Rp\ cannot be completely absent. Indeed, if only trilinear
couplings are present at some energy scale, they will generate bilinear
couplings at some other energy scale through renormalisation group
evolution\footnote{This is in fact a consequence of \SUSY\ breaking.
 In the limit of exact \SUSY, the $\mu_i H_u L_i$ terms generated radiatively
 can be removed from the superpotential by a rotation (\ref{eq:SU4_rotation}),
 eliminating any bilinear \Rp\ term from the Lagrangian. In the presence of
 soft \SUSY-breaking terms, however, the bilinear \Rp\ terms generated in
 the superpotential and in the scalar potential cannot be rotated away
 simultaneously.}~\cite{decarlosl,nardi} (see chapter~\ref{chap:evolution}
for details). Thus, in a consistent quantum field theory, one must include
bilinear \Rp\ terms as soon as trilinear terms are present. Still, since
the experimental and cosmological bounds on neutrino masses put strong
constraints on the tolerable amount of bilinear $R$-parity violation (see
next section), it makes sense to consider a situation in which the
phenomenology of trilinear \Rp\ interactions is not affected by the presence
of bilinear \Rp\ terms -- except for some specific phenomena such as neutrino
oscillations.
  \item[{\bf (ii)}] {\it explicit \Rp\ by bilinear terms}:
assuming that, in an appropriately chosen basis, $R$-parity is broken solely
by bilinear terms, the number of independent \Rp\ parameters reduces
to 9 (3 $\mu_i$ or $v_i$, 3 $B_i$ and 3 $\widetilde{m}^2_{di}$). Since
bilinear \Rp\ terms mix leptons with Higgs fields, however, trilinear \Rp\
interactions of the $\lambda$ and $\lambda'$ type (both triscalar and
Yukawa-like interactions) are generated upon rotating the initial weak
eigenstate basis to the mass eigenstate basis. Still those couplings can be
expressed in terms of the initial bilinear \Rp\ couplings and of the down-type
Yukawa couplings, and are not independent \Rp\ parameters.
Note that, due to the fact that $R$-parity breaking originates here
from $L$-number-violating interactions, no $B$-violating $\lambda''$-type
interactions are generated, thus avoiding the problem of a too fast
proton decay.
The advantage of such a scenario resides in its predictivity (albeit it is
difficult to motivate the absence of trilinear couplings by other
considerations than simplicity or predictivity); the main difficulty lies in
suppressing the neutrino masses induced by bilinear \Rp\ terms to an
acceptable level without fine-tuning (see chapter~\ref{chap:neutrinos}).
A detailed discussion of bilinear $R$-parity breaking is
given in section~\ref{sec:bilinear}.
  \item[{\bf (iii)}] {\it spontaneous \Rp}: a completely different option is
the spontaneous breaking of $R$-parity induced by the vacuum expectation value 
of an $R$-parity odd scalar (i.e. necessarily a scalar neutrino in the \MSSM).
Such a scenario can be implemented in various ways.
Common features of the models are a constrained pattern of \Rp\ couplings,
showing some analogy with the bilinear \Rp\ case, and - with the exception
of the models where lepton number is gauged - a variety of new
interactions involving a massless Goldstone boson (or a massive
pseudo-Goldstone boson in the presence of a small amount of explicit
lepton-number violation) associated with the breaking of the lepton number.
Spontaneous $R$-parity breaking is addressed in section~\ref{sec:spontaneous}.
\end{itemize}

Even within a restricted pattern of $R$-parity violation like (i) or (ii),
one is generally left with a very large number of arbitrary \Rp\ parameters.
For this reason, it is often necessary to make some assumptions on their
flavour structure; for example, the bounds coming from particular processes 
are generally derived under the assumption that a single coupling or
combination of couplings gives the dominant contribution (see
chapter~\ref{chap:indirect}). From a more theoretical perspective it is
interesting to study models that can constrain the flavour
structure of the \Rp\ couplings. In particular, the case of an abelian
flavour symmetry is addressed in section~\ref{sec:flavour}.
Furthermore, extensions of the \SSM\ with an enlarged gauge structure
may have a more restricted pattern of \Rp\ couplings than the \SSM\ itself.
$R$-parity violation in the context of \GUT\ is discussed in
section~\ref{sec:GUTs}.

Finally, it should be kept in mind that despite its simplicity, $R$-parity
may be viewed as having no clear theoretical origin, at least at this level. 
Thus, in the absence of $R$-parity, it is a valid option to consider other 
discrete or continuous symmetries sharing with $R$-parity the capability of 
protecting proton decay from renormalizable operators, while allowing some 
baryon or lepton-number-violating couplings.
Well-known examples of such symmetries are ``baryon parities'', which allow only
for lepton-number violation, and ``lepton parities'', which allow only for
baryon-number violation. Some of these symmetries are even 
more efficient than $R$-parity in suppressing proton decay from
non-renormalizable operators which may be generated from some fundamental
theory beyond the \SSM. These other symmetries ("alternatives" to $R$-parity) 
are discussed in section~\ref{sec:alternatives}.


\section{Effects of Bilinear {\boldmath{$R$}}-Parity violation}
\label{sec:bilinear}


As already argued in the previous section, bilinear \Rp\ terms are present
in any consistent pattern of $R$-parity violation, although they have often
been neglected in the literature; most of the studies assume
$R$-parity breaking by trilinear terms only. The purpose of this section
is to describe the effects associated with the presence of bilinear \Rp\
terms. The effects of trilinear \Rp\ terms will be discussed in chapters
\ref{chap:indirect} and \ref{chap:colliders}.
Let us stress that the following discussion does not only apply to the
bilinear and spontaneous $R_p$-breaking scenarios, but also to the most
general scenario in which both bilinear and trilinear \Rp\ couplings are
present. The phenomenology of bilinear $R$-parity violation has been first
investigated in Refs.~\cite{hall84,lee84,dawson85}.

\subsection{Distinguishing Between Higgs and Lepton Doublet Superfields}
\label{subsec:H_L_basis}

As we have already said, in the limit of unbroken \SUSY, the bilinear \Rp\
superpotential terms $H_u L_i$ can always be rotated away by a suitable
redefinition, as expressed by Eq.~(\ref{eq:SU4_rotation}) of the four
superfields $(H_d, L_{i=1,2,3})$.
This redefinition leaves a single bilinear term in the superpotential,
the $\mu$-term $\mu H_u H_d$, but generates new contributions to the
$\lambda_{ijk} L_i L_j E^c_k$ and $\lambda'_{ijk} L_i Q_j D^c_k$ terms
from the charged lepton and down quark Yukawa couplings.
So, as long as \SUSY\ is unbroken, $R$-parity violation can
be parametrized by trilinear couplings only. In the presence of soft terms,
however, the scalar potential and the superpotential provide {\it two 
independent sources of bilinear $R$-parity violation} that {\it cannot be 
simultaneously rotated away}~\cite{hall84}, unless some very particular conditions
to be discussed below are satisfied. Furthermore these conditions are not 
renormalization group invariant. Therefore bilinear \Rp\ terms, if absent at 
some energy scale, are always regenerated at other energy scales. 
One is then left with a \index{Mixing!higgs--lepton}
physically relevant mixing between the Higgs and lepton superfields, which
leads to specific signatures, as discussed in the next subsections.

The Higgs-lepton mixing \index{Mixing!higgs--lepton} associated with bilinear 
$R$-parity violationresults in an ambiguity in the definition of the $H_d$ and 
$L_i$ superfields,which carry the same gauge charges.
The values of the lepton-number-violating couplings, in particular, depend
on the choice of the ($H_d$, $L_i$) basis (see subsection \ref{subsec:basis}).
It is therefore crucial, in phenomenological studies of $R$-parity violation,
to specify in which basis one is working.
Of course any physical quantity one may compute will not depend on the
choice of basis made; but the formula that expresses this quantity as a function
of the lepton-number-violating couplings will. 
In this subsection, we shall introduce two basis-independent quantities
$\sin \xi$ and $\sin \zeta$ that control the size of the effects of bilinear
$R$-parity violation in the fermion and in the sfermion sectors, respectively.
%

Before doing so, let us rewrite the most general superpotential
and soft scalar potential, Eqs. (\ref{eq:W_Rp_even}) -- (\ref{eq:W_Rp_odd})
and (\ref{eq:V_Rp_even}) -- (\ref{eq:V_Rp_odd}), in a form that makes
apparent the freedom of rotating the $H_d$ and $L_i$ superfields.
For this purpose we group them into a 4-vector $\hat{L}_{\alpha}$,
$\alpha = 0,1,2,3$~\cite{banks95}. The renormalizable superpotential,
including all possible
\index{Superpotential!$R_p$ conserving} $R$-parity-preserving and 
\index{Superpotential!$R_p$ violating} -violating terms, then reads:
\begin{eqnarray}
  W & = & \mu_{\alpha}\, H_u \hat{L}_{\alpha}\ +\ \frac{1}{2}\,
  \lambda^e_{\alpha \beta k}\, \hat{L}_{\alpha} \hat{L}_{\beta} E^c_k\
  +\ \lambda^d_{\alpha j k}\,\hat{L}_{\alpha} Q_j D^c_k\  \nonumber \\
  & - & \lambda^u_{j k}\, H_u Q_j U^c_k\
  +\ \frac{1}{2}\, \lambda''_{i j k}\, U^c_i D^c_j D^c_k \ ,
\label{eq:W_general}
\end{eqnarray}
and the soft \SUSY-breaking terms in the scalar sector read:
\begin{eqnarray}
  V_{\rm soft} & = & \left(\, B_{\alpha}\,
  h_u \hat{\tilde L}_{\alpha}\, +\, \mbox{h.c.}\, \right)\
  +\ \widetilde m^2_{\alpha \beta}\, \hat{\tilde L}^{\dagger}_{\alpha}
  \hat{\tilde L}_{\beta}\ +\ \widetilde m^2_u\, h^{\dagger}_u h_u\
  +\ \ldots\  \nonumber \\
  & + & \left(\, 
  \frac{1}{2} A^e_{\alpha \beta k}\,
              {\tilde L}_{\alpha} {\tilde L}_{\beta} {\tilde l^c}_k\
  +\ A^d_{\alpha jk}\, {\tilde L}_{\alpha} {\tilde Q}_j {\tilde d^c}_k 
  +\ \frac{1}{2} A''_{ijk}\, {\tilde u^c}_i {\tilde d^c}_j {\tilde d^c}_k\
  +\, \mbox{h.c.}\, \right)\ ,
\label{eq:soft_general}
\end{eqnarray}
where the dots stand for the other soft (squark and "right-handed" slepton)
scalar mass terms.
In the presence of the bilinear soft terms in Eq. (\ref{eq:soft_general}),
the radiative breaking of the electroweak symmetry leads to a
vacuum expectation value for each of the neutral scalar components of the
$\hat{L}_{\alpha}$ doublet superfields,
$<\hat{\tilde \nu}_{\alpha}>\ \equiv v_{\alpha} / \sqrt{2}$.
The covariance of Eqs. (\ref{eq:W_general}) and (\ref{eq:soft_general}) under
$SU(4)$\index{Group symmetries!$SU(4)$ rotations} rotations of 
the $\hat{L}_{\alpha}$ superfields,
$\hat{L}_{\alpha} \rightarrow U_{\alpha \beta} \hat{L}_{\alpha}$, dictates
the transformation laws for the parameters $\mu_{\alpha}$,
$\lambda^e_{\alpha \beta k}$, $\lambda^d_{\alpha j k}$, $B_{\alpha}$,
$\widetilde m^2_{\alpha \beta}$, $A^e_{\alpha \beta k}$ and $A^d_{\alpha jk}$.

Up to now, we have not chosen a specific basis for the $\hat{L}_{\alpha}$
superfields. Several bases have been considered in the literature. A first
possibility \cite{hall84} is to define the down-type
Higgs superfield $H_d$ as the combination of the $\hat{L}_{\alpha}$
superfields that couples to $H_u$ in the superpotential. This choice implies
$\mu_i = 0$ for the orthogonal combinations $L_i$, i.e. all bilinear
\Rp\ terms are contained in the soft \SUSY-breaking contribution
to the scalar potential, but the lepton scalars have in general non-vanishing
\VEVs, $v_i \neq 0$.
A second, more physical possibility is to define $H_d$ as the combination of
the $\hat{L}_{\alpha}$ superfields whose vacuum expectation value breaks the
weak hypercharge\index{Hypercharge} \cite{banks95}, implying $v_i = 0$ for 
all three sneutrino fields. In the following, we choose the latter option and
define
\begin{equation}
  H_d\ =\ \frac{1}{v_d} \sum_{\alpha}\, v_{\alpha} \hat{L}_{\alpha} \ ,
\label{eq:H_d}
\end{equation}
where $v_d \equiv \left( \sum_{\alpha} v^2_{\alpha} \right)^{1/2}$ (here and
in the following, we neglect phases for simplicity, since taking them into
account is straightforward). The orthogonal combinations $L_i$, $i=1,2,3$
correspond to the usual slepton fields with vanishing \VEVs:
\begin{equation}
  \hat{L}_{\alpha}\ =\ \frac{v_{\alpha}}{v_d}\, H_d\ +\
        \sum_i\, e_{\alpha i} \, L_i \ ,
\label{eq:redefinition}
\end{equation}
where the 4-vectors $\vec{e}_i \equiv \{e_{\alpha i} \}_{\alpha = 0 \ldots 3}$
satisfy $\vec{v} . \vec{e}_i = 0$ and $\vec{e}_i . \vec{e}_j = \delta_{ij}$.
At this point one can still rotate freely the $L_i$ superfields. This freedom can be
removed by diagonalizing the charged lepton Yukawa couplings~\cite{bisset98}.
The advantage of this choice is that, in the physically relevant limit $\sin \xi \ll 1$,
to be discussed below, the charged leptons almost coincide with their mass
eigenstates, and the lepton flavour composition of the massive neutrino
(see below) can be parametrized in terms of the $\mu_i$.
Alternatively, one may require that a single lepton doublet superfield, say $L_3$,
couples to $H_u$ in the superpotential, i.e.
$\vec{\mu} . \vec{e}_1 = \vec{\mu} . \vec{e}_2 = 0$.  \cite{binetruy98}.
The latter choice allows one to rewrite the superpotential as (leaving aside
the last two terms in Eq. (\ref{eq:W_general}), which are not modified):
\begin{eqnarray}
  W & = & \lambda^e_{ik}\, H_d L_i E^c_k\
        +\ \lambda^d_{ik}\, H_d Q_i D^c_k\
        +\ \frac{1}{2}\, \lambda_{ijk}\, L_i L_j E^c_k\
        +\ \lambda'_{ijk}\, L_i Q_j D^c_k\  \nonumber  \\
        & & +\ \mu \cos \xi\, H_u H_d\ +\ \mu \sin \xi\, H_u L_3\ \ ,
\label{eq:W_redefined}
\end{eqnarray}
where $\mu \equiv \left( \sum_{\alpha} \mu^2_{\alpha} \right)^{1/2}$, $\xi$
is the angle between the 4-vectors $\vec{\mu}$ and $\vec{v}$, given by \cite{banks95}
\begin{equation}
 \cos \xi\ \equiv\ \frac{1}{\mu v_d}\, \sum_{\alpha} \mu_{\alpha}v_{\alpha}\ ,
\label{eq:cos_xi}
\end{equation}
and the physical Yukawa and trilinear \Rp\ couplings are given by:
\begin{eqnarray}
  \lambda^e_{ik}\ =\ \sum_{\alpha,\, \beta}\, \frac{v_{\alpha}}{v_d}\:
    e_{\beta i}\, \lambda^e_{\alpha \beta k}\ , & & \lambda^d_{ik}\
    =\ \sum_{\alpha}\, \frac{v_{\alpha}}{v_d}\: \lambda^d_{\alpha i k}\ ,
\label{eq:redefined_Yukawa} \\
  \lambda_{ijk}\ =\ \sum_{\alpha,\, \beta}\, e_{\alpha i}\, e_{\beta j}\,
    \lambda^e_{\alpha \beta k}\ , & & \lambda'_{ijk}\ =\ \sum_{\alpha}\,
    e_{\alpha i}\, \lambda^d_{\alpha j k} 
\label{eq:redefined_lambda'}
\end{eqnarray}
(similar relations hold for the associated $A$-terms $A^e_{ij}$, $A^d_{ij}$,
$A_{ijk}$ and $A'_{ijk}$). The residual term $H_u L_3$ in Eq.
(\ref{eq:W_redefined}) corresponds to a physical higgsino-lepton mixing 
\index{Mixing!higgsino-lepton} that
cannot be removed by a field redefinition, unless the vacuum expectation
values $v_{\alpha}$ are proportional to the $\mu_{\alpha}$, so that
$\sin \xi = 0$. Indeed, contrary to the $\mu_{\alpha}$ and $v_{\alpha}$ which
are basis-dependent quantities and do not have any intrinsic physical meaning,
their relative angle $\xi$ does not depend on the choice of basis for the
$\hat L_{\alpha}$ superfields. {\it This angle controls the
size of the effects of bilinear $R$-parity violation in the fermion
sector} (see section \ref{sec:bilinear}).

Similarly, the amount of bilinear $R$-parity violation in the scalar sector
is measured by the angle $\zeta$ formed by the 4-vectors $\vec{B}$ and
$\vec{v}$~\cite{haber2},
\begin{equation}
  \cos \zeta\ \equiv\ \frac{1}{B v_d}\, \sum_{\alpha} B_{\alpha}v_{\alpha}\ ,
  \label{eq:cosksi}
\end{equation}
where $B \equiv \left( \sum_{\alpha} B^2_{\alpha} \right)^{1/2}$. Indeed,
in the $v_i = 0$ basis, the \Rp\ scalar potential reduces to:
\begin{eqnarray}
  V^{(2)}_{\slash \!\!\!\! {R_p}} & = &
   \mu^\star\! \mu_i\, h^{\dagger}_d \tilde{L}_i\ +\ B_i\, h_u{\tilde L}_i\
   +\ \widetilde m^2_{di}\, h^{\dagger}_d\tilde{L}_i\
   +\ \mbox{h.c.}  \nonumber \\
  & = & B_i \left( h_u - \tan \beta\, h^{\dagger}_d \right) {\tilde L}_i\
   +\ \mbox{h.c.}\ ,
\label{eq:V_redefined}
\end{eqnarray}
where $\tan \beta \equiv v_u / v_d$, $B_i = \sum_{\alpha} B_{\alpha}
e_{\alpha i}$, and we have used the minimization condition \linebreak
$\widetilde m^2_{di} + \mu^\star\! \mu_i + B_i \tan \beta = 0$ to derive
the last equality. This condition ensures that the ${\tilde L}_i$
fields only couple to the combination of $h_u$ and $h_d$ that does not
acquire a \VEV, thus allowing for vanishing $v_i$'s. Since $\sum_i B^2_i =
B^2 \sin^2 \zeta$, Eq. (\ref{eq:V_redefined}) tells us that {\it the overall amount
of physical Higgs-slepton mixing \index{Mixing!higgs--slepton} is controlled by 
the angle $\zeta$}; in particular,it vanishes if and only if the vacuum 
expectation values $v_{\alpha}$ are proportional to the soft parameters 
$B_{\alpha}$, so that $\sin \zeta = 0$.In this case all bilinear soft terms are 
$R$-parity conserving in the$v_i=0$ basis, i.e. $\mbox{$V^{(2)}_{\rm soft} = 
(B\, h_u h_d + \mbox{h.c.})$}+ \widetilde m^2_d\, h^{\dagger}_d h_d + \widetilde 
m^2_u\, h^{\dagger}_u h_u+ \widetilde m^2_{ij}\, \tilde L^{\dagger}_i \tilde L_j 
+ \ldots$.Some phenomenological consequences of the Higgs-slepton mixing 
\index{Mixing!higgs--slepton} are 
discussedat the end of subsection \ref{subsec:implications} and later in section
\ref{sec:sneutrinos}.

In general the vacuum expectation values $v_{\alpha}$ are neither aligned
with the superpotential masses $\mu_{\alpha}$ nor with the soft parameters
$B_{\alpha}$. This results in a physical Higgs/lepton mixing 
\index{Mixing!higgs--lepton} both in the
fermion and in the scalar sectors, characterized by the two misalignment
\index{Misalignment} 
angles $\xi$ and $\zeta$, respectively. The case $\sin \xi = 0$
corresponds to the absence of mixing in the fermion sector: $\mu_i = v_i = 0$
in the same basis, and all effects of bilinear $R$-parity violation come
from the Higgs-slepton \index{Mixing!higgs--slepton}
mixing, Eq. (\ref{eq:V_redefined}). Conversely,
the case $\sin \zeta = 0$ corresponds to the absence of mixing in the scalar
sector: $B_i = v_i = 0$ in the same basis, and all effects of bilinear
$R$-parity violation come from the higgsino-lepton 
\index{Mixing!higgsino--lepton} mixing, Eq. (\ref{eq:W_redefined}). Finally, in 
the case where$\sin \xi = \sin \zeta = 0$, all bilinear \Rp\ terms can be 
simultaneously rotated away from the Lagrangian, and $R$-parity violation is 
purely of thetrilinear type. Again, the statement that $\sin \xi$ or $\sin 
\zeta$ vanishes is scale-dependent \cite{decarlosl,nardi}.

As we shall discuss in subsection \ref{subsec:H_L_mixing}, there are
particularly strong constraints on $\sin \xi$ coming from neutrino masses.
Therefore, it is legitimate to ask under which circumstances $\sin \xi$
vanishes.  To achieve this situation, it is sufficient (but not necessary)
to impose the following two conditions on the \Rp\ soft terms \cite{banks95}:
(i) the $B$ terms are proportional to the $\mu$-terms, $B_{\alpha} \propto
\nolinebreak \mu_{\alpha}$ \cite{hall84}; (ii) $\vec{\mu}$ is an
eigenvector of the scalar mass-squared matrix,
$\widetilde m^2_{\alpha \beta}\, \mu_{\beta}
= \widetilde m^2_d\, \mu_{\alpha}$.
Although these relations are not likely to hold exactly
at the weak scale, a strong correlation between the soft parameters and the
$\mu_{\alpha}$ may result from a flavour symmetry \cite{banks95} or from some
universality assumption at the GUT scale \cite{hempfling96,nilles97},
leading to an approximate alignment between the 4-vectors $\vec{\mu}$ and
$\vec{v}$, so that $\sin \xi \ll 1$ (see chapter~\ref{chap:evolution} for more
details). Actually conditions (i) and (ii) do not only imply $\sin \xi = 0$,
but also $\sin \zeta = 0$; therefore they are sufficient to ensure the
absence of bilinear $R$-parity violation at the scale at which they are
satisfied. Finally, when only condition (i) holds, one has
$\sin \xi = \sin \zeta$, hence bilinear $R$-parity violation is parametrized
by a single physical parameter $\xi$ (with $B_i = B \sin \xi\, \delta_{i3}$
in the basis where $\mu_i = \mu \sin \xi\, \delta_{i3}$).

\subsection{Trilinear Couplings Induced by Bilinear {\boldmath{\Rp}} Terms}
\label{subsec:induced}

Let us now comment on an interesting consequence of Eqs.
(\ref{eq:redefined_Yukawa})--(\ref{eq:redefined_lambda'})
in scenarios where, in some particular basis
$\hat{L}_{\alpha}$, the trilinear \Rp\ couplings vanish
(i.e. $\lambda^e_{ijk} = \lambda^d_{ijk} = 0$, while
$\lambda^e_{0jk} = - \lambda^e_{j0k}$ and $\lambda^d_{0jk}$ are
non-vanishing). This is the case, in particular, in the ``bilinear'' and
spontaneous $R_p$-breaking scenarios.
Then, in the ($H_d$, $L_i$) basis defined by Eq.
(\ref{eq:redefinition}), the trilinear \Rp\ couplings are related to the
ordinary Yukawa couplings $\lambda^e_{ij}$ and $\lambda^d_{ij}$ by
\begin{equation}
  \lambda_{ijk}\ =\ \frac{v_d}{v_0} \left( e_{0i}\, \lambda^e_{jk}\, -\,
        e_{0j}\, \lambda^e_{ik} \right)\ ,   \hskip 1cm
  \lambda'_{ijk}\ =\ \frac{v_d}{v_0}\ e_{0i}\, \lambda^d_{jk}\ ,
\label{eq:rotated_lambda}
\end{equation}
where $v_0$ and $e_{0i}$ are the $\alpha = 0$ components of the 4-vectors
$v_{\alpha}$ and $e_{\alpha i}$ that define the ($H_d$, $L_i$) basis.
The flavour dependence of the $\lambda_{ijk}$ and $\lambda'_{ijk}$ is then
determined by the charged lepton and down quark Yukawa couplings, respectively.

Eq. (\ref{eq:rotated_lambda}) has two obvious consequences.
First, unless there is a hierarchy of \VEVs\
$v_0 \ll v_i$ in the initial basis, the trilinear \Rp\
couplings are at most of the same order of magnitude as the down-type Yukawa
couplings. This suppression, which is stronger for smaller values of
$\tan \beta$, allows them to evade most of the direct and indirect bounds
that shall be discussed in chapters~\ref{chap:indirect} 
and~\ref{chap:colliders}
(some particularly severe bounds require an additional suppression of the
rotation parameters $e_{0i}$). Second, the contributions of the trilinear
\Rp\ couplings (\ref{eq:rotated_lambda}) to flavour-changing neutral current
(FCNC) processes either vanish or are strongly suppressed \cite{binetruy98}.
Indeed, the $\lambda'_{ijk}\, L_i Q_j D^c_k$ terms read, in the mass eigenstate
basis of the quarks (we put a bar on the trilinear \Rp\ couplings when
they are defined in the mass eigenstate basis for fermions):
\begin{equation}
  \bar \lambda'_{ijk}\, \left( N_i\, D_j\
  -\ E_i\, V^{\dagger}_{jp}\, U_p \right) D^c_k
  \hskip 1cm \mbox{with} \hskip 1cm
 \bar\lambda'_{ijk}\ =\ e_{0i}\, \frac{\sqrt{2}\, m_{d_j}}{v_0}\ \delta_{jk}\ ,
\label{eq:lambda'_bar}
\end{equation}
where $V$ is the CKM matrix. It follows that the $\lambda'$
couplings induced by bilinear \Rp\ vanish for quark-flavour-changing 
transitions $d_j \rightarrow d_k$, while $u_j \rightarrow d_k$ 
transitions are suppressed by the CKM angles. 
Thus most of the constraints on
$\lambda'$ couplings due to quark-flavour-violating processes
(see chapter~\ref{chap:indirect}) are trivially satisfied. 

The case of lepton-flavour violation is more subtle, both because
the relation between the $\lambda_{ijk}$ and $\lambda^e_{ij}$ couplings
is not an exact proportionality relation, and because the charged lepton masses
are not given by the eigenvalues of the Yukawa matrix, due to the
lepton-higgsino mixing \index{Mixing!higgsino--lepton}
(see discussion below). However, for small mixing
($\sin \xi \ll 1$), one can write, in the mass eigenstate basis of the charged
leptons:
\begin{equation}
  \bar \lambda_{ijk}\ N_i\, E_j\, E^c_k  \hskip 1cm \mbox{with}
  \hskip 1cm  \bar \lambda_{ijk}\ \simeq\ \frac{\sqrt{2}}{v_0}\,
  \left( R^{e \star}_{L il} e_{0l}\, m_{e_j}\, \delta_{jk}\
  -\ R^{e \star}_{L jl} e_{0l}\, m_{e_i}\, \delta_{ik} \right)\ ,
\label{eq:lambda_bar}
\end{equation}
where the matrix $R^e_L$ rotates the left-handed charged leptons fields to 
their corresponding mass
eigenstates\footnote{Strictly speaking, the couplings $\bar \lambda_{ijk}$
  are defined in the basis in which the charged lepton Yukawa couplings are
  diagonal, i.e. $R^e_L$ is defined by $R^e_L \lambda^e R^{e \dagger}_R =
  \mbox{Diag}\, (\lambda_{e_1}, \lambda_{e_2}, \lambda_{e_3})$. However, for
  $\sin \xi \ll 1$ this basis approximately coincides with the mass eigenstate
  basis, and therefore $\lambda_{e_i} v_d \simeq \sqrt{2}\, m_{e_i}$.
  In the limit $\sin \xi = 0$, both bases coincide and Eq. (\ref{eq:lambda_bar})
  is exact.}. These couplings violate lepton flavour, but there is a restricted
number of them, since $\bar \lambda_{ijk} \neq 0$ only if $i=k$ or $j=k$.
Furthermore, the non-vanishing couplings are suppressed by the small lepton
Yukawa couplings. This significantly reduces
their contributions to lepton-flavour-violating processes, especially in the
small $\tan \beta$ case.

The above conclusions hold when the trilinear \Rp\ couplings
vanish in the original $\hat{L}_{\alpha}$ basis, i.e.
$\lambda^e_{ijk} = \lambda^d_{ijk} = 0$. If this is not the case, the physical
trilinear \Rp\ couplings $\lambda_{ijk}$ and $\lambda'_{ijk}$ (defined in the
($H_d$, $L_i$) basis) receive an additional contribution from the initial
$\lambda^e_{ijk}$ and $\lambda^d_{ijk}$ couplings, which modifies Eqs.
(\ref{eq:rotated_lambda}), (\ref{eq:lambda'_bar}) and (\ref{eq:lambda_bar})
as well as the resulting conclusions for FCNC processes.

\subsection{Higgsino-Lepton Mixing}
\label{subsec:H_L_mixing}
\index{Mixing!higgsino-lepton|(}
As already mentioned, the main effect of bilinear $R$-parity violation is a
physical mixing between sleptons and Higgs bosons in the scalar sector, and
between leptons and neutralinos/char\-ginos in the fermion sector. We do not
discuss the details of the mixing in the scalar sector here, and refer the
interested reader to the literature (see e.g.
Refs.~\cite{campos95,diazrom98,chang00,davidson99,kong02}).
In the fermion sector, in an arbitrary basis $\hat{L}_{\alpha}$, the
superpotential mass parameters $\mu_{\alpha}$ mix the fermionic components
of the $\hat{L}_{\alpha}$ and $H_u$ superfields (i.e. the neutrino fields
$\hat{\nu}_{\alpha}$ with the neutral higgsino field $\tilde h^0_u$, and
the charged lepton fields $\hat{l}^-_{\alpha}$ with the conjugate of the charged 
higgsino field $\tilde h^+_u$), and the \VEVs\ $v_{\alpha}$ mix the neutrino fields
$\hat{\nu}_{\alpha}$ with the gaugino associated with
the $Z$ gauge boson. 

As a result the $4 \times 4$ neutralino mass matrix (resp.
the $2 \times 2$ chargino mass matrix) of the MSSM becomes a $7 \times 7$
neutralino-neutrino mass matrix (resp. a $5 \times 5$ chargino-\, charged lepton
mass matrix). With the notation
${\cal L}_{\rm mass} = - \frac{1}{2}\, \psi^{0 T} M_N \psi^0 + \mbox{h.c.}$,
the $7 \times 7$ neutralino-neutrino mass matrix $M_N$ reads, in the
$\psi^0 = (-i \lambda_{\gamma}, -i \lambda_Z, \tilde h^0_u,
\hat{\nu}_{\alpha})_{\, \alpha = 0 \cdots 3}$
basis\footnote{\label{footnote:two_comp}We use here a two-component notation
for spinors. With the conventions of Ref. \cite{haber85}, the two-component
spinors $\lambda_{\gamma}$ and $\lambda_Z$ are related to the four-component
Majorana spinors\index{Majorana fermions} $\tilde \gamma$ and $\tilde Z$ by
$\tilde \gamma = (-i \lambda_{\gamma}, i \bar \lambda_{\gamma})^T$ and
$\tilde Z = (-i \lambda_Z, i \bar \lambda_Z)^T$. Similarly,
the two-component spinors $\lambda^+$ and $\lambda^-$ are related to the
Dirac spinor\index{Dirac fermions} $\tilde W^+$ by
$\tilde W^+ = (-i \lambda^+, i \bar \lambda^-)^T$.}~\cite{banks95}:
\begin{equation}  \hskip -.5cm
  M_N\ =\ \left(
    \begin{array}{cccc}
    M_1 c^2_W + M_2 s^2_W & (M_2-M_1) s_W c_W & 0 & 0_{1 \times 4} \\
    (M_2-M_1) s_W c_W & M_1 s^2_W + M_2 c_W^2 &  \frac{g}{2 c_W}\, v_u
    & - \frac{g}{2 c_W}\, v_{\alpha} \\
    0 & \frac{g}{2 c_W}\, v_u & 0 & - \mu_{\alpha} \\
    0_{4 \times 1} & - \frac{g}{2 c_W}\, v_{\alpha} & - \mu_{\alpha}
    & 0_{4 \times 4} \\
    \end{array}  \right) ,
\label{eq:M_N_arbitrary}
\end{equation}
where $M_1$ and $M_2$ are the $U(1)_Y$\index{Group symmetries!$U(1)_Y$} and
$SU(2)_L$\index{Group symmetries!$SU(2)_L$} gaugino masses, $g$ is the 
$SU(2)_L$ gauge coupling, $c_W = \cos \theta_W$ and $s_W = \sin \theta_W$.

The $5 \times 5$ chargino-\, charged lepton mass matrix, defined by
${\cal L}_{\rm mass} = - \psi^{- T} M_C \psi^+ + \mbox{h.c.}$, reads, in the
$\psi^- = (-i \lambda^-, \hat{l}^-_{\alpha})_{\, \alpha = 0 \cdots 3}$ and
$\psi^+ = (-i \lambda^+, \tilde h^+_u, l^c_k)_{\, k = 1 \cdots 3}$ 
basis~\cite{banks95}:
\begin{equation}
  M_C\ =\ \left(
    \begin{array}{ccc}
    M_2 & g v_u / \sqrt{2} & 0_{1 \times 3} \\
    g v_{\alpha} / \sqrt{2} & \mu_{\alpha} &
    \lambda_{\alpha \beta k} v_{\beta} \\
    \end{array}  \right) ,
\label{eq:M_C_arbitrary}
\end{equation}
or, in a more explicit form:
\begin{equation}
  M_C\ =\ \left(
    \begin{array}{ccc}
    M_2 & g v_u / \sqrt{2} & 0_{1 \times 3} \\
    g v_0 / \sqrt{2} & \mu_0 & \lambda_{0 \beta k} v_{\beta} \\
    g v_i / \sqrt{2} & \mu_i & \lambda_{i \beta k} v_{\beta}
    \end{array}  \right) .
\end{equation}
In the $R$-parity conserving case, it is possible to find a ($H_d$, $L_i$) basis
in which $\mu_{\alpha} \equiv (\mu,0,0,0)$ and $v_{\alpha} \equiv (v_d,0,0,0)$.
Then Eqs. (\ref{eq:M_N_arbitrary})
and (\ref{eq:M_C_arbitrary}) reduce in the ``ino'' sector to the 
well-known ($4 \times 4$) neutralino and ($2 \times2$) chargino mass matrices 
of the MSSM with $R$-parity conservation (cf. Eq. (\ref{mwino})). 

In order to discuss the physical implications of the neutrino-neutralino and 
charged lepton-chargino mixings
caused by bilinear $R$-parity violation, we now turn to the basis defined by
Eqs. (\ref{eq:H_d}) and (\ref{eq:W_redefined}), in which $\mu_{\alpha} \equiv
(\mu \cos \xi, 0, 0, \mu \sin \xi)$ and $v_{\alpha} \equiv (v_d,0,0,0)$.
This choice of basis leaves only a physical mixing in
$M_N$ and $M_C$, proportional to the misalignment parameter $\sin \xi$.
The structure of both matrices is schematically:
\begin{equation}
  \footnotesize{\left( \mbox{\begin{tabular}{c|c}
    \begin{tabular}{c} MSSM \\ neutralino (chargino) \\ mass matrix
    \end{tabular} &
    \begin{tabular}{c} \Rp\ terms  \end{tabular} \\
    \noalign{\vskip .2cm}  \hline  \noalign{\vskip .2cm}
    \begin{tabular}{c} \Rp\ terms \end{tabular} & 
    \begin{tabular}{c} neutrino \\ (charged lepton) \\ mass matrix
    \end{tabular}
  \end{tabular}} \right)}\ .
\end{equation}

More precisely, $M_N$ reads:
\begin{equation}
  {\footnotesize M_N\ =\ \left(  \mbox{\begin{tabular}{c|c}
   $\begin{array}{cccc}
    M_1 c^2_W + M_2 s^2_W & (M_2-M_1) s_W c_W & 0 & 0 \\
    (M_2-M_1) s_W c_W & M_1 s^2_W + M_2 c_W^2 &  M_Z \sin \beta
    & -M_Z \cos \beta \\
    0 & M_Z \sin \beta & 0 & - \mu \cos \xi \\
    0 & -M_Z \cos \beta & - \mu \cos \xi & 0 \end{array}$
   & $\begin{array}{c} 0 \\ 0 \\ - \mu \sin \xi \\ 0 \end{array}$  \\
   \noalign{\vskip .2cm}  \hline  \noalign{\vskip .2cm}
   $\begin{array}{cccc} \hskip .5cm 0 \hskip 2.55cm & 0 \hskip 1.25cm
   & - \mu \sin \xi \hskip .8cm & 0 \end{array}$ & $0$
   \end{tabular}}  \right)}\ ,
\label{eq:M_N_xi}
\end{equation}
where $\tan \beta \equiv v_u/v_d$. As can also be seen from Eq.
(\ref{eq:W_redefined}), $\nu_1$ and $\nu_2$ decouple from the tree-level
neutrino-neutralino mass matrix, and only the mixing between
$\nu_3$ and the neutralinos remains (this is no longer the case at the
one-loop level, where all three neutrinos mix with the neutralinos, see
chapter~\ref{chap:neutrinos}). As for the chargino-charged lepton mass
matrix, it reads:
\begin{equation}
  \footnotesize{M_C\ =\ \left( \mbox{\begin{tabular}{c|c}
    $\begin{array}{cc}
    M_2 & g v_u / \sqrt{2} \\
    g v_d / \sqrt{2} & \mu \cos \xi \end{array}$ &
    $\begin{array}{c} 0_{1 \times 3} \\ 0_{1 \times 3} \end{array}$ \\
    \noalign{\vskip .2cm}  \hline  \noalign{\vskip .2cm}
    $\begin{array}{cc} \hskip .3cm 0_{3 \times 1} &
    \hskip .1cm \mu \delta_{i3} \sin \xi
    \end{array}$ & $\lambda^e_{ik} v_d / \sqrt{2}$
   \end{tabular}}
  \right)}\ .
\label{eq:M_C_xi}
\end{equation}

As already mentioned above, the mixing in $M_N$ and $M_C$, proportional to
$\sin \xi$, is suppressed when the \VEVs\ $v_{\alpha}$ are
approximately aligned with the $\mu_{\alpha}$; we shall see below that
experimental constraints on neutrino masses actually require a strong alignment,
i.e. a very small value of $\sin \xi$.
The matrix $M_C$ has five mass eigenstates; the
three lightest ones are identified with the charged leptons $l^-_i$ 
(which, due to the higgsino-lepton mixing, do not exactly coincide
with the eigenstates of the Yukawa matrix $\lambda^e_{ik}$, although
this mismatch can be neglected when $\sin \xi \ll 1$), and the two
heaviest ones with the charginos $\widetilde \chi^-_1$ and
$\widetilde \chi^-_2$. In addition to its two massless eigenstates $\nu_1$
and $\nu_2$, the matrix $M_N$ has five massive eigenstates;
the four heaviest ones are the neutralinos $\widetilde \chi^0_i$,
$i=1 \cdots 4\,$; the lightest one, which is mainly $\nu_3$ in the
$\sin \xi \ll 1$ case, can be identified with a Majorana
neutrino\index{Majorana fermions} with a mass~\cite{banks95}
\begin{equation}
  m_{\nu_{3}}\ =\ m_0 \tan^2 \xi\ ,
\label{eq:m_nu_tau}
\end{equation}
where $m_0$ is given, in the case
$\sin \xi \ll 1$, by the following expression~\cite{hempfling96}:
\begin{equation}
  m_0\ \simeq\ \frac{M_Z^2 \cos^2 \beta (M_1 c_W^2 + M_2 s_W^2)}
  {M_1 M_2 \mu \cos \xi - M_Z^2 \sin {2 \beta}
  ( M_1 c_W^2 + M_2 s_W^2)}\ \mu \cos \xi\ .
\label{eq:m_0}
\end{equation}
For a rough estimate, we can take:
\begin{equation}
  m_0\ \sim\ (100 \GeV)\, \cos^2 \beta\, \left( \frac{100 \GeV}{M_2}
  \right)\ .
\label{eq:m_0_estimate}
\end{equation}
The exact value of $m_0$ depends on the gaugino masses, $\mu$ and
$\tan \beta$, but Eq. (\ref{eq:m_0_estimate}) becomes a good approximation
for a heavy \susyq\ spectrum or for large values of $\tan \beta$.
Thus we see that the neutrino mass is proportional to the square of the \Rp\
angle $\xi$ (it could not have been linear in $\tan \xi$, since the neutrino
mass term is $R$-parity even), and therefore roughly measures
the overall amount of bilinear $R$-parity violation in the fermion sector.
The other two neutrinos remain massless at tree-level, but acquire 
masses at the one-loop level (see chapter~\ref{chap:neutrinos}).

Since $m_{\nu_{3}}$ is proportional to $\tan^2 \xi$, with a natural scale
in the $(1 - 100) \GeV$ range depending on the value of $\tan \beta$,
the known experimental and cosmological upper bounds on the heaviest neutrino
mass provide strong constraints on bilinear $R$-parity violation.
Indeed, the cosmological bound on neutrino masses
inferred from CMB \cite{WMAP} and large scale structure data \cite{2dFGRS}
($\sum_i m_{\nu_i} \lesssim 1 \eV$, where the sum runs over all neutrino
species) imposes a strong alignment of the $v_{\alpha}$ with the
$\mu_{\alpha}$, typically\footnote{Some scenarios in which the heaviest neutrino
has rapid enough decay modes could in principle evade this bound, but this
possibility looks rather unnatural in view of the experimental evidence for neutrino
oscillations (see e.g. the discussion at the end of section~\ref{sec:spontaneous}).}
$\sin \xi \lesssim 3 \times 10^{-6} \sqrt{1 + \tan^2 \beta}$.

We see that the presence of bilinear \Rp\ terms can only be tolerated at the 
expense of some significant amount of tuning in the parameters so as to keep
neutrinos sufficiently light. 
It is important to keep in
mind that the values of $\mu_i$ and $v_i$ by themselves are not a good
measurement of this tuning, since in an arbitrary weak eigenstate basis,
large values of the $\mu_i$ and $v_i$ can be compatible with a strong
alignment. However in the $v_i=0$ basis (resp. $\mu_i=0$ basis), the $\mu_i$
(resp. $v_i$) are constrained to be small, according to the estimate
$\mu_i \sim \mu \sin \xi$ (resp. $v_i \sim v_d \sin \xi$).

Finally, let us mention the fact that the neutrino sector also constrains
the tolerable amount of bilinear $R$-parity violation in the scalar sector,
yielding $\sin \zeta \lesssim (10^{-4} - 10^{-3})$ for the cosmological
bound (see section \ref{sec:problem}).
\index{Mixing!higgsino-lepton|)}

\subsection{Experimental Signatures of Bilinear {\boldmath{$R$}}-Parity Violation}
\label{subsec:implications}

The mixing of ordinary leptons with charginos and neutralinos also leads to
interactions that are characteristic of bilinear $R$-parity violation.
These are: {\it (i)} \Rp\ gauge interactions,
{\it (ii)} lepton-flavour-violating $Z$ couplings, and 
{\it (iii)} trilinear \Rp\ interactions distinct from those
generated from the superpotential $\lambda$ and $\lambda'$ terms.
These interactions are suppressed by at least one power of $\sin \xi$,
and are therefore very difficult to observe experimentally, with however a few exceptions.
We give a brief description of them below:
\begin{itemize}
  \item[{\it (i)}] Non-diagonal couplings of the $Z$ and $W$ bosons to a lepton
and a \susyq\ fermion (chargino or neutralino) appear when the currents are 
written in terms of the mass eigenstates:
$Z\, \widetilde \chi^{\pm}_i l^{\mp}_j$, $Z\, \widetilde \chi^0_i \nu_3$,
$W^- \widetilde \chi^+_i \nu_j$, $W^- \widetilde \chi^0_i l^+_j$.
These \Rp\ gauge couplings are proportional to $\sin \xi$,
and therefore correlated to the heaviest neutrino mass, $m_{\nu_3}$. 
They give rise to \Rp\ processes such as single production 
of charginos and neutralinos (e.g. through the decays 
$Z \rightarrow \widetilde \chi^{\pm}_i l^{\mp}_j$ and
$Z \rightarrow \widetilde \chi^0_i \nu_3$ at LEP) and decays
of the lightest neutralino into three standard fermions
($\widetilde \chi^0_1 \rightarrow \nu_3 f \bar f$ or
$\widetilde \chi^0_1 \rightarrow l_i f \bar f'$) 
or, if it is heavier than the gauge bosons, into $l_i W$ or $\nu_3 Z$.
Since the cross-section goes as $\sin^2 \xi$, single production
is unobservable in practice.
The two-body decays are characteristic of bilinear $R$-parity
violation, while the three-body decays are also induced by the trilinear
\Rp\ couplings (except however for
$\widetilde \chi^0_1 \rightarrow \nu_3 \nu \bar \nu$). Studies of the
corresponding signals at LEP and at hadron colliders can be found in
Refs.~\cite{bilinear_LEP} and~\cite{bilinear_hadron,bilinear_LSP_decay},
respectively. More recently, the \Rp\ decays of a neutralino LSP
have been discussed in Refs.~\cite{bilinear_LSP_decay,phrv00}.
See also Ref.~\cite{hp03} for the \Rp\ decays of a chargino LSP.
  \item[{\it (ii)}] Together with the previous \Rp\ gauge interactions,
bilinear \Rp\ also gives rise to flavour-violating couplings of the $Z$
boson to the leptons, $Z\, l^-_i l^+_j$.
These couplings, which contribute to
FCNC processes such as $\mu \rightarrow 3\, e$ \cite{bisset98}, are
proportional to $\sin^2 \xi$, and their effects are therefore extremely
difficult to observe experimentally. In particular, the flavour-violating
$Z$-boson decays $Z \rightarrow l^+_i l^-_j$ are suppressed by
$\sin^4 \xi$, well below the experimental upper limits,
$\mbox{BR}\, (Z \rightarrow l^+_i l^-_j) < (10^{-6} - 10^{-5})$~\cite{pdg04}.
  \item[{\it (iii)}] Finally, the couplings of neutralinos and charginos to
matter fields give rise, when written in terms of mass eigenstates, to
\Rp\ trilinear interactions proportional to $\sin \xi$
\cite{roy97}. The ones that originate from down-type higgsino couplings are
similar to interactions generated from the superpotential $\lambda$ and
$\lambda'$ couplings; they are suppressed both by the smallness of the Yukawa
couplings and by $\sin \xi$. The ones that originate from up-type higgsino or
gaugino couplings, on the other hand,
cannot arise from superpotential $\lambda$ or $\lambda'$ couplings.
Examples of such interactions are $\bar l_R d_L \widetilde{u}^\star_R$ and
$\bar \nu^c_R u_L \widetilde{u}^\star_R$.
\end{itemize}
Since all the above couplings are suppressed by at least one power of
$\sin \xi$, which is constrained to be very small by the cosmological bound
on neutrino masses, the corresponding experimental signatures are very
difficult to observe in practice, with a few exceptions like the decays
$\widetilde \chi^0_1 \rightarrow l_i f \bar f'$ when the lightest neutralino
$\widetilde \chi^0_1$ is the LSP.

Finally, bilinear $R$-parity violation introduces a mixing in the scalar
sector between the  Higgs bosons and the sleptons, 
\index{Mixing!higgs--slepton} which leads to \Rp\
decay modes of these scalars, such as
$h, H \rightarrow \widetilde \chi^0\, \nu,\, \chi^+\, l^-$ for neutral
$C P$-even Higgs bosons~\cite{davidson97,campos95,diazrom98},
or $\tilde \tau_1 \rightarrow l^- \nu,\, q \bar q'$ for the lightest stau~\cite{hprv02}.

\section{Spontaneous Breaking of {\boldmath{$R$}}-Parity}
\label{sec:spontaneous}

Spontaneous breaking of $R$-parity has been considered as an interesting
alternative to explicit $R$-parity breaking, because of its predictivity and
potentially rich phenomenology. Also, if spontaneous $R$-parity breaking
occurs below a few \TeV, the strong cosmological bounds on trilinear \Rp\
couplings associated with the requirement that \Rp\ interactions do not
erase any primordial baryon asymmetry (see section \ref{secxxx6c})
can be evaded.

The simplest possibility to break $R$-parity spontaneously is to give a
vacuum expectation value to a sneutrino
field~\cite{sneutrino_vev_1,ellis85,ross85},
e.g. $<\widetilde \nu_{\tau}>\, \neq 0$ (i.e. $v_3 \neq 0$ with our previous
notations). This may occur since the squared masses of the sneutrinos receive
negative contributions both at tree-level from the $D$-terms and from
radiative corrections. Due to the larger
third family Yukawa couplings, radiative corrections are expected to generate
a \VEV\ for the tau sneutrino only. However, since the conservation of lepton
number is associated with a global symmetry, its spontaneous breaking
would give rise to a massless Goldstone boson $J$, called
the \index{Majoron}Majoron~\cite{chikashige,gelmini},
together with a scalar $\rho$ having a mass of the order of
$<\widetilde \nu_{\tau}>$. The decay mode\footnote{Several constraints on
$<\widetilde \nu_{\tau}>$, e.g. the LEP limit on the tau neutrino mass which
requires $<\widetilde \nu_{\tau}>\, \lesssim 2 \GeV$
(as can be seen by adapting formulae (\ref{eq:m_nu_tau}) and (\ref{eq:m_0})
to the case $\mu_i = 0$, $v_3 \neq 0$), or the even stronger bound
$<\widetilde \nu_{\tau}>\, \lesssim 100 \keV$ coming from stellar
energy loss through Majoron\index{Majoron} emission in Compton scattering processes
$e \gamma \rightarrow e J$ \cite{Majoron_cooling}, ensure that
$m_{\rho} < M_Z$, so that this decay mode is indeed kinematically allowed.}
$Z \rightarrow J\, \rho$
would then contribute to the invisible decay width of the $Z$ boson like half
a neutrino flavour, which is excluded by experimental data.
There are several ways out: (i) introduce some amount of explicit lepton
number breaking into the MSSM, so as to give to the pseudo-Majoron a mass
$m_J > M_Z$ \cite{comelli94}; (ii) enlarge the gauge group to
include lepton number, so that the would-be Majoron\index{Majoron} becomes the longitudinal
degree of freedom of a new gauge boson~\cite{spontaneous_gauge}; (iii) break
$R$-parity through the \VEV\ of a right-handed sneutrino,
$<\widetilde \nu^c_{\tau}>\, \neq 0$, so that the Majoron\index{Majoron} is mainly an
electroweak singlet, and does not contribute sizeably to the $Z$ decay 
width~\cite{masiero90,romao92,spontaneous_nu_R_2}.

Let us illustrate approach (iii) by giving the main features of the model of
Ref.~\cite{masiero90}. The model contains, beyond the MSSM
superfields, the following $SU(2) \times U(1)$\index{Group symmetries!$SU(2)_L \times U(1)_Y$} 
singlets: three right-handed neutrinos $N^c_i$, three singlets $S_i$ with
lepton number $L=1$, and an additional singlet $\Phi$ whose r\^ole is to
generate the $\mu$-term -- as in the ``NMSSM'' (Next to Minimal \SSM).
The \index{Superpotential!spontaneous $R_p$ breaking}superpotential 
contains, beyond the quark and charged lepton Yukawa couplings, 
the following cubic terms~\cite{spontaneous_nu_R_radiative}:
\begin{equation}
  W\ =\ h_0\, H_u H_d \Phi\ +\ \lambda_{\Phi}\, \Phi^3\
  +\ h_{\nu ij}\, H_u L_i N^c_j\ +\ h_{ij}\, S_i N^c_j \Phi
\label{eq:W_spontaneous}
\end{equation}
This superpotential preserves both $R$-parity and total lepton number.
$R$-parity is spontaneously broken provided that the lepton singlets acquire
vacuum expectation values (in the following, we restrict ourselves to the
one-family case):
\begin{equation}
  <\widetilde \nu_{R \tau}>\, \ \equiv\ \frac{v_R}{\sqrt{2}}\ , \hskip .5cm
   <\widetilde S_{\tau}>\, \ \equiv\ \frac{v_S}{\sqrt{2}}\ .
\end{equation}
The most general minimum also involves, together with the Higgs \VEVs\ 
$v_u / \sqrt{2}$ and $v_d / \sqrt{2}$ required to break the electroweak 
symmetry, a sneutrino \VEV\ $<\widetilde \nu_{L \tau}>\, \equiv v_L / \sqrt{2}$,
assumed to be small. The Majoron\index{Majoron} $J$ is given by the imaginary part of the linear
combination:
\begin{equation}
  \frac{1}{\sqrt{v^2_R + v^2_S}}\ \left[\ \frac{v^2_L}{v^2}\:
  \left(\, v_u h_u\, -\, v_d h_d\, \right)\,
  +\, v_L\, \widetilde \nu_{\tau}\, +\, v_R\, \widetilde \nu^c_{\tau}\,
  +\, v_S\, \widetilde S_{\tau}\ \right] \, .
\end{equation}
The most stringent constraint on $J$ comes from astrophysics: in order to
avoid a too large stellar energy loss via Majoron\index{Majoron} emission, one must 
impose~\cite{masiero90} $v^2_L / v_R m_W \lesssim 10^{-7}$, 
i.e. $v_L \lesssim 100 \MeV$ for $v_R \sim 1 \TeV$. 
This small value of $v_L$ ensures that $J$ couples only very weakly to the 
$Z$ boson, and therefore does not affect its invisible decay width. 
Typical values for a viable $R_p$-breaking minimum of the scalar 
potential are $10 \GeV \lesssim v_R, v_S, v_{\Phi} \lesssim 1 \TeV$
and $v_L \lesssim (10-100) \MeV$. 
The hierarchy $v_L \ll v_R$ can be understood in terms of small 
Yukawa couplings, since $v_L$ is found to be proportional to the neutrino
Yukawa coupling constants $h_{\nu ij}$.

As can be seen from Eq. (\ref{eq:W_spontaneous}), a nonzero $v_R$ generates 
effective superpotential bilinear terms, $\mu_i = h_{\nu i3} v_R / \sqrt{2}$.
As a result, upon
a redefinition of the Higgs superfield $H_d$, small trilinear couplings
$\lambda$ and $\lambda'$ are generated with the same flavour structure
as in the previous section, and \Rp\ effects are induced in gauge and matter
interactions of neutralinos and charginos, as well as in slepton and Higgs
decays. The magnitude of these effects is related to the value of the heaviest
neutrino mass. However, there is a noticeable difference with the explicit
bilinear $R$-parity breaking discussed in the previous section, due to the
presence of the \index{Majoron}Majoron. This results in
new interactions, such as chargino and (invisible) neutralino decays
$\widetilde \chi^{\pm} \rightarrow \tau^{\pm} J$ and
$\widetilde \chi^0 \rightarrow \nu_i J$, invisible decay of the lightest
Higgs boson $h \rightarrow J J$ (which may be sizeable~\cite{invisible_Higgs}),
flavour-violating decays of charged leptons $e_i \rightarrow e_j J$, and
tau neutrino annihilation $\nu_{\tau} \nu_{\tau} \rightarrow J J$ and decay
$\nu_{\tau} \rightarrow \nu_{\mu} J$. It has been argued that the latter
processes could be large enough
to relax the energy density and nucleosynthesis constraints on the
tau neutrino mass, and allow it to be as large as the LEP limit of
$18.2$ MeV \cite{spontaneous_neutrino}. However such a possibility, quite
popular at a time where atmospheric neutrino oscillations were not
established on a very solid basis, does not look very appealing today since
it fails to accommodate both solar and atmospheric neutrino oscillations.

The feasibility of spontaneous \Rp\ along the lines of Ref. \cite{masiero90}
has been investigated by several authors; it has in particular been shown that
spontaneous $R$-parity breaking could be induced radiatively together with
electroweak symmetry breaking~\cite{spontaneous_nu_R_radiative}. 
Numerous studies of the experimental signatures of spontaneous \Rp\ can be 
found in the literature, see e.g. Refs.~\cite{invisible_Higgs,gonzalez90,
decampos99}.

\section{Constraining {\boldmath{\Rp}} Couplings from Flavour Symmetries}
\label{sec:flavour}

The trilinear terms in the \Rp\ superpotential of Eq. (\ref{eq:W_Rp_odd}) are
very similar to those associated with the quark and charged lepton Yukawa
couplings in Eq. (\ref{eq:W_Rp_even}), known from the pattern of
fermion masses and mixing angles to have a rather hierarchical structure. 
It is conceivable that the mechanism at the origin of this hierarchy 
also provides a hierarchical structure for \Rp\ couplings.

A possible simple explanation for the fermion mass hierarchy has been
provided long ago by Froggatt and Nielsen, who postulated the existence of
a spontaneously broken, flavour-dependent abelian symmetry. 
In this section, we show that such a symmetry may also naturally
generate a flavour  hierarchy between \Rp\ couplings
\cite{dreiner04,banks95,binetruy98,borzumatti96,BenHamo,binetruy96,ellis98}. 
For the case of a non-abelian flavour symmetry, see e.g. Ref.~\cite{bhatta98}.

Let us first explain how a family-dependent symmetry 
$U(1)_X$\index{Group symmetries!Family-dependent $U(1)_X$} 
constrains the Yukawa sector~\cite{binetruy96,froggatt79}. 
Consider a Yukawa coupling $H_u Q_i U^c_j$ and let us denote generically the
$X$-charge of a superfield $\Phi_i$ by the corresponding small letter 
$\phi_i$. Invariance under 
$U(1)_X$\index{Group symmetries!Family-dependent $U(1)_X$} implies 
that $H_u Q_i U^c_j$ appears in the superpotential only if its $X$-charge
vanishes, i.e. $q_i+u_j+h_u=0$.
To account for the large top quark mass, one assumes that this happens only
for the Yukawa coupling $H_u Q_3 U^c_3$; thus all fermions but the top quark
are massless before the breaking of this symmetry.
One further assumes that the flavour symmetry is broken by the \VEV\ 
of a \SM\ singlet $\theta$ with $X$-charge $-1$, and that the other Yukawa
couplings are generated from (gauge-invariant) interactions of the form
\begin{equation}
  y^u_{ij}\ H_u Q_i U^c_j \, \left( \frac{\theta}{M} \right)^{q_i+u_j+h_u}\ ,
\label{eq:nonrenormalizable}
\end{equation}
where $M$ is a mass scale, $y^u_{ij}$ is an unconstrained coupling of
order one, and $q_i+u_j+h_u>0$. Such non-renormalizable terms typically appear
in the low-energy effective field  theory of a fundamental theory with heavy
fermions of mass $M$ (one may also think of a string theory, in which case
$M \sim M_P$). If $U(1)_X$\index{Group symmetries!Family-dependent $U(1)_X$} 
is indeed broken below the scale $M$, $\epsilon =\, <\theta> / M$ is a
small parameter,  and (\ref{eq:nonrenormalizable}) generates an effective
Yukawa coupling
\begin{equation}
  \lambda^u_{ij}\ =\  y^u_{ij}\ \epsilon^{\, q_i+u_j+h_u} \ ,
\label{eq:Yu}
\end{equation}
whose order of magnitude is fixed by the values of the $X$-charges. Similarly
one has, for down quarks and charged leptons:
\begin{equation}
  \lambda^d_{ij}\ \sim\ \epsilon^{\, q_i+d_j+h_d}\ , \qquad
  \lambda^e_{ij}\ \sim\ \epsilon^{\, l_i+e_j+h_d}\ . 
\label{eq:Yd}
\end{equation}
Such a family-dependent symmetry thus naturally yields a hierarchy between
Yukawa couplings, and therefore fermion masses.
\vskip .3cm

For example, the charge assignment $q_1-q_3=3$, $q_2-q_3=2$, $u_1-u_3=5$,
$u_2-u_3=2$, $d_1-d_3=1$, $d_2-d_3=0$ yields quark Yukawa matrices of the
form:
\begin{equation} 
\lambda^u\ \sim\ 
\left(
\begin{array}{ccc} 
\epsilon^8 & \epsilon^5 & \epsilon^3 \\ 
\epsilon^7 & \epsilon^4 & \epsilon^2 \\
\epsilon^5 & \epsilon^2 & 1
\end{array}
\right)\ , \qquad 
\lambda^d\ \sim\ \epsilon^{q_3+d_3+h_d}
\left(
\begin{array}{ccc} 
\epsilon^4 & \epsilon^3 & \epsilon^3 \\ 
\epsilon^3 & \epsilon^2 & \epsilon^2 \\
\epsilon & 1 & 1
\end{array}
\right) \ ,
\label{eq:FN_matrices}
\end{equation}
where the symbol $\sim$ indicates that the entries are known up to factors
of order one only. Eq. (\ref{eq:FN_matrices}) holds at the scale at which
the abelian symmetry is spontaneously broken, usually taken to be
close to the Planck scale. With renormalisation group effects down to
the weak scale taken into account, these Yukawa matrices can accommodate the
observed quark masses and mixings if the small number $\epsilon$ is of the
order of the Cabibbo angle, i.e. $\epsilon \approx V_{us} \simeq 0.22$.
More generally, assuming that the $X$-charge associated with each Yukawa 
coupling is positive, only a few structures for $\lambda^u$ and $\lambda^d$,
which differ from Eq. (\ref{eq:FN_matrices}) by a $\pm 1$ change in the powers
of $\epsilon$, are allowed by the data. In the lepton sector
there is more freedom, as long as a mechanism for generating neutrino masses
is not specified. The combination of $X$-charges $q_3+d_3+h_d$,
related to the value of $\tan \beta$ by $m_t / m_b \sim \tan \beta\,
\epsilon^{-(q_3+d_3+h_d)}$, is actually constrained if one imposes gauge
anomaly cancellation conditions.
\vskip .3cm

\Rp\ couplings are then constrained by 
$U(1)_X$\index{Group symmetries!Family-dependent $U(1)_X$} exactly
as for Yukawa couplings. They are generated from the following
non-renormalizable superpotential terms:
\begin{equation}
L_i L_j E^c_k \left( \frac{\theta}{M} \right)^{l_i+l_j+e_k}\ , \
  \hskip .5cm  L_i Q_j D^c_k \left( \frac{\theta}{M} \right)^{l_i+q_j+d_k}
 \ . \label{eq:Rp_NR}
\end{equation}
To avoid unnaturally large values of the quark $X$-charges, we have assumed a
baryon parity that forbids the baryon-number-violating superpotential terms
$U^c D^c D^c$ as well as the dangerous dimension-5 operators discussed
in Section~\ref{sec:alternatives}, thus preventing proton decay.
One can see from Eq. (\ref{eq:Rp_NR}) that abelian
flavour symmetries yield a hierarchy between \Rp\ couplings 
that mimics (in order of magnitude) the down quark and charged lepton mass
hierarchies. Indeed, one has:
\begin{equation}
  \lambda_{ijk}\ \sim\ \epsilon^{\, l_i - h_d}\ \lambda^e_{jk}\ , \qquad
  \lambda'_{ijk}\ \sim\ \epsilon^{\, l_i - h_d}\ \lambda^d_{jk}\ .
\label{eq:lambda_lambda'}
\end{equation}
Provided that the Yukawa matrices $\lambda^d$ and $\lambda^e$ are known,
experimental limits on $\lambda$ and $\lambda'$ can be translated into a
constraint on $l_i - h_d$. We shall assume here that the $X$-charge carried by
each operator is positive, and take for $\lambda^d$ the structure of Eq.
(\ref{eq:FN_matrices}). In the lepton sector, the $\lambda^e_{ij}$ are less
constrained; however, it is possible to derive upper bounds on the 
$\lambda_{ijk}$ couplings from the three charged lepton masses.
Assuming a small value of $\tan \beta$ (corresponding to $q_3+d_3+h_d=3$), 
one finds that the experimental bounds on coupling products (including
the limit coming from $\epsilon_K$, $\mid \Im \left(
\lambda'_{i12}\ \lambda^{\prime \star}_{i21} \right)\! \mid\,
\leq 8 \times 10^{-12}$)
are satisfied as soon as~\cite{dudas98}:
\begin{equation}
  l_i-h_d\ \geq\ 2 - 3 \ . 
\label{eq:charge_constraint}
\end{equation}
For moderate or large values of $\tan \beta$, larger values of the $X$-charges
would be required.

The condition (\ref{eq:charge_constraint}) can now be used, together with Eq.
(\ref{eq:lambda_lambda'}), to derive, in the framework considered,
$\tan \beta$-independent upper bounds on the individual couplings
$\lambda$ and $\lambda'$. All of them
are well below the experimental limits. Thus, if abelian flavour symmetries
are responsible for the observed fermion mass spectrum along the lines
discussed above, one expects the first signals for broken $R$-parity to come
from FCNC processes~\cite{dudas98}. These conclusions, however, are not
completely generic for abelian flavour symmetries:
they would be modified if 
$U(1)_X$\index{Group symmetries!Family-dependent $U(1)_X$} were broken by a 
vector-like pair of singlets~\cite{ellis98}, or if we gave up the assumption
that the $X$-charge associated with each operator is positive.

Let us now consider the most general scenario of $R$-parity violation, with
both bilinear and trilinear \Rp\ couplings (see section \ref{sec:bilinear}
for the notations used in the following, where we closely follow the discussion
of Ref.~\cite{binetruy98}). Assuming that the bilinear terms
are generated through \SUSY\ breaking~\cite{giudice88} (which ensures that
the $\mu_{\alpha}$ are of the order of the weak scale, as required by
electroweak symmetry breaking), one finds:
\begin{equation}
  \mu_{\alpha}\ \sim\ \tilde{m}\, \epsilon^{\, \tilde{l}_{\alpha}} \ , 
        \hskip 1cm  v_{\alpha}\ \sim\ v_d\, \epsilon^{\,
        \tilde{l}_{\alpha} - \tilde{l}_0} \ , 
\end{equation}
where $\tilde{m}$ is the typical mass scale associated with the soft
\SUSY-breaking terms, $\tilde{l}_{\alpha}\ \equiv\ |l_{\alpha}+h_u|$, 
and the above estimates are valid for 
$0 \leq \tilde{l}_0 < \tilde{l}_i$, $i=1,2,3$. 
Thus the $v_{\alpha}$ are approximatively aligned along the $\mu_{\alpha}$ by
the flavour symmetry~\cite{banks95}, which implies (assuming with no loss of
generality $\tilde{l}_3 \leq \tilde{l}_{1,2}$):
\begin{equation}
  \sin^2 \xi\ \sim\ \epsilon^{\, 2\, (\tilde{l}_3 - \tilde{l}_0)} \ .
\label{eq:sin_xi}
\end{equation}
Furthermore, the redefinition (\ref{eq:redefinition}) is completely fixed by 
requiring $L_1 \simeq \hat{L}_1$ and $L_2 \simeq \hat{L}_2$, with
\begin{equation}
  \frac{v_{\alpha}}{v_d}\ \sim\ \epsilon^{\, \tilde{l}_{\alpha} -
        \tilde{l}_0}\ \ , \hskip 1cm\  e_{\alpha i}\ \sim\ \epsilon^{\,
        |\tilde{l}_{\alpha} - \tilde{l}_i|} \ . 
\end{equation}
Note that $H_d \simeq \hat{L_0}$, which allows us to define $h_d \equiv l_0$.

The low-energy \Rp\ couplings depend on the signs of the charges 
$l_{\alpha}+h_u$.
In all phenomenologically viable cases, the order of magnitude 
relations (\ref{eq:lambda_lambda'}) are modified to:
\begin{equation}
  \lambda_{ijk}\ \sim\ \epsilon^{\, \tilde{l}_i - \tilde{l}_0}\
   \lambda^e_{jk}\ , \qquad
  \lambda'_{ijk}\ \sim\ \epsilon^{\, \tilde{l}_i - \tilde{l}_0}\
    \lambda^d_{jk}\ .
\label{eq:lambda_lambda'_bis}
\end{equation}
By combining Eqs. (\ref{eq:m_nu_tau}), (\ref{eq:sin_xi}) 
and (\ref{eq:lambda_lambda'_bis}), we can write down a relation between the
mass of the tau neutrino, the \Rp\ couplings $\lambda'_{3jk}$ and the
down quark Yukawa couplings ($m_0$ is defined in Eq. (\ref{eq:m_0_estimate})),
\begin{equation}
 m_{\nu_3}\ \sim\ m_0 \left(\frac{\lambda'_{3jk}}{\lambda_{jk}^d}\right)^2\ ,
\label{eq:m_nu3_relation}
\end{equation}
which is a generic prediction of this class of models. Let us stress however
that, in Eqs. (\ref{eq:sin_xi}) and (\ref{eq:m_nu3_relation}), we have assumed
that the suppression of the misalignment angle $\xi$ is only due to the
abelian flavour symmetry. In this case the cosmological bound on neutrino
masses, $m_{\nu} \leq {\cal O} (1\, \mbox{eV})$, requires very large values
of the $\tilde{l}_{\alpha}$, so that $R$-parity violation should be very
suppressed and in practice unobservable. This conclusion can be evaded only
if some other mechanism provides the required alignment between the
$v_{\alpha}$ and the $\mu_{\alpha}$, in which case Eqs. (\ref{eq:sin_xi}) and
(\ref{eq:m_nu3_relation}) are no longer valid.

Let us now concentrate on the case $l_i+h_u \geq 0 > l_0+h_u$, which leads to
an enhancement of flavour-diagonal couplings relative to off-diagonal
couplings.  Indeed, the dominant terms in  Eq. (\ref{eq:redefined_lambda'})
correspond to $\alpha=0$ or $\beta=0$, which provides an alignment of the \Rp\
couplings along the Yukawa couplings:
\begin{equation}
  \lambda_{ijk}\ \simeq\ \left( e_{0i}\, \lambda^e_{jk}\, -\,
        e_{0j}\, \lambda^e_{ik} \right)\ , \qquad
  \lambda'_{ijk}\ \simeq\ e_{0i}\, \lambda^d_{jk}\ . 
\end{equation}
As a consequence, \Rp\ couplings are almost diagonal in the basis of fermion
mass eigenstates. Furthermore, they undergo an enhancement relative to the
naive power counting, since e.g.
\begin{equation}
  \lambda'_{ijk}\ \sim\ \epsilon^{\, \tilde{l_i}-\tilde{l_0}}\ \lambda^d_{jk}\
  \sim\ \epsilon^{\, -2\, \tilde{l_0}}\ \epsilon^{\, l_i+q_j+d_k} \ . 
\end{equation}
This opens the phenomenologically interesting possibility that $R$-parity
violation be sizeable while its contribution to FCNC processes is suppressed,
as required by experimental data -- provided however that the misalignment
angle $\xi$ is reduced to phenomenologically acceptable values by some
other mechanism than the suppression by large lepton $X$-charges.

Abelian flavour symmetries can also play a useful r\^ole in controlling the
proton decay rate in supersymmetric theories with unbroken $R$-parity.
Indeed, they can suppress the coefficients of the dangerous dimension-5
operators\footnote{Of course, this statement also holds in supersymmetric
theories without $R$-parity~\cite{BenHamo}, but in this case one may find
more appealing to invoke a discrete symmetry forbidding baryon
number violation from both dimension-4 and dimension-5 operators
(see Section~\ref{sec:alternatives}), as we did in the above discussion.}
$QQQL$ and $U^cU^cD^cE^c$~\cite{BenHamo,kaplan94}, which preserve
$R$-parity but violate baryon number and lepton number,
and are expected to be generated from unknown
Planck-scale physics (see Section~\ref{sec:alternatives}).
Flavour symmetry models that explain the fermion mass hierarchy and are compatible
with the experimental lower limit on the proton lifetime generally predict proton decay
close to the present experimental sensitivity~\cite{dreiner04,kaplan94,irges98}.

\section{{\boldmath{$R$}}-Parity Violation in \GUT\ }
\label{sec:GUTs}

Up to now we discussed $R$-Parity Violation in the framework of the \SSM\
(possibly augmented by some flavour symmetries). More sophisticated \susyq\
extensions of the \SM\ may lead to a different structure of \Rp\ couplings.
In particular, nontrivial constraints on the allowed \Rp\ couplings generally
result from an enlarged gauge structure, as in \GUT\ 
(see e.g. Refs.
\cite{hall84,hempfling96,smirnov96,sakai,brahm,tamvakis96,barbieri97,giudice97}).

In \GUT\ based on the $SU(5)$ gauge group, all trilinear \Rp\
superpotential couplings originate, at the renormalizable level, from
the same operator \cite{sakai}
\begin{equation}
  \frac{1}{2}\, \Lambda_{ijk}\: {\bf \bar 5}_i {\bf \bar 5}_j {\bf 10}_k\ ,
\end{equation}
antisymmetric under the exchange of ${\bf \bar 5}_i$ and
${\bf \bar 5}_j$, where the antifundamental representation ${\bf \bar 5}_i$
contains the $L_i$ and $D^c_i$ superfields, while the antisymmetric
representation ${\bf 10}_i$ contains the $Q_i$, $U^c_i$ and $E^c_i$
superfields\footnote{Ordinary quark and charged lepton Yukawa couplings are 
  generated, at the renormalizable level, from the superpotential terms 
  $\Lambda^d_{ij} {\bf 10}_i {\bf \bar 5}_j {\bf \bar 5}_d
   + \Lambda^u_{ij} {\bf 10}_i {\bf 10}_j {\bf 5}_u$,
  where the representations ${\bf \bar 5}_d$ and ${\bf 5}_u$ contain the
  doublet Higgs superfields $H_d$ and $H_u$, respectively. The first term
  leads to the relation $\lambda^d_{ij} = \lambda^e_{ji}$, which even after
  taking into account the renormalization effects is in gross contradiction
  with the measured fermion masses, and must be corrected by terms
  (renormalizable or not) involving higher-dimensional Higgs 
  representations.}. As a result, all three types of trilinear \Rp\ couplings
are simultaneouly present or absent, and related by 
\begin{equation}
  \lambda_{ijk}\ =\ \frac{1}{2}\,  \lambda'_{ikj}\ =\ \lambda''_{kij} \, ,
\label{eq:SU(5)_relations}
\end{equation}
with the resulting antisymmetry of the $\lambda'$ couplings, 
$\lambda'_{ijk} = - \lambda'_{kji}$.
We are then left with 9 superpotential couplings, as well as the 9 
corresponding $A$-terms satisfying the same relation. 
Since both $\lambda'$ and $\lambda''$ couplings are simultaneously present,
the experimental bound on the proton lifetime severely constrains
these couplings, with 
$|\Lambda_{ij1}|, |\Lambda_{123}| \lesssim 2 \times 10^{-13}$ at $M_{GUT}$
\cite{smirnov96}, where the constraint on the $\Lambda_{ij1}$ comes from
the familiar ($B-L$)-conserving operators, while the constraint on
$\Lambda_{123}$ comes from ($B+L$)-conserving operators generated by
diagrams involving a left-right mixing mass insertion of third generation
squarks~\cite{lee84}. 
The other $\Lambda_{ijk}$ couplings do not induce proton decay at tree 
level, but can contribute at the one-loop level, and must therefore satisfy 
$|\Lambda_{ijk} (M_{GUT})| \lesssim 3 \times 10^{-9}$ \,
\cite{smirnov96}.

However Eq. (\ref{eq:SU(5)_relations}) only applies
when the ${\bf \bar 5}_i {\bf \bar 5}_j {\bf 10}_k$ operator arises at the
renormalizable level, i.e. when the $\Lambda_{ijk}$ are field independent.
On the contrary when these couplings are induced by \VEVs\ responsible for 
GUT symmetry breaking, a different pattern for trilinear $R$-parity breaking
can be obtained (see e.g. Ref. \cite{tamvakis96} for a model leading to
$\lambda''_{ijk}$ couplings only, and Refs. \cite{barbieri97,giudice97}
for models leading to $\lambda'_{ijk}$ couplings only). $SU(5)$ \GUT\ also
potentially contain a bilinear \Rp\ superpotential operator
\begin{equation}
  M_i\, {\bf 5}_u {\bf \bar 5}_i\ ,
\label{eq:SU(5)_bilinear}
\end{equation}
where ${\bf 5}_u$ contains both the usual doublet Higgs superfield $H_u$ and
a (superheavy) Higgs colour triplet $T_u$. This yields, together with the
usual $H_u L_i$ terms, a baryon-number-violating term $T_u D^c_i$. While the
former is a source for $\lambda$ and $\lambda'$ couplings, the latter
is a source for $\lambda''$ couplings, which are the only surviving \BV\
couplings at low energy. However these are small if the ``doublet-triplet
splitting problem'' is solved (with
$m_{H_u} \sim M_{weak} \ll m_{T_u} \sim M_{GUT}$) and $M_i \lesssim M_{weak}$
\cite{hall84}. Then the effective low energy \BV\ couplings are suppressed
relative to \LV\ couplings by a typical factor of $M_{weak} / M_{GUT}$, and
one ends up with an approximate conservation of baryon number.

It is natural to ask whether one can avoid the generation of \Rp\ couplings
-- without imposing $R$-parity or any other global symmetry -- by considering
a unification group larger than $SU(5)$. It is well known that one of the
generators of the $SO(10)$ group\index{Group symmetries!$SO(10)$} acts as
$B-L$ on the \SSM\ fields,
with the matter superfields embedded into the spinorial representation
${\bf 16}_i$ (which decomposes under $SU(5)$ as
${\bf 16}_i = {\bf 10}_i \oplus {\bf \bar 5}_i \oplus {\bf 1}_i$, where
${\bf 1}_i$ corresponds to a right-handed neutrino), and the Higgs
doublet superfields embedded into the representation
${\bf 10} = {\bf 5} \oplus {\bf \bar 5}$.
Since $R$-parity can be rewritten as $(-)^{2S+3(B-L)}$, it follows that
\Rp\ operators are forbidden by the 
$SO(10)$\index{Group symmetries!$SO(10)$} gauge symmetry 
(at least as long as it is unbroken).
For instance, with the above field content,
the only cubic operator compatible with the $SO(10)$
gauge symmetry takes the form ${\bf 10}\, {\bf 16}\, {\bf 16}$ and 
preserves $R$-parity.

Still the process of gauge symmetry breaking down to
$SU(3)_C \times SU(2)_L \times U(1)_Y$
\index{Group symmetries!Standard Model}
may lead to a theory that does not conserve $R$-parity.
The question of whether $R$-parity remains exact
or not and of what types of \Rp\ couplings are generated crucially depends 
on the breaking scheme (for a discussion, see e.g. \cite{mohapatra98}).
If the $B-L$ symmetry is spontaneously broken by
the \VEV\ of (the right-handed neutrino-like component of) a Higgs boson in
the spinorial representation ${\bf 16}_H$, then $R$-parity gets broken in
the low-energy effective theory. More precisely, the renormalizable operator
\begin{equation}
  {\bf 10}_u {\bf 16}_H {\bf 16}_i\ ,
\label{eq:SO(10)_bilinear}
\end{equation}
%
gives rise, through $<\!\! {\bf 16}_H\!\! >\, \neq 0$, to both
baryon number and lepton-number-violating bilinear terms, contained in
the $SU(5)$ \Rp\ operator ${\bf 5}_u {\bf \bar 5}_i$. These are especially
dangerous since $<\!\! {\bf 16}_H\!\! >$ is generally much larger
than $M_{weak}$. Trilinear \Rp\ couplings are generated from
higher-dimensional operators like
\begin{equation}
  {\bf 16}_i {\bf 16}_j {\bf 16}_k {\bf 16}_H\ .
\label{eq:SO(10)_trilinear}
\end{equation}
If present, these operators generate trilinear \Rp\ couplings $\lambda$,
$\lambda'$ and $\lambda''$ satisfying the $SU(5)$ relations
(\ref{eq:SU(5)_relations}).
Alternatively, the $B-L$ symmetry can be broken by the \VEV\ of a Higgs
boson in the ${\bf 126}$ representation. In this case the selection rule
$\Delta (B-L) = 2$ holds, and $R$-parity is automatically conserved.

The breaking of $SO(10)$\index{Group symmetries!$SO(10)$} 
into an intermediate $SU(5)$\index{Group symmetries!$SU(5)$}
requires a ${\bf 16}_H$,
while if $SO(10)$ first breaks into a 
left-right symmetric\index{Group symmetries!Left-right symmetric} 
gauge group $SU(4)_C \times SU(2)_L \times SU(2)_R$ or
$SU(3)_C \times SU(2)_L \times SU(2)_R \times U(1)_{B-L}$, the breaking of
the $B-L$ gauge symmetry involves either a ${\bf 16}_H$ or a ${\bf 126}_H$.
(Actually supersymmetry demands a vector-like pair of Higgs representations,
e.g. ${\bf 16}_H \oplus {\bf \overline{16}}_H$.) We conclude
that the $SO(10)$ breaking chain with an intermediate $SU(5)$
gauge symmetry, or with an intermediate left-right symmetry spontaneously
broken by a ${\bf 16}_H$, leads to unacceptably large \Rp\ couplings,
unless the dangerous operators (\ref{eq:SO(10)_bilinear}) and
(\ref{eq:SO(10)_trilinear}) are forbidden by some symmetry or strongly
suppressed by some mechanism.

\section{\sloppy Restrictions on {\boldmath{$R$}}-Parity Violations from
Generalized Matter, Baryon or Lepton Parities}
\label{sec:alternatives} 

As already alluded to before in this chapter, there exist discrete
symmetries that can protect proton decay from renormalizable
operators as efficiently as $R$-parity, while allowing for some \Rp\ couplings.
Such symmetries therefore provide viable patterns of $R$-parity violations,
and it is useful to classify them, as we do in this section, following the discussion
of Ref.~\cite{ibanez92}.

Let us first note that $R$-parity actually does not forbid all dangerous
baryon-number-\index{Baryon number} and 
lepton-number-violating\index{Lepton number} couplings.
Indeed, it is highly probable that the \SSM\ is just an
effective theory, to be embedded in a more fundamental theory including
quantum gravity at some high energy scale $\Lambda$. Gauge-invariant,
higher-dimensional non-renormalizable operators are then generated in the
low-energy theory by integrating out massive particles. In particular the
effective superpotential is expected to contain the following quartic terms:
\index{Superpotential!alternative $B$ and $L$ violation}
\begin{eqnarray}
 \hskip -.5cm W_{n.r.}\ \ \ni\ \
 \frac{(\kappa_1)_{ijkl}}{\Lambda } (Q_iQ_j) (Q_kL_l)\
 +\ \frac{(\kappa_2)_{ijkl}}{\Lambda } (U^c_iU^c_j D^c_k) E^c_l\
 +\ \frac{(\kappa_3)_{ijk}}{\Lambda } (Q_iQ_j) (Q_kH_d) \nonumber \\
 +\ \ \frac{(\kappa_4)_{ijk}}{\Lambda } (Q_iH_d) (U^c_jE^c_k)\
 +\ \frac{(\kappa_5)_{ij}}{\Lambda } (L_iH_u) (L_jH_u)\
 +\ \frac{(\kappa_6)_i}{\Lambda } (L_iH_u) (H_dH_u)\ , \label{2}
\end{eqnarray}
where $\Lambda$ can be viewed as parametrizing the scale of new physics
(e.g. the string scale, the Planck scale or the GUT scale) beyond the \SSM. 
Other non-renormalizable \BV\ and \LV\ operators can be present in the K\"ahler
potential $K$, the function that defines the matter kinetic terms in a
non-renormalizable \susyq\ theory (for example, the kinetic terms for
complex scalar fields $\phi_i$ read ${\cal L}_{kin} = \left( \partial^2 K /
\partial \phi_i \partial {\bar \phi}_j \right)
\partial_{\mu} \phi_i \partial^{\mu} \bar \phi_j$) \cite{ibanez92,addprd69}:
\begin{eqnarray}
  K_{n.r.} & \ni & 
  \frac{(\kappa_7)_{ijk}}{\Lambda} (Q_i Q_j)  D^{c \dagger}_k\ +\  
  \frac{(\kappa_8)_i}{\Lambda} (H_u^{\dagger} H_d)  E^c_i  \nonumber \\
  & & +\ \ \frac{(\kappa_9)_{ijk}}{\Lambda} (Q_i L^{\dagger}_j) U^c_k\ +\
  \frac{(\kappa_{10})_{ijk}}{\Lambda} (U^c_i D^{c \dagger}_j) E^c_k\ ,
\label{3}
\end{eqnarray}
where we have kept only the trilinear terms, which correspond to dimension-5
operators in the Lagrangian density.

In Eqs. (\ref{2}) and (\ref{3}), we assumed the particle content of the MSSM;
other versions of the \SSM\ may allow for additional terms. It is easy to see
that the operators parametrized by $\kappa_1$, $\kappa_2$ and $\kappa_5$, while
compatible with $R$-parity, still violate the
baryon number\index{Baryon number} and lepton number\index{Lepton number}
symmetries. The operator $LH_uLH_u$ generates Majorana masses
for the neutrinos; as long as $\Lambda \gtrsim 5 \times 10^{13}$ GeV, its contribution
is compatible with the experimental constraints on neutrino masses,
and the $(\kappa_5)_{ij}$ are not required to be small.
The operators $QQQL$ and $U^c U^c D^c E^c$, on the other hand, induce
proton decay and their coefficients $\kappa_1$ and $\kappa_2$ are therefore
constrained to be small, even if the cutoff scale $\Lambda$ is taken to be as
large as the Planck mass (i.e. $\Lambda \sim 10^{19} \GeV$). In particular,
$(\kappa_1)_{112l} \leq {\cal O} (10^{-7})$ for typical superpartner masses
in the TeV range. Other operators in Eqs. (\ref{2}) and (\ref{3}) induce
proton decay in conjunction with the dimension-4 \Rp\ operators of Eq.
(\ref{eq:W_Rp_odd}), implying severe bounds on the products
$\kappa_4 \lambda^{''}$, $\kappa_9 \lambda^{''}$ and
$\kappa_{10} \lambda^{''}$. This provides an additional motivation for
searching for discrete symmetries protecting proton decay from renormalizable
operators, while allowing for some of the \Rp\ terms in
Eq. (\ref{eq:W_Rp_odd}). Some of these symmetries will
eventually turn out to be more efficient than $R$-parity in forbidding the
dangerous \BV\ and \LV\ nonrenormalizable operators in Eqs. (\ref{2}) and
(\ref{3}).

We are interested in discrete symmetries of the $R$-parity conserving
superpotential of Eq. (\ref{eq:W_Rp_even}) (i.e. symmetries compatible
with the presence of the $\mu$-term and of quark and lepton Yukawa couplings)
protecting proton decay from renormalizable operators.
These symmetries, which are either commuting with \SUSY\ or $R$-symmetries
also acting on the supersymmetry generator, can be divided into three 
general classes~\cite{ibanez92}:

i) {\it Generalized matter ($R$-)parities} (GMP) :  discrete ($R$-)symmetries 
protecting both baryon and lepton number from dimension-4 operators, i.e.
enforcing $\lambda = \lambda' = \lambda'' = 0$;

ii) {\it Generalized baryon ($R$-)parities} (GBP) :  discrete ($R$-)symmetries
protecting baryon number from dimension-4 operators and allowing for 
lepton-number violations, i.e. enforcing $\lambda'' = 0$;

iii)  {\it Generalized lepton ($R$-)parities} (GLP) :  discrete ($R$-)
symmetries protecting lepton number from dimension-4 operators and allowing
for baryon-number violations, i.e. enforcing $\lambda = \lambda' = 0$.

For simplicity, we restrict our discussion to flavour-blind
symmetries (the case of flavour-dependent symmetries has been illustrated
in section \ref{sec:flavour}), and we assume the particle content of the MSSM.
To start with, let us consider \index{Discrete symmetries}discrete 
$Z_N$ symmetries commuting with \SUSY, which act on the chiral superfields
as ($k = 0 \cdots N-1$):
\begin{eqnarray}
 Q & \rightarrow & e^{2k\pi i \ q/N} \ Q \quad , \quad
 U^c\ \rightarrow\ e^{2k\pi i \ u/N} \ U^c \quad , \quad
 D^c\ \rightarrow\  e^{2k\pi i \ d/N} \ D^c \quad , \nonumber \\
 L & \rightarrow & e^{2k\pi i \ l/N} \ L \quad , \quad
 E^c\ \rightarrow\ e^{2k\pi i \ e/N} \ E^c \quad , \quad
 H_{u,d}\ \rightarrow\ e^{2k\pi i \ h_{u,d}/N} \ H_{u,d} \quad ,
\label{5} 
\end{eqnarray}
where $q, \cdots, h_d, h_u$ denote here the $Z_N$ charges of the MSSM
superfields, defined modulo $N$. Restricting the scan to 
$Z_2$\index{Discrete symmetries!$Z_2$} and 
$Z_3$\index{Discrete symmetries!$Z_3$} symmetries,
one finds the generalized 
parities\index{Discrete symmetries!Generalized parities} listed
in Table~\ref{tab:znsym} \cite{ibanez92}.

\begin{table}[h]
\begin{center}
\begin{tabular}{l|c|c|c|c|c|c|c}   \hline
{Generalized parities} \quad & \quad {$q$} \quad & \quad {$u$} \quad &
\quad {$d$} \quad & \quad
{$l$} \quad & \quad {$e$} \quad & \quad {$h_d$} \quad  & \quad {$h_u$}
\quad  \\ \hline\hline
\quad  $Z_2^M$ \quad , \quad $Z_3^M$ \quad &  \quad 0
\quad  & \quad -1 \quad & \quad 1 \quad & \quad 0 \quad & \quad
 1 \quad & \quad -1 \quad & \quad 1 \quad \\ \hline
\quad $Z_3^M$ &
\quad 0 \quad & \quad -1 \quad & \quad 1 \quad & \quad 1 \quad & \quad 
0 \quad & \quad -1 \quad  & \quad 1 \quad \\ \hline\hline
\quad $Z_2^L$ \quad , \quad $Z_3^L$ &
\quad 0 \quad & \quad 0 \quad & \quad 0 \quad & \quad -1 \quad  & \quad 
1 \quad & \quad 0 \quad & \quad 0 \quad \\ \hline
\quad $Z_2^B$ &
\quad 0 \quad & \quad -1 \quad & \quad 1 \quad & \quad -1 \quad  & \quad 
0 \quad & \quad -1 \quad  & \quad 1 \quad   \\ \hline\hline
\quad $Z_3^B$ &
\quad 0 \quad & \quad -1 \quad & \quad 1 \quad & \quad -1 \quad & \quad
-1 \quad & \quad -1 \quad & \quad 1 \\ 
\hline
\end{tabular}
\caption{{\it Matter, lepton and baryon generalized parities in the MSSM.
Superscript indices distinguish between matter (M), baryon (B) and lepton (L)
generalized parities. The letters $q$, $u$, $d$, $\cdots$ denote the $Z_N$
charges referring to the action of the generalized parities as expressed
in Eq. (\ref{5}).}}
\label{tab:znsym}
\end{center}
\end{table}

The $Z_2$\index{Discrete symmetries!$Z_2$} GMP ($Z^M_2$) is actually equivalent, after
a weak hypercharge\index{Hypercharge} rotation, to a symmetry $X$ acting 
on MSSM superfields as $X (Q_i,U^c_i,D^c_i,L_i, E^c_i)=
-(Q_i,U^c_i,D^c_i,L_i,E^c_i)$, $X (H_d,H_u)=+(H_d,H_u)$. This symmetry is
identical to the matter parity symmetry already considered in chapter
\ref{chap:intro}, and is therefore equivalent to $R$-parity.
Contrary to the other generalized parities listed in Table~\ref{tab:znsym},
the $Z_2$\index{Discrete symmetries!$Z_2$} and $Z_3$\index{Discrete symmetries!$Z_3$}
GBP's do not forbid the bilinear lepton number violating operators $L_i H_u$.
Some of the symmetries listed in Table~\ref{tab:znsym} are actually not better
than $R$-parity in forbidding the dangerous non-renormalizable dimension-5 operators
in Eqs. (\ref{2}) and (\ref{3}). Indeed, the first two GMP's (including the
usual matter parity) allow for such terms. 

A similar analysis can be done for discrete $R$-symmetries, i.e. discrete
versions of the continuous $R$-symmetries discussed in chapter
\ref{chap:intro}.
Since $R$-symmetries do not commute with \SUSY,
the transformations (\ref{5})
must be supplemented with a corresponding action on Grassmann variables 
$\theta$, which defines also the total charge of the superpotential (see
Eq. (\ref{eq:R_weight})),
\begin{equation}
 W (x, \theta)\ \rightarrow\ e^{- 2k\pi i/N}\
 W (x, e^{k\pi i/N} \theta)\ .
\label{6} 
\end{equation}
Scanning over all flavour-blind $Z_2$\index{Discrete symmetries!$Z_2$} and 
$Z_3$\index{Discrete symmetries!$Z_3$} discrete $R$-symmetries, one
finds the generalized 
$R$-parities~\cite{ibanez92}\index{Discrete symmetries!Generalized $R$-parity}
listed in Table~\ref{tab:rsymm}.
Like the generalized parities of Table~\ref{tab:znsym}, the generalized
$R$-parities do not necessarily forbid all dangerous dimension-5 operators.
Actually only one generalized matter $R$-parity satisfies the requirement
of forbidding all operators in Eqs. (\ref{2}) and (\ref{3}) but $LH_uLH_u$,
the second $Z^M_3$ symmetry listed in Table~\ref{tab:rsymm}.
The baryon and lepton $R$-parities listed in Table~\ref{tab:rsymm} are at this
stage all safe, since they forbid at least one dangerous coupling appearing
in experimentally constrained products of couplings.

\begin{table}[h]
\begin{center} 
\begin{tabular}{l|c|c|c|c|c|c|c}   \hline
{Generalized $R$-parities} \quad & \quad {$q$} \quad & \quad {$u$} \quad
& \quad {$d$} \quad & \quad
{$l$} \quad & \quad {$e$} \quad & \quad {$h_d$} \quad  & \quad {$h_u$}
\quad  \\ \hline\hline
\quad  $Z_2^M$  \quad &  \quad 0
\quad  & \quad 0 \quad & \quad 1 \quad & \quad 1 \quad & \quad
 0 \quad & \quad 0 \quad & \quad 1 \quad \\ \hline
\quad $Z_3^M$ &
\quad 0 \quad & \quad 0 \quad & \quad -1 \quad & \quad -1 \quad & \quad 
0 \quad & \quad 0 \quad  & \quad -1 \quad \\ \hline
\quad $Z_3^M$ &
\quad 0 \quad & \quad 1 \quad & \quad 1 \quad & \quad 0 \quad  & \quad 
1 \quad & \quad 1 \quad & \quad 1 \quad \\ \hline
\quad $Z_3^M$ &
\quad 0 \quad & \quad 1 \quad & \quad 1 \quad & \quad -1 \quad  & \quad 
-1 \quad & \quad 1 \quad  & \quad 1 \quad   \\ \hline\hline
\quad $Z_2^L$ &
\quad 0 \quad & \quad -1 \quad & \quad 0 \quad & \quad 0 \quad & \quad
0 \quad & \quad -1 \quad & \quad 0 \\ \hline
\quad $Z_3^L$ &
\quad 0 \quad & \quad -1 \quad & \quad 0 \quad & \quad 1 \quad & \quad
-1 \quad & \quad -1 \quad & \quad 0 \\ \hline
\quad $Z_3^L$ &
\quad 0 \quad & \quad -1 \quad & \quad 0 \quad & \quad 0 \quad & \quad
0 \quad & \quad -1 \quad & \quad 0 \\ \hline\hline
\quad $Z_2^B$ &
\quad 0 \quad & \quad 0 \quad & \quad 1 \quad & \quad 0 \quad & \quad
1 \quad & \quad 0 \quad & \quad 1 \\ \hline
\quad $Z_3^B$ &
\quad 0 \quad & \quad 1 \quad & \quad 1 \quad & \quad 1 \quad & \quad
0 \quad & \quad 1 \quad & \quad 1 \\ \hline
\quad $Z_3^B$ &
\quad 0 \quad & \quad 0 \quad & \quad -1 \quad & \quad 0 \quad & \quad
-1 \quad & \quad 0 \quad & \quad -1 \\ \hline
\end{tabular}
\caption{{\it Matter, lepton and baryon generalized $R$-parities in the MSSM.
Superscript indices distinguish between matter (M), baryon (B) and lepton (L)
generalized $R$-parities. The letters $q$, $u$, $d$, $\cdots$ denote the $Z_N$
charges referring to the action of the generalized $R$-parities as expressed
in Eq. (\ref{5}).}}
\label{tab:rsymm}
\end{center}
\end{table}
 
One may wonder whether the discrete symmetries considered above are likely
to be exact at low energy.
Indeed, any global symmetry, continuous or discrete, might be broken
by quantum gravity effects. Even if this happens only at the Planck scale,
as noticed above, the strong constraints coming from proton decay rule
out the corresponding symmetry~\cite{gilbert}.
It is however well known that a discrete symmetry originating from 
the spontaneous breaking of some continuous gauge symmetry is protected
against quantum-gravity violations.
The original gauge quantum numbers are of course very constrained by
cancellation of triangle gauge anomalies, as well as by mixed
gauge-gravitational anomalies. 
As noticed in~\cite{ibanez92,ibanez91}, there are remnants of these conditions,
called ``discrete gauge anomaly cancellation conditions'' in the low-energy 
theory after spontaneous symmetry breaking. Discrete symmetries
that respect these conditions are therefore safe with
respect to anomalies. Among the generalized parities
of Table~\ref{tab:znsym}, only two are discrete anomaly free in the MSSM,
namely the standard $Z_2$\index{Discrete symmetries!$Z_2$} matter parity ($Z^M_2$), 
actually equivalent to $R$-parity, and the $Z_3$\index{Discrete symmetries!$Z_3$}
generalized baryon parity ($Z^B_3$).
If in addition we require the absence of the dimension-5 operators of Eqs.
(\ref{2}) and (\ref{3}), we are left with the $Z_3$ generalized baryon
parity only, a remarkable degree of uniqueness.
 
Let us finally note that the standard matter parity $X$ could also originate
from a spontaneouly broken anomalous 
$U(1)_X$\index{Group symmetries!Family-dependent $U(1)_X$} 
gauge symmetry under which
all matter superfields have charge one and the two Higgs superfields have
charge zero (however the Yukawa couplings are not invariant under this
symmetry, and should therefore be generated by a Froggatt-Nielsen mechanism,
as in section \ref{sec:flavour}). Indeed, compactifications of the heterotic
string often contain
a seemingly anomalous abelian gauge symmetry, whose mixed gauge anomalies are
compensated for by the Green-Schwarz mechanism \cite{green84}. For this
mechanism to work, the following condition must be satisfied:
$A' g^{\prime\, 2} = A\, g^2 = A_3\, g^2_3$, where $A'$, $A$, $A_3$ are the
coefficients of the mixed gauge anomalies
$[U(1)_Y]^2 U(1)_X$, $[SU(2)_L]^2 U(1)_X$ and $[SU(3)_C]^2 U(1)_X$, and
$g'$, $g$, $g_3$ are the coupling constants of the gauge groups
$U(1)_Y$, $SU(2)_L$ and $SU(3)_C$. This condition can be understood as a
relation determining the relative normalization of the generators associated
with different gauge groups in terms of the anomaly coefficients. Now the
above charge assigment yields $A_3 = A = -12$ and $A' = -20$, implying
$g^{\prime\, 2} / g^2 = A / A' = 3 / 5$ (or simply $g_1 = g = g_3$
with $g_1 = \sqrt{5 / 3}\, g'$) and therefore
\cite{ibanez93}
\begin{equation}
  \sin^2 \theta_W\ =\ \frac{g^{\prime\, 2}}{g^{\prime\, 2}+g^2}\
    =\ \frac{3}{8}\ ,
\label{9}
\end{equation}
a relation known to be successful at the grand unification scale. We conclude
that a gauge continuous version of the standard matter parity discussed
in chapter \ref{chap:intro}, which is equivalent to the 
$Z_2$\index{Discrete symmetries!$Z_2$}
matter parity of Table~\ref{tab:znsym} and has the same effect as $R$-parity,
may be a good string symmetry.
This provides a further motivation for $R$-parity conservation.
Still the arguments developed earlier in this section as
well as in section \ref{sec:flavour} motivate possible violations of the
$R$-parity symmetry, with a hierarchy of \Rp\ couplings that could make them
well compatible with experimental data.


\vskip .5cm

In this chapter, we studied from a theoretical point of view the possible
violations of the $R$-parity symmetry introduced in chapter 1
in connection with baryon and lepton-number conservation.
We first gave the most general form of the \Rp\ terms that may
be present in the superpotential and in the soft \SUSY-breaking scalar
potential, and addressed issues associated with the choice of a basis for the
Higgs and lepton superfields $H_d$ and $L_i$, in the presence of bilinear
$R$-parity violation.

Then we classified the patterns of $R$-parity
breaking that are consistent at the quantum level, according to which types
of \Rp\ terms are present in the Lagrangian density. We also discussed three
scenarios of $R$-parity violation often considered in the literature, namely
explicit \Rp\ by trilinear terms, explicit \Rp\ by bilinear terms and
spontaneous $R$-parity breaking. We then discussed the effects of the
Higgs-lepton mixing \index{Mixing!higgs--lepton} associated with the presence of 
bilinear \Rp\ terms, which leads to a potentially very large (order $M_Z$) 
neutrino mass and to specific\Rp\ signatures at colliders. To suppress the 
neutrino mass to aphenomenologically acceptable value, either a strong 
fine-tuning of bilinear\Rp\ parameters or an ``alignment'' mechanism is 
required. We also presented ascenario of spontaneous $R$-parity breaking 
involving a singletMajoron\index{Majoron}, thus
avoiding conflict with the measured invisible decay width of the $Z$ boson.

Finally, we discussed possible microscopic origins for terms violating
lepton and baryon numbers. We focussed essentially on abelian flavour
symmetries, Grand Unified gauge symmetries and discrete symmetries
generalizing $R$-parity. The latter appear for example in string theories.
The phenomenological constraints on the lepton- and baryon-number violations,
which are crucial tests of any extension of the Standard Model, become
therefore a window into the structure of the underlying high energy theory.

\cleardoublepage
\chapter{RENORMALIZATION GROUP SCALE EVOLUTION 
OF {\boldmath{\Rp}} COUPLINGS}
\label{chap:evolution}  

One of the most interesting indications for supersymmetry
is the unification of gauge couplings at a high scale, obtained
by renormalization group evolution of the measurements done by LEP\index{LEP}
at the ``low'' scale of $M_Z$ \cite{marcianunif}. The renormalization
group allows one to evolve couplings and mass parameters between two energy 
scales, thus providing a way to test at the available energies physical 
assumptions postulated at a higher scale, or vice-versa to translate available 
experimental data into quantities at high energy. 

In this chapter, we shall focus on the renormalization group
evolution of constraints for the \Rp\ interactions and
cover in particular those associated with the perturbative unitarity
or the so-called ``triviality bounds'', the infrared fixed points and the 
tests of grand unification schemes. The r\^ole of supersymmetry-breaking 
effects will also be discussed. In the limit of unbroken supersymmetry the 
bilinear \Rp\ interactions can be recast only in terms of the $\mu$-term 
by a suitable redefinition of the four superfields ${\hat L}_\alpha 
\equiv$($H_d$, $L_1$, $L_2$, $L_3$), therefore $R$-parity violation can be 
parametrized by trilinear couplings only. In the presence of soft terms, as 
explained in chapter \ref{chap:theory}, the superpotential and the 
scalar potential contain two independent sources of bilinear $R$-parity 
violation which, in general, cannot be simultaneously rotated away by field 
redefinitions. In turn, the Higgs-lepton mixing due to bilinear $R$-parity 
violation leaves an arbitrariness in the choice of the  ${\hat L}_\alpha$ 
basis. It is therefore crucial, when dealing with numerical results, to 
specify the choice of basis. A detailed discussion is given in section 
\ref{subsec:H_L_basis}. In the following we shall keep the discussion 
general, and, when referring to numerical results, the reader will be 
guided to cited papers for the choice of basis and detailed assumptions.

Our discussion of renormalization group\index{Renormalization group} 
studies in presence of \Rp\ will mostly concern the so-called supergravity 
framework. 
In this framework, the soft \SUSY-breaking terms are 
supposed to be generated through gravity in the limit in which the
gravitational coupling constant ($\kappa = \sqrt{8 \pi G_N}$) is taken
to be small.
Since this is closely tied with a grand unified theory
approach, one is led to consider the
scale evolution of parameters up to large energy scales of the order
of the grand unification scale $M_X$ (or the compactified string
theory scale $M_C$, or the Planck scale $M_P$). Conversely to this
bottom-up type scale evolution, one may also envisage a top-down type scale
evolution, by assigning boundary condition values for the (fewer in
number) unified parameters at the large scale and using the
renormalization group equations (RGEs) to evolve the entire set of
\MSSM\ parameters values down to the electroweak symmetry breaking 
mass scale. 

\section{Renormalization Group Equations}
\label{sec:rengr0}
As remarked in chapter~\ref{chap:theory}, in the absence of $R$-parity 
and lepton-number conservation laws, there is no {\it a priori} distinction 
between the $H_d$ Higgs and $L_i$ lepton superfields, as they have the same
gauge quantum numbers. 
Only after electroweak symmetry breaking, the mass eigenstate basis defines
which combinations of the $H_d$ and $L_i$ component fields correspond to the
physical leptons and sleptons.
The magnitude of \Rp\ couplings depends in particular on which direction in
the space of weak doublet superfields with weak hypercharge\index{Hypercharge} 
$-1$ ultimately corresponds to the Higgs.
One possible strategy consists in constructing combinations of coupling 
constants that are invariant under these basis redefinitions 
\cite{davidson97,davidson98,ferrandis99}, and parametrize the 
\Rp\ content of the Lagrangian density in a way similar to the 
Jarlskog invariants for $C P$ violations \cite{Jarlskog:1985cw}.

In order to express the superpotential and the renormalization group 
equations\index{Renormalization group!equations} (RGEs) in a compact  
way, we rewrite Eqs. (\ref{eq:W_Rp_even}) and (\ref{eq:W_Rp_odd}) of section
\ref{subsec:couplings} in the form of Eq.~(\ref{eq:W_general}):
\begin{equation}
W_{R_p}\ + W_{R\!\!\! \slash_p}\ =\ \mu_\alpha\, H_u {\hat L}_\alpha\
+\ \frac{1}{2}\, \lambda^e_{\alpha \beta k}\, {\hat L}_\alpha {\hat L}_\beta 
E^c_k + \lambda^d_{\alpha jk}\, {\hat L}_\alpha Q_j D^c_k\
  + \lambda'' \; {\mathrm {and}} \; \lambda^u_{ij} \; {\mathrm {terms}} ,
\label{eq:ch3wrwrp}
\end{equation}
where $\alpha \equiv (0,i) = (0,1,2,3)$, $\beta \equiv (0,j) = (0,1,2,3)$.
Here we use the following notation:
\begin{equation}
H_d \equiv {\hat L}_0 \, , \;\;\;\;\;\;\;\;\; \mu \equiv \mu_0 \, ,
\end{equation}
\begin{equation}
\lambda^e_{0jk}\equiv \lambda^e_{jk} \, , \;\;\;\;\;\;\;\;
\lambda^d_{0jk}\equiv \lambda^d_{jk} \, .
\end{equation}
This allows us to write the $SU(4)$ transformation of 
Eq.~(\ref{eq:SU4_rotation}) and following equations as:
\begin{eqnarray}
{\hat L}_\alpha &\to& U_{\alpha \beta} \; {\hat L}_\beta \; ,\\
\mu_\alpha &\to& U^\star_{\alpha \beta} \; \mu_\beta \; ,\\
{\lambda}^e_{\alpha \beta k} &\to& U^\star_{\alpha \gamma} \, 
U^\star_{\beta \delta} \; \lambda^e_{\gamma \delta k} \; ,\\
{\lambda}^d_{\alpha jk} &\to& U^\star_{\alpha \beta} \;
\lambda^d_{\beta jk} \; ,
\end{eqnarray}
where $U$ is the $SU(4)$ matrix with entries $U_{\alpha \beta}$ associated
with the basis rotation. 
It is clear from the above equations that the lepton-number-violating 
couplings are basis-dependent. For a detailed discussion see
chapter \ref{chap:theory}.

In the following, explicit expressions for the $\Rp$ RGEs up to the two 
loop order will be written using the above notations. 
To facilitate a comparison with the existing literature 
we give in Table~\ref{tab:allanach} the correspondence between our 
notations and the ones of \cite{allner}.
\begin{table} [h]
\begin{center}
\begin{tabular}{c|c}
\hline
Our Notation&Notation of \protect{\cite{allner}} \\ \hline \hline
$H_d$ &$H_1$ \\
$H_u$ &$H_2$ \\
\hline
$\lambda^e_{ijk}$ & $(\Lambda_{E^k})_{ij} $ \\
$\lambda^e_{jk} \equiv \lambda^e_{0jk}$ & $-(Y_E)_{jk}$ \\
$\lambda^d_{ijk}$ & $(\Lambda_{D^k})_{ij} $ \\
$\lambda^d_{jk} \equiv \lambda^d_{0jk}$ & $-(Y_D)_{jk}$ \\
$\lambda''_{ijk}$ & $(\Lambda_{U^k})_{ij} $ \\
$\lambda^u_{jk}$ & $(Y_U)_{jk}$ \\
\hline
\end{tabular}
\caption{Correspondence among our notation and the one of
\protect{\cite{allner}}.} 
\end{center}
\label{tab:allanach}
\end{table}

\subsection{Evolution of the Bilinear {\boldmath{$\mu$}} Terms}

Following the general equations given in \cite{marv} the RGEs for the
bilinear $\mu$ terms\index{Renormalization group!bilinear terms} including 
all \Rp\ effects can be written as:
\begin{equation}
\frac{d}{dt}\mu_\alpha=\mu_\alpha \; \Gamma_{uu}+\mu_\beta
\; \Gamma_{\alpha \beta},
\label{eq:bilcompact}
\end{equation}
where $t=\log{q^2}$ and $\Gamma$ are the anomalous dimensions. 
The index $\alpha$ is defined as in the previous section 
($\alpha\equiv (0,i)$). The notation $\Gamma_{uu}$ is 
a shorthand for the anomalous dimension \index{Anomalous dimension} 
$\Gamma_{H_u \, H_u}$ for the $H_u$ superfield. In a similar way $\Gamma_{00}$ 
stands for $\Gamma_{H_d \, H_d}$ and $\Gamma_{ij}$ means $\Gamma_{L_i \, L_j}$.
Expanding Eq.~(\ref{eq:bilcompact}) into its components one obtains
\begin{eqnarray}
\frac{d}{dt}\mu_0&=&\mu_0 \, \Gamma_{uu} + \mu_0 \, \Gamma_{00}+
\mu_i \, \Gamma_{0i} \, ,\\
\frac{d}{dt}\mu_i&=&\mu_i\, \Gamma_{uu} + \mu_0 \, \Gamma_{i0}+
\mu_j \, \Gamma_{ij} \; .
\end{eqnarray}
This set of equations implies that even if we start with all $\mu_i=0$, 
non-zero $\mu_i$ will be generated through the RGEs via a non-zero $\mu_0$ and 
vice-versa. 

The bilinear terms do not appear in the equations for the evolution of the 
Yukawa couplings or the gauge couplings. Thus, they do not directly 
affect the unification of the latter.
The anomalous dimensions\index{Anomalous dimension} $\Gamma$ transform as 
follows under the $SU(4)$ rotation of the fields:
\begin{eqnarray}
{\Gamma}_{uu}&\to& \Gamma_{uu} \, ,\\
{\Gamma}_{\alpha \beta}&\to& U_{\beta \gamma} U^\star_{
\alpha \tau} \Gamma_{\tau \gamma} \; .
\label{gammatransf}
\end{eqnarray}
The anomalous dimensions\index{Anomalous dimension} are given by 
\begin{equation}
\Gamma_{i j}=\frac{1}{16 \pi^2} \gamma_{i j}^{(1)}+
\frac{1}{(16 \pi^2)^2} \gamma_{i j}^{(2)} + \dots 
\end{equation}
where $\gamma^{(1)}$,  $\gamma^{(2)}$, $\dots$ are 1--loop, 2--loop, $\dots$ 
contributions:
\begin{eqnarray}
\gamma_{ij}^{(1)}&=& \frac{1}{2} Y_{imn}Y^\star_{jmn}-2\, \delta_{ij}
\sum_ag_a^2C_a(i)\; ,\\ 
\gamma_{ij}^{(2)}&=&-\frac{1}{2} Y_{imn}Y^\star_{npq}Y_{pqr}Y^\star_{mrj}
+Y_{imn}Y^\star_{jmn} \sum_a g_a^2\, \left[ 2C_a(p)-C_a(i) \right] \nonumber \\
&+&2\, \delta_{ij}\sum_a g_a^2 \left[g_a^2 C_a(i) S_a(R)+2
\sum_b g_b^2 C_a(i) C_b(i) - 3g_a^2C_a(i)C(G_a)\right] \, . 
\end{eqnarray} 
$Y_{ijk}$ is a generic Yukawa coupling (it stands for $\lambda^e$, 
$\lambda^d$ or $\lambda''$), $C_a(f)$ is the quadratic Casimir of the  
representation $f$ 
of the gauge group $G_a$. $C(G)$ is an invariant of the adjoint representation 
of the gauge group $G$ and $S_a(R)$ is the second invariant of the 
representation $R$ in the gauge group $G_a$. Explicitly if $t^A$ are the 
representation matrices of a group $G$ one has: 
\begin{equation}
(t^A t^A)_{ij}= C(R) \delta_{ij} \, ,
\end{equation}
for $SU(3)$ triplets $Q$ and $SU(2)$ doublets $L$:
\begin{equation}
C_{SU(3)}(Q)=\frac{4}{3}, \; \; \;  C_{SU(2)}(L)=\frac{3}{4} \, ,
\end{equation}
and for the $U(1)$ weak hypercharge\index{Hypercharge} embedded in 
$SU(5)$\index{Group symmetries!$SU(5)$}:
\begin{equation}
C(\phi)=\frac{3}{5} y^2(\phi),
\end{equation}
where $y(\phi)$ is the weak hypercharge\index{Hypercharge} of the field $\phi$. The factor $3/5$
comes from the $SU(5)$ grand unified normalisation of the weak 
hypercharge\index{Hypercharge} generator.
For the adjoint representation:
\begin{equation}
C(G)\; \delta^{AB}=f^{ACD}f^{BCD},
\end{equation}
where $f^{ABC}$ are the structure constants.\index{Structure constant} For the
groups under study: \begin{equation}
C(SU(3)_C)=3,\quad C(SU(2)_L)=2,\quad C(U(1)_Y)= 0,\quad C(SU(N))=N
\; .
\end{equation}
The Dynkin index\index{Dynkin index} is defined by
\begin{equation}
\mbox{Tr}_R(t^A t^B) = S(R)\; \delta^{AB}.
\end{equation}
For the fundamental representations $f$ we have
\begin{eqnarray}
SU(3),\, SU(2) \to \quad &&S(f)=\frac{1}{2} \, ,\\
U(1)_Y \to \quad&&S(f)=\frac{3}{5}y^2(f)\, .
\end{eqnarray}

\subsection{Evolution of the Trilinear {\boldmath{\Rp}} Yukawa Couplings}
As already stated in chapter \ref{chap:theory}, scenarios in which $R$-parity 
is broken only by trilinear interaction terms are in general not consistent 
since bilinear \Rp\ terms are generated by quantum corrections and cannot be
rotated away through a 
$SU(4)$\index{Group symmetries!$SU(4)$ rotations} 
redefinition of the superfields, due to the
presence of soft \SUSY-breaking terms. 
One has therefore to consider all the couplings when studying the
renormalization group evolution.
The general RGEs for the Yukawa
couplings\index{Renormalization group!Yukawa couplings} 
$Y_{ijk}$ (where $Y$ stands for $\lambda^e_{\alpha \beta k}$, or 
$\lambda^d_{\alpha jk}$, $\lambda^u_{ik}$ or 
$\lambda''_{ijk}$ of Eq.~(\ref{eq:ch3wrwrp})) are given by \cite{marv}: 
\begin{equation}
\frac{d}{dt} Y_{ijk}=Y_{ijl} \; \Gamma_{lk} \; +\; 
(k \leftrightarrow j)\; + \; (k \leftrightarrow i) \;.
\end{equation}
For the Yukawa couplings $\lambda^e_{\alpha \beta k}$ and 
$\lambda^d_{\alpha jk}$ this gives
\begin{eqnarray}
\frac{d}{dt} \lambda^e_{\alpha \beta k}&=&\lambda^e_{\alpha \beta l}\; 
\Gamma_{E_l\, E_k} +
\lambda^e_{\alpha \delta k} \; \Gamma_{\delta \beta} 
+ \lambda^e_{\gamma \beta k} \; \Gamma_{\gamma\alpha}\; ,\\
\frac{d}{dt} \lambda^d_{\alpha jk}&=&\lambda^d_{\alpha jl}\; 
\Gamma_{D_l\, D_k} + \lambda^d_{\alpha lk} \; \Gamma_{Q_l\, Q_j} + 
\lambda^d_{\gamma jk} \; \Gamma_{\gamma\alpha}\; ,
\end{eqnarray}
while for the Yukawa couplings
$\lambda''_{ijk}$ and $\lambda^u_{jk}$ we have
\begin{eqnarray}
\frac{d}{dt} \lambda^u_{ij}&=&\lambda^u_{ik}\; \Gamma_{U_j\, U_k} +
\lambda^u_{ij} \; \Gamma_{uu} + \lambda^u_{kj} \; 
\Gamma_{Q_i \; Q_k} \; ,\\
\frac{d}{dt} \lambda''_{ijk}&=&\lambda''_{ilk}\; \Gamma_{D_j\,
D_l} + \lambda''_{ljk} \; \Gamma_{U_i\, U_l} + \lambda''_{ijl} \; 
\Gamma_{D_k \; D_l} .
\end{eqnarray}

Of course the two-loop RGEs for the Yukawa
couplings preserve ${\lambda''}_{ijk}=0$ (for all $i,j,k$) at all scales 
if they are zero at some scale (i.e. baryon parity is conserved).
The same is true if lepton parity is imposed at some scale for 
$\lambda^e_{ijk}$ and $\lambda^d_{ijk}$.
If however one imposes only one coupling to be non-zero at some
scale, this remains in general not true at all scales. Take for example
only $\lambda^d_{111}\neq 0$ at some scale. Then through the CKM
mixing the other terms $\lambda^d_{1ij}$ will be generated by the 
RGEs.

\subsection{Evolution of the Gauge Couplings}
The RGEs for the \SM\ 
gauge couplings\index{Renormalization group!gauge couplings} 
$g_1$ [for $U(1)_Y$\index{Group symmetries!$U(1)_Y$}, with $g_1=\sqrt{5/3}~g'$ using for
example the $SU(5)$\index{Group symmetries!$SU(5)$} GUT normalisation], 
$g_2$ [for $SU(2)_L$\index{Group symmetries!$SU(2)_L$}] and 
$g_3$ [for $SU(3)_c$\index{Group symmetries!$SU(3)_C$} can be written:  
\begin{eqnarray}
&&\frac{d}{dt}g_a = \frac{g_a^3}{16\pi^2} B_a^{(1)}
+\frac{g_a^3}{(16\pi^2)^2} \Big[ \sum_{b=1}^3 B_{ab}^{(2)} g_b^2
-C_a^e \,\mbox{Tr}\, (\lambda^{e\dag}_{jk} \lambda^e_{jk})
-C_a^d\, \mbox{Tr}\, (\lambda^{d\dag}_{jk} \lambda^{d}_{jk})
\\
&&-C_a^u \,\mbox{Tr}\, (\lambda^{u\dag}_{jk} \lambda^{u}_{jk})
+A_a^e \sum_{k=1}^3\, \mbox{Tr}\, (\lambda^{e\dag}_{ijk} \lambda^e_{ijk})
+A_a^d \sum_{k=1}^3\, \mbox{Tr}\, (\lambda^{d\dag}_{ijk} 
\lambda^d_{ijk})
+A_a^u \sum_{k=1}^3\, \mbox{Tr}\, (\lambda^{\prime\prime\dag}_{ijk} \lambda^{\prime\prime}_{ijk})
\Big]\nonumber \, .
\end{eqnarray}
In the evolution equation for the gauge couplings, the equations are
coupled only starting with the two-loop term, while up to one loop each
coupling has an independent evolution.

The coefficients $B_a,\,B_{ab},$ and $C_a^x$ are calculated
in \cite{bjorkjones}:
\begin{eqnarray}
B_a^{(1)}&=&(\frac{33}{5},1,-3),\\
B_{ab}^{(2)}&=&\left( \begin{array}{ccc}
199/25 & 27/5 & 88/5 \\
9/5 & 25 & 24 \\
11/5 & 9 & 14
\end{array}
\right),\\
C_{a}^{u,d,e}&=&\left( \begin{array}{ccc}
26/5 & 14/5 & 18/5 \\
6 & 6 & 2 \\
4 & 4 & 0
\end{array}
\right)\; ,
\end{eqnarray}
where the index $u,\, d,\, e$ refers to the lines of the matrix.
The \Rp\ contributions to the running\index{Renormalization group!equations} 
of the gauge couplings appear only at two-loops. 
They are given in~\cite{allner}:
\begin{equation}
A_{a}^{u,d,e}=\left( \begin{array}{ccc}
12/5 & 14/5 & 9/5 \\
0 & 6 & 1 \\
3 & 4 & 0
\end{array}
\right)\; .
\end{equation}
In the \MSSM\ there is a relation between the running of the top-quark mass 
and the ratio of \VEVs\ of the two Higgs doublets, $\tan \beta$. Given the 
measure of the top-quark mass there is a restriction for the allowed
$\tan \beta$ range. 
In the presence of \Rp, however, there is no such a restrictive prediction. 
Furthermore, allowing the bilinear lepton-number-violating 
terms~\cite{rengroup,allner,nilles97,diaz98,diazrom98}, bottom-tau 
Yukawa unification can occur for any value of $\tan \beta$.

\section{Perturbative Unitarity Constraints}
\label{sec:rengr1}

It is possible to derive upper bounds on the \Rp\ coupling constants, without 
the need of specifying further input boundary conditions, simply by imposing
the requirement that the ultraviolet scale evolution remains perturbative
\index{Renormalization group!perturbative unitarity} up to the large 
unification scale, 
\begin{equation}
\frac{ Y^2_{ijk} (M_X)}{(4\pi)^2} < 1 \; ,
\end{equation}
where $Y_{ijk}$ is a generic trilinear \Rp\ coupling constant. In more general 
terms, the unitarity limits concern the upper bound constraints on the 
coupling constants imposed by the condition of a scale evolution
between the electroweak and the unification scales, free of
divergences or Landau poles for the entire set of coupling constants.
The principal inputs here are the \SM\  gauge coupling constants, the
superpartner spectrum together with the ratio of Higgs bosons \VEVs\
parameter, $ \tan \b =v_u/v_d$, and the quark and lepton
mass spectra, as described by the Yukawa coupling constants, 
$ \l ^{u, d, e}_{ij}$.  Since the third generation Yukawa coupling constants,
$\l_t = \l^u _{33}, \ \l_b =
\l^d _{33}, \ \l_\tau = \l^e_{33}, $ are predominant, the influential  
\Rp\ coupling constants are expected to be those containing the maximal
number of third generation indices, namely $ \l^e_{233},\ \l^d_{333}$, 
$\lambda''_{313}, \ \l ''_{323}$.

The first study developing perturbative unitarity bounds is due to Brahmachari 
and Roy~\cite{brahmarchi}. The derived bounds for the baryon-number-violating 
interactions, $ [\l ''_{313} , \ \l '' _{323}] < 1.12$, turn out to be very 
weakly dependent on the input value for $\tan \b $.  These bounds increase 
smoothly with the input value of $m_t$, diverging at $ m_t \approx 185 \ 
\text{GeV}$.
Bounds for the other configurations of the generation indices of $\l
'' _{ijk} $ have been obtained on the same basis by Goity and
Sher~\cite{goity}, $ \l ''_{mjk} <1.25, \ [m=1,2]$. 
Allanach et al.~\cite{allner} carry out a systematic analysis of the 
renormalization flow equations, up to the two-loop order, for the lepton
and baryon-number-violating interactions $\lambda^e_{ijk}$, $\lambda^d_{ijk}$
and $\lambda''_{ijk}$.
The resulting coupling constant bounds read 
$\l^e_{323} (m_t)<0.93, \ \l^d_{333} (m_t)<1.06, \ \l '' _{323} (m_t)<1.07,$
at $\tan \b =5$.
Choosing a higher $\tan \b $ lowers these bounds slightly.

\section{Quasi-fixed points analysis for {\boldmath{\Rp}} couplings}
The RGEs describing the evolution of the Yukawa
couplings down from a large scale $M_X$ may have fixed 
points\index{Renormalization group!fixed points} which give
information on the couplings. 
The existence of infrared fixed points (IRFP) for the third generation Yukawa
coupling constants and the relevant \Rp\ coupling constants, is signalled by
vanishing solutions for the beta-functions describing the scale
evolution for the ratios of the above Yukawa coupling constants to the
gauge interaction coupling constants.
In principle, one seeks fixed point solutions~\cite{pendleton} 
characterized by the exact absence in the infrared regime of a
renormalization group flow for ratios such as, for instance,
$\l^2_t/ g^2_3 $ or $ {\l^e_{323}}^2/g^2_3 $.
In practice however, this fixed point regime may be inaccessible since
it would set in at a scale much lower than the electroweak scale,
making it irrelevant.
In that case the values of the Yukawa couplings are determined by
quasi-fixed points (QFP)\index{Renormalization group!quasi--fixed points}
\cite{hill} describing the actual asymptotic behaviour of the couplings. 
In such a case, the values at the weak scale are essentially independent of
their values at the large scale, provided the initial values are large.
For an analytical study see \cite{kazakov00}.

As an example let us consider a simplified renormalization group equation
for the top-quark Yukawa coupling $\lambda_t$ at one loop in the Standard
Model:
\begin{equation}
16 \pi^2\frac{d\lambda_t}{dt}=\frac{\lambda_t}{2} \left(9 \lambda_t^2 -16 
g_3^2\right) \, ,
\label{eq:3-ren}
\end{equation}
where $t=\log q^2$ and we have neglected the contributions from the lighter 
quarks and the electroweak contributions. By forming the difference of the 
previous equation with the one for the evolution of the QCD strong interaction 
coupling:
\begin{equation}
16 \pi^2 \frac{d g_3}{dt}=g_3^3 \left( \frac{2}{3} N_f -11 \right)
\end{equation}
where $N_f$ is the number of flavours, one obtains:
\begin{equation}
16 \pi^2 \frac{d}{dt} \log (\lambda_t/g_3) =\frac{9}{2} \lambda_t^2 +g_3^2 
\left( 3-\frac{2}{3} N_f\right) \; .
\label{eq:3-evo}
\end{equation}
When the value
\begin{equation}
\lambda_t^2=\frac{2}{9} g_3^2 \left(\frac{2}{3} N_f -3 \right)
\end{equation}
is reached, Eq.~(\ref{eq:3-evo}) has zero on the right-hand side 
which implies a constant ratio of the two couplings for subsequent
decreasing values of the scale $t$.
This is the Pendleton-Ross fixed point.
However this behaviour would only set in at a very low scale, of the order
of 1 GeV, while the region of interest is the one in which the scale
is around $m_t$.
The reason why the fixed point is important only at a very low scale
is that in Eq.~(\ref{eq:3-ren}) the 
strong coupling constant $g_3$ becomes large only below 1 GeV.
In the intermediate region $\lambda_t$ evolves with the lowering of the
mass scale until the $g_3$ coupling becomes of the same order, i.e.~:
\begin{equation}
\frac{9}{2} \lambda_t^2 \simeq 8 g_3^2\; .
\label{eq:condition}
\end{equation} 
In this intermediate region the right-hand side of equation (\ref{eq:3-ren}) 
is close to zero, which in turn implies that $\lambda_t$ must remain 
relatively constant.
The previous argument can be made more precise by a detailed calculation or a 
graph of $\lambda_t(q)$ versus $\lambda_t(M_X)$ where $M_X$ is the high scale
and $q$ is in the intermediate range. In both cases the asymptotic behaviour 
is ascribed to the condition (\ref{eq:condition}) which is termed a 
quasi-fixed point.

Before considering the effect of $\Rp$, let us consider the quasi-fixed point 
regime for the Minimal Supersymmetric Standard Model. 
In this case the renormalization group flow points 
towards the value of the top quark coupling constant $ \l_t (m_t) \simeq 1.1$, 
which establishes a correlation between the top mass and $\tan \b $, described 
by the relation
\begin{equation}
m_t ({\mathrm{pole}}) =\frac{v\sin \beta}{\sqrt{2}}\, 
\lambda_t({\mathrm{pole}})\; . 
\end{equation}
Substitution of the physical top mass $m_t$ as an input value singles out a 
discrete range for $\tan \b$.
 
When the third generation \Rp\ interactions are switched on, individually or
collectively, solutions of the quasi-fixed point type continue to exist. These 
fixed point values of the coupling constants provide theoretical bounds under 
the assumption that the theory remains perturbative. By requiring a lower 
bound on the top mass, say, $ m_t > 150 \ \text{GeV}$, they would lead to 
excluded domains in the parameter space of $\l_t$ and the \Rp\ Yukawa coupling 
constants~\cite{brahmarchi}. Looking for a simultaneous quasi-fixed point in 
$\l_t$ and/or $\l_b$ and in the \Rp\ coupling constants one at a time, one 
obtains~\cite{fp2,fp1}: $ \l_t \simeq 0.94, \ \l ''_{323} \simeq 1.18$, 
$\l^d_{333} \simeq 1.07$,
and $\l_t \simeq 1.16$, $\l^e_{233}  \simeq 0.64$, at small $\tan \b$ and  
$\l_t \simeq 0.92$, $\l_b \simeq 0.92$, $\l ''_{323} \simeq 1.08$, 
at large $\tan \beta$. As the \Rp\ couplings $ \l^e $, $\l^d $, $\l '' $ are
successively switched on, the regular top Yukawa coupling varies as, 
$\l_t \simeq 1.06 \to 1.06 \to 0.99 $, respectively in the small $\tan \b $
regime. In the large $\tan \b\simeq m_t/m_b  \approx 35$ regime, 
the solution for the quasi-fixed point  
predictions are modified as, $ \l _t \simeq 1.00 \to 1.01 \to 0.87 $,
respectively and $\l_b \simeq 0.92 \to 0.78 \to 0.85 $, respectively, the
corresponding fixed point values for the \Rp\ coupling constants being 
$\l^d_{333} \simeq 0.71 , \ \l ''_{323} \simeq 0.92$~\cite{fp2,fp1}. 
Further discussions of the fixed point physics in connection with \Rp\
can be found in~\cite{rengroup}.

The stability condition with respect to small variations of the parameters 
for a renormalization group fixed point requires that the matrix of 
derivatives of the beta-functions with  respect to the coupling constants has 
all its eigenvalues of the same fixed (positive in our conventions) sign. 
Discussion of the stability issue motivated by the supersymmetric models 
can be found in~\cite{allan97,Abel:1997va,jack98}. The above stability
condition  is actually never satisfied in the \MSSM\, 
even for the trivial fixed point at which the \Rp\ coupling
constants tend to zero. Once one includes the \Rp\ interactions, a
stable infrared quasi-fixed point does exist, but only if one considers
simultaneously the third generation regular Yukawa coupling constants,
$\l_t, \ \l_b $, along with $\l ''_{332}$  \cite{ananpandita}. In particular
there is no simultaneous $B$- and $L$-violating infrared fixed point. Note 
that the validity of these results is based on the extent of what variation 
of the parameters is considered ``small''.

The quasi-fixed points are reached for large initial
values of the couplings at the GUT scale, therefore they reflect the
assumption of  perturbative unitarity of the corresponding couplings.
Under this assumption, the quasi-fixed points provide upper bounds on the 
relevant Yukawa couplings, especially the $B$-violating
coupling $\lambda_{332}''$ \cite{pandi01}. 

In the \MSSM\ the coupling constants $g_1$, $g_2$ and $g_3$ unify at a certain 
scale $M_{GUT}$ thanks to $R$-parity conservation. 
The scale evolution of the gauge couplings leads to a successful 
unification with the values of the unified coupling constant and the 
unification scale given by, $ \a_X(M_X)\simeq 1/24.5 =
0.041, \ [g_X(M_X)\simeq 0.72], \ M_X \simeq 2.3
\times 10^{16}\ \text{GeV}$.
Besides gauge coupling unification, GUT theories reduce the number of free
parameters in the Yukawa sector.  
\Rp\ affects this picture: the feed-back effects of the \Rp\ trilinear 
interactions on the regular Yukawa interactions may have significant 
implications for the constraints set by grand unification on the \MSSM\ 
parameters.      

In the context of \GUT\ one could consider a unification of 
the \Rp\ parameters. However, if the \Rp\ interactions arise from a $SU(5)$ 
invariant term there would be a relation between $B$- and $L$-violating 
terms and this in turn would imply non-zero contributions to the proton 
decay, either directly or at one-loop level through flavour mixing, 
therefore limiting the \Rp\ couplings to very small values
as already discussed in section~\ref{chap:theory}.
The situation for the widely used hypothesis of Yukawa coupling
unification $\l_b=\l_\tau$ is analogous, even if there is no direct
link between the two sorts of coupling unification.
Analyses avoiding the assumption of $\l_b=\l_\tau$ unification 
have been performed and the quasi-fixed point values for the \Rp\
coupling constants are found as~\cite{fp2,fp1},  
$\l^e_{233} = 0.90, \ \l^d_{333} = 1.01 , \ \l ''_{323}  
= 1.02$, for $\tan \b < 30 $.
  
As a simplifying assumption one can take a hierarchy similar to the one
between \SM\ Yukawa couplings, and therefore consider only one
coupling at a time. 
Solving the two-loop RGEs, Allanach et al.~\cite{allner} find that by 
turning on any one of the three relevant \Rp\
third generation related coupling constants, from zero to their
maximally allowed values, the unification coupling constant,
$\a_X$, is insignificantly affected by less than $ 5 \% $, 
while the unification scale, $M_X$, can be reduced by up to $ 20\% $.
Note also that for large values of the $R$-parity violating coupling, 
the value of $\alpha_s(M_Z)$ predicted from unification can be reduced 
by 5\% with respect to the $R$-parity conserving case.

\section{Supersymmetry Breaking}  
\label{sec:rengr3}
  
The renormalization group studies
in the presence of the soft \SUSY-breaking become far less
tractable.  A proper treatment of $R$-parity
violation must also include \Rp\ soft terms, therefore a large number of
additional parameters arise which all have a mutual influence on one another.
Within the MSSM, these additional terms are given in Eq.~(\ref{eq:V_Rp_odd})
and introduce 51 new \Rp\ parameters:
3 $B_i$ associated with the bilinear superpotential terms,
$45\ $ \Rp\ $A$-terms with the same antisymmetry properties as
the corresponding trilinear superpotential couplings, and 3 \Rp\ soft 
mass terms $\widetilde{m}^2_{di}$ mixing the down-type Higgs boson and
slepton fields. 

The inclusion of $R$-parity violation in the superpotential allows the 
generation of lepton-Higgs mixing which leads to sneutrino \VEVs\ and hence 
neutrino masses as discussed in chapter \ref{chap:theory}.
The indirect generation of sneutrino \VEVs\ through the running of the RGEs 
for the soft terms can lead to large effects. 
They induce finite sneutrino \VEVs\ $v_i$, 
via the renormalization group evolution of the \Rp\ trilinear
interactions from the grand unification scale to the electroweak scale, 
as discussed in~\cite{decarlosl}.  
A renormalization group analysis, including the soft \SUSY-breaking 
parameters, is developed within a supergravity framework \cite{addprd69},
where the \Rp\ trilinear interactions are specified at the grand
unification scale, $ \l^e_{ijk} (M_X) $, and one performs at each energy
scale the requisite field transformation aimed at removing away the
bilinear interactions in the superpotential, $\mu_i (q) =0$.  
Since finite $ v_i$ contribute, via the mixing with neutralinos, to the
neutrino Majorana masses (see chapter \ref{chap:neutrinos}), the condition
that the experimental limits on these masses are satisfied leads to the 
following qualitative bounds at the unification scale, 
$ \l^e_{i33} < (10^{-2} - 10^{-3}) $ and $\l^d_{i33} <(10^{-2} - 10^{-3})$. 
The \Rp\ interactions also initiate,
through the renormalization group evolution, indirect contributions to
flavour changing soft mass and \Rp\ parameters.  An application to the
prototype process $\mu \to e+\g$ indicates that these indirect
effects turn out to dominate over the direct effects associated with
the explicit contributions from the one-loop diagrams 
discussed in chapter~\ref{chap:indirect}. However, the
situation cannot be described in terms of quantitative
predictions, owing to the large number of free parameters and the
occurrence of strong cancellations amongst contributions from
different sources.
  
In another study~\cite{decarlosq} the r\^{o}le of the \Rp\ interactions in
driving certain superpartner mass squared to negative values is examined.  
The sneutrinos are most sensitive
to this vacuum stability constraint because of the weaker experimental
bounds on their masses.  The attractive contribution
from the \Rp\ interactions reads $ \delta m ^2_{ \tilde \nu } \simeq - |\l
_{ijk} ^d (M_X)|^2 (13 m_0^2 + 49 M_{\ud } ^2 -1.5 M_\ud A -12 A^2 )$, 
where $m_0$, $M_{\ud}$ and $A$ stand for the unification values of the soft
scalar masses, the gaugino masses and the $A$-terms respectively, assumed
to be universal.    
Invoking the experimental constraint on  $ m_{ \tilde \nu }$
from LEP, one may derive bounds on the \Rp\ coupling constants at the unification
scale, valid for all flavour configurations, such as
$\l^d_{ijk} (M_X)< 0.15 $, which translate into bounds at the electroweak scale,
of the form $\l^d_{ijk} \approx \l^d_{ijk} (M_Z) < 0.3$.
The indirect effects of the \Rp\ interactions on the flavour changing
parameters are also examined for the process $b \to s + \gamma$.
These contributions appear to dominate over the direct perturbative contributions
from the one-loop diagrams. However, because of the large number of relevant
parameters and the complicated dependence on the observables, one
can again only infer conclusions of a qualitative nature.  

In conclusion, the renormalization group evolution is a powerful tool to link 
theoretical hypotheses and experimental data, by allowing the comparison  
of quantities such as coupling constants at different scales. It is however 
difficult to draw general conclusions on the bounds that one can 
obtain as they may strongly depend on assumptions, in a range of energy that is 
still to be explored. Nonetheless if one allows to include the general picture 
of grand unification and supersymmetry a number of interesting results can
be obtained.


\cleardoublepage

\chapter{COSMOLOGY AND ASTROPHYSICS}

\label{chap:cosmology}


One of the first merit of a conserved $R$-parity is to provide, naturally,
a stable lightest supersymmetric particle (LSP). If $R$-parity is 
absolutely conserved, the LSP is absolutely stable \,-- none of its
possible decay channels being kinematically allowed --\ and therefore it
constitutes a possible dark matter candidate\index{Dark matter}.

A broken $R$-parity supersymmetry could have important implications on
this issue of the dark matter\index{Dark matter} of the universe.
The LSP can then decay through
\Rp\ interactions into Standard Model particles only. Such an unstable LSP
can still remain, however, a possible dark matter\index{Dark matter}
candidate, provided its lifetime is sufficiently long; the corresponding
\Rp\ couplings are then required to be extremely small. On the other hand
a short-lived LSP, irrelevant to the dark matter problem, is required to
decay sufficiently quickly so as not to affect the successful predictions
of Big Bang nucleosynthesis.

A second important issue concerns the cosmological baryon
asymmetry\index{Baryon asymmetry}, i.e. the fact that there is no
significant amount of antibaryons in the universe. Understanding the
origin of the observed baryon-to-photon ratio
$n_B / n_\g = (6.1^{+ 0.3}_{- 0.2}) \times 10^{-10}$ \cite{WMAP},
or in other terms of the cosmological baryon asymmetry
$\eta_B \equiv (n_B - n_{\bar B}) / n_\g$, raises the question of how and
when this baryon-antibaryon asymmetry was generated, and what the protection
of this asymmetry against subsequent dilution by
baryon-number-violating\index{Baryon number!violation} interactions over 
the history of the universe requires. 
Different solutions for the creation of the cosmological baryon asymmetry
have been proposed in the literature, either directly through
$B$-violating interactions, or indirectly through 
$L$-violating interactions (the resulting lepton number asymmetry
being turned into a baryon asymmetry through $\,B-L\,$ conserving but 
$\,B+L\,$ violating processes known as sphalerons).
This also requires that the corresponding interactions should
violate, in addition to baryon and/or lepton number, the $C$ and $CP$
symmetries between particles and antiparticles.

Supersymmetric theories with broken $R$-parity have the interesting
feature of providing the baryon and/or lepton-number non-conservation
needed for baryogenesis.
However, while these \Rp\ interactions may generate a baryon
or lepton asymmetry all by themselves, in reverse, they might also dilute a
pre-existing baryon asymmetry.

\section{Constraints from the lifetime of the 
Lightest Supersymmetric Particle}

\index{Relic particles|(}


\label{secxxx6a}

\subsection{Decays of the Lightest Supersymmetric Particle}


\label{subsec:mc5}

In \susyq\ extensions of the \SM\ with unbroken $R$-parity,
the LSP plays a fundamental r\^ole as the sole \susyq\ relic
from the Big Bang, and may then provide the non-baryonic
component of the dark matter\index{Dark matter} of the universe.

In principle the LSP could be any \susyq\ particle, such as the lightest
neutralino or chargino, a sneutrino, a charged slepton, a squark or a gluino.
There are however strong arguments in favour of an electrically neutral and
uncoloured (stable) LSP \cite{ellis84}. Stable, electrically charged and
uncoloured particles would combine with electrons (if they have charge $+1$)
or with protons or nuclei (if they have charge $-1$) to form superheavy
isotopes of the hydrogen or of other elements. 
Stable coloured particles would first bind into new hadrons (such 
as $(\tilde t u d)^+$ or $(\tilde t d d)^0$ in the case of a stop LSP), 
which would then combine with electrons (in the case of a stable,
charge $+1$ heavy hadron) or with nuclei to form superheavy isotopes
of the hydrogen or of other elements.
The relic number densities of such
massive stable particles have been evaluated to be
$n_X /n_B \simeq 10^{-6} \ (m_X / 1 \ \GeV)$ for an electrically
charged, uncoloured particle \cite{wolfram79} such as a charged slepton LSP,
and $n_X /n_B \simeq 10^{-10}$, independently of the hadron mass, for a
coloured particle \cite{wolfram79,dover79}, such as a squark or a gluino LSP.
However, terrestrial experiments searching for anomalously heavy protons
or superheavy isotopes have placed stringent upper limits on the relic
abundances of electrically charged 
or coloured stable particles\index{Hadrons!Coloured stable particles}
(for a review see Ref. \cite{Perl2001}).
For example ``heavy proton'' experimental
searches yield the limit $n_X /n_B < 10^{-21}$ for $m_X < 350 \GeV$
\cite{smith79}; heavy isotopes searches, $n_X /n_B < (2 \times 10^{-16}
- 7 \times 10^{-9})$, depending on the element, for
$10^2 \GeV < m_X < 10^4 \GeV$ \cite{hemmick90}; and searches for superheavy
isotopes of hydrogen in water, $n_X /n_B < 10^{-28}$ \cite{smith82},
$3 \times 10^{-20}$ \cite{hemmick90} and $6 \times 10^{-15}$ \cite{verkerk}
in the mass ranges $(10 - 10^3) \GeV$, $(10^2 - 10^4) \GeV$ and
$(10^4 - 10^8) \GeV$, respectively.
The comparison of these negative experimental results with the above
predicted relic abundances almost certainly rules out charged or coloured
superparticles as suitable (stable) LSP candidates.

Among the possible electrically neutral and uncoloured LSPs, the
lightest neutralino $\tilde \chi^0_1$ appears to be the best candidate for
the non-baryonic dark matter of the universe.
The gravitino remains a possible dark matter candidate, but it 
generally suffers from an abundance excess problem,
while the possibility of a sneutrino LSP has been excluded, in the \MSSM,
by direct dark matter searches in underground experiments
\cite{beck94,falk94}. The requirement that the relic density of the
lightest neutralino falls within the range allowed by observations,
$\Omega_{CDM} = 0.23 \pm 0.04$ \cite{WMAP}, where $\Omega_{CDM} \equiv
\rho_{CDM} / \rho_c$ is the ratio of the present cold dark matter (CDM)
energy density to the critical energy density, puts strong constraints
on the parameters of the \SSM. But the fact that satisfactory values of the
relic abundance can be obtained constitutes one of the important motivations
for $R$-parity conservation in \susyq\ extensions of 
the \SM\footnote{If the strong $CP$ problem is solved by the Peccei-Quinn
                 mechanism~\cite{peccei77}, the supersymmetric partner of the
		 axion, the axino\index{Axion!axino}, could also be a viable
		 dark matter candidate (see e.g. Ref. \cite{covi99},
                 assuming primordial axinos to have been diluted by 
		 inflation).}.

The above state of affairs gets drastically modified in the case
of a broken $R$-parity.  The most important effect of \Rp\ interactions
is the resulting instability of the LSP. An unstable LSP can
still be a dark matter\index{Dark matter} component of the universe
provided it is sufficiently long-lived, so as to retain most of its
primordial abundance until the present time -- but this requires extremely
small values of the \Rp\ couplings. A LSP with lifetime shorter
than a fraction of the age of the universe, on the other hand, would now
have disappeared almost completely and can no longer play a r\^ole as
a dark matter\index{Dark matter} component of the universe.  
In this case however, the constraints associated with experimental searches
for anormalously heavy protons or superheavy isotopes no longer apply,
and the LSP can be any superpartner -- not necessarily an electrically
neutral and uncoloured particle. We shall though restrict ourselves to the case
of a neutralino LSP in the following.

Depending on the lifetime $\tau_\tchi^0$ of the LSP (i.e. depending on the
strength of the \Rp\ couplings responsible for its decay), different types
of cosmological constraints apply. The decays of a long-lived LSP, with a
lifetime comparable to, or slightly larger than, the present age $t_0$
of the universe, $\tau_\tchi^0 \gtrsim t_0$, can produce an excess of
particles such as antiprotons or positrons in our galaxy at a level
incompatible with observations. To avoid this problem, one must require
$\tau_\tchi^0 \gg t_0$, i.e. extremely small values of the trilinear
\Rp\ couplings, at the level of ${\cal O} (10^{-20})$ or below.
These very strong constraints do not
apply, of course, when the LSP lifetime is shorter than the age of the
universe. In this case, the LSP must decay sufficiently quickly so that
its late decays do not modify the light element abundances successfully
predicted by Big-Bang nucleosynthesis. This constraint results in an
upper bound on $\tau_\tchi^0$, or equivalently on a lower bound on trilinear
\Rp\ couplings of the order of ${\cal O} (10^{-12})$ ~\cite{kim98}.
For comparison, these couplings are required to be larger than
${\cal O} (10^{-8})$ for the LSP to decay inside a laboratory detector.

We now present a more detailed discussion of the constraints originating 
from nucleosynthesis. The decay of an unstable relic particle after the
nucleosynthesis epoch would have produced electromagnetic
and/or hadronic showers that could have either dissociated or created light
nuclei \cite{lindley}. Hence, in order not to destroy the
predictions of Big-Bang nucleosynthesis, the LSP lifetime, if not
greater than the age of the universe, must not exceed some upper limit.
Focusing on the constraints arising from deuterium photo-dissociation,
Kim et al.~\cite{kim98} estimate the maximal allowed lifetime to be
$(\tau _{\tchi^0})_{max} \simeq 2.24 \times 10^{7} \text{s}\, /
\left[\, 4.92 + \ln\, (m_{\tchi^0} / 1 \GeV) - \ln\, (n_B /n_\g )\, \right]$. 
Imposing that the neutralino LSP lifetime associated to its decays via
trilinear \Rp\ couplings is shorter than the above value leads to a lower
bound of the order of $10^{-12}$ on a weighted sum of squared couplings.
As an example, for a $60 \GeV$ photino-like neutralino, assuming
a universal sfermion mass of $1 \TeV$, the constraint reads: 
\begin{eqnarray}
  & 0.12 \sum_{i,j,k} \vert \l_{ijk} \vert^2\,
  +\, 0.31 \sum_{i,j \neq 3,k} \vert \l'_{ijk} \vert^2\,
  +\, 0.04 \sum_{i,k} \vert \l'_{i3k} \vert^2 &  \nonumber \\
  & +\, 0.23 \sum_{i<j, k \neq 3} \vert \l''_{ijk} \vert^2\
  >\  7.7 \times 10^{-24}\ . &
\end{eqnarray}

Let us now discuss the constraints applying to a neutralino LSP with a
lifetime greater than the age of the universe ($\tau_\tchi^0 > t_0$).
A first set of constraints comes from the production of antiprotons through
LSP decays mediated by the $\lambda'_{ijk}$ and $\lambda''_{ijk}$
couplings~\cite{baltz98}. The observed flux of cosmic rays antiprotons
places a strong bound on such decays, resulting in stringent upper limits
on the corresponding \Rp\ couplings:
\begin{equation}
  \lambda'_{ijk},\, \lambda''_{ijk}\ <\
  \left( 10^{-24} - 10^{-19} \right)\ ,
\end{equation}
for all generation indices, exclusive of $\l''_{3jk}$ in the case where
the LSP is lighter than the top quark.
The upper bound on a given coupling strongly depends on the model parameters
(especially on the neutralino and squark masses), but is always
smaller by some 3 orders of magnitude than the upper bound corresponding
to the condition that the LSP lifetime is greater than the age of the
universe, $\tau_\tchi^0 > t_0$ (see Ref. \cite{baltz98} for details).

A very long-lived LSP neutralino can also produce positrons
through the three-body decays $\tchi^0 \to e^+ + 2\, \mbox{fermions}$,
which can be induced both by trilinear and bilinear \Rp\ interactions.
The experimentally measured positron flux in our
galaxy imposes the following bound on the corresponding partial lifetime
of the neutralino~\cite{berezinsky91}:
$\tau (\tchi^0 \to e^+ + 2\, \mbox{fermions}) / t_0 > 6 \times 10^{10}\,
h\, (m_{\tchi^0} / 100 \GeV)^\ud \ ({\tilde m} / 100 \GeV)^\ud$, where
all sfermion masses have been set to $\tilde m$, and $h$ is the reduced
Hubble parameter defined by $H_0 = 100 \ h\,\text{km/s/Mpc}$.
This leads to stringent upper bounds on all trilinear and bilinear \Rp\
superpotential couplings~\cite{berezinsky98}:
\begin{eqnarray}
  \lambda_{ijk},\, \lambda'_{ijk},\, \lambda''_{ijk} & < &
    4 \times 10^{-23}\ N_{1l}^{-1}\
    \left( \frac{m_{\tilde f}}{100 \GeV} \right)^2
    \left( \frac{m_{\tchi^0}}{100 \GeV} \right)^{-9/8}
    \left( \frac{1 \GeV}{m_f} \right)^{1/2}\ ,  \nonumber \\
  \mu_i & < & 6 \times 10^{-23 }\ N_{1l}^{-1}\
    \left( \frac{m_{\tchi^0}}{100 \GeV} \right)
    \left( \frac{\tilde m}{100 \GeV} \right)^{-7/4} \GeV\ ,
\end{eqnarray}
where $ m_f$ is the emitted fermion mass and the $N_{1l} \ [l=3,4] $
parametrize the amount of the higgsino components in the neutralino.

\subsection{Gravitino Relics}

\label{subsec:mc6}

It is well known that supergravity theories are plagued with a cosmological
gravitino problem. Indeed, since the gravitino interacts only gravitationally,
it has a very small annihilation cross-section and tends to overclose the
universe; or, if it is unstable, to destroy the successful predictions of
Big-Bang nucleosynthesis through its late decays.
In the case of a stable gravitino (or a quasi-stable one with a lifetime
longer than the age of the universe), the annihilation rate is too weak to
prevent the relic energy
density of heavy gravitinos from exceeding the critical energy density
\cite{weinberg83,fayetmrnd82}. Then gravitinos must be very light,
$m_{3/2} \lesssim 1$ keV \cite{pagels}, in order for their relic abundance
not to overclose the universe\footnote{Because the effective strength of
its couplings are fixed by the ratio $G_N / m_{3/2}^2$, a gravitino heavier
than a few eV would have extremely small interaction cross-sections and
decouple very early, allowing for its residual abundance to be higher than
that of a neutrino with the same mass \cite{fayet79,fayetmrnd81}. The upper
limit on the mass of such a light gravitino, obtained by demanding that its
relic energy density be less that the critical density, is then increased
(as compared to the corresponding limit for neutrinos), up to $\,\sim 1 $ keV
\cite{pagels} \,--  its precise value depending on the number of particle
species in thermal equilibrium at the gravitino decoupling time.}.
In the case of an unstable gravitino, its decay must occur sufficiently
early so as not to affect nucleosynthesis. Indeed, if the gravitino decays
after nucleosynthesis, its decay products will either dissociate or create
light nuclei and modify their relative abundances, thus destroying the
agreement between Big-Bang nucleosynthesis predictions and observations.
Furthermore the entropy release subsequent to gravitino decays will wash
out the baryon asymmetry and spoil the concordance between the observed
baryon-to-photon ratio and the light nuclei abundances. The second problem
can be evaded if the gravitino is heavier than about $10^4$ GeV
\cite{weinberg83}. This lower bound assumes that the gravitino is not the
LSP, so that it can decay to lighter \susyq\ partners; if the gravitino is
the LSP and decays via \Rp\ channels, it should be even heavier -- but
then all superpartners should be extremely heavy. To summarize, there is no
cosmological gravitino problem if $m_{3/2} \lesssim 1$ keV, or if the
gravitino is unstable and heavy ($m_{3/2} \gtrsim 10$ TeV if it is not
the LSP).

The above constraints, however, were derived within standard cosmology
(without inflation), and can be relaxed if there is an inflationary phase
which dilutes the gravitino abundance \cite{khlopov84,ellis84_bis,moroi93}.
In this case one still has to face a cosmological gravitino problem
associated with the gravitinos produced during the reheating phase after
inflation, whose abundance is essentially proportional to the reheating
temperature $T_R$.
If the gravitino is stable, requiring that its relic energy density is less
than the critical density therefore results in an upper bound on $T_R$.
If it is unstable, its decays should not affect the light nuclei abundances
successfully predicted by Big-Bang nucleosynthesis, which requires
values of $T_R$ lower than in the case of a stable gravitino.
One typically finds $T_R \lesssim 10^7$ GeV for $m_{3/2} \sim 100$ GeV if
the gravitino is not the LSP \cite{cyburt03}. Such a stringent upper bound
is problematic for standard inflationary models \cite{linde90,lyth99}, which
generally predict much higher values of the reheating temperature.

Let us consider in greater detail the case of an unstable gravitino.
All decay rates of the gravitino are proportional to Newton's constant
$G_N = 1 / M^2_P$, and may be expressed as
$\,\G_ {\tilde G } \approx \,\a_ {\tilde G }\ \,m_{3/2 }^3/ M_P^2 \,$,
where $\a_ {\tilde G }\,$ is a dimensionless coefficient.
The fastest possible decay modes, for which the coefficient $\a_{\tilde G}$
is of order one, are the $R$-parity conserving two-body decay modes, such as
${\tilde G} \to \tilde \chi^\mp +W^\pm$,
${\tilde G} \to \tilde \chi^0+ \g (Z)$ and
${\tilde G} \to \tilde l^\mp + l^\pm$.
These are allowed only if the gravitino is not the LSP. In the presence
of \Rp\ interactions, the gravitino can also decay into channels solely
comprising the ordinary ($R$-even) particles \cite{farrar83}, but with
much smaller rates than the $R$-parity conserving modes
($\a_ {\tilde G } \ll 1$) due to the smallness of the \Rp\ couplings.
As a results, the \Rp\ decay channels are relevant only for the case of
a gravitino LSP, on which we shall concentrate now.

The case of bilinear $R$-parity violation has been discussed in
Ref. \cite{takayama2000}.
Assuming that the lightest neutralino is essentially bino-like,
the dominant decay mode of the gravitino LSP is ${\tilde G } \to \nu \gamma$,
for which $\a_ {\tilde G } \simeq \frac{1}{32 \pi}\, \cos^2 \theta_W m_\nu
/ m_{\tilde \chi^0_1}$, where $m_\nu$ is the neutrino mass generated at
tree level by the bilinear \Rp\ terms (see chapter \ref{chap:neutrinos}).
The experimental and cosmological constraints on neutrino masses then
imply that the gravitino lifetime is much longer than the age of the universe,
even for a gravitino mass as high as $100$ GeV. The gravitino
relic abundance and mass are further constrained by the requirement that the
photon flux produced in gravitino decays does not exceed the observed diffuse
photon background; for a relic abundance in the relevant range for dark matter
and $m_\nu \sim 0.07$ eV, this implies $m_{3/2} \lesssim 1$ GeV. Thus, in the
presence of bilinear $R$-parity violation (at the level required to explain
atmospheric neutrino data), a gravitino LSP can constitute the dark matter
of the universe only if it is lighter than about $1$ GeV, assuming
in addition that the reheating temperature is low enough for the
gravitino relic density to fall in the range relevant for dark matter.
The case of trilinear $R$-parity violation has been discussed in Ref.
\cite{moreau2002}. Assuming standard cosmology (without an inflationary
phase), a gravitino LSP which decays via trilinear \Rp\ couplings
$\lambda_{ijk}$, $\lambda'_{ijk}$ or $\lambda''_{ijk}$ can evade the
relic abundance problem, but it is excluded by nucleosynthesis constraints,
unless the gravitino mass is unnaturally large. This abundance problem,
however, can be solved by inflation. To conclude, $R$-parity violation
does not seem to provide a natural solution to the cosmological gravitino
problem.

Still a gravitino with \Rp\ decay can have interesting implications in
astrophysics and cosmology. As mentioned above, nucleosynthesis severely
constrains the possibility of a late decaying massive particle, and the
constraint is particularly strong for an unstable gravitino. However one
can consider an alternative scenario to Big-Bang nucleosynthesis which
relies
on such a particle \cite{dimo88}. In this scenario, light element 
production takes place when the hadronic decay products interact with the
ambient protons and $^{4}He$. In order to reproduce the observed abundances,
very specific properties of the decaying particle are required; it must in
particular decay after nucleosynthesis and have a small baryonic branching
ratio, $r_B \sim 10^{-2}$. The candidate proposed in Ref. \cite{halltasi}
is a massive, not LSP gravitino decaying to hadrons predominantly via
the $L$-violating trilinear \Rp\ couplings. One can arrange for
the required small baryonic branching ratio in gravitino decays by
considering a sneutrino LSP and assuming non-vanishing \Rp\ couplings
$\l_{131}$, $\l_{232}$ and $\l'_{3jk}$ with $\l'_{3jk} / \l_{131} \sim 0.1$
and $\l'_{3jk} / \l_{232} \sim 0.1$. The gravitino then undergoes the
following cascade decays: $ {\tilde G} \to \nu \bar {\tilde \nu}, \
\tilde \nu \to ( e\bar e, \mu \bar \mu , \cdots ) + (q \bar q ) $.
The predicted abundances of $D$, $^{4}He$ and
$^{7}Li$ can be made to match the observations even for a universe closed
by baryons \cite{dimo88,jedamzik00} (i.e. with $\Omega_B \simeq 1$), but
the scenario overproduces $^{6}Li$ and is therefore disfavoured
\cite{sarkar02}.
\index{Relic particles|)}

\section{Cosmological Baryon Asymmetry}


\label{secxxx6b}

\index{Baryon asymmetry|(}

\subsection{Baryogenesis from {\boldmath{$R$}}-Parity-Violating Interactions}


\label{secxxx6d}

\index{Baryogenesis|(}

Generating the observed baryon asymmetry of the universe is one
of the challenges of particle physics. In order for a baryon-antibaryon
asymmetry to be dynamically generated in an expanding universe, three
necessary conditions, known as Sakharov's conditions \cite{sakharov67},
must be met: (i) baryon-number violation; (ii) $C$ and $CP$ violation;
(iii) departure from thermal equilibrium. In principle, all three
ingredients are already present in the \SM, where baryon number is
violated by nonperturbative processes known as sphalerons
\cite{manton,klinkhamer84,kuzmin85} (which violate $B+L$ but preserve $B-L$ and
are in thermal equilibrium above the electroweak scale), and the departure
from thermal equilibrium could be due to the electoweak phase transition.
This leads to the standard electroweak baryogenesis scenario, which however
has been excluded as a viable mechanism in the \SM \ \cite{farrar93}, and works only in a
restricted portion of the \MSSM\ parameter space \cite{quiros01}. Other
mechanisms, such as leptogenesis \cite{fukugita86}, in which a lepton
asymmetry is generated by out-of-equilibrium decays of heavy Majorana
neutrinos and then partially converted into a baryon asymmetry by sphaleron
transitions, or Affleck-Dine baryogenesis \cite{affleck85}, offer possible
alternatives to the standard scenario. In this section, we review several
attempts to generate the observed baryon asymmetry from $R_p$-violating
interactions.

A first class of scenarios uses the trilinear \Rp\ couplings
$\lambda''_{ijk}$ and their associated $A$-terms $A''_{ijk}$
as the source of baryon-number violation. The $\lambda''_{ijk}$
couplings induce decays of a squark (resp. an antisquark) into
two antiquarks (resp. two quarks), which violate baryon number by
one unit. The differences between the various scenarios that rely on this
process reside in the way departure from equilibrium is realized,
and in the mechanism that produces squarks.

In the scenario proposed by Dimopoulos and Hall~\cite{halldimo},
squarks are produced far from thermal equilibrium at the end of inflation
as decay products of the inflaton field.
Their subsequent decays into quarks and antiquarks
induced by the \Rp\ couplings $\lambda''_{ijk}$ generate a baryon asymmetry
directly proportional to the $CP$ asymmetry in these decays,
$\Delta \Gamma_{\tilde q} =
(\Gamma (\tilde q_R \rightarrow \bar q_R \bar q_R) -
\Gamma (\tilde q^c_L \rightarrow q_R q_R)) /
(\Gamma_{\tilde q_R} + \Gamma_{\tilde q^c_L})$. The
dominant contribution to this $CP$ asymmetry comes from the interference
between tree-level and two-loop diagrams involving the $CP$-violating
phases present in the $A$-terms (in the convention in which gaugino mass
parameters are real).
In order for this scenario to work, the reheating temperature $T_R$ must be
extremely low (typically $T_R \lesssim 1$ GeV) so that scattering processes
induced by the $\lambda''_{ijk}$ couplings, which could dilute the baryon
asymmetry created in squark decays, are suppressed. 

In the scenario considered by Cline and Raby \cite{cline91}, the departure
from thermal equilibrium is provided by the late decays of the gravitino,
and the baryon asymmetry is produced in two steps. First an asymmetry
in the number densities of squarks and antisquarks is produced by
$CP$-violating decays of the neutral gauginos produced in the out-of
equilibrium decays of the gravitino, or by $CP$-violating decays
of the gravitino itself, but the
total baryon number remains conserved due to an opposite asymmetry in the
number densities of quarks and antiquarks. As in the previous model, the
source of $CP$ violation is the relative phase between the gaugino mass
parameters and the $A$-terms, but now the $CP$ asymmetry arises from the
interference between tree-level and one-loop diagrams. In the case of
gluons, the asymmetry induced by non-vanishing $\lambda''_{323}$ and
$A''_{323}$ couplings reads:
\begin{equation}
  \Delta \Gamma_{\tilde g}\ \equiv\
  \frac{\Gamma (\tilde g \rightarrow t \tilde t^c) -
  \Gamma (\tilde g \rightarrow \bar t \tilde t)}{\Gamma_{\tilde g}}\
  \approx\ \frac{\lambda''_{323}}{16 \pi}\,
  \frac{\Im (A^{\prime \prime \star}_{323} m_{\tilde g})}
  {|m_{\tilde g}|^2}\ ,
\end{equation}
where $\Im$ denotes the imaginary part.
In the second step, this asymmetry gets partially converted into a baryon
asymmetry by the $B$-violating decays of the (anti)squarks
induced by the $\lambda''_{ijk}$ couplings -- it is assumed here that the
squarks are lighter than the gauginos, so that their only relevant tree-level
decay mode is into two quarks. At the time where the squark decays occur, the
scattering processes that could erase the $CP$ asymmetry are highly
suppressed by low particle densities. For this scenario to work,
the reheating temperature must be high enough for the
required gravitino abundance to be regenerated after inflation, i.e. typically
$T_R \gtrsim 10^{15}$ GeV. In addition, the gravitino should be heavy enough
($m_{3/2} \gtrsim 50$ TeV) so that its decay products do not affect
nucleosynthesis.

To generate the observed baryon asymmetry, the previous two
scenarios require the $CP$ asymmetry to be close to its maximal allowed
value, i.e. the $CP$-violating phase should be close to the upper bound
associated with the electric dipole moment of the neutron, and the
dominant $B$-violating coupling should be of order one. The
source of baryon-number violation must then be a coupling
that is not constrained by $\Delta B = 2$ processes such as
neutron-antineutron oscillations and heavy nuclei decays (see chapter
\ref{chap:indirect}), e.g. $\lambda''_{323}$.

A variant of the scenario considered by Cline and Raby, which also works
for smaller values of the $CP$ asymmetry and for lower reheating temperatures,
has been proposed by Mollerach and Roulet \cite{moller92}. In this scenario,
a large, out-of-equilibrium population of gluinos is created in the decays
of heavy axinos\index{Axion!axino} ($\tilde a$) and/or 
saxinos\index{Axion!saxino} ($s$), the
fermionic superpartners and scalar partners of the pseudoscalar
axions\index{Axion!axion}, respectively.
This requires $m_{\tilde a} > m_{\tilde g}$ ($m_s > 2 m_{\tilde g}$), so that
the decay channel $\tilde a \rightarrow g \tilde g$
($s \rightarrow \tilde g \tilde g$) be kinematically allowed. The axinos
(saxinos)\index{Axion!axino}\index{Axion!saxino} decay at a temperature 
around $1$ GeV and thus do not interfere with nucleosynthesis.
The baryon asymmetry is then generated in two steps
from gluino decays, like in the scenario of Cline and Raby, but the present
scenario is much more efficient and the $CP$ asymmetry is not required to be
close to its maximal value. The $CP$-violating phase can thus be small, and
the $B$-violating coupling can be smaller
than one. In the presence of an inflationary phase, 
the observed amount of baryon asymmetry can be obtained for
reheating temperatures as low as $10^{4}$ GeV (in the case of a large
$CP$ asymmetry), due to the fact that 
(s)axinos\index{Axion!axino}\index{Axion!saxino}
are regenerated much more efficiently than gravitinos.

In another scenario studied by Adhikari and Sarkar~\cite{adhikari},
the baryon asymmetry is generated in out-of-equilibrium decays of the
lightest neutralino induced by the $\lambda''_{ijk}$ couplings,
$\tilde \chi^0_1 \rightarrow u_{iR}\, d_{jR}\, d_{kR}$, rather than
in squark decays. The $CP$ asymmetry in these decays arises at the one-loop
level, and can be large even for small values of the $B$-violating
couplings, which are required by the out-of-equilibrium condition. Unlike
in the previous scenarios, $CP$ violation is due to the complexity of the
$\lambda''_{ijk}$ couplings. The sfermions are assumed to be much heavier
than the lightest neutralino, so that the former have already decayed at
the time where the latter decays, and their \Rp\ decay modes do not erase
the generated baryon asymmetry. The other processes that could dilute the
baryon asymmetry, such as the $\Delta B = 1$ scattering processes
$u_{iR}\, d_{jR} \rightarrow \bar d_{kR}\, \tilde \chi^0_1$, must be out of
equilibrium. This requires rather small values of the $\lambda''_{ijk}$
couplings; still Adhikari and Sarkar estimate that it is possible
to generate the observed baryon asymmetry of the
universe for values of the $\lambda''_{ijk}$ couplings in the
$10^{-4} - 10^{-3}$ range (see however the footnote below).

In a second class of scenarios, the \Rp\ couplings $\lambda_{ijk}$ and
$\lambda'_{ijk}$, which violate lepton number, are used to create a
lepton asymmetry at the electroweak scale. This one is then partially
converted into a baryon asymmetry by the sphaleron processes\footnote{It
has been noted in Ref. \cite{hambye01} that the lepton or baryon asymmetry
generated at the electroweak scale from the decay of gauge non-singlet
particles can be strongly suppressed due to the efficiency of the
annihilation of these particles into two gauge bosons. This effect had
been overlooked or underestimated in the scenarios discussed below, as well
as in the scenario of Ref. \cite{adhikari}.}.

In the scenario considered by Masiero and Riotto~\cite{masiero92}, the
lepton asymmetry is generated through the \Rp\ (and $CP$-violating) decays 
of the LSP, which is produced out of equilibrium in bubble collisions
during the electroweak phase transition. The origin of $CP$ violation is
simply the presence of phases in the \Rp\ couplings $\lambda_{ijk}$ and
$\lambda'_{ijk}$. Assuming that the LSP is the
lightest neutralino, the dominant decay channel is expected to be
$\tilde \chi^0_1 \rightarrow l^-_i t \bar d_k$
($l^+_i \bar t d_k$), and the $CP$ asymmetry is given by:
\begin{equation}
  \epsilon\ =
    \frac{\Gamma (\tilde \chi^0_1 \rightarrow l^- t \bar d)
    - \Gamma (\tilde \chi^0_1 \rightarrow l^+ \bar t d)}
    {\Gamma (\tilde \chi^0_1 \rightarrow l^- t \bar d)
    + \Gamma (\tilde \chi^0_1 \rightarrow l^+ \bar t d)}\ \simeq\
  \frac{1}{16 \pi}\ \frac{\sum_{i,k,l,m,n} \Im \,
    (\lambda^{\prime \star}_{inl} \lambda'_{mnl} \lambda^{\prime \star}_{m3k}
    \lambda'_{i3k})} {\sum_{i,k} |\lambda'_{i3k}|^2}\ .
\end{equation}
For this mechanism to work, the phase transition must be first order,
i.e. it must proceeds by nucleation of bubbles of true vacuum
in the unbroken phase.
Furthermore, contrary to what is required in the standard electroweak
baryogenesis scenario, the sphaleron processes must remain in equilibrium
until the temperature drops below the critical temperature.
This can indeed be the
case if the Higgs sector of the \SSM\ is extended by the addition of
one or more singlet superfields. Finally, the \Rp\ interactions responsible
for the generation of the lepton asymmetry must be in equilibrium at the
electroweak scale, while the $\Delta L = 1$ scattering processes
$l_{iL} \bar d_{kR} \rightleftharpoons \tilde \chi^0_1\, \bar t$, which could
wash it out, must be out of equilibrium. The first condition turns into 
a lower bound on the $\lambda'_{i3k}$ couplings:
\begin{equation}
  |\lambda'_{i3k}|\ \gtrsim\ 5.3 \times 10^{-5}
    \left( \frac{500\, \mbox{GeV}}{m_{\tilde \chi^0_1}} \right)
    \left( \frac{T_0}{150\, \mbox{GeV}} \right)
    \left( \frac{m_{\tilde t_L}}{1\, \mbox{TeV}} \right)^2\ ,
\label{eq:MR_condition1}
\end{equation}
and the second one into an upper bound on the same couplings:
\begin{equation}
  |\lambda'_{i3k}|\ \lesssim\ 2.4 \times 10^{-4}
    \left( \frac{500\, \mbox{GeV}}{m_{\tilde \chi^0_1}} \right)
    \left( \frac{150\, \mbox{GeV}}{T_0} \right)^{1/2}
    \left( \frac{m_{\tilde t_L}}{1\, \mbox{TeV}} \right)^2\ .
\label{eq:MR_condition2}
\end{equation}
where $T_0$ is the critical temperature of the electroweak phase transition. 
For values of the \Rp\ couplings
in the range delimited by Eqs. (\ref{eq:MR_condition1}) and
(\ref{eq:MR_condition2}), this scenario can generate the observed baryon
asymmetry, provided that the top squark is heavy. Indeed, relatively high
values of $\lambda'_{i3k}$ are needed to provide a large enough
$CP$ asymmetry (typically $|\lambda'_{i3k}| \sim 10^{-2}$, which requires
$m_{\tilde t_L} \gtrsim 6\, (3)$ TeV if
$m_{\tilde \chi^0_1} \simeq 500\, (100)$ GeV).

Adhikari and Sarkar~\cite{sarkarba} noticed that, in the presence of
flavour-violating couplings of the neutralinos, the generation of the lepton
asymmetry can be much more efficient.
This is especially so if some sfermions are not
much heavier than the lightest neutralino, so that they can be produced
in bubble collisions and contribute to the lepton asymmetry through their
\Rp\ decays $\tilde \nu_{iL} \rightarrow d_{kR}\, \bar d_{jL}$,
$\tilde d_{jL} \rightarrow d_{kR}\, \bar \nu_{iL}$,
$\tilde d_{kR} \rightarrow d_{jL}\, \bar \nu_{iL}$, \dots.

Hambye, Ma and Sarkar \cite{hambye00} consider another scenario
in which a lepton asymmetry is created in out-of-equilibrium decays of
the lightest neutralino into a charged Higgs boson and a lepton singlet,
$\tilde \chi^0_1 \rightarrow l^{\pm}_R\, h^{\mp}$. The $CP$ asymmetry is
proportional to the square of the parameter $\xi$ that accounts for the
mixing \index{Mixing!higgs--slepton}
between the slepton singlets and the charged Higgs boson, and the
$CP$-violating phase comes from the neutralino mass matrix.
The same parameter $\xi$ controls the out-of-equilibrium condition
for $\Delta L = 2$ processes that could erase the generated lepton
asymmetry. For this mechanism to produce enough baryon asymmetry,
$\xi$ must be close to the upper bound associated with this condition.
This can be achieved by introducing non-holomorphic \Rp\ soft terms of the
form $H^+_u H_d\, \tilde l^c$, which contrary to the standard \Rp\ soft
terms are not constrained by experimental data.

\index{Baryogenesis|)}

\subsection{Survival of a Baryon Asymmetry in the Presence of 
{\boldmath{\Rp}} Interactions}


\label{secxxx6c}

We have seen in the previous subsection that $R$-parity violation may
be at the origin of the baryon asymmetry of the universe. In general,
however, \Rp\ interactions are considered as a danger since they can
erase a baryon asymmetry that would be present before the
electroweak phase transition. In order to avoid this, one
requires \Rp\ interactions to be out of equilibrium above the
critical temperature of the electroweak phase transition,
which results in strong upper bounds on the \Rp\ couplings. 

Let us first review the standard conditions for a baryon asymmetry generated
during the thermal history of the universe to be preserved until today,
in the absence of $B$- and $L$-violating interactions.
Above the critical temperature $T_c \sim 100$ GeV and up to temperatures
of the order of $10^{12}$ GeV, nonperturbative processes which violate
$B+L$ but preserve $B-L$ are in thermal equilibrium
\cite{klinkhamer84,kuzmin85}. These processes, known as sphalerons,
tend to erase any $B+L$ asymmetry present in the high temperature phase,
in such a way that a baryon asymmetry can persist only if it corresponds
to a $B-L$ asymmetry. More precisely, the existence of sphaleron processes
in thermal equilibrium with the interactions of the \SM\ (or of the \MSSM)
leads to the following proportionality relation between the $B$ and
$B-L$ asymmetries \cite{khlebnikov88,harvey90}:
\begin{equation}
  Y_B\ =\ \frac{24 + 4 N_H}{66 + 13 N_H}\ Y_{B-L}\ ,
\label{eq:Y_B}
\end{equation}
where $Y_B \equiv (n_B - n_{\bar B}) / s$
($Y_L \equiv (n_L - n_{\bar L}) / s$), with $n_B$ ($n_L$) the baryon
(lepton) number density and $s$ the entropy density of the universe,
and $N_H$ is the number of Higgs doublets. For the \SM\ with one Higgs
doublet, one has $Y_B / Y_{B-L} = 28/79$, while for the \MSSM\ one has
$Y_B / Y_{B-L} = 8/23$. As a result, there are only two viable possibilities
for generating the baryon asymmetry of the universe: one can generate it
at or after the electroweak phase transition (when sphaleron transitions
are suppressed), or above the electroweak phase transition in the form
of a $B-L$ asymmetry.

The above discussion must be modified in the presence of interactions
that violate $B-L$, such as \Rp\ interactions
\cite{salati,campbell91,fischler,campbell92,dreiross,inui94}.
Assuming that a $B-L$ asymmetry is generated by some mechanism above the
electroweak phase transition, the only possibility for it to be preserved
is that the $B-L$-violating interactions be out of equilibrium, i.e. that
their characteristic timescale be longer than the age of the universe.
This condition can  be written as $\Gamma_{B-L} < H$, where $\Gamma_{B-L}$
is the rate of a typical $B-L$-violating process, and $H$ is the Hubble
parameter. In the case of $R$-parity violation, this yields strong upper
bounds on \Rp\ parameters. For the trilinear couplings, the processes that
yield the best bounds are the decays of squarks and sleptons into two
fermions or sfermions, and the corresponding rates are given by
\cite{campbell91,campbell92}:
\begin{equation}
  \Gamma_{\hat \lambda}\ \simeq\ 1.4 \times 10^{-2}\,
  |\hat \lambda|^2\, \frac{m^2_\Phi}{T}\ , \qquad
  \Gamma_{\hat A}\ \simeq\ 1.4 \times 10^{-2}\,
  \frac{|\hat A|^2}{T}\ ,
\end{equation}
where $m_\Phi$ is the mass of the decaying sfermion, $\hat \lambda$ stands
for any of the couplings $\lambda_{ijk}$, $\lambda'_{ijk}$ or
$\lambda''_{ijk}$, and $\hat A$ for $A_{ijk}$, $A'_{ijk}$ or
$A''_{ijk}$. Since the temperature dependence of the Hubble parameter
is given by $H \simeq 1.66\, g^{1/2}_* T^2 / M_P$ (where $g_*$ is the
number of effectively massless degrees of freedom at the temperature $T$),
the out-of equilibrium condition $\Gamma < H$ is more easily satisfied
at high temperature, and the best bounds are obtained for $T \sim T_c$.
Assuming $m_\Phi \sim T_c \sim 100$ GeV, one obtains \cite{campbell92}:
\begin{eqnarray}
  |\lambda_{ijk}|,\ |\lambda'_{ijk}|,\ |\lambda''_{ijk}|
  & \lesssim & 10^{-7}\ ,
\label{eq:lambda_bound} \\
  \qquad |A_{ijk}|,\ |A'_{ijk}|,\ |A''_{ijk}| & \lesssim
  & 10^{-5}\, \mbox{GeV}\ .
\label{eq:A_bound}
\end{eqnarray}
For heavier sfermions, these constraints are slightly weakened, e.g. for
$T \sim m_\Phi \sim 1$ TeV the upper bounds (\ref{eq:lambda_bound}) are
increased by a factor of $3$. Also, the bounds (\ref{eq:lambda_bound})
and (\ref{eq:A_bound}) were derived under the assumption that the
decays are kinematically allowed, which is not necessarily the case for
the decays induced by $A$-terms. The bounds on the \Rp\ $A$-terms associated
with scattering processes can be estimated to be one order of magnitude
weaker \cite{davidson98_talk}. 

For bilinear \Rp\ couplings, the relevant interaction rates are 
\cite{campbell92}:
\begin{eqnarray}
  & \Gamma_{\mu_i}\ \simeq\ 1.4 \times 10^{-2}\, g^2\,
  \frac{|\mu_i|^2}{\tilde m^2}\, T\ , \qquad
  \Gamma_{B_i}\ \simeq\ 1.4 \times 10^{-2}\, g^2\,
  \frac{|B_i|^2}{\tilde m^4}\, T\ , & \nonumber \\
  & \Gamma_{\tilde m^2_{di}}\ \simeq\ 1.4 \times 10^{-2}\, g^2\,
  \frac{|\tilde m^2_{di}|^2}{\tilde m^4}\, T\ ,
\end{eqnarray}
where $\tilde m$ stands for the relevant scalar or supersymmetric fermion
mass. As usual, the bilinear \Rp\ parameters $\mu_i$, $B_i$ and
$\tilde m^2_{di}$ are expressed in the ($H_d$, $L_i$) basis in which the
sneutrino \VEVs\
$< \tilde \nu_i >$ vanish and the charged lepton Yukawa couplings are
diagonal (see subsection \ref{subsec:H_L_basis}).
Assuming $\tilde m \sim T_c \sim 100$ GeV, one obtains the bounds:
\begin{equation}
  |\mu_i|\  \lesssim\ 2 \times 10^{-5}\, \mbox{GeV}, \qquad
  |B_i| ,\ |\tilde m^2_{di}|\ \lesssim\ 2 \times 10^{-3}\, \mbox{GeV}^2\ .
\label{eq:m2_bound}
\end{equation}
There is however a little subtlety in deriving these bounds,
due to the fact that the thermal mass eigenstate basis for the $H_d$ and
$L_i$ fields above the electroweak phase transition is not the same as
the zero temperature mass eigenstate basis \cite{davidson97,davidson98}.
For a discussion of the cosmological bounds on \Rp\ couplings in terms
of basis-independent quantities, see Ref. \cite{davidson98}.

As discussed in section \ref{sec:alternatives}, baryon- and 
lepton-number violation may also proceed through non-renormalizable
operators, which are expected to be generated from some more fundamental
theory than the \SSM. Writing a generic
non-renormalizable operator of dimension $4+n$ as
${\cal O}_{4+n} / M^n_{4+n}$, where all couplings have been absorbed
in the definition of the mass scale $M_{4+n}$, one can estimate the
rate of the $B-L$-violating processes induced by this operator to be
$\Gamma \sim 10^{-3} T^{2n+1} / M^{2n}_{4+n}$. The requirement that
these processes are out of thermal equilibrium at the temperature $T$
yields a lower bound on the mass scale $M_{4+n}$:
\begin{equation}
  M_{4+n}\ \gtrsim\ 10^{2+ \frac{6}{n}}\, \mbox{GeV}
  \left( \frac{T}{100\, \mbox{GeV}} \right)^{1 - \frac{1}{2n}}\ .
\label{eq:nr_bound}
\end{equation}
The interaction rates for non-renormalizable operators increase faster
with temperature than the Hubble parameter; therefore, the strongest
bounds on $M_{4+n}$ are obtained at high temperature. In
principle\footnote{For operators that do not involve the right-handed
electron field, it may be enough to require that the corresponding interactions
are out of equilibrium up to $T_{e_R} \sim (10^4-10^5)$ GeV
\cite{campbell92_bis,cline94}.
Indeed, above $T_{e_R}$, the electron Yukawa coupling is out of equilibrium,
so that an asymmetry stored in $e_R$ cannot be transferred to other
particle species. In baryogenesis scenarios that generate an $e_R$
asymmetry, this can protect the baryon asymmetry down to the temperature
$T_{e_R}$.}, the bound (\ref{eq:nr_bound}) should be applied at the
temperature at which the baryon asymmetry has been created, but it is
no longer valid above $T \sim 10^{12}$ GeV, where sphaleron processes
are out of equilibrium. It should be noted
that the bound (\ref{eq:nr_bound}) applies not only to \Rp\ operators,
but also to the $R$-parity conserving operators which violate $B-L$,
such as the superpotential term $\frac{1}{M_5} LH_uLH_u$, which induces
Majorana masses for the neutrinos. The bound on $M_5$ is
$M_5 \gtrsim 10^{8}\, \mbox{GeV}\, (T / 100\, \mbox{GeV})^{1/2}$,
which is compatible with the cosmological bound on neutrino masses
$\sum_i m_{\nu_i} \lesssim 1$ eV \cite{WMAP},
even if the bound on $M_5$ applies up to $T \sim 10^{12}$ GeV.

In the above, we did not make any distinction between couplings
which violate different lepton flavours. However,
since the sphaleron processes preserve each one of the
three combinations $B/3 - L_i$, $i=1,2,3$, a $B-L$ asymmetry generated
before the electroweak phase transition will survive as soon as the
processes violating one of the $B/3 - L_i$ are out of equilibrium,
even if the other two are violated by processes in thermal equilibrium.
This means that the bounds (\ref{eq:lambda_bound}) to
(\ref{eq:nr_bound}) must be satisfied by all
baryon-number-violating couplings and by the couplings that violate,
say $L_e$, while the \Rp\ couplings that violate $L_\mu$ or $L_\tau$
can be arbitraly large -- provided however that the sources of 
lepton-flavour violation are out of equilibrium  \cite{dreiross}.
Explicitly, the conditions for preserving a $B/3 - L_e$ asymmetry 
generated before the electroweak phase transition read (for superpartner
masses of the order of $T_c \sim 100$~GeV):
\begin{eqnarray}
  |\lambda_{1jk}|,\ |\lambda_{ij1}|,\ |\lambda'_{1pq}|,\ |\lambda''_{npq}|
  & \lesssim & 10^{-7}\ ,
\label{eq:lambda_L_e_bound}  \\
  |A_{1jk}|,\ |A_{ij1}|,\ |A'_{1pq}|,\ |A''_{npq}| 
  & \lesssim & 10^{-5}\, \mbox{GeV}\ ,
 \label{eq:A_L_e_bound}  \\
 |\mu_1|\  \lesssim\ 2 \times 10^{-5}\, \mbox{GeV}, \qquad
  |B_1| ,\ |\tilde m^2_{d1}| & \lesssim & 2 \times 10^{-3}\, \mbox{GeV}^2\ ,
\label{eq:m2_L_e_bound}
\end{eqnarray}
where $i, j, k \neq 1$.
The \Rp\ couplings that violate $L_\mu$ or $L_\tau$ can be much larger
if the off-diagonal slepton soft terms are small enough, so that they do
not induce lepton-flavour-violating processes at thermal equilibrium
\cite{dreiross,davidson98_talk}:
\begin{equation}
  \left| \frac{(m^2_{\tilde L})_{1j}}{(m^2_{\tilde L})_{11}} \right|\, ,\
  \left| \frac{(m^2_{\tilde l^c})_{1j}}{(m^2_{\tilde l^c})_{11}} \right|\
  \lesssim\ 5 \times 10^{-2}\ ,
  \qquad |A^e_{1j}|\ \lesssim\ 10^{-5}\, \mbox{GeV}\ ,
\end{equation}
where $j = 2$ or $3$. 
%
%
The constraints (\ref{eq:lambda_L_e_bound}) to (\ref{eq:m2_L_e_bound})
can be summarized by saying that baryon-number violation, as well as
lepton-number violation in at least one generation, must be strongly
suppressed.

The constraints presented in this subsection should be regarded as sufficient
conditions for the baryon asymmetry of the universe not to be erased
by \Rp\ interactions, rather than strict bounds. First of all, they do
not apply if the baryon asymmetry of the universe is generated at or after
the electroweak phase transition, like in the standard electroweak
baryogenesis scenario, or in all baryogenesis scenarios discussed
in the previous subsection, where the \Rp\ interactions act as the
source of the baryon asymmetry. Indeed, the sphaleron processes come
out of equilibrium just below the critical temperature, so that they
can no longer erase a $B+L$ asymmetry. Furthermore, there are several
loopholes in the cosmological arguments used to derive the bounds on
\Rp\ couplings displayed above, and these (or some of them)
can be evaded in several baryogenesis scenarios,
even if the baryon asymmetry is generated above the electroweak
phase transition (see e.g. Refs. \cite{dreiross,davids97}).

\index{Baryon asymmetry|)}


\vskip .5cm

In this chapter, we studied the implications of a broken $R$-parity
in cosmology and astrophysics. The most dramatic change with respect
to the $R$-parity conserving \SSM\ is the unstability of the LSP,
which rules it out as a natural candidate for the 
non-baryonic dark matter of the universe,
unless \Rp\ couplings are unrealistically small. An even more damaging
effect of $R$-parity violation is the potential erasure of the baryon
asymmetry of the universe by \Rp\ interactions, if it has been generated
before the electroweak phase transition.
These could be two good reasons to stick to a conserved $R$-parity.
On the other hand, $R$-parity violation can help solving the cosmological
gravitino problem, although with some difficulty, and provide new
mechanisms for generating the baryon asymmetry of the universe at or
after the electroweak phase transition.

\cleardoublepage                                                         %
\chapter{NEUTRINO MASSES AND MIXINGS}                                    %
\label{chap:neutrinos}                                                   %

$R$-parity forbids lepton-number ($L$) violation from renormalizable
interactions. Allowing for violation of $L$-conservation law has several
important effects in the neutrino sector.
The most dramatic implication of non-vanishing couplings
$\lambda_{ijk}$, $\lambda'_{ijk}$ and/or bilinear \Rp\ parameters is the
automatic generation of neutrino masses and mixings. As a consequence,
the possibility that the atmospheric and solar neutrino data, 
now interpreted in terms of neutrino oscillations, be
explained by \Rp\ interactions has motivated a large number of studies
and models. Besides neutrino masses, $R$-parity violation in the lepton
sector leads to neutrino transition magnetic moments, new contributions to
neutrinoless double beta decays, neutrino-flavour transitions in matter
and sneutrino-antisneutrino oscillations. 
In this chapter, we shall concentrate on the question of neutrino masses
and mixings in \susyq\ models without $R$-parity, and on the related issues
of neutrino transition magnetic moments, \Rp-induced neutrino-flavour
transitions in matter and sneutrino-antisneutrino mixing 
\index{Sneutrino!mixing}. Neutrinoless double beta decay will be discussed in
Chapter \ref{chap:indirect}.


\section{{\boldmath{\Rp}} Contributions to Neutrino Masses and Mixings}
\label{sec:problem}
\index{Neutrino!mass|(}
\index{Neutrino!mixing|(}
\subsection{{\boldmath{$R$}}-Parity Violation as a Source of Neutrino Masses}

In order to account for nonzero neutrino masses and mixings, the \SM\
(with two-component, left-handed neutrino fields) has to be supplemented
with additional particles. The simplest possibility is to add right-handed
neutrinos, which either leads to Dirac neutrinos\index{Dirac mass} or, if
the right-handed neutrinos have heavy Majorana masses\index{Majorana mass},
to Majorana neutrinos\index{Majorana mass} through the well-known
seesaw mechanism \cite{seesaw}. Other mechanisms, which directly generate a
Majorana mass\index{Majorana mass} term for the standard two-component
neutrino fields, do not involve right-handed neutrinos but require an
enlarged Higgs sector, involving additional 
$SU(2)_L$\index{Group symmetries!$SU(2)_L$}
triplet \cite{gelmini}
or singlet and doublet~\cite{zee,babu88} Higgs fields.

A new possibility arises in the \SSM. Indeed, in the absence of $R$-parity,
lepton-number-violating couplings induce Majorana masses\index{Majorana mass}
for neutrinos without the need of right-handed neutrinos or exotic Higgs
fields. In other words, {\it \SUSY\ without $R$-parity automatically
incorporates massive neutrinos}. This may be regarded both as an
appealing feature of \Rp\ models and as a potential problem, since the
contribution of \Rp\ couplings may exceed by several orders of magnitude
the experimental bounds on neutrino masses.

One can distinguish between two types of contributions to
neutrino masses and mixings in \susyq\ models without $R$-parity\footnote{For
early discussions on neutrino masses in supersymmetric models with
explicitly or spontaneously broken $R$-parity, see Refs.
\cite{hall84,lee84,dawson85,sneutrino_vev_1,ellis85}.} \cite{hall84}:
(i) a tree-level contribution arising from the neutrino-neutralino mixing
due to bilinear $R$-parity violation;
(ii) loop contributions induced by the trilinear \Rp\ couplings
$\lambda_{ijk}$ and $\lambda'_{ijk}$ and/or bilinear \Rp\ parameters.
These contributions, if present, are generally expected to be
large and potentially in conflict with the experimental limits on neutrino
masses \cite{pdg04},
\begin{equation}
  m_{\nu_e} < 3 \eV\ , \qquad m_{\nu_\mu} < 190 \keV\ , \qquad
  m_{\nu_\tau} < 18.2 \MeV\ ,
\end{equation}
and with the cosmological bound on stable neutrinos,
$\sum_i m_{\nu_i} \lesssim 1 \eV$ \cite{WMAP}. The effective Majorana
masses\index{Majorana mass} generated through mechanisms (i) and (ii)
can be written as (with the neutral gauginos and higgsinos integrated out)
\begin{equation}
  -\, \frac{1}{2}\, M^{\nu}_{ij}\: \bar \nu_{Li}\, \nu^c_{Rj}\
  +\ \mbox{h.c.}\ ,
\end{equation}
where $i,j = 1,2,3$ are generation indices, and the $3 \times 3$ matrix
$M^{\nu}$ is symmetric by virtue of the properties of the charge conjugation
matrix. The relative rotation between charged lepton and neutrino mass
eigenstates defines a lepton mixing matrix\index{Lepton!mixing matrix} $U$, 
$3 \times 3$ and unitary, the Pontecorvo-Maki-Nakagawa-Sakata (PMNS) matrix
\cite{pontecorvo57,MNS}, responsible for neutrino oscillations.
With the conventions $R^e_L M^e R^{e+}_R =
\mbox{Diag}\, (m_{e_1}, m_{e_2}, m_{e_3})$ and $M^{\nu} =
R^{\nu}\, \mbox{Diag}\, (m_{\nu_1}, m_{\nu_2}, m_{\nu_3})\, R^{\nu T}$,
the lepton mixing matrix reads $U = R^e_L R^{\nu}$, and the weak eigenstate
neutrinos $\nu_{\alpha = e, \mu, \tau}$ \, -- i.e. the 
$SU(2)_L$\index{Group symmetries!$SU(2)_L$}
partners of the mass eigenstate charged leptons -- \, are related to the mass
eigenstate neutrinos
$\nu_{i=1,2,3}$ by $\nu_{\alpha} = \sum_i U_{\alpha i}\, \nu_i$.
Here $M^e$ is an effective mass matrix obtained after integrating out the
charged gauginos and higgsinos, which mix with charged leptons through
bilinear $R$-parity violation (see subsection \ref{subsec:H_L_mixing}). In
general it is not simply proportional to the charged lepton Yukawa matrix,
although $M^e = \lambda^e v_d / \sqrt{2}$ remains a good approximation in the
phenomenologically relevant limit of small bilinear $R$-parity violation.

In the following, we shall neglect $C P$ violation in the lepton sector, and
therefore assume the charged lepton Yukawa couplings $\lambda^e_{ij}$ and the
\Rp\ couplings $\mu_i$, $\lambda_{ijk}$ and $\lambda'_{ijk}$ to be real. 
The MNS matrix is then real and can be parametrized by three
angles $\theta_{12}$, $\theta_{13}$ and $\theta_{23}$, responsible for solar,
reactor and atmospheric neutrino oscillations respectively:
\begin{equation}
  U\ =\ \left(
    \begin{array}{ccc}
    c_{13} c_{12} & c_{13} s_{12} & s_{13}  \\
    - c_{23} s_{12} - s_{13} s_{23} c_{12} &
    c_{23} c_{12} - s_{13} s_{23} s_{12} & c_{13} s_{23}  \\
    s_{23} s_{12} - s_{13} c_{23} c_{12} &
    - s_{23} c_{12} - s_{13} c_{23} s_{12} & c_{13} c_{23}
    \end{array} \right) .
\label{eq:parametrization_MNS}
\end{equation}

Let us finally notice that since we are assuming non-vanishing 
$L$-violating couplings, we should make sure that the $B$-violating
couplings $\lambda''_{ijk}$ are absent from the theory -- otherwise the
proton would decay at a much too rapid rate. This can be ensured by imposing,
instead of $R$-parity, the $Z_3$\index{Discrete symmetries!$Z_3$} baryon parity
discussed in section \ref{sec:alternatives}.
The advantage of this symmetry, which is discrete anomaly free in the MSSM, is
that it forbids not only the $\lambda''$ couplings, but also the dangerous
dimension-5 operators leading to proton decay.
\index{Neutrino!mass|)}


\subsection{Tree-Level Contribution Generated by Neutrino-Neutralino Mixing}
\label{subsec:tree-level}
\index{Neutrino!mass!tree--level|(}

Let us first consider the tree-level contributions. As discussed in
section \ref{subsec:H_L_mixing}, bilinear $R$-parity violation induces
a mixing between neutrinos and neutralinos, which yields a single massive
neutrino state at tree level. This can be understood as a kind of seesaw
mechanism, in which the neutral gauginos and higgsinos play the r\^ole of the
right-handed neutrinos. Indeed, in the limit of a small neutrino-neutralino
\index{Mixing!neutrino--neutralino}
mixing, the $7 \times 7$ neutrino-neutralino mass matrix $M_N$ has a
``seesaw'' structure, with a strong hierarchy\footnote{As in section \ref{sec:bilinear},
we are working in a ($H_d$, $L_i$) basis in which the sneutrino \VEVs\ vanish,
$v_i=0$. The seesaw structure may be less obvious in an arbitrary basis
where both $\mu_i \neq 0$ and $v_i \neq 0$, since large values of $\mu_i$
and $v_i$ are in principle compatible with a
small physical neutrino-neutralino mixing (see subsection
\ref{subsec:H_L_mixing}).} between the gaugino-higgsino diagonal $4 \times 4$
block $M_{\chi}$ and the off-diagonal $3 \times 4$ block $m$:
\begin{equation}
  M_N |_{tree}\ =\ \left(
    \begin{array}{cc}
    M_{\chi} & m^T \\
        m    &  0_{3 \times 3}
    \end{array} \right)\ .
\label{eq:M_N_tree}
\end{equation}
The effective mass matrix obtained by integrating out the neutralinos,
which yields the neutrino mass and mixing angles, is given by
$M^{\nu}_{tree} \simeq -\, m\, M^{-1}_{\chi}\, m^T$. The fact that only one
neutrino becomes massive is most easily seen in a basis in which the
superpotential \Rp\ mass parameters $\mu_1$ and $\mu_2$ (together
with the sneutrino \VEVs\ $v_i$) vanish; then both $L^0_1$ and $L^0_2$
decouple from $M_N |_{tree}$, as can be seen from Eq. (\ref{eq:M_N_xi}).
Of course, when quantum corrections are included, all three neutrinos acquire
a mass, as explained in the next sections.

In order to determine the flavour composition of the single neutrino that
acquires a mass at this level, one has in principle to diagonalize the
chargino mass matrix. Indeed, since bilinear $R$-parity violation mixes
charged leptons with charginos, the physical charged leptons are the three
lightest eigenstates of the $5 \times 5$ charged lepton-chargino mass matrix
$M_C$, Eq. (\ref{eq:M_C_xi}). In general these do not coincide with the
eigenstates of the Yukawa matrix $\lambda^e_{ij}$. However, in the limit of a
small charged lepton-chargino mixing we are interested in, one can identify
the two bases to a good approximation. We therefore choose to write $M_N$
and $M_C$ in the basis in which the sneutrino \VEVs\ $v_i$ vanish and the
charged lepton Yukawa couplings $\lambda^e_{ij}$ are diagonal. In this basis,
the effective neutrino mass matrix reads \cite{joshinowa95_first}:
\begin{equation}
  M^{\nu}_{tree}\ \simeq\ -\, m\, M^{-1}_{\chi}\, m^T\ =\
   -\, \frac{m_{\nu_{tree}}}{\sum_i \mu^2_i}\ \left(
    \begin{array}{ccc}
    \mu^2_1 & \mu_1 \mu_2 & \mu_1 \mu_3  \\
    \mu_1 \mu_2 & \mu^2_2 & \mu_2 \mu_3  \\
    \mu_1 \mu_3 & \mu_2 \mu_3 & \mu^2_3
    \end{array} \right)\ ,
\label{eq:M_nu_tree}
\end{equation}
where $m_{\nu_{tree}}$ is given by Eqs. (\ref{eq:m_nu_tau}) and
(\ref{eq:m_0}),
\begin{equation}
 m_{\nu_{tree}}\ \simeq\ \frac{M_Z^2\, \cos^2\!\beta\, (M_1 c_W^2+M_2 s_W^2)\,
  \mu \cos \xi}{M_1\, M_2\, \mu \cos \xi - M_Z^2\, \sin {2 \beta}\,
  (M_1 c_W^2 + M_2 s_W^2)}\ \tan^2 \xi\ .
\label{eq:m_nu_tau_bis}
\end{equation}
The misalignment angle $\xi$, defined in subsection \ref{subsec:H_L_basis},
is a basis-independent quantity that
controls the size of the bilinear \Rp\ effects in the fermion sector (in particular,
assuming small neutrino-neutralino and charged lepton-chargino mixings
amounts to require $\sin \xi \ll 1$). 
In the basis in which we are working,
it is given by $\sin \xi = \sqrt{\sum_i \mu^2_i} / \mu$.
As already noticed in section \ref{subsec:H_L_mixing}, phenomenologically relevant
values of $m_{\nu_{tree}}$ require
a strong alignment\index{Alignment} between the 4-vectors
$\mu_{\alpha} \equiv (\mu_0, \mu_i)$ and $v_{\alpha} \equiv (v_0, v_i)$,
$\sin \xi \sim 3 \times 10^{-6} \sqrt{1 + \tan^2 \beta}\, \sqrt{m_{\nu_{tree}} / 1 \eV}$.
Several authors \cite{hempfling96,nilles97} have studied the possibility of
obtaining the desired amount of alignment\index{Alignment} from GUT-scale
universality in the soft terms. With this assumption the LEP \index{LEP} bound 
on the tauneutrino mass can be satisfied in a relatively large region of the 
parameterspace, but a significant amount of fine-tuning in the bilinear \Rp\
parameters is necessary in order to reach the \eV\ region (one would
typically need $\mu_i / \mu \sim 10^{-4}$ at the GUT scale). Another
possibility is to invoke an abelian flavour symmetry \cite{banks95}; however
rather large values of the associated charges are necessary to yield the
desired alignment\index{Alignment} (see section \ref{sec:flavour}).

The massive neutrino is mainly a superposition of the electroweak neutrino
eigenstates, and its flavour composition is given, in the basis we are
considering, by the superpotential \Rp\ mass parameters $\mu_i$
\cite{joshinowa95_first}:
\begin{equation}
  \nu_3\ \simeq\ \frac{1}{\sqrt{\sum_i \mu^2_i}}\
  \left( \mu_1 \nu_e\, +\,  \mu_2 \nu_{\mu}\, +\, \mu_3 \nu_{\tau} \right)\ .
\label{eq:flavour_nu_tree}
\end{equation}
In terms of mixing angles, this gives the relations
\begin{equation}
  \sin \theta_{13}\ =\ \frac{\mu_1}{\sqrt{\sum_i \mu^2_i}}\ ,  \hskip 1cm
  \sin \theta_{23}\ =\ \frac{\mu_2}{\sqrt{\mu^2_2 + \mu^2_3}}\ ,
\end{equation}
while $\sin \theta_{12}$ is undetermined.
\index{Neutrino!mass!tree--level|)}
\index{Neutrino!mixing|)}


\subsection{One-Loop Contributions Generated by Trilinear {\boldmath{\Rp}} Couplings}
\label{subsec:loop}

At the one-loop level\index{Neutrino!mass!one--loop|(}, a variety of
diagrams involving the trilinear \Rp\ couplings $\lambda$ and $\lambda'$
and/or insertions of bilinear \Rp\ masses contribute to the
neutralino-neutrino mass matrix, thus correcting Eq. (\ref{eq:M_nu_tree}). 
In this subsection, we concentrate on the diagrams involving
trilinear \Rp\ couplings only. These diagrams represent the dominant
one-loop contribution to neutrino masses and mixings when bilinear $R$-parity
violation is strongly suppressed (i.e. when $\sin \xi \simeq 0$
and $\sin \zeta \simeq 0$ in the language of subsection
\ref{subsec:H_L_basis}, where the angle $\zeta$ formed by the 4-vectors
$B_{\alpha} \equiv (B_0, B_i)$ and $v_{\alpha} \equiv (v_0, v_i)$
controls the Higgs-slepton mixing \index{Mixing!higgs--slepton}). 
The one-loop diagrams involving 
bilinear\Rp\ masses will be discussed in the next subsection.

\begin{figure}[t]
\begin{center}
\mbox{\epsfxsize=\textwidth
       \epsffile{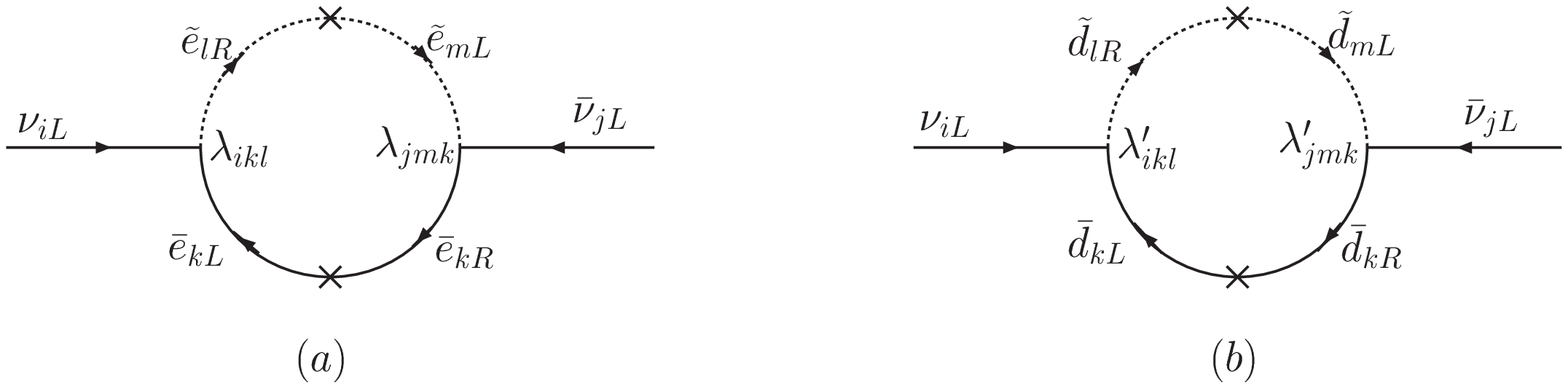}}
\caption{{\it One-loop contributions to neutrino masses and mixings
induced by the trilinear \Rp\ couplings $\lambda_{ijk}$ (a) and
$\lambda'_{ijk}$ (b). The cross on the sfermion line indicates the
insertion of a left-right mixing mass term. The arrows on external legs
follow the flow of the lepton number.}}
\label{fig:trilin_loop}
\end{center}
\end{figure}

The trilinear \Rp\ couplings $\lambda_{ijk}$ and $\lambda'_{ijk}$
contribute to each entry of the neutrino mass matrix through the
lepton-slepton and quark-squark loops of Fig. \ref{fig:trilin_loop},
yielding \cite{hall84,babu90}
\begin{equation}
  M^{\nu}_{ij} |_{\lambda}\ =\ \frac{1}{16 \pi^2} \sum_{k,l,m}
  \lambda_{ikl} \lambda_{jmk}\, m_{e_k}\
  \frac{(\tilde{m}^{e\, 2}_{\scriptscriptstyle{LR}})_{ml}}
  {m^2_{\tilde e_{Rl}} - m^2_{\tilde e_{Lm}}}\,
  \ln \left( \frac{m^2_{\tilde e_{Rl}}}{m^2_{\tilde e_{Lm}}} \right)
  +\ (i \leftrightarrow j)\ ,
\label{eq:m_nu_lambda_1}
\end{equation}
\begin{equation}
  M^{\nu}_{ij} |_{\lambda'}\ =\ \frac{3}{16 \pi^2} \sum_{k,l,m}
  \lambda'_{ikl} \lambda'_{jmk}\, m_{d_k}\
  \frac{(\tilde{m}^{d\, 2}_{\scriptscriptstyle{LR}})_{ml}}
  {m^2_{\tilde d_{Rl}} - m^2_{\tilde d_{Lm}}}\,
  \ln \left( \frac{m^2_{\tilde d_{Rl}}}{m^2_{\tilde d_{Lm}}} \right)
  +\ (i \leftrightarrow j)\ ,
\label{eq:m_nu_lambda'_1}
\end{equation}
Here the couplings $\lambda_{ijk}$ (resp. $\lambda'_{ijk}$) and the
left-right slepton mixing matrix\index{Slepton!mixing matrix} 
$\tilde{m}^{e\, 2}_{\scriptscriptstyle{LR}} =
(A^e_{ij} - \mu \tan \beta\, \lambda^e_{ij})\, v_d / \sqrt{2}$
(resp. the left-right squark mixing matrix
$\tilde{m}^{d\, 2}_{\scriptscriptstyle{LR}} =
(A^d_{ij} - \mu \tan \beta\, \lambda^d_{ij})\, v_d / \sqrt{2}$)
are expressed in the basis in which the charged lepton masses (resp. the
down quark masses) as well as the mass matrices for the associated doublet
and singlet scalars are diagonal.
The above expressions simplify if, as is customary, one assumes that the
sfermion masses are approximately degenerate, and that the $A$-terms are
proportional to the Yukawa couplings, $A^e_{ij} = A^e \lambda^e_{ij}$ and
$A^d_{ij} = A^d \lambda^d_{ij}$. Then Eqs. (\ref{eq:m_nu_lambda_1}) and
(\ref{eq:m_nu_lambda'_1}) reduce to:
\begin{equation}
  M^{\nu}_{ij} |_{\lambda}\ \simeq\ \frac{1}{8 \pi^2}\
  \frac{A^e - \mu \tan \beta}{\overline{m}^2_{\tilde e}}\
  \sum_{k,l} \lambda_{ikl} \lambda_{jlk}\, m_{e_k} m_{e_l}\ ,
\label{eq:m_nu_lambda_2}
\end{equation}
\begin{equation}
  M^{\nu}_{ij} |_{\lambda'}\ \simeq\ \frac{3}{8 \pi^2}\
  \frac{A^d - \mu \tan \beta}{\overline{m}^2_{\tilde d}}\
  \sum_{k,l} \lambda'_{ikl} \lambda'_{jlk}\, m_{d_k} m_{d_l}\ ,
\label{eq:m_nu_lambda'_2}
\end{equation}
where $\overline{m}_{\tilde e}$ (resp. $\overline{m}_{\tilde d}$) is
an averaged scalar mass parameter, and
the couplings $\lambda_{ijk}$ (resp. $\lambda'_{ijk}$) are
now expressed in the superfield basis corresponding to the 
charged lepton (resp. down quark) mass eigenstate basis.
Even after those approximations, the neutrino mass matrix still depends on
a large number of trilinear \Rp\ couplings (9 $\lambda_{ijk}$ and 27
$\lambda'_{ijk}$). To obtain a more predictive scheme, one has
to make assumptions on the generational structure of the trilinear \Rp\
couplings.

One may for instance assume that, for a given generation index $i$, there is
no strong hierarchy among the couplings $\lambda_{ijk}$ and $\lambda'_{ijk}$,
or that their flavour structure in the indices $j$ and $k$ is linked to the
fermion mass hierarchy \cite{banks95,borzumatti96,binetruy98}. 
The second assumption is natural in models where the fermion mass hierarchy 
is explained by flavour symmetries
(see section \ref{sec:flavour}). In both cases the contributions with
$k,l = 2 \mbox{ or } 3$ dominate in Eqs. (\ref{eq:m_nu_lambda_2}) and
(\ref{eq:m_nu_lambda'_2}), and one obtains
\begin{equation}
  M^{\nu}_{ij} |_{\lambda}\ \simeq\ \frac{1}{8 \pi^2} \left\{
  \lambda_{i33} \lambda_{j33}\, \frac{m^2_{\tau}}{\tilde m}\
  +\ (\lambda_{i23} \lambda_{j32} + \lambda_{i32} \lambda_{j23})\,
  \frac{m_{\mu} m_{\tau}}{\tilde m}\ +\  \lambda_{i22} \lambda_{j22}\,
  \frac{m^2_{\mu}}{\tilde m}\, \right\} ,
\label{eq:m_nu_lambda_3}
\end{equation}
\begin{equation}
  M^{\nu}_{ij} |_{\lambda'}\ \simeq\ \frac{3}{8 \pi^2} \left\{
  \lambda'_{i33} \lambda'_{j33}\, \frac{m^2_b}{\tilde m}\
  +\ (\lambda'_{i23} \lambda'_{j32} + \lambda'_{i32} \lambda'_{j23})\,
  \frac{m_s m_b}{\tilde m}\ +\ \lambda'_{i22} \lambda'_{j22}\,
  \frac{m^2_s}{\tilde m}\, \right\} ,
\label{eq:m_nu_lambda'_3}
\end{equation}
where we have set all sfermion mass parameters equal to $\tilde m$.
The term proportional to $m^2_\tau$ in Eq. (\ref{eq:m_nu_lambda_3}) comes
from the tau-stau loop and gives
$M^{\nu}_{ij} |_{\lambda} \sim \lambda_{i33} \lambda_{j33}\,
(4 \times 10^5 \eV)\, (100 \GeV / \tilde m)$; similarly, the
term proportional to $m^2_b$ in Eq. (\ref{eq:m_nu_lambda'_3}) comes from the
bottom-sbottom loop and gives
$M^{\nu}_{ij} |_{\lambda'} \sim \lambda'_{i33} \lambda'_{j33}\,
(7.7 \times 10^6 \eV)\, (m_b / 4.5 \GeV)^2\, (100 \GeV / \tilde m)$.

This shows that trilinear \Rp\ couplings of order 1 would lead to large
entries in the neutrino mass matrix, grossly conflicting with
experimental data. This in turn puts strong constraints on the trilinear
\Rp\ couplings. The most stringent upper bound comes from the non-observation
of neutrinoless double beta decay, whose rate is directly related to
the $(i,j)=(11)$ element of $M^{\nu}$ \cite{bhattacharyya99}:
\begin{equation}
  \vert \lambda_{133} \vert \; \leq \; 9.4 \times 10^{-4}\
  \left(\frac{<\! m_{\nu}\! >}{0.35 \eV}\right)^{\frac{1}{2}}\
  \left(\frac{\tilde m}{100 \GeV}\right)^{\frac{1}{2}}\ ,
\label{eq:m_nu_lambda_bound}
\end{equation}
\begin{equation}
  \vert \lambda'_{133} \vert \; \leq \; 2.1 \times 10^{-4}\
  \left(\frac{<\! m_{\nu}\! >}{0.35 \eV}\right)^{\frac{1}{2}}\
  \left(\frac{4.5 \GeV}{m_b}\right)\
  \left(\frac{\tilde m}{100 \GeV}\right)^{\frac{1}{2}}\ ,
\label{eq:m_nu_lambda'_bound}
\end{equation}
where $<\! m_{\nu}\! >$ is the effective neutrino mass, bounded by
neutrinoless double beta decay experiments. From the other terms in
Eqs. (\ref{eq:m_nu_lambda_3}) and (\ref{eq:m_nu_lambda'_3}) one can also
extract (weaker) bounds on the couplings $\lambda_{1kl}$ and $\lambda'_{1kl}$,
$(k,l) \neq (3,3)$. From a different perspective it is quite remarkable
that a small amount of $R$-parity violation through trilinear couplings,
with $\lambda_{ijk}$ and $\lambda'_{ijk}$ comparable in strength with the
charged lepton and down quark Yukawa couplings, can induce neutrino masses
in the phenomenologically interesting range, namely
$10^{-3} \eV \lesssim m_{\nu} \lesssim 1 \eV$.

Let us now discuss the flavour structure of the neutrino mass matrix.
Assuming further that the $\lambda$-type couplings are not greater than
the $\lambda'$-type couplings, the leading contribution to
$M^{\nu}_{loop} \equiv M^{\nu} |_{\lambda} + M^{\nu} |_{\lambda'}$ comes
from the bottom-sbottom loop. Then
\begin{equation}
  M^{\nu}_{loop}\ =\ \frac{m_{\nu_{loop}}}{\sum_i \lambda^{\prime 2}_{i33}}\
    \left( \begin{array}{ccc}
    \lambda^{\prime 2}_{133} & \lambda'_{133} \lambda'_{233}
    & \lambda'_{133} \lambda'_{333}  \\
    \lambda'_{133} \lambda'_{233} & \lambda^{\prime 2}_{233}
    & \lambda'_{233} \lambda'_{333}  \\
    \lambda'_{133} \lambda'_{333} & \lambda'_{233} \lambda'_{333}
    & \lambda^{\prime 2}_{333}
    \end{array} \right)\ +\ \cdots\ ,
\label{eq:m_nu_structure}
\end{equation}
where
\begin{equation}
  m_{\nu_{loop}}\ =\ \frac{3 m^2_b}{8 \pi^2 \tilde m}\ 
\sum_i \lambda^{\prime 2}_{i33}
\label{eq:m_nu_loop}
\end{equation}
and the dots stand for corrections of order 
\[
  \frac{m^2_{\tau}}{8 \pi^2 \tilde m}\,
  \lambda_{i33} \lambda_{j33}\ , \qquad
  \frac{3 m_b m_s}{8 \pi^2 \tilde m}\,
  (\lambda'_{i23} \lambda'_{j32} + \lambda'_{i32} \lambda'_{j23})\ , \qquad
  \frac{m_{\mu} m_{\tau}}{8 \pi^2 \tilde m}\,
  (\lambda_{i23} \lambda_{j32} + \lambda_{i32} \lambda_{j23})\ \cdots
\]
(with $i,j \neq 3$ for the
$m^2_{\tau}$ corrections, and $(i,j) \neq (2,2), (3,3)$ for the
$m_{\mu} m_{\tau}$ corrections).
The structure (\ref{eq:m_nu_structure}) generally leads to a hierarchical
mass spectrum. Indeed, only one neutrino becomes massive at leading order,
with a mass $m_{\nu_3} = m_{\nu_{loop}}$ and a flavour composition given
by the mixing angles $\sin \theta_{13} = \lambda'_{133} /
\sqrt{\sum_i \lambda^{\prime 2}_{i33}}$
and $\sin \theta_{23} = \lambda'_{233} / \sqrt{\lambda^{\prime 2}_{233}
+ \lambda^{\prime 2}_{333}}$
($\theta_{12}$ remains undetermined at dominant order). Once sub-dominant
contributions to $M^{\nu}$ are included, all three neutrinos become massive.
Let us stress again that the above discussion does not take into account
the contributions of bilinear \Rp\ parameters to the neutrino mass matrix.


\subsection{One-Loop Contributions Generated by both Bilinear and
Trilinear {\boldmath{\Rp}} Couplings}
\label{subsec:general}

In the above, we discussed one-loop contributions to neutrino masses in the
limit where bilinear $R$-parity violation can be neglected. However this
is generally not a valid approximation, since bilinear \Rp\ terms are
always generated radiatively from trilinear \Rp\ interactions, and
the presence of bilinear \Rp\ terms drastically changes the discussion of
one-loop neutrino masses. First of all, the neutrino mass matrix receives
contributions already at tree level, as discussed in section
\ref{subsec:tree-level}. Secondly, in addition to the lepton-slepton and
quark-squark loops already encountered, one-loop diagrams involving
insertions of bilinear \Rp\ masses or slepton \VEVs\ must be considered
\cite{hall84,hempfling96,haber1,haber2,kaplan00,chun00,cheung00,kong00,
hirsch00,davidson00,davidson00_bis,kang02}. 
Here we shall only present briefly the
main classes of loop contributions, and comment on the level of suppression
of bilinear \Rp\ parameters required by neutrino mass constraints.
We refer the interested reader to Ref. \cite{davidson00_bis} for a detailed
classification and evaluation of the various diagrams.

\begin{figure}[t]
\begin{center}
\mbox{\epsfxsize=\textwidth
       \epsffile{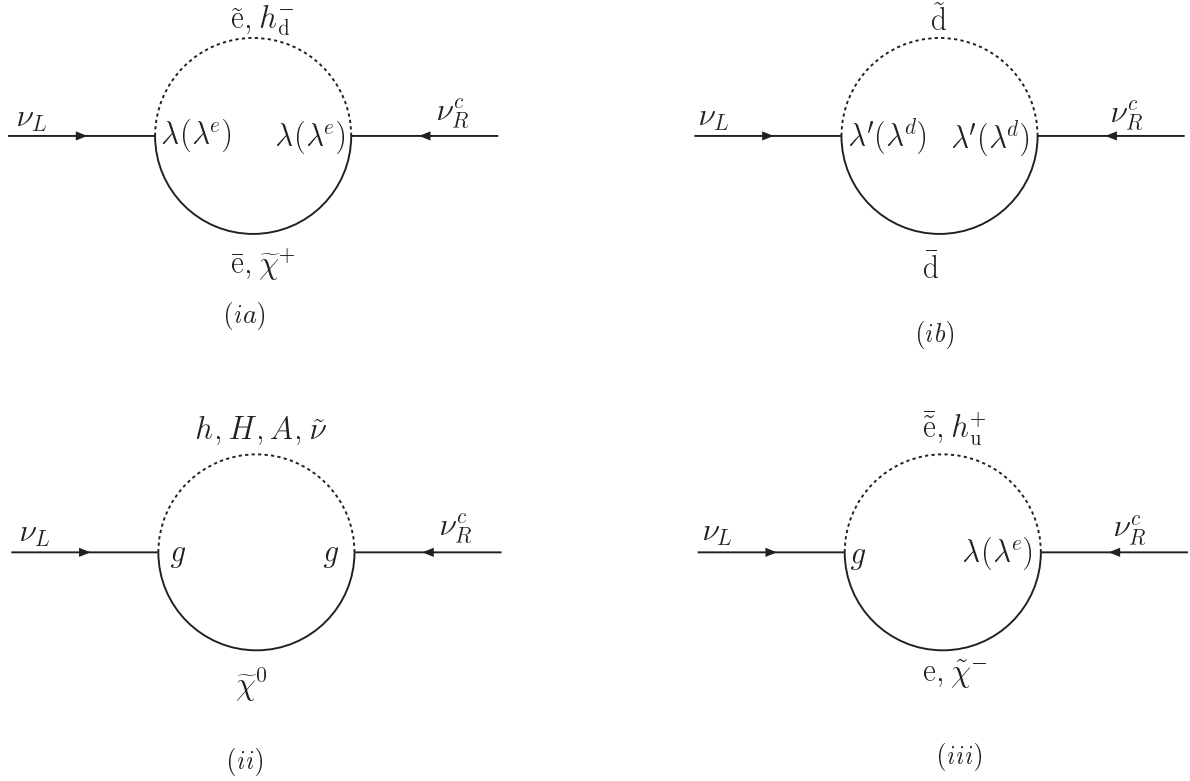}}
\caption{{\it Schematic description of the one-loop diagrams contributing
to neutrino masses and mixings, divided into three classes as described in
the text. \Rp\ mass insertions on internal and/or external lines are not
shown.  The arrows on external legs follow the flow of the lepton number.}}
\label{fig:loop_nu}
\end{center}
\end{figure}

Let us first notice that there are two ways of computing the one-loop neutrino
masses and mixing angles. The first one is to compute one-loop corrections to
the full $7 \times 7$ neutralino-neutrino mass matrix, Eq. (\ref{eq:M_N_tree})
\cite{hempfling96,chun00,hirsch00}. The second one is to compute one-loop
corrections to the tree-level effective $3 \times 3$ neutrino mass matrix,
Eq. (\ref{eq:M_nu_tree}) \cite{davidson00,davidson00_bis}; in this case
the Feynman rules are written in terms of tree-level MSSM mass eigenstates
(i.e. the tree-level mass matrices are diagonalized neglecting $R$-parity
violation) and the bilinear \Rp\ masses are included in the diagrams as mass
insertions, both on internal and external lines.
The second approach is more suitable for a discussion of the various
contributions to neutrino masses and mixings. Leaving aside gauge boson loops
and diagrams with two \Rp\ mass insertions on the external legs, which
renormalize the tree-level neutrino mass, one can divide the one-loop
contributions to the neutrino mass matrix into three classes (see Fig.
\ref{fig:loop_nu}), depending on which couplings appear at the two
vertices (a diagram with two couplings $\lambda_1$ and $\lambda_2$ at the
vertices will be denoted by ($\lambda_1$, $\lambda_2$)) \cite{davidson00_bis}:
\begin{enumerate}
  \item[{\bf (i)}] diagrams involving trilinear \Rp\ couplings and/or Yukawa
couplings at the vertices, with
charged fermions and scalars in the loop; in addition to the
($\lambda$, $\lambda$) and ($\lambda'$, $\lambda'$) diagrams discussed above,
there are ($\lambda$, $\lambda^e$) and ($\lambda'$, $\lambda^d$) diagrams
with one \Rp\ mass insertion, and ($\lambda^e$, $\lambda^e$) and
($\lambda^d$, $\lambda^d$) diagrams with two \Rp\ mass insertions;
  \item[{\bf (ii)}] diagrams involving two gauge couplings, with a neutralino
and neutral scalars in the loop \cite{haber1,haber2}; these diagrams have two
\Rp\ mass insertions;
  \item[{\bf (iii)}] diagrams involving a trilinear \Rp\ coupling or a Yukawa
coupling at one vertex and a gauge coupling at the other vertex, with a
chargino and charged fermions and
scalars in the loop; the ($g$, $\lambda$) and ($g$, $\lambda'$) diagrams have
one \Rp\ mass insertion, while the ($g$, $\lambda^e$) and ($g$, $\lambda^d$)
diagrams have two \Rp\ mass insertions.
\end{enumerate}
Each of these diagrams contains two \Rp\ interactions, which can be trilinear
($\lambda$ and $\lambda'$ couplings), mass insertions on lepton or higgsino
lines ($\mu_i$ mixing parameters or slepton \VEVs\ $v_i$), LR mixing mass
insertions on scalar lines (slepton \VEVs\ $v_i$) or soft \Rp\ mass insertions
on scalar lines ($B_i$ and $\tilde m^2_{di}$ parameters). Note that the
mass insertion approximation is valid only in a basis in which \Rp\ masses are
indeed small, e.g. in the $v_i=0$ basis. We adopt this basis for the rest of
the section.

\begin{figure}[t]
\begin{center}
\mbox{\epsfxsize=0.45\textwidth
       \epsffile{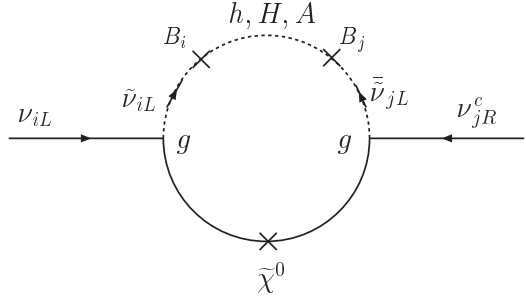}}
\caption{{\it Neutral loop with gauge couplings at the vertices and two
\Rp\ mass insertions on the scalar line. The cross on the neutralino line
indicates a Majorana mass insertion.  The arrows on external legs
follow the flow of the lepton number.}}
\label{fig:GH}
\end{center}
\end{figure}

Depending on the relative size of the various \Rp\ parameters,
some diagrams can be neglected. For example, assuming
that the trilinear couplings always give a significant contribution, one can
consider three representative cases \cite{davidson00_bis} (it may also
happen that the contribution of the bilinear terms in the loops dominates
over the contribution of the trilinear couplings, see Ref.
\cite{grossman04}): \\ 
\, (a) the contribution of bilinear terms in the loops is negligible. 
       This is the situation discussed in the previous subsection, where only 
       the ($\lambda$, $\lambda$) and ($\lambda'$, $\lambda'$) diagrams were 
       considered; \\
\, (b) the bilinear soft terms, but not the bilinear superpotential masses, 
       induce sizeable loop contributions. In addition to the previous 
       diagrams, the neutral loop from class (ii) with two \Rp\ soft mass 
       insertions on the scalar line (see Fig.~\ref{fig:GH}) must be
       taken into account; \\
\, (c) both the bilinear soft terms and the bilinear superpotential masses 
       induce sizeable loop contributions. In this case, all diagrams listed
       above must be considered a priori.

The contribution of the diagrams of Fig.~\ref{fig:GH} can be estimated
to be \cite{davidson00_bis},
\begin{equation}
  \Delta M^{\nu}_{ij}\ \sim\ \frac{g^2}{64 \pi^2}\,
  \frac{B_i B_j}{\tilde m^3}\ \frac{\epsilon_H}{\cos^2 \beta}\ ,
\label{eq:m_nu_GH}
\end{equation}
where $\tilde m$ is the typical mass of the particles in the loop,
and $\epsilon_H$ takes into account the cancellation between the
$h$, $A$ and $H$ loops. As noticed in Ref. \cite{grossman04}, the
different Higgs loops tend to cancel partially, and the
cancellation becomes stronger when the pseudoscalar Higgs boson
becomes heavier ($\epsilon_H \rightarrow 0$ in the decoupling limit
$M_A \rightarrow \infty$). Furthermore, $\epsilon_H$ decreases when
$\tan \beta$ increases, which softens the dependence of
Eq. (\ref{eq:m_nu_GH}) on $\tan \beta$.
Assuming $\tilde m \sim 100$ GeV and $\epsilon_H / \cos^2 \beta
\sim 0.1$, one can see from Eq. (\ref{eq:m_nu_GH}) that soft \Rp\ masses
$\sqrt{B_i} \sim 1$ GeV are enough to generate neutrino mass matrix entries
of the order of the cosmological bound ($\Delta M^{\nu}_{ij} \sim 1 \eV$). 
Thus, neutrino masses do not only constrain the misalignment angle in the
fermion sector ($\sin \xi \lesssim 3 \times 10^{-6} \sqrt{1 + \tan^2 \beta}$
for $m_\nu \leq 1 \eV$), but also the misalignment angle in the
scalar sector. Namely, one has $\sin \zeta \lesssim (10^{-4} - 10^{-3})\,
(\tilde m / 100 \GeV)^{3/2}\, (100 \GeV / \sqrt{B})^2$ for
$m_\nu \leq 1 \eV$, assuming a partial cancellation between the different
Higgs loops in the range $\epsilon_H / \cos^2 \beta = 0.01 - 1$.


\section{Explicit Models of Neutrino Masses}  
\label{sec:models}
\index{Neutrino!mass|(}

We have seen in the previous section how the violation of $R$-parity
improves our understanding of the generation of neutrino masses. The
questions that immediately arise is how well \Rp\ models can account
for the observed neutrino oscillation parameters, and whether these models
lead to specific experimental signatures that could allow to test them.
Numerous studies have been devoted to these questions, and we shall
not attempt to give an exhaustive account of the existing literature
on the subject. Rather we would like to stress the main characteristics of
\Rp\ models of neutrino masses through a detailed discussion of some
representative examples.

Before doing so, let us summarize the experimental status of neutrino
masses and mixings.

\subsection{Experimental Constraints on Neutrino Masses and Mixings}
\label{subsec:experiments}

Atmospheric neutrino data \cite{SK_atm,Soudan,MACRO}
strongly suggest oscillations of atmospheric $\nu_\mu$'s into
$\nu_\tau$'s, with a squared mass difference $\Delta m^2_{atm}
\equiv m^2_{\nu_3} - m^2_{\nu_2} \simeq (1.5-3.9) \times 10^{-3} \eV^2$,
and a large-to-maximal mixing angle, $\tan^2 \theta_{23} = (0.45-2.3)$,
both at the $3 \sigma$ level \cite{gonzalez-garcia03}. The results of
the K2K long-baseline neutrino oscillation experiment \cite{K2K}
are consistent with these oscillation parameters.

Solar neutrino data \cite{Homestake,gallium,Kamiokande_solar,SK_solar,SNO}
combined with the results of the  KamLAND experiment \cite{KamLAND}
provide evidence for oscillations of solar $\nu_e$'s into
$\nu_\mu$'s and $\nu_\tau$'s. Before the SNO and KamLAND results,
four different solutions to the solar neutrino deficit
were allowed, out of which three are now excluded. We nevertheless list
them for future reference (the following pre-SNO allowed ranges of parameters
are taken from Ref. \cite{gonzalez-garcia01}):
(i) a small mixing angle solution (SMA) in the MSW regime, in which
neutrino oscillations inside the sun are enhanced by matter effects
(known as the Mikheev-Smirnov-Wolfenstein, or MSW effect
\cite{wolfenstein78,mikheev85}), with a squared mass difference
$\Delta m^2_{\odot} \equiv m^2_{\nu_2} - m^2_{\nu_1}
\simeq (4 \times 10^{-6} - 10^{-5}) \eV^2$ and a mixing angle
$\tan^2 \theta_{12} \simeq (10^{-4} - 2 \times 10^{-3})$;
(ii) a large mixing angle MSW solution (LMA) with
$\Delta m^2_{\odot} = (10^{-5} - 5 \times 10^{-4}) \eV^2$ and
$\tan^2 \theta_{12} \simeq (0.2-1.)$;
(iii) a solution with low squared mass difference (LOW), which extends to
quasi-vacuum oscillations, with
$\Delta m^2_{\odot} = (4 \times 10^{-10} - 4 \times 10^{-7}) \eV^2$
and $\tan^2 \theta_{12} \simeq (0.2-4.)$;
(iv) a tower of regions in the vacuum oscillation regime (VO), in which
the oscillations occur during the propagation of the neutrinos from the sun
to the Earth, with
$\Delta m^2_{\odot} \sim (10^{-11}- 5 \times 10^{-10}) \eV^2$ and a
large mixing angle. After the results from the SNO and KamLAND collaborations,
LMA is the only allowed solution (see e.g.
Refs. \cite{fogli02,valle02,bahcall02,smirnov02}), with smaller allowed
regions in the oscillation parameter space:
$\Delta m^2_{\odot} \simeq (5.4 - 10.) \times 10^{-5} \eV^2$ and
$\Delta m^2_{\odot} \simeq (14. - 19.) \times 10^{-5} \eV^2$ (high-LMA region),
$\tan^2 \theta_{12} \simeq (0.29 - 0.82)$ at the $3 \sigma$ level
\cite{gonzalez-garcia03}.

The CHOOZ reactor experiment \cite{CHOOZ} provides an upper limit on the
mixing angle $\theta_{13}$ that connects the solar and atmospheric neutrino
sectors, $|U_{e3}| = |\sin \theta_{13}| < 0.2$ (90\% CL).

One should also mention the LSND experiment \cite{LSND}, which claims
evidence for $\nu_\mu \leftrightarrow \nu_e$ oscillations with parameters
$\Delta m^2 = (0.2-2) \eV^2$ and
$\sin^2 2 \theta = (3 \times 10^{-3} - 3 \times 10^{-2})$.
However this result, which cannot be accounted for together with the other
neutrino oscillation data within the standard scheme of three-neutrino
oscillation, is still controversial.

Finally, upper bounds on the absolute neutrino mass come from direct mass
measurements, $m_{\nu_e} < 3 \eV$ \cite{pdg04}; CMB \cite{WMAP} and large
scale structure data \cite{2dFGRS}, $\sum_i m_{\nu_i} \lesssim 1 \eV$; and
from neutrinoless double beta decay experiments which are sensitive to the
effective neutrino mass $<\! m_{\nu}\! >\, \equiv \sum_i m_{\nu_i} U^2_{ei}$,
found to verify
$<\! m_{\nu}\! >\, \leq (0.35-1.05) \eV$ (90\% CL) \cite{heidelberg-moscow}.

\subsection{Classification of Models}
\label{subsec:classification}

One can classify the \Rp\ neutrino mass models according to the pattern of
$R$-parity violation that they assume. We shall distinguish between models
with trilinear couplings only, models with both bilinear and trilinear
couplings, and models with bilinear couplings only. In the first class of
models, bilinear $R$-parity violation
is neglected, and neutrino masses and mixings arise at the one-loop level.
A more realistic variant assumes the absence of bilinear $R$-parity violation
at the GUT scale, and takes into account the tree-level contribution of the
bilinear \Rp\ terms generated from the renormalization group evolution.
In the second class of models, both bilinear and trilinear \Rp\ terms are
present, and the neutrino mass matrix receives both tree-level and loop
contributions. In the third class of models, $R$-parity violation can be
parametrized in terms of bilinear \Rp\ couplings only.
The neutrino mass matrix receives both tree-level and loop
contributions, as in the second class of models.

\vskip .5cm

\noindent \addtocounter{sss}{1}
{\bf \thesss Models with Trilinear Couplings only} 
 \addcontentsline{toc}{subsection}{\hspace*{1.2cm} \alph{sss})
             Models with Trilinear Couplings only}
\vskip .1cm

In the limit where bilinear $R$-parity violation can be neglected,
the only contributions to the neutrino mass matrix come from the
($\lambda$, $\lambda$), or lepton-slepton, and ($\lambda'$, $\lambda'$),
or quark-squark loop diagrams, and are given by Eqs. (\ref{eq:m_nu_lambda_1})
and (\ref{eq:m_nu_lambda'_1}) (cf. subsections~\ref{subsec:loop}
and~\ref{subsec:general}).
These expressions simplify to Eqs.
(\ref{eq:m_nu_lambda_2}) and (\ref{eq:m_nu_lambda'_2}) under the (very
common) assumptions of proportionality of the $A$-terms to the Yukawa
couplings and flavour-independence of sfermion masses. Despite the
large number of arbitrary parameters involved in those formulae,
trilinear $R$-parity violation leads to an interesting structure
for the neutrino mass matrix, given by Eq. (\ref{eq:m_nu_structure}),
when the hierarchy among trilinear couplings
is mild or linked to the fermion mass hierarchy, and the $\lambda$-type
couplings are not greater than the $\lambda'$-type couplings. Indeed the
structure (\ref{eq:m_nu_structure}) can account for both the large
atmospheric mixing angle (with $\lambda'_{233} \approx \lambda'_{333}$)
and the hierarchy of oscillation frequencies
$\Delta m^2_{\odot} \ll \Delta m^2_{atm}$ (with $\Delta m^2_{\odot}$
determined by the subdominant contributions to $M^{\nu}_{loop}$, 
governed by the charged lepton and down quark mass hierarchies).
The correct scale of atmospheric
neutrino oscillations is obtained for values of the \Rp\ couplings that
could give rise to FCNC processes (see chapter \ref{chap:indirect}) and to
observable signals at colliders (see chapter \ref{chap:colliders}).

As a prototype of model trying to explain atmospheric and solar neutrino
oscillations in terms of trilinear \Rp\ couplings, let us discuss a
model by Drees et al. \cite{drees98}. Besides the standard assumptions on
soft terms, the authors require that all \Rp\ trilinear couplings that are
not forbidden by a symmetry be of comparable magnitude. While theoretically
not well motivated, this hypothesis gives them some control on the subdominant
contributions to the neutrino mass matrix.
Note that the only way
to account for the smallness of $|U_{e3}|$ consistently with their hypothesis
is to set $\lambda'_{133} = 0$. Then, leaving aside the contribution of
$\lambda$ couplings, one obtains at leading order:
\begin{equation}
  M^{\nu}_{loop}\ \sim\ \frac{3 m^2_b}{8 \pi^2 \tilde m}\
  \lambda^{\prime 2}_{333}\ \left(
    \begin{array}{ccc}
    \frac{m_s}{m_b} & \frac{m_s}{m_b} & \frac{m_s}{m_b}  \\
    \frac{m_s}{m_b} & 1 & 1  \\
    \frac{m_s}{m_b} & 1 & 1
    \end{array} \right)\ ,
\label{eq:m_nu_drees}
\end{equation}
where each entry in Eq. (\ref{eq:m_nu_drees}) should be multiplied by an
arbitrary factor of order one, and the determinant of the lower right
$2 \times 2$ sub-matrix is of order $m_s / m_b$. This structure is not
altered when the contribution of $\lambda$-type couplings, which involves
the charged lepton masses, is included.
One typically obtains
$m_{\nu_2} / m_{\nu_3} \sim m_s / m_b \sim 0.04$ (taking $m_s = 200 \MeV$
and $m_b = 5 \GeV$), yielding a $\Delta m^2_{\odot}$ roughly in the MSW
range, and a moderate to large solar mixing angle.
This points towards the large mixing angle solution, which is precisely
the only allowed solution to the solar neutrino problem after the KamLAND
results.
Such a scenario would require $\lambda'$ couplings in the
$(5 \times 10^{-5} - 10^{-4})\, (\tilde m / 100 \GeV)^{1/2}$ range, in order
for $\Delta m^2_{atm} \simeq m^2_{\nu_3}$ to fall in the allowed interval.
Due to the hierarchy $m_{\nu_2} \ll m_{\nu_3}$ (a frequent
feature of \Rp\ neutrino mass models), no measurable signal is expected
for neutrinoless double beta decay. The required  value of \Rp\ trilinear
couplings, on the other hand, may give rise to sizeable FCNC decays
such as $\mu \rightarrow e \gamma$ and $K^0 \rightarrow \mu e$, depending
on the model. Signals at colliders are also expected, due to the decay
of the LSP inside the detector -- probably without an observable displaced
vertex (see also Ref. \cite{barger01}).
In order to motivate the structure of Eq. (\ref{eq:m_nu_drees}), the authors
of Ref. \cite{drees98} try to find a discrete flavour symmetry allowing
for the desired couplings, while forbidding $\lambda'_{133}$ as well as
the $B$-violating couplings $\lambda''_{ijk}$ and the bilinear
\Rp\ parameters $\mu_i$. They are able to identify such a $Z_3$
\index{Discrete symmetries!$Z_3$}symmetry, which however must be explicitly 
broken by the strange quark Yukawa coupling.
This problem is generic for abelian discrete symmetries and makes their
model less natural.

As discussed in sections \ref{subsec:patterns} and \ref{sec:rengr3},
neglecting bilinear $R$-parity violation in the presence of trilinear
\Rp\ couplings is not always a valid assumption, since the latter induce
all types of bilinear \Rp\ parameters at the one-loop level
\cite{decarlosl,nardi}. In particular, even if only trilinear \Rp\ couplings
are present at the GUT scale, the renormalization group induced bilinear
\Rp\ terms generally give the dominant contribution to the heaviest neutrino
mass. This is the situation studied by Joshipura and Vempati in Ref.
\cite{joshipura99_bis}. For simplicity only the $\lambda'_{ijk}$ couplings and
their associated $A$-terms are considered, and universality among soft
terms is assumed at the GUT scale. At the weak scale, bilinear \Rp\ terms are
generated and give a tree-level contribution to the neutrino mass matrix,
which upon neglecting the scale-dependence of the soft parameters in the
renormalization group evolution takes the form
$(M^{\nu}_{tree})_{ij} = m_0\, a_i a_j$, where
$a_i \equiv \sum_k \lambda'_{ikk} m_{d_k} / v_d$ in the $\mu_i = 0$ basis,
and $m_0$ is determined by solving the renormalization group equations.
At the one-loop level, the trilinear couplings give an additional contribution
$(\Delta M^{\nu}_{loop})_{ij} = m_1 \sum_{k,l} \lambda'_{ikl}
\lambda'_{jlk} m_{d_k} m_{d_l} / v^2_d$, see Eq. (\ref{eq:m_nu_lambda'_2})
(other loop contributions are neglected). Depending on the MSSM parameters,
the ratio $m_1 / m_0$ varies between typical values of $10^{-3}$
and $10^{-1}$; in some regions of the parameter space cancellations in
$m_0$ can lead to $m_1 / m_0 > 1$. For
$m_1 / m_0 \sim (10^{-2} - 10^{-1})$ and
$\lambda'_{ijk} \sim 10^{-4}$ (for small values of $\tan \beta$), one
naturally obtains a neutrino mass spectrum compatible with both atmospheric
neutrino data and the -- now excluded --
vacuum oscillation solution of the solar neutrino problem. MSW solutions
can also be obtained provided $m_1 / m_0 \sim 1$, which happens
in a particular region of the MSSM parameter space.

Other examples of neutrino mass models based on trilinear $R$-parity
breaking can be found in Refs. \cite{chun00,joshipura01,koide02,borzumati02}.
The last reference also discusses two-loop contributions induced by
both superpotential and soft trilinear \Rp\ couplings.


\vskip .5cm

\noindent \addtocounter{sss}{1}
{\bf \thesss Models with both Bilinear and Trilinear Couplings} 
 \addcontentsline{toc}{subsection}{\hspace*{1.2cm} \alph{sss})
             Models with both Bilinear and Trilinear Couplings}
\vskip .1cm

In the presence of all types of \Rp\ couplings, the neutrino mass
matrix receives a tree-level contribution from bilinear \Rp\ terms,
Eq. (\ref{eq:M_nu_tree}), and one-loop corrections involving both bilinear
and trilinear \Rp\ couplings, as explained in section \ref{subsec:loop}.
In practive however most studies have omitted the loop diagrams containing
bilinear \Rp\ mass insertions (such as the neutral loop diagrams of Fig.
\ref{fig:GH}), thus keeping only the $(\lambda, \lambda)$ and
$(\lambda', \lambda')$ loop contributions, Eqs. (\ref{eq:m_nu_lambda_1}) and
(\ref{eq:m_nu_lambda'_1}) (see however Ref. \cite{cheung00}).


Since flavour symmetries can constrain \Rp\ couplings, it is interesting
to study their predictions for \Rp\ models of neutrino masses. The case
of a $U(1)_X$ flavour symmetry has been considered by Borzumati et al.
\cite{borzumatti96} and by several other authors
(see e.g. Refs. \cite{binetruy98,grossman04,choi99}). In this framework the
order of magnitude of each \Rp\ coupling is determined by the $X$-charge of
the corresponding operator
(see section \ref{sec:flavour} for details and notations):
$\lambda_{ijk} \sim \epsilon^{\tilde{l}_i - \tilde{l}_0}\, \lambda^e_{jk}$,
$\lambda'_{ijk} \sim \epsilon^{\tilde{l}_i - \tilde{l}_0}\, \lambda^d_{jk}$,
$\widetilde m^2_{\alpha \beta} \sim \tilde{m}^2
\epsilon^{|l_{\alpha} - l_{\beta}|}$ and $B_{\alpha} / \tilde{m}
\sim \mu_{\alpha} \sim \tilde{m}\, \epsilon^{\tilde{l}_{\alpha}}$, where
$\tilde m$ is the typical scale associated with the soft \SUSY-breaking terms,
$\epsilon \simeq V_{us} = 0.22$, $\tilde{l}_{\alpha} \equiv |l_{\alpha}+h_u|$,
and the generations of leptons are labelled in such a way that
$\tilde l_0 < \tilde l_3 \leq \tilde l_{1,2}$. The flavour symmetry ensures
an approximate alignment of the doublet \VEVs\ $v_{\alpha}$ along the
superpotential mass parameters $\mu_{\alpha}$, resulting in a misalignment
angle $\sin^2 \xi \sim \epsilon^{2(\tilde l_3 - \tilde l_0)}$; still large
lepton $X$-charges are necessary in order for neutrino masses to reach the
phenomenologically interesting range. Keeping only the ($\lambda$, $\lambda$)
and ($\lambda'$, $\lambda'$) diagrams at the one-loop level, one finds the
following structure for the neutrino mass matrix,
\begin{equation}
  M^{\nu}_{ij}\ \sim\ \left( m_0 \delta_{i3} \delta_{j3} + m_1 \right)
    \epsilon^{\tilde l_i + \tilde l_j - 2 \tilde l_0}\ ,
\end{equation}
where $m_0 \sim (100 \GeV) \left( 100 \GeV / \tilde{m} \right)
\epsilon^{2 \tilde l_0}$ is associated with the tree-level contribution,
and $m_1 \sim (5 \keV) \left( m_b / 4.5 \GeV \right)^4
\left( 100 \GeV / \tilde{m} \right) \epsilon^{- 2 \tilde l_0}$
with the one-loop contribution.
The neutrino masses and mixing angles are then given by
$m_{\nu_3} \sim m_0\, \epsilon^{2(\tilde l_3 - \tilde l_0)}$,
$m_{\nu_2} \sim m_1\, \epsilon^{2(\tilde l_2 - \tilde l_0)}$,
$m_{\nu_1} \sim m_1\, \epsilon^{2(\tilde l_1 - \tilde l_0)}$
and $\sin \theta_{ij} \sim \epsilon^{|l_i-l_j|}$ ($i \neq j$). The mass
spectrum is characterized by a large hierarchy $m_{\nu_2} \ll m_{\nu_3}$.
For this model to account for atmospheric and solar neutrino oscillations,
one would need both large values of the lepton charges, with
$\tilde l_3 = 9$ or $10$, and $\tilde l_0 \geq 2$, which corresponds to
$\tan \beta \gtrsim 20$ (indeed $\tan \beta \sim \epsilon^{- \tilde l_0}$).
Since the large atmospheric mixing angle favours $\tilde l_2 = \tilde l_3$,
the -- now excluded -- low $\Delta m^2_{\odot}$  solution is selected
for $\tilde l_0 = 2$. Then, in order to account for the large solar mixing
angle, $|U_{e3}|$ should be close to its present limit.
The large mixing angle solution can be accommodated for
$\tilde l_2 = \tilde l_3 + 1$, which is marginally compatible with the large
atmospheric mixing angle, and $\tilde l_0 = 3$. This discussion neglects
the contribution of the diagrams with \Rp\ mass insertions,
however. As noticed in Ref. \cite{grossman04}, the diagrams of
Fig. \ref{fig:GH} dominate over the ($\lambda'$, $\lambda'$) diagrams
in a large portion of the parameter space. If one assumes
a moderate cancellation between the different Higgs loops, it becomes
possible to accommodate the large mixing angle solution to the solar
neutrino problem together with the large atmospheric mixing angle.

Other examples of neutrino mass models based on both trilinear and
bilinear $R$-parity breaking can be found in Refs.
\cite{cheung00,chun99,kong99,SYchoi99,haug00}.

\vskip .5cm

\noindent \addtocounter{sss}{1}
{\bf \thesss Models with Bilinear Couplings only} 
 \addcontentsline{toc}{subsection}{\hspace*{1.2cm} \alph{sss})
             Models with Bilinear Couplings only}
\vskip .1cm
\label{sec:ModBilinear}

The above discussion clearly shows that models of neutrino masses based on
the most general \Rp\ mass terms and couplings suffer from a lack of
predictivity. This led several authors \cite{hempfling96,nilles97,kaplan00,
chun00,hirsch00,joshipura99,takayama00,joshipura02,chun02}
to consider the so-called ``bilinear $R$-parity
breaking'' scenario, in which one assumes that the only seed of $R$-parity
violation resides in the bilinear superpotential and soft terms (see section
\ref{subsec:patterns}). This scenario yields only one massive neutrino
at tree level with the flavour composition of Eq. (\ref{eq:flavour_nu_tree}).
Radiative corrections to the neutralino-neutrino mass matrix then generate
the other two masses and the solar mixing angle, while slightly modifying
the heaviest neutrino state. In order to accommodate the atmospheric
neutrino mass scale, these studies generally assume universality of the soft
terms at the GUT scale \cite{hempfling96,nilles97,hirsch00} (or at the
messenger scale in the context of gauge-mediated \SUSY\ breaking
\cite{kaplan00,chun00}) together with the smallness of the \Rp\ parameters,
with typically $\mu_i / \mu\, (M_{GUT}) \leq 10^{-3}$.
The second assumption is crucial for obtaining $m_{\nu_{tree}} =
\sqrt{\Delta m^2_{atm}} \sim 0.05\, \mbox{eV}$. Indeed, while the universality
conditions lead to an exact alignment between the superpotential mass
parameters $\mu_{\alpha}$ and the doublet \VEVs\ $v_{\alpha}$ at the GUT
scale, radiative corrections
induce some amount of non-universality among the soft terms at the
weak scale, which spoils this alignment and induces a nonzero neutrino mass.
For $\mu_i\, (M_{GUT}) \sim \mu\, (M_{GUT})$, however, the resulting neutrino
mass would lie in the range $100\, \mbox{eV} \leq m_{\nu_{tree}} \leq 100\,
\mbox{MeV}$ \cite{hempfling96,nilles97}, well above the atmospheric neutrino
scale. This is the reason why $\mu_i \ll \mu$ is required at $M_{GUT}$.
Note that the universality assumption reduces the number of independent
\Rp\ parameters to only 3, which one can choose to be the three \Rp\ 
supersymmetric masses at the GUT scale, 
$\epsilon_i \equiv \mu_i\, (M_{GUT})$.

A detailed study of this model has been presented in Ref. \cite{hirsch00}.
The authors define an alignment vector
$\Lambda_i \equiv (\mu v_i - v_d \mu_i) |_{M_{weak}}$, which parametrizes
the misalignment induced at the weak scale by the renormalization group
evolution of the soft terms (in the $\epsilon_i \ll \mu$ limit that we
are considering, this alignment vector is related to the misalignement
angle $\xi$ defined in subsection \ref{subsec:H_L_basis} by
$\sum_i \Lambda^2_i \simeq \mu^2 v^2_d \sin^2 \xi$).
The tree-level neutrino mass matrix is given by Eq. (\ref{eq:M_nu_tree}),
with the replacement $\mu_i \rightarrow \Lambda_i$. In the regime where
$\sum_i \epsilon^2_i \ll \sqrt{\sum_i \Lambda^2_i}$ and
$\epsilon_2 \Lambda_2 / \epsilon_3 \Lambda_3 < 0$,
the one-loop corrections are small and do not spoil the
structure of the tree-level neutrino mass matrix . Thus the atmospheric
neutrino parameters are essentially determined by the $\Lambda_i$, with
$m_{\nu_3} \sim \sum_i \Lambda^2_i / \mu^2 M_2$,
$\tan \theta_{23} \approx \Lambda_2 / \Lambda_3$ and
$|U_{e3}| = \sin \theta_{13} \approx \Lambda_1 / \sqrt{\sum_i \Lambda^2_i}$.
Consistency with experimental data requires
$\Lambda_1 \ll \Lambda_2 \approx \Lambda_3$ and
$\sqrt{\sum_i \Lambda^2_i} \sim 0.1\, \mbox{GeV}^2$ (the second constraint
depends on the values of the supersymmetry parameters). As for the solar
neutrino parameters, they are determined by the one-loop corrections to the
neutralino-neutrino mass matrix, controlled by the ratios
$\epsilon_i / \mu\, $: $\Delta m^2_{21}$ is a function of
$\sqrt{\sum_i \epsilon^2_i} / \mu$, while $\theta_{12}$ depends on
$\epsilon_1 / \sqrt{\epsilon^2_2 + \epsilon^2_3}$. In the case of universal
boundary conditions at $M_{GUT}$, the $\Lambda_i / \Lambda_j$ are correlated
with the $\epsilon_i / \epsilon_j$, so that  the CHOOZ limit on $|U_{e3}|$
constrains the solar mixing angle to be small. A departure from the
universality hypothesis is therefore necessary to accommodate the large
mixing angle solution (see Ref. \cite{diaz03} for a more recent analysis
of this model).

Interestingly, this scenario can be checked at colliders such as the LHC or
a future linear collider \cite{bilinear_LSP_decay,phrv00,hirsch00}. Indeed,
for the required values of the $\epsilon_i$, the lightest neutralino
(assumed to be the LSP) should decay within the detector. Furthermore, the
ratios of branching ratios for semi-leptonic decays into different charged
leptons show some correlation with the lepton mixing angles. In particular,
$\mbox{BR}\, (\widetilde \chi^0_1 \rightarrow \mu q \bar q') /
\mbox{BR}\, (\widetilde \chi^0_1 \rightarrow \tau q \bar q')$ is strongly
correlated with $\tan \theta_{23}$, irrespective of the lightest neutralino
mass~\cite{bilinear_LSP_decay,hirsch00}.
The experimentally allowed range for the atmospheric mixing angle
indicate that this ratio should be of order
one. Similarly, $\mbox{BR}\, (\widetilde \chi^0_1 \rightarrow e q \bar q') /
\mbox{BR}\, (\widetilde \chi^0_1 \rightarrow \tau q \bar q')$ and to a
smaller extent $\mbox{BR}\, (\widetilde \chi^0_1 \rightarrow e \tau \nu_i) /
\mbox{BR}\, (\widetilde \chi^0_1 \rightarrow \mu \tau \nu_i)$
are correlated with $|U_{e3}|$ and $\tan \theta_{12}$,
respectively~\cite{phrv00}.
Since the solar mixing angle is large,
$\mbox{BR}\, (\widetilde \chi^0_1 \rightarrow e \tau \nu_i) /
\mbox{BR}\, (\widetilde \chi^0_1 \rightarrow \mu \tau \nu_i)$ should be
of order one. All above branching ratios, except
$\mbox{BR}\, (\widetilde \chi^0_1 \rightarrow e q \bar q')$, are
larger than $(10^{-4} - 10^{-3})$ and it should be possible to measure
them. Other collider signatures of this scenario are discussed in
Refs. \cite{hp03,hprv02,rpv01}.

The weakness of this scenario is that there is no a priori
reason for the absence of trilinear \Rp\ couplings, nor for the smallness
of the $\epsilon_i$. To cure this problem, one may invoke spontaneous
$R$-parity breaking, or an abelian flavour symmetry \cite{mira00}.

\vskip .5cm

\noindent
{\bf A model with soft bilinear couplings only} 


In Ref. \cite{abada01}, another option has been explored, namely the
possibility that $R$-parity is broken by soft bilinear terms only. This
scenario has 6 parameters (3 $B_i$ and 3 $\widetilde{m}^2_{di}$, or
equivalently 3 $B_i$ and 3 $v_i$) but does not assume anything about the
structure of the soft terms, contrary to the previous ``bilinear $R$-parity
breaking'' scenario, whose predictivity relies on the universality
assumption at the GUT scale. In addition to the tree-level contribution,
the neutrino mass matrix receives contributions from the neutral loop
diagrams from class (ii), in the classification of subsection
\ref{subsec:general}. The other loop contributions, class (i) and
class (iii), are negligible for low values of $\tan \beta$.
The tree-level and loop contributions are governed by the quantities
$\delta^i_{\mu} \equiv v_i / v_d$ and
$\delta^i_B \equiv (B v_i - B_i v_d) / v_d \sqrt{B^2 + \sum_i B^2_i}$,
respectively. To make the connection with the misalignment angles $\xi$
and $\zeta$ defined in subsection \ref{subsec:H_L_mixing}, note that
$\sum_i (\delta^i_{\mu})^2 = \sin^2 \xi$ and
$\sum_i (\delta^i_B)^2 = \sin^2 \zeta$.

In the limit where the subleading loop diagrams are neglected, only two
neutrinos are massive. Depending on the values of the various parameters,
the heaviest neutrino mass is determined either by the tree-level or the
loop contribution. Both atmospheric and solar neutrino data can be
accommodated, but the required values of the soft bilinear \Rp\ parameters
are very small, with $|\delta^i_{\mu}| \leq 8 \times 10^{-7}$ and
$|\delta^i_B| \leq 3 \times 10^{-5}$. In the absence of
a specific mechanism that would explain the weakness of $R$-parity violation
in the soft terms, it is difficult to motivate such small values.

 
\section{Neutrino Transition Magnetic Moments}
\label{sec:magnetic}
\index{Neutrino!magnetic moment|(}

Massive neutrinos can have magnetic dipole moments (and also electric
dipole moments if $C P$ is violated). Since the magnetic moment of a neutrino
is induced by loop diagrams involving a chirality flip, it is generally
proportional to its mass and therefore very tiny. For instance, a Dirac
neutrino with no interaction beyond the \SM\ has
$\mu_\nu = 3 e G_F m_\nu / 8 \sqrt{2} \pi^2
\simeq 3.2 \times 10^{-19}\, (m_\nu / \eV)\, \mu_B$ \cite{lee77,marciano77},
where $\mu_B \equiv e / 2 m_e$ is the Bohr magneton, to be compared with
a laboratory limit of $\mu_\nu < 1.0 \times 10^{-10}\, \mu_B$
at $90 \%$ C.L. \cite{pdg04}.
In the case of Majorana neutrinos -- relevant for \SUSY\
without $R$-parity -- only transition dipole moments $\mu_{\nu_i\nu_j}$,
$i \neq j$ are allowed (the coefficients $\mu_{\nu_i\nu_j}$ of the effective
operators $\bar \nu_i \sigma^{\mu \nu} \nu_j F_{\mu \nu}$ are antisymmetric
in $(i,j)$ due to the Majorana nature of the neutrino).
These correspond to transitions between a left-handed
neutrino and a right-handed (anti)neutrino with different flavours, and
mediate radiative neutrino decays $\nu_i \rightarrow \nu_j \gamma$.
Neutrino transition magnetic moments have several implications in
astrophysics; in particular, they can induce spin-flavour transitions such as
$\nu_{e L} \rightarrow \nu^c_{\mu R}$ or $\nu_{\mu L} \rightarrow \nu^c_{e R}$
in the solar magnetic field \cite{akhmedov88,lim88}. However this possibility
is strongly constrained by solar neutrino and KamLAND data, from which the bound
$\mu_\nu \lesssim (10^{-12} - 10^{-11})\, \mu_B$, similar to other astrophysical
bounds, has been derived \cite{miranda04}. Still obtaining such a large value
for $\mu_\nu$ while keeping small neutrino masses in explicit models
is a theoretical challenge.

In \Rp\ models, transition magnetic moments are generated from
the trilinear couplings $\lambda_{ijk}$ and $\lambda'_{ijk}$ via
lepton-slepton and quark-squark loops \cite{babu90,barbnum1}.
The corresponding diagrams only differ from the ones responsible for
neutrino masses by an additional photon vertex attached to an internal line.
As a consequence, an upper bound on $\mu_{\nu_i \nu_j}$
depending on the neutrino masses can be derived. Barring accidental
cancellations between different contributions to the neutrino mass matrix,
and assuming a conservative upper bound of $10 \eV$ for each
$M^{\nu}_{ij}$, one finds \cite{barbnum1}
$|\mu_{\nu_e \nu_\mu}| \leq {\cal O}\, (10^{-13} \mu_B)$ for light sleptons
(squarks), much above the \SM\ value. This bound can
be saturated only if the mass scales involved in atmospheric and solar
neutrino oscillations result from accurate degeneracies among neutrino masses.
%

An interesting mechanism for avoiding the constraining proportionality
between the mass and magnetic moment of a Dirac neutrino is to postulate
an approximate $SU(2)_\nu$\index{Group symmetries!Approximate $SU(2)_\nu$} 
symmetry under which $\nu_L$ and
$\nu^c_L$ (the $C P$ conjugate of $\nu_R$) form a doublet~\cite{voloshyn88}.
The Lorentz structure of interaction terms is then such that the
electromagnetic dipole moment operator $\nu_L^T C \s^{\mu \nu } \nu^c_L$ is
antisymmetric under $\nu_L \leftrightarrow \nu^c_L$ and transforms as a
singlet of $SU(2)_\nu$, while the Dirac mass\index{Dirac mass} operator
$\nu_L^T C \nu^c_L$ is symmetric and transforms as a triplet. Thus, in the
$SU(2)_\nu$ symmetric limit, $\mu_\nu$ can be nonzero while $m_\nu$ vanishes.

Babu and Mohapatra~\cite{babun} have generalized this mechanism to Majorana
neutrinos by replacing $SU(2)_\nu$ with an horizontal 
$SU(2)_H$\index{Group symmetries!Horizontal $SU(2)_H$}
flavour symmetry acting on the first two generations of leptons. In a \Rp\
model~\cite{babu90}, they consider a discrete version of this symmetry,
namely the $Z_2$\index{Discrete symmetries!$Z_2$} flavour group acting on 
the electroweak doublet and singlet lepton fields of the first two
generations as
$(L_e, L_\mu) \to (L_\mu, - L_e)$, $(e^c_L, \mu^c_L) \to (\mu^c_L, - e^c_L)$,
with all other fields left invariant. In combination with the assumption of
conservation of the lepton number difference $L_e-L_\mu$, this leads to a
large transition magnetic moment $\mu_{\nu_e \nu_\mu}$ together with
vanishing masses for the first two generation neutrinos. At the same time,
however, the $Z_2$\index{Discrete symmetries!$Z_2$} symmetry yields
$m_e = m_\mu$. 
This can be cured by assuming a soft breaking of $Z_2$ in the slepton sector.
Then a contribution to the transition moment as large as
$\mu_{\nu_e \nu_\mu } \approx (10^{-11} - 10^{-10})\, \mu_B$, consistent
with light neutrino masses, $m_{\nu_e }, m_{\nu_\mu } < 10 \eV$, and with
the observed value of the splitting between the electron and muon masses,
$m_\mu - m_e$, can be achieved at the price of a fine-tuning in the slepton
mass matrices.

Motivated by the desire of avoiding an unnatural fine-tuning, Barbieri
et al.~\cite{barbnum1} consider the continuous 
$SU(2)_H$\index{Group symmetries!Horizontal $SU(2)_H$}
flavour symmetry to be broken solely by the Yukawa couplings of the
electron and the muon. The mismatch $\lambda^e_{11} \neq \lambda^e_{22}$
results in the splitting
$(\tilde m^{e2}_{\scriptscriptstyle{LR}})_{11}
- (\tilde m^{e2}_{\scriptscriptstyle{LR}})_{22} \neq 0$
necessary to generate masses for the first two generation neutrinos, while
a large transition magnetic moment $\mu_{\nu_e \nu_\mu}$ can be obtained
by requiring
$(\tilde m^{e2}_{\scriptscriptstyle{LR}})_{11}
\simeq (\tilde m^{e2}_{\scriptscriptstyle{LR}})_{22} \gg
|(\tilde m^{e2}_{\scriptscriptstyle{LR}})_{11}
- (\tilde m^{e2}_{\scriptscriptstyle{LR}})_{22}|$.
However $(\tilde m^{e2}_{\scriptscriptstyle{LR}})_{11}$ is bounded by its
contribution to the one-loop corrections to the electron mass, resulting
in the upper bound $\mu_{\nu_e \nu_\mu} \leq {\cal O}\, (10^{-12} \mu_B)$.
Therefore, even with the help of suitable symmetries, the neutrino transition
magnetic moment generated by \Rp\ couplings happens to be at least 2 orders
of magnitude smaller than the present laboratory upper bound.
\index{Neutrino!magnetic moment|)}


\section{Neutrino Flavour Transitions in Matter Induced by {\boldmath{\Rp}} Interactions}

The oscillations of neutrinos in matter are affected
by their interactions with the medium. The most familiar illustration of this
phenomenon is the Mikheev-Smirnov-Wolfenstein (MSW) mechanism
\cite{wolfenstein78,mikheev85}, i.e. the enhancement of neutrino oscillations
inside the sun due to their coherent forward scatterings on electrons and
nucleons. Since the charged current interactions only contribute to
scatterings of electron neutrinos, the electron neutrinos on one side, and
the muon and tau neutrinos on the other side have different scattering
amplitudes on electrons, which results in oscillation parameters in matter
different from the oscillation parameters in vacuum.

Similarly, any non-standard interaction of neutrinos with the charged
leptons and down quarks, such as \Rp\ interactions \cite{roulet91_bis,guzzo91},
modifies neutrino oscillations in matter. For instance
the couplings $\lambda'_{131}$ and $\lambda'_{331}$ contribute to the
scattering processes $\nu_e d \rightarrow \nu_e d$ and
$\nu_\tau d \rightarrow \nu_\tau d$, respectively, and the corresponding
amplitudes are different as soon as $\lambda'_{131} \neq \lambda'_{331}$.
Moreover these couplings, if both present, induce the flavour-changing
scatterings $\nu_e d \rightarrow \nu_\tau d$. It follows that \Rp\
interactions can induce flavour transitions of neutrinos inside the sun
or the Earth, even if neutrinos are massless and do not oscillate in vacuum.

Several authors have studied the possibility of accounting for the
solar and atmospheric neutrino data with \Rp-induced flavour transitions,
or more generally with non-standard neutrino interactions.
While flavour-changing non-standard neutrino interactions can only play
a subleading r\^ole with respect to oscillations in the atmospheric
neutrino sector \cite{bergmann00}, several authors have found that they
could be responsible for the solar neutrino deficit (for recent studies,
see Refs. \cite{bergmann00_bis,gago02,guzzo02,adhikari02,dreiner02}),
although the case of pure \Rp-induced flavour transitions is strongly 
disfavoured by the SNO data \cite{dreiner02}. However, the KamLAND
results have showed that neutrinos in the solar neutrino energy range
oscillate in vacuum with parameters consistent with the large mixing angle
MSW solution, leaving only the possibility that \Rp-induced flavour
transitions contribute as a subdominant effect.


\section{\sloppy {\boldmath{$\Delta L = 2$}} Sneutrino Masses and Sneutrino-Antisneutrino
Mixing}
\label{sec:sneutrinos}
\index{Sneutrino!mass|(}

Supersymmetry breaking $\Delta L = 2$ sneutrino mass terms, parametrized by 
the Lagrangian terms $- \ud\, (\tilde m^2_{\Delta L = 2})_{ij}\,
\tilde \nu_i \tilde \nu_j + \mbox{h.c.}$ \cite{haber1,hirschosc}, are
expected in any \susyq\ model with nonzero neutrino Majorana masses.
These terms induce a mass splitting and a mixing between the sneutrino and
the antisneutrino of a same generation, which gives rise to characteristic
experimental signatures such as sneutrino-antisneutrino oscillations.
In the one-generation case, the sneutrino mass splitting is given by
$\Delta m^2_{\tilde \nu} \equiv m^2_{\tilde \nu_2} - m^2_{\tilde \nu_1}
= 2\, \tilde m^2_{\Delta L = 2}$, where the mass eigenstates $\tilde \nu_1$
and $\tilde \nu_2$ are linear combinations of $\tilde \nu$ and $\tilde \nu^c$.

In \susyq\ models with bilinear $R$-parity violation, the sneutrinos mix
with the neutral Higgs bosons, which leads to \LV\ sneutrino-antisneutrino 
mixing at the tree level. 
Under the assumption of $C P$ conservation, it is convenient
to deal with the sneutrino $C P$ eigenstates rather than with the lepton number
eigenstates $\tilde \nu$ and $\tilde \nu^c$. The $C P$-even sneutrinos
$\nu_{+i} \equiv (\tilde \nu_i + \tilde \nu^c_i) / \sqrt 2$ mix with the
$C P$-even Higgs bosons $h$ and $H$, and the $C P$-odd sneutrinos
$\nu_{-i} \equiv - i (\tilde \nu_i - \tilde \nu^c_i) / \sqrt 2$ with the
$C P$-odd Higgs boson $A$ and with the Goldstone boson that in the absence of
$R$-parity violation is absorbed by the $Z$ boson. As a result the mass
degeneracy between the $C P$-even and $C P$-odd sneutrinos of each 
generation is
broken. The sneutrino mass splitting within each generation reads, if
one neglects flavour mixing \cite{haber2}:
\begin{equation}
  \Delta m^2_{\tilde \nu_i}\ \equiv\ m^2_{\tilde \nu_{+i}}
  - m^2_{\tilde \nu_{-i}}\ =\ \frac{4\, B^2_i\, M^2_Z\, m^2_{\tilde \nu_i}
  \sin^2 \beta}{(m^2_{\tilde \nu_i} - m^2_H) (m^2_{\tilde \nu_i} - m^2_h)
  (m^2_{\tilde \nu_i} - m^2_A)}\ ,
\label{eq:snu_mass_splitting}
\end{equation}
where $m^2_{\tilde \nu_i} = (M^2_{\tilde L})_{ii} + \mu^2_i - \frac{1}{8}
(g^2 + g^{\prime 2}) (v^2_u - v^2_d)$, and as usual we are working in a
($H_d$, $L_i$) basis in which the sneutrino \VEVs\ $v_i$ vanish. In this
basis $\Delta m^2_{\tilde \nu_i}$ is proportional to the square of the
\Rp\ soft terms $B_i$; therefore the contribution of bilinear $R$-parity
violation to the sneutrino-antisneutrino mixing 
\index{Sneutrino!mixing} is controlled by the
misalignment\index{Misalignment} between the 4-vectors
$B_{\alpha} \equiv (B_0, B_i)$ and $v_{\alpha} \equiv (v_0, v_i)$
(see subsection \ref{subsec:H_L_basis}).
At the one-loop level, $\Delta m^2_{\tilde \nu_i}$ receives
additional contributions from the trilinear \Rp\ couplings $\lambda$ and
$\lambda'$ and from their associated soft parameters $A$ and $A'$.
In practice $\Delta m^2_{\tilde \nu_i} \ll m^2_{\tilde \nu_i}$
and one can write $\Delta m_{\tilde \nu_i} \equiv m_{\tilde \nu_{+i}}
- m_{\tilde \nu_{-i}} \simeq \Delta m^2_{\tilde \nu_i}\, /\,
2 m_{\tilde \nu_i}$.

Sneutrino-antisneutrino mixing \index{Sneutrino!mixing}
and neutrino masses are closely
linked at the one-loop order. A Majorana\index{Majorana mass} neutrino mass
term induces radiatively a sneutrino mass splitting term and vice-versa.
Taking these effects into account, one finds \cite{haber2} that for generic
model parameters the sneutrino mass splitting to neutrino mass ratio
$\frac{\Delta m_{\tilde \nu}}{m_\nu}$ falls in the interval
\begin{equation}
  10^{-3}\ \lesssim\ \frac{\Delta m_{\tilde \nu}}{m_\nu}\
  \lesssim\ 10^{3}\ .
\end{equation}
Cancellations between the tree-level and one-loop contributions to $m_\nu$
may enhance this ratio, thus allowing for larger sneutrino mass splittings
at the price of a fine-tuning. Furthermore there arise strong pairwise
correlations, of nearly linear character, between the contributions to the
Majorana neutrino masses, the $\Delta L = 2$ sneutrino masses and the
neutrinoless double beta decay amplitudes. Hirsch et al.~\cite{klapdoretal}
find in the framework of the $R$-parity conserving \SSM\ that the induced
effect of $\tilde m_{\Delta L = 2}$ on neutrinoless double beta decay imposes
the bound $\tilde m_{\Delta L = 2} < 2 \GeV \ (\tilde m / 100 \GeV)^{3/2}$,
resp. $\tilde m_{\Delta L = 2} < 11 \GeV\ (\tilde m / 100 \GeV)^{7/2}$, in
the extreme case where the lightest neutralino is a pure bino, resp.
a pure higgsino (all superpartner masses assumed equal to
$\tilde m$). Another bound, $\tilde m_{\Delta L = 2} < (60 - 125) \MeV \
(m_\nu / 1 \eV)^\ud$, is associated with the one-loop contribution to
the neutrino mass induced by $\tilde m_{\Delta L = 2}$. These bounds,
which can be converted into bounds on $\Delta m_{\tilde \nu}$ via
$\Delta m_{\tilde \nu} \simeq \tilde m_{\Delta L = 2} / m_{\tilde \nu}$,
also apply in the \Rp\ case.

The phenomenological implications of the sneutrino-antisneutrino
mass splittings and mixings have been examined in recent
works~\cite{haber1,haber2,hirschosc}. For large mass splittings, 
$\Delta m_{\tilde \nu } > 1 \GeV$, the sneutrino pair
production at colliders could be tagged through the leptonic decays of the
sneutrinos resulting in characteristic charged
dilepton final states~\cite{haber1} through the decay modes
$\tilde \nu \to e^\pm \tchi^\mp$. Interesting signals could also
arise from the resonant sneutrino or antisneutrino production~\cite{feng98}
in $e^+e^-$ or $q\bar q$ collisions. The corresponding off-shell sneutrino
or antisneutrino exchange processes could also be observed via
the fermion-antifermion pair production reactions,
$e^+e^- \to \tilde \nu,\, \tilde \nu^c \to f \bar f$,
at high energy lepton colliders and similarly at hadron
colliders~\cite{haber2,shalom}. 

For small mass splittings, $\Delta m_{\tilde \nu } << 1 \GeV$,
sneutrino-antisneutrino oscillations could rise to a measurable level
provided that the oscillation time is shorter than the sneutrino lifetime,
corresponding to
$x_\nu \equiv \frac{\Delta m _{\tilde \nu}}{\G_{\tilde \nu}} > 1$.
By analogy with the $ B-\bar B$ system, the
production of a sneutrino-antisneutrino pair would be signaled by
characteristic like-sign dileptons, provided that the branching ratios
of the decay modes $\tilde \nu \to e^\pm \tchi^\mp$
are appreciable~\cite{haber1}.

Bar-Shalom et al.~\cite{shalom} have developed an interesting  test 
for the \Rp -induced resonant
production of sneutrinos at leptonic and hadronic colliders.  Based on
the current sensitivity reaches, a mere observation of deviations with
respect to the \SM\  predictions for the tau-antitau lepton pair
production reactions $e^+e^- \to \tilde \nu \to \tau^+ \tau^-$, resp.
$p\bar p \to \tilde \nu \to \tau^+ \tau ^- + X$,   would
lead to bounds on coupling products of the form
$\l_{232} \l'_{311} < 0.003$, resp. $\l_{232} \l'_{322} < 0.011$.
In the presence of $C P$ violation among \Rp\ couplings,
the $\tilde \nu$-$\tilde \nu^c$ mixing could contribute to the $C P$-odd
double spin correlation observables associated with the spin polarisation
of the $\tau^+ \tau^-$ pair. Nonzero and large contributions arise
already at the tree level, thanks
to the complex phase dependence provided by the spatial azimuthal angle.  
Analogous  $C P$-even double spin correlation observables may
also be induced through the same type of \Rp\ interactions. 
\index{Sneutrino!mass|)}

\vskip .5cm

Altogether, we have seen how the violation of $R$-parity
in \susyq\ theories naturally leads to massive neutrinos.
The fact that \Rp\ models automatically
incorporate massive neutrinos, while allowing for observable signals at
colliders for the values of \Rp\ couplings suggested by neutrino data, is
certainly an appealing feature of $R$-parity violation. However, the smallness
of neutrino masses dictates strong constraints on \Rp\ couplings, especially
on bilinear \Rp\ couplings, while \Rp\ \susyq\ models also suffer from a lack
of predictivity in the neutrino sector. This motivates the study of restricted
scenarios such as the ``bilinear $R$-parity breaking'' scenario, which are
more predictive. We have also seen that, given the acceptable neutrino masses,
the neutrino transition magnetic moments generated from \Rp\ couplings
generally lie well below the current laboratory upper limit. Finally,
\Rp\ couplings induce $\Delta L = 2$ sneutrino mass terms,
which leads to mass splittings and mixings between sneutrinos and
antisneutrinos and gives rise to sneutrino-antisneutrino oscillations.
\index{Neutrino!mass|)}

\cleardoublepage                                                          %
\chapter{INDIRECT BOUNDS ON {\boldmath{$R$}}-parity ODD INTERACTIONS}                  %
\label{chap:indirect}                                                     %
\index{Bounds on \Rp\ interactions|(}  
The assumption of a broken $R$-parity introduces in the \SSM\ new  
interactions between ordinary particles and \susyq\ particles, which 
can contribute to a variety of low, intermediate and high energy 
processes that do not involve the production of  
superpartners in the final state. Requiring that the \Rp\ contribution to  
a given observable does not exceed the limit imposed by the precision of  
the experimental measurement and by the theoretical uncertainties on the  
\SM\ prediction (or, for a process that has not been observed,   
that it does not exceed the experimental upper limit), yields upper bounds  
on the \Rp\ couplings involved. In addition to the bounds associated  
with renormalization group effects and with astrophysics and cosmology,  
discussed in chapters \ref{chap:evolution} and \ref{chap:cosmology},   
and with the direct limits extracted from \SUSY\ searches at colliders,   
to be discussed in chapter \ref{chap:colliders}, these indirect bounds   
provide useful information on the possible patterns of $R$-parity violation.  
  
In this chapter, we give a comprehensive review of the indirect bounds  
on \Rp\ interactions coming from low, intermediate and high energy particle  
phenomenology, as well as from nuclear and atomic physics observables.  
This complements and updates other existing reviews, e.g. Refs.  
\cite{barger89,dreiner1,hinchliffe,reviewsm,reviews,reviews2,allanach99}.  
  
The chapter is organised in five main sections.  
In section~\ref{sec:introind}, the assumptions under which the bounds on   
\Rp\ couplings have been extracted are presented, and issues concerning   
quark and lepton superfield bases are addressed, in particular in connection   
with the single coupling dominance hypothesis. A basis-independent   
parametrization of $R$-parity violation is also presented,  
as an alternative to the choice made in this review.  
Section~\ref{sec:bilcons} deals with the constraints on bilinear \Rp\   
parameters, both for an explicit breaking and in the case of a   
spontaneous $R$-parity breaking.  
Section~\ref{sec:tricons} reviews the indirect bounds on trilinear \Rp\  
couplings associated with fundamental tests of the Standard Model   
in charged current and neutral current interactions, with $C P$ violation  
and with high precision measurements of electroweak observables.  
In section~\ref{sec:decaycons} the constraints on trilinear \Rp\  
interactions coming from a variety of hadron flavour or lepton   
flavour violating processes, and $B$ or $L$ violating processes  
are presented.  
Finally, section~\ref{sec:indcons} provides a list of the indirect bounds  
on trilinear \Rp\ couplings presented in chapters \ref{chap:evolution} and  
\ref{chap:neutrinos} and in this chapter. The robustness and significance  
of these bounds are discussed.

\section{Assumptions and Framework}  
\label{sec:introind}  
  
Our main focus in this chapter will be on the $R$-parity odd  
renormalizable superpotential couplings $\mu_i$, $\lambda_{ijk}$,  
$\lambda'_{ijk}$ and $\lambda''_{ijk}$, defined in Eq. (\ref{eq:W_Rp_odd}).  
The simultaneous presence of bilinear and trilinear \Rp\ terms results  
in an ambiguity in the choice of a basis for the down-type Higgs and lepton  
doublet superfields ($H_d$, $L_i$), an issue already addressed  
in subsections \ref{subsec:basis} and \ref{subsec:H_L_basis}. In this chapter  
we adopt a superfield basis in which the \VEVs\ of the sneutrino fields  
vanish and, unless otherwise stated, the charged lepton Yukawa couplings  
are diagonal (see the discussion at the beginning of subsection  
\ref{subsec:bilinear_bound}). 
  
\subsection{The Single Coupling Dominance Hypothesis}  
\label{subsecmeth}  
    
Most of established indirect bounds on the trilinear couplings  
$\lambda_{ijk}$, $\lambda'_{ijk}$ and $\lambda''_{ijk}$  
have been derived under the so-called single  
coupling dominance hypothesis, where a single \Rp\ coupling is  
assumed to dominate over all the others.  This useful working  
hypothesis can be rephrased by saying that each of the \Rp\ couplings  
contributes one at a time~\cite{barger89,dimopoulos88}.  The  
perturbative constraints arising from contributions at loop order $l$  
then involve upper bounds on combinations of \Rp\ couplings and  
superpartner masses of the form ${\hat  \l ^p \over \tilde m^q } ({\hat e^2  
\over (4\pi )^2 } )^l $, where $ \hat \l , \ \hat e , \ \tilde m $ are  
generic symbols for \Rp\ couplings, gauge couplings and superpartner masses,  
and the power indices $p \ge  2  , \ q  \ge 2$ depend on the   
underlying  mechanism.  Apart from  
a few isolated cases, the bounds derived under the single coupling  
dominance hypothesis are in general  
moderately strong.  The typical orders of magnitude are $\l ,\ \l'  
,\ \l '' < ( 10^{-2} - 10^{-1}) { \tilde m\over 100 \ \text{GeV} } $,  
involving generically a linear dependence on the superpartner mass.  
  
The constraints based on the single coupling dominance  
hypothesis deal with a somewhat restricted set of applications, such  
as charged current or neutral current gauge interactions, neutrinoless  
double beta decay and neutron-antineutron oscillation. By contrast, a much  
larger fraction of the current constraints on \Rp\ interactions are  
derived from extended hypotheses where the dominance is postulated for  
quadratic or quartic products of couplings. The literature abounds with  
bounds involving a large variety of flavour configurations for quadratic  
products of the coupling constants. The processes that yield constraints  
on products of \Rp\ couplings can be divided into four main classes:  
(i)  hadron flavour changing processes, such as oscillations of neutral  
flavoured mesons, and leptonic or semileptonic decays of $K$ or $B$ mesons  
like $K\to e_i  \bar e_j$ and $K\to \pi \nu \bar \nu $;  
(ii)  lepton flavour changing processes, such as $\mu ^-\to e^-$ conversion  
in nuclei, or radiative decays of charged leptons;  
(iii) $L$-violating processes, such as neutrinoless double beta  
decay, neutrino Majorana masses and mixings (cf. chapter  
\ref{chap:neutrinos}), or three-body decays of charged  
leptons, $l_l^\pm \to l^\pm l_n^- l^+_p $;  
(iv)  $B$-violating processes, such as nucleon decay,  
neutron-antineutron oscillations, double nucleon decay or  
some rare decays of heavy mesons.  
  
\subsection{Choice of the Lepton and Quark Superfield Bases}  
\label{secchoice}  
 
When discussing specific bounds on coupling constants, it is necessary  
to choose a definite basis for quark and lepton superfields, especially  
if the single coupling dominance hypothesis is used. Two obvious basis  
choices can be made for quark superfields: the current (or weak eigenstate)  
basis, in which left-handed quarks have flavour diagonal couplings to the  
$W$ gauge boson, and the ``super-CKM'' basis, in which quark mass matrices  
are diagonal. A similar distinction between weak eigenstate basis and mass  
eigenstate (or ``super-MNS'') basis exists for leptons when neutrino masses  
are taken into account. 
In most studies, it is tacitly understood that the single coupling dominance  
hypothesis applies in the mass eigenstate basis. It may appear more  
natural, however, to apply this hypothesis in the weak eigenstate basis   
when dealing with models in which the hierarchy among (weak-eigenstate-basis) 
couplings originates from some flavour theory. In this case, a single  
process may allow to constrain several couplings, provided one has some  
knowledge of the rotations linking the weak eigenstate and mass eigenstate  
bases~\cite{agashe}.  
  
It is useful to write down the trilinear \Rp\ superpotential terms in  
the two superfield bases. Let us first consider the $L Q D^c$ terms.  
We denote by $\hat \lambda'_{ijk}$ the corresponding couplings expressed in  
the weak eigenstate basis, and by $\lambda^{'A}_{ijk}$ and  
$\lambda^{'B}_{ijk}$ the same couplings in two useful representations  
of the mass eigenstate basis:  
\begin{eqnarray}  
W^ {(\l ')} &=&  
\hat \l '_{ijk} (\hat{N}_i \hat{D}_j - \hat{E}_i \hat{U}_j) \hat{D}^c_k=  
\l^{'A}_{ijk} (\hat{N}_i D'_j - \hat{E}_i U_j)D^c_k=  
\l^{'B}_{ijk} (\hat{N}_i D_j - \hat{E}_i U'_j)D^c_k \  
, \cr \l^{'A}_{ijk}&=& \hat \l '_{imn}(V_L^{u\dagger })_{mj} (V_R^{dT  
})_{nk}\ , \ \ \ D'_j =V_{jl} D_l\ ,\ \cr \l^{'B}_{ijk}&= & \hat \l  
'_{imn}(V_L^{d\dagger })_{mj} (V_R^{dT })_{nk}\ , \ \ \ U'_j=  
V^\dagger_{jl} U_l\ .  
\label{eqmp0}  
\end{eqnarray}  
In (\ref{eqmp0}), hatted superfields are in the weak eigenstate  
basis and unhatted superfields in the mass eigenstate  
basis; $V^u_L$, $V^d_L$ and $V^d_R$ are the matrices that rotate,  
respectively, the left-handed up quarks, the left-handed down quarks  
and the right-handed down quarks to their mass eigenstate basis (one has  
e.g. $U_i = \sum_{j} (V^u_L)_{ij} \hat{U}_j$); and $V = V_{CKM} =  
V_L^uV_L^{d\dagger}$ is the Cabibbo-Kobayashi-Maskawa (CKM) matrix.  
Since the three sets of couplings $\l^{'A}$, $\l^{'B}$ and $\hat \l'$  
are related by unitary matrices, the following sum rules hold:  
$\sum_{jk} \vert \l ^{'A}_{ijk}\vert ^2 =\sum_{jk} \vert \l ^{'B}_{ijk}  
\vert ^2 = \sum_{jk} \vert \hat \l'_{ijk}\vert ^2$.  
The hypothesis of a single dominant \Rp\ coupling, when applied to the  
coupling sets $\{ \l^{'A}_{ijk} \}$ or $\{ \l^{'B}_{ijk} \}$, may allow  
for flavour changing transitions in the down quark or up quark sectors,  
respectively, but not in both simultaneously.  
The flavour mixing may be formally obtained by replacements of the form   
$b \to b' = V_{33} b + V _{32} s +V_{31} d$ in the former case, and  
$t \to t' = V^\dagger_{33} t + V^\dagger_{32} c +V^\dagger _{31} u$ in the  
latter case (with similar relations for the first two generations).  
  
Analogous expressions can be written for the $L L E^c$ interaction terms:  
\begin{eqnarray}  
W^{(\l )} &=&\ud \hat \l_{ijk} (\hat{N}_i \hat{E}_j - \hat{E}_i \hat{N}_j)  
\hat{E}^c_k=\ud \l^A_{ijk}  
(N_i E'_j -E'_i N_j)E^c_k=\ud \l^B_{ijk} ( N'_i E_j -E_i  
N'_j)E^c_k\ , \cr \l^A_{ijk}&=& \hat \l_{lmn} (V_L^{\nu \dagger })_{li}  
(V_L^{\nu \dagger })_{mj}(V_R^{e T })_{nk}\ ,\ \ E'_j=V'_{jm} E_m\ , \cr  
\l^B_{ijk}&=& \hat \l_{lmn} (V_L^{e \dagger })_{li} (V_L^{e \dagger  
})_{mj}(V_R^{e T })_{nk}\ ,\ \ N '_i=V^{'\dagger }_{il} N _l\ ,  
\label{eqmn0}  
\end{eqnarray}  
where $V^\nu_L$, $V^e_L$ and $V^e_R$ are the rotation matrices for  
left-handed neutrinos, left-handed charged leptons and right-handed  
charged leptons, respectively, and the Maki-Nakagawa-Sakata matrix is  
$U_{MNS} = V^{' \dagger} = V_L^e V_L^{\nu \dagger}$. For the $U^c D^c D^c$  
interactions, the mass eigenstate basis couplings $\l''_{ijk}$ are  
related to the current basis couplings $\hat \l''_{ijk}$ by $\l''_{ijk}  
= \hat \l''_{lmn} (V_R^{uT})_{li}(V_R^{dT})_{mj}(V_R^{dT})_{nk}$. 
  
As mentioned above, if the single coupling dominance hypothesis applies  
in the weak eigenstate basis, several bounds may be derived from  
a single process. Indeed, let us assume that, in the weak eigenstate basis,  
the single \Rp\ coupling $\hat \l'_{IJK}$ is dominant. Then, in the mass  
eigenstate basis, this coupling generates the operator  
$\l^{'A}_{IJK} E_I U_J D_K^c$ and, due to the flavour mixing, subdominant  
operators $\l^{' A}_{ijk} E_i U_j D_k^c$, $(i,j,k) \neq (I,J,K)$, with  
couplings $\l^{' A}_{ijk}$ suppressed relative to $\l^{'A}_{IJK}$ by fermion  
mixing angles. Explicitly, we have  
$\hat \l'_{IJK} \hat{E}_I \hat{U}_J \hat{D}_K^c\, \simeq\,  
\hat \l'_{IJK} E_I U_J D_K^c\,  
+\, \sum_{i \neq I} (V^{e \star}_L)_{iI} \hat \l'_{IJK}\, E_i U_J D_K^c\,  
+\, \sum_{j \neq J} (V^{u \star}_L)_{jJ} \hat \l'_{IJK}\, E_I U_j D_K^c\,  
+\, \sum_{k \neq K} (V^d_R)_{kK} \hat \l'_{IJK}\, E_I U_J D_k^c\,  
+\, \cdots$, where we have assumed that the rotation matrices  
$V_{L,R}^{(e,u,d)}$ are close to the unit matrix. Since  
$\l^{'A}_{IJK} \simeq \hat \l'_{IJK}$, the following relations among  
couplings in the mass eigenstate basis   
hold (from now on, we drop the upper  
index $A$ in the mass eigenstate basis couplings $\l^{'A}_{ijk}$.):  
$\l'_{iJK} \simeq (V^{e \star}_L)_{iI} \l'_{IJK}$,  
$\l'_{IjK} \simeq (V^{u \star}_L)_{jJ} \l'_{IJK}$,  
$\l'_{IJk} \simeq (V^d_R)_{kK} \l'_{IJK}$, etc.  
It may then happen that the most severe bound on the  
dominant coupling $\l'_{IJK}$ comes from a process involving a subdominant  
operator $E_i U_j D_k^c$. For instance an upper bound on the coupling  
$\l'_{IjK}$, $|\l'_{IjK}| < |\l'_{IjK}|_{upper}$, yields  
$|\l'_{IJK}| < |\l'_{IjK}|_{upper}\, /\, |(V^{u \star}_L)_{jJ}|$.  
This constraint may be stronger than bounds extracted from processes  
involving the dominant operator $E_I U_J D_K^c$. 
 
More generally, one can derive a sequence of bounds on \Rp\ couplings 
from a single process without any reference to the single coupling dominance  
hypothesis, provided that one has some knowledge of the rotation matrices  
$V_{L,R}^{(e,u,d)}$. For instance, in the model of Ref. \cite{ellis98},  
in which the quark rotation matrices are controlled by an abelian flavour  
symmetry, the constraints $\l'_{111} < 0.002$ and $\l'_{133} < 0.001$  
(associated with the non-observation of neutrinoless double  
beta decay and with the upper limit on the electron neutrino mass,  
respectively, and derived assuming superpartner masses of $200 \GeV$)  
are used to derive upper bounds on all $\l'_{1jk}$ couplings. Combining  
the latter with the bounds extracted from other observables, one obtains  
$\l'_{1jk} < (10^{-3} - 2 \times 10^{-2})$, depending on the coupling.  
Such bounds obtained from flavour mixing arguments, however, are  
model-dependent. In Ref.~\cite{addprd69}, upper bounds on the $\l'_{ijk}$ 
couplings are derived from the cosmological neutrino mass bound 
under various assumptions regarding the quark rotation matrices 
and $(V_{L,R}^e)_{ij} = \delta_{ij}$. 
  
The same arguments allow to transform a bound on a product of \Rp\  
couplings into a bound on a single coupling. For instance, in the model  
of Ref. \cite{ellis98}, the contraint   
$\vert  \l ^{'\star }_{i13}\l '_{i31} \vert < 3.2 \times 10^{-7}$,  
associated with $B - \bar B$ mixing, yields $|\l '_{i13}| < 6 \times 10^{-4}$  
(again for sfermion masses of $200 \GeV$). 

\subsection{A Basis-Independent Parametrization of {\boldmath{$R$}}-Parity Violation}  
\label{sec:invariants}  
\index{Parameters for \Rp\ models!basis--independent|(}  
As explained in subsection \ref{subsec:basis}, there is no a priori  
distinction between the ($Y=-1$) Higgs superfield and the lepton superfields  
in the absence of $R$-parity and lepton number. The choice of a basis  
for $H_d$ and the $L_i$ becomes even a delicate task if bilinear \Rp\  
terms and/or sneutrino \VEVs\ are present.  
As a result, the definition of the \Rp\ parameters is ambiguous  
(except for the baryon number violating couplings $\lambda''_{ijk}$, which  
are not discussed here). This ambiguity is removed either by  
choosing a definite ($H_d$, $L_i$) basis, as explained in subsection  
\ref{subsec:H_L_basis}, or by parametrizing $R$-parity violation by a  
complete set of basis-invariant quantities instead of the original \Rp\  
Lagrangian parameters.  
  
In this subsection, we briefly present the second approach, as an  
alternative to the choice made in this 
review~\footnote{Although we frequently use the basis-independent quantities 
                 $\sin \xi$ and $\sin \zeta$ to measure the total amount of 
		 bilinear $R$-parity violation in the fermion and scalar 
                 sector, respectively  
		 (cf. subsection~\ref{subsec:H_L_basis}),  
                 we do not make a systematic use of invariants 
                 to parametrize $R$-parity violation.}. We use  
the notations of subsection \ref{subsec:H_L_basis}, with the 4 doublet  
superfields $H_d$ and $L_i$ grouped into a 4-vector $\hat L_\alpha$,  
$\alpha = 0,1,2,3$. The renormalizable, baryon number conserving  
superpotential then reads  
$W = \mu_\a H_u \hat{L}_\a + \frac{1}{2} \l^e _{\a \b k }  
\hat{L}_\a \hat{L}_\b E_k^c + \l^d_{\a pq} \hat{L}_\a Q_p D_q^c +  
\l^u _{pq} Q_p U^c_q H_u$. The couplings  
$\l^e_{\a \b k} = (\l _e ^k)_{\a \b}$, $\l^d_{\a pq} =(\l_d ^{pq})_\a$  
and $\mu_\a$ define 3 matrices and 10 vectors in the $\hat L_\alpha$ field  
space (each value of $k$ defines a $4 \times 4$ matrix, and each value of  
($p$, $q$) a 4-vector). In addition, the \Rp\ soft terms $B_\a$,  
$\widetilde m^2_{\a \b}$, $A^e_{\a \b k} = (A_e^k)_{\a \b}$ and  
$A^d_{\a pq} = (A_e^{pq})_\a$ define 4 matrices and 10 vectors.  
At this stage the ``ordinary'' Yukawa couplings and the trilinear \Rp\  
interactions cannot be disentangled; the distinction  
between lepton number conserving and lepton number violating interactions  
arises once a direction in the $\hat{L}_\a$ field space has been chosen  
for the Higgs superfield.   
  
However, an intrinsic definition of $R$-parity violation is possible with  
the help of basis-invariant products of the \Rp\ parameters  
\cite{davidson97, davidson98}. These invariants, being independent of the  
basis choice in the $\hat L_\a$ field space, can be defined in a geometrical  
way. Let us give an example of an invariant involving only superpotential  
parameters, together with its geometrical interpretation. We first notice  
that each of the 10 vectors $(\l_d ^{pq})_\a$ and $\mu_\a$ defines a  
(would-be) Higgs direction in the $\hat L_\a$ field space (if  
$(H_d)_\a \propto (\l_d ^{pq})_\a$, all $\lambda'_{ijk}$ vanish, and if  
$(H_d)_\a \propto \mu_\a$, all $\mu_i$ vanish).  
If the directions defined by $\mu_\a$ and one of the $(\l_d ^{pq})_\a$  
differ, $R$-parity is violated. Another, slightly different criterion  
for $R$-parity violation is when one of the $(\l_d ^{pq})_\a$ has a nonzero  
component along one of the vectors $\sum_\b \mu^\star_\b (\l_e ^k)_{\b \a}$  
(to be interpreted as the three lepton directions, labeled by the flavour  
index $k$, the direction for the Higgs superfield being defined by $\mu_\a$).  
The projection of the vector $(\l_d ^{pq})_\a$ on the direction  
$\sum_\b \mu^\star_\b (\l _e ^k)_{\b \a}$ is then a measure of $R$-parity  
violation in the $k^{\rm th}$ lepton number. The corresponding invariant can  
be defined as \cite{davidson97}:  
\begin{eqnarray}    
  \delta_1^{kpq}\ =\ \frac{|\mu^\dagger \l_e^k \l_d^{pq \star}|^2}  
  {|\mu|^2 |\l_e^k|^2 |\l_d^{pq}|^2}\ ,   
\label{eq:basis_invariant}  
\end{eqnarray}    
where $\mu^\dagger \l_e^k \l_d^{pq \star}  
\equiv \sum_{\a, \b} \mu^\star_\a (\l_e^k)_{\a \b} (\l_d^{pq})^\star_\b$,  
$|\mu|^2 \equiv \sum_\a |\mu_\a|^2$,  
$|\l_e^k|^2 \equiv \sum_{\a, \b} |(\l_e^k)_{\a \b}|^2$ and  
$|\l_d^{pq}|^2 \equiv \sum_\a |(\l_d^{pq})_\a|^2$.  
If $\delta_1^{kpq} \neq 0$ for some ($p$, $q$), $R$-parity and the   
$k^{\rm th}$ lepton number are violated (summing $\delta_1^{kpq}$  
over the right-handed lepton index $k$ provides a measure of the breaking of  
total lepton number). However, $\delta_1^{kpq} = 0$ for all  
($k$, $p$, $q$) does not imply the absence of $R$-parity violation from  
the superpotential, since other invariants involving the \Rp\ parameters  
$(\l_e^k)_{\a \b}$, $(\l_d^{pq})_{\a}$ and $\mu_\a$ can be constructed.  
  
The invariants that may be constructed out of the \Rp\ Lagrangian  
parameters are much more numerous than the \Rp\ parameters themselves,  
but not all of them are independent.  
After removing the redundancies, one finds a set of  
36 independent invariants parametrizing $R$-parity violation from the  
superpotential. Once soft \SUSY\ breaking terms are included, one obtains  
a total number of 78 independent invariants, in agreement with the counting  
of independent \Rp\ parameters that do not break baryon number (3 $\mu_i$,  
9 $\lambda_{ijk}$ and 27 $\lambda'_{ijk}$ in the superpotential; 3 $B_i$,  
3 $\widetilde m^2_{di}$, 9 $A_{ijk}$ and 27 $A'_{ijk}$ in the scalar  
potential; but due to the freedom of redefining the ($H_d$, $L_i$) basis,  
only 6 among the 9 bilinear \Rp\ parameters are physical).  
  
The basis-independent approach has been used for the derivation of  
cosmological bounds on $R$-parity violation \cite{davidson97,davidson98}  
and for the computation of neutrino masses and mixings in \Rp\ models  
\cite{davidson00,davidson00_bis,abada01}. 
\index{Parameters for \Rp\ models!basis--independent|)} 
\subsection{Specific Conventions Used in this Chapter}  
\label{sec:specif}  
  
In presenting numerical results for coupling constants, we need at  
times to distinguish between the first two families and the third.  
For this purpose, when quoting numerical bounds only, we assume the  
following conventions for the alphabetical indices: $ l, m, n \in [1,2]  
$ and $ i, j , k \in [1,2,3]$. The mass of superpartners are fixed  
at the reference value of $\tilde m= 100 \ \text{GeV}$ unless  
otherwise stated.  A notation like $\tilde d_{kR}^p$ in a numerical  
relationship, such as $\l'_{ijk} < 0.21\, \tilde d_{kR}^p$,  
stands for the expression, $\l'_{ijk} < 0.21\, (\frac{m_{\tilde  
d_{kR}}}{100 \text{GeV}})^p$.  
  
The following auxiliary parameters   
for the \Rp\ coupling constants arise in the main body of the   
text~\cite{barger89}:   
\begin{eqnarray}  
 r_{ijk} (\tilde l)= {M_W^2\over g_2^2 m^2_{\tilde l} } \vert  
 \l_{ijk}\vert^2  
 \qquad , \qquad  
 r'_{ijk} (\tilde q)= {M_W^2\over g_2^2 m^2_{\tilde q} } \vert  
 \l'_{ijk}\vert^2 \qquad .  
\end{eqnarray}  
  
Finally, unless otherwise stated, we also rely on the Review of Particle   
Physics from the Particle Data Group~\cite{pdg04} as a source for the  
experimental data information as well as for short reviews on the main   
particle physics subjects. 
Our notations and conventions are given in appendix~\ref{chap:appendixA}.  
We also assume familiarity with the standard textbooks and reviews,   
such as~\cite{rosstext} for general theory  
and~\cite{haber85,palmoha,bargerphil} for phenomenology.  
\index{Bounds on \Rp\ interactions|)} 

\section{Constraints on Bilinear {\boldmath{\Rp}} Terms and on Spontaneously Broken  
{\boldmath{$R$}}-Parity}  
\label{sec:bilcons}  
\index{Bounds on \Rp\ interactions!Bilinear Terms|(}  
In this section, we summarize the main constraints on bilinear \Rp\  
parameters, both for an explicit breaking (cf. section  
\ref{sec:bilinear}) and for a spontaneous breaking of $R$-parity  
(cf. section \ref{sec:spontaneous}). 
The spontaneous breaking of $R$-parity is  
characterised by an $R$-parity invariant Lagrangian leading to non-vanishing  
\VEVs\ for some $R$-parity odd scalar field, which in turn generates \Rp\  
terms. Since such a spontaneous breakdown of $R$-parity generally also  
entails the breaking of the global continuous symmetry associated lepton 
number conservation, this scenario is distinguished by a non-trivial scalar 
sector including a massless Goldstone boson, the Majoron\index{Majoron}, and a 
light scalar field, partner of the pseudo-scalar Majoron.   
Some scenarios of spontaneous $R$-parity breaking also   
involve a small amount of explicit lepton number breaking, in which case  
the Majoron\index{Majoron} becomes a massive pseudo-Goldstone boson.  
By contrast, the explicit $R$-parity breaking case may lead to finite  
sneutrino VEVs $<\tilde \nu_i>\, \equiv v_i / \sqrt{2}$, but the  
Lagrangian density always includes terms that violate $R$-parity 
intrinsically.  

\subsection{Models with Explicit {\boldmath{$R$}}-Parity Breaking}  
\label{subsec:bilinear_bound}  
   
The bilinear \Rp\ parameters consist of  
3 \Rp\ superpotential masses $\mu_i$ and 6 soft \SUSY\ breaking parameters  
\index{Mixing!higgs--slepton}
mixing the Higgs bosons with the sleptons (3 \Rp\ $B$-terms $B_i$  
associated with the $\mu_i$, and 3 \Rp\ soft mass parameters   
${\tilde m}^2_{di}$).  
In the presence of these parameters, the sneutrinos generally acquire  
\VEVs\ $v_i$, which in turn induce new bilinear \Rp\ interactions.  
However the $v_i$ are not independent parameters, since they can be  
expressed in terms of the Lagrangian parameters -- or, alternatively,  
they can be chosen as input parameters, while 3 among the 9 bilinear \Rp\  
parameters $\mu_i$, $B_i$ and ${\tilde m}^2_{di}$ are functions of the  
$v_i$ and of the remaining \Rp\ parameters.  
  
As explained in subsection \ref{subsec:H_L_basis}, in the presence of  
bilinear $R$-parity violation, the ($H_d$, $L_i$) superfield basis in which  
the Lagrangian parameters are defined must be carefully specified.  
Indeed, due to the higgsino-lepton \index{Mixing!higgsino--lepton}
and Higgs-slepton \index{Mixing!higgs--slepton} mixings induced by  
the bilinear \Rp\ terms, there is no preferred basis for the $H_d$ and $L_i$  
superfields, and a change in basis modifies the values  
of all lepton number violating parameters, including the trilinear couplings  
$\lambda_{ijk}$ and $\lambda'_{ijk}$ (cf. subsection \ref{subsec:basis}).  
In practice the most convenient choice is the basis in which the \VEVs\ of  
the sneutrino fields vanish and the Yukawa couplings of the charged leptons  
are diagonal. In the (phenomenologically relevant) limit of small bilinear  
$R$-parity violation, this superfield basis is very close to the fermion mass  
eigenstate basis, and therefore allows for comparison with the indirect bounds  
on trilinear \Rp\ couplings derived later in section \ref{sec:tricons}.  
>From now on we shall assume that this choice of basis has been made.  
Therefore:  
\begin{equation}  
  v_i\ =\ 0 \quad \mbox{and} \quad  
  \lambda^e\ =\ \mbox{Diag} (m_e, m_\mu, m_\tau)\ ,  
\end{equation}  
and the bilinear \Rp\ parameters $\mu_i$, $B_i$ and ${\tilde m}^2_{di}$,  
as well as the trilinear \Rp\ couplings $\lambda_{ijk}$, $\lambda'_{ijk}$  
and their associated $A$-terms, are unambiguously defined.  
  
Let us now turn to the bounds that can be put on bilinear \Rp\ parameters,   
or equivalently on the induced mixings between leptons and neutralinos/charginos,   
and between sleptons and Higgs bosons.  
  
In the fermion sector, the neutralino-neutrino and chargino-charged lepton  
mixings lead to a variety of characteristic signatures (cf. subsection  
\ref{subsec:implications}) which in principle can be used to constrain the  
superpotential \Rp\ masses $\mu_i$.  
In practice however, the strongest bounds on these parameters come  
from the neutrino sector. Indeed, the neutralino-neutrino mixing induces  
a tree-level neutrino mass, given by Eqs. (\ref{eq:m_nu_tau}) and  
(\ref{eq:m_0}). In the absence of a fast decay mode of the corresponding  
neutrino, this mass is subject to the cosmological bound  
$m_\nu \lesssim 1 \eV$ \cite{WMAP}, which in turn requires a strong  
suppression of bilinear $R$-parity violation in the fermion sector:  
\begin{equation}  
  \sin \xi\ \lesssim\ 3 \times 10^{-6} \sqrt{1 + \tan^2 \beta}\ ,  
\label{eq:xi_cosmo}  
\end{equation}  
where $\xi$ is the misalignment angle introduced in subsection  
\ref{subsec:H_L_basis} to quantify in a basis-indepen\-dent way the size  
of the neutralino-neutrino and chargino-charged lepton mixings. Since we  
are working in a basis where $v_i = 0$, $\sin \xi$ is related to the \Rp\  
superpotential mass parameters $\mu_i$ by  
$\sin^2 \xi = \sum_i \mu^2_i / \mu^2$. The bound (\ref{eq:xi_cosmo}) is  
strong enough to suppress the experimental signatures of bilinear \Rp\  
violation in the fermion sector below observational 
limits~\footnote{This conclusion would have been different if the heaviest 
                 neutrino mass could have been as large as the $\nu_{\tau}$ 
		 LEP \index{LEP} limit of $18.2 \MeV$ \cite{pdg04}, as was often 
                 assumed in the early literature on bilinear \Rp.   
                 Indeed, in most models of spontaneous \Rp\, the tau 
		 neutrino was unstable enough to evade the cosmological 
		 energy density bound.   
                 Since then many of these models have been excluded by the 
		 invisible decay width of the $Z$ boson, and the scenario 
		 of a heavy decaying neutrino has become less attractive 
		 after the discovery of atmospheric and solar neutrino  
		 oscillations 
		 (see also subsection~\ref{subsec:spontaneous_bound}).}.  
   
In the scalar sector, the strongest contraints on the Higgs-slepton mixing  
\index{Mixing!higgs--slepton}
comes again from the neutrino sector. Indeed the \Rp\ soft masses contribute  
to the neutrino mass matrix at the one-loop level, through the diagrams  
discussed in subsection \ref{subsec:general}. According to  
Eq. (\ref{eq:m_nu_GH}), the cosmological bound on neutrino masses yields  
the following upper limit on bilinear \Rp\ in the scalar sector:  
\begin{equation}  
  \sin \zeta\ \lesssim\ 10^{-4}\ ,  
\label{eq:zeta_cosmo}  
\end{equation}  
where $\zeta$ is the misalignment angle in the scalar sector introduced in  
Eq. (\ref{eq:cosksi}) of subsection \ref{subsec:H_L_basis}.   
Since we are working in a basis where $v_i = 0$, $\sin \zeta$ is related  
to the \Rp\ $B$-terms $B_i$ by $\sin^2 \zeta = \sum_i B^2_i / B^2$.  
  
Finally, one should mention that some physical quantities receive  
contributions involving simultaneously bilinear and trilinear \Rp\  
parameters, especially in the neutrino sector (cf. subsections  
\ref{subsec:general} and \ref{subsec:L_number}). This results in upper  
bounds on products of \Rp\ parameters like $\l_{ijk}\, \mu_i$ or  
$\l'_{ijk}\, \mu_i$.  

\subsection{Models with Spontaneous {\boldmath{$R$}}-Parity Breaking}  
\label{subsec:spontaneous_bound}  
  
The main constraints on models of spontaneously broken $R$-parity are  
essentially due to the existence of a Goldstone  
boson (or, in the presence of interactions breaking explicitly lepton number,  
of a pseudo-Goldstone boson) associated with the breakdown of lepton number,  
the Majoron\index{Majoron} $J$. The first constraint comes from the invisible decay width  
of the $Z$ boson, which excludes a massless doublet or triplet 
Majoron\index{Majoron}.  
In viable models, the Majoron must be either mainly an electroweak  
singlet (i.e. it contains only a very small doublet or triplet component)  
like in the model of Ref. \cite{masiero90}, or a massive pseudo-Goldstone  
boson like in the model of Ref. \cite{comelli94}. In the second case, $m_J$  
should be large enough for the decay $Z \rightarrow J \rho$, where $\rho$  
is the scalar partner of the Majoron, to be kinematically 
forbidden; in practice $m_J \gtrsim M_Z$ is required. A third option,  
not discussed here, is to gauge the lepton number, in which case the  
Majoron\index{Majoron} disappears from the mass spectrum by virtue of the 
Higgs mechanism.  
  
Due to its electroweak non-singlet components, the Majoron possesses  
interactions with quarks and leptons of the form ${\cal L}_{ffJ}\, =\,  
g_{ffJ}\, \bar f \gamma_5 f J\, +\, \mbox{h.c.}$, where the coupling $g_{ffJ}$  
is model-dependent but related to the electroweak non-singlet \VEV\ $v_L$  
involved in the breaking of lepton number -- generally the \VEV\ of a  
sneutrino field. In the case of a light Majoron\index{Majoron}, these  
couplings induce  
physical processes that can be used to put upper bounds on $v_L$, such as  
exotic semileptonic decay modes of $K$ and $\pi$ mesons, like  
$K^+ \to l^+\bar \nu J$ \cite{barger82}; neutrino-hadron deep inelastic  
scattering with Majoron emission initiated by the subprocess  
$\nu_\mu u \to l^+ d J$ \cite{barger82}; or lepton flavour violating decays  
of charged leptons, like $e \to \mu J$.  
In practice however, the strongest constraints on $v_L$ come from  
astrophysical considerations. Indeed light Majorons\index{Majoron} can be 
produced inside  
the stars via processes such as the Compton scattering $e + \gamma \rightarrow  
e + J$ \cite{Majoron_cooling}. Being weakly coupled, Majorons, once produced,  
escape from the star, carrying some energy out. The requirement that the  
corresponding energy loss rate should not modify stellar dynamics beyond  
observational limits puts a severe bound on the couplings $g_{ffJ}$,  
therefore on $v_L$. The strongest bounds come from red giant stars:  
\begin{equation}  
  g_{eeJ}\ \lesssim\ 5 \times 10^{-13}\ ,  
\label{eq:g_eeJ}  
\end{equation}  
if the Majoron\index{Majoron} mass does not exceed a few times the characteristic  
temperature of the process, $m_J \lesssim 10 \keV$ \cite{raffelt95}.  
In the model of Ref. \cite{masiero90}, where $J$  
is mainly an electroweak singlet, this bound translates into  
\begin{equation}  
  \frac{v^2_L}{v_R\, M_{W}}\ \lesssim\ 10^{-7}\ ,  
\end{equation}  
where $<\widetilde \nu_{\tau}>\, \equiv v_L / \sqrt{2}$, and  
$<\widetilde \nu_{R\tau}>\, \equiv v_R / \sqrt{2}$ is the \VEV\  
of the right-handed sneutrino field  
involved in the spontaneous breaking of $R$-parity (cf. section  
\ref{sec:spontaneous} for details). For $v_R \sim 1 \TeV$,  
this is satisfied as soon as $v_L \lesssim 100 \MeV$.  
Models involving a doublet or triplet $\mbox{(pseudo-)}$Majoron are not  
subject to the constraint (\ref{eq:g_eeJ}), since such Majorons\index{Majoron} 
are too heavy to be produced in stars.  
  
  
Finally, since spontaneous $R$-parity breaking involves the \VEV\ of a  
left-handed sneutrino field and/or generates bilinear \Rp\ terms through  
the \VEV\ of a right-handed sneutrino field, the constraints on models  
with explicit bilinear $R$-parity breaking also apply here.  
In particular, a single neutrino becomes massive at tree level.  
This neutrino, if cosmologically stable, is subject to the bound  
$m_\nu \lesssim 1 \eV$ \cite{WMAP}, which in turn requires a strong  
suppression of the misalignment angle $\xi$ as expressed by   
Eq. (\ref{eq:xi_cosmo}).  
The misalignment angle $\xi$, defined by Eq. (\ref{eq:cos_xi}), can be  
expressed in terms of the parameters of the model. In the model of Ref.  
\cite{comelli94}, $\sin \xi = v_L / v_d$ and the constraint (\ref{eq:xi_cosmo})  
translates into  
\begin{equation}  
  < \widetilde \nu_\tau >\ \equiv\ \frac{v_L}{\sqrt{2}}\ \lesssim 500 \keV\ .  
\label{eq:v_L_cosmo}  
\end{equation}  
This bound is  
independent~\footnote{In principle models with a massless or light 
                      Majoron\index{Majoron} 
                      can evade such a cosmological bound, hence the  
                      constraint (\ref{eq:xi_cosmo}), since the heavy neutrino 
		      can decay into a lighter one plus a Majoron, as  
		      originally suggested in Ref.~\cite{masiero90}. 
		      However such a scenario, quite popular at a time where 
		      oscillations of atmospheric neutrinos were not 
		      established on a firm basis, no longer appears to be 
		      very appealing, since it cannot be reconciled with both 
		      solar and atmospheric neutrino oscillations.  
                      }  
\index{Bounds on \Rp\ interactions!Bilinear Terms|)}
of $\tan \beta$, due to the fact that $v_d = v \cos \beta$, which turns out to   
be a strong constraint on such models.

\section{Constraints on the Trilinear {\boldmath{\Rp}} Interactions}  
\label{sec:tricons}  
\index{Bounds on \Rp\ interactions!Trilinear Terms|(}  
In this section we discuss the subset of the charged and neutral  
electroweak current phenomena which forms the basis for the high  
precision measurements. We also consider applications at the interface  
of $C P$ violation and $R$-parity violation and review some miscellaneous  
topics associated with high precision observables (anomalous magnetic  
moments or electric dipole moments).  Unless otherwise stated, the  
various numerical results  quoted in this section employ Standard Model    
predictions which include either tree and/or one-loop level  
contributions. 
The limits on the \Rp\ coupling constants quoted in this section are  
$2 \s $ bounds unless otherwise stated.  
 
\subsection{Charged Current Interactions}  
\label{secxxx2a}  
  
Two important issues associated  with the Standard Model charged current   
interactions are: (1) the universality with respect to the $ W^\pm $ gauge boson  
couplings to quarks and leptons, and between the couplings of different   
lepton families; (2) the relations linking the independent renormalised physical   
parameters of the Standard Model at the quantum level.  

\subsubsection{Charged Current Universality in Lepton Decays}  
  
\begin{figure}[htb]  
  \begin{center}  
     \mbox{\epsfxsize=0.95\textwidth  
       \epsffile{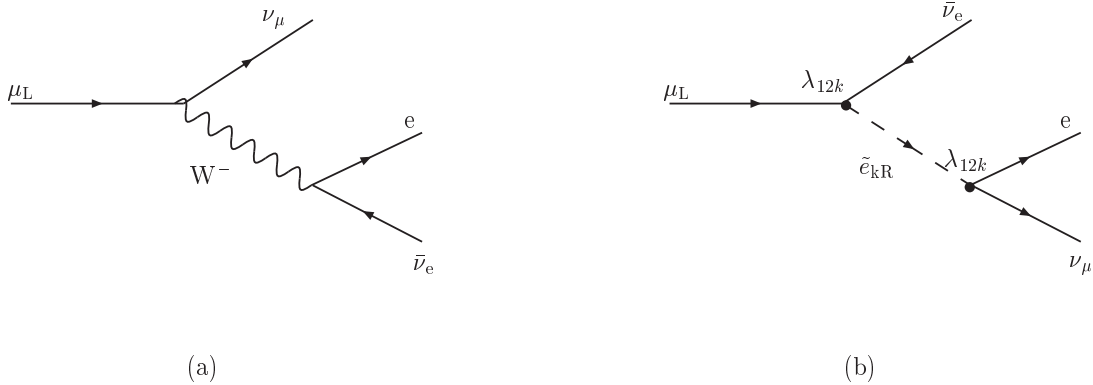}}  
  \end{center}  
 \vspace*{-0.5cm} 
  
 \caption[]{ \label{fig:muondecay_lle}  
 {\it{ Contributions to $G_{F}$ from (a) the standard model  
       and (b) an \Rp\ operator $ L_1 L_2 E^c_k $.  
 }}}  
\end{figure}

%
The presence of a   
$ L_1 L_2 E^c_k$  
operator leads to the additional contribution to the muon decay  
shown in Fig.~\ref{fig:muondecay_lle}b.   
%
The effective tree-level  
Fermi coupling $G_{F}$ which determines the $\mu$ lifetime   
becomes:  
\begin{eqnarray}  
 {G_{F} \over \sqrt 2} = {g^2\over 8M_W^2} (1+r_{12k}(\tilde e_{kR})) ,  
 \label{eqn:gmu_c6}  
\end{eqnarray}  
the auxiliary parameter $ r_{ijk}$ being defined in  
section~\ref{sec:specif}.  
The direct measurement of $G_{F}$ together with the tree-level  
relation~(\ref{eqn:gmu_c6}) cannot be used, however, to set conservative   
constraints on $\lambda_{12k}$, due to the large effects induced by   
radiative corrections. 
A study of the one-loop quantum relations linking the basic   
set of renormalized  input parameters $\a ,\ G_{F} ,\ M_Z $,   
with the weak angle  and/or the  $W$ boson mass parameter \MW    
~has to be performed. For an estimate see \cite{dimopoulos88}.   
We shall examine here two  different versions  
associated with the off-shell  ${\overline {MS} }$ and on-shell   
regularization schemes, respectively~\cite{langarad}: 
\begin{eqnarray}  && \text{off-shell} \ (\overline {MS}) :\   
M_W^2 =  { \pi \a (1+r_{12k} (\tilde e_{kR} ))\over \sqrt 2 G_F   
\sin^2\t_W (M_Z)\vert_ {\overline {MS} } (1-\Delta r(M_Z) \vert_{\overline  
{MS} } )} , \cr && \text{on-shell}:\ 
\sin ^2 \t _W \equiv 1-({ M_W \over M_Z} ) ^2=  { \pi \a (1+r_{12k}  
(\tilde e_{kR} ))\over \sqrt 2 G_F  M_W^2 (1-\Delta r ) } . 
\label{eqx5}  
\end{eqnarray}  
The quantities  labeled by $\overline{MS}$  refer to the   
modified  minimal subtraction scheme and those without a label refer to the  
on-shell renormalization scheme. 
The off-shell scheme relation can be interpreted as a prediction for the   
$W$ boson mass \MW\ depending on the weak interaction parameter   
$ \sin^2\t_W (M_Z)\vert_ {\overline {MS} }$,   
and the on-shell scheme relation as a prediction for this weak interaction   
parameter, linked to the $W$ mass to all orders of perturbation theory by   
$\sin ^2 \t _W = 1- M_W ^2/M_Z ^2$.   
The auxiliary parameters  in these two schemes,   
$\Delta  r, \ \Delta  r (M_Z) \vert _{\overline {MS} }  
$ are calculable renormalization scheme  
dependent  functions which depend on the basic input parameters  and  
the  Standard Model mass spectrum.   
  
We evaluate both relations by using the experimental values for the  
input parameters~\cite{erler98}. The parameters common to both relations are   
set as:  
\begin{eqnarray} &&  
\a = {1/137.035} ,\ G_F =1.16639 \times10^{-5} \ \text{GeV} ^{-2} ,\cr && M_Z =  
91.1867 \pm 0.0020 \ \ \text{GeV} , \ M_W= 80.405  
\pm 0.089 \ \  \text{GeV}  \end{eqnarray} . 
The weak angle in the off-shell $\overline {MS} $ relation is  
set as $\sin ^2 \t _W (M_Z) \vert _{\overline {MS} } = 0.23124 \pm  
0.00017 $, while in the on-shell it is in principle determined in  
terms of the W mass by $\sin ^2 \t _W = 1- M_W ^2/M_Z ^2$.  For  
the auxiliary parameters, we use the values: $\Delta r = 0.0349 \pm 0.0019  
\pm 0.0007$, and $\Delta r (M_Z) \vert _{\overline {MS} } = 0.0706  \pm   
0.0011$.  
  
Let us now quote the results of the calculations. We find that the  
off-shell scheme relation tends to rule out the existence of $\l _{12k}$. 
However, taking into account the uncertainties on the input parameters  
leaves still the possibility of inferring bounds on the \Rp\ coupling  
constants.  The uncertainty in $ M_W$ dominates by far all the other  
uncertainties. A calculation at the $1\s $ level leads to the  
coupling constant bound $\l _{12k} <0.038 \ \tilde e_{kR}$. 
 
For the on-shell scheme relation, we still find that this tends to rule   
out $\l_{12k}$, but yields the $1\s $ level bound   
$\l_{12k} <0.046 \ \tilde e_{kR}$.    
To illustrate the importance of the uncertainties in  
the $W$ boson mass in this context, we consider the alternative  
calculation in the on-shell scheme where we use the experimental value  
for the on-shell renormalized weak angle   
$ \sin ^2 \t _W =0.2260 \pm 0.0039 $ and evaluate the $W$ mass from the  
relation $ M_W ^2 = M_Z ^2 (1-\sin ^2 \t _W )$.  This prescription is  
now found to yield definite values for the \Rp\ coupling constants given  
by $ \l _{12k} = 0.081 \ \tilde e_{kR}$ .  
  
The main conclusion here is that the constraints for the coupling  
constants $\l _{12k}$ extracted from the \Rp\ correction to $G_F$  
depend sensitively on the input value of the $W$ boson mass.  The  
comparison of results obtained with the off-shell and on-shell  
regularization schemes serves, however, as a useful consistency check.  
  
New contributions to the $\mu$ decay can be probed by   
comparing the measurement of the ratio of rates:   
$$ R_{\tau \mu} =   
 \Gamma(\tau \rightarrow \mu \nu \bar{\nu}) /  
 \Gamma(\mu \rightarrow e \nu \bar{\nu}) $$  
to its SM expectation $R_{\tau \mu}^{SM}$.  
This was first considered in~\cite{barger89}  
where it was shown that the small experimental value  
for $R_{\tau \mu}$ reported in~\cite{rtaumu_Amaldi}  
could be accounted for by \Rp\ muon decays for  
coupling values $\lambda_{12k} \sim 0.15$.  
An updated analysis~\cite{ledroit} using more precise measurements of  
$R_{\tau \mu}$~\cite{pdg02} and $O(\alpha)$ values for  
$R_{\tau \mu}^{SM}$ now yields the bound  
\begin{eqnarray}  
 \l _{12k} < 0.07 \, \tilde e_{kR} .  
 \label{eqn:l12k_mudecay}  
\end{eqnarray}  
  
More generally, the ratio $R_{\tau \mu}$ can also be affected by   
$L_2 L_3 E^c_k$ operators modifying the $\tau$ leptonic decays via   
$\tilde{e}_{kR}$ exchanges, similarly to the process shown in   
Fig.~\ref{fig:muondecay_lle}b.  
The expression of $R_{\tau \mu}$ reads as:  
\begin{equation}  
R_{\tau \mu } \simeq   
  R_{\tau \mu } ^{SM} [1+2 (r_{23k}(\tilde e_{kR}) -r_{12k}(\tilde e_{kR})) ]  
  \ ,   
\end{equation}  
while the ratio of both leptonic $\tau$ decay widths is:  
\begin{equation}  
R_\tau ={\G (\tau \to e \bar \nu_e \nu_\tau )  
 \over  
 \G (\tau \to \mu \bar \nu_\mu \nu_\tau ) }   
  \simeq   
 R_\tau ^{SM} [1+2 (r_{13k}(\tilde e_{kR}) -r_{23k}(\tilde e_{kR}))]  
 \ .   
\end{equation}  
The comparison of the experimental measurements with the SM values  
yields the following bounds~\cite{ledroit} on the coupling constants  
$\lambda_{i3k}$:  
\begin{equation}  
 \l_{13k} < 0.07 \, \tilde e_{kR} \, [R_\tau ] ; \quad  
 \l_{23k} < 0.07 \, \tilde e_{kR} \, [R_\tau ] ; \quad  
 \l_{23k} < 0.07 \, \tilde e_{kR} \, [R_{\tau \mu }] .  
 \label{eqn:rtau_bounds}  
\end{equation}  
  
Aside from the  muon lifetime, the energy and angular distributions of the  
charged lepton emitted in muon decay offer useful observables  in  
order to test the Lorentz covariant structure of the charged current   
interactions, through the presence of either non $V-A$ couplings or   
tensorial couplings.  The information is encoded in terms of the Michel  
parameter $\rho $ and the analogous parameters, $ \eta , \ \xi , \  
\delta $, functions of the independent Fermi $S, V,T$ invariant   
couplings, which enter the differential (energy and angle) muon decay   
distributions~\cite{gerber}.   
The $ \tilde e_{kR}$ exchange depicted in Fig.~\ref{fig:muondecay_lle}b   
and induced by the $\l_{12k} $ coupling alone  
initiates \Rp\ contributions to the Lorentz vector and axial  
vector couplings. However the corresponding corrections in this case are   
masked by a  
predominant Standard Model contribution of the same structure, so that no 
useful constraints can be inferred.    
In contrast, the  
tree level exchange of a stau $\tilde \tau_{L}$ initiates  
corrections to the scalar Lorentz coupling, yielding the bound:  
$\vert \l^\star _{232} \l _{131} \vert < 0.022\ \tilde \tau_L ^2   
$~\cite{cheung98}.  
While the above quadratic bound actually  
turns out to be weaker than those deduced by combining tentatively   
the individual  
bounds on the coupling constants $ \l_{13k}$ and $\l_{23k}$  
given in equation~(\ref{eqn:rtau_bounds}), it  
has the advantage of providing a more robust bound,  
not exposed to invalidating cancellations.  
  
\subsubsection{Charged Current Universality in {\boldmath{$\pi$}} 
               and {\boldmath{$\tau$}}  Decays}  
  
\begin{figure}[tbh]  
  \begin{center}  
     \mbox{\epsfxsize=0.95\textwidth  
       \epsffile{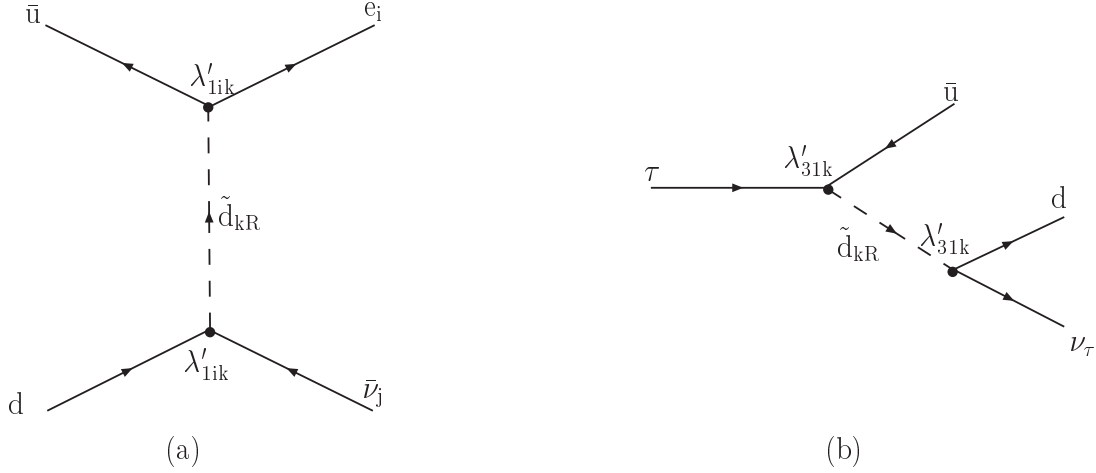}}  
  \end{center}  
 \vspace*{-0.5cm} 
  
 \caption[]{ \label{fig:pidecays}  
 {\it{ Contributions of \Rp\ interactions to (a) the leptonic  
       $\pi$ decays and (b) the $\tau \rightarrow \pi \nu_{\tau}$  
       decay.  
 }}}  
\end{figure}

Leptonic decays of the $\pi$ as well as   
$\tau^- \rightarrow \pi^- \nu_{\tau}$ can be mediated at the  
tree level by \Rp\ interactions, as shown by the  
diagrams of Fig.~\ref{fig:pidecays}.  
At low energies, these \Rp\  contributions can be represented  
by four fermion interactions  
between pairs of quarks and leptons,  
$(\bar l \G l )(\bar q \G q )$.  
Writing the effective interactions in this form allows a systematic  
calculation of the \Rp\ induced meson leptonic decays and $\tau$  
semi-hadronic decays.  
For a general and complete study of bounds  
from meson decays we refer to~\cite{phdherz}. 
In comparing with the experimental data for the $\pi  
$-meson  decay width $ \G (\pi^-  
\to \mu^- \bar \nu_\mu ) $, it is advantageous to eliminate the  
dependence on the pion decay coupling constant, $ F_\pi $, by  
considering the ratio~\cite{barger89}:  
\begin{eqnarray}  
 R_\pi & =&{\G (\pi^- \to e^- \bar \nu_e ) \over \G (\pi^- \to \mu^-  
 \bar \nu_\mu ) }= R_\pi ^{SM} [1+{2\over V_{ud}} (r'_{11k}(\tilde  
 d_{kR}) -r'_{21k}(\tilde d_{kR}))]\ ,  
\end{eqnarray}  
where the auxiliary parameters $r'_{ijk}$ are defined in  
section~\ref{sec:specif}. The inferred  coupling constant bounds are  
\cite{phdherz}:   
\begin{equation}  
 \l '_{21k} < 0.059\  \tilde d_{kR}, \  
 \l'_{11k} < 0.026\  \tilde d_{kR} \ .  
 \label{eqn:rpi_bounds}  
\end{equation}  
The closely related   
two-body decay $ \tau^- \to \pi^- \nu_\tau $ also offers an  
additional useful test of the lepton universality~\cite{bhattachoud}. 
A model-independent analysis based on a comparison  
with the experimental results for the ratio of $\tau $ lepton  
and $\pi $  meson  decay widths,  
\begin{eqnarray}  
 R_{\tau \pi } &=& { \G (\tau^- \to \pi^-  \nu _\tau )   
 \over \G (\pi^- \to \mu^- \nu_\mu ) } =  
 R_{\tau \pi } ^{SM} { \vert V_{ud} + r'_{31k} (\tilde d_{kR}) \vert ^2   
 \over  
 \vert V_{ud} + r'_{21k} (\tilde d_{kR}) \vert ^2 }  
\end{eqnarray}  
yields the coupling constant bound:  
\begin{eqnarray}  
 \l '_{31k} < 0.12 \ \tilde d_{kR}  & , &  
 \l'_{21k} < 0.08\  \tilde d_{kR} \quad .  
 \label{eqn:rtaupi_bounds}  
\end{eqnarray}

\subsubsection{Charged Current Universality in the Quark Sector}  
  
In the quark sector, the presence of a \LQD\ operator leads to  
additional contributions to quark semileptonic decays,  
via processes similar to that shown in Fig.~\ref{fig:pidecays}   where  
the incoming antiquark line is reversed.  
The experimental value of the CKM matrix element $V_{ud}$,  
determined by comparing the nuclear $\beta$ decay to the muon decay,  
is then modified according to:  
\begin{equation}  
 \vert V_{ud} \vert^2 =  
 {\vert V^{SM} _{ud} +r'_{11k}(\tilde d_{kR}) \vert ^2 \over  
 \vert 1+r_{12k}(\tilde e_{kR}) \vert ^2 } \qquad.  
\end{equation}  
Similarly, the rates for $s \rightarrow u  l  \bar{\nu}_l$ and  
$b \rightarrow  u  l \bar{\nu}_l$, measured in $K$ and charmless   
$B$ decays respectively, are affected by \Rp\ corrections induced by  
$\lambda'$ couplings.   
The values of $V_{us}$ and $V_{ub}$ extracted from these  
rates depend again on $r_{12k}$ via the effect of  
$\l_{12k}$ couplings on $G_{F}$.  
Summing over the down quark generations yields~\cite{barger89}:  
\begin{eqnarray}  
\sum_{j=1} ^ 3\vert V_{ud_j}\vert^2   
& = &  
  { 1 \over \vert 1+r_{12k}(\tilde e_R) \vert ^2 } \   
  \ [\ \ \vert V^{SM} _{ud} +r'_{11k}(\tilde d_{kR}) \vert ^2 \    
\cr & +&   
  \vert V^{SM} _{us} +[r'_{11k}(\tilde  
  d_{kR}) r'_{12k}(\tilde d_{kR}) ]^ {1\over 2}  \vert ^2  
\cr & +&  
  \vert V^{SM} _{ub} +[r'_{11k}(\tilde  
  d_{kR}) r'_{13k}(\tilde d_{kR}) ]^ {1\over 2}\vert ^2\ ] \\  
&= &  
 1 - 2 r_{12k}(\tilde e_{kR}) V^{SM} _{ud}  
   + 2 r'_{11k}(\tilde d_{kR}) V^{SM} _{ud}   
\cr & +&  
   2 [r'_{11k}(\tilde d_{kR}) r'_{12k}(\tilde d_{kR}) ]^ {1\over 2} V^{SM} _{us}  
 + 2 [r'_{11k}(\tilde d_{kR}) r'_{13k}(\tilde d_{kR}) ]^ {1\over 2} V^{SM} _{ub}  
   \, ,  
\label{eqx3}  
\end{eqnarray}  
the last equality resulting from the unitarity of the CKM   
matrix~\footnote{We have corrected the formula for the unitarity   
constraint used in the work by Ledroit and Sajot~\cite{ledroit} by   
noting that the \Rp\ corrections  to the flavour mixing matrix   
elements $V_{us}$ and $V_{ub}$  are given by quadratic products   
of the coupling  constants.}.  
At the lowest order in the \Rp\ corrections into which we specialize,   
it is consistent to identify the flavour mixing matrix elements   
appearing in the right-hand side with the measured CKM matrix  
elements, $V_{ud_j}^{SM} \simeq  V_{u d_j} $.  
Setting the  various CKM matrix elements $V_{ud_j} $ and in the sum   
$\sum_j \vert V_{u d_j} \vert^2$ at their measured values~\cite{pdg02},  
the following bounds are obtained in the single   
and quadratic coupling constant dominance hypothesis, respectively:  
\begin{eqnarray}  
\begin{array}{lll}  
\l_{12k} < 0.05 \ \tilde e_{kR}, & \l '_{11k} < 0.02\  \tilde d_{kR}, &  \\   
& (\l ^{'\star } _{11k} \l'_{12k})^{1/2}  < 0.04\ \tilde d_{kR},    
& (\l ^{'\star } _{11k}  \l '_{13k})^{1/2}  < 0.37 \ \tilde d_{kR} .  
\end{array}  
 \label{eqn:vud_bounds}  
\end{eqnarray}  
For more bounds on leptonic meson decays see \cite{dreiner02_bis}. 
A consideration of the unitarity constraints on the other sums of CKM  
matrix elements, \\   
$\sum _{j=1,2,3} \vert V _{cd_j} \vert ^2$  and   
$\sum _{j=1,2,3} \vert V _{cd_j} \vert ^2$,   could  also  be used  
with the same prescriptions as in the above comparison,   
to derive bounds on the single  
coupling constants $ \l _{i2k} $ and $ \l _{i3k} $, respectively.  
  
\subsubsection{Semileptonic and Leptonic Decays of Heavy Quark Hadrons}  
  
\begin{figure}[htb]  
  \begin{center}  
     \mbox{\epsfxsize=0.7\textwidth  
       \epsffile{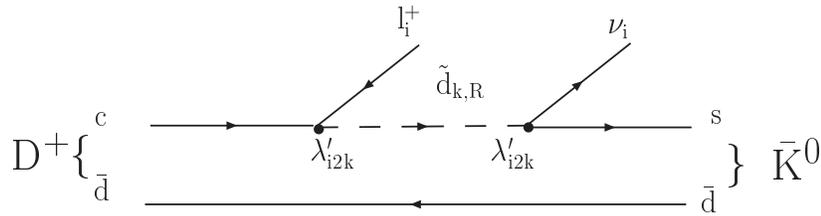}}  
  \end{center}  
 \caption[]{ \label{fig:ddecay}  
 {\it{ \Rp\ contributions to the semileptonic decay of a charmed meson.  
 }}}  
\end{figure}

An experimental information on the three-body decay channels of  
charmed mesons is available for the following three classes of  
semileptonic processes, differing in the  final state   
by the lepton generation or the  strange meson type:  
 $D^+\to \bar K^0 l_i^+ \nu_i , \ D^+ \to \bar K^  
{0\star } l_i^+ \nu_i , \ D^0 \to K^- l_i^+ \nu_i ,\ [l_i= e, \mu;\  
\nu_i = \nu_e, \nu_\mu ] $.    
These decays could be enhanced by \Rp\ contributions involving  
a $\lambda'_{i2k}$ coupling as shown in Fig.~\ref{fig:ddecay}.  
%
%
Denoting the branching fraction ratios   
$B( D \rightarrow \mu \nu_{\mu} K^{(*)}) \ / \   
 B (D \rightarrow e \nu_e K^{(*)})$  
by $R^{(*)}_{D^+},   
 \ R_{D^0}$ respectively, one can write the \Rp\  
corrections as~\cite{altarelli}:  
\begin{eqnarray}  
{ R_{D^+} \over (R_{D^+})^{SM} }& = & { R_{D^+}^\star \over (R^\star  
_{D^+})^{SM} } = { R_{D^0} \over (R_{D^0})^{SM} } = {\vert  
1+r'_{22k}(\tilde d_{k R}) \vert ^2 \over \vert 1+r ' _{12k}(\tilde  
d_{kR} ) \vert ^2 }.  
\label{eqx4}  
\end{eqnarray}  
Following~\cite{altarelli}, we use  
$(R^{(*)}_D)^{SM} = 1/1.03$ to account for the phase space suppression  
in the muon channel.   
>From the experimental values of  
$R^{(*)}_D$ given in~\cite{pdg02}, one deduces the  following    
$2 \s $  coupling constant bounds:  
\begin{eqnarray}  
\begin{array}{lll}  
| \l '_{12k}| < 0.44 \  \tilde d_{kR} ,    
&  
| \l '_{22k}| < 0.61 \  \tilde d_{kR}  
&  
 [R_{D^+} ] ;  \cr  
| \l '_{12k}|<  0.23 \  \tilde d_{kR} ,  
&  
| \l '_{22k}| < 0.38\  \tilde d_{kR}  
&  
 [R^\star _{D^+} ]; \cr  
| \l '_{12k}|<  0.27 \  \tilde d_{kR} ,  
&  
| \l '_{22k}| <  0.21 \ \tilde d_{kR}   
&  
 [R_{D^0}].  
\end{array}  
 \label{eqn:ddecays_bounds}  
\end{eqnarray}  

By invoking the existence of a flavour mixing in the up-quark sector,  
within the current basis single coupling constant dominance  
hypothesis, the \Rp\ contribution to the inclusive semileptonic $B$ meson  
inclusive decay process, $ B^- \to X_q\tau^- \bar \nu ,$ may be  
expressed solely in terms of the single coupling constant   
$ \l ' _{333} .$ The comparison with  
experiment yields the following estimate for the  
bound~\cite{grossman}:  
\begin{eqnarray}  
| \l '_{333}| & < & 0.12 \ \tilde b_{R}.  
 \label{eqn:grossman}  
\end{eqnarray}  
The same process as considered in~\cite{erler}   
leading to $\l '_{333} < 0.32 \ \tilde b_{R}$.  
The different predictions furnish an indication of the dependence on  
the input hadronic parameters.  
 
The two-body leptonic decay channels of the charmed  
quark mesons, $ D^-_s \to l^- \nu $, also serve a good use in testing  
the lepton universality.  A comparison with the experimental results  
for the ratios of $\tau  $ to $\mu $ emission,  
\begin{equation}   
 R_ {D_s} (\tau \mu ) = {B(D_s \to \tau \nu_\tau )  
 \over B(D_s \to \mu \nu_\mu ) } = { \vert V_{cs} + r'_{32k} (\tilde  
 d_{kR}) \vert ^2 \over \vert V_{cs} + r'_{22k}(\tilde d_{kR}) \vert ^2 }  ,   
\end{equation}  
yields the  coupling constant bounds~\cite{ledroit}:   
\begin{eqnarray}  
| \l '_{22k}| < 0.65 \ \tilde d_{kR},\  
| \l '_{32k}| < 0.52 \ \tilde d_{kR} \quad .  
 \label{eqn:rdstaumu_bounds}  
\end{eqnarray}  
  
The $\Delta S=1$ decays of strange baryons, e.g.   
$ \L \to p  l^-  \bar \nu_e , \cdots \ [l=e, \mu ] $ , provide  
bounds on quadratic products of the $\l ' $ interactions.  
We quote the $2 \sigma$ bounds obtained by Tahir et al.~\cite{tahir}:  
\begin{eqnarray}  
\begin{array}{lllll}  
\vert \l ^{'\star }  _{11k} \l '_{12k} \vert & <&  1.3 \ \times \ 10^{-1} \   
(5.3 \ \times \ 10^{-3})  
\ \tilde d^2_{kR} & [ \L \to  p  l^-  \bar{\nu}_l ] &;  \\  
\vert \l^{'\star }  _{11k} \l '_{12k}\vert  & < & 8.5 \ \times \ 10^{-2} \   
(1.6 \ \times \ 10^{-2})  
 \ \tilde d^2_{kR}  
 & [\S^- \to  n  l^-  \bar{\nu}_l ] &; \\  
\vert \l^{'\star }  _{11k} \l '_{12k} \vert & < & 1.2 \ \times \ 10^{-1}  \   
(5.0 \ \times \ 10^{-2})  
 \ \tilde d^2_{kR}  
 & [\Xi ^- \to  \L  l^-  \bar{\nu}_l ] &,  
\end{array}  
\label{eqn:tahir_bounds}  
\end{eqnarray}  
from the upper limits on the branching ratios   
of the indicated decays with $l=e$ ($l=\mu$).  
 
\subsection{Neutral Current Interactions}   
\label{secxxx2b}  
  
\subsubsection{Neutrino-Lepton Elastic Scattering and Neutrino-Nucleon   
                   Deep Inelastic Scattering}  
    
Most of the experimental information on neutrino interactions with  
hadron  targets or with leptons is accessed via experiments using  
$\nu_\mu $ and $ \bar \nu_\mu $ beams.  One may consult  
\cite{panman,perrierlanga} for reviews.      
The elastic scattering $\nu_\mu e \to \nu_\mu e , \ \bar  
\nu_\mu e \to \bar \nu_\mu e $ has been studied by the  
CHARM II experiment, which provides measurements  
for the ratio  
$ R = \s (\nu_\mu)  / \s (\bar \nu_\mu)$,  
where $\sigma ( \nu_\mu (\bar{\nu}_\mu) )$ denotes the  
cross-section  
$\sigma ( \nu_\mu (\bar{\nu}_\mu) e \rightarrow  
          \nu_\mu (\bar{\nu}_\mu) e) $.  
Neutral current (NC) and charged current (CC) deep  
inelastic scattering on nucleons or nuclei,  
$\nu _\mu  N (A) \to \nu_\mu  X $ and  
$\nu _\mu  N (A) \to \mu  X $,  
has been studied by the CDHS and CHARM experiments at CERN, and   
by the CCFR Collaboration  
at Fermilab.   
These experiments measure the following rates:  
\begin{equation}  R_\nu ={ <\s ( \nu N) ^{NC} >  
\over <\s ( \nu N) ^{CC} >}, \quad R_{\bar \nu } ={ <\s (\bar \nu N)    
^{NC} > \over <\s (\bar \nu N) ^{CC} > } ,   
\end{equation}    
where the brackets stand  
for an average over the neutrino beam energy flux distribution. Useful  
information is also collected through the  elastic scattering of neutrinos on  
a proton target~\cite{mannlanga}.  Each of these observables presents  
its own specific advantages by providing highly sensitive measurements  
of the Standard Model   parameters.  
    
\begin{figure}[htb]  
  \begin{center}  
     \mbox{\epsfxsize=0.95\textwidth  
       \epsffile{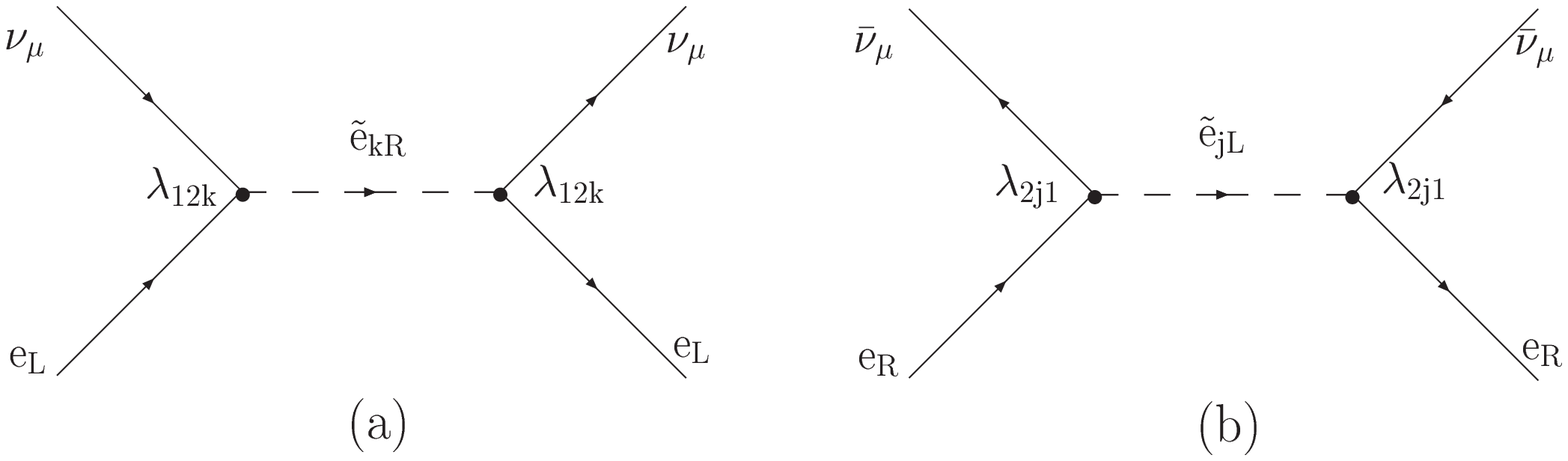}}  
  \end{center}  
 \vspace*{-0.3cm} 
  
 \caption[]{ \label{fig:nuscat}  
 {\it{ Examples of \Rp\ contributions to (a) $\nu - e$ and (b)  
       $\bar{\nu}_\mu -e$ scattering.  
       Other contributions, as coming from the $t$-channel   
       exchange of a selectron, are not represented.  
 }}}  
\end{figure}

The neutrino $\nu_\mu $ scattering on quarks and charged leptons is  
described by the $ s$-channel $Z$ boson exchange diagram.  At energies  
well below $M_Z$, the relevant neutral current couplings are encoded  
in the parameters $ g_{L,R} ^{\nu f} $  for charged leptons, and $  
\e_{L,R} (f) $ for quarks, as defined in terms of the effective  
Lagrangian density,  
\begin{eqnarray}    
{\cal{L}}&=&- {4G_{F} \over \sqrt 2} (\bar \nu_L\g_\mu \nu_L) [\sum_{l=e, \mu  
} g_L^{\nu f } ( \bar l_L\g^\mu l_L)+ g_R^{\nu l } ( \bar l_R\g^\mu  
l_R) \cr &+ & \sum_{q=u,d} \e _L( f ) ( \bar q_L\g^\mu q_L) + \e _R  
(f)  (\bar q_R\g^\mu q_R) ].    
\end{eqnarray}    
The \Rp\ contributions to neutrino elastic scattering arise at the tree  
level order.    
Examples are shown in Fig.~\ref{fig:nuscat} in the case of  
$\nu_\mu e$ and $\bar{\nu}_\mu e$ scattering. Similar contributions   
induced by a $\lambda'_{21k}$ ($\lambda'_{2j1}$) coupling  
affect the $\nu_\mu  d$ ($\bar{\nu}_\mu)  d$ scattering  
via the exchange of a $\tilde{d}_{kR}$ ($\tilde{d}_{jL}$) squark.  
%
%
%
The  results for the combined Standard Model   and \Rp\ contributions  
read~\cite{barger89}:  
\begin{eqnarray}    
g^{\nu e} _L&=&(-\ud +x_W)(1-r_{12k}(\tilde e_{kR}) )-r_{12k}(\tilde  
e_{kR}), \cr g^{\nu e} _R&=& x_W(1-r_{12k}(\tilde e_{kR}) )+r_{211}(\tilde  
e_{L}) +r_{231}(\tilde \tau_{L}), \cr \e _L(d)&=&(-\ud +{1\over 3}  
x_W)(1-r_{12k}(\tilde e_{kR}) )-r'_{21k}(\tilde d_{kR}),  
\cr  \e _R (d)&=& {x_W\over 3}(1-r_{12k}(\tilde e_{kR})    
)+r'_{2j1}(\tilde d_{jL}),  
\end{eqnarray}    
with $x_W = \sin^2 \theta_W$.  
Note that although a $\lambda_{12k}$ coupling does not lead  
to sfermion exchange contributing to $\nu q \rightarrow \nu q$   
scattering, it affects $\e_R$ and $\e_L$ via its effects  
on $G_{F}$ (see equation~(\ref{eqn:gmu_c6})).  
>From these relations and using the experimental values for $g^{\nu e}_L$ and  
$g^{\nu e}_R$ given in~\cite{pdg02},   
which rely on the $\sigma ( \nu_\mu (\bar{\nu}_\mu) )$ measurements   
from the CHARM II experiment~\cite{charm}, one obtains  
the bounds: 
\begin{eqnarray}  
 \l_{12k} < 0.14 \ \tilde e_{kR},  
 & \l_{231} < 0.11 \ \tilde \tau_{L},  
 & \l_{121} < 0.13 \ \tilde e_{L} \quad .  
 \label{eqn:nue_lambda_bounds}  
\end{eqnarray}  
The fits to the  
current data from the CDHS and CCFR Collaborations~\cite{ccfr}  
determine the numerical values for the parameters $ \e _{L,R}$~\cite{pdg02}.     
Comparing with the Standard Model values, suitably including the radiative   
corrections, yields the following limits on the \Rp\ coupling   
constants~\cite{ledroit}:  
\begin{eqnarray}  
 \l_{12k} < 0.13 \ \tilde e_{kR},  
& \l'_{21k} < 0.15 \ \tilde d_{kR},   
& \l'_{2j1} < 0.18 \ \tilde d_{jL} \quad .  
 \label{eqn:nue_lambdap_bounds}  
\end{eqnarray}  

The elastic scattering of $\nu_\mu $ and $\bar \nu_\mu $ on a proton 
target is known to furnish an independent sensitive means to measure the 
weak angle, $ \sin ^2\t_W$~\cite{mannlanga}. Although this case has been 
included in the global studies of the electron quark four fermion 
contact interactions~\cite{bargerz}, the present experimental accuracy 
and the  theoretical uncertainties on the nucleon weak form factors do 
not warrant a detailed study of the constraints on the \Rp\ interactions. 
  
\subsubsection{Fermion-Antifermion Pair Production   
                   and {\boldmath{$Z$}}-Boson Pole Observables}  
\label{sec:fcnczpole}  
    
The fermion pair production reactions, $e^+ e^-\to f \bar f, \  
[f=l,q] $ have been studied over a wide range of incident  energies at  
the existing leptonic colliders, PEP, PETRA, TRISTAN, LEP \index{LEP} and SLC.  
These reactions provide sensitive tests of the Standard Model ~\cite{panman}. 
For the high energy regime, the data for the $Z$ boson pole resonant 
production, $e^+ e^- \to Z^0 \to f \bar f$, as collected at the CERN LEP 
and the SLC colliders, have provided a wealth of experimental measurements   
of the Standard Model parameters.  
    
At low energies, the basic parameters are conventionally defined in  
terms of the following parametrisation for the effective Lagrangian  
density,  
\begin{eqnarray}    
{\cal{L}}&=&- {4G_{F} \over \sqrt 2}\sum_{f=l, q} \bar e \g ^\mu (g_L^e P_L +  
g_R^e P_R )e \ \bar f \g^\mu (g_L^f P_L + g_R^f P_R )f.  
\label{eq7p}    
\end{eqnarray}    
The high energy scattering regime, $ \sqrt s \ge M_Z $, is described  
by analogous transition amplitudes, differing in form only by the insertion  
of an energy dependent $Z$ boson resonance propagator factor. The $Z$-pole   
measurements provide information on a large   
set of observables.    
Of particular interest here are the forward-backward asymmetries  
$$ A^f _{FB}= \left. { (\s )_> -(\s )_< \over   
            (\s )_> +(\s )_< }  
             \right \vert_{e\bar e\to f\bar f} $$  
which, in the Standard Model, are related to the vector and axial  
couplings of fermions to the $Z$ boson via~:  
$$ A^{f} _{FB} =  {3\over 4} \cala ^e \cala ^{f}  
   \qquad {\mbox{ where }} \qquad  
  \cala ^{f} = 2g_V^fg_A^f/(g_V^{f2}+ f_A^{f2})  
   \stackrel{\mbox{ tree level}}{=}  
  -T_{3L}^f  \qquad . $$  
 
%
\begin{figure}[htb]  
  \begin{center}  
   \begin{tabular}{ccc}  
     \mbox{\epsfxsize=0.30\textwidth  
       \epsffile{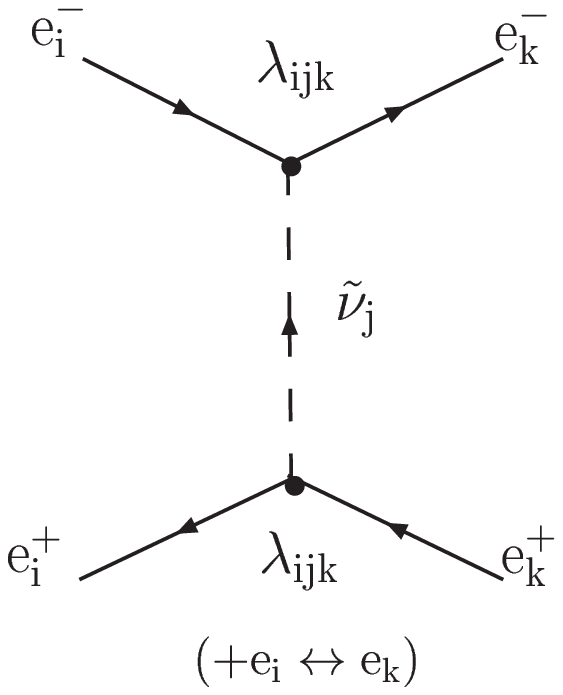}}  
    & \mbox{\epsfxsize=0.30\textwidth  
       \epsffile{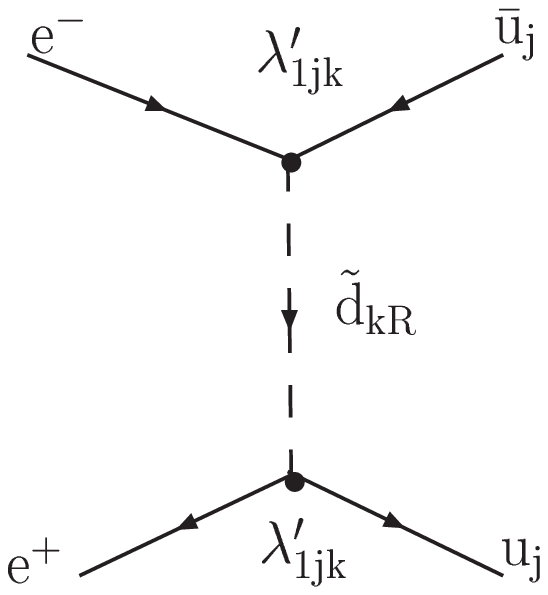}}  
    & \mbox{\epsfxsize=0.30\textwidth  
       \epsffile{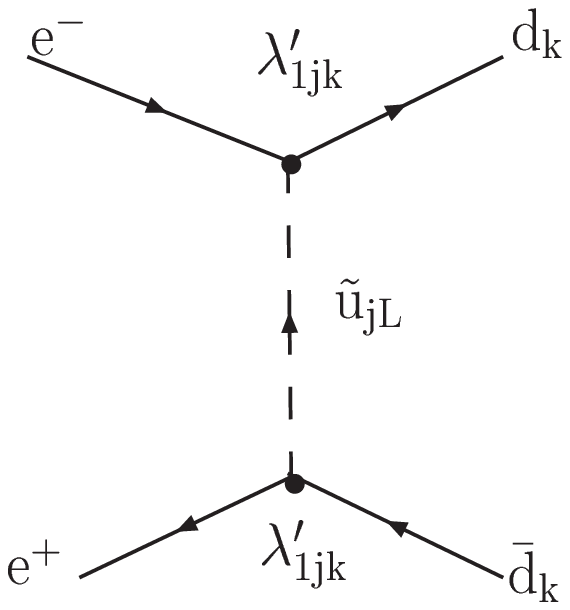}}  
    \end{tabular}  
  \end{center}  
 \caption[]{ \label{fig:asym}  
 {\it{ \Rp\ contributions to the forward-backward asymmetries.  
 }}}  
\end{figure}  
%
%
The tree level $t$-channel exchange of a sneutrino (squark)   
induced by a $\lambda_{ijk}$ ($\lambda'_{ijk}$) coupling affects  
the forward-backward asymmetries $A^l_{FB}$  
($A^q_{FB}$) as shown in Fig.~\ref{fig:asym}.  
Note that the $s$-channel exchange of a sneutrino also affects  
the Bhabha cross-section, but leaves $A^e_{FB}$ unchanged since  
the sneutrino decays isotropically in its rest frame.  
The SM values of $A^f_{FB}$ are modified according to:  
\begin{eqnarray}  
 \begin{array}{llll}  
  (A^{l_k}_{FB})_{SM} / A^{l_k}_{FB} & = &  
    \vert 1 + r_{1jk}(\tilde{\nu}_j) \vert^2  
   &  
    l_k = e \ : \  ijk=121,131 \\  
 & & & l_k = \mu \ : \  ijk=121, 122, 132, 231 \\  
 & & & l_k = \tau \ : \  ijk=123,133,131,231 \\  
  (A^{u_j}_{FB})_{SM} / A^{u_j}_{FB} & = &  
    \vert 1 + r'_{1jk}(\tilde{d}_{kR}) \vert^2   
  \\  
  (A^{d_k}_{FB})_{SM} / A^{d_k}_{FB} & = &  
    \vert 1 + r'_{1jk}(\tilde{u}_{jL}) \vert^2   
   \quad .  
 \end{array}  
\end{eqnarray}  
Taking from~\cite{pdg02} the experimental values for  
$A^f_{FB}$, as well as the SM predictions which include  
radiative corrections, the following $2\sigma$ bounds are obtained:  
\begin{eqnarray}  
 \begin{array}{cccl}  
 \l_{ijk} < 0.37 \ \tilde \nu _{jL} &;& \ (ijk)= (121), (131)   
  & [ A^e_{FB} ] \\  
\l_{ijk} < 0.25 \ \tilde \nu _{jL} &;& \ (ijk)= (122), (132),  
   (211), (231)   
  & [ A^{\mu}_{FB} ] \\  
\l_{ijk} < 0.11 \ \tilde \nu_{jL} &;& \ (ijk)= (123), (133), (311), (321)    
  & [ A^{\tau}_{FB} ] \\  
\l '_{12k} < 0.21 \ \tilde d_{kR}  & &   
  & [ A^c_{FB} ] \\  
\l ' _{1j2} < 0.28 \ \tilde u_{jL} & &   
  & [ A^s_{FB} ] \\  
\l ' _{1j3} < 0.18 \ \tilde u_{jL} & &   
  & [ A^b_{FB} ]   
 \end{array}  
 \label{eqn:afb_bounds}  
\end{eqnarray}

\begin{figure}[bth]  
  \begin{center}  
     \mbox{\epsfxsize=0.9\textwidth  
       \epsffile{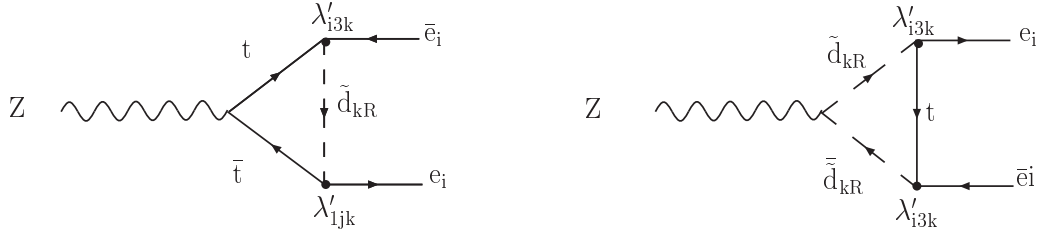}}  
  \end{center}  
 \caption[]{ \label{fig:rl}  
 {\it{ \Rp\ contributions to the leptonic $Z$ decay width.  
       The (subleading) self-energy diagrams are not  
       represented.  
 }}}  
\end{figure}

At the one-loop order, \Rp\ interactions lead to  
$Z \bar{f} f  \ [f=q,l]$ vertex corrections.   
The diagrams of the dominant \Rp\ processes contributing  
at the one-loop order to the leptonic $Z$ decay  
width $\Gamma_l$ are shown in Fig.~\ref{fig:rl}.  
The $\lambda'$ couplings which lead to these contributions  
also affect the $Z \rightarrow b \bar{b}$ decay width  
$\Gamma_b$,  
via loop processes propagating internal top and slepton lines.  
The subsequent change of the hadronic decay width of the $Z$  
has thus to be taken into account when calculating  
the \Rp\ induced corrections to  
$R_l = \Gamma_h \ / \Gamma_l$.   
%
Those corrections read~\cite{ellishar}:  
$$  
\delta R_l \equiv  {R_l \over R_l^{SM} } -1= -R_l^{SM} \Delta_l  
+R_l^{SM}  
R_b^{SM} \Delta_b $$  
where  
$$ R_b \equiv \Gamma_b \ / \Gamma_h,  
\qquad \Delta_f \equiv { \G (Z \to f \bar f) \over \G_{SM} (Z \to f \bar  
f) } -1,  
$$  
%
%
The comparison with the CERN LEP-I  \index{LEP}  
measurements~\cite{pdg02} of $R_l^Z \ [l=e, \ \mu , \ \tau ]$    
leads to the   
following $ 2\s $  
bounds~\cite{ledroit}, valid for $m(\tilde{d}_{kR}) = 100 \GeV$:    
\begin{eqnarray}  
 \l '_{13k} < 0.47, &  
 \l '_{23k}< 0.45, &  
 \l'_{33k} < 0.58 .  
 \label{eqn:rl_bounds}  
\end{eqnarray}  
The dependence of these bounds on  the $\tilde{d}_{kR}$ mass  
is not trivial and can be found in~\cite{ellishar}.  
 
The \Rp\ $\l ''$ interactions affect $\Gamma_h$ via diagrams similar 
to those shown in Fig.~\ref{fig:rl}, and thus affect $R_l$ and $R_b$ 
as well~\cite{sridhar}. 
From the measurement of the average leptonic decay width 
$R_l = 20.795 \pm 0.040$, the following bounds are 
obtained assuming squark masses of 100 GeV~\cite{sridhar}: 
\begin{eqnarray}  
 [ \l ''_{313}, \ \l ''_{323}] < 1.46 (0.97) & , &  
 \l''_{312} < 1.45 (0.96),  
 \label{eqn:rl_bounds_lpp}  
\end{eqnarray}  
the first (second) number corresponding to the $2 \sigma$ ($1 \sigma$)  
upper bound.  
Less stringent upper limits on the same $\lambda''$ couplings were also 
obtained in~\cite{sridhar} from an old measurement of $R_b$.  
These are out-of-date since the measured $R_b$ is now in  
good agreement with its SM value. 
  
Lebedev et al.~\cite{lebedev99} have performed a global statistical 
fit of the one-loop level $R$-parity violating contributions to 
the experimental data for the asymmetry parameters 
$ A^b , \ A_{FB} ^b$ gathered at the CERN LEP \index{LEP} and Stanford SLC 
colliders.   The pattern of signs from the \Rp\ corrections is such that certain 
  contributions enter with a sign opposite to the one that would be required  
by observations.   
This circumstance lead these authors to conclude that out the whole set of   
coupling constants   
$ \l '_{i31}, \ \l '_{i32}, \  \l ''_{321}  $     
and  
$\l '_{i33}, \  \l ''_{331} $   
was ruled out at the $1 \s $ and $2 \s $ level respectively.  
  
The pair-production of leptons with different flavours  
at the $Z$ pole provides quadratic bounds on a  
large subset of the \Rp\ coupling constants, 
$ \l _{ijk} , \ \l '_{ijk}$. 
Lepton-flavour violating (LFV) decays of the $Z$, with branching ratios  
defined by:  
\begin{eqnarray}    
B_{ii'} & \equiv & {\G (Z \to e_i \bar e_{i'} )+ \G (Z \to e_{i'} \bar  
e_i ) \over \G (Z \to all) } , \ [i \ne i']   
\label{eqx8}  
\end{eqnarray} 
may occur through \Rp\ induced one-loop processes similar to those 
shown in Fig.~\ref{fig:rl}, whose amplitude is proportional to 
quadratic products of the relevant \Rp\ coupling constants. 
The contribution of such processes induced by \LQD\ operators to the 
LFV decays of the $Z$ was studied by Anwar Mughal et al. 
in~\cite{anwar}, focusing again on the dominant diagrams involving a 
top quark in the internal loop.  
A comparison with the current experimental limits 
on $B_{e \mu}$, $B_{e \tau}$ and $B_{\mu \tau}$~\cite{pdg02} 
yields the bounds: 
\begin{eqnarray}   \sum_k \vert \l^{'\star } _{13k} \l '_{23k} \vert  &<&  
6.2 \ \times 10^{-2} \quad [Z\to e\mu ]; \quad    
\sum_k \vert \l^{'\star } _{33k} \l '_{13k} \vert < 1.5 \  
\times 10^{-1} \quad  [Z\to e\tau ]; \cr   
\sum_k \vert \l^{'\star } _{23k} \l '_{33k}\vert  &<&  1. 7\  
\times 10^{-1} \quad  [Z\to \mu \tau ]  
\end{eqnarray}     
established assuming $m( \tilde{d}_{kR}) = 200 \GeV$.  
The non trivial dependence of these bounds on $m( \tilde{d}_{kR})$   
can be found in~\cite{anwar}.  
  
A more general study of the LFV $Z$ decays induced by  
\Rp\ process can be found in~\cite{chemtobm}.  
This analysis is not restricted to diagrams involving  
a top quark in the loop, and $\l$ couplings are  
also considered.  
When LFV decays are mediated by $\l$ couplings,  
it is shown that   
$ B_{JJ'}\approx ( \vert {\l_{ijJ} \l_{ijJ'}^\star \vert /  
   0.01 })^2 \  4.  \times 10^{-9}$.  
Under the hypothesis of a  
pair of dominant coupling constants, one deduces from the current  
experimental limits:   
$\vert \l_{ijJ}\l_{ijJ'}^\star \vert < [0.46,\ 1.1, \ 1.4 ]$  
for $[(J \ J')= (12),\ (23),\ (13)] $ and   
$\tilde{m} = 100 \GeV$, $\tilde{m}$ being the common mass  
of the sfermions involved in the contributing loops.  
LFV decays mediated by $\l '$ interactions  
are enhanced   
by an extra colour factor and by the  
possibility of accommodating an internal top quark in the  
loops. An approximate estimate reads:  
$B_{JJ'} \approx ({ \vert \l ^{'\star } _{Jjk} \l'_{J'jk}\vert  / 0.01 })^2  
1.17 \times 10^{-7}$.   
The comparison with the experimental limits yields the  
coupling constant bounds:  
$\vert \l ^{'\star } _{Jjk} \l'_{J'jk} \vert < [3.8 \times 10^{-2} ,\ 9.1   
\times 10^{-2} , \  
1.2 \times 10^{-1} ]$, for $[(J \ J')= (12),\ (23),\ (13)] $ 
and $\tilde{m} = 100 \GeV$. 
  
These $e^+e^-$ collider bounds are not yet competitive with bounds   
obtained from fixed target experiments but they are expected to be  
improved in the context of future $e^+e^-$ linear colliders.  
The $\Rp$ contributions to Flavour Changing Neutral Currents at  
$e^+e^-$ colliders for centre of mass energies above the   
$Z$ boson pole will be discussed in chapter~\ref{chap:colliders}.  
  
\subsubsection{Atomic Parity Violation and Polarisation Asymmetries}  
\index{Atomic parity violation|(}  
Atomic Parity Violation (APV)  has been observed via 
the $6S \rightarrow 7S$ transitions of 
$^ {133} _{55}\text{Cs} $~\cite{wood}. 
In the SM, parity violating transitions between particular atomic levels 
occur via $Z$-exchange between the nucleus and the atomic electrons. 
The underlying four fermion contact interactions, flavour diagonal  
with respect to the leptons and quarks, are conventionally represented  
by the effective Lagrangian:  
\begin{eqnarray}  
{\cal{L}}= {G_{F} \over \sqrt 2} \sum_{i=u,d} \left[ C_1(i) (\bar e \g_\mu 
\g_5 e) (\bar q_i \g^\mu q_i) +C_2(i) (\bar e \g_\mu e) (\bar q_i \g^\mu  
\g_5q_i) \right] 
\end{eqnarray}  
with  
\begin{eqnarray}  
 \begin{array}{lll}  
 C_1 (u) = 2 g^e_A g^u_V = -{1\over 2} + {4\over 3} x_W &,&  
 C_1 (d) = 2 g^e_A g^d_V ={1\over 2} -{2\over 3} x_W     
 \\  
 C_2 (u) = 2 g^e_V g^u_A =-{1\over 2} + 2 x_W &,&  
 C_2 (d) = 2 g^e_V g^d_A ={1\over 2} - 2 x_W   
 ,  
 \end{array}  
\end{eqnarray}  
and $x_W = \sin^2 \theta_W$.   
%
%
In the presence of a $\lambda'_{11k}$ ($\lambda'_{1j1}$) coupling, the 
$s$-channel  
exchange of a $\tilde{d}_{kR}$ ($\tilde{u}_{jL}$) between an electron   
and a $u$ ($d$) quark in the nucleus, as shown by the crossed processes  
depicted in Fig.~\ref{fig:asym}b(c), leads to additional parity  
violating interactions.  
Moreover, the coefficients $C_1$ and $C_2$ are affected by  
a non vanishing $\lambda_{12k}$ coupling via its effect on  
$G_{F}$ given by equation~(\ref{eqn:gmu_c6}).  
The expression of $C_1$ and $C_2$ in the presence of \Rp\ interactions  
reads as~\cite{barger89}:  
\begin{eqnarray}   
  C_1(u)&=&(-\ud +{4\over 3} x_W )(1-r_{12k}(\tilde e_{kR}))  
  -r'_{11k}(\tilde d_{kR}), \cr C_2(u)&=& (-\ud +2x_W  
  )(1-r_{12k}(\tilde e_{kR}))-r'_{11k}(\tilde d_{kR}), \cr  
  C_1(d)&=&(\ud -{2\over 3} x_W )(1-r_{12k}(\tilde e_{kR}))  
  +r'_{1j1}(\tilde u_{jL}),\cr C_2(d)&=& (\ud -2x_W )(1-r_{12k}(\tilde  
  e_{kR})) -r'_{1j1}(\tilde u_{jL}).  
\label{eq10}  
\end{eqnarray} 
  
Instead of using the $C$ coefficients directly,   
one can use the measurement of the weak charge $Q_W$, defined as  
$ Q_W= -2[(A+Z) C_1(u)+(2A-Z) C_1(d)]$ where $Z$ ($A$) is the number of protons  
(nucleons) in the considered atom, or its difference  
$\delta Q_W = Q_W - Q_W^{SM}$ to its SM value.  
In $\text{Cs}$ atoms \cite{revginges}:  
$$ \delta (Q_W(\text{Cs})) = -2 [72.07\  r_{12k}(\tilde e_{kR}) +376\    
r'_{11k}(\tilde  
d_{kR}) -422\ r'_{1j1}(\tilde u_{jL}) ] \quad .$$  
Comparison with the experimental measurement :  
$$ \delta (Q_W(\text{Cs})) = 0.45 \pm 0.48$$  
in the single coupling dominance hypothesis leads to the $2-\sigma$   
bounds:  
\begin{eqnarray}  
| \l_{12k}|< 0.05 \ \tilde e_{kR}, \   
|\l '_{11k}|< 0.02 \ \tilde d_{kR}, \  
|\l '_{1j1}|< 0.03\ \tilde u_{jL} \quad .  
 \label{eqn:apv_bounds}  
\end{eqnarray}  
  
In $\text{Tl}$ atoms,  
$\delta (Q_W(\text{Tl}))= -2 [116.89\  r_{12k}(\tilde e_{kR}) +570\    
r'_{11k}(\tilde d_{kR}) -654\ r'_{1j1}(\tilde q_{jL}) ] $.  
However the resulting bounds on the \Rp\ couplings   
are less stringent than the ones given above, due to the  
large error on the measurement of $\delta (Q_W(\text{Tl}))$.  
  
Closely related to APV measurements in atoms, polarisation  
asymmetries in elastic and inelastic scattering of longitudinally  
polarised electrons on proton or nuclear targets  
can also be used to constrain \Rp\ interactions~\cite{marcianolanga}. 
Below, we shall use the summary of  
experimental results for the parameters $ C_{1,2} (q)$ as quoted   
in \cite{bargerz}.   
  
A relevant observable is the asymmetry with respect to  
the initial lepton longitudinal polarisation for the elastic  
scattering on scalar $ J^P= 0^+$ nuclear targets,  
\begin{equation}   
\cala_{pol} = { d\s _{R} - d\s _{L} \over d\s _{R} +d\s _{L} } ;  
\quad [\cala _{pol} ={G_F q^2 \over \sqrt 2 \pi \a } {3\over 2 }  
(C_1(u) +C_1(d)) (1 +R(q^2) ) ] .   
\end{equation}    
The elastic electron scattering $ e^- + ^{12} C \to   e^- + ^{12} C  $  
is studied at the BATES accelerator.  
The difference between the experimental  
measurement and the Standard Model prediction~\cite{souder,souderlanga} is   
given  
by $ \delta (C_1(u) +C_1(d)) = (0.137 \pm 0.033) -(0.152 \pm 0.0004)  
= -0.015 \pm 0.033$.  Fitting this to the \Rp\  
contribution yields the  coupling constant bounds  
$ \l_{12k} < 0.255 \, \tilde e_{kR} , \ \l '_{11k} < 0.10 \, \tilde d_{kR} $,  
and the $1\s $ level bound   
$\l '_{1j1} < 0.11 \tilde q_{jL}$.  
  
The polarisation asymmetry of inelastic electron scattering on deuteron  
$e +d \to e ' +X$, as measured by the SLAC experiment~\cite{prescott}, is  
described by,  
\be\cala_{pol} = {3 G_F Q^2 \over 5 \sqrt 2 \pi \a }  [(C_1(u) -\ud C_1 (d) )   
+ (C_2(u) -\ud C_2 (d) ){ 1 -(1-y)^2 \over 1  
+(1-y)^2} ]  \quad  
[y= {E_e - E'_e\over E_ e} ] . \end{equation}    
The differences between the experimental values and the Standard Model   
predictions are, \\  
$ \delta (2C_1(u) -C_1 (d) ) = (-0.22\pm 0.26),   
\ \delta (2C_2(u) -C_2 (d) ) = (-0.77\pm 1.23)$.    
Comparing with the \Rp\ contribution for the first quantity yields the   
coupling constant bounds   
$ \l '_{11k} < 0.29 \tilde d_{kR}, \   
  \l '_{1j1} < 0.38 \tilde q_{jL} $,   
and the $1\s $ level bound,   
$ \l_{12k} < 0.20 \tilde e_{kR} $.    
Comparing for the second quantity yields   
$ \l_{12k} < 2.0 \ \times \tilde e_{kR}, \ \l '_{1j1} < 0.71 \tilde  
  q_{jL} $   
and at the $1\s $ level,   
$ \l '_{11k} < 0.39 \tilde d_{kR}.$  
  
The electron polarisation asymmetry, as  measured for the $ e_{L,R} +  
^9Be \to p +X $  
quasi-elastic scattering Mainz accelerator experiment~\cite{mainz},  
exhibits a discrepancy with respect to the Standard Model   prediction, ${\cal  
{A}}_{Mainz} = 12.7\  C_1(u) -0.65 \ C_1(d) +2.19 \ C_2(u) -2.03 \ C_2(d) =  
(-0.065 \pm 0.19)$.  Comparing with the \Rp\ contribution yields the  
coupling constant bounds $ \l '_{11k} < 0.93\ \times  10^{-1} \tilde d_{kR},$   
and at the $1\s $ level, $ \l_{12k} < 3.0 \ \times 10^{-1} \tilde e_{kR}, \  
\l '_{1j1} < 2.4 \ \times 10^{-1} \tilde q_{jL}$.  
The above results clearly show that the strongest bounds from $\g - Z$  
interference effects are those emanating from the APV experiments.  
\index{Atomic parity violation|)}   

\subsection{Anomalous Magnetic Dipole Moments}  
\index{Anomalous magnetic dipole moment|(}  

The anomalous magnetic dipole ($M1$) moments of the quarks and leptons  
represent valuable observables that may be accessed in both low and energy  
experiments.    
For  the light leptons, these observables are determined with   
high precision  thanks to  the  high sensitivity currently   
attained by the experimental  measurements.   
Other moments such as the  $Z$ boson current  
anomalous  magnetic dipole moments  for  the $\tau $-lepton $a_{\tau  
}(m_Z^2)$ or  the $b $-quark $ a_{b}(m_Z^2)$    
are currently accessed in experiments at the  leptonic and hadronic colliders.   
  
The discrepancies with respect to the Standard Model expectations for the   
leptons or quarks, defined as $ \delta  a_l =a_l^{EXP}-a_l^{SM}, $    
reflect the corrections arising from the perturbative higher-loop  
orders electroweak contributions, the virtual hadronic corrections and  
possibly the Minimal Supersymmetric Standard Model loop corrections.    
The  comparison between theory and experiment    
is expected to provide a sensitive test for new physics~\cite{marciano}.    
The electron  moment $a _e$ is a basic   
data for the purpose of extracting the experimental value   
of the hyperfine constant  $\a $.    
A small finite discrepancy is present for the electron   
in the difference between the  
experimental and Standard Model value   given by  $ \d  
a_e \equiv a_e^{EXP}-a^{SM}_e= 1. \ \times \   10^{-11}. $    
  
The measurement of the muon anomalous magnetic  
moment~\cite{brown01} exhibits a finite deviation from the Standard  
Model prediction,  
$\delta  a_\mu = a_ \mu ^{EXP}-a^{SM}_ \mu  = 33.7 \ (11.2) \  
\times \ 10^{-10} $, based on $e^+e^-$ data,   
or $\delta  a_\mu = a_ \mu ^{EXP}-a^{SM}_ \mu  = 9.4 \ (10.5) \  
\times \ 10^{-10} $, based on $\tau$ data. The theoretical value   
$(a_\mu )_{SM}$ includes the contribution  from the hadronic radiative   
effects~\cite{davier98}.   
The above quoted deviation with respect to the  
Standard Model is deduced on the basis of calculations of multi-loop  
diagrams which have not been verified in totality. Recent  
works by Knecht et al.~\cite{knecht01}    
have resolved the long-standing problem associated with the sign in the   
muon anomalous magnetic moment~\cite{hayakino01} for the    
non-perturbative contribution from   
the pion-pole term in the light-by-light scattering amplitude.   
We refer to the review cited in \cite{knecht01} for more details.  
Although the exact size of the hadronic contributions  to   
$a_\mu ^ {SM}$ remains an unsettled  problem, the comparison of  the various  
existing calculations~\cite{hadronmu98}   
indicates that the corresponding uncertainties do   
not affect significantly the Standard Model  prediction.    
  
An early study  by Frank and Hamidian~\cite{hamidian} of the \Rp\  
contribution to the leptons anomalous $M1$  moments  indicated that the   
resulting constraints on the \Rp\ coupling constants were relatively   
insignificant.   
The recently reported  measurement  of the muon anomalous magnetic  
moment~\cite{brown01} has stimulated two detailed studies of the \Rp\  
effects~\cite{kimk01,adhir01} focused on the muon anomalous $M1$   
moment.    
  
The study by Kim et al.,~\cite{kimk01} is performed within the so-called  
effective supersymmetry framework. One  retains the contributions from   
the third generation sfermions only, based on the assumption that the   
first and second generation  sfermions  decouple as having   
large masses of order $m _{\tilde {l} } = O(20)\  \text{TeV} , \  [l=1,2]$.   
Such a radical hypothesis would, of course, relax significantly   
the various bounds on the \Rp\ coupling constants   
with  superpartner indices associated to the first two generations.   
The one-loop diagram contributions to the  anomalous $M1$   
moment   $a_{f_J} $ of  a fermion  $f_J$ enter in two  types depending   
on whether the required chirality flip  between the external  
fermions takes place on the external lines themselves (giving   
an external fermion mass overall factor $ m_J$) or on the internal   
fermion and sfermion  lines (giving an  overall factor  
$m_{f_J} \tilde m^2_{\scriptscriptstyle{LR}} /  m_{\tilde f }^2 $).   
For the muon case, it turns out that the contributions of the first type,   
with a chirality flip on the external line, are the predominant ones.   
The  bounds inferred by assuming the single coupling constant dominance  
hypothesis are:   
\begin{eqnarray} [\l _{32k}, \  \l _{3j2},\ \l _{i23},\ \l _{2j3} ]   
< 0.52 {m_{\tilde f } \over 100 GeV} .   
\end{eqnarray}  
Alternatively, if one focuses solely on the coupling constant $\l _{322}$,  
based on the observation that this is   
the least constrained  of all the coupling constants involved, then the   
predicted value of the anomalous moment  associated with the \Rp\ effects,   
\begin{eqnarray}(a_\mu )_{RPV} \simeq 34.9 \ \times \ 10^{-10} ({100 \  
GeV\over \tilde m} )^2 \vert \l _{322} \vert ^2, \end{eqnarray} is seen by   
comparison with the experimental result to be compatible   
with the perturbativity bound on the corresponding coupling constant.   
  
A comparison of  the \Rp\ effects on the muon magnetic moment and    
the neutrino masses is of interest, despite the fact that the   
one-loop contributions to the neutrino masses are of the type involving    
chirality flip mass insertion  terms for the fermion and sfermion   
internal lines. Indeed, some   
correlation still exists between the one-loop contributions  in these  
two cases owing to the identity of the \Rp\ coupling constant   
factors.  Adhikari and Rajasekaran~\cite{adhir01}   
observe that in order to get   \Rp\    
contributions to  the muon anomalous  magnetic  moment and neutrino   
mass of the size  required by the current experiments,   
$a _\mu =  O(10^{-9}),\  m_{\nu_\mu } = O(1) $ eV,    
one needs to suppress in some way   
the  one-loop contribution to the neutrino mass. This can be achieved   
by postulating for the chirality flip slepton mass parameters  
$(\tilde m^{e2}_{\scriptscriptstyle{LR}})_{ij}$ either reduced values or  
a degeneracy with respect to the first two generations,   
$(\tilde m^{e2}_{\scriptscriptstyle{LR}})_{11} \simeq  
(\tilde m^{e2}_{\scriptscriptstyle{LR}})_{22}$.  
A natural resolution  of this issue can be achieved in a model using  
a discrete symmetry acting on the lepton sector.   
\index{Anomalous magnetic dipole moment|)}  
  
\subsection{{\boldmath{$C P$}} Violation}  
\label{secxxx2c}  
\index{Violation of $C P$}  
 
The existence of a possible connection between $C P$ violation and  
$R$-parity violation has received increased attention in the  
literature~\cite{shalom,grossmanworah,kaplan,abel,guetta,jang,mohanta00,chen,  
shalom2,handoko,bhatta99}.   
In this section, we discuss a non-exhaustive set  
of physical applications which lie at the interface between $R$-parity  
violation and $C P$ or $T$ violation.   
The topics to be adressed include discussions on observables in the   
neutral $ K \bar K $ system, the electric dipole moment (EDM) of leptons   
and quarks, the $C P$-odd asymmetries in $B$ meson hadronic decay rates and  
the  $C P$-odd asymmetries in $Z$ boson decay rates into fermion-antifermion  
pairs.  
A study of the $C P$-violation   
effects in association with sneutrino flavour oscillations 
\index{Violation of $C P$!Sneutrino flavour oscillations}   
was carried out in chapter~\ref{chap:neutrinos}.  

\subsubsection{General Considerations on {\boldmath{$C P$}} Violation}  
  
As is known, the violation of $C P$ symmetry is revealed in the  
context of field theories by complex  
phases present in VEVs of scalar fields, in particles mass parameters  
or in Yukawa interaction coupling constants which cannot be removed by  
field redefinitions.  
  
It is conventional to distinguish between soft and hard $C P$ violation,  
depending on whether the dimensionality of the $C P$-odd operators in the  
effective lagrangian is $\le 3$ and $\ge 4$ respectively. 
The spontaneous\index{Violation of $C P$!Spontaneous} $C P$ violation 
case, as characterized by the presence of $C P$-odd complex phases  
in scalar fields VEVs resulting from a $C P$ conserving lagrangian, 
falls naturally within the soft violation category. 
The distinction between soft and hard $C P$ violations is motivated by the  
different impact that the quantum and thermal fluctuations effects  
have in the two cases.  The soft 
$C P$ violation interactions cannot renormalize the hard interactions,  
unlike the hard 
$C P$ violation  
interactions which indeed can renormalize the operators of lower  
dimensions.  The soft $C P$ violation parameters may also be suppressed  
by thermal fluctuations, eventually leading to a restoration of $C P$  
symmetry at high temperatures.  By contrast, the thermal effects do  
not affect significantly the coupling constant parameters of operators  
of dimension $\ge 4$, a fact which makes the hard $C P$ violation  
mechanisms more robust candidates for generating the baryon or lepton  
asymmetry in the early Universe.  
  
The Standard Model includes two sources of hard $C P$  
violation 
One is the complex phase in the CKM matrix with three quark  
generations. The other is the QCD theta-vacuum angle.  
For supersymmetric models, new sources of soft $C P$ violation appear  
with the $\mu $ term\index{Violation of $C P$!$\mu$ term} 
and soft supersymmetry breaking\index{Violation of $C P$!Soft SUSY breaking} 
interactions. 
With the known structure of the constrained MSSM classical action,  
assuming fully universal soft supersymmetry breaking,  
the unremovable $C P$-odd phases are restricted to a pair of phases given  
by the relative complex phases $\phi _{A} = \arg (A M_\ud ^\star )$ and  
$ \phi _B = \arg (B \mu ^\star ). $ The experimental constraints  
give strong individual bounds on these phases i.e. $ \phi _{A,B} < O(10 ^  
{-3}) $.  However several other additional phases arise once one  
relaxes the universality hypothesis for the supersymmetry breaking parameters  
for the quark and lepton generations or the different gaugino mass  
parameters.  It has been found~\cite{bhrlik99} that in cases where some built-in  
correlations between the various phases are included, the evaluation  
of physical observables includes such strong cancellation effects that  
the experimental bounds relax to values $O(1)$.  
  
With broken $R$-parity, new sources of $C P$ violation can contribute  
through complex phases in the parameters $\mu , \ \mu _i $ and  
$B, \ B_i$ for the bilinear interactions\index{Violation of $C P$!$\mu$ term} 
and in the  
parameters $\l _{ijk} , \ \l ' _{ijk} , \ \l ''_{ijk} $ and $ A _{ijk}  
, \ A ^ {'}_{ijk} , \ A ^ {''} _{ijk} $ for the trilinear  
interactions.  Each of these coupling constants can carry a complex  
phase although only the subset of these phases invariant under the  
fields rephasing is physical.  For products of the trilinear \Rp\  
coupling constants only, a $C P$-odd phase invariant under complex phase  
redefinitions of the fields can be defined starting from the  
quartic order.  Examples encountered in the calculations of scattering  
amplitudes for processes involving four fermion fields such as  
$ e^++e^- \to f_J + \bar f_{J'}$ are given by:    
$\arg ( \l _{i1J} \l ^\star _{i1J'} \l^\star _{i'jJ} \l _{i'jJ'} ), $  
and $\arg ( \l _{ijJ} \l ^\star _{ijJ'} \l^\star _{i'j'J} \l _{i'j'J'}  
)$.   
  
Although basis-independent studies have been performed   
for specific cases~\cite{davidson97,davidson97b,grossman00}  
a full systematic discussion of a basis independent  
parametrisation of $C P$ violation for the \Rp\ interactions would be  
useful in characterizing the natural size of the relevant parameters  
as can be emphasized from the example described in~\cite{huitu}.    
  
Concerning the context of bilinear $R$-parity violation,  
a partial study of the  parametrisation of $C P$ violation  
is performed in~\cite{joshi95}. For that part of the scalar  
potential which determines the Higgs bosons VEVs, the $C P$-odd phase is  
included through the coupling constant product  
$\mu ^\star \mu _i B_i B$. Including the trilinear terms, one  
characteristic $C P$ violation condition can be identified in terms of  
the phase in the coupling constant product of $\l $ interactions and  
regular Yukawa interactions given by: $\Im ( \l ^{\star }_{nmk} \l  
_{ijk} \l ^e _{nl} \l ^{e\star } _{il} \l ^e _{mp} \l ^{e\star } _{jp} ) \ne  
0$.  Another interesting conclusion concerning a spontaneous violation  
of $C P$\index{Violation of $C P$!Spontaneous} 
in the presence of \Rp\  
interactions is that complex valued sneutrino \VEVs\ can occur in a natural 
way without fine-tuning of the parameters~\cite{joshi95}. 

\subsubsection{Neutral {\boldmath{$K \bar K $}} System}  
\index{Violation of $C P$!$K \bar K$ system}  
  
The possibility of embedding a $C P$-odd complex phase in the \Rp\  
coupling constants has been envisaged at an early stage in a work by Barbieri  
and Masiero~\cite{barbieri86} and then further discussed 
in~\cite{choudhury96,carlson,decarlosq,slavich,kundu}  
and references therein.  
A complex relative phase present in a  
quadratic product of $\l''_{ijk}$ coupling constants can contribute to  
the $K_S-K_L$ mass difference.  
  
\begin{figure}[htb]  
 \begin{center}  
     \mbox{\epsfxsize=0.9\textwidth  
       \epsffile{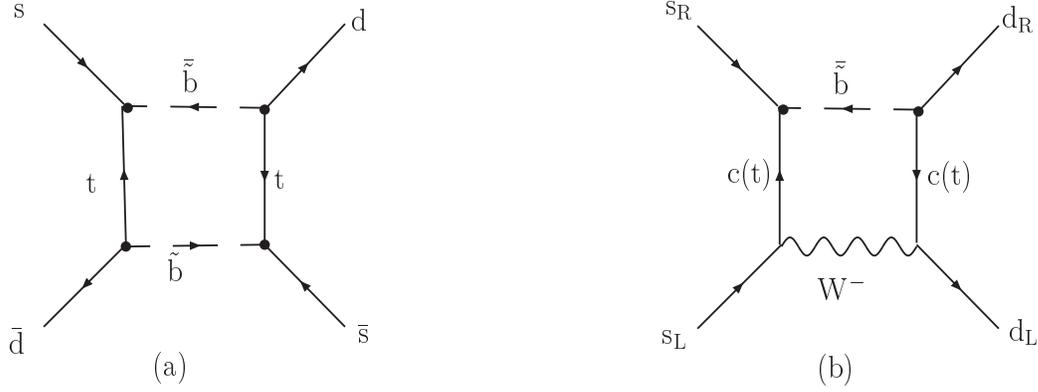}}  
  \end{center}  
 \caption[]{\it{ Box diagrams leading to $K \bar{K}$ mixing induced  
       by $\l''$ couplings.}}  
\label{fig:cp1box}  
\end{figure}  
  
  
\begin{figure}[htb]  
\begin{center}  
\epsfig{file=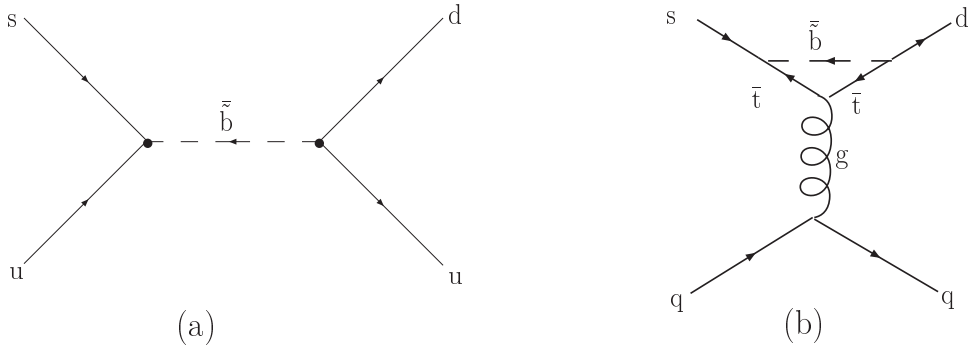,width=0.85\textwidth}   
\end{center}  
 \caption{\it{Tree level diagram (a) and gluonic penguin one-loop  
              diagram (b) contributing to the direct  
	      $\Delta S =1 $ $C P$ violation involving $\l ''$ couplings.}}  
\label{fig:cp2box}  
\end{figure}  
  
The most general $ \Delta S =2 $ effective lagrangian 
${\cal{L}}^{\Delta S =2}_{eff}$  
including contributions of charginos and charged Higgs boson neglected in 
earlier works~\cite{barbieri86,carlson,decarlosq} has been considered 
in~\cite{slavich} (see also~\cite{kundu}).  
Its contribution to the $K_S-K_L$ mass difference is related to the matrix 
element $< K^o \vert \bar {\cal{L}}^{\Delta S =2}_{eff} \vert \bar {\bar K}^o >$ 
and involves the products of $\l ''_{313} \l'' _{323}$ couplings (see for 
example Fig.~\ref{fig:cp1box}) as well as CKM matrix 
elements~\footnote{In~\cite{barbieri86} the charm contribution and in  
                   consequence the $\l''_{232} \l'' _{213}$ products have also 
		   been considered where the t-quark in the loop is replaced 
		   by a c-quark.}.   
This \Rp\ coupling's contribution to the $K_S-K_L$ mass difference  
has been calculated in~\cite{slavich} using NLO QCD evolution of Wilson  
coefficient also included in 
${\cal{L}}^{\Delta S =2}_{eff}$~\cite{ciuchini}  
as well as lattice calculations for long-distance hadronic processes which 
cannot be evaluated pertubatively and also contribute to the above matrix 
element.  
Requiring that this contribution to the $K_S-K_L$ mass difference is not 
larger than the experimental  
value~\cite{pdg02}~\footnote{Actually the upper bound derived in~\cite{slavich} 
                    comes from the experimental value published in~\cite{pdg00}  
                    on the $K_S-K_L$ mass difference. However the difference 
		    with the published value in~\cite{pdg02} being marginal 
		    for the present purpose, the conclusion of the analysis  
                    presented in~\cite{slavich} 
		    on $\l''_{232} {\l''}^{\star}_{213}$ is unchanged.}   
allows one to set an upper limit~\cite{slavich}:   
\begin{eqnarray}  
\l''_{313} {\l''}^{\star}_{323} < O(0.033)  
\end{eqnarray}  
by performing a general scan  
over the parameter space on the minimimal supersymmetric extension of the  
standard model at the weak scale and taking into account the contraints from  
direct searches for supersymmetric particles.     
  
The $ \l ''$ interactions contribute also at the tree level to the  
direct $ \Delta S =1 $ $C P$ violation 
(see Fig.~\ref{fig:cp2box}(a)),  
as described by the observable parameter $\e '$.   
The \Rp\ contribution to $\e '$ is described by the 
relation~\cite{barbieri86}:   
\begin{eqnarray}  
{\Im} ( \l '' _{123} \l ^{ ''\star } _{113} )  
\approx \vert \e '\  \vert \  10 ^{1} \ \tilde q^2 .   
\end{eqnarray}  
The gluonic penguin one-loop diagram, see Fig.~\ref{fig:cp2box}(b),   
provides a competitive contribution to that of the box diagram due  
to the existence of a logarithmic enhancement factor in the amplitude.  
The resulting bound reads~\cite{barbieri86}:   
\begin{eqnarray}   
\Im( \l '' _{313} \l ^{'' \star } _{323} )  
\approx \vert \e ' \vert \  10 ^{-2}\    \tilde q^2 .   
\end{eqnarray}  
In order to match the  currently observed  value for $\e ' =  
O(10^{-6})$ one should require values for the quadratic   
coupling constants  
$ {\Im} ( \l '' _{123} \l ^{ ''\star } _{113} )$  
(respectively $\ {\Im} ( \l '' _{313} \l ^{''\star } _{323}$))  
of $O(10^{-5} ) \ \tilde q^2$ 
(respectively $O(10^{-8} ) \ \tilde q^2 $).  
  
Since different generational  configurations of the  
coupling constants  contribute to the  $C P$ violation parameters $\e  
$ and $\e '$,  one concludes that the \Rp\ interactions   $\l ''$   
might be relevant candidates for milliweak type $C P$  
violation contributing solely to the  
indirect $C P$ violation. 
  
\begin{figure}[htb]  
\begin{center} 
 
\vspace*{0.3cm} 
 
\epsfig{file=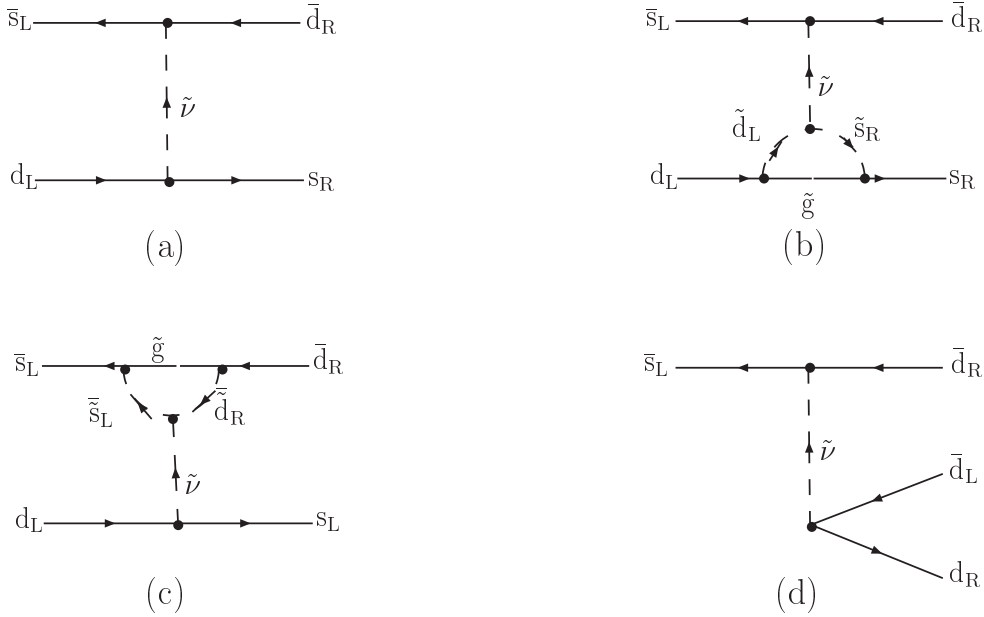,width=0.85\textwidth}   
\end{center}  
 
\vspace*{-0.2cm} 
 
 \caption{\it{Contributions to $\Delta S =2 $ (a, b and c)   
and $\Delta S =1 $ (d) observables involving  $ \l '$ couplings.}}  
\label{fig:cp9box}  
\end{figure}  
  
An interesting proposal~\cite{abel} is to incorporate the $C P$-odd phase  
in the scalar superpartner interactions corresponding to the soft  
supersymmetry breaking \Rp\ interactions of the form   
\mbox{${V_{soft} = A^{ '}_{ijk} \tilde L_i \tilde Q_j \tilde D_k^c +\  
\mbox{h.c.}} \ $}.   
The contribution  
to the imaginary part of the $ \Delta S =2$ mass shift is estimated  
qualitatively as   
$ \e \simeq  10^{-2} \ \Im( A ^{'} _{i21}  -   
A ^{'} _{i12}  ) /m_{\tilde g} $.   
The comparison with the experimental value   
indicates that some cancellation between the above two flavour  
non-diagonal configurations i.e. $ A ^{'} _{i21} and \ A ^{'}  
_{i12}$,  
may be required to take place.  A contribution to the four quark  
interaction $ \bar s_R d_L \bar d_R s_L$ arises from a sneutrino  
exchange penguin type diagram involving a one-loop correction to the  
$\tilde \nu ds $ vertex as shown in  
Fig.~\ref{fig:cp9box}(b) and~\ref{fig:cp9box}(c).    
The predicted effect on the direct $C P$  
violation $ \Delta S =1 $ observable $ \e '$, see Fig.~\ref{fig:cp9box}(d), 
reads~\cite{abel}:   
\begin{eqnarray}  
\vert { \e ' \over \e } \vert \simeq 10^{-7} \ { {\l '} _{i11} \over  
{\l '} _{i12} } > 10^2 \ {\l '} _{i11} {\l '} _{i21} , \end{eqnarray}   
where the inequality  obtained  at the second step uses   
the  bounds on the coupling constant products,   
$ {\l ^{' \star } } _{i12} {\l '} _{i21} <  10^{-9} \tilde \nu_i ^2 . $   

\subsubsection{Asymmetries in Hadron Decay Rates   
                   and Polarisation Observables}  
\label{sec:Aszpole}  
  
The polarisation of the muon emitted in the $ K$-meson three-body  
semileptonic decay $ K^+\to \pi ^0 +\nu +\mu^+ $ ($K_{\mu 3}$) or in  
the radiative decay $ K^+\to \mu^+ +\nu +\g $ ($K_{\mu 2\g }$)  
constitute useful observables  for testing $ T $ and/or   
$C P $ violation.   
  
The transverse muon polarization $ P_T( K_{\mu 2\g })$ can be 
related~\footnote{Various expressions for $P_T(K_{\mu 3}$) can be found 
		  in~\cite{belanger} in the context of leptoquark models.} 
to $\vert \Im (\l ^\star _{2i2} \l ' _{i12} ) \vert  / m^2_{\tilde e_{iL} }$   
as shown in~\cite{chen}.  
  
Under various simplifying assumptions~\cite{chen}, rough estimates on upper 
bounds on $ P_T( K_{\mu 2\g })$ may be derived from bounds on the  
branching ratio of $\mu \rightarrow e \gamma$ and from the measured  
value~\footnote{At the time of~\cite{chen} only upper bounds on   
$ BR(K^+  \rightarrow  \pi^+ \nu \bar \nu) $  
were known.} of $ BR(K^+  \rightarrow  \pi^+ \nu \bar \nu) $.  
The $K_{\mu 2\g }$ decay has been measured in~\cite{aliev}  
(see also~\cite{adler}) and should provide useful handles in constraining 
these \Rp\ couplings.  
  
The transverse muon polarization $P_T(K_{\mu 3}$ is O($10^{-10}$) in the  
standard model~\cite{valencia}.  
It can be related to   
$(\l ^\star _{232} \l ' _{312} )   / m^2_{{\tilde {\tau}}_{iL} }$  
and $ (\l ^\star _{122} \l ' _{112} )  / m^2_{\tilde e_{iL} }$ as shown   
in~\cite{herczegb99} (see also~\cite{fabbrichesi,wung}). These contributions 
from \Rp\ couplings can be as large as the present experimental   
limits~\footnote{The limits used in~\cite{herczegb99} come from~\cite{abetal1}  
                 and should be replaced by those published in~\cite{abetal2}. 
		 However the general conclusion drawn in~\cite{herczegb99} 
		 should not be significantly affected.}.  
on $P_T(K_{\mu 3}$~\cite{abetal1,abetal2} (see also~\cite{anisimov,blatt}).  
  
As discussed in~\cite{diwan} future  
projects may reach uncertainties approaching the O($10^{-4}$ ) level for the 
measurement of $P_T(K_{\mu 3})$ thus allowing also to test \Rp\ couplings 
further.  
  
\subsubsection{{\boldmath{$Z$}} boson partial decay}  
\label{sec:zpartialdec}  
    
The $Z$ boson partial decay channels into fermion-antifermion (up-quark,  
down-quark, charged lepton) pairs of different flavours,   
$Z \to f_J +\bar f_{J'} \,  [J\ne J' ; \ f=u,  
\ d, \ l ] $,  may exhibit  potentially observable  $C P$  
violating decay asymmetries. These are defined by the normalised  
differences of flavour non-diagonal, spin-independent rates,   
\begin{eqnarray} {\cal  
A}_{JJ'}={B_{JJ'}-B_{J'J}\over B_{JJ'}+B_{J'J} },  
\end{eqnarray}  
where the   branching ratios $ B_{JJ'}$ are  defined in  
equation (\ref{eqx8}). A finite contribution to the flavour decay asymmetry is   
rendered possible by the existence of a finite $C P$-odd complex phase,  
$\psi $, embedded in the \Rp\ coupling constants.  The asymmetries  
\cite{chemtobm} are proportional to ratios of the coupling constants    
of the form:   
\begin{eqnarray}  
\Im ( { \l ^{'\star } _{iJk } \l '_{iJ'k} \over \l ^{'\star  
} _{1Jk'} \l '_{1J'k'} } ) \propto  \sin \psi .    
\end{eqnarray}  
Should the \Rp\ coupling constants exhibit generational hierarchies,   
one could then expect large enhancement or suppression of the   
asymmetries, depending on the flavour of the emitted fermions.    
Assuming that the above ratios of \Rp\ coupling constants products take  
values of order unity, the resulting asymmetries for the emission of  
charged leptons, down-quarks and up-quarks are found to be of the order of  
${\cal A}_{JJ'} \approx (10^{-1}\ -\ 10^{-3}) \sin \psi $.

\subsubsection{Neutron Electric Dipole Moment}    
\label{sec:neutronedm}  
  
\begin{figure}[htb]  
\begin{center}  
\epsfig{file=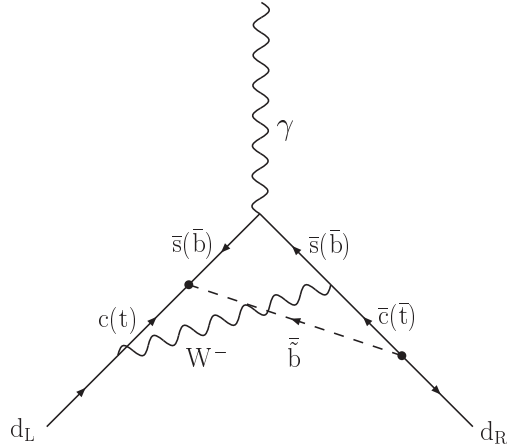,width=0.45\textwidth}   
\end{center}  
 \caption{\it{Contributions to the neutron electric  
dipole moment involving  $ \l ''$ couplings.}}  
\label{fig:cp6box}  
\end{figure}

The \Rp\ interactions $ \l ''$ may contribute to the neutron electric  
dipole moment $d^\g_n$~\cite{barbieri86} via a quark electric dipole  
moment described by a two-loop vertex Feynman diagram involving the  
crossed exchange of $W $ and $ \tilde d $ internal particle lines   
as seen in Fig.~\ref{fig:cp6box}.    
Note that no contributions from the \Rp\ interactions to  
the neutron dipole moment can arise at the one-loop order  
level~\cite{godbole99,abel99}.  
 
The needed suppression  to account for a sufficiently small   
electric dipole moment is provided in part by the light quark mass   
factors  reflecting the chirality flip selection rules  of the \Rp\   
couplings. A relative $C P$-odd complex phase embedded   
in a pair product of $\l ''$  
coupling constants is required in order to obtain a finite  
contribution to the electric dipole moment.  The contribution is  
maximised by choosing third generation $ b-\tilde b$ quarks  
configuration for the internal fermion and sfermion particles.  The  
results derived by Barbieri and Masiero~\cite{barbieri86} on the basis of  
the double coupling constant dominance hypothesis read:   
\begin{eqnarray}  \Im (\l  
''_{213} \l _{232}^{''\star } ) \simeq 10^{-2} ({d ^\g _n\over 10^{-25} \, e  
\times  \text{ cm}  })  
\tilde q ^2  
\label{eq:EDM_neutron_1} 
\end{eqnarray}  
\begin{eqnarray} \Im (\l    
''_{312} \l _{332}^{''\star } ) \simeq 10^{-1} ({d^\g _n\over 10^{-25} \, e  
\times  \text{ cm}  }) \tilde q ^2  
\label{eq:EDM_neutron_2} 
\end{eqnarray}  
for the internal charm quark and top quark cases,  respectively.   
Using the current experimental bound on the neutron electric dipole moment,  
$d_n^\g < 1.2 \ \times \ 10^{-25} \, e   \text{ cm} $ from~\cite{kfsmith}  
one obtains a bound  
for the  top quark  box diagram contribution only which is  
$ \ \vert \Im (\l ''_{312} \l _{332}^{''\star  
} ) \vert \, < 0.12 \tilde q^2 $.  
\par 
The discussion on systematic uncertainties concerning 
the more recent bound $d_n^\g < 6.3 \ \times \ 10^{-26} \, e \cdot  \text{ cm} $     
recently published in~\cite{pgharris} has been criticized in~\cite{golub}.  
  
Contributions from products of $\l'$ couplings such as $\Im({\l'}^{\star}_{i33} \l'_{i11})$ are discussed in   
\cite{changetal}.  
  
New contributions involving both bilinear and trilinear couplings can lead to a neutron  
electric dipole moment as discussed in~\cite{choichun,keum}.  
  
\subsubsection{Electron Electric Dipole Moment}    
\label{sec:electronedm}  
  
Even if one assumes purely real $C P$ conserving \Rp\ coupling constants,  
a non-vanishing $C P$ violating contribution could possibly be induced by  
invoking the existence of other possible sources of complex phases  
present in the (minimal) Supersymmetric Standard Model.  
Thus the $\l_{ijk} $ and $\l '_{ijk}$ \Rp\ couplings may induce one-loop  
contributions to the electric dipole moment of leptons (and quarks)  
through an interference with a complex valued $C P$-odd soft   
supersymmetry breaking parameters $A^{u}_{ij} and \ A^{d}_{ij} $   
associated with the regular Yukawa interactions.  
A non-vanishing amplitude associated with the Feynman diagram with a pair  
of sfermion and fermion internal lines requires the presence of a  
$L-R$ chirality flip mass-mixing insertion  
$\tilde m^2_{\scriptscriptstyle{LR}}$ for the  
internal sfermion.  Stated equivalently, it requires a mass splitting between  
the opposite chiralities sfermion eigenstates.   
The bounds are  
strongest for the electron electric dipole moment  
$d_e^\g $ and can lead to several strong individual coupling constant bounds as  
shown in~\cite{hamidian}.   
A representative subset is tentatively summarized by   
$\l '_{111} < 5.5 \ \times \ 10^{-5} , \ \l  
'_{121} < 8.7 \ \times \ 10^{-6} , \ \l '_{213} < 9.5 \ \times \ 10^{-2} , \   
\l '_{233} < 1.5 \ \times \ 10^{-2}$ using $d_e^{\gamma} = (3 \pm 8) \times 10^{-27} e \text{ cm}$  
~\cite{pdg96}  
which sould be superseded by the recently~\cite{regan}   
measured value $d_e^{\gamma} = (6.9 \pm 7.4) \times 10^{-28} e \text{ cm}$  
i.e. $ \vert d_e^{\gamma} \vert = 1.6 \times 10^{-27} e \text{ cm}$  
  
Focusing on the contribution from a complex \Rp\  
coupling constant $\l '_{133} =\vert \l '_{133}\vert e^{i\b } $,  
interfering with a complex soft coupling constant $ A ^q = \vert A ^q \vert  
e^{i\a _A }$, the current bound on the experimental electron electric dipole moment  
can admit solutions with large values for both of the above $C P$-odd  
phases $ \b $ and $  \a_A $ as shown in~\cite{adhi99}.  
  
\begin{figure}[htb]  
\begin{center}  
\epsfig{file=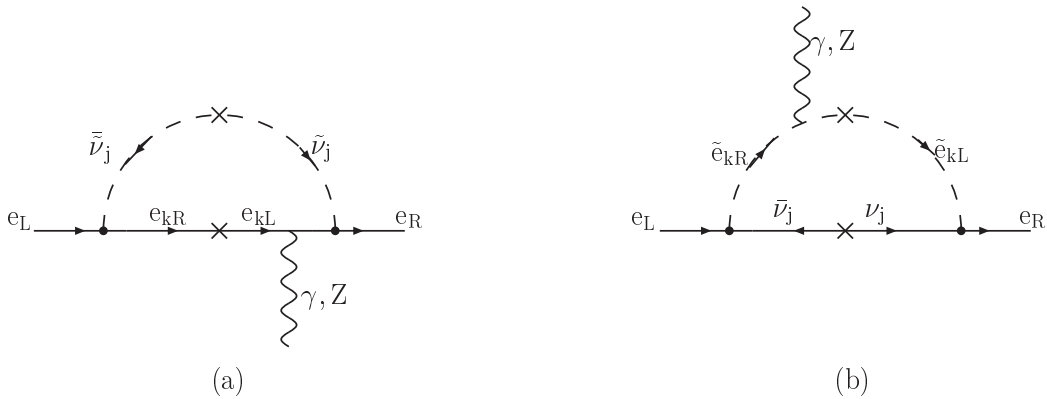,width=0.9\textwidth}   
\end{center}  
 
\vspace*{-0.2cm} 
 
 \caption{\it{One-loop diagrams contribution  to the electron   
dipole moment involving  $ \l $ couplings.}}  
\label{fig:edm3box}  
\end{figure}

However the conclusions drawn from the above two studies  
\cite{hamidian,adhi99} have been recently challenged by the  
observation that no contributions from the \Rp\ interactions to  
the electric dipole moment can arise at the one-loop order  
level~\cite{godbole99,abel99}.   
Due  
to the chirality selection rules, an $ \tilde e_R $ particle line can  
never be emitted nor absorbed.  For similar reasons, a $\tilde d_R -  
\tilde d_L $ mass insertion  on the squark line must     
be accompanied by a neutrino Majorana mass insertion, resulting in a strongly  
suppressed contribution to the electric dipole moment which exactly  
vanishes in  the zero neutrino mass limit.  A one-loop  
contribution could only be possible through a sneutrino Majorana mass  
term, $\tilde m_{ij} \tilde \nu _{iL} \tilde \nu _{jL} $ as can be   
seen in   
Fig.~\ref{fig:edm3box}.    
  
Similar chirality  
selection rules would also apply for the analogous chirality flip  
observables, such as involved in the contributions to the   
neutrino Majorana mass, the  
neutrino $ M1 $ or $ E1 $ diagonal or off-diagonal moments, or the  
charged fermions $ M1 $ transition moments.  
On the other hand, $E1$ transition  
moments for Majorana neutrinos may possibly be initiated by the \Rp\   
interactions   
at the one-loop order. The  above observations figure in~\cite{godbole99,abel99}.   
\begin{figure}[htb]  
\begin{center}  
  \begin{tabular}{c}  
  \hspace*{-0.4cm} \epsfig{file=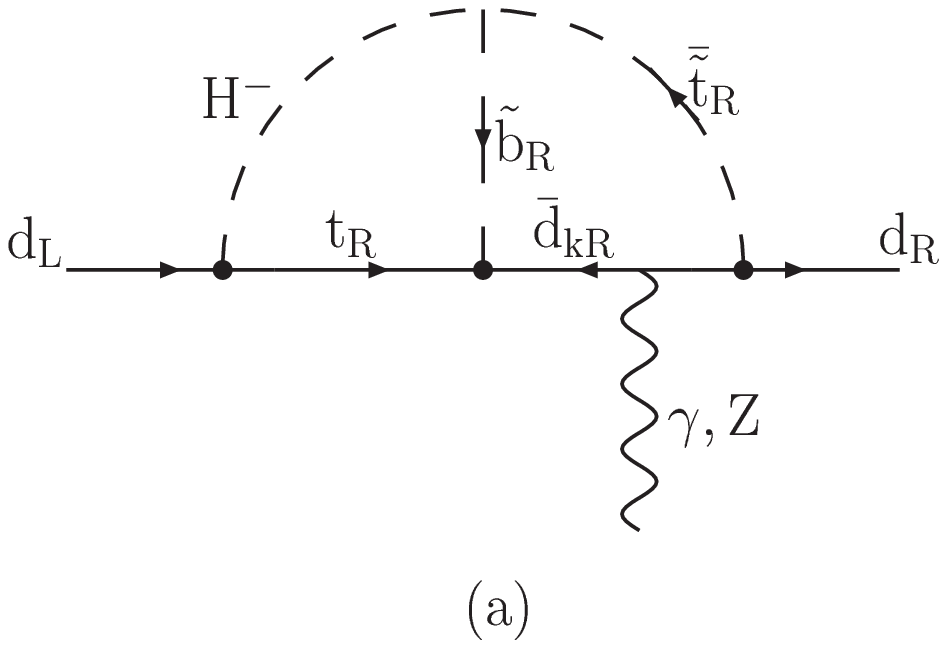,width=0.45\textwidth}   
  \hspace*{-0.8cm} \epsfig{file=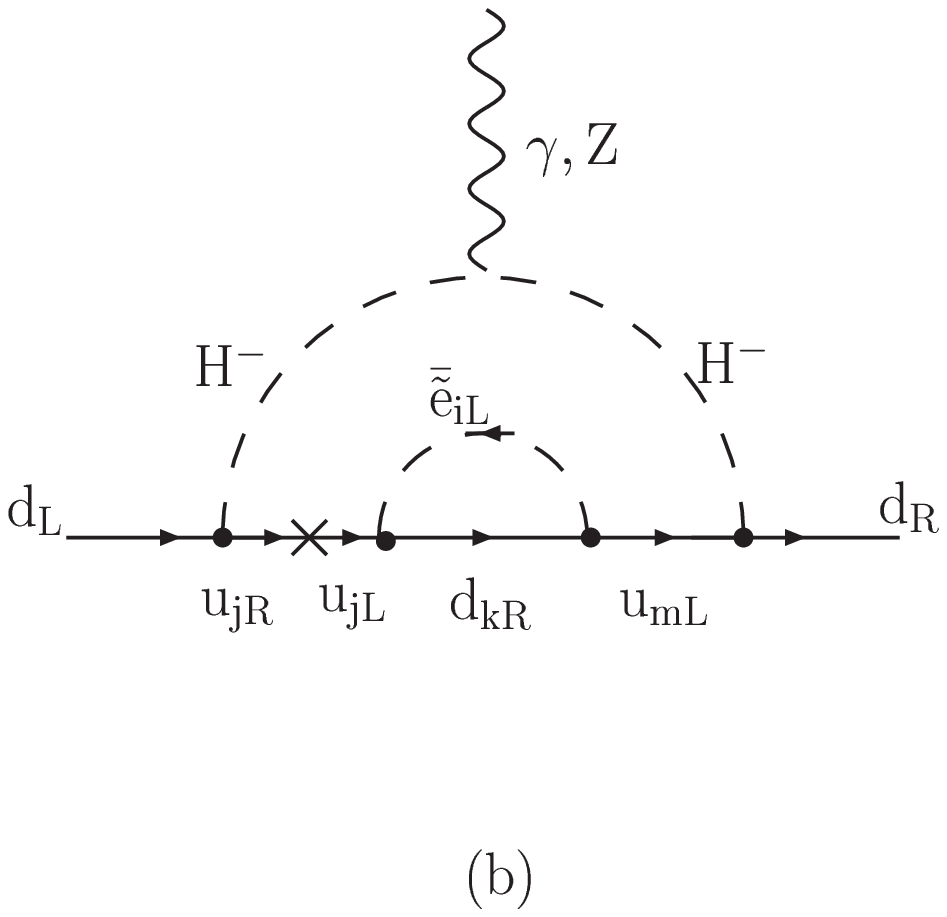,width=0.45\textwidth}  
  \end{tabular}  
  
\end{center}  
 
\vspace*{-0.2cm} 
  
\caption{{\it Examples of  two-loops diagrams contribution  
to the electric dipole moment of the electron with   
Higgs and sfermions exchanges   
$ \l ' $   
couplings (a) and $ \l $ and $ \l ''$  couplings (b). }}  
  
\label{fig:edm4et5box}  
\end{figure}  
\begin{figure}[htb]  
\begin{center}  
  \begin{tabular}{c}  
  \hspace*{-0.4cm} \epsfig{file=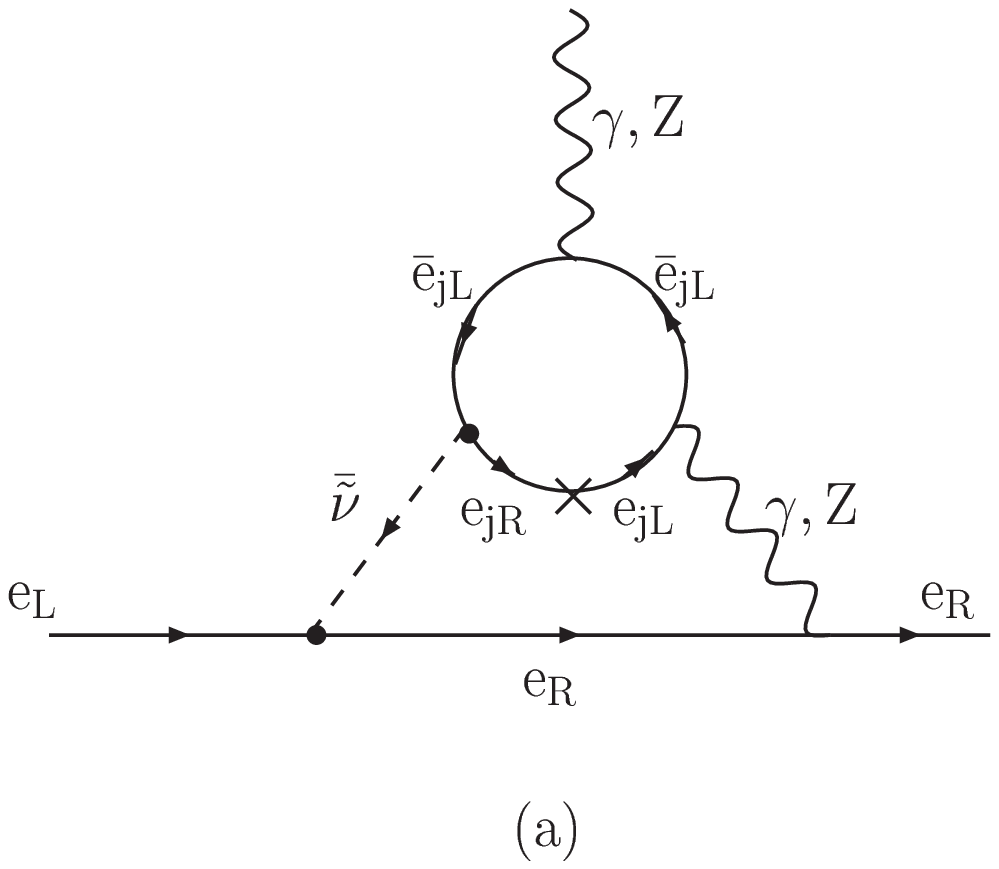,width=0.45\textwidth}   
  \hspace*{-0.8cm} \epsfig{file=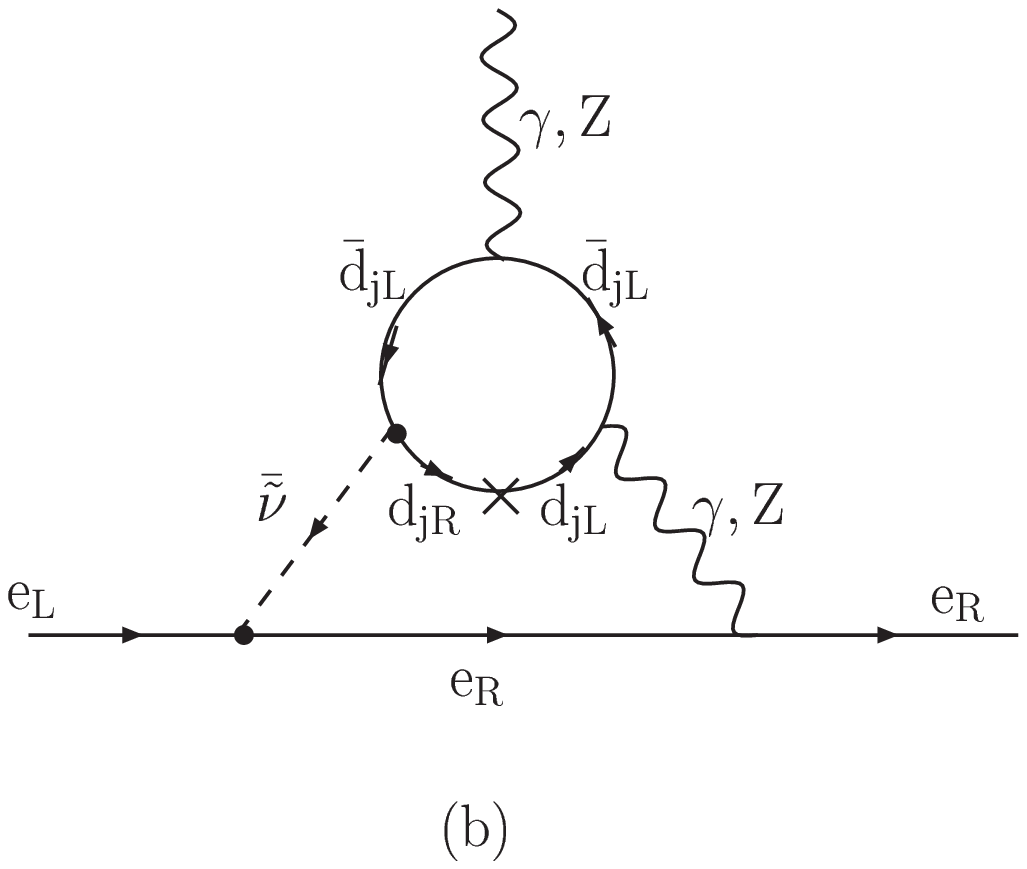,width=0.45\textwidth}  
  \end{tabular}  
  
\end{center}  
  
\vspace*{-0.2cm}  
  
\caption{{\it Examples of  two-loops diagrams contribution  
to the electric dipole moment of the electron involving $ \l $   
couplings (a) and $ \l $ and $ \l '$  couplings (b). }}  
  
\label{fig:edm1et2box}  
\end{figure}  
\begin{figure}[htb]  
\begin{center}  
\epsfig{file=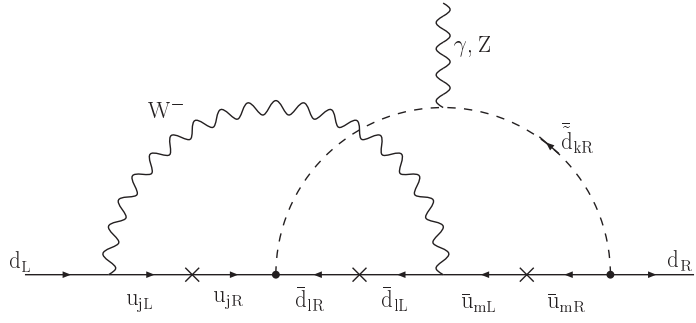,width=0.6\textwidth}   
\end{center}  
 \caption{\it{Example of two-loop diagram contribution  to the electron   
dipole moment with gauge boson and sfermion exchange  
involving  $ \l ''$ couplings.}}  
\label{fig:edm7box}  
\end{figure}  
  
As discussed in~\cite{godbole99,abel99}, at   
the two-loop order  many possible mechanisms can contribute to the  
electric dipole moment.  We have already discussed above the  
contributions from the $\l ''$ interactions.  The two-loop diagrams with  
overlapping or crossed exchanges of Higgs bosons and sfermions  
(see Fig.~\ref{fig:edm4et5box}), or of  
gauge boson and sfermions  
(see Fig.~\ref{fig:edm1et2box} and Fig.~\ref{fig:edm7box}),   
yield contributions proportional to  
quadratic products of the \Rp\ coupling constants times quadratic  
products of the CKM matrix elements, entering in appropriately rephased  
invariant flavour configurations.    
  
New contribution involving both bilinear and trilinear couplings can also lead to electron  
electric dipole moment as discussed in~\cite{choichun,keum}.  
  
\subsubsection{Atomic Electric Dipole Moment}    
\label{sec:atomedm}  
  
The electric dipole moment of atoms also present a strong potential interest~\cite{Khrip97,bern91}   
owing to the high experimental sensitivity   
that can be attained in the experimental measurements of   
the  electric dipole moment of atoms such as $^{133}$~Cs or $^{205}$~Tl.    
The \Rp\ contributions to the  
electron electric dipole moment from the other  mechanism  described  
by the two-loop diagram with one  
fermion closed loop 
leads to the~coupling~constant   
bounds~\cite{godbole99,abel99,herczeg99}~(for~J=1,2,3):  
\begin{eqnarray}  
  \vert \Im(\l_{1J1} ^\star \l '_{J33} ) \vert & < & 6.  \ \times \ 10^{-7}    
 \label{eqn:nedm}  
\end{eqnarray}  
  
These bounds should be revisited in view of new experimental results~\cite{regan}.  
  
The \Rp\ contributions from the two-loop  
diagram with two crossed sfermionic loops attached to the external  
line, yield bounds on quartic coupling constant products of the  
form~\cite{abel99}:    
\begin{equation}   
{m_l \over m_\tau } \vert \Im (\l _{1mn} \l ^\star _{jln}  
\l ^\star  _{iml}   \l _ {ij1} ) \vert < 10^{-6 } \ \       
\end{equation}  
\begin{equation}  
{m_l \over m_t} \vert \Im (\l  _{1mn} \l ^{\star } _{jln}  
\l ^{ \star } _{iml}   \l  _ {ij1} ) \vert <3. \times  10 ^  {-6}   
\end{equation}      
using experimental bounds on the electric dipole moment of both the electron  
and the neutron.  
  
These bounds on quartic coupling constant products  
should also be revisited in view of the new experimental result~\cite{regan}  
on the electric dipole moment of the electron and the discussion  
published in~\cite{golub} on the experimental value of the electric dipole moment   
of the neutron.

Finally, useful information on the $P$ and $T$ violating $ e-N$ interactions as  
parametrised by the effective Lagrangian:  
\begin{equation}    
{\cal{L}}= -{G_F\over \sqrt 2} (C_{Sp} \bar e i\g_5 e \bar pp +i C_{Tp}  
\bar e \s _{\a \b }  e \bar p  \tilde  \s ^{\a \b  }  p )     
+ (p\leftrightarrow n),  
\end{equation}  
with:  
\begin{equation}    
{\tilde {\s}}_{\a \b} = \ud \e^{\a \b \g \delta } \s^{\g \delta },  
\end{equation}     
can be obtained from the experimental limits for the electric dipole  
moment of atoms such as currently available for the $ ^{133} \text{Cs} $ or   
$  \ ^{205} \text{Tl} $ atoms~\cite{Khrip97}.    
The comparison can result for example in the bound  
\cite{herczeg99}   
$ \vert \Im(\l _{1I1} ^\star \l '_{I11} ) \vert < 1.7 \ \times \ 10^{-8}  
\tilde \nu_I ^2 $ (I=2,3).

\subsubsection{Hadronic {\boldmath{$B$}} Meson Decay Asymmetries}  
  
The formalism for $B$ meson physics as well as $C P$ violation in the 
$B$ meson system can be found in~\cite{pdg02}. 
  
The $ \Delta b=1$ non-leptonic decay transition  
amplitudes arise from tree level diagrams and, when these are  
forbidden, from one-loop box type and penguin type diagrams associated  
with the quark subprocess $ b \to d_i \bar q' q'', \ [q' , q'' = (u,  
c, d, s)]$. The relevant effective Lagrangian consists of $10$  
independent operators quartic in the quark fields.   
  
The tree level  
contribution from the \Rp\ interactions having the specific form:  
\begin{equation}   
 {\cal{L}} = \sum_i {\l'  
_{ijk} \l ^{'\star } _{ij'k'} \over m_{\tilde e_iL} ^2 } (\bar d_{kR}  
d_{jL} )(\bar d_{j'L} d_{k'R} ).  
\end{equation}  
  
Assuming the mixing decay amplitudes to consist of the two additive  
contributions from the Standard Model   and \Rp\ interactions, one can   
parametrise the off-diagonal elements of the mass matrix $M_{12}$ of 
the neutral $B$ mesons  
and the decay amplitude $A$ in terms of real parameters and complex  
phases  $r_X$  and $\t _X$ where $X$ stands either for $M$ (mixing) or $D$   
(decay) as:  
\begin{equation}    
M_{12}= M_{12} ^{ SM} (1+ r_M e^ {i\t_M}  )  
\end{equation}  
\begin{equation}   
 A = A_{SM} (1 +r_D e^{i\t _D } )   
\end{equation}    
  
Under the simplifying approximation where the final  
state $C P$-even strong interactions phase is the same for all the  
additive terms in the decay amplitudes, the ratio \mbox{${ \bar A /A =  
e^{-2i\phi_D} }$}, where $\bar A$ is the $C P$ mirror conjugate decay amplitude   
becomes a pure complex phase, so that one expresses the  
basic asymmetry parameter as:  
\begin{equation}   
 { r_{f(C P)} = e^{-2i(\phi _M + \phi_D) }}  
\end{equation}   
The \Rp\ corrections may be represented in terms of shifts in the  
mixing and decay complex phases $ \phi_X = \phi_X ^{SM} +\delta \phi  
_X $ (with X= M, D) such that:  
\begin{equation}   
  \delta \phi _X = \tan ^{-1}{ r_X \sin  
\t_X \over1 + r_X \cos \t_X }.   
\end{equation}  
For illustration, note that  in the  
Standard Model  the mixing phase for the $ B_d$ system is described by  
\mbox{${\phi_M=  
-\ud \arg(V_{tb}^\star V_{td})}$}, and the decay mode to the final state  
$ f(C P)= J/\Psi \pi ^0$ by:  
\begin{equation}    
r_{f(C P)} = \exp i[ \arg(V_{tb}^\star V_{td}) +\ \arg(V_{cs}^\star  
V_{cb}) + \arg(V_{cd}^\star V_{cs}) ] = e^{-2i\b }   
\end{equation}    
  
The decay  
channels such as $B \to K^0\bar K^0, \ \phi \pi^0 , \ \phi K_S $ are of special  
interest since their associated quark subprocesses,   
$ \bar b \to \bar d d\bar d, \  
\bar d d \bar s, \  \bar s s\bar s$, respectively, are tree level  
forbidden in the Standard Model~\footnote{See for example~\cite{pdg02,lp03} for experimental results.}.   
  
The \Rp\ contribution to the mixing parameter reads   
$ r_M \simeq  \ 10^8  
 \ \l ^{' \star } _{i13} \l '_{ i31}  
\tilde \nu ^{ -2}$~\cite{guetta}.    
Predictions for the \Rp\ contributions to the $C P$-odd asymmetry  
parameter $ r_D$ in the various decay channels are provided in  
refs.~\cite{kaplan,guetta,jang}.     
  
Using bounds on quadratic products of the $\l ' $ and $\l '' $ coupling 
constants from experimental constraints in~\cite{pdg96}, one can  
observe~\cite{kaplan} that the predicted bounds on $ r_D$  
follow different patterns for the heavy meson decay channels such as  
$ B\to J/\Psi K_S, \ B\to D^+ D^- $, in comparison to   
the light meson decay channels such  
as $ B\to  \phi K_S, \ \phi \pi ^0, K_S K_S ,$ the latter generally  
yielding more favourable signals with \mbox{${ (1+ r_D )\simeq \vert  
A_{ \, \not \hspace{-0.06cm} R_p} /R_{SM} \vert >> 1}$}.   
One can also incorporate systematically the mixing effects and  
updated values for the Wilson coefficients of the operators~\cite{jang}.    
The contributions to the asymmetry parameters $ r_D$ in the various decays  
from the $\l ' $ interactions are typically of order, \mbox{${ 10^{-3} -  
10^{-4}}$}.  By contrast, those from the $\l '' $ interactions turn out to 
arise with  a more interesting order  of magnitude i.e.  $O( 1 \  -\ 10^{-1})$.  
  
The important decay modes with the final states   
\mbox{${f= \phi K_S}$} and \mbox{${J/\Psi K_S} $} have been considered 
in~\cite{guetta}. 
The Standard Model predicts equal decay phases  
$\phi_D$ along with controlled uncertainties for the difference of  
phases~\cite{worah2}:   
\begin{equation}  
\Delta \phi_D  = \vert \phi_D  (B_d \to  
\phi K_S ) - \phi_D  (B_d \to J/\Psi K_S ) \vert < O( 10^{-1} ).   
\end{equation}    
Anticipating the possibility that the experimental errors may reach a  
sensitivity at this level of accuracy or higher, an important question is 
the expected size for the \Rp\ contributions.  
These are found to be~\cite{guetta}:  
  
\begin{eqnarray}   &&   r_D (B_d \to \phi K_S ) \simeq 8. \ \times \ 10^ 2   
\vert \l ^{ '\star }  
 _{i23} \l ' _{i22} + \l ^{ '\star }_{i22} \l '_{ i32} \vert   
({m_W \over m_ {\tilde \nu _i } })^2  \cr &&    
r_D (B_d \to J/\Psi K_S ) \simeq 2. \ \times \ 10^ 2 \vert \l^{  
 '\star }_{i23} \l ' _{i22} \vert   
({m_W \over m_ {\tilde e _{iL} } })^2  \end{eqnarray}  
Using the  current bounds on the \Rp\ coupling constants, especially those  
coming from $BR(b \rightarrow X_s \nu {\bar {\nu}})$~\cite{barate}, yields an 
encouraging prediction for the above difference of phases, 
$ \Delta \phi _D  \simeq O(1)$~\cite{guetta}.     
 
Let us finally note that analogous methods have been developed to  
extract experimental information for neutral or charged $B$ meson decays  
into non-pure $C P$ channels.  The $C P$ decay rate asymmetries are obtained by  
forming differences between the decay rates for the $C P$ conjugate  
transitions, \mbox{${B^0 \to f ,\  
\bar B^0 \to \bar f  }$}. Interesting signals from the \Rp\  
contributions are also expected for  
the charged $B$ meson decays $C P$ asymmetries in the transitions \mbox{$B^+  
\to f$} and \mbox{$B^- \to \bar f  $}~\cite{kaplan} such as, for example,   
\mbox{$B_d  
\to J/\Psi \rho ^0, \ D^\pm \pi ^\mp , K^+ \pi ^-$ or $ B_d^+ \to  
J/\Psi K^+ , \ \pi^+\pi^0 $}.   
The leptonic or semileptonic decay modes  
of the $ B$ mesons also deserve a due consideration.  
  
For the decay mode  $ B^\pm \to \pi^\pm K$, the \Rp\ induced amplitude   
$ A_{ \, \not \hspace{-0.06cm} R_p} \propto [\l '_{i23} \l ^{'\star }  
 _{i12} /\tilde m ^2] ( b \bar s) (d\bar d)$ could yield a nearly   
 100~\% contribution to the $C P$-odd asymmetry much  larger than the Standard Model    
contributions which  are  expected not to exceed a   
 40~\% effect~\cite{bhatta99}.  
       
The $C P$ violating asymmetries in the decay  and polarization observables   
of hyperon nonleptonic weak decay 
\index{Violation of $C P$!Hyperon weak decays} 
modes $ \L _b \to p + \pi ^-$ are examined in~\cite{mohanta00}. 

 %
\section{Trilinear {\boldmath{\Rp}} Interactions in Flavour Violating Processes
and in {\boldmath{\BV}} and {\boldmath{\LV}} Processes}
\label{sec:decaycons}

A very large number of bounds for the trilinear \Rp\ couplings have been
deduced from studies of low and intermediate energy processes.
In particular, rare decays involving either hadron flavour
violation or lepton flavour violation (LFV), or both combined, constitute
a nearly inexhaustible source of constraints on the trilinear \Rp\ couplings.
Processes that violate lepton number or baryon number also
provide strong constraints on trilinear $R$-parity violation.
To review the results obtained in the
current literature, we shall organise the discussion into four
subsections, where we discuss in succession hadron flavour violating
processes, lepton flavour violating processes, lepton number violating
processes and baryon number violating processes.
   
\subsection{Hadron Flavour Violating Processes}   
\label{sec:hfccons}   
   
\subsubsection{Mixing of Neutral Mesons}   
\index{Mixing!neutral mesons|(}   
The contribution from \Rp\ coupling to the mass difference and mixing 
observables  for the neutral $B\bar{B}$ meson system (i.e. $[\Delta b=2]$), 
has been considered  in~\cite{decarlosq,bhattaray} and further updated 
in~\cite{kundua,kundu}.  
 
\begin{figure}[htb] 
  \begin{center} 
     \mbox{\epsfxsize=0.9\textwidth 
       \epsffile{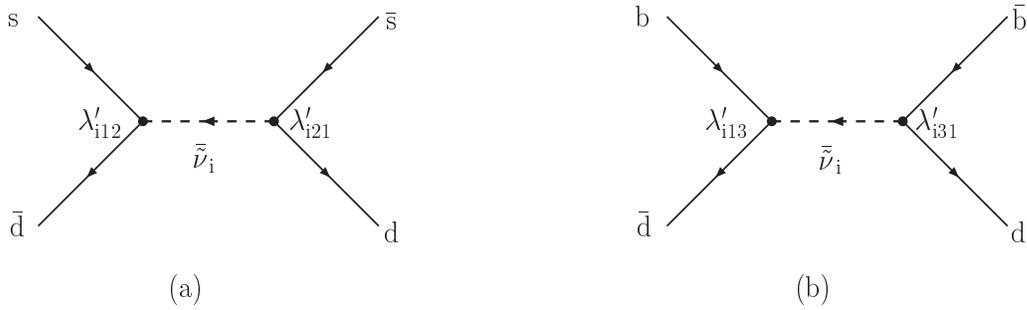}} 
  \end{center} 
 \vspace*{-0.2cm}

 \caption[]{ \label{fig:kkbar_tree} 
 {\it{ \Rp\ contributions to (a) $K \bar{K}$ and (b) $B \bar{B}$ mixing 
       involving sneutrino in the $s$-channel. 
 }}} 
\end{figure}

Sneutrino exchange can contribute to the $B \bar{B}$ mixing 
as well as $K \bar{K}$ mixing through two $\lambda'$ couplings 
via the tree level diagrams shown in Fig.~\ref{fig:kkbar_tree}a,b. 
 
%
\begin{figure}[htb] 
  \begin{center} 
     \mbox{\epsfxsize=0.5\textwidth 
       \epsffile{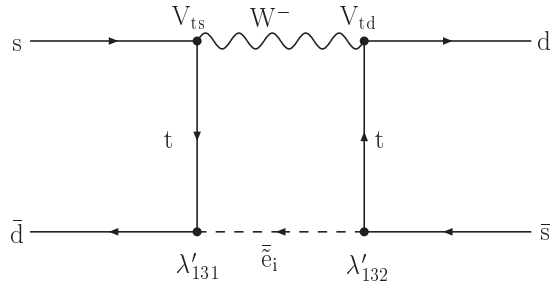}} 
  \end{center} 
 \caption[]{ \label{fig:kkbar_box} 
 {\it{ Box diagram leading to $K \bar{K}$ mixing induced 
       by $\l'$ couplings. 
 }}} 
\end{figure} 
 
Individual coupling constant bounds involving the $\l '$ interactions 
alone, based on the single coupling constant dominance hypothesis in 
the current basis for quark fields can be obtained~\cite{agashe}. 
Applying the transformation from current to the mass   
basis, one may express the transition amplitudes so that only a single \Rp\ 
coupling constant appears. The bounds deduced in~\cite{agashe} 
deserve an update in view of the experimental results published 
in~\cite{pdg02} including the results from the BABAR and BELLE collaborations. 
These bounds would however depend on the absolute mixing in the 
quark sector and would be valid if the relative mixing of the up and down 
quark sectors was entirely due to the absolute mixing 
in the down sector. However in this case, no $D\bar D$ mixing 
can be induced by a single \Rp\ coupling. 
Alternatively, if the CKM mixing comes only from the mixing 
in the up quark sector then $D \bar{D}$ mixing can provide a very 
stringent bound on $\lambda'_{ijk}$.
\index{Mixing!neutral mesons|)}
\subsubsection{Rare Leptonic Decays of Mesons}   
   
The study of rare leptonic decay modes of the $K$ and $B$ 
mesons offers distinctive probes for new physics beyond the
Standard Model \cite{litt98}.   
We shall consider in this subsection the leptonic two-body decay 
channels corresponding to final states with a charged 
lepton-antilepton pair, $M^0 \to l^-_i l^+_j$ (with
$M^0 = K^0_L$, $K^0_S$, $B^0_d$ or $B^0_s$), as well as charged
$B$ meson decays into a charged lepton and a (anti)neutrino,
$B^- \to l^- \bar \nu$.

The decays $K^0, B^0 \to l^-_i l^+_j$ arise via the underlying quark flavour
violating subprocess $ d_k + \bar d_l \to e_i +\bar e_j \ (k \neq l)$. 
In the Standard Model, the transitions that preserve lepton flavour
($i=j$), such as $K_L \to \mu^+ \mu^-$ or $B^0 \to \mu^+ \mu^-$,
arise through loop diagrams and are strongly suppressed, while
the transitions that violate lepton flavour ($i \neq j$), such as
$K_L \to \mu^+ e^-$, are unobservable due to the smallness of
neutrino masses. 
On the experimental side, the decays $K_L \to e^+ e^-$
and $K_L \to \mu^+ \mu^-$ have been measured, while only upper bounds
are available on the corresponding $K_S$ decays. In the $B$ meson
sector, the experimental upper bounds on the decays
$B^0_{d,s} \to \mu^+ \mu^-$ and $B^0_{d,s} \to e^+ e^-$ are still
several orders of magnitude above the Standard Model predictions, 
while $B^0_{d,s} \to \tau^+ \tau^-$ is yet unconstrained.

\begin{figure}[htb] 
  \begin{center} 
     \mbox{\epsfxsize=0.9\textwidth 
       \epsffile{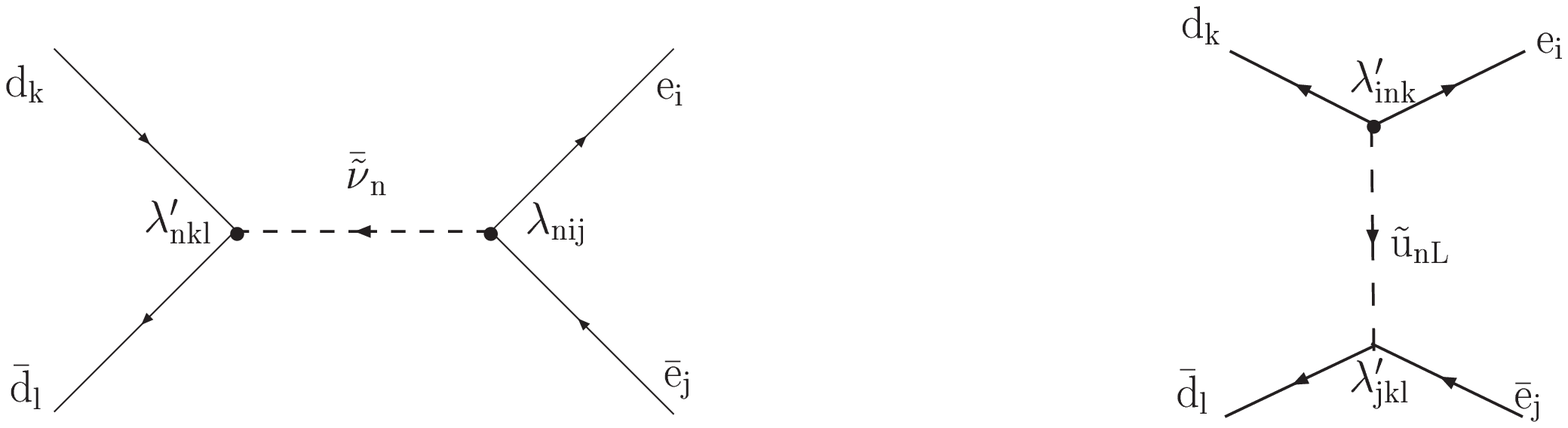}} 
  \end{center} 
 \caption[]{ \label{fig:dkdleiej} 
 {\it{ \Rp\ contributions to the process $d_k + \bar d_l \to e_i +\bar e_j$. 
 }}} 
\end{figure}

\Rp\ interactions contribute to the subprocess
$d_k + \bar d_l \to l_i +\bar l_j$ via tree-level sneutrino and up
squark exchange, as shown in Fig.~\ref{fig:dkdleiej}. This allows
to extract significant bounds on quadratic products of \Rp\ couplings
from rare leptonic meson decays. Specifically,
the decay $M^0 \to l^-_i l^+_j$, where $M^0 = d_k \bar d_l$,
constrains the following quantities:
\begin{equation}   
  A^{kl}_{ij}\ \equiv\ \sum_{n,p,q} V_{np} V^\dagger_{qn}\,
    \frac{\l^{\prime \star}_{ipk} \l'_{jql}}{m^2_{\tilde u_{nL}}}\ , \qquad
  B^{kl}_{ij}\ \equiv\ \sum_{n,p,q} U^\dagger_{np} U_{qn}\,
    \frac{\l^\star_{pij} \l'_{qkl}}{m^2_{\tilde \nu_{nL}}}\ ,
\label{eq:ABklij}
\end{equation} 
where $A^{kl}_{ij}$ and $B^{kl}_{ij}$ are associated with up squark and
sneutrino exchange, respectively.
In Eq. (\ref{eq:ABklij}), the couplings $\l_{ijk}$ and $\l'_{ijk}$ are
expressed in the mass eigenstate bases of down quarks and charged leptons,
which explains the presence of the CKM and MNS mixing angles (see
subsection~\ref{secchoice}), and the sfermion mass matrices are assumed
to be diagonal in the mass eigenstate basis of their fermion partners.
In the following, we shall further assume that the masses
of the exchanged sfermions are
degenerate, i.e. $m_{\tilde u_{nL}} \equiv m_{\tilde u_L}$ and
$m_{\tilde \nu_{nL}} \equiv m_{\tilde \nu_L}$; then Eq. (\ref{eq:ABklij})
reduces to $A^{kl}_{ij} = \frac{1}{m^2_{\tilde u_L}}
\sum_p \l^{\prime \star}_{ipk} \l'_{jpl}$ and
$B^{kl}_{ij} = \frac{1}{m^2_{\tilde \nu_L}} \sum_p \l^\star_{pij} \l'_{pkl}$.

For mesons that have wave functions of the form
$M^{kl} = (d_k \bar d_l \pm d_l \bar d_k) / \sqrt{2}$, like $K_L$
and $K_S$ in the limit where $C P$ violation is neglected,
$A^{kl}_{ij}$ and $B^{kl}_{ij}$ must be replaced by
$(A^{kl}_{ij} \pm A^{lk}_{ij}) / \sqrt{2}$ and
$(B^{kl}_{ij} \pm B^{lk}_{ij}) / \sqrt{2}$, respectively. One then defines,
for the kaon system \cite{choudhury96}:
\begin{eqnarray}   
  A^L_{ij}\ \equiv\ \frac{1}{m^2_{\tilde u_L}}
    \sum_p ( \l^{\prime \star}_{ip1} \l'_{jp2} 
    - \l^{\prime \star}_{ip2} \l'_{jp1} )\ ,  \quad
  B^L_{ij}\ \equiv\ \frac{1}{m^2_{\tilde \nu_L}}
    \sum_p \l^\star_{pij} ( \l'_{p12} - \l'_{p21} )\ ,  \\
  A^S_{ij}\ \equiv\ \frac{1}{m^2_{\tilde u_L}}
    \sum_p ( \l^{\prime \star}_{ip1} \l'_{jp2} 
    + \l^{\prime \star}_{ip2} \l'_{jp1} )\ ,  \quad
  B^S_{ij}\ \equiv\ \frac{1}{m^2_{\tilde \nu_L}}
    \sum_p \l^\star_{pij} ( \l'_{p12} + \l'_{p21} )\ ,
\end{eqnarray}   
where $A^L_{ij}$, $B^L_{ij}$ are relevant for $K_L$ decays,
and $A^S_{ij}$, $B^S_{ij}$ are relevant for $K_S$ decays.
Since $A^L_{ji} = - A^{L \star}_{ij}$, $A^L_{11}$ and $A^L_{22}$
vanish for real \Rp\ couplings. As a result, the lepton flavour
conserving decays $K_L \to e^+ e^-$ and $K_L \to \mu^+ \mu^-$
only constrain the imaginary part of the products
$\l^{\prime \star}_{ip1} \l'_{ip2}$ 
($i = 1, 2$)\footnote{This conclusion remains true when the exchanged sfermions
are not degenerate in mass, which is the case considered
in Ref.~\cite{choudhury96}. In this case however, the products
$\l^{\prime \star}_{ip1} \l'_{iq2}$ and $\l^{\prime \star}_{ip2} \l'_{iq1}$
($p \neq q$) also contribute to the decays $K_L \to l^+_i l^-_i$,
but their contribution is suppressed by CKM mixing angles. For the
imaginary part of these products, the order of magnitude of the suppression
is given by $|V_{pq}|$ (and is therefore rather mild for
$\Im (\l^{\prime \star}_{i11} \l'_{i22})$ and
$\Im (\l^{\prime \star}_{i21} \l'_{i12})$, but the latter is much more
constrained by $C P$ violation in the 
$K \bar K$ system, see
subsection~\ref{secxxx2c}),
while it can be much stronger for the real part.
This results in weaker bounds on $\Im (\l^{\prime \star}_{ip1} \l'_{iq2})$,
and especially on $\Re (\l^{\prime \star}_{ip1} \l'_{iq2})$, than on
$\Im (\l^{\prime \star}_{ip1} \l'_{ip2})$.}.
By requiring that the \Rp\ contribution itself does not exceed 
the $2 \sigma$ upper bound on the branching ratios of
$K_L \rightarrow e^+ e^-$ and $K_L \rightarrow \mu^+ \mu^-$,
measured to be $(9^{\ +\, 6}_{\ -\, 4}) \times 10^{-12}$ and
$(7.25 \pm 0.16) \times 10^{-9}$~\cite{pdg02}, respectively, one obtains
the following bounds, which update the ones given in Ref.~\cite{choudhury96}:
$\vert B^L_{11} \vert < 1.0 \times 10^{-8}\, \tilde \nu_L^2$,
$\vert B^L_{22} \vert < 2.2 \times 10^{-7}\, \tilde \nu_L^2$,
$\vert \Im (A^L_{11}) \vert < 8.1 \times 10^{-5}\, \tilde u_L^2$,
$\vert \Im (A^L_{22}) \vert < 7.8 \times 10^{-6}\, \tilde u_L^2$.
Under the double coupling dominance hypothesis, these bounds yield:
\begin{equation} 
\begin{array}{llll} 
  \vert \l^\star_{121} \l'_{212} \vert ,\
  \vert \l^\star_{121} \l'_{221} \vert 
  & < & 1.0 \times 10^{-8}\ \tilde \nu_L^2 & [K_L \to e^+ e^-]\ , \\ 
  \vert \l^\star_{131} \l'_{312} \vert ,\
  \vert  \l^\star_{131} \l'_{321} \vert 
  & < & 1.0 \times 10^{-8}\ \tilde \nu_L^2 & [K_L \to e^+ e^-]\ ,  \\ 
  \vert \l^\star_{122} \l'_{112} \vert ,\
  \vert \l^\star_{122}  \l'_{121} \vert
  & < & 2.2 \times 10^{-7}\ \tilde \nu_L^2 & [K_L \to \mu^+ \mu^-]\ ,  \\ 
  \vert \l^\star_{232} \l'_{312} \vert ,\
  \vert \l^\star_{232} \l'_{321} \vert
  & < & 2.2 \times 10^{-7}\ \tilde \nu_L^2 & [K_L \to \mu^+ \mu^-]\ ,
\end{array} 
\label{eqn:kdecays_ee_mumu_ll'} 
\end{equation} 
\begin{equation} 
\begin{array}{llll} 
  \vert  \Im (\l^{\prime \star}_{1j1} \l'_{1j2}) \vert
  & < & 8.1 \times 10^{-5}\ \tilde u_L^2 & [K_L \to e^+ e^-]\ ,  \\
  \vert  \Im (\l^{\prime \star}_{2j1} \l'_{2j2}) \vert
  & < & 7.8 \times 10^{-6}\ \tilde u_L^2 & [K_L \to \mu^+ \mu^-]\ .
\end{array} 
\label{eqn:kdecays_ee_mumu_l'l'} 
\end{equation} 
The bounds associated with the lepton flavour violating decay
$K_L \to e^\pm \mu^\mp$ have been derived in Ref.~\cite{choudhury96}
and updated in Ref.~\cite{dreiner02_bis} with the $90\%$~CL
experimental limit $B (K_L \to e^\pm \mu^\mp) < 4.7 \times 10^{-12}$
given in Ref.~\cite{pdg02}: 
\begin{eqnarray} 
\begin{array}{lll} 
  \vert \l^\star_{122} \l'_{212} \vert ,\
  \vert \l^\star_{122} \l'_{221} \vert
  & < & 6 \times 10^{-9}\ \tilde \nu_L^2\ ,  \\ 
  \vert \l^\star_{132} \l'_{312} \vert ,\
  \vert \l^\star_{132} \l'_{321} \vert 
  & < & 6 \times 10^{-9}\ \tilde \nu_L^2\ , \\ 
  \vert \l^\star_{121} \l'_{112} \vert ,\
  \vert \l^\star_{121} \l'_{121} \vert
  & < & 6 \times 10^{-9}\ \tilde \nu_L^2\ ,  \\ 
  \vert \l^\star_{231} \l'_{312} \vert ,\
  \vert \l^\star_{231} \l'_{321} \vert 
  & < & 6 \times 10^{-9}\ \tilde \nu_L^2\ , \\
  \vert  \l^{\prime \star}_{1j1} \l'_{2j2} \vert ,\
  \vert  \l^{\prime \star}_{1j2} \l'_{2j1} \vert
  & < & 3 \times 10^{-7}\ \tilde u_L^2\ .
 \end{array} 
 \label{eqn:kdecays_emu} 
\end{eqnarray} 
Significantly better bounds are obtained for $\lambda \lambda'$-type
products of couplings; the reason for that is that the contribution
of $B^L_{ij}$ to the decay amplitude is enhanced with respect to the
contribution of $A^L_{ij}$ by a factor $2 m^2_{K^0} / m_l (m_d + m_s)$,
where $m_l = m_\mu$ or $m_e$.
In updating the bounds (\ref{eqn:kdecays_ee_mumu_ll'}),
we have used the central values of the estimated ranges for
$m_s$ and $m_s / m_d$ given in Ref. \cite{pdg02}, $m_s = (80-155) \MeV$
and $m_s / m_d = (17-22)$.

The bounds derived from $K_S$ leptonic decays are less stringent
than the ones derived from $K_L$ leptonic decays, due to the weaker
experimental sensitivity to $K_S$ decays, and we do not list them here.
We just mention that, since $A^S_{ji} = A^{S \star}_{ij}$,
the decays $K_S \rightarrow e^+ e^-$ and $K_S \rightarrow \mu^+ \mu^-$
provide bounds on the real part of the products
$\l^{\prime \star}_{ip1} \l'_{ip2}$ ($i = 1, 2$), while the decays
$K_L \rightarrow l^+_i l^-_i$ only constrain their imaginary part.
Stronger bounds on $\Re (\l^{\prime \star}_{ip1} \l'_{ip2})$
can however be derived from $K \bar K$ mixing 
\index{Mixing!neutral mesons} by considering the
contribution of box diagrams with an internal $W$ boson, charged
Higgs or charged Goldstone boson~\cite{bhattaray}.

The decays $B^0 \to l^-_i l^+_j$ provide bounds on the coupling
products $\l^\star_{pij} \l'_{pkl}$, $\l^\star_{pji} \l'_{plk}$
and $\l^{\prime \star}_{ipk} \l'_{jpl}$ with $(k,l) = (1,3), (3,1)$
($B^0_d$ decays) and $(k,l) = (2,3), (3,2)$ ($B^0_s$ decays).
Since leptonic $B$ meson decays are less constrained experimentally
than leptonic kaon decays, the bounds on coupling products associated
with the former are less stringent than those associated with the latter,
Eqs. (\ref{eqn:kdecays_ee_mumu_ll'}) -- (\ref{eqn:kdecays_emu}).
Nevertheless leptonic $B$ meson decays provide the best bounds
(with some exceptions) on coupling products of the form
$\l^\star_{pij} \l'_{pkl}$, with $k = 3$ or $l = 3$.
These bounds were derived in Ref.~\cite{jangkim} and updated in
Refs~\cite{dreiner02_bis,kundu02} with the $90\%$~CL experimental limits
given in Ref.\cite{pdg02}. We list below the bounds on the $\l \l'$-type
coupling products given in Ref.~\cite{kundu02}:
\begin{equation}
\begin{array}{rlll}
  \vert \l^\star_{i11} \l'_{i13} \vert ,\
  \vert \l^\star_{i11} \l'_{i31} \vert 
  & < & 1.7 \times 10^{-5}\ \tilde \nu^2_L
  & [B^0_d \to e^+ e^-]\ ,  \\ 
  \vert \l^\star_{i22} \l'_{i13} \vert ,\
  \vert \l^\star_{i22} \l'_{i31} \vert
  & < & 1.5 \times 10^{-5}\ \tilde \nu^2_L
  & [B^0_d \to \mu^+ \mu^-]\ ,  \\
  \vert \l^\star_{i12} \l'_{i13} \vert, \
  \vert \l^\star_{i12} \l'_{i31} \vert, \
  \vert \l^\star_{i21} \l'_{i13} \vert, \
  \vert \l^\star_{i21} \l'_{i31} \vert
  & < & 2.3 \times 10^{-5}\ \tilde \nu^2_L
  & [B^0_d \to e^\pm \mu^\mp]\ ,  \\
  \vert \l^\star_{i13} \l'_{i13} \vert, \
  \vert \l^\star_{i13} \l'_{i31} \vert, \
  \vert \l^\star_{i31} \l'_{i13} \vert, \
  \vert \l^\star_{i31} \l'_{i31} \vert
  & < &  4.9 \times 10^{-4}\ \tilde \nu^2_L
  & [B^0_d \to e^\pm \tau^\mp]\ ,  \\
  \vert \l^\star_{i23} \l'_{i13} \vert, \
  \vert \l^\star_{i23} \l'_{i31} \vert, \
  \vert \l^\star_{i32} \l'_{i13} \vert, \
  \vert \l^\star_{i32} \l'_{i31} \vert
  & < &  6.2 \times 10^{-4}\ \tilde \nu^2_L
  & [B^0_d \to \mu^\pm \tau^\mp]\ ,  \\
  \vert \l^\star_{i11} \l'_{i23} \vert ,\
  \vert \l^\star_{i11} \l'_{i32} \vert 
  & < & 1.4 \times 10^{-4}\ \tilde \nu^2_L
  & [B^0_s \to e^+ e^-]\ ,  \\ 
  \vert \l^\star_{i22} \l'_{i23} \vert ,\
  \vert \l^\star_{i22} \l'_{i32} \vert
  & < & 2.7 \times 10^{-5}\ \tilde \nu^2_L
  & [B^0_s \to \mu^+ \mu^-]\ ,  \\
  \vert \l^\star_{i12} \l'_{i23} \vert, \
  \vert \l^\star_{i12} \l'_{i32} \vert, \
  \vert \l^\star_{i21} \l'_{i23} \vert, \
  \vert \l^\star_{i21} \l'_{i32} \vert
  & < & 4.7 \times 10^{-5}\ \tilde \nu^2_L
  & [B^0_s \to e^\pm \mu^\mp]\ .
\end{array} 
\label{eqn:bdecays_li_lj_ll'} 
\end{equation}
By contrast, the bounds on $\l' \l'$-type coupling products associated
with rare leptonic decays of $B$ mesons are generally weaker than the
products of bounds on individual couplings. We nevertheless list
the bounds on the coupling products $\l^{\prime \star}_{ipk} \l'_{jpl}$
given in Ref.~\cite{kundu02} (there is no significant bound associated
with the decay modes $B^0_{d,s} \to e^+ e^-$):
\begin{equation}
\begin{array}{rlll}
  \vert \l^{\prime \star}_{2j1} \l'_{2j3} \vert
  & < & 2.1 \times 10^{-3}\ \tilde u^2_L
  & [B^0_d \to \mu^+ \mu^-]\ ,  \\
  \vert \l^{\prime \star}_{1j1} \l'_{2j3} \vert, \
  \vert \l^{\prime \star}_{1j3} \l'_{2j1} \vert
  & < & 4.7 \times 10^{-3}\ \tilde u^2_L
  & [B^0_d \to e^\pm \mu^\mp]\ ,  \\
  \vert \l^{\prime \star}_{1j1} \l'_{3j3} \vert, \
  \vert \l^{\prime \star}_{1j3} \l'_{3j1} \vert
  & < &  5.9 \times 10^{-3}\ \tilde u^2_L
  & [B^0_d \to e^\pm \tau^\mp]\ ,  \\
  \vert \l^{\prime \star}_{2j1} \l'_{3j3} \vert, \
  \vert \l^{\prime \star}_{2j3} \l'_{3j1} \vert
  & < &  7.3 \times 10^{-3}\ \tilde u^2_L
  & [B^0_d \to \mu^\pm \tau^\mp]\ ,  \\
  \vert \l^{\prime \star}_{2j2} \l'_{2j3} \vert
  & < & 3.9 \times 10^{-3}\ \tilde u^2_L
  & [B^0_s \to \mu^+ \mu^-]\ ,  \\
  \vert \l^{\prime \star}_{1j2} \l'_{2j3} \vert, \
  \vert \l^{\prime \star}_{1j3} \l'_{2j2} \vert
  & < & 9.6 \times 10^{-3}\ \tilde u^2_L
  & [B^0_s \to e^\pm \mu^\mp]\ .
\end{array} 
\label{eqn:bdecays_li_lj_l'l'} 
\end{equation}
The bounds (\ref{eqn:bdecays_li_lj_ll'}) and (\ref{eqn:bdecays_li_lj_l'l'})
have been derived using $f_{B_d} = f_{B_s} = 200 \MeV$; they scale as
$(200 \MeV / f_{B_{d,s}})$. Also $m_b + m_d \approx M_{B^0_d}$ and
$m_b + m_s \approx M_{B^0_s}$ have been used for the bounds
(\ref{eqn:bdecays_li_lj_ll'}).
 
No dedicated search for the decays $B^0 \rightarrow \tau^+ \tau^-$
has yet been carried out explicitely at the existing colliders. 
However, such decays would manifest themselves at the LEP \index{LEP} 
experiments as $b \bar{b}$ events associated with large missing energy, 
due to the neutrinos emerging from the $\tau$ decays. 
Such events were studied at LEP to set constraints on 
the branching ratio for $B^{-} \rightarrow \tau \bar{\nu}$, 
and from an analysis of the same data Grossman et al. infer
the upper bounds $B (B^0_d \to \tau^+ \tau^-) < 0.015$ and
$B (B^0_s \to \tau^+ \tau^-) < 0.05$~\cite{grossman2}, four orders
of magnitude above the \SM\ predictions. These results in the
following bounds on the coupling products $\l^\star_{i33} \l'_{ikl}$,
with $k = 3$ or $l = 3$~\cite{grossman2}:
\begin{eqnarray}
  \vert \l^\star_{i33} \l'_{i13} \vert ,\
  \vert \l^\star_{i33} \l'_{i31} \vert
  & < & 6.4 \times 10^{-3}\ \tilde \nu^2_L
  \quad [B_d \to \tau^+ \tau^-]\ ,  \\ 
  \vert \l^\star_{i33} \l'_{i23} \vert ,\
  \vert \l^\star_{i33} \l'_{i32} \vert
  & < & 1.2 \times 10^{-2}\ \tilde \nu^2_L
  \quad [B_s \to \tau^+ \tau^-]\ .
\end{eqnarray}

For completeness, we also mention the bounds that have been derived from
the non-observa\-tion of
the lepton flavour violating neutral pion decay $\pi^0 \to \mu^+ e^-$
in Ref.~\cite{dreiner02_bis}, using the $90\%$ CL experimental upper limit
$B (\pi^0 \to \mu^+ e^-) < 3.8 \times 10^{-10}$ \cite{pdg02}. The following
bounds are better than previous bounds:
\begin{equation}
  \vert \l^\star_{312} \l'_{311} \vert ,\
  \vert \l^\star_{321} \l'_{311} \vert\
  <\ 3 \times 10^{-3}\ \tilde \nu_L^2\ .
\end{equation}

We now consider leptonic decays of charged $B$ mesons,
$B^- \to l^- \bar \nu$. In the \SM, these decays are suppressed by
the CKM angle $V_{ub}$ and by charged lepton masses, and the
experimental upper bounds on their branching ratios
are still well above the theoretical predictions, except for
the decay mode $B^- \to \tau^- \bar \nu$. \Rp\ interactions contribute
to these decays via similar tree-level diagrams to those of
Fig.~\ref{fig:dkdleiej}, with the exchanged sneutrino (resp. up squark)
replaced by a charged slepton (resp. down squark). Specifically, the
decay $B^- \to l^-_i \bar \nu$ constrains the following
quantities~\cite{baek99}:
\begin{equation}
  C_{ij}\ \equiv\ \sum_{n,p} V_{1p}\,
    \frac{\l^{\prime \star}_{ipn} \l'_{j3n}}{m^2_{\tilde d_{nR}}}\ , \qquad
  D_{ij}\ \equiv\ \sum_{n,p} V_{1p}\,
    \frac{\l^{\prime \star}_{np3} \l_{nji}}{m^2_{\tilde e_{nL}}}\ ,
\label{eq:CDij}
\end{equation} 
where, as in Eq. (\ref{eq:ABklij}), the couplings $\l_{ijk}$ and $\l'_{ijk}$
are expressed in the mass eigenstate bases of down quarks and charged leptons,
and the sfermion mass matrices are assumed to be diagonal. Due
to the impossibility of distinguishing experimentally the flavour of the
neutrino produced, a single decay mode $B^- \to l^-_i \bar \nu$ constrains
the six quantities $C_{ij}$ and $D_{ij}$, $j=1,2,3$. Then, from a single
constraint $|D_{ij}| < {\cal B}$, one can derive, under the double
coupling dominance hypothesis, the following set of bounds ($n=1,2,3$):
$\l^{\prime \star}_{n13} \l_{nji} < {\cal B}\, \tilde d^2_{nR}$,
$\l^{\prime \star}_{n23} \l_{nji} < ({\cal B} / V_{us})\, \tilde d^2_{nR}$ and
$\l^{\prime \star}_{n33} \l_{nji} < ({\cal B} / V_{ub})\, \tilde d^2_{nR}$.
A similar statement holds for the bounds on the coupling products
$\l^{\prime \star}_{i1n} \l'_{j3n}$, $\l^{\prime \star}_{i2n} \l'_{j3n}$,
and $\l^{\prime \star}_{i3n} \l'_{j3n}$ derived from $|C_{ij}| < {\cal B}'$.

The bounds on quadratic products of \Rp\ couplings associated with the
$90\%$~CL experimental upper limit on $B(B^- \to l^-_i \bar \nu)$
have been derived in Ref.~\cite{baek99}. Ref.~\cite{dreiner02_bis}
obtained weaker bounds from a more careful analysis based on a
conservative treatment of the experimental errors. We list below
the bounds that are not weaker than bounds associated with other processes
or products of individual bounds~\cite{dreiner02_bis}:
\begin{equation}
\begin{array}{cl}
  \vert \l^{\prime \star}_{i13} \l_{i31} \vert\
  <\ 6 \times 10^{-4}\ \tilde l^2_{iL}
  & [B^- \to e^- \bar \nu]\ ,  \\
  \vert \l^{\prime \star}_{i13} \l_{i32} \vert\
  <\ 7 \times 10^{-4}\ \tilde l^2_{iL}
  & [B^- \to \mu^- \bar \nu]\ ,  \\
  \vert \l^{\prime \star}_{313} \l_{233} \vert\
  <\ 2 \times 10^{-3}\ \tilde l^2_{3L}
  & [B^- \to \tau^- \bar \nu]\ ,  \\
  - 6 \times 10^{-4}\ \tilde l^2_{2L}\
  <\ \Re (\l^{\prime \star}_{213} \l_{233})\
  <\ 1 \times 10^{-3}\ \tilde l^2_{2L}
  & [B^- \to \tau^- \bar \nu]\ .
\end{array} 
\label{eqn:bdecays_l_nu} 
\end{equation}
The bounds on $\l' \l'$-type products associated with leptonic decays
of charged $B$ mesons are not competitive, since the contribution of
$C_{ij}$ to the decay amplitude is suppressed by a factor of
$m_{l_i} / m_{B^\pm}$ with respect to the contribution of $D_{ij}$.

\subsubsection{Rare Semileptonic Decays of Mesons} 
 
The rare semileptonic FCNC decay $K^+ \to \pi^+ \nu \bar \nu$ is often
regarded as a hallmark for tests of the \SM\ and searches for new physics.
Indeed, this process is theoretically very clean, since the hadronic
matrix element can be extracted from the well-measured decay
$K^+ \rightarrow \pi^0 e^+ \nu$, and the long-distance hadronic physics
contributions are known to be small~\cite{litt93,bigi}. The present
experimental value, $B (K^+\to \pi^+\nu \bar \nu)
= 1.6 ^{+1.8} _{-0.8} \times 10^{-10}$~\cite{pdg04}, is compatible
with the SM predictions, but has still large errors.
 
%
\begin{figure}[htb] 
  \begin{center} 
     \mbox{\epsfxsize=0.9\textwidth 
       \epsffile{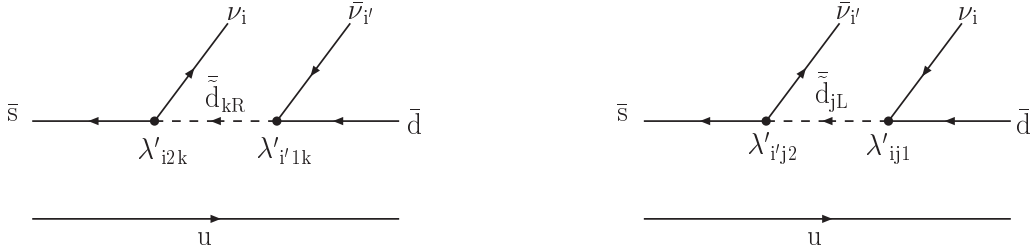}} 
  \end{center} 
 \caption[]{ \label{fig:ktopinunu} 
 {\it{ \Rp\ contributions to $K^+ \rightarrow \pi^+ \nu \bar{\nu}$. 
 }}} 
\end{figure} 
%

The \Rp\ interactions can contribute to the process $K^+\to \pi^+ \nu \bar \nu$
through the tree-level diagrams shown in Fig.~\ref{fig:ktopinunu}, which
involve a $\tilde d_{kR}$ or a $\tilde d_{kL}$ exchange.   
The dependance of the branching ratio on the \Rp\ couplings is encapsulated
in the auxiliary parameters ${\cal E }_{ii'}$~\cite{choudhury96}:
\begin{equation}   
  {\cal E}_{ii'}\ =\ \sum_k \l^{\prime \star}_{i2k} \l'_{i'1k}\
  \tilde d^{-2} _{kR} - \sum_j \l^{\prime \star}_{ij1} \l'_{i'j2}\
  \tilde d^{-2} _{jL}\ . 
\end{equation}   
In Ref.~\cite{choudhury96}, an experimental upper limit
was used to put an upper bound
on $\sum_{ii'} \vert {\cal E}_{ii'} \vert^2$, neglecting the \SM\
contribution. A warning is in order concerning this method. Since the
present experimental value $B(K^+\to \pi^+\nu \bar \nu)=(1.47\
^{+1.3}_{-0.8})\ 10^{-10}$ \cite{KPINUNUexp} is close
to the expected \SM\ value, it is no longer legitimate to neglect
the \SM\ contribution when deriving constraints on \Rp\ couplings.

Since then a detailed analysis, including all relevant contributions, was 
performed \cite{dow04}, yielding the upper bound
$\sum_{ii'} \vert {\cal E}_{ii'} \vert^2 < 4.45 \times 10^{-10}$.

One can use the bound on $\sum_{ii'} \vert {\cal E}_{ii'} \vert^2$ to
infer bounds on products of $\l'$-type couplings. Applying the double
coupling dominance hypothesis in the mass eigenstate basis, one obtains 
\cite{dow04}:
\begin{eqnarray}
\begin{array}{lll}
  \vert \l^{\prime \star}_{i2k} \l'_{i'1k} \vert
  & < &  2.11 \times 10^{-5}\ \tilde d_{kR}^2\ ,  \\
  \vert \l^{\prime \star}_{ij1} \l'_{i'j2} \vert
  & < & 2.11 \times 10^{-5}\ \tilde d_{jL}^2\ .
\end{array}
  \qquad  [K^+ \to \pi^+ \nu \bar \nu]
\label{eqn:kpinunu_bounds} 
\end{eqnarray}
Bounds on individual coupligs may also be obtained if, instead of
applying the double coupling dominance hypothesis in the mass
eigenstate basis, it is assumed that a single coupling is nonzero
in the weak eigenstate basis~\cite{agashe}, i.e. $\hat \l'_{ipq} \neq 0$
in the notation of subsection~\ref{secchoice}. Then, upon rotating
the down quarks to their mass eigenstate basis, several couplings
$\l'_{ijk} = (V^{d \dagger}_L)_{pj} (V^{d T}_R)_{qk} \hat \l'_{ipq}$
are generated, and one has $\sum_k \l^{\prime \star}_{i2k} \l'_{i'1k}
= \delta_{i'i} (V^d_L)_{2p} (V^{d \star}_L)_{1p} |\hat \l'_{ipq}|^2$.

In a similar manner, bounds on the coupling products
$\l^{\prime \star}_{i3k} \l'_{i'2k}$ and
$\l^{\prime \star}_{ij2} \l'_{i'j3}$ can be extracted from the
non-observation of the rare semileptonic $B$ meson decay
$B \to X_s \nu \bar \nu$~\cite{grossman}. We have updated the result
of Ref.~\cite{grossman} with the $90\%$~CL experimental upper limit
$B (B \to X_s \nu \bar \nu) < 7.7 \times 10^{-4}$~\cite{ALEPH96},
which lies an order of magnitude above the \SM\ prediction:
\begin{equation}
\begin{array}{lll}
  \vert \l^{\prime \star}_{i3k} \l'_{i'2k} \vert
  & < & 1.5 \times 10^{-3}\ \tilde d^2_{kR}\ ,  \\
  \vert \l^{\prime \star}_{ij2} \l'_{i'j3} \vert
  & < & 1.5 \times 10^{-3}\ \tilde d^2_{jL}\ .
\end{array}
  \qquad  [B \to X_s \nu \bar \nu]
\label{eqn:bsnunu}
\end{equation} 

Finally, the rare semileptonic decays $B \to X_s l^+_i l^-_j$ can also
be used to set bounds on $\l' \l'$-type and $\l \l'$-type coupling
products \cite{jang98}.

\subsubsection{Rare Hadronic Decays of the {\boldmath{$B$}} Mesons} 
 
The hadronic $B$ meson decays that do not proceed through a
$b \to c$ transition are suppressed in the \SM, and offer
potentially promising constraints on the \Rp\ interactions.
Unlike the rare leptonic and semileptonic decays discussed before,
however, these processes are plagued with large hadronic uncertainties
and the bounds on products of \Rp\ couplings presented below should be
considered as indicative.

In Ref.~\cite{carlson}, the decays $B^+ \to \bar K^0 K^+$ and
$B^+ \to K^0 \pi^+$ have been used to set bounds on
the products of baryon number violating couplings
$\l''_{i23} \l^{\prime \prime \star}_{i12}$ and
$\l''_{i13} \l^{\prime \prime \star}_{i12}$, which contribute
to these processes via tree-level exchange of an up squark.
We have updated the bound estimates of Ref.~\cite{carlson} by using
the $90\%$~CL experimental upper limit
$B (B^+ \to \bar K^0 K^+) < 2.4 \times 10^{-6}$~\cite{pdg02}
and by requiring that the \Rp\ contribution to $B^+ \to K^0 \pi^+$
does not exceed by more than $2 \sigma$ the measured value of
the branching ratio, $B (B^+ \to K^0 \pi^+)
= (1.73^{\, +\, 0.27}_{\, -\, 0.24}) \times 10^{-5}$~\cite{pdg02},
and found:
\begin{equation}
\begin{array}{llll}
  \vert \l''_{i23} \l^{\prime \prime \star}_{i12} \vert
  & < &  1.7 \times 10^{-3}\ \tilde u^2_{iR} & [B^+ \to \bar K^0 K^+]\ ,  \\
  \vert \l''_{i13} \l^{\prime \prime \star}_{i12} \vert
  & < &  6.4 \times 10^{-3}\ \tilde u^2_{iR} & [B^+ \to K^0 \pi^+]\ .
\end{array}
\label{eqn:btokpi} 
\end{equation}   

Ref.~\cite{chakraverty01} improves the results of Ref.~\cite{carlson}
by considering a large sample of hadronic decay modes of the $B$ mesons,
which receive contributions of the baryon number violating \Rp\
interactions through tree-level exchange of either a down squark
or an up squark. Assuming naive factorization of the hadronic matrix
elements, Ref.~\cite{chakraverty01} obtains the following allowed
ranges at $90\%$~CL (we give only the stronger constraints):
\begin{equation}
\begin{array}{cl}
%
  - 1.1 \times 10^{-3}\ \tilde d^2_{1R}\
  <\ \l''_{113} \l''_{112}\
  <\ 7.8 \times 10^{-4}\ \tilde d^2_{1R}
  & \scriptstyle{[B^0 \to \pi^0 K^{0 \star}\, B^+ \to \pi^0 K^+]}\ ,  \\
  - 1.2 \times 10^{-3}\ \tilde d^2_{2R}\
  <\ \l''_{123} \l''_{212}\
  <\ 1.4 \times 10^{-3}\ \tilde d^2_{2R}
  & \scriptstyle{[B^+ \to \pi^+ \bar D^0,\, \rho^+ \bar D^0;\,
    B^0 \to \bar D^0 \pi^0]}\ ,  \\
  - 1.4 \times 10^{-2}\ \tilde d^2_{1R}\
  <\ \l''_{213} \l''_{112}\
  <\ 2.0 \times 10^{-2}\ \tilde d^2_{1R}
  & \scriptstyle{[B^+ \to D^+_s \pi^0]}\ ,  \\
  - 7.9 \times 10^{-4}\ \tilde u^2_{iR}\
  <\ \l''_{i13} \l''_{i12}\
  <\ 1.2 \times 10^{-3}\ \tilde u^2_{iR}
  & \scriptstyle{[B^+ \to \pi^+ K^0,\, \pi^0 K^+,\, \pi^+ K^{0 \star}]}\ , \\
  - 1.9 \times 10^{-3}\ \tilde u^2_{iR}\
  <\ \l''_{i23} \l''_{i12}\
  <\ 2.8 \times 10^{-3}\ \tilde u^2_{iR}
  & \scriptstyle{[B^0 \to K^0 \bar K^0]}\ .
\end{array}
\end{equation}

Ref.~\cite{barshalom03} considers the decay mode $B^- \to \phi \pi^-$,
which, using QCD factorization, they estimate to be suppressed at the level
of $B (B^- \to \phi \pi^-) = (2.0^{\, + 0.3}_{\, -0.1}) \times 10^{-8}$
in the \SM. From the $90\%$~CL upper limit
$B (B^- \to \phi \pi^-) < 1.6 \times 10^{-6}$~\cite{pdg02}, they derive
the following upper bounds:
\begin{equation}
\begin{array}{lll}
  \vert \l^{\prime \prime \star}_{i23} \l''_{i12} \vert
  & < & 6 \times 10^{-5}\ \tilde u^2_{iR}\ ,  \\
  \vert \l'_{i32} \l^{\prime \star}_{i12} \vert
  & < & 4 \times 10^{-4}\ \tilde \nu^2_{iL}\ ,  \\
  \vert \l^{\prime \star}_{i23} \l'_{i21} \vert
  & < & 4 \times 10^{-4}\ \tilde \nu^2_{iL}\ .  \\
\end{array}
  \qquad [B^- \to \phi \pi^-]
\label{eqn:B_phi_pi}
\end{equation}

\subsubsection{FCNC Top Quark Decays} 
 
The \fcnc\ decays of the top quark, which will be best constrained
in future Tevatron experiments at Fermilab and at the CERN LHC, might
provide bounds on products of \Rp\ couplings that are not constrained
by other processes. Ref.~\cite{yang98} considered the FCNC top quark
decays $t \to c + V$ ($V = Z, \g, g$), for which an experimental
sensitivity of ($10^{-5} - 10^{-3}$) is expected, depending on the
decay mode. These decays, which are negligible in the \SM, are induced
at the one-loop level by the \Rp\ couplings $\l'_{ijk}$ and $\l''_{ijk}$;
however, given the constraints
from other processes on the $\l'_{ijk}$ couplings, only the $\l''_{ijk}$
couplings are likely to give an observable contribution.
The corresponding branching ratios are estimated to be, for 
squark masses $m_{\tilde d_{kR}} \lesssim 170 \GeV$~\cite{yang98}:
\begin{equation} 
  B(t \to c + [Z, \g, g])\ =\ [3.6 \times 10^{-5},\ 9 \times 10^{-7},\
  1.6 \times 10^{-4}]\ \ \vert \sum_{j < k}
  \l^{\prime \prime \star}_{3jk} \l''_{2jk} \vert^2\ , 
\end{equation} 
and scale as $1 / m^4_{\tilde d_{kR}}$ for larger squark masses.
Although modest, the bounds that might be inferred from the expected
experimental sensitivities are complementary with the other bounds
on the $\l''_{ijk}$ couplings discussed in this chapter.

\subsubsection{The Rare Decay {\boldmath{$b \to s \gamma$}}} 

\begin{figure}[htb] 
  \begin{center} 
     \mbox{\epsfxsize=0.6\textwidth 
       \epsffile{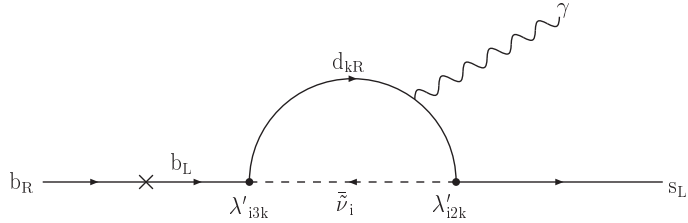}} 
  \end{center} 
 \caption[]{ \label{fig:btosgamma} 
 {\it{ Example of the \Rp\ contributions to the decay 
       $b \rightarrow s \gamma$. 
 }}} 
\end{figure} 

The measured inclusive $b \to s \gamma$ rate,
$B (B \to X_s \gamma) = (3.3 \pm 0.4) \times 10^{-4}$ \cite{pdg04},
is in good agreement with the \SM\
prediction~\cite{chetyrkin97,buras97,kagan99}. This
constrains new physics contributions, and in particular implies
restrictions on the \SSM\ spectrum and on some combinations of
\Rp\ couplings. Ref.~\cite{decarlosq} considered both the direct
contribution of the \Rp\ interactions, which can mediate
$b \to s \gamma$ through one-loop diagrams such as the one shown
in Fig.~\ref{fig:btosgamma}, and their indirect contribution through
the renormalization group evolution of the soft \SUSY\ breaking masses.
Indeed, the \Rp\ couplings enter the renormalization
group equations for the \susyq\ parameters, and can generate large
flavour violating entries in the squark mass matrices which then
induce the decay $b \to s \gamma$. This indirect contribution can
enhance the branching ratio by up to an order of magnitude with
respect to the direct contribution, but it is difficult to derive
bounds on the \Rp\ couplings from this effect due to its
complicated dependence on the \susyq\ mass spectrum. The direct
contribution yields the following upper bounds~\cite{decarlosq}:
\begin{equation}
\begin{array}{lll}
  \vert \l'_{i3k} \l^{\prime \star}_{i2k} \vert
  & < & 0.09\ (2 \tilde \nu_{iL}^{-2} - \tilde d_{iR}^{-2})^{-1}\ ,  \\
  \vert \l^{\prime \star }_{ij3} \l'_{ij2} \vert
  & < & 0.035\  (\tilde l_{iL}^{-2} - \tilde d_{jL}^{-2})^{-1}\ ,  \\
  \vert \l^{\prime \prime \star }_{i3k} \l''_{i2k} \vert
  & < & 0.16\  \tilde q_R^{2}\ . 
\end{array}
\label{eqn:btosgamma} 
\end{equation}
Since the experimental value of $B (B \to X_s \gamma)$ has
significantly changed with respect to the one given in Ref.~\cite{pdg96}
that was used in Ref.~\cite{decarlosq}, these bounds should be considered
as indicative. Using more recent data, Ref.~\cite{chakraverty01_bis}
derives a weaker bound on the coupling products
$\l^{\prime \prime \star}_{33k} \l''_{32k}$.
An update of their results,
taking into account the reduction of the experimental error, yields
the $2 \sigma$ upper bound:
\begin{equation}
  \vert \l^{\prime \prime \star }_{33k} \l''_{32k} \vert\
  < \ 0.35\  \tilde d_{iR}^{2}\ . 
\end{equation}

\subsection{Lepton Flavour Violating Processes} 
\label{secxxx3b}

In the \SM, lepton flavour violating (LFV) processes occur at a negligible
rate due to the smallness of neutrino masses. They are therefore
very sensitive probes of new physics, and can be used to place bounds
on \Rp\ couplings. In order to disentangle the effect of \Rp\ interactions
from the effect of possible flavour non-universalities in the slepton
sector, we shall assume in this subsection that the slepton mass matrices
are diagonal and proportional to the identity matrix, i.e.
$m^2_{\tilde l_{R_i}} \equiv m^2_{\tilde l_R}$,
$m^2_{\tilde l_{L_i}} \equiv m^2_{\tilde l_L}$,
$m^2_{\tilde \nu_{L_i}} \equiv m^2_{\tilde \nu_L}$.

\subsubsection{Lepton Flavour Violating Radiative Decays of Charged Leptons} 

The \Rp\ interactions can induce LFV radiative decays of charged leptons,
$l_j \to l_i + \gamma$ ($i \neq j$), through one-loop diagrams analogous
to the one shown in Fig.~\ref{fig:btosgamma}. The most constrained of
these decays, $\mu \to e \gamma$, yields the following upper bounds on
$\l \l$ and $\l' \l'$-type coupling products (the bounds given in
Refs.~\cite{decarlosl} and~\cite{chaichian} have been updated in
Ref.~\cite{gouvea01} using the $90\%$~CL experimental upper limit
$B (\mu \to e \gamma) < 1.2 \times 10^{-11}$~\cite{pdg00}):
\begin{equation}
\begin{array}{lll}
  \vert \l^\star_{ij2} \l_{ij1} \vert & < & 8.2 \times 10^{-5}\
  (2 \tilde \nu_L^{-2} - \tilde l_L^{-2})^{-1}\ ,  \\
  \vert \l_{23k} \l^\star_{13k} \vert & < & 2.3 \times 10^{-4}\
  (2 \tilde \nu_L^{-2} - \tilde l_R^{-2})^{-1}\ ,  \\
  \vert \l'_{2jk} \l^{\prime \star}_{1jk} \vert
  & < & 7.6 \times 10^{-5}\ \tilde d_{kR}^{2} \quad (j=1,2)\ .
\end{array}
\label{eqn:mutoegamma} 
\end{equation}
Due to the large top quark mass, the bound on
$\vert \l'_{23k} \l^{\prime \star}_{13k} \vert$ does not scale as
$m^2_{\tilde d_{kR}}$. Indeed, updating the bounds of
Ref.~\cite{chaichian}, one obtains:
\begin{equation}
\begin{array}{lll}
  \vert \l'_{23k} \l^{\prime \star}_{13k} \vert
  & < & 1.3 \times 10^{-3}\ ,\ 2.0 \times 10^{-3}\ ,\
  9.9 \times 10^{-3}  \quad (k=1,2)\ ,  \\
  \vert \l'_{233} \l^{\prime \star}_{133} \vert
  & < & 1.7 \times 10^{-3}\ ,\ 2.0 \times 10^{-3}\ ,\
  9.9 \times 10^{-3}\ ,
\end{array}
\label{eqn:mutoegamma_bis} 
\end{equation}
for $m_{\tilde d_{kR}} = m_{\tilde t_L} = 100 \GeV$, $300 \GeV$ and
$1 \TeV$, respectively. In Eqs.~(\ref{eqn:mutoegamma}) and
(\ref{eqn:mutoegamma_bis}), left-right mixing in the squark and
charged slepton mass matrices has been neglected.
The bounds that can be inferred from $\tau \to \mu \gamma$ and from
$\tau \to e \gamma$ are much weaker.

Ref.~\cite{decarlosl} also investigates the indirect contribution of
\Rp\ interactions to $\mu \to e \gamma$ through their effect on the
renormalization group evolution of the slepton masses. The indirect
contribution often dominates over the direct contribution discussed
above; however, due to its complicated dependence on the supersymmetric
parameters, it is not possible to derive bounds on \Rp\ couplings
from this effect.

\subsubsection {Lepton Flavour Violating Decays of {\boldmath{$\mu$}} and
                {\boldmath{$\tau$}} into three Charged Leptons} 
 
The lepton flavour violating decay 
$l^-_m \to l^-_i +l^-_j+l^+_k$, where  
$l_m = \mu$ or $\tau$, can be mediated by tree-level $t$- and $u$-channel 
sneutrino exchange when the involved leptons have nonzero 
$\lambda$-type couplings. The non-observation of these processes
yield bounds on products of \Rp\ couplings of the form
$\l_{nmi} \l^\star_{njk}$, $\l^\star_{nim} \l_{nkj}$,
$\l_{nmj} \l^\star_{nik}$ and $\l^\star_{njm} \l_{nki}$
\cite{hinchliffe,choudhury96}. We have updated the bounds of
Ref.~\cite{choudhury96} using the $90\%$~CL experimental upper
limits on $B (l^-_m\to l^-_i +l^-_j+l^+_k)$ given in
Ref.~\cite{pdg02}:
\begin{equation}
\hskip -0.3cm
\begin{array}{rll}
  \vert \l_{321} \l^\star_{311} \vert\ ,
  \vert \l^\star_{i12} \l_{i11} \vert
  & < & 6.6 \times 10^{-7}\ \tilde \nu^2_L \quad [\mu \to eee]\ ,  \\
  \vert \l_{231} \l^\star_{211} \vert\ ,
  \vert \l^\star_{i13} \l_{i11} \vert
  & < & 2.7 \times 10^{-3}\ \tilde \nu^2_L \quad [\tau \to eee]\ ,  \\
  \vert \l_{231} \l^\star_{212} \vert\ ,
  \vert \l^\star_{313} \l_{321} \vert
  & < & 2.0 \times 10^{-3}\ \tilde \nu^2_L
  \quad [\tau^- \to \mu^+ e^- e^-]\ ,  \\
  \vert \l_{232} \l^\star_{211} \vert\ ,
  \vert \l^\star_{323} \l_{311} \vert\ ,
  \vert \l_{131} \l^\star_{121} \vert\ ,
  \vert \l^\star_{i13} \l_{i12} \vert
  & < & 2.1 \times 10^{-3}\ \tilde \nu^2_L
  \quad [\tau^- \to \mu^- e^+ e^-]\ ,  \\
  \vert \l_{132} \l^\star_{121} \vert\ ,
  \vert \l^\star_{323} \l_{312} \vert
  & < & 2.0 \times 10^{-3}\ \tilde \nu^2_L
  \quad [\tau^- \to e^+ \mu^- \mu^-]\ ,  \\
  \vert \l_{131} \l^\star_{122} \vert\ ,
  \vert \l^\star_{313} \l_{322} \vert\ ,
  \vert \l_{232} \l^\star_{212} \vert\ ,
  \vert \l^\star_{i23} \l_{i21} \vert
  & < & 2.1 \times 10^{-3}\ \tilde \nu^2_L
  \quad [\tau^- \to e^- \mu^+ \mu^-]\ ,  \\
  \vert \l_{132} \l^\star_{122} \vert\ ,
  \vert \l^\star_{i23} \l_{i22} \vert
  & < & 2.2 \times 10^{-3}\ \tilde \nu^2_L \quad [\tau \to \mu\mu\mu]\ .
\end{array}
\label{eqn:l_to_three_l}
\end{equation}

The decays $l^-_m \to l^-_i +l^-_j+l^+_j$ ($k=j$) can also be induced
through photon penguin diagrams by the same $\l \l$- and $\l' \l'$-type
coupling products as the radiative decays $l_m \to l_i \gamma$.
In the case of $\mu \to eee$, the associated bounds are stronger than
the ones extracted from the non-observation of $\mu \to e \gamma$.
We list below the bounds given in Ref.~\cite{gouvea01}:
\begin{equation}
\begin{array}{lll}
  \vert \l^\star_{232} \l_{231} \vert < 4.5 \times 10^{-5}\ ,
 & \vert \l_{232} \l^\star_{132} \vert < 7.1 \times 10^{-5}\ ,
 & \vert \l_{233} \l^\star_{133} \vert < 1.2 \times 10^{-4}\ , \\
  \vert \l'_{211} \l^{\prime \star}_{111} \vert < 1.3 \times 10^{-4}\ ,
 & \vert \l'_{212} \l^{\prime \star}_{112} \vert < 1.4 \times 10^{-4}\ ,
 & \vert \l'_{213} \l^{\prime \star}_{113} \vert < 1.6 \times 10^{-4}\ , \\
  \vert \l'_{221} \l^{\prime \star}_{121} \vert < 2.0 \times 10^{-4}\ ,
 & \vert \l'_{222} \l^{\prime \star}_{122} \vert < 2.3 \times 10^{-4}\ ,
 & \vert \l'_{223} \l^{\prime \star}_{123} \vert < 2.9 \times 10^{-4}\ .
\end{array}
\label{eqn:l_to_three_l_penguin} 
\end{equation}
In Eq. (\ref{eqn:l_to_three_l_penguin}), the bounds on $\l \l$-type
(resp. $\l' \l'$-type) coupling products have been derived assuming that
all slepton (resp. squark) masses are degenerate and equal to
$\tilde m = 100 \GeV$ (resp. $\tilde m = 300 \GeV$)
and neglecting left-right mixing in sfermion mass matrices. These bounds
do not simply scale as $\tilde m^2$.

 \subsubsection{Muon to Electron Conversion in Nuclei} 

$\mu^- \rightarrow  e^-$ conversion in a nucleus can be induced
by $\l \l'$- and $\l' \l'$-type coupling products via the
exchange of a sneutrino (resp. a squark) in the $t$-channel
(resp. $s$- and $u$-channels)~\cite{kim}.
Experimentally, stringent bounds are set on the rate of 
$\mu -e$ conversion in a nucleus $A$ relative to the ordinary muon 
capture, $R_{\mu e} \equiv \G (\mu^- + A \to e^- + A) /
\G (\mu^- \mbox{capture in}\, A)$. Using the $90\%$~CL upper limit
$R_{\mu e} < 6.1 \times 10^{-13}$ obtained by the SINDRUM II experiment
on a $^{48} \rm{Ti}$ target~\cite{sindrumii}, the following bounds
are deduced, updating those given in Ref.~\cite{kim}:
\begin{equation}
\begin{array}{rll}
  \vert \l^\star_{i12} \l'_{i11} \vert\ ,
  \vert \l_{i21} \l^{\prime \star}_{i11} \vert
  & < & 2.1 \times 10^{-8}\ \tilde \nu^2_L\ ,  \\
  \vert \l'_{2j1} \l^{\prime \star}_{1j1} \vert
  & < & 4.3 \times 10^{-8}\ \tilde u^2_{jL} \quad (j=2,3)\ ,  \\
  \vert \l'_{21k} \l^{\prime \star}_{11k} \vert
  & < & 4.5 \times 10^{-8}\ \tilde d^2_{kR} \quad (k=2,3)\ .
\end{array}
\label{eqn:mutoe}
\end{equation}
The combination $\l'_{211} \l^{\prime \star}_{111}$ can also induce
$\mu -e$ conversion in $^{48} \rm{Ti}$, but cancellations may occur
between the up squark and the down squark contributions, resulting
in a weaker bound, $\vert \l'_{211} \l^{\prime \star}_{111} \vert
< 4.3 \times 10^{-8}\, (\tilde u^{-2}_L
- \frac{70}{74} \tilde d^{-2}_R)^{-1}$.

\Rp-induced $\mu^- \rightarrow  e^-$ conversion in a nucleus can also
proceed through photon penguin diagrams~\cite{huitu98}, in the same
way as the LFV decay $\mu \to eee$. The associated bounds are stronger
than the ones extracted from the non-observation of $\mu \to e \gamma$
and $\mu \to eee$, if the latter does not occur at tree level.
We list below the bounds given in Ref.~\cite{gouvea01}:
\begin{equation}
\begin{array}{lll}
 \vert \l^\star_{122} \l_{121} \vert < 6.1 \times 10^{-6}\ ,
 & \vert \l^\star_{132} \l_{131} \vert < 7.6 \times 10^{-6}\ ,
 & \vert \l^\star_{232} \l_{231} \vert < 8.3 \times 10^{-6}\ , \\
  \vert \l_{231} \l^\star_{131} \vert < 1.1 \times 10^{-5}\ ,
 & \vert \l_{232} \l^\star_{132} \vert < 1.3 \times 10^{-5}\ ,
 & \vert \l_{233} \l^\star_{133} \vert < 2.3 \times 10^{-5}\ , \\
%
%
   \vert \l'_{222} \l^{\prime \star}_{122} \vert < 4.3 \times 10^{-5}\ ,
 & \vert \l'_{223} \l^{\prime \star}_{123} \vert < 5.4 \times 10^{-5}\ .
\end{array}
\label{eqn:mutoe_penguin}
\end{equation}
In Eq. (\ref{eqn:mutoe_penguin}), the bounds on $\l \l$-type
(resp. $\l' \l'$-type) coupling products have been derived assuming that
all slepton (resp. squark) masses are degenerate and equal to
$\tilde m = 100 \GeV$ (resp. $\tilde m = 300 \GeV$)
and neglecting left-right mixing in sfermion mass matrices. These bounds
do not simply scale as $\tilde m^2$.

\subsubsection{Muonium to Antimuonium Conversion} 

\begin{figure}[htb] 
  \begin{center} 
   \begin{tabular}{cc} 
     \mbox{\epsfxsize=0.45\textwidth 
       \epsffile{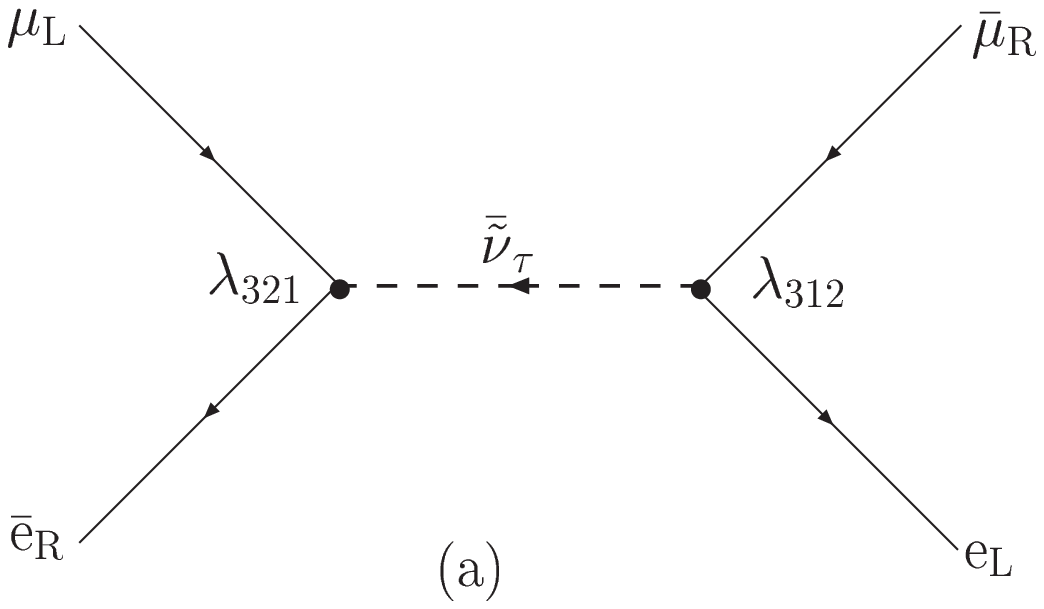}} 
    & \mbox{\epsfxsize=0.30\textwidth 
       \epsffile{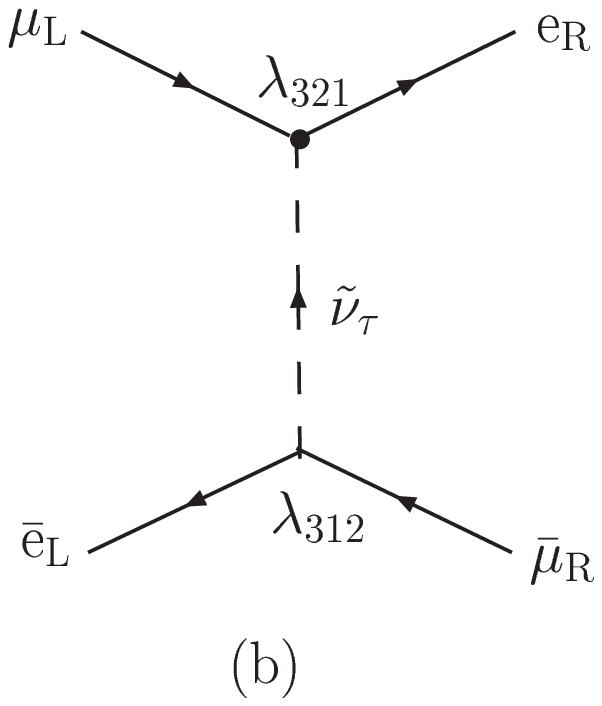}} 
    \end{tabular} 
  \end{center} 
 \caption[]{ \label{fig:mueconv} 
 {\it{\Rp\ contributions to muonium-antimuonium conversion. 
      Similar diagrams involving an incoming $\mu_R$ are not shown. 
 }}} 
\end{figure}

The conversion reaction of a muonium atom into an 
antimuonium atom, $M (\mu^+e^-)\to \bar M( \mu^-e^+)$,  
has been initially proposed as a test of a multiplicative lepton number 
symmetry~\cite{feinberg} which would forbid $\Delta L_{\mu} = \pm 1$ 
transitions, but would allow for $\Delta L_{\mu} = \pm 2$ transitions
such as $M \rightarrow \bar{M}$. Experimental limits on this
process are conventionally expressed in terms of an effective coupling
$G_{M \bar M}$ defined by~\cite{feinberg0}:
\begin{equation} 
  {\cal L}_{eff} (M \to \bar M)\ =\ \frac{4G_{M \bar M}}{\sqrt 2}\, 
 (\bar \mu_L \g ^\mu e_L ) (\bar \mu_L \g_\mu e_L)\ +\ \mbox{h.c.}\ .
\label{eq:L_M_Mbar}
\end{equation} 
The current $90\%$~CL experimental limit~\cite{M_Mbar98} is
$G_{M \bar M} < 3.0 \times 10^{-3}\, G_F$.
In the presence of $R$-parity violation, muonium to antimuonium
conversion can be mediated by tree-level exchange of a tau sneutrino
in the $s$- or the $u$-channel as shown in Fig.~\ref{fig:mueconv}.  
%
The associated effective interaction is of the $(V-A)(V+A)$ form (after
a Fierz transformation) and is described by an effective coupling
$\tilde G_{M \bar M}$ distinct from the coupling $G_{M \bar M}$ defined
by Eq. (\ref{eq:L_M_Mbar}). Explicitly, one has~\cite{kim,halprin,moha92}:
\begin{equation} 
  \frac{\tilde G_{M \bar M}}{\sqrt 2}\ =\ 
  \frac{\l_{312} \l^\star_{321}}{8 m^2_{\tilde \nu_L}}\ .
\end{equation}
The $90\%$~CL experimental limit on muonium-antimuonium
conversion~\cite{M_Mbar98}, in the case of a $(V-A)(V+A)$ interaction,
translates into $\tilde G_{M \bar M} < 2.0 \times 10^{-3}\, G_F$. This
yields the following upper bound on the coupling product
$\l_{312} \l^\star_{321}$, which updates the bound given in
Ref.~\cite{kim}:
\begin{equation}
  \vert \l_{312} \l^\star_{321} \vert\
  <\ 1.9 \times 10^{-3}\ \tilde \nu^2_L\ .
\label{eqn:muonium} 
\end{equation}

\subsubsection{Lepton Flavour Violating Semileptonic Decays of the 
{\boldmath{$\tau$}}} 
 
The $\tau-$lepton decay modes include a variety of lepton flavour 
violating processes which yield constraints on several products of
\Rp\ couplings. Of special interest are the two-body decay modes
into pseudoscalar and vector mesons, $\tau \to l + P^0$ and
$\tau \to l + V^0$, with $l = e, \mu$, $P=\pi^0, \eta , K^0$ and
$V=\rho^0, \omega, K^{\star 0}, \phi$. 
The \Rp\ interactions contribute to these processes via tree-level  
sneutrino or squark exchange, induced by $\l \l'$-type or $\l' \l'$-type
coupling pairs, respectively~\cite{kim}. For sneutrinos or up-type
squarks, the corresponding diagrams are the time-reversed of the
diagrams shown in Fig.~\ref{fig:dkdleiej}; the exchange of a down-type
squark, which is not shown, corresponds to the subprocess
$e_i + \bar e_j \to u_k + \bar u_l$.

The sneutrino exchange mediates tau decays into pseudoscalar mesons only;
hence the $\l \l'$-type coupling products are only constrained
by these decays. Using the $90\%$~CL experimental upper limits on
$B (\tau \to l + P^0)$ given in Ref.~\cite{pdg02}, one obtains
the following bounds, which update the bounds of Ref.~\cite{kim}:
\begin{equation}
\begin{array}{llll}
  \vert \l_{i31} \l^{\prime \star}_{i11} \vert\ ,
  \vert \l^\star_{i13} \l'_{i11} \vert
  & < & 1.6 \times 10^{-3}\ \tilde \nu_{iL}^2 
  & \left[\tau^- \to e^- + \eta^0 \right] ,  \\
  \vert \l_{i31} \l^{\prime \star}_{i22} \vert\ ,
  \vert \l^\star_{i13} \l'_{i22} \vert
  & < & 1.6 \times 10^{-2}\ \tilde \nu_{iL}^2 
  & \left[\tau^- \to e^- + \eta^0 \right] , \\
  \vert \l_{i31} \l^{\prime \star}_{i12} \vert\ ,
  \vert \l^\star_{i13} \l'_{i21} \vert
  & < & 8.5 \times 10^{-2}\ \tilde \nu_{iL}^2  
  & \left[\tau^- \to e^- + K^0 \right] ,  \\
  \vert \l_{i32} \l^{\prime \star}_{i11} \vert\ ,
  \vert \l^\star_{i23} \l'_{i11} \vert
  & < & 1.7 \times 10^{-3}\ \tilde \nu_{iL}^2 
  & \left[\tau^- \to \mu^- + \eta^0 \right] ,  \\
  \vert \l_{i32} \l^{\prime \star}_{i22} \vert\ ,
  \vert \l^\star_{i23} \l'_{i22} \vert
  & < & 1.7 \times 10^{-2}\ \tilde \nu_{iL}^2 
  & \left[\tau^- \to \mu^- + \eta^0 \right] ,  \\
  \vert \l_{i32} \l^{\prime \star}_{i12} \vert\ ,
  \vert \l^\star_{i23} \l'_{i21} \vert
  & < & 7.6 \times 10^{-2}\ \tilde \nu_{iL}^2  
  & \left[\tau^- \to \mu^- + K^0 \right] .
\end{array} 
\label{eqn:lfvtau_meson_ll'} 
\end{equation} 

$\l' \l'$-type coupling products induce both decays $\tau \to l + P^0$
and $\tau \to l + V^0$, but the latter are more constrained
experimentally and therefore provide stronger bounds than the former.
Using the $90\%$~CL experimental upper limits on
$B (\tau \to l + V^0)$ given in Ref.~\cite{pdg02}, one obtains
the following bounds, which update the bounds of Ref.~\cite{kim}:
\begin{equation}
\begin{array}{llll}
  \vert \l'_{3j1} \l^{\prime \star}_{1j1} \vert
  & < & 2.4 \times 10^{-3}\ \tilde u_{jL}^2
  & \left[\tau^- \to e^- + \rho^0 \right] ,  \\
  \vert \l'_{3j1} \l^{\prime \star}_{1j2} \vert
  & < & 2.7 \times 10^{-3}\ \tilde u_{jL}^2
  & \left[\tau^- \to e^- + K^{\star 0} \right] ,  \\
  \vert \l'_{3j1} \l^{\prime \star}_{2j1} \vert
  & < & 4.4 \times 10^{-3}\ \tilde u_{jL}^2
  & \left[\tau^- \to \mu^- + \rho^0 \right] ,  \\
  \vert \l'_{3j1} \l^{\prime \star}_{2j2} \vert
  & < & 3.4 \times 10^{-3}\ \tilde u_{jL}^2
  & \left[\tau^- \to \mu^- + K^{\star 0} \right] ,  \\
 \vert \l'_{31k} \l^{\prime \star}_{11k} \vert
  & < & 2.4 \times 10^{-3}\ \tilde d_{kR}^2
  & \left[\tau^- \to e^- + \rho^0 \right] ,  \\
  \vert \l'_{31k} \l^{\prime \star}_{21k} \vert
  & < & 4.4 \times 10^{-3}\ \tilde d_{kR}^2
  & \left[\tau^- \to \mu^- + \rho^0 \right] .
\end{array} 
\label{eqn:lfvtau_meson_l'l'} 
\end{equation} 

\subsection{Lepton Number Non-Conserving Processes } 
\label{subsec:L_number} 
 
\subsubsection{Neutrinoless Double Beta Decay} 
\index{Neutrinoless double beta decay|(} 
Searches for neutrinoless double beta decay  
($\b \b_{0 \nu}$) of nuclei 
($(Z,N) \to (Z+2,N-2)+l^-_i+l^-_j$) are performed 
using 
$^ {76} {\rm{Ge}},\ ^ {48} {\rm{Ca}}, \ ^ {82} {\rm{Se}}, \ ^ {100} {\rm{Mo}}$. 
They are mainly carried out in underground laboratories 
and make use of various detection techniques. 
The current experimental information and some of the 
promising future prospects are reviewed in Ref.~\cite{expbb}.

\begin{figure}[htb] 

\vspace*{0.2cm} 
 
  \begin{center} 
     \mbox{\epsfxsize=0.9\textwidth 
       \epsffile{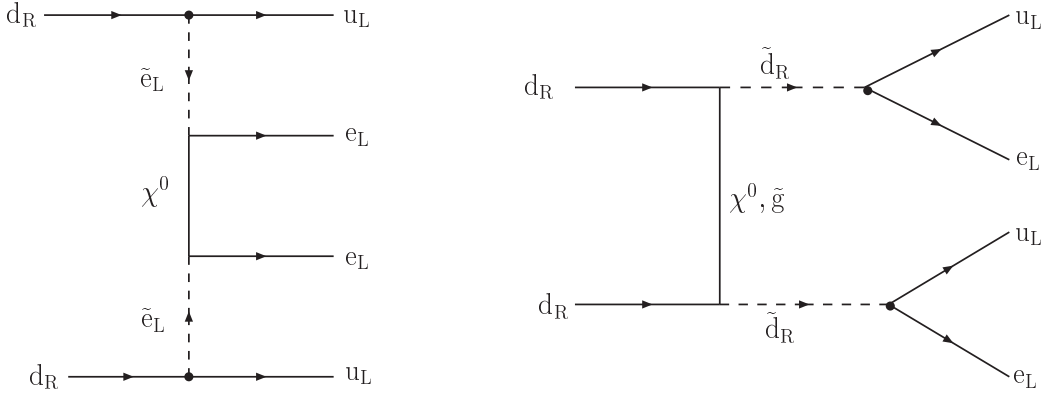}} 
  \end{center} 
 \caption[]{ \label{fig:betadecay} 
 {\it{ Contributions to the neutrinoless double beta decay 
       induced by a $\lambda'_{111}$ coupling. 
 }}} 
\end{figure}

The nucleon level transition, $ n +n \to p+p + e^- + e^-$ is 
induced at the quark level by the subprocess $ d+d \to u +u +e+e$.  
The \Rp\ operator {$ L_1 Q_1 D^c_1$} 
would allow such a transition to occur at the tree level, 
via processes involving  
the sequential $t$-channel exchange of two sfermions and a gaugino, 
where the sfermion may be a slepton or a squark, $\tilde e_L $ or $ 
\tilde u_{L}, \tilde d_{R} $, and the gaugino, a neutralino or a 
gluino~\cite{mohapatra86,hirsch}.  
Corresponding diagrams are shown in Fig.~\ref{fig:betadecay}. 
%
%
In the limit of large masses for the exchanged sparticles, 
these mechanisms can be described in terms of point-like six fermion 
effective interactions.
Relying on such an effective Lagrangian and using 
an approximate evaluation of the nuclear operator matrix element, an 
early  study by Mohapatra   
led to the bounds~\cite{mohapatra86}: 
$|\l '_{111}| <0.48\ \times \ 10^{-9/4} \tilde f^2 \tilde g ^\ud$,
$\vert \l '_{111} \vert <   2.8   \ \times \ 10^{-9/4} \tilde f^2 \tchi ^\ud $. 
Meanwhile, detailed calculations of the $\b \b_{0 \nu}$  amplitudes 
have been performed including all contributing 
diagrams, and the relevant nuclear matrix elements have 
been calculated in the proton-neutron Quasiparticle 
Random Phase Approximation (QRPA)~\cite{hirsch}. 
>From the lower limit on the half-life of $^{76} {\rm{Ge}}$ measured 
by the Heidelberg-Moscow experiment~\cite{heidelberg}: 
$$ T^{\b \b_{0 \nu}}_{1/2} (^{76} {\rm{Ge}})  
   > 1.1 \times \ 10^{25} {\rm{yr}} $$ 
the following bound is obtained in the minimal  
supergravity framework~\cite{hirsch1}: 
\begin{eqnarray} 
 \vert \l'_{111} \vert & < & 3.3\times \ 10^{-4}\ {\tilde q} ^2 {\tilde g}^{\ud }. 
  \label{eqn:betabeta_bounds} 
\end{eqnarray}  
Since the upper bound on $\l'_{111}$ scales with  
$(T^{\b \b_{0 \nu}}_{1/2})_{lim}^{-1/4}$, most recent bounds 
on the half-life of $^{76} {\rm{Ge}}$ do not significantly improve 
the above result. 
Slightly more serere bounds on $\l'_{111}$ have been 
obtained in Ref.~\cite{pionexchange} by including  
the pion-exchange contributions to the \Rp\ induced 
$\b \b_{0 \nu}$ decay. 
   
\begin{figure}[htb] 
  \begin{center} 
     \mbox{\epsfxsize=0.55\textwidth 
       \epsffile{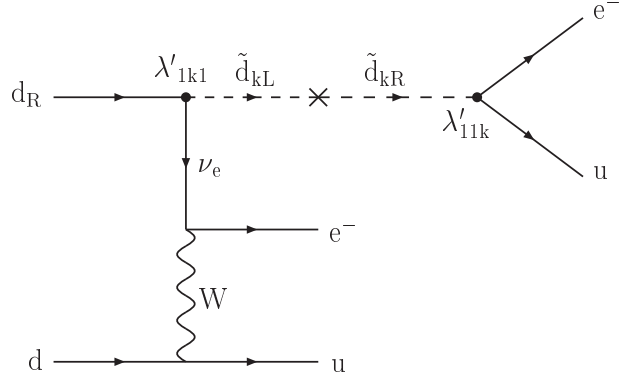}} 
  \end{center} 
 \caption[]{ \label{fig:babu} 
 {\it{ Contributions to the neutrinoless double beta decay 
       induced by squark and $W$ exchanges. 
 }}} 
\end{figure}

Babu and Mohapatra~\cite{babu} identified another \Rp\ 
contribution to $\b \b_{0 \nu}$, based on the 
$t$-channel scalar-vector type exchange of a sfermion and a charged 
$W$ boson linked together through an intermediate internal  
neutrino exchange. 
The corresponding diagram is shown in Fig.~\ref{fig:babu}. 
%
The amplitude for this process is closely related to that of the 
familiar SM neutrino exchange, except for the 
important fact that no chirality flip is required for the intermediate 
internal neutrino line  propagation. 
The strong suppression factor arising  within the Standard Model 
contribution  from the 
neutrino propagator factor is replaced as, ${ m_\nu / q^2 } \to { 
1 / g \cdot q } = { \g 
\cdot q / q^2 }$, where $q$ is the  intermediate neutrino four momentum.   
The chirality flip  penalty  is  
transferred instead to the exchanged down-squark, as seen in 
Fig.~\ref{fig:babu}. 
The contribution shown in Fig.~\ref{fig:babu} thus disappears 
in case of a vanishing mixing between $\tilde{d}_{kR}$ and 
$\tilde{d}_{kL}$. 
The bound on $(T^{\b \b_{0 \nu}}_{1/2})$ leads then
to upper limits on the products $\lambda'_{1k1} \l'_{11k}$ 
which scale with $m^4 _{\tilde{d}_{kR} } \ / \ (A_k - \mu \tan \beta)$, 
the denominator determining the left-right mixing in the $\tilde{d}_k$ 
sector. The resulting bounds for  the third, second and first generations,
as quoted in Ref.~\cite{hirsch_babu}, read: 
\begin{eqnarray} 
 \begin{array}{llll} 
\vert  \l'_{113}\l'_{131} \vert & < &  3.8 \times 10^{-8} 
  &  \left(  \tilde m /  100 \GeV \right)^3 , \\ 
\vert  \l'_{112}\l'_{121}\vert  & < & 1.1 \times 10^{-6} 
  &  \left(  \tilde m /  100 \GeV \right)^3, 
 \end{array} 
\label{eqn:babu_bounds} 
\end{eqnarray} 
assuming the input values for the down squark mass parameters  
$m_{\tilde{d}_{kR}} \simeq (A_k - \mu \tan \beta) \equiv \tilde m$. 
   
A systematic discussion including both the bilinear and trilinear \Rp\ 
interactions has been given by Faessler et al. in Ref.~\cite{faessler98}. 
It includes a detailed study of the validity of the 
different approximation schemes in the determination of the 
relevant nuclear matrix elements.
The bilinear \Rp\ terms give rise to several contributions to $\b \b_{0 \nu}$,
either alone or in combination with trilinear \Rp\ interactions.
The dominant contribution turns out to be the neutrino exchange
diagram controlled by the effective neutrino mass parameter
induced by bilinear $R$-parity violation,
$<m_\nu> \equiv \sum_i m_{\nu_i} U^2_{ei}\ \propto\ (v_1 \mu -v_d \mu _1)^2$.
The comparison of their  predicted results with the
experimental limit  for $\b \b_{0 \nu}$  yields the bounds (assuming a common
superpartner mass parameter $\tilde m = 100 \GeV$ and $\tan \beta = 1$):  
\begin{eqnarray}  
  |\mu_1|\ < 470 \keV\ , & & |\mu_1 \l'_{111}|\ <\ 100 \eV\ ,  \label{eq:faessler_mu1}  \\
  |v_1|\  <\ 840 \keV\ , & & |v_1 \l'_{111}|\ <\ 55  \eV\ .  \label{eq:faessler_v1}
\end{eqnarray}  
Strictly speaking, the bounds (\ref{eq:faessler_mu1})
(resp. the bounds (\ref{eq:faessler_v1})) apply in a ($H_d$, $L_i$) basis
in which $v_i \equiv\, < \tilde \nu_i >\, = 0$ (resp. $\mu_i = 0$).
 
In Ref.~\cite{hirsch99}, Hirsch studied the contribution of bilinear
$R$-parity violation to $\b \b_{0 \nu}$ both at the tree-level and at the
one-loop level. The consideration of the tree-level contribution led
him to exclude values for $\mu_1$ or $ v_1$ in the interval
${\cal O} (0.1)   -   {\cal O} (1) \MeV$, in agreement with
Faessler et al.~\cite{faessler98}.
He further observed that even in the case of a perfect alignment
between the  $\mu_1$ and $v_1$ parameters (such that
$v_1 \mu -v_d \mu _1 = 0$, hence bilinear \Rp\ violation does not contribute
to $ <m_\nu > $ at the tree level), there can  occur finite contributions to
$<m_\nu >$ arising at  the  one-loop level, which leads to the bound
$\vert \mu_1/\mu \vert < 0.01$.

A lepton number violating process which is closely related to  
the $\b \b_{0 \nu }$  reaction concerns the $ \mu^+ \to e^- $ conversion   
reaction 
taking place in atomic nuclei via the atomic orbit capture reaction of 
muons~\cite{palmoha}, $\mu^+ +(Z,N) \to e^- +(Z+2, N-2) $. The 
numerical result for the predicted branching fraction,  
$ (B(\mu ^- \to e^+)/ 10^{-12}) \simeq  \vert 
\l'_{213} \l '_{131} \vert /  2.3 \ \times \ 10^{-2} $,  as represented 
by scaling with respect to the current ${\cal O} (10^{-12})$  
experimental sensitivity,  
indicates the extent to which future improvement in the 
measurements of $ \mu^+ \to e^- $ conversion 
could bring useful information on the \Rp\ interactions. 
\index{Neutrinoless double beta decay|)} 

\subsection{Baryon Number Non-Conserving Processes} 
\label{secxxx3d}   
 
\subsubsection{Single Nucleon Decay} 
 
\begin{figure}[htb] 
  \begin{center} 
     \mbox{\epsfxsize=0.55\textwidth 
       \epsffile{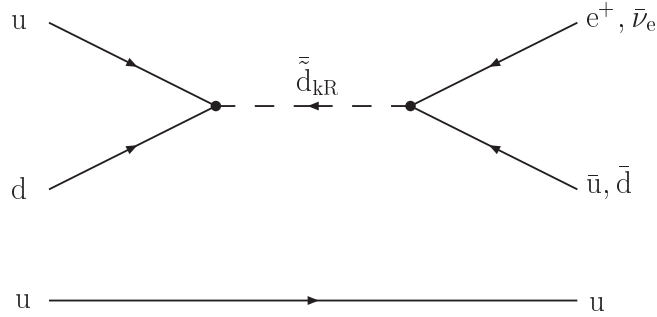}} 
  \end{center} 
%
 \caption[]{ \label{fig:pdecay} 
 {\it{ \Rp\ contributions to the proton decay. 
 }}} 
\end{figure}

Matter instability, as would be  implied by a  non-conservation
of baryon number, is a well-documented subject thanks
to the extensive research developed in  connection with
\GUT~\cite{weinberg82,langacker,masiero1,weinbergs,weinberg80}. 
The combined contributions of the $\l ' $ and $ \l ''$ interactions 
lead to an effective interaction of the form
$ {\cal L} = ( \l '' \l ^{\prime \star } / 
m^2_{\tilde d_{R}} ) [(u^cd^c)^\dagger (\nu d) -(u^cd^c) ^\dagger (eu)] 
+ \ h. \ c. $, which we have written using a two-component Weyl 
spinor representation  for the fermion fields. 
This effective interaction is obtained by contracting a pair of down-squark fields
as $(d^c)^\dagger d^c $, and therefore yields a $(B-L)$-conserving amplitude.
This is illustrated by the the tree-level $\tilde d_{kR}$ squark  $s$-channel exchange 
diagrams shown in Fig.~\ref{fig:pdecay}. 
%
The comparison with the experimental limits on nucleon 
decays yields extremely severe bounds on the \Rp\ coupling
products $\l'_{imk} \l^{\prime \prime \star}_{11k}$ ($i,k=1,2,3$, $m=1,2$). Adapting
to the \Rp\ case the computation of the proton decay rate
from dimension-6 operators done in the context of \GUT\
in Ref.~\cite{hisano00}, and using the experimental lower
bounds on partial nucleon lifetimes given in Ref.~\cite{pdg04},
we obtain:
\begin{equation} 
\begin{array}{llll} 
\vert  \l '_{l1k} \l ^{\prime \prime \star }_{11k} \vert & \lesssim &  
 (2-3) \times 10^{-27}\ {\tilde d_{kR}}^2\ \ (l = 1,2) & [ p \to \pi^0 l^+ ]\  , \\ 
\vert  \l '_{31k} \l ^{\prime \prime \star }_{11k} \vert & \lesssim &  
 7 \times 10^{-27}\ {\tilde d_{kR}}^2 & [ n \to \pi^0 \bar \nu ]\  , \\ 
\vert  \l '_{i2k} \l ^{\prime \prime \star }_{11k} \vert & \lesssim & 
 3 \times 10^{-27}\ {\tilde d_{kR}}^2 & [ p \to K^+ \bar \nu]\ ,
\end{array} 
\label{eqn:nucleon_decay} 
\end{equation} 
which update the estimate given in Ref.~\cite{hinchliffe}.
The $\tilde d_{kR}$ squark exchange can also occur in the
$t$-channel, yielding a bound on the \Rp\ coupling products
$\l'_{l1k} \l^{\prime \prime \star}_{12k}$ ($l=1,2$, $k=1,2,3$):
\begin{equation} 
\begin{array}{llll} 
\vert  \l '_{l1k} \l ^{\prime \prime \star }_{12k} \vert & \lesssim &  
 (6-7) \times 10^{-27}\ {\tilde d_{kR}}^2 & [ p \to K^0 l^+ ]\  . 
\end{array} 
\label{eqn:nucleon_decay_bis} 
\end{equation} 

One could also alternatively contract the down-squark fields  
as $d - d^{c \dagger }$ by including a left-right mixing mass insertion term
$\tilde m^{d\, 2}_{\scriptscriptstyle{LR}}$, which 
then yields a $(B+L)$-conserving amplitude~\cite{vissani95}.   
The bounds derived from the experimental limits on the nucleon
decay channels $p\to K^+\nu$ and $n \to \pi^0 \nu$~\cite{pdg04}
read ($i, j = 1,2,3$): 
\begin{equation} 
\begin{array}{llll} 
\vert  \l '_{ij1} \l''_{11j} \vert & \lesssim &  
 7 \times 10^{-27}\ {\tilde d_{jL}}^2  
 \left( \frac{m^2_{\tilde d_{jR}}}{(\tilde m^{d\, 2}_{\scriptscriptstyle{LR}})_{jj}} \right)
 & [ n \to \pi^0 \nu ]\ ,  \\
 \vert  \l '_{ij2} \l''_{11j} \vert & \lesssim &  
 3 \times 10^{-27}\ {\tilde d_{jL}}^2  
 \left( \frac{m^2_{\tilde d_{jR}}}{(\tilde m^{d\, 2}_{\scriptscriptstyle{LR}})_{jj}} \right)
 & [ p \to K^+ \nu ]\ ,  \\
 \vert  \l '_{i31} \l''_{123} \vert & \lesssim &  
 3 \times 10^{-27}\ {\tilde b_L}^2  
 \left( \frac{m^2_{\tilde b_R}}{(\tilde m^{d\, 2}_{\scriptscriptstyle{LR}})_{33}} \right)
 & [ p \to K^+ \nu ]\ .
\end{array} 
\label{eqn:nucleon_decay2} 
\end{equation} 

These bounds are less stringent than the previous ones due to the
presence of the left-right mass term
$(\tilde m^{d\, 2}_{\scriptscriptstyle{LR}})_{jj} =  (A^d - \mu \tan \beta) m_{d_j}$
in the denominator; the best bounds are obtained for $j = 3$
($\tilde b_L - \tilde b_R$ exchange). The exchanged scalar field
can also be an up-squark, with the insertion of a left-right
mass term $(\tilde m^{u\, 2}_{\scriptscriptstyle{LR}})_{jj} =
(A^u - \mu \cot \beta) m_{u_j}$~\cite{vissani95}.
The bounds derived from the experimental limits on the neutron
decay modes $n \to K^+ l^-$~\cite{pdg04} read ($l=1,2$, $j = 1,2,3$): 
\begin{equation} 
\begin{array}{llll} 
\vert  \l '_{lj1} \l''_{j12} \vert & \lesssim &  
 10^{-26}\ {\tilde u_{jL}}^2  
 \left( \frac{m^2_{\tilde u_{jR}}}{(\tilde m^{u\, 2}_{\scriptscriptstyle{LR}})_{jj}} \right)
 & [ n \to K^+ l ]\ .
\end{array} 
\label{eqn:nucleon_decay2_bis} 
\end{equation} 
Again the best bounds are obtained for $j = 3$ ($\tilde t_L - \tilde t_R$ exchange). 
 
The above very  stringent bounds concern $ \l' \l^{\prime \prime (\star)}$
products involving dominantly the first two light generation indices.  It 
was observed by Smirnov and Vissani~\cite{smirnov} that an 
appropriate extension of the above analysis to the one-loop level 
could be used to set strong bounds on coupling products  
for all possible configurations of the generation indices. 
The contributions come from one-loop diagrams obtained from the 
above tree-level diagrams by adding a vertex diagram dressing for the 
coupling $\nu_i d_{jL} \tilde d^\star_{kR}$ or for the coupling 
$u_{i'R} d_{j'R} \tilde d_{k'R}$, or a box diagram dressing for both 
couplings, 
%
%
where the internal lines propagating in the loops are charged or neutral 
Higgs bosons, winos or sfermions. The loop and 
flavour  mixing  suppression  factors in the transition 
amplitudes result in much weaker bounds than
Eqs.~(\ref{eqn:nucleon_decay})--(\ref{eqn:nucleon_decay2_bis}).
Assuming squark masses around $1 \TeV$, one obtains the
following conservative bound on any product of $\l$-
and $\l''$-type couplings~\cite{smirnov}: 
\begin{equation} 
\vert  \l'_{ijk} \l^{\prime \prime \star}_{i'j'k'} \vert\ <\ {\cal O} (10 ^{-9})\ . 
 \label{eqn:nucleon_decay3} 
\end{equation} 
For squark masses around $100 \GeV$, this bound would be
${\cal O} (10 ^{-12})$.

Single nucleon decays can also be induced by products of $\l$-type and
$\l''$-type couplings, through tree-level diagrams involving the sequential
exchange of a squark, a neutralino or chargino, and a
slepton, or through one-loop or two-loop diagrams obtained
from the dressing of the former tree-level
diagrams~\cite{carlson,long98,bhattapal99}.
Bhattacharyya and Pal~\cite{bhattapal99}
consider proton decay mediated by diagrams involving the exchange
of a neutralino. Assuming a common superpartner mass $\tilde m = 1 \TeV$,
they obtain the following bounds on the $\l \l^{\prime \prime \star}$ products
involving a coupling $\l''_{112}$:
\begin{equation} 
\begin{array}{llll} 
\vert  \l_{231} \l^{\prime \prime \star}_{112} \vert,\
\vert  \l_{132} \l^{\prime \prime \star}_{112} \vert & \lesssim &  
10^{-16} & [ p \to K^+ e^\pm \mu^\mp \bar \nu ]\ ,  \\
\vert  \l_{123} \l^{\prime \prime \star}_{112} \vert & \lesssim &  
10^{-14} & [ p \to K^+ \nu \bar \nu \bar \nu ]\ ,  \\
\vert  \l_{121} \l^{\prime \prime \star}_{112} \vert,\
\vert  \l_{131} \l^{\prime \prime \star}_{112} \vert & \lesssim &  
10^{-17} & [ p \to K^+ \bar \nu ]\ ,  \\
\vert  \l_{122} \l^{\prime \prime \star}_{112} \vert,\
\vert  \l_{232} \l^{\prime \prime \star}_{112} \vert & \lesssim &  
10^{-20} & [ p \to K^+ \bar \nu ]\ ,  \\
\vert  \l_{133} \l^{\prime \prime \star}_{112} \vert,\
\vert  \l_{233} \l^{\prime \prime \star}_{112} \vert & \lesssim &  
10^{-21} & [ p \to K^+ \bar \nu ]\ .
\end{array} 
\label{eqn:nucleon_decay4} 
\end{equation} 
The bounds obtained from four-body decay modes could actually
be relaxed by about two orders of magnitude, due to phase space
factors. The constraints on products involving any other $\l''_{ijk}$ coupling
are much weaker, since the corresponding vertex must be dressed
by a loop with a charged Higgs boson in order to induce
proton decay. The resulting bounds read, assuming
$m_{H^+} = \tilde m = 1 \TeV$ ($(i,j,k) \neq (1,1,2)$)~\cite{bhattapal99}:
\begin{equation} 
\begin{array}{llll} 
\vert  \l_{231} \l^{\prime \prime \star}_{ijk} \vert,\
\vert  \l_{132} \l^{\prime \prime \star}_{ijk} \vert & \lesssim &  
(10^{-7}\ -\ 10^{-5}) & [ p \to \pi^+ (K^+) e^\pm \mu^\mp \bar \nu ]\ ,  \\
\vert  \l_{123} \l^{\prime \prime \star}_{ijk} \vert & \lesssim &  
(10^{-5}\ -\ 10^{-3}) & [ p \to \pi^+ (K^+) \nu \bar \nu \bar \nu ]\ ,  \\
\vert  \l_{121} \l^{\prime \prime \star}_{ijk} \vert,\
\vert  \l_{131} \l^{\prime \prime \star}_{ijk} \vert & \lesssim &  
(10^{-8}\ -\ 10^{-6}) & [ p \to \pi^+ (K^+) \bar \nu ]\ ,  \\
\vert  \l_{122} \l^{\prime \prime \star}_{ijk} \vert,\
\vert  \l_{232} \l^{\prime \prime \star}_{ijk} \vert & \lesssim &  
(10^{-11}\ -\ 10^{-9}) & [ p \to \pi^+ (K^+) \bar \nu ]\ ,  \\
\vert  \l_{133} \l^{\prime \prime \star}_{ijk} \vert,\
\vert  \l_{233} \l^{\prime \prime \star}_{ijk} \vert & \lesssim &  
(10^{-12}\ -\ 10^{-10}) & [ p \to \pi^+ (K^+) \bar \nu ]\ .
\end{array} 
\label{eqn:nucleon_decay4_bis}
\end{equation} 
Considering nucleon decay mediated by tree-level diagrams involving
a chargino exchange, Carlson et al.~\cite{carlson} find stronger bounds
than Eqs~(\ref{eqn:nucleon_decay4_bis}) for the $\l \l''$ products
involving the couplings $\l''_{113}$, $\l''_{123}$, $\l''_{212}$
and $\l''_{312}$:
%
\begin{equation}
\vert \l_{ijk} \l''_{113} \vert \lesssim 10^{-13},\ \
\vert \l_{ijk} \l''_{123} \vert \lesssim 10^{-12},\ \
\vert \l_{ijk} \l''_{212} \vert \lesssim 10^{-13},\ \
\vert \l_{ijk} \l''_{312} \vert \lesssim 10^{-12}\ . 
\label{eqn:nucleon_decay4_ter} 
\end{equation}
These bounds correspond to $(B+L)$-conserving decays such as
$p \to l^ +_k  \nu_i  \nu_j$. The bounds on $\l \l''$ products
involving the other $\l''_{ijk}$ couplings are weaker than
Eqs~(\ref{eqn:nucleon_decay4_bis}),
since a loop dressing of the corresponding vertex is necessary
in order to induce nucleon decay; we therefore do not give them here.

Other  independent bounds resulting from the combined effects of the trilinear 
interactions $\l''$ and bilinear interactions $\mu_i$ are obtained by 
Bhattacharyya and Pal~\cite{bhattapal98}, based on a tree-level 
mechanism involving the intermediate action of the Yukawa 
interaction of quarks with the up-type Higgs boson.  The relevant effective 
Lagrangian is $(B+L)$-conserving and contributes to the proton
decay channels $p \to K^+ \nu$  and $p \to K^+  \pi ^+ l^-$. 
The bound associated with the channel $p \to K^+ \nu$ reads
($i=1,2,3$)~\cite{bhattapal98}:  
%
\begin{equation}
  \vert \l ''_{112}\, \frac{\mu_i}{\mu} \vert\ \lesssim\ 10^{-23}\, \tilde u^2_{ R }
   \quad  [ p \to K^+ \nu ]\ .
\end{equation}
At the one-loop level, the same type of Higgs or gaugino dressing
as described above can be invoked to deduce analogous bounds  
involving also the heavy quark generations.  The associated  
mechanism requires the intermediate action of the 
Yukawa interactions of quarks with Higgs bosons, which results 
in an extra suppression factor $(\l ^d)^2$.  The associated upper
bounds on $\l''_{ijn} \mu_{i'} / \mu$, $n = 1,2$, $(i,j,n) \neq (1,1,2)$, 
vary inside the following range~\cite{bhattapal98}:
\begin{equation}  
 \vert \l''_{ijn}\, \frac{\mu_{i'}}{\mu} \vert\ \lesssim\ (10^{-16}\ - \ 10^{-12})\,
   \tilde d^2_{nR}  \quad  [ p \to \pi^+ \nu,\ p \to K^+ \nu ]\ .
\end{equation} 
We quote below a representative  subset  of the  derived bounds:  
\begin{equation}   
\vert \l''_{321} \frac{\mu_{i'}}{\mu} \vert\ \lesssim\ 10^{-16}\, \tilde d^2_R\, , \ \
\vert \l''_{331} \frac{\mu_{i'}}{\mu} \vert\ \lesssim\ 10^{-15}\, \tilde d^2_R\, , \ \
\vert \l''_{332} \frac{\mu_{i'}}{\mu} \vert\ \lesssim\ 10^{-16}\, \tilde s^2_R\, . 
\end{equation} 

If the production of superpartner particles in single nucleon decays 
were energetically allowed, additional exotic decay modes  
could arise from the baryon number violating 
interactions alone.  A familiar example is furnished, for the case of a very 
light neutralino, $ m_{\tilde \chi^0 } << m_p - m_{K^+}$, by the exotic 
proton decay channel $ p \to \tilde \chi ^0 K ^+ $, which can 
proceed via $ \tilde s $ tree-level exchange. This decay mode sets the 
bound $\vert \l ''_{112} \vert \lesssim 10^{-15}$~\cite{chang96}. 
 
For the case of an ultralight gravitino $\tilde G $ or  
axino\index{Axion!axino} $\tilde a$~\cite{kimrep,rajagopal}, as
characteristically arises in the low-energy 
gauge-mediated supersymmetry breaking approach, additional single 
nucleon decay channels may appear where the \Rp\ interactions initiate 
processes involving the emission of a light strange meson accompanied 
by an $R$-parity odd gravitino or axino\index{Axion!axino}.
The tree-level $\tilde s$ 
exchange graph for the relevant subprocesses, $ud \to \bar s  \tilde G$
and $ud \to \bar s \tilde a$,  leads to the bounds~\cite{choit},
\begin{eqnarray} 
 \vert \l''_{112} \vert & \lesssim & 6 \times 10^{-17}\
 \tilde s^2_R \left( \frac{m_{3/2}}{1 \eV} \right) 
 \qquad \qquad \quad [ p \to K^+  \tilde G ]\ ,
 \label{eq:nucleon_gravitino}  \\
 \vert \l''_{112} \vert & \lesssim & 8 \times 10^{-17}\
 C_q^{-1}\ \tilde s^2_R \left( \frac{F_a}{10^{10} \GeV} \right) 
 \quad  [ p \to K^+  \tilde a ]\ ,
 \label{eq:nucleon_axino}
\end{eqnarray} 
applying to the gravitino and axino emission cases, respectively.  
For the axino\index{Axion!axino} case, $F_a$ designates the axionic 
symmetry breaking mass scale, and the parameter $C_q$,
which describes the model dependence of the axino\index{Axion!axino} couplings
to quark and lepton fields, is assigned an order one value or
values in the range $C_q^{-1} = {\cal O} (10^{2}\ - \ 10^{3})$,
depending on the type of axino\index{Axion!axino} considered.  
 
Pursuing along the same lines as  above  with  the study of the two-body 
single nucleon decay modes at the one-loop level, one can derive 
strong bounds on all couplings $\l''_{ijk}$. Accounting 
approximately for the loop and flavour suppression factors 
associated with the one-loop dressing of the previous tree-level
diagrams, Choi et al. obtain bounds in the ranges~\cite{choil}:
\begin{eqnarray}
 \vert \l''_{ijk} \vert & \lesssim & \left( 10^{-11}\, \tilde m^3 \ - \
 10^{-8}\, \tilde m^2 \right)  \left( \frac{m_{3/2}}{1 \eV} \right) ,
  \label{eq:nucleon_gravitino2}  \\
 \vert \l''_{ijk} \vert & \lesssim & \left( 10^{-11}\, \tilde m^3 \ - \
 10^{-8}\, \tilde m^2 \right)  C_q^{-1}  \left( \frac{F_a}{10^{10} \GeV} \right) , 
  \label{eq:nucleon_axino2}  
\end{eqnarray}
%
%
where $\tilde m$ denotes a common superpartner mass, for the gravitino
and axino\index{Axion!axino} emission  cases, respectively.  
A representative subset of these bounds reads:  
%
\begin{eqnarray}
 \left[ \frac{\vert \l''_{113} \vert}{\tilde m^3}, \frac{\vert \l''_{212} \vert}{\tilde m^2},  
 \frac{\vert \l''_{323} \vert}{\tilde m^2} \right] & \lesssim &
 \left[ 2 \times 10^{-11}, 3 \times 10^{-9}, 6 \times 10^{-9} \right]
 \left( \frac{m_{3/2}}{1 \eV} \right) ,  \\  
 \left[ \frac{\vert \l''_{113} \vert}{\tilde m^3}, \frac{\vert \l''_{212} \vert}{\tilde m^2},
 \frac{\vert \l''_{323} \vert}{\tilde m^2} \right] & \lesssim &
 \left[ 3 \times 10^{-11}, 4 \times 10^{-9}, 8 \times 10^{-9} \right]
 C_q^{-1} \left( \frac{F_a}{10^{10} \GeV} \right) ,  \hskip 1cm 
\end{eqnarray}  
for the gravitino and axino\index{Axion!axino} emission cases, respectively. 
 
\subsubsection{Nucleon-Antinucleon Oscillations and Double Nucleon Decay} 
\index{Nucleon--antinucleon oscillations|(} 
The $n\to \bar n$ transition is governed by the effective Lagrangian, 
\begin{equation}
{\cal{L}}= -(\bar n \ \bar n^c ) 
\left( \begin{array}{cc}  m & \delta m \\ \delta m ^\star & m 
         \end{array} \right) \left( \begin{array}{cc}  n \\  n^c  
         \end{array} \right) ,
\end{equation}
where the inputs needed to 
determine the mass shift parameter $\delta m$, involve the couplings
$ \l ''$, the superpartner mass parameters, and the 
hadronic matrix elements of the relevant $\cddd =9$ local operators
$d_R d_R d_R u_R q_L q_L$ and $d_R d_R q_L q_L q_L q_L$.
While the neutron-antineutron oscillation time, defined approximately
by $\tau _{osc} \simeq 1/\delta m$, is strongly hindered by the nuclear
interactions, one hopefully anticipates to find observable manifestations
of $\Delta B =2$ baryon number violation under the guise of nuclear
two-nucleon disintegration processes,
$N + (A -1) \to \bar N + (A -1) \to X + (A-2) $,
where $X $ denotes the possible decay channels for the 
nucleon-antinucleon pair annihilation reaction
$n \bar n \to X$,  $n \bar p \to X$ and $p \bar p \to X$, 
with $X= \pi, \ 2\pi , \ 3\pi, \ 2K, \ \cdots $~\cite{moha80}.
The r\^ole of the hadronic and nuclear structure effects in the estimation 
of the dimension-9 operators matrix elements is discussed  
in~ Refs.~\cite{pasupathy,gal83,alberico}. 
 
\begin{figure}[htb] 
  \begin{center} 
     \mbox{\epsfxsize=0.65\textwidth 
       \epsffile{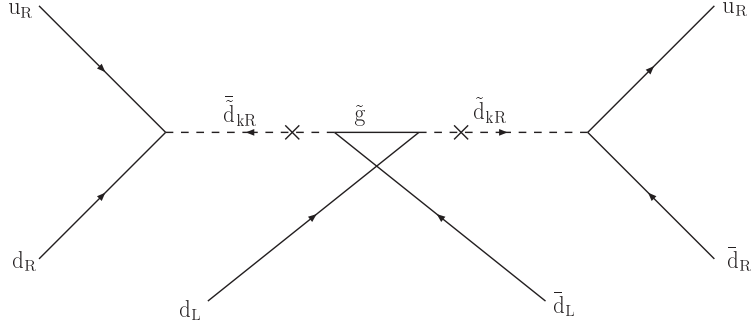}} 
  \end{center} 
 \caption[]{ \label{fig:nosci} 
 {\it{ \Rp contribution to nucleon-antinucleon oscillation.  
 }}} 
\end{figure}

Two competitive tree-level mechanisms for the \Rp\ contributions were 
originally discussed in an initiating  study by Zwirner~\cite{zwirner83}.  
The dominant process is shown in Fig.~\ref{fig:nosci}. 
%
The bounds inferred by comparison with the experimental limit 
due to the non-observation of $n$-$\bar n$ oscillations, using the tentative  
estimate $\vert \psi_{N} (0) \vert ^4 \approx 10^{-4} \ \text{GeV} ^{6}$
for the  wave function, are 
rather strong~\cite{halldimo,hinchliffe,zwirner83}:  
\begin{equation}
\vert \l''_{11k} \vert\ \lesssim\ (10^{-8} \ - \ 10^{-7})\ \frac{10^8\, \text{s}}{\tau_{osc}}
\left( \frac{\tilde m}{100 \GeV} \right)^{5/2} .
\label{eq:nn_bar}
\end{equation}
An alternative estimate reads: $\vert \l''_{11k} \vert \lesssim
(0.3 - 1.7 ) \times 10^{-10}\, \tilde g^{1/2} \tilde d_{kR}^2$.
However, these  bounds should be taken 
as indicative only, since an unknown suppression factor from the 
flavour off-diagonal entries of the left-right mixing squark 
mass matrix was ignored.  The second 
mechanism discussed by Zwirner~\cite{zwirner83} is described by an 
intermediate vertex at which three sfermions, which are emitted by 
quark lines via $\l''$ interactions, jointly annihilate via a 
soft supersymmetry breaking interaction of the type
$A''_{ijk} \tilde u^c_i \tilde d^c_j \tilde d^c_k$~\cite{zwirner83}. 
This contribution faces the same problem regarding the unknown input 
for the interaction trilinear in the squark fields.  Being of order 
$\l^{\prime \prime 4}$, it should be subdominant compared to the above one. 
 
Goity and Sher~\cite{goity} have challenged the view that $n$-$\bar n$ 
oscillations do actually constrain the coupling constant $ \l 
''_{121}$, in view of the uncertain information on the input supersymmetry 
breaking mass parameters.  They argue that one can identify a 
competitive mechanism, with a fully calculable transition amplitude, 
which sets a bound on $ \l ''_{131}$.  This alternative 
mechanism~\cite{goity} is based on the sequence of reactions
%
$u_R d_R + d_L \to \tilde b^\star_R + d_L \to \tilde b^\star_L + d_L 
\to \bar d_L + \tilde b_L \to \bar d_L + \bar u_R \bar d_R$,
where the intermediate transition $\tilde b^\star_L + d_L \to 
\bar d_L + \tilde b_L$ is due to a 
$W$ boson and gaugino exchange box diagram~\cite{goity}.  The choice of 
intermediate bottom squarks is the most favourable one in order to 
maximise factors such as $ m^2_{d} /M_W $, which arise from the 
electroweak interactions of d-quarks in the box diagram amplitude. 
The resulting bound  must be evaluated numerically  and  lies in the 
wide interval $\vert \l''_{131} \vert \lesssim (2 \times 10^{-3} - 10^{-1})$, for 
squark masses varying in the range $m _{\tilde q} = (200 - 600) \GeV$. 
The bound on $\l''_{121} $ is a factor $m_s/m_b \approx 4 \times 10^ {-2}$
weaker, $\vert \l''_{121} \vert \lesssim (5 \times 10^{-2} - 2.5)$, and is 
of marginal physical interest~\cite{goity}.
Chang and Keung~\cite{chang96} observe that the above 
mechanism actually includes three other analogous one-loop box 
diagrams involving the exchange of gaugino-$W$ boson and
quark-squark pairs. A single one of these dominates and yields
bounds  for the associated couplings of the form~\cite{chang96},  
\begin{equation}
\begin{array}{lll}
\vert \l''_{321} \vert & \lesssim & [2.1 \times 10^{-3},\ 1.5 \times 10^{-2}]
  \left( \frac{m_s}{200 \MeV} \right)^{-2}\ ,  \cr
\vert \l''_{331} \vert & \lesssim & [2.6 \times 10^{-3},\ 2 \times 10^{-2}]\ ,
\end{array}
\label{eq:nn_bar_2}
\end{equation}
where the two numbers inside brackets  
are in correspondence with the two input values 
used for the squark mass,  
$m_{\tilde q} = [100, 200] \GeV$, and $m_s$ is the 
strange quark mass.  
 
The generational structure of the $\l ''_{ijk}$ couplings imposes non-diagonal 
flavour configurations for the $ d^c$ quarks, which disfavours the 
strangeness conserving $n\to \bar n$ transition.  Based on an 
observation by Dimopoulos and Hall~\cite{halldimo}, Barbieri and 
Masiero~\cite{barbieri86} propose  instead to apply the same mechanism to the 
$\Delta S =2$ process $udd \to u^c s^c s^c$, which contributes 
to the transition $ n \to \bar \Xi $.  One avoids in 
this way the penalty of two flavour off-diagonal left-right mixing squark 
mass insertions, but the counterpart of this advantage is an 
energetically suppressed, off-shell $ n \to \bar \Xi $ transition. 
Another   hadronic physics aspect in the  comparison  
between the $ n \to \bar \Xi $ and $ n \to \bar n $ 
systems resides in their different short distance hadronic 
interactions.  An $ n - \bar {\Xi } $ oscillation initiates
$\Delta B = \Delta S =2$ double nucleon decay processes
such as $p+p \to K^+ + K^+$ or $n+n \to K^0 + K^0$, 
which could become the predominant channel for a double nucleon 
induced nuclear decay.   
Application to the nuclear decay reaction  
$^{16}\rm{O} \to ^{14}\rm{C} + K^{+} + K^{+}$ yields a relationship for 
the double nucleon decay lifetime, $ \tau_{NN} \simeq (10^{-2} \  \text 
{ y } ) \ \l ^{''-4}_{112} \  (\tilde q ^4  \ \tilde g)^2 $,  
which results in the bound~\cite{barbieri86}: 
\begin{equation}  
\vert \l''_{121} \vert\ \lesssim\ 10^{-8.5}\ \tilde g^\ud\ \tilde q ^2
\left( \frac{\tau _{NN}}{10^{32}\, \text{yr}} \right)^{-1/4}
\left( \frac{10^{-6} \GeV^6}{\lesssim \bar N \vert ududss \vert \Xi >} \right)^{1/2} .
\end{equation} 
An alternative treatment of the nuclear decay process is proposed by 
Goity and Sher~\cite{goity}, where one bypasses the intermediate step 
of the $n\to \bar \Xi $ transition by dealing directly with the 
transition $ NN \to KK$.  These authors  identify a mechanism where the \Rp\ 
interactions contribute through a Feynman diagram involving the 
$s$-channel production of a pair of squarks mediated by the 
$t$-channel exchange of a gluino, based on the reaction scheme
%
%
$(q_iq_j) (q_lq_m) \to \tilde q^\star_k \tilde q^\star_n 
\to \bar q_k \bar q_n$, where ($q_i$, $q_j$, $q_k$) is a permutation of 
($u_R$, $d_R$, $s_R$), and similarly for ($q_l$, $q_m$, $q_n$).  
The decay amplitude for the 
nuclear reaction $ ^{16}\rm{O} \to ^{14}\rm{C}+ K^++K^+$ is evaluated within 
an impulse approximation nuclear Fermi gas model for the nuclei, where 
the nuclear momentum integral contains the folded product of the 
elementary process cross-section with the nuclear momentum 
distributions of nucleons.  The resulting bound reads~\cite{goity}:    
\begin{eqnarray}  
\vert \l''_{121} \vert\ \lesssim\ 10^{-15}\, {\cal R}^{-5/2}\ ,  \quad
{\cal R}\ \equiv\ \frac{\tilde \L}{(m_{\tilde g} m_{\tilde q}^4)^{1/5}}\ ,
\label{eq:OCKK}
\end{eqnarray} 
where the parameter 
$\tilde \L $  in the  overall scale  factor parameter ${\cal R}$ 
describes a hadronic scale representing the dimensional 
analysis estimates of the hadronic and nuclear matrix elements. 
Varying ${\cal R}$ inside the range $(10^{-3} - 10^{ -6} )$, 
one finds a bound spanning a wide interval:
$\vert \l ''_{121} \vert \lesssim (10^{-7} - 10^{0})$.
In spite of the strong dependence on the hadronic and 
nuclear structure inputs, the preferred estimates are quoted  
in Ref.~\cite{goity} as $\vert \l''_{121} \vert \lesssim 10^{-6} $ and
$\vert \l ''_{131} \vert \lesssim 10^{-3}$, for 
the choice of a common superpartner mass $\tilde m = 300 \GeV$. 
\index{Nucleon--antinucleon oscillations|)} 
\section{General Discussion of Indirect Trilinear Bounds}
\label{sec:indcons}

In this section we shall attempt to assess from a more global perspective
the current situation regarding the indirect bounds on the trilinear \Rp\
coupling constants.

\subsection{Summary of Main Experimental Bounds}
\label{secin4a}

We have collected together a sizeable  
subset of the strongest bounds  available in the literature.   
Table~\ref{singletab} displays the results for single coupling
constants and Table~\ref{table62} to~\ref{table65}
for quadratic coupling constant products.   
These lists recapitulate bounds already encountered in the preceding sections.
The bounds derived from the perturbative unitarity conditions following  
the discussion of the renormalisation group constraints in 
chapter~\ref{chap:evolution} have also been included, as well as the bounds
inferred from the neutrino mass discussed in chapter~\ref{chap:neutrinos}. 
Self-evident abbreviations (described in the caption of Table~\ref{singletab}) 
are used to identify the associated observable processes from which the
bounds are inferred. 
 
\begin{table} 
\begin{center} 
\vskip 0.1cm 
%
%
\begin{tabular}{l|l|l|l} \hline 
& {\bf Charged } & {\bf Neutral} & {\bf Other Processes} \\ 
 & {\bf Current } & {\bf Current} & \\ \hline \hline 
  $\l_{12k}$  
& $ \ 0.05 \ \tilde e_{kR} \ [V_{ud}]$ (\ref{eqn:vud_bounds}) 
& $ \ 0.14 \ \tilde e_{kR} \ [\nu_\mu e ] $ 
           (\ref{eqn:nue_lambda_bounds})  & \\  
& & $ \ 0.13 \ \tilde e_{k=1L}\ [\nu_\mu e ] $ 
           (\ref{eqn:nue_lambda_bounds})  & \\ 
& $ \ 0.07 \ \tilde e_{kR} [R_{\mu \tau}](\ref{eqn:l12k_mudecay}) $  
& $ \ [0.37, \ 0.25, \ 0.11]  \tilde \nu $ & \\  
& & $\qquad k=1,2,3 \ [A_{FB}] $ (\ref{eqn:afb_bounds}) & \\  
& & $ \ 0.05 \tilde e_{kR} \ [Q_W (\text{Cs}) ] $  
           (\ref{eqn:apv_bounds}) & \\ 
& & $ \ 0.13 \ \tilde e_{kR} \ [\nu_\mu q ] $  
       (\ref{eqn:nue_lambdap_bounds}) & \\ \hline 
 
  $\l_{13k}$  
& $ \ 0.07 \ \tilde e_{kR}\ [R_{\tau }]$ (\ref{eqn:rtau_bounds}) 
& $ \ [0.37, \ 0.25, \ 0.11] \tilde \nu $ & \\ 
& & $ \qquad k=1,2,3 \ [A_{FB}] $ (\ref{eqn:afb_bounds}) 
& \\ \hline 
 
  $\l_{23k}$  
& $ \ 0.07 \ \tilde e_{kR} \ [R_{\tau }]$ (\ref{eqn:rtau_bounds}) 
& $ \ 0.11\ \tilde \tau_L \ [\nu_\mu e ]$   
           (\ref{eqn:nue_lambda_bounds}) & \\ 
& $ \ 0.07 \ \tilde e_{kR}\ [R_{\tau \mu }]$ (\ref{eqn:rtau_bounds}) 
& $ \qquad k=1 $ & \\ \hline 
 
$\l_{233}$ & & & $ \ 0.90 \ [RG]$ \\ \hline 
 
$\l_{i22}$ & && $ 2.7 \times 10^{-2} \ \tilde{\mu} \tilde{m}^{-\ud} \ 
[m_\nu < 1 \eV]$ \\ 
& & & $\qquad (\tilde{m}^{e\, 2}_{\scriptscriptstyle{LR 22}} = \tilde{m} m_\mu)$
(\ref{eq:m_nu_lambda_2}) \\ 
\hline 
 
 
$\l_{i33}$ & && $ 1.6 \times 10^{-3} \ \tilde{\tau} \tilde{m}^{-\ud} \ 
[m_\nu < 1 \eV]$ \\ 
& & & $\qquad (\tilde{m}^{e\, 2}_{\scriptscriptstyle{LR 33}} = \tilde{m} m_\tau)$
(\ref{eq:m_nu_lambda_2}) \\ 
\hline \hline


  $\l'_{11k}$ &  
$ \ 0.02 \ \tilde d_{kR} \ [ V_{ud}]$ (\ref{eqn:vud_bounds})  
& $ \ [0.28, 0.18] \ \tilde u_L \ $
\\
& & $\qquad k=2,3 \ [A_{FB}]$ (\ref{eqn:afb_bounds})  
\\ 
& $ \ 0.03 \ \tilde d_{kR} \ [R_{\pi }] (\ref{eqn:rpi_bounds}) $ 
& $ \ 0.02 \tilde d_{kR} \ [Q_W (\text{Cs})]  
      (\ref{eqn:apv_bounds}) $ 
\\ \hline 
  
$\l'_{111}$ &&& $3.3\times 10^{-4}\, \tilde q^2 \tilde g^\ud\ [\b \b {0\nu }]$
(\ref{eqn:betabeta_bounds}) \\ \hline

$ \l'_{12k}$ 
& $ \ 0.44\ \tilde d_{kR} [R_{D^+}]$ (\ref{eqn:ddecays_bounds})   
& $ \ 0.21\ \tilde d_{kR} \ [A_{FB}]$ (\ref{eqn:afb_bounds}) 
\\  
& $ \ 0.27 \ \tilde d_{kR} [R_{D^0} ]$ (\ref{eqn:ddecays_bounds}) 
& $\ [0.28, \ 0.18] \ \tilde{c}_L$   
\\ 
& $ \ 0.23 \ \tilde d_{kR} [R_{D^+} ^\star ]$ (\ref{eqn:ddecays_bounds}) 
& $\qquad k=2,3 \ [A_{FB}]$ (\ref{eqn:afb_bounds})  
\\ \hline 
 
$ \l'_{13k}$ &  
& $ \ [0.28, \ 0.18] \ \tilde{t}_L $  
\\ 
& & $ \qquad k=2,3 \ [A_{FB}]$ (\ref{eqn:afb_bounds}) & \\ 
& & $ \ 0.47 \ [R_{e}] $ (\ref{eqn:rl_bounds}) &   \\ 
& & $ \qquad (m(\tilde{d}_{kR}) = 100 \GeV)$  & \\ 
\hline 
 
$ \l'_{1j1}$ 
& & $ \ 0.03\ \tilde u_{jL} \ [Q_W(\text{Cs})]$ (\ref{eqn:apv_bounds}) & \\ 
\hline 

$ \l'_{2j1}$ & & $ \  0.18 \ \tilde d_{jL}\ [\nu_\mu q]$  
(\ref{eqn:nue_lambdap_bounds}) \\ \hline 

  $ \l'_{21k}$  
& $ \ 0.06 \ \tilde d_{kR} \ [R_\pi ]$ (\ref{eqn:rpi_bounds})  
& $ \ 0.15 \ \tilde d_{kR} \ [\nu_\mu q] $ 
        (\ref{eqn:nue_lambdap_bounds}) 
\\ 
& $ \ 0.08 \ \tilde d_{kR} \ [R_{\tau \pi }]$ (\ref{eqn:rtaupi_bounds}) 
& \\ \hline 

  $ \l'_{22k}$  
& $ \ 0.61 \ \tilde d_{kR} [R_{D^+}]$ (\ref{eqn:ddecays_bounds}) 
& \\ 
& $ \ 0.38 d_{kR} \ [R_{D^+} ^\star ]$ (\ref{eqn:ddecays_bounds}) &
\\ 
& $ \ 0.21 \tilde d_{kR} \ [R_{D^0} ]$ (\ref{eqn:ddecays_bounds}) 
& &   
\\ 
& $ \ 0.65 \ \tilde d_{kR} [R_{D_s} (\tau \mu)]$ (\ref{eqn:rdstaumu_bounds}) 
& & \\ \hline

  $\l'_{23k}$  
&& $ \ 0.45 \ [ R_{\mu}]$ (\ref{eqn:rl_bounds}) & \\ 
& &  $\qquad (m_{\tilde{d}_{kR}} = 100 \GeV)$  & \\ \hline 

\end{tabular} 
\end{center}
\end{table} 


\begin{table} 
\begin{center} 
\begin{tabular}{l|l|l|l} \hline 
 & {\bf Charged } & {\bf Neutral} & {\bf Other Processes} \\ 
 & {\bf Current } & {\bf Current} & \\ \hline \hline 
 
  $ \l'_{31k}$  
& $ \ 0.12 \ \tilde d_{kR} \ [R_{\tau \pi } ]$ 
      (\ref{eqn:rtaupi_bounds})  
& \\ \hline 
 
  $ \l'_{32k}$  
& $ \ 0.52 \ \tilde d_{kR}\ [R_{D_s} ( \tau \mu )]$ 
           (\ref{eqn:rdstaumu_bounds}) 
& \\ \hline 

  $\l'_{33k}$  
& & $ \ 0.58\  \ [R_{\tau}] $ (\ref{eqn:rl_bounds})  
\\
& &  $\qquad (m_{\tilde{d}_{kR}} = 100 \GeV)$
\\ \hline 

$\l'_{333}$ & $ 0.32\ \tilde b_R \ [B \to \tau \nu X] $  
      (\ref{eqn:grossman})  & & $ \ 1.06 \ [RG]$ \\ 
\hline 
 
$ \l'_{i11} $ & & & $ \ 0.2 \ \tilde{d} \tilde{m}^{-\ud} \ 
[m_\nu < 1 \eV]$  \\ 
& & & $\qquad (\tilde{m}^{d\, 2}_{\scriptscriptstyle{LR 11}} = \tilde{m} m_d)$
(\ref{eq:m_nu_lambda'_2}) \\ 
\hline 
 
$ \l'_{i22} $ & & & $ 10^{-2} \ \tilde{s} \tilde{m}^{-\ud} \ 
[m_\nu < 1 \eV]$  \\ 
& & & $\qquad (\tilde{m}^{d\, 2}_{\scriptscriptstyle{LR 22}} = \tilde{m} m_s)$
(\ref{eq:m_nu_lambda'_2}) \\ 
\hline 
 
$ \l'_{i33} $ & & & $ 4 \times 10^{-4} \ \tilde{b} \tilde{m}^{-\ud} \ 
[m_\nu < 1 \eV]$  \\ 
& & & $\qquad (\tilde{m}^{d\, 2}_{\scriptscriptstyle{LR 33}} = \tilde{m} m_b)$
(\ref{eq:m_nu_lambda'_2}) \\ 
\hline \hline

 
$ \l'' _{11k}$ & & & $(10^{-8} - 10^{-7}) (10^8 \text{s} / \tau_{osc}) \tilde m^{5/2}$ \\
&&& $\qquad [n\bar n ]$ (\ref{eq:nn_bar}) \\ \hline 

$\l''_{112}$ & & & $10^{-6} \ [NN] \ (\tilde m = 300 \GeV)$ (\ref{eq:OCKK}) \\
& & & $6 \times 10^{-17}\, \tilde s^2_R\, (m_{3/2} / 1 \eV)$ \\
& & & $\qquad [ p \to K^+  \tilde G ]$ (\ref{eq:nucleon_gravitino}) \\ 
& & & $8 \times 10^{-17}\, C_q^{-1}\, \tilde s^2_R\, (F_a / 10^{10} \GeV)$ \\ 
& & & $\qquad  [ p \to K^+  \tilde a ]$ (\ref{eq:nucleon_axino}) \\ \hline 
 
$\l''_{113}$ & & & $10^{-3} \ [NN] \ (\tilde m = 300 \GeV)$ (\ref{eq:OCKK}) \\
\hline
 
$ \l'' _{123}$ & & & $ \ 1.25 \ [RG] $ \\ \hline 
 
$ \l'' _{212}$ & & & $ \ 1.25 \ [RG] $ \\ \hline 
 
$ \l'' _{213} $ & & & $ \ 1.25 \ [RG] $ \\ \hline 
 
$ \l'' _{223}$ & & & $ \ 1.25 \ [RG] $ \\ \hline

$ \l ''_{312}$ & & $ \ 1.45 \ [R_l]$  
     (\ref{eqn:rl_bounds_lpp}) & $\ 4.28 \ [RG]$ \\  
&& $ \qquad (\tilde{m} = 100 \GeV)$  & $\ 2.1 \times 10^{-3} \ [n \bar n] $
(\ref{eq:nn_bar_2}) \\ \hline 
 
$ \l ''_{313}$ & & $ \ 1.46 \ [R_l]$  
      (\ref{eqn:rl_bounds_lpp})  & $\ 1.12 \ [RG]$ \\  
& & $ \qquad (\tilde{m} = 100 \GeV)$   
& $\ 2.6\times 10^{-3} \ [n\bar n] $ (\ref{eq:nn_bar_2}) \\  
\hline 
 
$ \l ''_{323}$ & & $ \ 1.46 \  [R_l]$ (\ref{eqn:rl_bounds_lpp}) & $ \ 1.12 \ 
[RG]$ \\  
& & $ \qquad (\tilde{m} = 100 \GeV)$ & \\ \hline 
 
$\l''_{ijk}$ & & & $(10^{-11}\, \tilde m^3 - 10^{-8}\, \tilde m^2 )$ \\
& & & $\times (m_{3/2 } / 1 \eV)\ [p \to K^+ \tilde G]$
(\ref{eq:nucleon_gravitino2}) \\
& & & $\times (F_a / 10^{10} \GeV)\ [p \to K^+ \tilde a]$
(\ref{eq:nucleon_axino2}) \\ \hline 
\end{tabular} 
\end{center}

\caption{{\it Single bounds for the \Rp\ coupling constants at the  
   2$\sigma$ level.  We use the notation $ V_{ij} $ for the CKM matrix, 
   $ R_l , \ R_{l\ l '}, \ R_D , \ R_l ^Z $ 
   for various branching fractions or ratios of branching 
   fractions as defined in the text, $ Q_W$ for the weak 
   charge, $\nu q , \ \nu l$ for the neutrino elastic scattering on 
   quarks and leptons, $ m_\nu $ for the neutrino Majorana mass, $ RG$ 
   for the renormalisation group, $A_{FB}$  for forward-backward asymmetry,  
   $Q_W (\text{Cs})$  
   for atomic physics parity violation, $n\bar n$ for neutron-antineutron 
   oscillation and $ NN$ for two nucleon nuclear decay, $[ K\bar
   K]$, for $K^0 - \bar K^0$ mixing \index{Mixing!neutral mesons}.
   The generation 
   indices denoted $ i, j, k$ run over the three generations while those 
   denoted $l, m, n$ run over the first two generations.  The dependence 
   on the superpartner mass follows the notational convention $\tilde m^p 
   = ({\tilde m \over 100 \ \text{GeV}} )^p$.
   Aside from a few cases associated with one-loop effects, we use the
   reference value $\tilde m = 100 \ \text{GeV}$. 
   The quoted equation labels refer to equations in the text.}} 

\label{singletab}

\end{table}
 
%
%
 
\begin{table} 
\begin{center} 
\vskip 1 cm 
\begin{tabular}{l|l|l|l} \hline 
 & {\bf Lepton } & {\bf Hadron } & {\bf L and/or B } \\ 
 & {\bf Flavour } & {\bf Flavour } & {\bf violation} \\ 
\hline \hline 
 

$\vert \l^\star_{ij2} \l_{ij1} \vert$ & $ \ 8.2 \times 10^{-5}
(\tilde \nu_L^2, \tilde l_L^2) \ [\mu \to e \g]$
(\ref{eqn:mutoegamma})& & \\  
\hline

$\vert \l_{23k} \l^\star_{13k} \vert$ & $ \ 2.3 \times 10^{-4}
(\tilde \nu_L^2, \tilde l_R^2) \ [\mu \to e \g]$
(\ref{eqn:mutoegamma})& & \\  
\hline

$ \vert \l_{312} \l^\star_{321} \vert $ & $ \ 1.9 \times 10^{-3}
\tilde \nu_L^2$ && \\ 
& $[\mu^+ e^- \to \mu^- e^+ ]$ (\ref{eqn:muonium}) && \\ \hline 
 
$\vert \l^\star_{i12} \l_{i11} \vert$ & $ \ 6.6 \times 10^{-7}
\tilde \nu_L^2 \ [\mu \to 3e]$ (\ref{eqn:l_to_three_l}) && \\ 
$\vert \l_{321} \l^\star_{311} \vert$ & $ \ 6.6  \times 10^{-7}
\tilde \nu_L^2 \ [\mu \to 3e]$ (\ref{eqn:l_to_three_l}) && \\
\hline 
 
$\vert \l^\star_{i23} \l_{i22} \vert $ & $ \ 2.2 \times 10^{-3}
\tilde \nu_L^2 \ [\tau \to 3 \mu ]$ (\ref{eqn:l_to_three_l})  && \\  
$\vert \l_{132} \l^\star_{122} \vert $ & $ \ 2.2 \times 10^{-3}
\tilde \nu _L^2 \ [\tau \to 3 \mu ]$ (\ref{eqn:l_to_three_l})  &&  \\
\hline
 
$\vert \l_{i12} \l_{j21} \vert$ 
& & & $0.15 \ \tilde{l}^2 \tilde{m}^{-1} \ [m_\nu < 1 \eV]$ \\ 
$\vert \l_{i13} \l_{j31} \vert$ 
& & & $8.7\times 10^{-3} \ \tilde{l}^2 \tilde{m}^{-1} \ [m_\nu < 1 \eV]$ \\ 
$\vert \l_{i22} \l_{j22} \vert$ 
& & & $7\times 10^{-4} \ \tilde{\mu}^2 \tilde{m}^{-1} \ [m_\nu < 1 \eV]$ \\ 
$\vert \l_{i23} \l_{j32} \vert$ 
& & & $4.2\times 10^{-5} \ \tilde{l}^2 \tilde{m}^{-1} \ [m_\nu < 1 \eV]$ \\ 
$\vert \l_{i33} \l_{j33} \vert$ 
& & & $2.5\times 10^{-6} \ \tilde{\tau}^2 \tilde{m}^{-1} \ [m_\nu < 1 \eV]$ \\
& & & $\quad (\tilde{m}^{e\, 2}_{\scriptscriptstyle{LR}} = \tilde{m}\, M^e)$
(\ref{eq:m_nu_lambda_2}) \\
\hline \hline


$\vert \l^\star_{i12} \l'_{i11} \vert$ & $ \ 2.1 \times 10^{-8}
\tilde \nu^2_L$ & & \\
& $\quad [\mu \to e\, ({\rm Ti})]$ (\ref{eqn:mutoe}) & & \\
$\vert \l_{i21} \l^{\prime \star}_{i11} \vert$ & $ \ 2.1 \times 10^{-8}
\tilde \nu^2_L$ & & \\
& $\quad [\mu \to e\, ({\rm Ti})]$ (\ref{eqn:mutoe}) & & \\
\hline

$\vert \l^{\star }_{1j1} \l'_{j33} \vert $ & $ {\cal I} \ 6 \times 10^{-7}  
\tilde \nu _j ^2 \ [d_e ^\g ]$ (\ref{eqn:nedm}) & & \\
\hline 

$\vert \l_{i31} \l^{\prime \star}_{i11} \vert $ & $ \ 1.6 \times 10^{-3}  
\tilde \nu_{iL}^2 \ [\tau \to e \eta]$ (\ref{eqn:lfvtau_meson_ll'}) && \\ 
$\vert \l^\star_{i13} \l'_{i11} \vert $ & $ \ 1.6 \times 10^{-3}  
\tilde \nu_{iL}^2 \ [\tau \to e \eta]$ (\ref{eqn:lfvtau_meson_ll'}) && \\
\hline
 
$\vert \l_{i32} \l^{\prime \star}_{i11} \vert $ & $ \ 1.7 \times 10^{-3}
\tilde \nu_{iL}^2 \ [\tau \to \mu \eta]$ (\ref{eqn:lfvtau_meson_ll'}) && \\
$\vert \l^\star_{i23} \l'_{i11} \vert $ & $ \ 1.7 \times 10^{-3}
\tilde \nu_{iL}^2 \ [\tau \to \mu \eta]$ (\ref{eqn:lfvtau_meson_ll'}) && \\
\hline 

\end{tabular}
\end{center}
 
\caption{{\it Quadratic coupling constant product bounds.
           We use the same conventions as in the preceding
	   table for the single coupling constant bounds. 
	   The presence of a symbol ${\cal I}$ means that the bound applies 
           to the imaginary part  of the coupling constant products.}}
\label{table62}
\end{table} 
 
 
\begin{table} 
\begin{center} 
\vskip 0.3cm 
\begin{tabular}{l|l|l|l} \hline 
 & {\bf Lepton } & {\bf Hadron } & {\bf L and/or B } \\ 
 & {\bf Flavour } & {\bf Flavour } & {\bf violation} \\ 
\hline \hline 
$ \vert \l^{\star }_{122} \l' _{112} \vert $ & & $ \ 2.2 \times 10^{-7}
\tilde  \nu_{L}^2 \ [ 
 K_L \to \mu ^+\mu ^- ]$ (\ref{eqn:kdecays_ee_mumu_ll'}) & \\  
$ \vert \l^{\star }_{122} \l' _{121} \vert $ & & $ \ 2.2 \times 10^{-7} 
 \tilde \nu _{L}^2  
 \ [ K_L \to \mu  ^+\mu  ^-]$ (\ref{eqn:kdecays_ee_mumu_ll'}) & \\ \hline 
 
$\vert \l^{\star }_{121} \l' _{212} \vert $ && $ \ 1.0 \times 10^{-8}
\tilde \nu_{L}^2 \ [K_L \to e ^+ e^-]$ (\ref{eqn:kdecays_ee_mumu_ll'}) & \\  
$\vert \l^{\star }_{121} \l' _{221} \vert $ &   
& $ \ 1.0 \times 10^{-8} \tilde \nu_{L}^2 \ [ K_L \to e^+ e ^- ]$ 
    (\ref{eqn:kdecays_ee_mumu_ll'}) &  \\ \hline 
 
$\vert \l^\star_{i12} \l'_{i12} \vert $ && $ \ 6 \times 10^{-9}
\tilde \nu_{L}^2 \ [ K_L \to e^\pm \mu^\mp ]$ (\ref{eqn:kdecays_emu}) & \\  
$\vert \l^\star_{i12} \l'_{i21} \vert $ & & $ \ 6 \times 10^{-9}
\tilde \nu_{L}^2 \ [ K_L \to e^\pm \mu^\mp ]$ (\ref{eqn:kdecays_emu}) & \\
$\vert \l^\star_{i21} \l'_{i12} \vert $ && $ \ 6 \times 10^{-9}
\tilde \nu_{L}^2 \ [ K_L \to e^\pm \mu^\mp ]$ (\ref{eqn:kdecays_emu}) & \\  
$\vert \l^\star_{i21} \l'_{i21} \vert $ & & $ \ 6 \times 10^{-9}
\tilde \nu_{L}^2 \ [ K_L \to e^\pm \mu^\mp ]$ (\ref{eqn:kdecays_emu}) & \\
\hline 
 
$\vert \l_{i31} \l^{\prime \star}_{i13} \vert$ && $ \ 6 \times 10^{-4}
\tilde l^2_{iL} \ [ B^- \to e^- \bar \nu ]$ (\ref{eqn:bdecays_l_nu}) & \\
$\vert \l_{i32} \l^{\prime \star}_{i13} \vert$ && $ \ 7 \times 10^{-4}
\tilde l^2_{iL} \ [ B^- \to \mu^- \bar \nu ]$ (\ref{eqn:bdecays_l_nu}) & \\
$\vert \l_{233} \l^{\prime \star}_{313} \vert$ && $ \ 2 \times 10^{-3}
\tilde l^2_{3L} \ [ B^- \to \tau^- \bar \nu ]$ (\ref{eqn:bdecays_l_nu}) & \\
\hline
 
$\vert \l^{\star }_{i11} \l'_{i13} \vert $ && $ \ 1.7 \times 10^{-5}
\tilde \nu^2_L \ [ B_d^0 \to e^+ e^- ]$ (\ref{eqn:bdecays_li_lj_ll'}) & \\  
$\vert \l^{\star }_{i11} \l'_{i31} \vert $ && $ \ 1.7 \times 10^{-5}
\tilde \nu^2_L \ [ B_d^0 \to e^+ e^- ]$ (\ref{eqn:bdecays_li_lj_ll'}) & \\
\hline 
 
$ \vert \l^{\star }_{i22} \l'_{i13}\vert $ && $ \ 1.5 \times 10^{-5}
\tilde \nu^2_L \ [ B_d^0 \to \mu ^+ \mu ^- ]$
(\ref{eqn:bdecays_li_lj_ll'}) & \\
$ \vert \l^{\star }_{i22} \l'_{i31}\vert $ && $ \ 1.5 \times 10^{-5}
\tilde \nu^2_L \ [ B_d^0 \to \mu ^+ \mu ^- ]$
(\ref{eqn:bdecays_li_lj_ll'}) & \\ \hline 
 
$ \vert \l^{\star}_{i12} \l'_{i13}\vert $ && $ \ 2.3 \times 10^{-5}
\tilde \nu^2_L \ [ B_d^0 \to e^\pm \mu^\mp ]$
(\ref{eqn:bdecays_li_lj_ll'}) & \\
$ \vert \l^{\star}_{i12} \l'_{i31}\vert $ && $ \ 2.3 \times 10^{-5}
\tilde \nu^2_L \ [ B_d^0 \to e^\pm \mu^\mp ]$
(\ref{eqn:bdecays_li_lj_ll'}) & \\
$ \vert \l^{\star}_{i21} \l'_{i13}\vert $ && $ \ 2.3 \times 10^{-5}
\tilde \nu^2_L \ [ B_d^0 \to e^\pm \mu^\mp ]$
(\ref{eqn:bdecays_li_lj_ll'}) & \\
$ \vert \l^{\star}_{i21} \l'_{i31}\vert $ && $ \ 2.3 \times 10^{-5}
\tilde \nu^2_L \ [ B_d^0 \to e^\pm \mu^\mp ]$
(\ref{eqn:bdecays_li_lj_ll'}) & \\ 
\hline \hline

 
$\vert  \l_{231} \l^{\prime \prime \star}_{112} \vert$
& & & $10^{-16} \ [ p \to K^+ e^\pm \mu^\mp \bar \nu ]$ \\
$\vert  \l_{132} \l^{\prime \prime \star}_{112} \vert$
& & & $10^{-16} \ [ p \to K^+ e^\pm \mu^\mp \bar \nu ]$ \\
$\vert  \l_{123} \l^{\prime \prime \star}_{112} \vert$
& & & $10^{-14} \ [ p \to K^+ \nu \bar \nu \bar \nu ]$ \\
$\vert  \l_{i11} \l^{\prime \prime \star}_{112} \vert$
& & & $10^{-17} \ [ p \to K^+ \bar \nu ]$ \\
$\vert  \l_{i22} \l^{\prime \prime \star}_{112} \vert$
& & & $10^{-20} \ [ p \to K^+ \bar \nu ]$ \\
$\vert  \l_{i33} \l^{\prime \prime \star}_{112} \vert$
& & & $10^{-21} \ [ p \to K^+ \bar \nu ]$ \\
& & & $\quad (\tilde m = 1 \TeV)$ (\ref{eqn:nucleon_decay4}) \\
\hline 

$\vert \l_{ijj} \l^{\prime \prime \star}_{i'j'k'} \vert$ & & & $(10^{-12} - 10^{-6})
\ [ p \to \pi^+ (K^+) \bar \nu ]$ \\
& & & $\quad (m_{h^+} = \tilde m = 1 \TeV)$ (\ref{eqn:nucleon_decay4_bis}) \\
\hline 
 
$\vert \l_{ijk} \l''_{113} \vert$ & & & $10^{-13} \ [p\to l^+ \nu \nu]$
(\ref{eqn:nucleon_decay4_ter}) \\
$\vert \l_{ijk} \l''_{123} \vert$ & & & $10^{-12} \ [p\to l^+ \nu \nu]$
(\ref{eqn:nucleon_decay4_ter}) \\
$\vert \l_{ijk} \l''_{212} \vert$ & & & $10^{-13} \ [p\to l^+ \nu \nu]$
(\ref{eqn:nucleon_decay4_ter}) \\
$\vert \l_{ijk} \l''_{312} \vert$ & & & $10^{-12} \ [p\to l^+ \nu \nu]$
(\ref{eqn:nucleon_decay4_ter}) \\
\hline \hline
 
\end{tabular}  
\end{center} 

\caption{{\it Quadratic coupling constant product bounds. We use the same 
              conventions as in the preceding table.}}
\label{table63}
\end{table} 
 

\begin{table} 
\begin{center} 
\vskip 0.3cm 
\begin{tabular}{l|l|l|l} \hline 
 & {\bf Lepton } & {\bf Hadron } & {\bf L and/or B } \\ 
 & {\bf Flavour } & {\bf Flavour } & {\bf violation} \\ 
\hline \hline 
 
$\vert \l^{'\star }_{i21} \l' _{i12} \vert $
& & $4.5 \times 10^{-9} \ \tilde \nu _{iL}^2 \ [K \bar K ]$ 
&\\ $\vert \l '_{i31} \l^{'\star}_{i22} \vert $ && $1. \times 10^{-4}
\ [K \bar K ] \ (\tilde{m} = 100 \GeV) $ & \\

$ \vert \l^{'\star }_{i31} \l' _{i32} \vert $ && $7.7 \times 10^{-4}  
 \ [K \bar K ] \ (\tilde{m} = 100 \GeV) $ & \\ 
\hline 
 
$\vert \l^{\prime \star}_{i2k} \l'_{i'1k} \vert$ && $2.11 \times 10^{-5}\   
\tilde d_{kR}^2 \ [K^+ \to \pi^+ \nu \bar \nu ]$
(\ref{eqn:kpinunu_bounds}) &\\  
$\vert \l^{\prime \star}_{ij1} \l'_{i'j2} \vert$ && $2.11 \times 10^{-5}\  
\tilde d _{jL}^2 \ [K^+ \to \pi^+ \nu \bar \nu ]$  
(\ref{eqn:kpinunu_bounds}) &\\ \hline 

$\vert \l^{'\star }_{i31} \l' _{i13} \vert $ && $3.3 \times 10^{-8}  
\tilde \nu_{iL}^2 \ [B \bar B ]$ &  
\\ $\vert \l '_{i31} \l^{'\star }_{i33} \vert $ &&$1.3 \times 10^{-3}  
 \ [B \bar B ]$ & \\ \hline 
 
$\vert \l^{\prime \star}_{i3k} \l'_{i'2k} \vert$ && $1.5 \times 10^{-3}  
\tilde d_{kR}^2 \ [B \to X_s \nu \bar \nu ]$ (\ref{eqn:bsnunu}) & \\
$\vert \l^{\prime \star}_{ij2} \l'_{i'j3} \vert$ && $1.5 \times 10^{-3}  
\tilde d_{jL}^2 \ [B \to X_s \nu \bar \nu ]$ (\ref{eqn:bsnunu}) & \\
 \hline 
 
$\vert \l'_{2mk} \l^{\prime \star}_{1mk} \vert$ & $ \ 7.6 \times 10^{-5}
\tilde d_{kR}^{2} \ [\mu \to e \g]$
(\ref{eqn:mutoegamma})& & \\  
$\vert \l'_{23k} \l^{\prime \star}_{13k} \vert$ & $ \ 2.0 \times 10^{-3}
\ [\mu \to e \g]$
(\ref{eqn:mutoegamma_bis})& & \\  
& $\quad (m_{\tilde d_{kR}} = m_{\tilde t_L} = 300 \GeV)$ & & \\
\hline

$\vert \l^{\prime \star }_{1j1} \l'_{1j2} \vert $ &&
${\cal I} \ 8.1 \times 10^{-5} \tilde u^2_L \ [ K_L \to e^+ e^- ] $ 
(\ref{eqn:kdecays_ee_mumu_l'l'}) & \\
$\vert \l^{\prime \star }_{2j1} \l'_{2j2} \vert $ &&
${\cal I} \ 7.8 \times 10^{-6} \tilde u^2_L \ [ K_L \to \mu^+ \mu^- ]$
(\ref{eqn:kdecays_ee_mumu_l'l'}) & \\ 
\hline 

$\vert \l^{\prime \star}_{1j1} \l'_{2j2} \vert$ & &
$ \ 3 \times 10^{-7} \tilde u^2_L \ [K_L \to e^\pm \mu^\mp]$
(\ref{eqn:kdecays_emu}) & \\
$\vert \l^{\prime \star}_{1j2} \l'_{2j1} \vert$ & &
$ \ 3 \times 10^{-7} \tilde u^2_L \ [K_L \to e^\pm \mu^\mp]$
(\ref{eqn:kdecays_emu}) & \\
\hline

$\vert \l^{\prime \star}_{2j1} \l'_{2j3} \vert$ & &
$ \ 2.1 \times 10^{-3} \tilde u^2_L \ [B^0_d \to \mu^+ \mu^-]$
(\ref{eqn:bdecays_li_lj_l'l'}) & \\
\hline

$\vert \l^{\prime \star}_{1j1} \l'_{2j3} \vert$ & &
$ \ 4.7 \times 10^{-3} \tilde u^2_L \ [B^0_d \to e^\pm \mu^\mp]$
(\ref{eqn:bdecays_li_lj_l'l'}) & \\
$\vert \l^{\prime \star}_{1j3} \l'_{2j1} \vert$ & &
$ \ 4.7 \times 10^{-3} \tilde u^2_L \ [B^0_d \to e^\pm \mu^\mp]$
(\ref{eqn:bdecays_li_lj_l'l'}) & \\
\hline

$\vert \l'_{2j1} \l^{\prime \star}_{1j1} \vert $ & $4.3 \times 10^{-8}
\tilde u_{jL}^2 \ [\mu \to e\, ({\rm Ti}) ]$ (\ref{eqn:mutoe}) & & \\ 
\hline 

$\vert \l'_{3j1} \l^{\prime \star}_{1j1} \vert $ & $2.4 \times 10^{-3}
\tilde u_{jL}^2 \ [\tau \to e \rho]$
(\ref{eqn:lfvtau_meson_l'l'}) & & \\
\hline
 
$\vert \l^{'\star }_{11k} \l '_{12k} \vert $ & & $5.3 \times 10^{-3} \tilde  
d^2_{kR}\ [\L   \to \ p  l^-  \bar{\nu}_l ]$   (\ref{eqn:tahir_bounds})  & \\ \hline 
 
$\vert \l'_{21k} \l^{\prime \star}_{11k} \vert $ & $4.5 \times 10^{-8}
\tilde d_{kR}^2 \ [\mu \to e\, ({\rm Ti}) ]$ (\ref{eqn:mutoe}) & & \\ 
\hline

$\vert \l'_{31k} \l^{\prime \star}_{11k} \vert $ & $2.4 \times 10^{-3}
\tilde d_{kR}^2 \ [\tau \to e \rho]$
(\ref{eqn:lfvtau_meson_l'l'}) & & \\
\hline 
\end{tabular} 
\end{center} 

\caption{{\it Quadratic coupling constant product bounds. 
              We use the same conventions as in the preceding 
	      table.}} 
\label{table64}
\end{table} 
 
\begin{table} 
\begin{center} 
\vskip 0.3cm 
\begin{tabular}{l|l|l|l} \hline 
 & {\bf Lepton } & {\bf Hadron } & {\bf L and/or B } \\ 
 & {\bf Flavour } & {\bf Flavour } & {\bf violation} \\ 
\hline \hline 
$\vert \l'_{113}\l '_{131} \vert $ & && $3.8 \times 10^{-8}\ [ \b \b {0\nu } ]$ 
(\ref{eqn:babu_bounds}) \\ \hline 
$\vert \l'_{112}\l '_{121} \vert $ &&& $1.1 \times 10^{-6}\ [ \b \b {0\nu } ]$ 
(\ref{eqn:babu_bounds}) \\ \hline 
 
$\vert \l'_{i3k} \l^{\prime \star}_{i2k} \vert $ & & $0.09 \ (\tilde \nu  
_{iL}^2 , \tilde 
d_{iR}^2) \ [B \to K \g ] $ (\ref{eqn:btosgamma}) & \\  
$\vert \l^{\prime \star}_{ij3} \l '_{ij2} \vert $ & & $0.035 
\ (\tilde e _{iL}^2 , \tilde d_{jL}^2) \  
[B \to K \g ] $ (\ref{eqn:btosgamma}) & \\ \hline 
 
$\vert \l'_{i11} \l'_{j11} \vert$ 
& & & $5\times 10^{-2} \ \tilde{d}^2 \tilde{m}^{-1} \ [m_\nu < 1 \eV]$ \\
$\vert \l'_{i12} \l'_{j21} \vert$ 
& & & $3\times 10^{-3} \ \tilde{q}^2 \tilde{m}^{-1} \ [m_\nu < 1 \eV]$ \\ 
$\vert \l'_{i13} \l'_{j31} \vert$ 
& & & $8\times 10^{-5} \ \tilde{q}^2 \tilde{m}^{-1} \ [m_\nu < 1 \eV]$ \\ 
$\vert \l'_{i22} \l'_{j22} \vert$ 
& & & $2\times 10^{-4} \ \tilde{s}^2 \tilde{m}^{-1} \ [m_\nu < 1 \eV]$ \\ 
$\vert \l'_{i23} \l'_{j32} \vert$ 
& & & $5\times 10^{-6} \ \tilde{q}^2 \tilde{m}^{-1} \ [m_\nu < 1 \eV]$ \\ 
$\vert \l'_{i33} \l'_{j33} \vert$ 
& & & $10^{-7} \ \tilde{b}^2 \tilde{m}^{-1} \qquad [m_\nu < 1 \eV]$ \\
&&& $\quad (\tilde{m}^{d\, 2}_{\scriptscriptstyle{LR}} = \tilde{m}\, M^d)$
(\ref{eq:m_nu_lambda'_2}) \\  
\hline \hline 
 

%
$\vert \l'_{l1k} \l^{\prime \prime \star}_{11k} \vert$
& & & $(2-3) \times 10^{-27} {\tilde d_{kR}}^2\ [ p \to \pi^0 l^+ ]$
(\ref{eqn:nucleon_decay}) \\ 
$\vert \l'_{31k} \l^{\prime \prime \star}_{11k} \vert$
& & & $7 \times 10^{-27} {\tilde d_{kR}}^2\ [ n \to \pi^0 \bar \nu ]$
(\ref{eqn:nucleon_decay}) \\ 
$\vert \l'_{i2k} \l^{\prime \prime \star}_{11k} \vert$ 
& & & $3 \times 10^{-27} {\tilde d_{kR}}^2\ [ p \to K^+ \bar \nu]$
(\ref{eqn:nucleon_decay}) \\ \hline 
 
$\vert  \l '_{l1k} \l ^{\prime \prime \star}_{12k} \vert$ & & &
$(6-7) \times 10^{-27} {\tilde d_{kR}}^2$
$\ [ p \to K^0 l^+ ]$ (\ref{eqn:nucleon_decay_bis}) \\ \hline

$\vert  \l '_{ij1} \l''_{11j} \vert$ & & &  
$7 \times 10^{-26}\ {\tilde d_{jL}}^2  
( m^2_{\tilde d_{jR}} / (\tilde m^{d\, 2}_{\scriptscriptstyle{LR}})_{jj} )$ \\
$\vert  \l '_{ij2} \l''_{11j} \vert$ & & &  
$3 \times 10^{-27}\ {\tilde d_{jL}}^2  
( m^2_{\tilde d_{jR}} / (\tilde m^{d\, 2}_{\scriptscriptstyle{LR}})_{jj} )$ \\
$\vert  \l '_{i31} \l''_{123} \vert$ & & &  
$3 \times 10^{-27}\ {\tilde b_L}^2  
( m^2_{\tilde b_R} / (\tilde m^{d\, 2}_{\scriptscriptstyle{LR}})_{33} )$ \\
& & & $\quad [ n \to \pi^0 \nu,\ p \to K^+ \nu ]$ (\ref{eqn:nucleon_decay2}) \\ \hline

$\vert  \l '_{lj1} \l''_{j12} \vert$ & & &  
$10^{-26}\ {\tilde u_{jL}}^2  
( m^2_{\tilde u_{jR}} / (\tilde m^{u\, 2}_{\scriptscriptstyle{LR}})_{jj} )$ \\
& & & $\quad [ n \to K^+ l ]$ (\ref{eqn:nucleon_decay2_bis}) \\ \hline

$\vert \l'_{ijk} \l^{\prime \prime \star}_{i'j'k'} \vert $ &&&  
$10^{-9}\ \ [p \to X \bar \nu (X \nu) ]$ \\
& & & $\quad (\tilde m = 1 \TeV)$ (\ref{eqn:nucleon_decay3}) \\ \hline
\hline
%



 
$\vert \l ^{''\star }_{232} \l'' _{231} \vert $ & & $ \ {\rm Min}[6. \times  
10^{-4} \tilde c , \ 2. \times 10^{-4} \tilde c ^2] $  & \\ 
& & $ \qquad  [K\bar K]$ &  
\\ $\vert \l ^{''\star }_{332} \l'' _{331} \vert $ && $ 
{\rm Min} [ 6. \times 10^{-4} \tilde t , \ 3. \times 10^{-4} \tilde t ^2]$ & 
\\ 
& & $\qquad [K\bar K]$& \\ 
\hline 
  
$\vert \l''_{i13} \l^{\prime \prime \star}_{i12} \vert $ &&
$ \ 6.4 \times 10^{-3} \tilde u_{iR}^2 \ [B^+ \to K^0 \pi^+ ]$
(\ref{eqn:btokpi}) & \\ \hline 

$\vert \l^{\prime \prime \star}_{i23} \l''_{i12} \vert $ &&
$ \ 6 \times 10^{-5} \tilde u_{iR}^2 \ [B^- \to \phi \pi^- ]$
(\ref{eqn:B_phi_pi}) & \\ \hline
 
$\vert \l''_{213} \l^{\prime \prime \star}_{232} \vert $
& & $ {\cal I} \ 10^{-2} \tilde q^2 \ [d_n ^\g]$
(\ref{eq:EDM_neutron_1}) & \\ \hline 

$\vert \l''_{312} \l^{\prime \prime \star}_{332} \vert $
& & $ {\cal I} \ 10^{-1} \tilde q^2 \ [d_n ^\g]$
(\ref{eq:EDM_neutron_2}) & \\ \hline 

$\vert \l^{\prime \prime \star}_{i3k} \l''_{i2k} \vert $ && $ 0.16
\tilde q_R^2 \ [ B \to K \g ] $ (\ref{eqn:btosgamma}) & \\ \hline 
\end{tabular} 
\end{center}

\caption{{\it Quadratic coupling constant product bounds.
           We use the same conventions as in the preceding
	   table.}} 

\label{table65} 
\end{table} 
 
It appears clearly from the tables that the low energy phenomenology 
is a rich and valuable source of information on the \Rp\ interactions. 
The most robust cases include the single nucleon decay 
channels, neutrinoless double beta decay, double nucleon decay, the neutral 
$ K, \ B$ meson mixings  \index{Mixing!neutral mesons}
and rare leptonic or semileptonic decays, the lepton 
number and/or flavour violating decays of leptons.   
The strongest single coupling constant bounds arise, in order of decreasing 
strength, from the baryon number violating processes of $n-\bar n$  
oscillation and $NN$ decay, from neutrinoless double beta decay, 
neutrino masses, semileptonic decays of $K$ mesons, and from 
neutral current (APV) and charged current lepton universality.   
For the quadratic coupling constant bounds, a similar classification ordered 
with respect to decreasing strength, places baryon number violating 
processes in first position, followed by $K$ and $B$ meson 
\index{Mixing!neutral mesons} mixing, 
$\mu \rightarrow e$ conversion, leptonic or semileptonic decays 
of $K$ and $B$ mesons, 
neutrinoless double beta decay, three-body lepton decays, 
and neutrino masses. 
 
The consideration of the loop level contributions is a very effective 
way to deduce complementary bounds on coupling constants involving the 
heavier generations of quarks and leptons.  For instance, the single 
nucleon stability bounds 
$ \vert \l'_{ijk} \l^{\prime \prime \star}_{i'j'k'} \vert < {\cal O} (10^{-9}) $, 
which are valid for all the generation indices, have far-reaching 
implications.  Should a single lepton number violating coupling 
constant $\l ' _{ijk} $ be sizeable, then one would 
conclude  a strong suppression of the full set of baryon number 
violating $ \l '' _{i'j'k'} $ coupling constants.  An analogous 
converse statement would hold for all the $\l ' _{ijk} $ if a single $ 
\l '' _{i'j'k'} $ were sizeable.  
 
 
%
\subsection{Observations on the Bound Robustness and Validity} 

\label{subsec:mcd1} 
 
Before discussing the impact on supersymmetric model building of the 
indirect bounds, we start with some general preliminary observations 
aimed at appreciating their potential usefulness.   
 
\noindent \addtocounter{sss}{1} 
{\bf \thesss Natural Order of Magnitude for \Rp\ Couplings} 
 \addcontentsline{toc}{subsection}{\hspace*{1.2cm} \alph{sss})  
             Natural Order of Magnitude for \Rp\ Couplings} 
 
First, it is 
important to ask what might be considered as natural values for the 
dimensionless $R$-parity violating trilinear interactions coupling constants.
(For convenience, we shall denote these collectively as $\hat \l_{ijk}$.) 
In the absence of any symmetry, the anticipated natural values are
${\cal O} (1)$ or ${\cal O} (g)$.  If one assumed instead a hierarchical structure 
with respect to the quarks and leptons generations, analogous to that 
exhibited by the regular $R$-parity conserving Yukawa interactions, an 
educated guess could be, for instance, $\hat \l_{ijk} = {\cal O} ((m_im_jm_k 
/v ^3)^{1/3}) $, where $ m_j$ denote the $q, l $ masses.  A variety  
of alternative forms for the generation dependence are 
suggested by considerations based on the physics of discrete 
symmetries or grand unified and string theories.  Just on the basis of 
the existing experimental constraints, one can check that the 
individual coupling constant bounds, as given in Table~\ref{singletab}
fall in an interval of values ${\cal O} (10^{-1}) \ - \ {\cal O} (10^{-2} )$ which 
interpolates between the above two extreme estimates.

\noindent \addtocounter{sss}{1} 
{\bf \thesss Impact of the SUSY Masses on the Bounds} 
 \addcontentsline{toc}{subsection}{\hspace*{1.2cm} \alph{sss})  
             Impact of the SUSY Masses on the Bounds} 
 
A second observation concerns the dependence of the indirect bounds 
with respect to the superpartner mass parameters.  Our reference value 
for the supersymmetry breaking mass parameter is set uniformly at $ 
\tilde m = 100 \GeV$, apart from a  very  few exceptions.
The tree level mechanisms have, of 
course, a transparent dependence on the superpartners masses, such 
that the single or quadratic  coupling constant bounds scale linearly or  
quadratically 
with $\tilde m$.  Several tree level dominated observables involve a 
single superpartner species, which then allows us to identify the 
relevant sfermion by indicating explicitly its particle name.  The 
bounds associated with one-loop effects have, in general, a weaker 
mass dependence over the mass interval, $ \tilde m = 100 \ - \ 500
\GeV$.  With increasing values of the superpartner masses, both the 
tree and loop level bounds gradually get weaker.  In fact, if the 
supersymmetry breaking mass scale happened to reach  the
${\cal O} (1) \TeV$
extrem  limit  for the  Standard Model  naturalness, a 
large number of the existing individual indirect bounds would become 
useless.  For the so-called "More Minimal Supersymmetric Standard  
Model"~\cite{cohen} where the third generation squarks or sleptons constitute 
the lightest scalar superpartners and the first and second generation 
sfermion masses are raised up to the TeV scale, one would be led to 
strongly weakened bounds for the first two generations of sfermions. 
Several quadratic bounds would still remain of interest, especially 
those associated with a simultaneous $ B $ and $L$ number violation, 
due to the extreme severity of the nucleon stability bounds. 
 
There are two other important exceptions to the suppression effect 
of the bounds from large sfermions masses.  
The first concerns the process independent bounds derived from 
the renormalisation group considerations, 
as was discussed in chapter~\ref{chap:evolution}.   
Because the perturbative unitarity or 
quasi-fixed points bounds originate from indirect effects associated 
with the resummation of large logarithms, they are practically 
insensitive to the value of the supersymmetry breaking scale as long 
as this does not extend  beyond the TeV decade.  The second exception concerns  
the 
class of observables governed by dimension nine operators, such as the 
amplitudes for the $\b \b _{0\nu }$ or $n-\bar n$ processes, where 
several contributions from different intermediate states, associated 
with sfermions and gauginos, compete with one another.  The 
destructive interference of these contributions renders the inferred 
bounds sensitive to the supersymmetry breaking mass spectrum as a 
whole. 

\noindent \addtocounter{sss}{1} 
{\bf \thesss Validity of the Assumption of one or two Dominant 
  Couplings} 
 \addcontentsline{toc}{subsection}{\hspace*{1.2cm} \alph{sss})  
             Validity of the Assumption of one or two Dominant 
  Couplings} 
 
A third observation concerns the validity of the single or 
double coupling constant dominance hypotheses.  When applied to the 
\Rp\ interactions, the dominance hypotheses rest on the premise that 
some hierarchy exists either between the $ B$ and $L$ number 
violations or between the different quark and lepton generations. 
The conclusions from certain studies might be altered if the dominance 
hypotheses were not justified or if certain unexpected finely tuned 
cancellations were at work.  One could imagine, for example, that a 
subset of the coupling constants exhibited generational degeneracies 
that would induce cancellations between the contributions of different 
component interactions part of the predominant subset.  An indirect 
evidence for a possible correlation between different coupling 
constants is furnished by the observation that the strongest 
constraints arise for quadratic products rather than the individual 
coupling constant bounds. This is clearly not surprising since the 
latter entail less demanding model-dependent assumptions.  
 
 An examination of Table~\ref{singletab} reveals that a few amongst the 
charge current and neutral current single coupling constant bounds 
are immune to invalidating cancellation effects.  Examples of robust 
bounds comprise for the charged current interactions,  
those deduced   from  the renormalised 
observable parameters $ G _\mu $ and $ M_W$, and for the neutral 
current interactions, those deduced from the  
forward-backward asymmetry parameter $ 
A_{FB}$, and the auxiliary parameter, $ C_2 (d)$.  Partly responsible 
for this state of affairs is the  use of ratios of rates or branching 
fractions. While the comparison with 
experimental data for such ratios removes 
the dependence on some poorly known hadronic matrix elements 
parameters, this has the drawback of introducing cancelling 
contributions.  As a result, these ratios obtain corrections from 
different \Rp\ interactions which often combine together 
destructively,  with opposite 
signs.  The quadratic  coupling constant bounds are exposed to a much lesser  
extent to 
cancellations since they are often derived for observables where the 
contributions from different sets of coupling constants add up 
incoherently. 
 
\noindent \addtocounter{sss}{1} 
{\bf \thesss Bound Robustness in Regards to Model Dependence} 
 \addcontentsline{toc}{subsection}{\hspace*{1.2cm} \alph{sss})  
 Bound Robustness in Regards to Model Dependence} 
 
As a final remark we should emphasise that not all  coupling constant  
bounds are to be 
treated indiscriminately.  One must exercise a critical eye on the 
model-dependent assumptions.  It is important to keep track of the 
superpartner generation index in light of the possibility of a large 
splitting between the sfermion generations.  The generational 
structure of the sfermion chirality-flip mass matrices 
$\tilde m^2_{\scriptscriptstyle{LR}}$ is a crucial input  for 
the one-loop contributions to the neutrinos Majorana masses or the 
$n-\bar n$ oscillation amplitude.  Deviations from a generation 
universality yield large off-diagonal contributions 
$(\tilde m^2_{\scriptscriptstyle{LR}})_{ij}$ which could modify 
the ensuing predictions. 
 
\subsection{Phenomenological Implications of Bounds} 
\label{subsec:mc3}

What implications on theoretical models beyond the \SM\ 
can be drawn from the existing bounds?  
As discussed in chapitre~\ref{chap:evolution}, works have been
done on model building using renormalization group equations
(RGE) to get bounds at the $M_{GUT}$ scale~\cite{allner}.
One might ask whether the results hint at any of the known 
alternatives, especially the  gauged horizontal continuous  
or discrete symmetries. The existence of some correlations with the 
flavour symmetries is clearly suggested. Indeed, it is generally the 
case that the coupling constants in the first and second generations 
are more constrained, although this might just reflect the lack of 
direct experimental data for the heavy flavoured hadrons or leptons. 
The fact that the strongest bounds are for products $ \l ' \l '' $ and 
$\l \l ''$ hints at an incompatibility between simultaneous $ B$ and 
$L$ number violations.  Separate $ B$ or $L$ number violation, as 
described by interactions governed by the coupling constant products 
$ 
\l \l , \ \l \l' , \ \l' \l' $ or $\l'' \l'' $, 
may also be disfavoured but in a way which depends on flavour, the first 
and second generations being those most strongly constrained. 
Ready solutions to prevent a coexistence of $B$ and $ 
L$ number non-conservation are offered by the $B$ and $L$ parities or 
the corresponding generalised discrete symmetries versions. 
 
The existence of a strong hierarchical structure in the \Rp\ coupling 
constants does not exclude the presence of certain unexpected 
degeneracies with respect to the quarks and leptons generations, as 
would be implied by the presence of unbroken discrete symmetries.  To establish 
this possibility one would need global studies of the \Rp\ interactions 
effects encompassing a large body of experimental data.  One way to 
infer robust bounds in cases where one suspects cancellation effects 
to be at work, is by choosing suitable observables which depend 
selectively on fixed products of the coupling constants.  Such an 
example was encountered with the \Rp\ induced contributions to the non 
$V-A$ charged current interactions (see section~\ref{secxxx2a}).  
Another attractive idea would be 
to fit a selected subset of the \Rp\ interactions coupling constants to 
a correspondingly selected subset of experimental constraints.  While 
global studies along these lines are routinely performed in the 
context of the contact interactions physics~\cite{bargerz} or the 
mirror fermions physics~\cite{londonlanga} their application to the 
\Rp\ physics appears problematical in view of the proliferation of the 
coupling constants.  
Interesting partially global  studies of the \Rp\ interactions  have 
been recently reported  in the literature  regarding fits to the 
APV observables~\cite{barger00}  
 or the $Z$ boson partial decay width  observables~\cite{lebedev99}.  
 The recent accumulated experimental information 
on the neutrino oscillations has  also  allowed to implement in part such a 
program by envisaging global fits to the data for the neutrino
Majorana masses based on the \Rp\ contributions.  Even if these studies 
must appeal to some assistance from theory, through certain 
specialised assumptions on the generational structure of the sfermion 
mass parameters, they have yielded a wealth of useful, although 
model-dependent, information on the \Rp\ interactions.  Of course, one 
should keep in mind the alternative options to explain the neutrino 
physics experimental data, which include in fact the \Rp\ mechanism of 
flavour changing neutral current neutrino interactions with  quarkonic 
and leptonic matter. 
 
The model-independent studies devoted to the four fermions contact 
interactions may be of some use in establishing the existence of 
possible cancellation effects amongst different \Rp\ interactions. 
Special attention, motivated by searches of 
compositeness~\cite{contact,eichten,hagiwara}, has been devoted in 
recent years to the flavour diagonal contact interactions.  Along with 
the low energy neutral and charged current interactions (neutrino or 
(polarised) electron elastic and inelastic scattering data) the 
current fits~\cite{bargerz,zarnecki} include the high energy data for 
the Drell-Yan dilepton production and large $p_T$ jet production at 
the Fermilab Tevatron collider, the dijet production (e.g., $e^-e^+\to 
s\bar s $) at the CERN LEP \index{LEP} collider, and the deep inelastic electron 
and positron scattering at the  HERA collider at DESY.  Based on the 
initial studies~\cite{contact}, one generally restricts the
consideration to the dominant $\cddd =6$ Lorentz vector interactions, 
using an effective Lagrangian of the general form, 
\begin{eqnarray}  
 {\cal{L}}_{NC}&=& \sum_{[(i,j)=(L,R); q=(u,d)] } {4\pi 
\eta_{ij}^q\over \L _{ij}^{q 2} } \bar e_i\g_\mu e_i \bar q_j \g_\mu 
q_j , 
\label{eqaz1} 
\end{eqnarray} 
 where a sum over the fermions flavour and chirality is understood and 
$\eta ^q =\pm 1$ stand for a sign phase.  One may express the 
relationship between the different scale parameters, denoted generically by $\L 
$, and the \Rp\ interactions coupling constants by the order of 
magnitude relation, $\hat \l ^2 /\tilde m^2 \approx 4\pi / \L ^2 $. 
More quantitatively, an identification with the neutral current 
interactions, for instance, yields~\cite{wilczhera},  
$$ 
 C_1(q) ={\sqrt 2 \pi \over G_F} 
( {\eta_{RL} ^q\over \L _{RL}^{q2} } - {\eta_{LL} ^q\over \L _{LL}^{q 
2} } - {\eta_{LR} ^q\over \L _{LR}^{q2} }+ {\eta_{RR} ^q\over \L 
_{RR}^{q2} }). 
$$ 
An important observation here is that the high energy collider experimental   
data 
favour low values of the energy scales.  Some currently quoted 
experimental bounds are,  
\mbox{$\L_{[LR,RL]}^{[-,+]d}> [1.4 ,$} \ \mbox{$  1.6] \ \ \text{TeV} $ },  
from the dijet production data~\cite{lepcol}, 
\mbox{$\L_{[LR,RL]}^{[+,+]u}>[ 2.5 ,\ 2.5] \ \ \text{TeV}  $ } 
from the Drell-Yan production data~\cite{abe} and  
\mbox{$\L \approx 1 \ \text{TeV}  $} from the 
anomalous deep inelastic scattering events 
data~\cite{altarelli,hera,alta1}.  In contrast to these results, it 
appears that the low energy experimental data consistently favours 
larger values of 
the energy scales.  This is most explicit in the Cesium atom APV data, 
where assuming that no cancellations occur between the different terms 
in $ C_1 (q)$, leads to the strong bound, $ \L > 10 \ \text{TeV}  $. More 
quantitatively, the simultaneous fits of the flavour diagonal $ eeqq$ 
contact interactions to both low and high energy experimental data, as 
completed by incorporation of the HERA high $ Q^2$ 
data~\cite{bargerz}, infer large values of the scale parameters with a 
non-trivial trend of relative signs between the different 
interactions. Quoting from ~\cite{bargerz} one finds the fitted 
values:  
$$ 
\L_{LL} ^{-eu} = 12.4 {+50.6 \choose - 34.8} \ \ \text{TeV}  ,\ \ 
\L_{LR} ^{+eu} = 3.82 {+0.93 \choose - 1.62}  \  \ \text{TeV}   ,\ \   
\L_{RL} ^{+eu} = 5.75  {+5.06 \choose - 6.88}  \  \ \text{TeV}   .\ \   
$$ 
These quantitative analyses indicate that cancellation effects are 
taking place at low energies between the contributions from different 
interactions.  Such cancellations would clearly pass unnoticed within 
analyses based on a single coupling constant dominance hypothesis.  Tentative 
explanations have been sought in terms of a short distance parity 
conserving interaction~\cite{wilczhera}, implying the relations $ { 
\eta_{iL}^q \over (\L ^q_{iL } )^2} =- { \eta_{iR}^q \over (\L ^q_{iR 
} )^2} $, or an extended global flavour symmetry group~\cite{nelsoncc}. 
Applied to the \Rp\ interactions, the implications would be in the existence  
of degeneracies amongst the subset of relevant \Rp\ coupling constants. 
 
The Lorentz vector component of the charged current (CC) electron-quark four  
fermions contact interactions appears also to lead to similar conclusions.   
The fits to the leptonic or hadronic colliders data based on the conventional  
parametrisation of the effective Lagrangian, 
$$ 
 {\cal{L}}_{CC}= {4\pi \eta \over \L _{CC}^{\eta 2} }  
 \bar e_L\g_\mu \nu_L \bar u_L  
  \g_\mu d'_L  
$$ 
yield the typical bounds, $\L^-_{CC} > 1.5 \ \ \text{TeV}  $.  
The recent deep inelastic scattering events observed at HERA also favour  
low scales, $\L _{CC} = (0.8\ - \ 1 ) \ \text{TeV}  $~\cite{altarelli,alta1}.   
By contrast, the fits to the low energy experimental data associated with  
the leptons universality or for meson decays favour larger  
scales~\cite{alta1}, $\L_{CC} \approx (10. \ - \ 30. ) \ \text{TeV}  $.   
A recent study of the non $ V-A$ charged current interactions, based on  
the high precision measurements of the muon decay rate differential  
distributions, also predicts strong bounds for the  
scales~\cite{hagiwara97}, $ \L^{\pm ll} _{CC } > [7.5 , \ 10.2]  \ \text{TeV}   
, $ for the  
four lepton interactions and $ \L^{\pm lq} _{CC } > [5.8 , \ 10.1] \  
\text{TeV}  $ 
for the two lepton two quark interactions. 
\index{Bounds on \Rp\ interactions!Trilinear Terms|)} 
%

  
\cleardoublepage                                                         %
\chapter{PHENOMENOLOGY AND SEARCHES AT COLLIDERS}                        %
\label{chap:colliders}                                                   %

\section{Introduction}
\label{sec:collintro}
The search for \Rp\ supersymmetry processes has been a major analysis 
activity at high energy colliders over the past decade, and is likely 
to remain so at existing and future colliders unless the idea of 
supersymmetry itself somehow becomes falsified.

We have seen in chapters~\ref{chap:intro} and~\ref{chap:theory} that
on the theory side, \Rp\ is (and will) remain a central issue  since gauge
invariance and renormalizability do not ensure lepton- and baryon-number 
conservation in supersymmetric extensions of the Standard Model. 
A consequence is that a general superpotential allows for trilinear terms 
corresponding to \Rp\ fermion-fermion-sfermion interactions involving 
$\lambda, \lambda'$ or $\lambda''$ Yukawa couplings.
It moreover possibly allows for additional {\it explicit} (bilinear)
or {\it spontaneous} sources of lepton-number violation.

The presence of \Rp\ interactions could have important consequences on the
phenomenology relevant for supersymmetry searches at high energy colliders.
This is because \Rp\ entails a fundamental instability of 
supersymmetric matter, thus opening up new decay channels for sparticles.
Especially crucial in this respect will be the fate of the lightest 
supersymmetric particle (LSP). 
Even for relatively weak \Rp\ interaction strengths, the decay of the 
LSP will lead to event topologies departing considerably from the
characteristic "missing momentum" signal of $R_p$ conserving theories.
But \Rp\ could be more than a mere observational complication.
It could also enlarge the discovery reach for supersymmetry itself 
as it allows for the creation or exchange of single sparticles.

In this chapter, essential ingredients of the phenomenology and search
strategies for \Rp\ physics at colliders are presented. Extensive 
references to related detailed studies for specific supersymmetry models
are provided. Existing experimental constraints established at LEP 
\index{LEP} $e^+e^-$, HERA $ep$ and Tevatron $p\bar{p}$ colliders are reviewed 
and discovery prospects in future collider experiments are discussed. 
\begin{table}
   \begin{center}
   \begin{tabular}{c|c|r|c|c}
   \hline
     Collider & Beams  & $\sqrt{s}$ \,\,\, & $\int {\cal{L}} dt$ & Years \\
   \hline
    LEP 1   & $e^+ e^- $ & $M_Z$ 
                              & $\sim 160 \picob^{-1} \otimes 4 $ & 1989-95 \\ 
    LEP 2   & $e^+ e^- $  & $> 2 \times M_W $ 
                             & $\sim 620 \picob^{-1} \otimes 4 $ & 1996-00 \\ 
   \hline
    HERA~Ia & $e^{-} p$ & $300 \GeV$ 
                               & ${\cal{O}} (1 \picob^{-1}) \otimes 2$ & 1992-93  \\ 
    HERA~Ib & $e^{\pm} p$ & $\lsim 320 \GeV$ 
                     & ${\cal{O}} (100 \picob^{-1}) \otimes 2$ & 1994-00 \\ 
   \hline
    Tevatron Run Ia & $p \bar{p}$ & $1.8 \TeV$  
                             & ${\cal{O}} (10 \picob^{-1})$   & 1987-89  \\
    Tevatron Run Ib & $p \bar{p}$ & $1.8 \TeV$  
                    & ${\cal{O}} (100 \picob^{-1}) \otimes 2$ & 1992-96  \\ 
   \hline \hline 
    HERA~II & $e^{\pm}_{L,R} p$ & $\sim 320 \GeV$ 
                            & $\sim 1 \femtob^{-1} \otimes 2$  & $\geq$ 2002 \\
   \hline
    Tevatron Run II & $p \bar{p}$ & $\sim 2.0 \TeV$  
                                  & $1 - 10 \femtob^{-1} \otimes 2$
                                                           & $\geq$ 2002 \\
   \hline
    LHC     & $pp$ & $14.0 \TeV$  & 
                        $10 - 100 \femtob^{-1} \otimes 2$ & $\gsim$ 2007 \\
   \hline \hline 
   Future LC     &  $e^+ e^- $ & $\sim 0.5 - 1.0 \TeV$  & 
                                 $50 \femtob^{-1}$  & $\ldots$ NLC~\cite{NLCworks} \\
                 &             & $\sim 0.5 - 1.0 \TeV$  & 
                                 $500 \femtob^{-1}$ & $\ldots$ TESLA~\cite{TESLAworks} \\
   \hline
   Future $\mu$C &  $\mu^+ \mu^- $ & $\sim 0.35 - 0.5 \TeV$  & 
                                  $10 \femtob^{-1}$ & $\ldots$  FMC~\cite{FMCNMCworks} \\
                 &                 & $\sim 1.0 - 3.0 \TeV$  & 
                                  $1000 \femtob^{-1}$ & $\ldots$ NMC~\cite{FMCNMCworks} \\
   \hline
   \end{tabular}
   \end{center}
   \caption{\label{tab:colliders}
            {\it Main contemporary and future collider facilities which are
	    considered in the search analyses and prospective studies described
	    in this chapter. The facilities are listed together with the nature
	    of the colliding beams, the available centre-of-mass energies $\sqrt{s}$, 
	    and the integrated luminosities $\int {\cal{L}} dt$ accumulated 
	    (or the range of $\int {\cal{L}} dt$ expected) per experiment.
	    The multiplicative factors after the $\otimes$ sign denotes 
	    the number of multi-purpose collider experiments operating 
	    (or expected to be operating) simultaneously around each 
	    collider.}}
\end{table}
The analyses and prospective studies in the literature have generally been
carried in the context of a given existing of future collider project.
The Table~\ref{tab:colliders} gives a list of the machine parameters 
considered in the studies reviewed in this chapter.

\section{Interaction Strength and Search Strategies}
\label{sec:collstrat}
The way supersymmetry could become manifest at colliders crucially 
depends both on the structure and parameters of the model followed by
Nature and on the a priori unknown magnitudes (individual and relative) 
of the new \Rp\ couplings. The weakest \Rp\ coupling values are likely to 
be felt mostly through the decay of sparticles otherwise pair produced 
via gauge couplings. The strongest \Rp\ coupling values could contribute 
to direct or indirect single sparticle production.
The best search strategy at a given collider will ultimately depend on 
the specific signal and background environment.

In the absence of definite theoretical predictions for the values 
of the 45 independent trilinear Yukawa couplings $\Lambda$
($\lambda_{ijk}$, $\lambda'_{ijk}$ and $\lambda''_{ijk}$), and 
facing the formidable task of testing $2^{45}-1$ possible 
non-vanishing coupling combinations, it is necessary in practice 
to assume a strong hierarchy among the couplings.
For the "hierarchy" between different types of couplings this is an 
arbitrary choice since $\lambda$, $\lambda'$ and $\lambda''$ appear
fundamentally independent.
Empirically, it can be partially justified by the fact that 
indirect bounds are particularly stringent on non-vanishing 
coupling products involving a \LV\ and a \BV\ coupling as was
seen in section~\ref{secxxx3d}. For example, the lower limit on the 
proton lifetime translates~\cite{smirnov} into very stringent bounds 
on the $\l' \times \l'' < O(10 ^{-9} )$ applicable to all possible 
flavour combinations.
Restrictions on combinations of couplings of a given type can be 
legitimized by analogy with the strong hierarchy of the Higgs 
Yukawa couplings structure in the Standard 
Model~\cite{dreiner1,dimopoulos88}. 
It may also be empirically justified by the fact that indirect 
bounds (chapter~\ref{chap:indirect}) are generally more stringent 
on the product of two different couplings than on the square of 
individual $\l$, $\l'$ or $\l''$ couplings. 

Thus, a reasonable simplifying assumption for the search strategy at
colliders is to postulate the existence of a single (dominant) 
\Rp\ coupling. Most of the prospective studies on \Rp\ and actual
search analyses at colliders rely on this assumption, i.e. that only one
\Rp\ coupling exists which can connect sleptons or squarks to ordinary
fermions. By doing so, it is in addition assumed implicitly or 
explicitly (through some mixing angles~\cite{dreiner1} connecting 
the squark current and mass basis) that flavour mixing relating
various couplings (see section~\ref{subsec:basis}) is suppressed.

Having chosen (somehow) a single dominant coupling $\Lambda$, the next
question is that of the range of coupling values relevant for collider 
physics. As to what concerns lower bounds, cosmology considerations do 
not provide much help.
As discussed in section~\ref{subsec:mc5}, a lightest supersymmetric 
particle (LSP) can no more be considered as a cold dark matter 
candidate in presence of a single non-vanishing \Rp\ Yukawa coupling 
with values even as small as ${\cal{O}}(10^{-20})$.
Strengthened lower bounds of 
$[ \lambda, \lambda',\lambda'' ] > {\cal{O}}(10^{-12})$
are obtained from the argument (section~\ref{subsec:mc5}) that an 
unstable LSP ought to decay fast enough in order not to disrupt 
nucleosynthesis. But even these still lie many orders of magnitude 
below the sensitivity reach of collider experiments.
For $\Lambda$ coupling values below ${\cal{O}}(10^{-8}-10^{-6})$ 
(depending in detail on model parameters), the lifetime 
of the LSP is so large that it is likely to completely escape 
detection in a typical ${\cal{O}}(10)$m diameter collider experiment.
An immediate consequence is that for a wide range of coupling 
values, the phenomenology at colliders would appear indistinguishable
from that of $R_p$ conserving theories. 
Only a discovery that the LSP turns out to be coloured or charged,
a fact forbidden by cosmological constraints for a stable LSP,
could be an indirect hint of the existence of \Rp\ interactions 
beyond the collider realm. 
Otherwise, there exist no known direct observational tests for such 
{\it very long-lived} LSPs~\cite{reviews2}. This inaccessible coupling
range will not be discussed further in this chapter. 

In case a non-vanishing \Rp\ coupling does exist with a magnitude 
leading to distinct phenomenology at colliders, the optimal search 
strategy will then depend on the absolute coupling value and
the relative strength of the \Rp\ and gauge interactions, 
as well of course on the nature of the supersymmetric model 
considered (sparticle spectrum and parameter
space depending on the supersymmetry breaking mechanism, etc.).
Sparticle direct and indirect $\Rp$ decay topologies will be discussed 
on general grounds in section~\ref{sec:sdecays}.
Anticipating this discussion, a direct sensitivity to a {\it long-lived}
LSP might be provided by the observation of displaced vertices in an 
intermediate range of coupling values up to ${\cal{O}}(10^{-5}-10^{-4})$.

For even larger $\Lambda$ values, the presence of \Rp\ supersymmetry 
could become trivially manifest through the decay of {\it short-lived}
sparticles pair produced via gauge couplings. 
A possible search strategy in such cases consists of neglecting
\Rp\ contributions at production (in non-resonant processes).
This is valid provided that the \Rp\ interaction strength 
remains sufficiently small compared to electromagnetic or weak 
interaction strengths, i.e. for $\Lambda$ values typically below 
${\cal{O}}(10^{-2}-10^{-1})$.
Such a strategy has been thoroughly explored at existing colliders to 
study how the experimental constraints on basic model parameters in specific 
supersymmetry models be affected by the presence of \Rp\ interactions.
This and the question of whether and how different types of couplings
could be distinguished at colliders in such a scenario will be discussed
in section~\ref{sec:spair}.

In a similar range of (or for larger) coupling values, $\Rp$ could
manifest itself most strikingly at colliders via single resonant or 
non-resonant production of supersymmetric particles. 
Single sparticle production involving \Rp\ couplings and how it
allows the extension of the discovery mass reach for supersymmetric 
matter at a given collider is discussed in section~\ref{sec:sproduct}.

Coupling values corresponding to interactions stronger than the 
electromagnetic interaction might still be allowed for sufficiently large
masses. For masses beyond the kinematic reach of a given collider, 
\Rp\ could contribute to observable processes through virtual sparticle
exchange. This is discussed in section~\ref{sec:virtuprod}.

Realistic search strategies at colliders must take into account the 
upper bounds on the $\Lambda$ couplings derived from indirect processes.
As was seen in chapter~\ref{chap:indirect}, these bounds all become 
weaker with increasing scalar masses but each possibly with a specific 
functional mass-dependence and each depending on a specific
type of scalar~\cite{reviews2}. 


Ultimately, the question of whether or not a given \Rp\ process is truly 
allowed by existing constraints must be answered at each point of the parameter 
space of a given supersymmetry model. 
A review of the huge number of publications dealing with specific aspects 
of \Rp\ in specific supersymmetry models would clearly be beyond the scope 
of this chapter. In the following, essential aspects of the phenomenology
will be discussed and references to detailed studies provided.

\vspace*{1.0cm}


\section{Decay of Sparticles Involving {\boldmath{$\Rp$}} Couplings}
\label{sec:sdecays}

In the scenario with \Rp\ due to the trilinear terms, the supersymmetric 
particles are allowed to \index{Decays involving \Rp\}decay 
into standard particles through one \Rp\ coupling. 
For sparticles other than the LSP, these $\Rp$-decays will in general 
compete with "cascade decays" initiated by standard gauge couplings.
The review of possibly allowed direct and cascade $\Rp$-decays for 
sfermions and gauginos-higgsinos is presented in this section.
This will later on allow us to easily characterize the essential 
event topologies expected in $\Rp$-SUSY searches.
Direct decays are discussed in subsections~\ref{sec:dirsfdecays} 
and~\ref{sec:dirghdecays}.
Indirect cascade decays are discussed in subsection~\ref{sec:cascdec}.
For completeness, decays involving bilinear interactions are discussed 
in~\ref{sec:mixdec}.

\subsection{Direct {\boldmath{\Rp}} Decays of Sfermions}
\label{sec:dirsfdecays}


The \LLE, \LQD, or \UDD\ couplings allow for a \Rp\ direct 
\index{Decays involving \Rp\ ! sfermions direct}decay 
into two standard fermions of, respectively, sleptons, sleptons 
and squarks, or squarks.
The allowed decays become evident when considering the Lagrangian 
for the trilinear Yukawa interactions written in expended notations 
in Eq.~(\ref{eq:laglambda}) to Eq.~(\ref{eq:laglambdapp}) and 
discussed in more details in Appendix B.
For convenience the corresponding list of decay channels is given
in Table~\ref{tab.sec6.rpv.sfermions.channels}.
The \Rp\ decay of a sfermion of a particular family will be possible 
only for specific indices $i,j,k$ of the relevant Yukawa coupling. 
%
\begin{table}[htb]
\begin{center}
\begin{tabular}{|c|c|c|c|}
\hline
 Supersymmetric & \multicolumn{3}{|c|}{Couplings} \\
 particles & 
 \multicolumn{1}{c}{\lam} & \multicolumn{1}{c}{\lamp} &
                                               \multicolumn{1}{c|}{\lampp} \\
\hline
 $\snu_{i,L}$ & $\ell^+_{j,L} \ell^-_{k,R} $ & $\bar{d}_{j,L} d_{k,R}$ & \\
 $\slep^-_{i,L}$ & $\bar{\nu}_{j,L} \ell^-_{k,R}$ & $\bar{u}_{j,L} d_{k,R}$ & \\
 $\snu_{j,L}$ & $\ell^+_{i,L} \ell^-_{k,R}$ &  & \\
 $\slep^-_{j,L}$ & $\bar{\nu}_{i,L} \ell^-_{k,R}$ & &  \\
 $\slep^-_{k,R}$ & $\nu_{i,L} \ell^-_{j,L}$ , $\ell^-_{i,L}
 \nu_{j,L}$ & &  \\ \hline 
 $\squ_{i,R}$ & & & $\bar{d}_{j,R} \bar{d}_{k,R}$ \\
 $\squ_{j,L}$ & & $\ell^+_{i,L} d_{k,R}$ & \\
 $\sqd_{j,L}$ & & $\bar{\nu}_{i,L} d_{k,R}$ & \\ 
 $\sqd_{j,R}$ & & & $\bar{u}_{i,R} \bar{d}_{k,R}$ \\ 
 $\sqd_{k,R}$ & & $\nu_{i,L} d_{j,L}, \ell^-_{i,L} u_{j,L}$ &
 $\bar{u}_{i,R} \bar{d}_{j,R}$\\ 
\hline
\hline
\end{tabular}
\caption{{\it Direct decays of sleptons and squarks via trilinear 
         $\Rp$ operators \lam\LiLjEk, \mbox{\lamp\LiQjDk} 
         and \mbox{ \lampp\UiDjDk}.}} 
\label{tab.sec6.rpv.sfermions.channels}
\end{center}
\end{table}

The partial widths of $\snu_i$'s when decaying via \mbox{$\lam$\LLE } or 
\mbox{$\lamp$\LQD} are given (neglecting lepton and quark masses)
by~\cite{Barger95}:
\begin{eqnarray}
\Gamma (\snu_i \rightarrow \ell^+_j \ell^-_k) = {1 \over 16 \pi} \lam ^2
{\rm m_{\snu_i}} \,\, , \label{larg1}\\
\Gamma (\snu_i \rightarrow \bar{d}_j d_k) = {3 \over 16 \pi} \lamp ^2
{\rm m_{\snu_i}} \,\, , \label{larg2}
\end{eqnarray}
where the factor 3 between relations~\ref{larg1} and~\ref{larg2} is a 
colour factor.
For sneutrinos undergoing direct decays, the mean decay length $L$ in 
centimetres can then be numerically estimated from:
\begin{eqnarray}
 L(\rm cm) = 10^{-12} (\beta \gamma) 
  \left({ 1 {\rm GeV} \over \rm{m}_{\snu_i} } \right)
 {1 \over \lam ^2 , 3 \lamp ^2} \, .
\end{eqnarray}
Similar formulae hold for charged sleptons in absence of mixing.

In the case of squarks or sleptons of the third generation, a possible 
mixing between supersymmetric partners of left- and right-handed fermions
has to be taken into account.
For instance, the width of the lightest stop $\sqt_1$ writes~\cite{Kon94}:
\begin{eqnarray} 
\Gamma ( \sqt_1 \rightarrow \ell^+_i d_k) = {1 \over 16 \pi} \lamp ^2
                               cos^2 (\theta_t) {\rm m}_{\sqt_1} \, ,
\end{eqnarray}
where $\theta_t$ is the mixing angle of top squarks. 
The case of the stop is somewhat special. 
The typical decay time of a $100 \GeVcc$ stop via a \Rp\ decay 
mode is roughly $3 \times 10^{-23}$~s for a coupling value of 10$^{-1}$,
and $3 \times 10^{-21}$~s for a coupling value of 10$^{-2}$. 
So the stop \Rp\ decay time is of the same order or even greater 
than its hadronization time which from the strong interaction is
${\cal{O}}(10^{-23})$~s. Thus, the stop may hadronize before it decays.

\subsection{Direct {\boldmath{\Rp}} Decays of Gauginos-Higgsinos}
\label{sec:dirghdecays}


In a direct \Rp\ decay, the neutralino (chargino) decays into a fermion 
and a virtual sfermion with this virtual sfermion subsequently 
\index{Decays involving \Rp\ ! gauginos-higgsinos direct}decaying to standard 
fermions via a \Rp\ coupling.
Thus, direct \Rp\ decays of gauginos-higgsinos are characterized by three 
fermions in the final state with the fermion type depending on the dominant 
coupling. The possible decays are listed in 
Table.~\ref{tab.sec6.rpv.gauginos.channels}.
\begin{table}[htb]
\begin{center}
\begin{tabular}{|c|c|c|c|}
\hline
Supersymmetric & \multicolumn{3}{|c|}{Couplings} \\
particles & \multicolumn{1}{c}{\lam} & \multicolumn{1}{c}{\lamp} &
                                       \multicolumn{1}{c|}{\lampp} \\
\hline
\XO & $\ell^+_i \bar{\nu}_j \ell^-_k~,~\ell^-_i \nu_j
 \ell^+_k,$ & $\ell^+_i \bar{u}_j d_k~,~\ell^-_i u_j
 \bar{d}_k , $ & $\bar{u}_i \bar{d}_j \bar{d}_k~,
u_i d_j d_k$ \\
 & $\bar{\nu}_i \ell^+_j \ell^-_k~,~\nu_i \ell^-_j
 \ell^+_k $ &  $ \bar{\nu}_i \bar{d}_j d_k~,~\nu_i
 d_j \bar{d}_k$ & \\
\hline
\XP & $\ell^+_i \ell^+_j  \ell^-_k~,~\ell^+_i \bar{\nu}_j
 \nu_k$ & $\ell^+_i \bar{d}_j d_k~,~\ell^+_i \bar{u}_j u_k$ & $u_i d_j
 u_k~,~u_i u_j d_k $ \\
 &  $\bar{\nu}_i \ell^+_j \nu_k~,~\nu_i \nu_j \ell^+_k$ & $\bar{\nu}_i
\bar{d}_j u_k~,~\nu_i u_j \bar{d}_k$ & $\bar{d}_i \bar{d}_j
 \bar{d}_k$\\ 
\hline 
\end{tabular}
\caption{{\it Direct decays of neutralinos and charginos with
 trilinear $\Rp$ operators \lam\LiLjEk,
 \mbox{\lamp\LiQjDk} and \mbox{ \lampp\UiDjDk}.}} 
\label{tab.sec6.rpv.gauginos.channels}
\end{center}
\end{table}
The corresponding diagrams are shown for the $L_{i}L_{j}E^c_{k}$ interactions
in Figs.~\ref{fig:chi0direct} and~\ref{fig:chipdirect}.
\begin{figure}[htb]
\begin{center}

\begin{tabular}{ccc}
\vspace*{-1.0cm}

\hspace*{-0.5cm} \epsfig{file=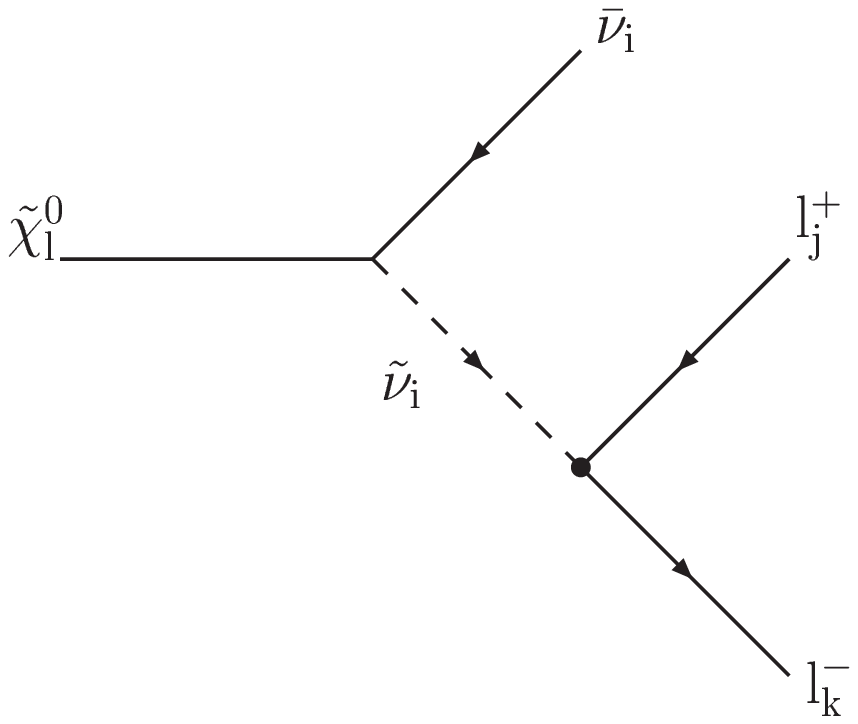,width=0.35\textwidth} &
\hspace*{-1.0cm} \epsfig{file=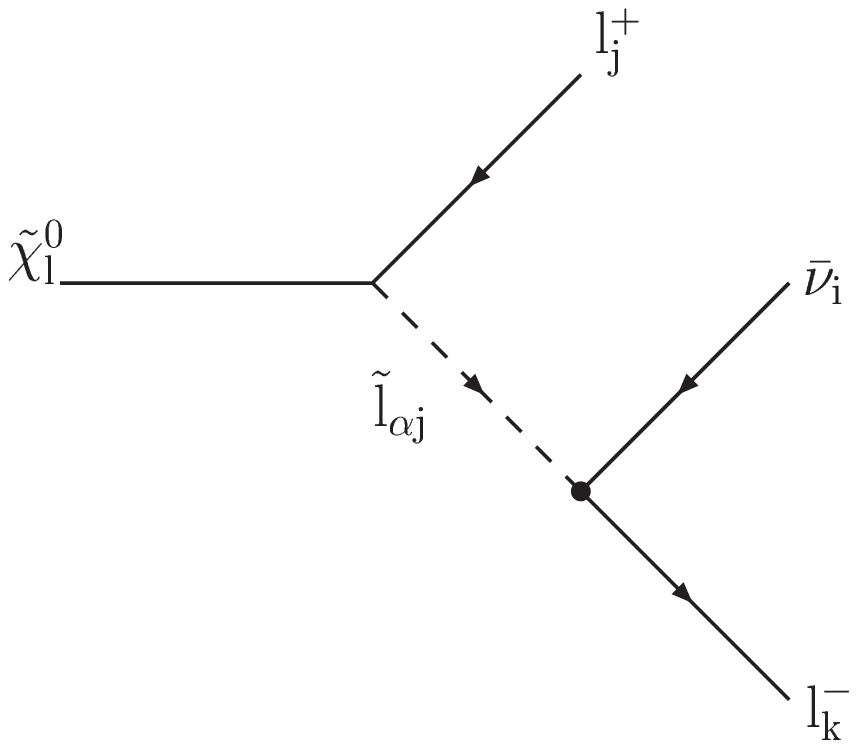,width=0.35\textwidth} &
\hspace*{-1.0cm} \epsfig{file=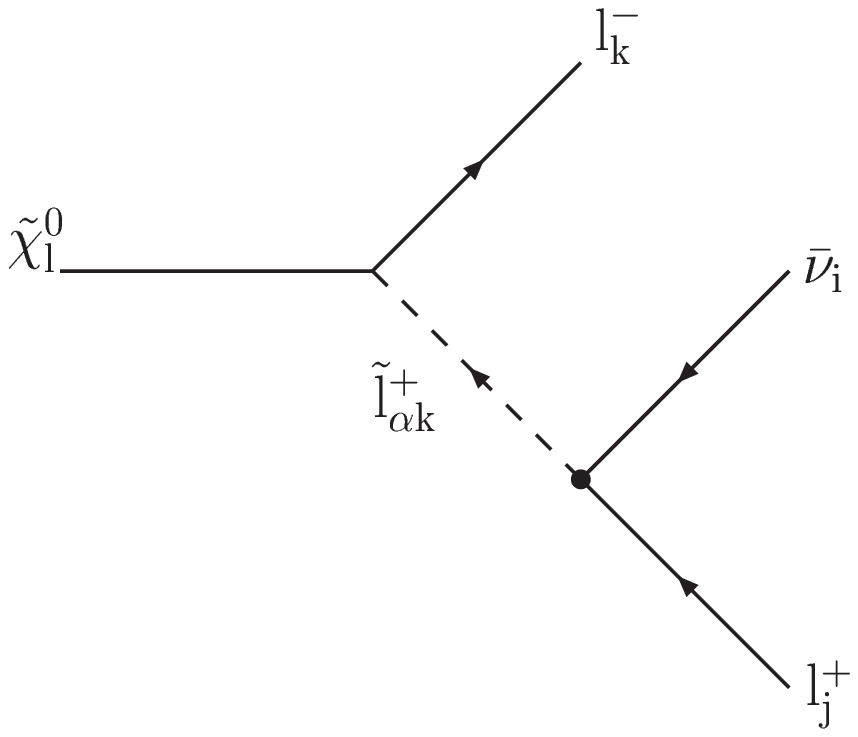,width=0.35\textwidth} \\
\end{tabular}
\vspace*{0.7cm}

 \caption{{\it Diagrams for the direct decays of the neutralino $\tilde{\chi}^0_l$
               via the coupling $\l_{ijk}$ of the $\Rp$ trilinear  
	       $L_{i}L_{j}E^c_{k}$ interaction. The index $l = 1 \ldots 4$ 
	       determines the mass eigenstate of the neutralino. The indices
	       $i,j,k = 1,2,3$ correspond to the generation. Gauge invariance
	       forbids $i=j$. The index $\alpha = 1,2$ gives the slepton mass
	       eigenstate (i.e. the chirality of the SM lepton partner in absence
	       of mixing).}} 
 \label{fig:chi0direct}
 \end{center}
 \end{figure}
%
\begin{figure}[htb]
\begin{center}

\begin{tabular}{ccc}
\vspace*{-1.0cm}

\hspace*{-0.5cm} \epsfig{file=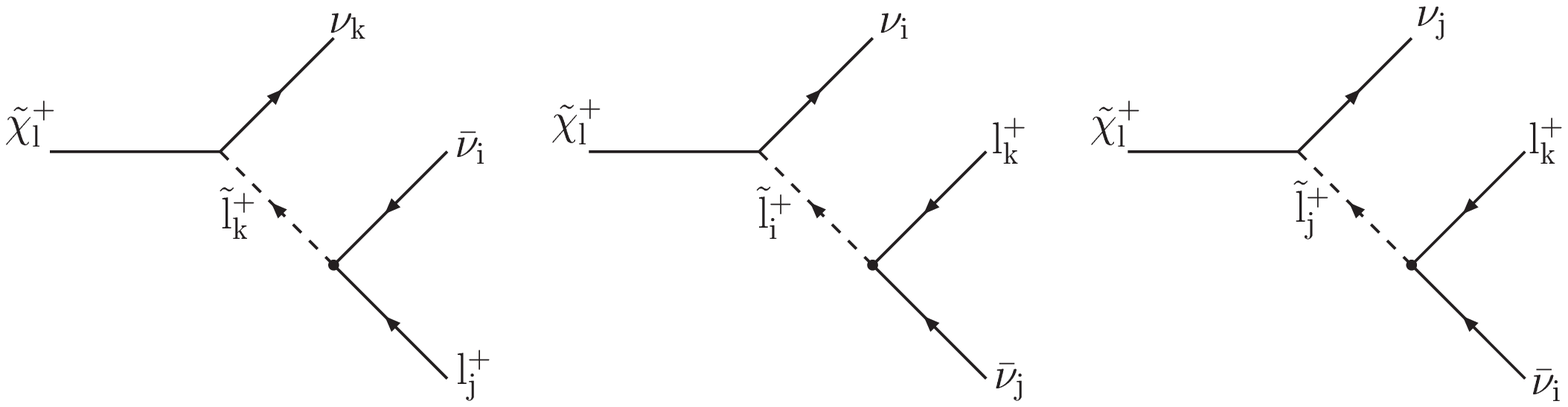,width=1.0\textwidth} \\
\end{tabular}

\vspace*{1.0cm}

\begin{tabular}{cc}

\hspace*{-1.5cm} \epsfig{file=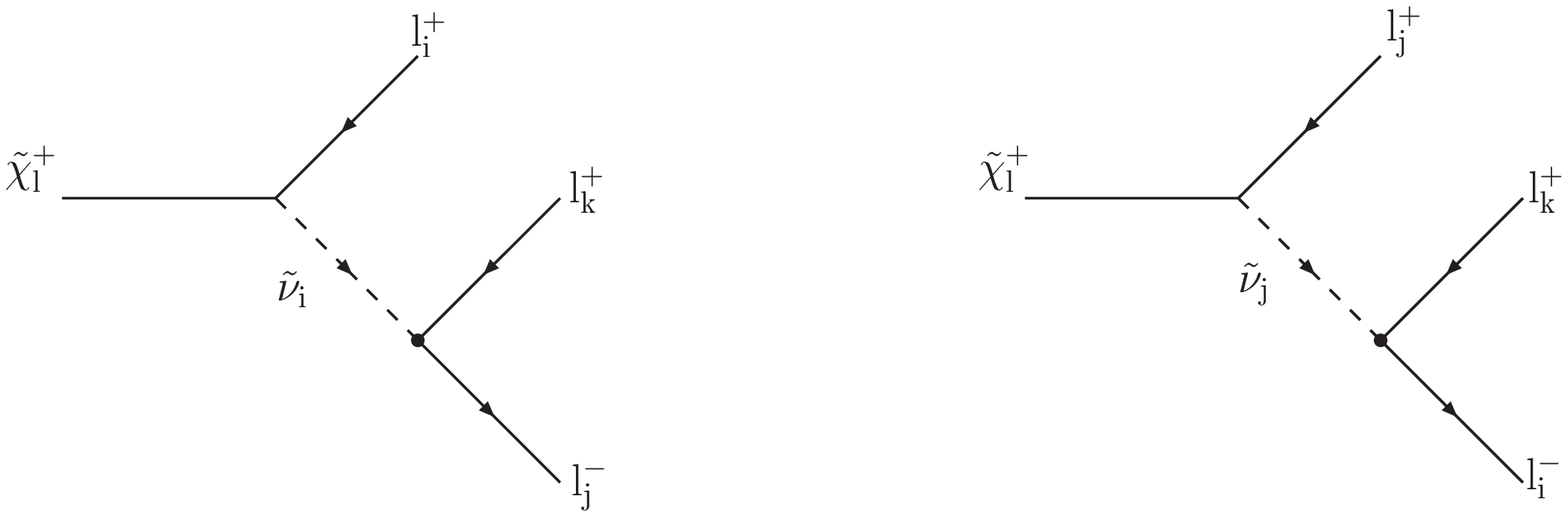,width=0.80\textwidth} \\
\end{tabular}


 \caption{{\it Diagrams for the direct decays of the chargino $\tilde{\chi}^+_l$
               via the coupling $\l_{ijk}$ of the $\Rp$ trilinear  
	       $L_{i}L_{j}E^c_{k}$ interaction. The index $l = 1 \ldots 4$ 
	       determines the mass eigenstate of the neutralino. The indices
	       $i,j,k = 1,2,3$ correspond to the generation. Gauge invariance
	       forbids $i=j$. The index $\alpha = 1,2$ gives the slepton mass
	       eigenstate (i.e. the chirality of the SM lepton partner in absence
	       of mixing).}} 
 \label{fig:chipdirect}
 \end{center}
 \end{figure}

A collection of general expressions for three-body decays and matrix
elements entering in the calculation of partial widths can be found 
in appendix~\ref{chap:appendixC}. In the case of a pure photino neutralino 
decaying with \lam\ , the expression for the partial width
simplifies~\cite{dawson85} to
\begin{eqnarray}
 \Gamma = \lam ^2 {\alpha \over 128 \pi ^2} { \rm m_{ \achia }^5 \over
          \rm m_{\widetilde{f}}^4 } 
\end{eqnarray}
with $\rm m_{\widetilde{f}}$ the mass of the virtual slepton in the decay. 
Further details can be found in Ref.~\cite{baltzgond}. 

In practice, the LSP lifetime is a crucial observable when discussing 
the final state topology to be expected for supersymmetric events.
The experimental sensitivity of collider experiments is often optimal 
if the LSP has a negligible lifetime so that the production and decay 
vertices coincide. 
Otherwise the LSP decay vertex is displaced. If the lifetime is sufficiently
large, the LSP decays may occur outside the detector, giving rise to final 
states characteristic of $R_p$ conserving models.

The mean decay length $L$ in centimetres for the lightest neutralino
can be numerically estimated~\cite{dreiner1} from:
\begin{eqnarray}
 L({\rm cm}) = 0.3 (\beta \gamma) \left( {\rm m_{\widetilde{f}} \over
  100 \GeVcc
  } \right)^4 \left( {1 \GeVcc  \over {\rm m}_{ \achia }} \right)^5
  [ {1 \over \lam ^2}  ,\ { 3 \over \lamp ^2},\ { 3 \over \lampp ^2} ] \, .
\label{gaulife}
\end{eqnarray}
Figure~\ref{fig:sec6life} illustrates the behaviour of the LSP 
\begin{figure}[htb]
\begin{center}
  \begin{tabular}{c}
 \hspace*{-0.4cm} \epsfig{file=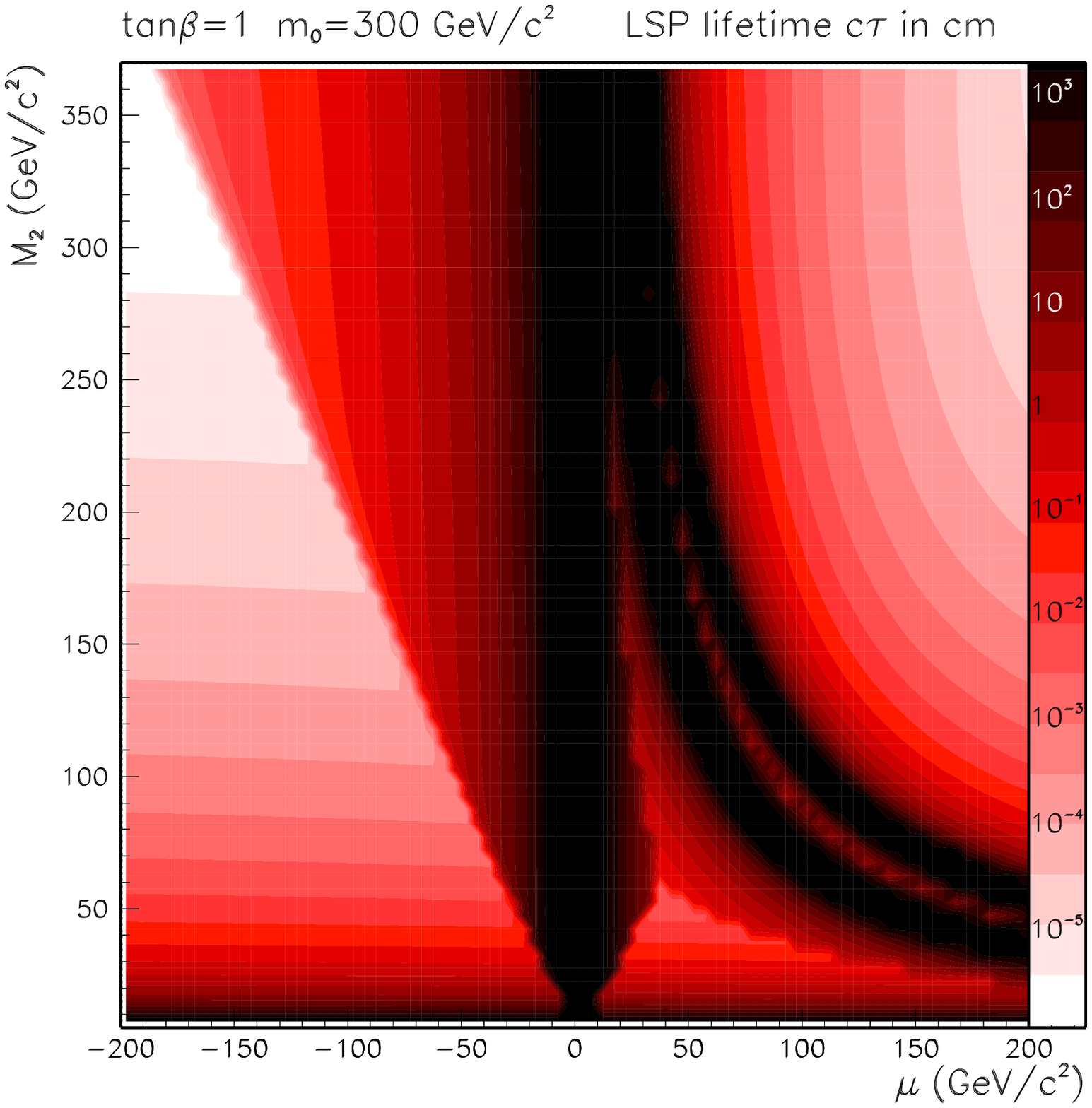,width=0.54\textwidth} 
 \hspace*{-0.8cm} \epsfig{file=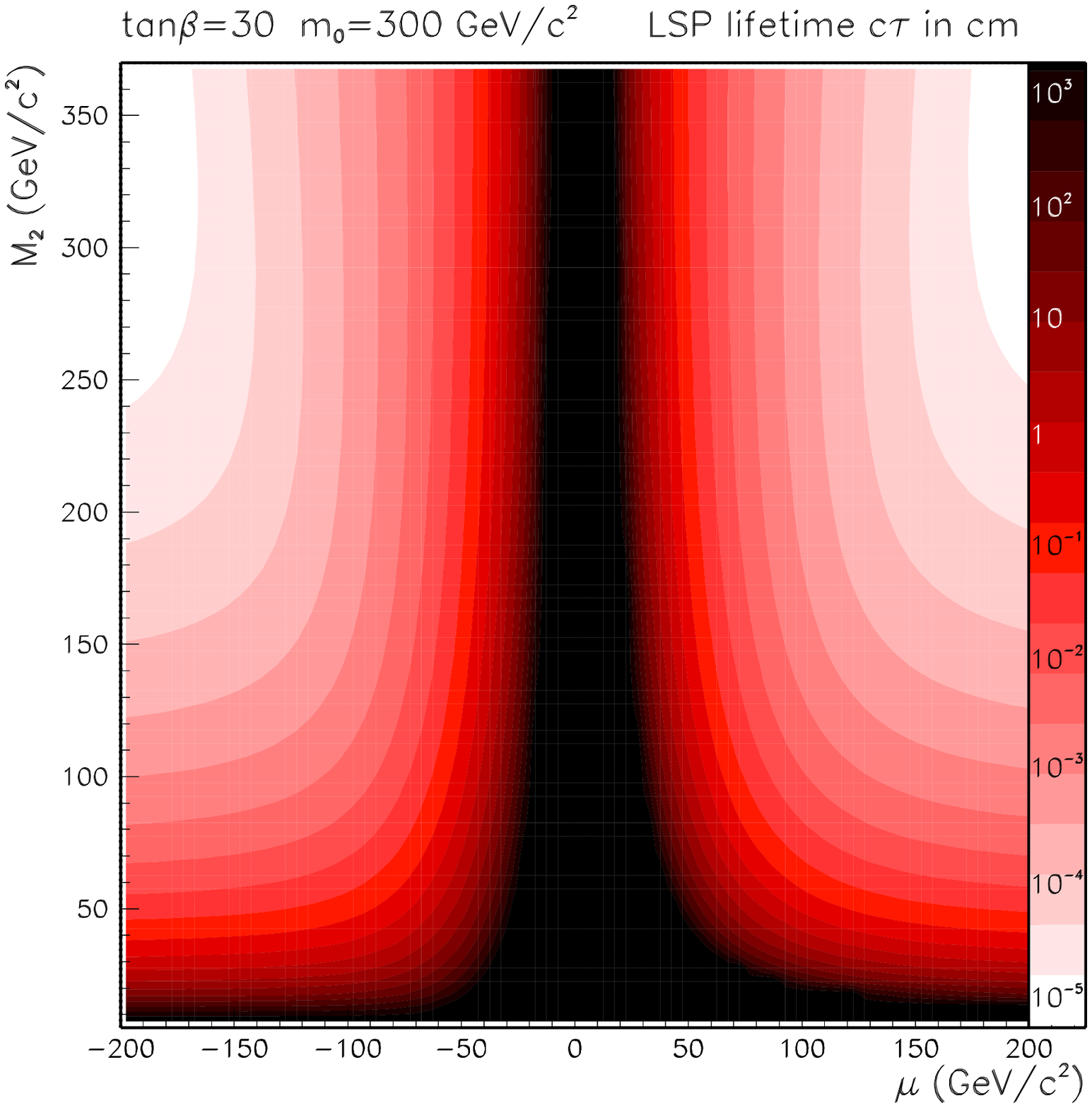,width=0.54\textwidth}
  \end{tabular}

\end{center}

\caption{{\it LSP lifetime for different values of the MSSM parameters,
         and with a dominant \Lacc\ coupling;
	 For this illustration, the coupling has been set to
	 \Lacc\ $= 0.004$. }}
\label{fig:sec6life}
\end{figure}
lifetime as presented in M$_2$ versus $\mu$ planes for different values 
of $\tan \beta$ and m$_0$, and considering a dominant \Lacc\ coupling.
A translation in terms of $L$ as a function of $m_{\achia}$ for a fixed
$m_{\tilde{f}}$ is shown in Fig.~\ref{fig:alephx1cm}.
\begin{figure}[!htb]
\begin{center}
\epsfig{file=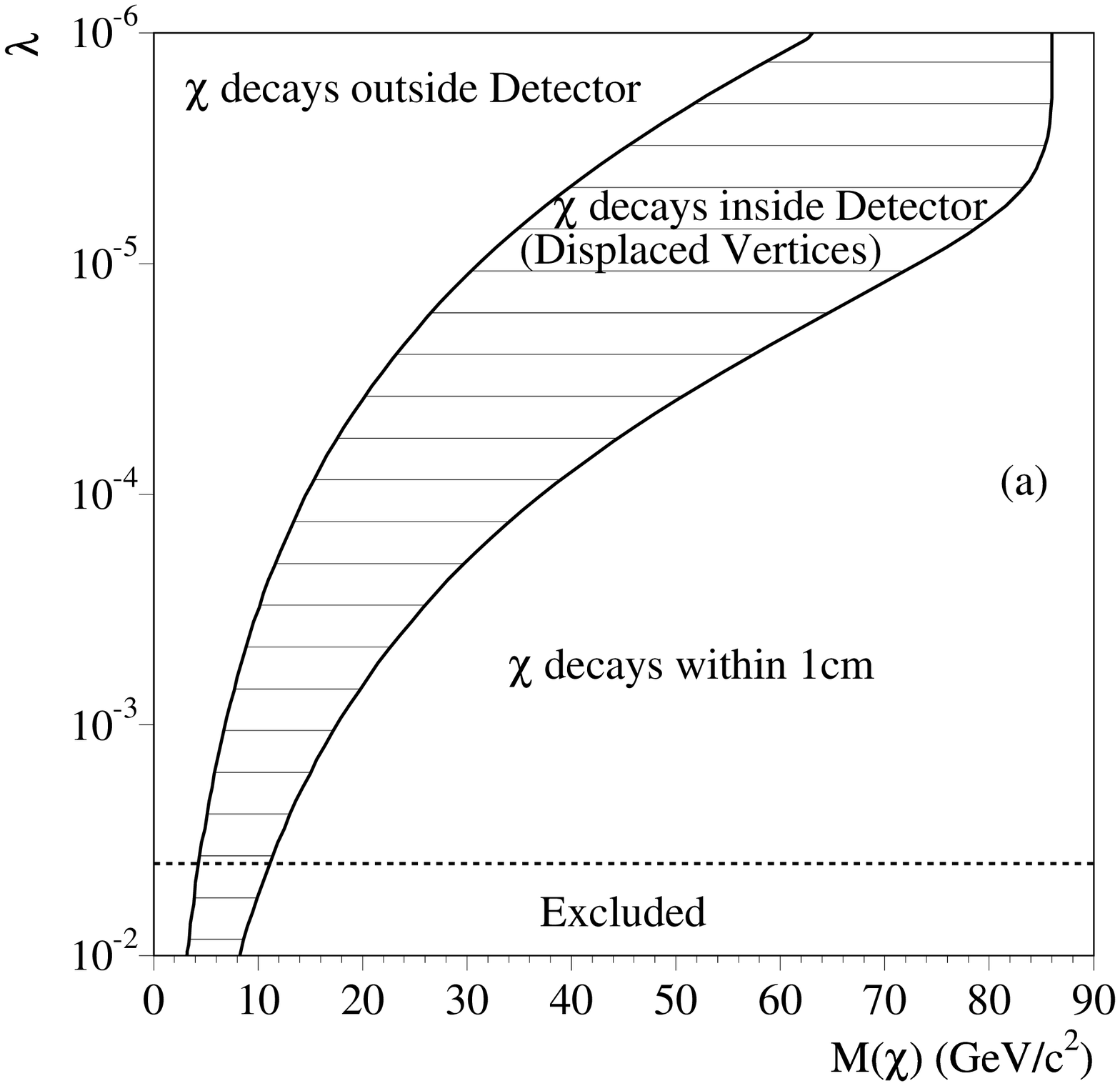,width=6cm}  
\hspace{0.4cm}
\epsfig{file=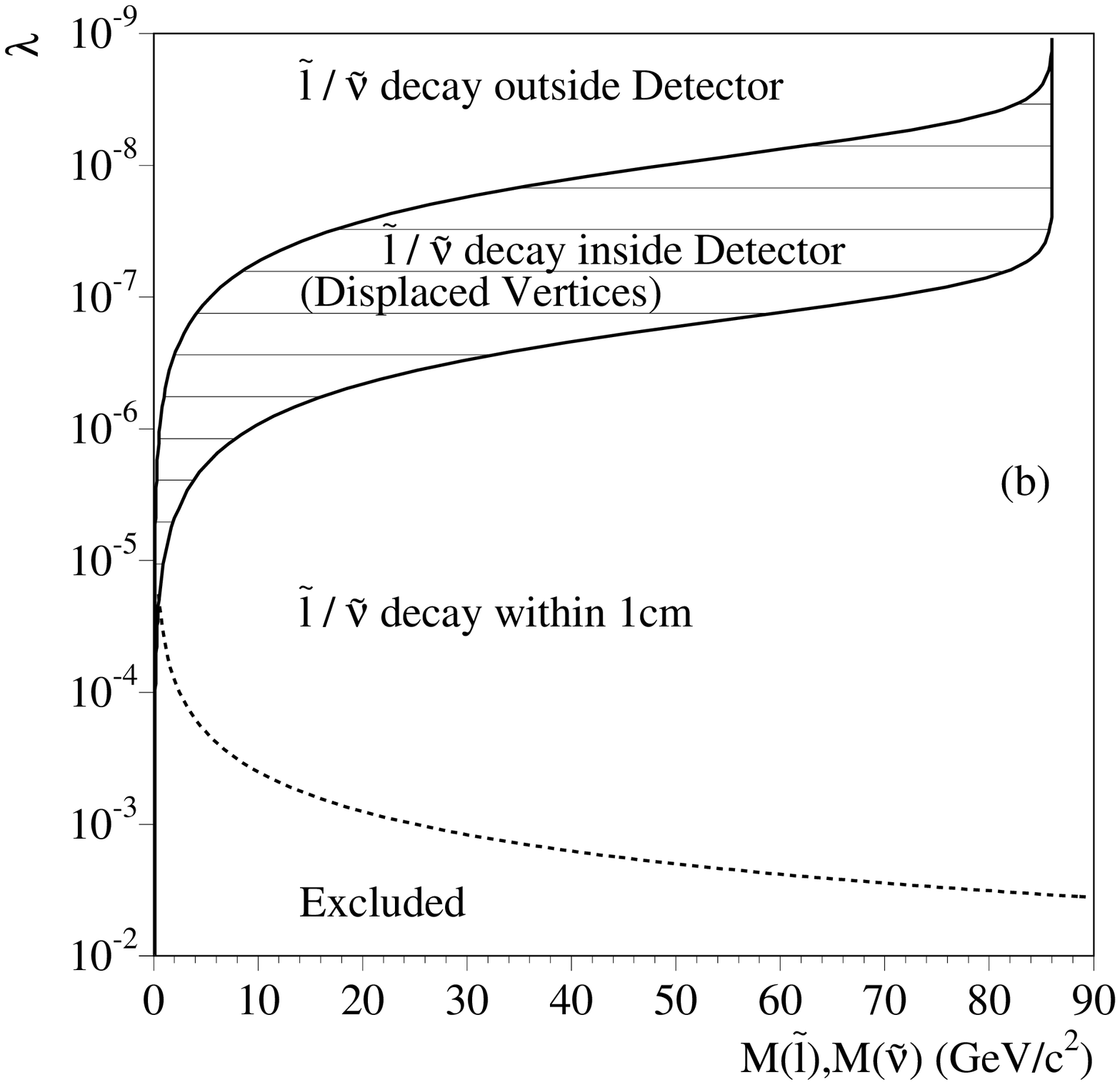,width=6cm}  
\caption{ 
{ \it 
      Regions in the $\lambda$ versus  sparticle mass plane where
      the sparticle has a mean decay length of  $L < 1 \cm$, 
      $ 1 < L < 3 $m (displaced vertices), and $L > 3$m (decay outside 
      a typical HEP detector) for a) $\achia$ assuming 
      $m_{\tilde{f}} = 100 \GeV$ and b) sleptons and sneutrinos.
      The dashed lines show an indirect limit on 
      $\lambda_{133}$~\cite{aleph_decay_X}.} }
\label{fig:alephx1cm}
\end{center}
\end{figure}
Measurements of \Rp\ coupling values can be performed 
through displaced vertex associated to the \Rp\ decay of the LSP. 

The sensitivities on the \Rp\ couplings obtained via a displaced vertex
depend of course on the specific detector geometry and performances.
Let us estimate the largest values of the \Rp\ coupling constants 
that can be measured via the displaced vertex analysis. The LSP
is assumed to be the lightest neutralino ($\tilde \chi^0_1$). 
Since a displaced vertex analysis is an experimental challenge at
hadron colliders, the performance typically achievable at a future $e^+e^-$ 
linear collider is considered here.
Assuming that the minimum distance between two vertices necessary to 
distinguish them experimentally is of order ${\cal{O}}(2 \times 10^{-5})$m,
it can be seen from Eq.~(\ref{gaulife}) that the \Rp\ couplings can be 
measured up to the values,   
\begin{eqnarray}
 \L < 1.2 \times 10^{-4} \gamma^{1/2} ({ {\rm m_{\widetilde{f}}} \over 100 \GeVcc })^2
       ({100 \GeVcc \over {\rm m}_{ \achia }})^{5/2}.
\label{LSP}
\end{eqnarray}
where $\L=\l$, $\l'/\sqrt{3}$ or $\l''/\sqrt{3}$, and $\gamma$ is the Lorentz
boost factor.

There is a gap between these values and the sensitivity of low-energy 
experiments which requires typically \Rp\ coupling values in the range 
$\L \sim {\cal{O}}(10^{-1}-10^{-2})$ for superpartners masses of 
$100 \GeVcc$. However, the domain above the values of Eq.~(\ref{LSP})
can be tested through the study of the single production of 
supersymmetric particles as will be discussed in 
section~\ref{sec:sproduct}.
Indeed, the cross-sections of such reactions are directly proportional
to a power of the relevant \Rp\ coupling constant(s), which allows
the determination of the values of the \Rp\ couplings.
Therefore, there exists a complementarity between the displaced vertex 
analysis and the study of singly produced sparticles, since these
two methods allow to investigate different ranges of values of the \Rp\ 
coupling constants.

\subsection{Cascade Decays Initiated by Gauge Couplings}
\label{sec:cascdec}

In an indirect decay, the supersymmetric particle first decays through a 
$R_p$ conserving vertex (i.e. through gauge couplings) to an 
on-shell supersymmetric particle, thus initiating a 
\index{Decays involving \Rp\ ! cascades}cascade which continues till 
reaching the LSP. 
The LSP then decays as described above via one \Rp\ coupling.

The sfermions may for example decay indirectly (i.e. undergo first a gauge 
decay) into a fermion plus a \XOI\, if the lightest neutralino \XOI\ is 
the LSP, as shown for example in the case of sleptons in 
Fig.~\ref{fig:indecay}a and b.
\begin{figure}[htb]
\begin{center}
\epsfig{file=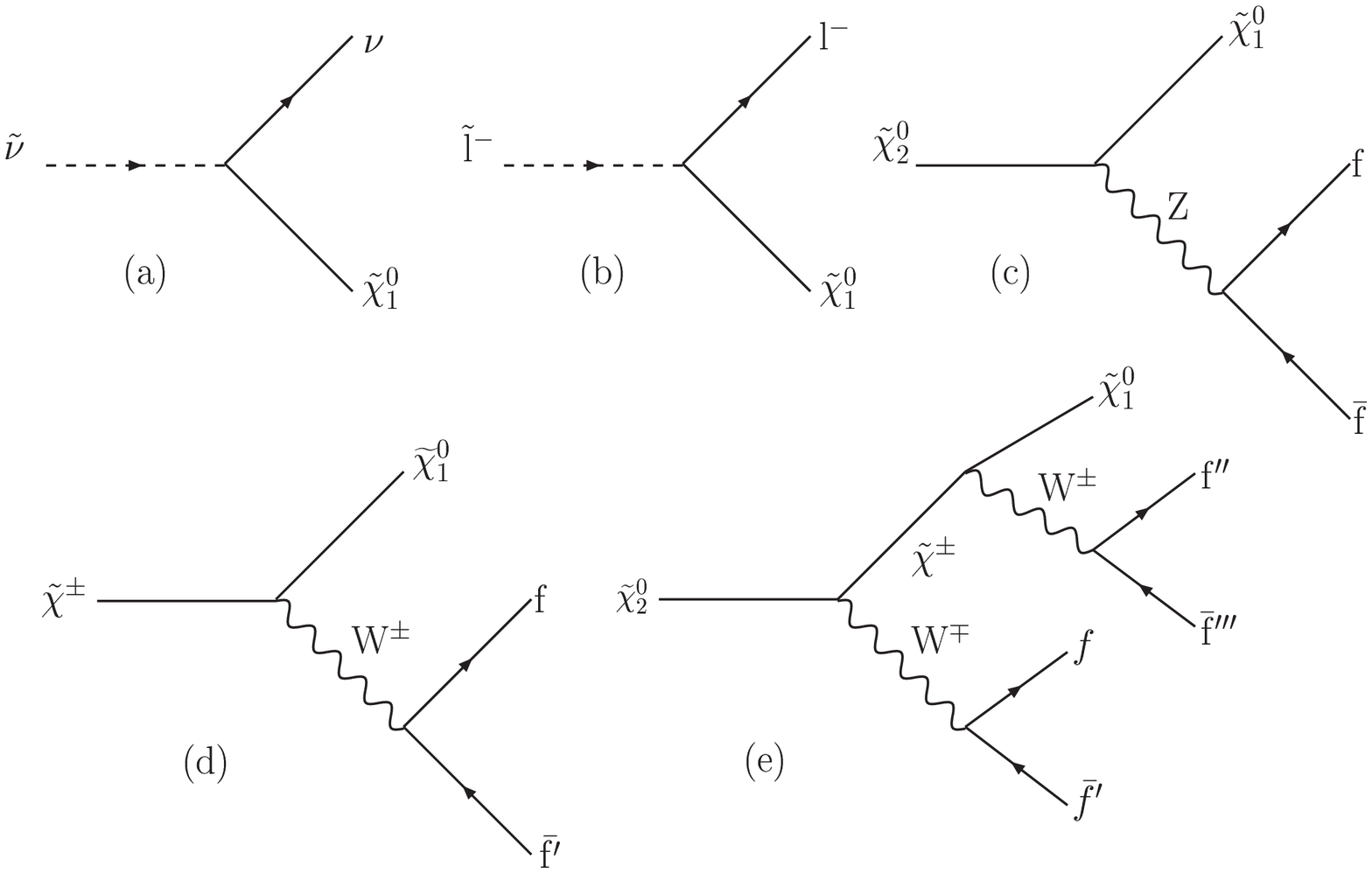,width=15cm}
\end{center}
\vspace*{-0.5cm}

\caption{{\it Slepton (a, b) and gauginos (c, d, e) indirect decay diagrams.}}
\label{fig:indecay}
\end{figure}
The \XOI\ will subsequently undergo a \Rp\ decay via one of the trilinear 
couplings.
In the squark sector, such decays have mainly been considered for the stop
and sbottom in actual searches at colliders 
as they possess a mass eigenstate which can be among the lightest for squarks.
If the lightest stop mass eigenstate $\sqt_1$ is not the LSP but the lightest 
charged supersymmetric particle, the cascade will be initiated through a decay
to $\sqt_1$ \Ra c\XOI. 
If m$_{\sqt_1} > $m$_{\tilde{\chi}^\pm} + $m$_{\rm {b}}$ 
then the decay $\sqt_1$ \Ra b\XPM\ is possible. 
In the case of the sbottom, the indirect decay $\sqb_1$ \Ra b\XOI\ 
is generally treated as the dominant one.

In the gaugino-higgsino sector, the heavy neutralino and chargino mass
eigenstates can decay, depending on their mass difference with the \XOI, 
either directly into three standard fermions, or indirectly to \XOI\ via 
a virtual $Z$ or $W$, as illustrated in Fig.~\ref{fig:indecay}c, d and e.

Assuming a small value for the $\Rp$ coupling, the indirect decay 
mode will generally dominate as soon as there is enough phase space 
available between ``mother'' and ``daughter'' sparticles.
For searches at existing colliders this happens when the mass difference 
between these two sparticles is larger than about $5$ to $10 \GeVcc$.
As an example, the Fig.~\ref{fig:staubrpv} shows the $\Rp$ decay branching 
fraction of the \staur\ via the \Lacc\ coupling as a function of the stau mass, 
for different values of the neutralino mass. If the slepton is lighter
than the neutralino, only the \Rp\ mode is opened.
As soon as the indirect decay mode is possible, it dominates.
\begin{figure}[htb]

\begin{center}
\epsfig{file=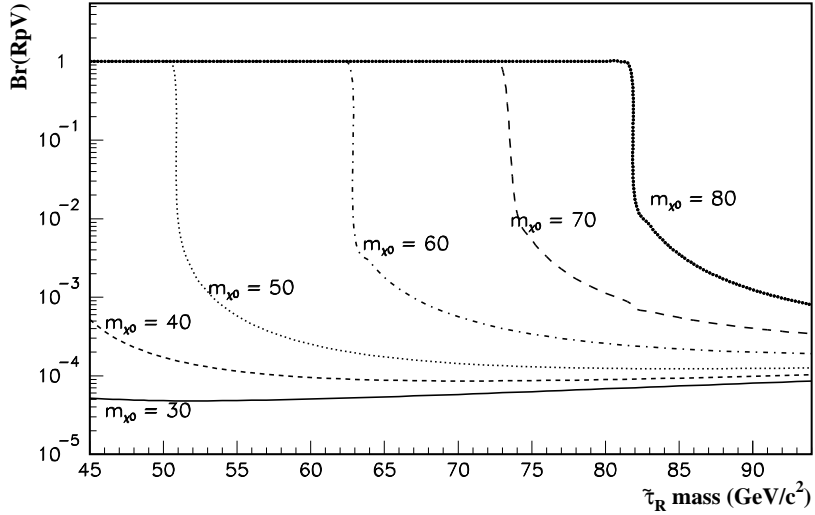,width=12cm}
\caption{{\it Branching ratio of the $\Rp$ decay \staur\ \Ra ee$ , $e$\tau ,
\tau\tau + $\Emiss\ as a function of the mass of the \staur\
and for different values of the neutralino mass.}}
\label{fig:staubrpv}
\end{center}
\end{figure}

Nevertheless, there exist regions of the SUSY parameter space where the
$R_p$ conserving decay (initiating the cascade) suffers from a 
``dynamic'' suppression. 
This is the case for example if the field component of the two lightest
neutralinos is mainly the photino, in which case the indirect decay
\mbox{\XOII\ \Ra \XOI\ Z$^*$} is suppressed. 
In these regions, even if the sparticle is not the LSP it will
decay through a direct \Rp\ mode.

It should be emphasized here that the indirect decays lead to final state
topologies which differ strongly from direct \Rp\ decays.
For example, the direct decay of a neutralino via a \LLE\ coupling
leads to a purely leptonic final state, while the indirect decay adds 
(mainly) jets to such final state. The allowed indirect decays and 
their branching ratios depend on the parameters values for the specific
supersymmetry model (e.g. MSSM, mSUGRA, ...) as well as on the value
of the \Rp\ coupling considered. 

Detailed strategies for practical searches at colliders will be discussed 
starting in section~\ref{sec:spair}.

\subsection{Decays Through Mixing Involving Bilinear Interactions}
\label{sec:mixdec}

For completeness, the effects of the $\mu_i L_i H_u$ 
bilinear terms in the superpotential which violate lepton-number 
conservation must be discussed in the context of collider physics.
The theoretical motivations for the appearance of such terms in the 
superpotential of supersymmetric models were discussed in detail
in chapter~\ref{chap:theory}.
They appear for instance as a consequence of non-zero right handed 
sneutrino vacuum expectation values in models with spontaneous $R$-parity 
violation, or in models with explicitly broken $R$-parity.

As discussed in section~\ref{sec:bilinear}, there are strong
constraints on the bilinear \Rp\ parameters from the neutrino sector.
For this reason, the $\mu_i$ are often considered to be
suppressed~\cite{reviews2} and thus neglected in collider
analyses, with the notable exception of the LSP decays,
where the LSP is not necessarily a neutralino in this context. 
Indeed, values of the $\mu_i$ suggested by the solar and atmospheric
neutrino experiments can lead to observable decays of the LSP at colliders,
with correlations between the neutrino mixing 
\index{Neutrino!mixing} angles and the
branching ratios into different lepton 
flavours~\cite{bilinear_LSP_decay,phrv00,hp03,hprv02,rpv01}
(see section~\ref{sec:ModBilinear}).

However, exotic scenarios with non-negligible $\mu_i$ parameters
have been considered, in particular in the context of spontaneous
$R$-parity violation~\cite{masiero90,romao92,spontaneous_nu_R_2}.
Although these scenarios are disfavoured by neutrino oscillation data,
we mention for completeness the expected effect of e.g. a non-negligible
$\mu_3$. Such a term introduces a tau component in the chargino 
mass eigenstate. As a consequence, the tree level decay
$Z \rightarrow {\tilde {\chi}}^- \tau^+$ becomes for instance 
possible.
In the top sector, bilinear terms could rise to additional decay modes 
for top-quarks and stop-squarks such as $t \rightarrow \stau_{1}^{+} + b$ 
or $\tilde{t}_{1} \rightarrow \tau^{+} + b$.
Most of these new decay modes result, through cascade decays, in final 
states with two $\tau$'s and two b-quarks plus the possibility of 
additional jets and leptons. 
B-tagging and $\tau$ identification are therefore important tools for the 
analysis. Note that the $\tilde{t} \rightarrow \tau + b$ decay could also
occur via a trilinear $\lambda^\prime_{333}$ coupling.

Spontaneous $R$-parity violation has been studied in the context 
of a single tau-lepton-number-violating bilinear term at the
Tevatron $p \bar{p}$ collider in Ref.~\cite{biteva} and the  
$\tilde{t} \rightarrow \tau + b$ decay was found to be competitive 
with the decay in c\XOI\ for a discovery of the $\tilde{t}$.
For the LHC $p p$ collider, the possibility to observe spontaneous 
$R$-parity violation through multilepton and same sign dilepton signatures 
in gluino pair-production has been considered in Ref.~\cite{bilhc}.

\section{$\Rp$ Phenomenology from Pair Produced Sparticles}
\label{sec:spair}


In this section, we are interested in the way the collider phenomenology 
for Supersymmetric models is affected by the presence of an individual
\Rp\ coupling with a value $\Lambda^2 \ll 4 \pi \alpha$ such that
\Rp\ contributions can be neglected at production. 

In such a configuration, a first question of interest for searches at
colliders is whether or not the nature of a specific non-vanishing 
\Rp\ coupling can be identified (within a range of allowed values) 
starting from the characteristics of the observed final states. 
Assuming that the presence of specific \Rp\ interactions is eventually 
established, a next important question is to understand whether and 
how the sensitivity to the fundamental parameters of a given supersymmetric 
theory is affected.

\subsection{Gaugino-Higgsino Pair Production} 
\label{sec:pairgg}
\index{Sparticle production!Gaugino-Higgsino pair|(}
 \noindent \addtocounter{sss}{1}
 {\bf \thesss Production and final states}   
  \addcontentsline{toc}{subsection}{\hspace*{1.2cm} \alph{sss}) 
              Production and Final States}

Gaugino-higgsino pair production via standard gauge couplings at colliders has 
been thoroughly studied in the literature and a detailed review is clearly 
outside the scope of this paper. Here, only the key ingredients shall be 
summarized.
Otherwise we concentrate on the phenomenology associated with the 
presence of \Rp\ interactions.


At $l^+l^-$ lepton colliders, the neutralinos are produced by pairs via 
$s$-channel $\mathrm{Z}$ exchange (provided they have a higgsino component), 
or via $t$-channel $\tilde{l}^{\pm}$ exchange (provided they have a gaugino 
component). The charginos are produced by pairs in the $s$-channel via 
$\gamma$ or $\mathrm{Z}$ exchange, or in the $t$-channel via sneutrino 
($\tilde{\nu}_l$) exchange if the charginos have a gaugino component.
Of course, the $t$-channel contributions are suppressed for high 
slepton masses. 
           
In the case of neutralinos, the $t$-channel exchange contributes to 
an enhancement which can be significant for slepton masses typically 
below $\sqrt{s_{ll}}/2$ (i.e. $m_{\sel} \lsim 100 \GeV$ in the
case of selectron exchange at LEP~2 \index{LEP} $e^+e^-$ collider).
In constrast, the chargino pair production cross-section can decrease 
due to destructive interference between the $s$- and $t$-channel amplitudes
(i.e. between \sel\ and \snue\ exchange at a $e^+e^-$ collider) 
if the $\tilde{l}^{\pm}$ and $\tilde{\nu}_l$ masses are comparable.

As an example, one can consider pair production in the framework
of the MSSM, assuming in addition that scalars have a common 
mass m$_0$ at the GUT scale. In such a case m$_0$ determines the slepton 
masses at EW scale and the relevant MSSM parameters are 
M$_2$, $\mu$, \tanb\, and m$_0$. 
In such a framework the production cross-sections are generally
found to be large.
If the dominant component of neutralinos and charginos is the 
higgsino ($|\mu| \ll $M$_2$) the production cross-sections are also
insensitive to slepton masses.
Over a wide range of MSSM parameter values, the pair production
cross-sections at LEP~2 for $\sqrt{s_{ee}} \simeq 200 \GeV$ 
is found to vary typically from 0.1 to $10$~pb.
Investigations of gaugino pair production in a similar constrained 
MSSM framework and in presence of \Rp\ have been performed
for the case of a future 500~\GeV\ \ee\ collider in Ref.~\cite{Godbole}. 


At $p \bar{p}$ and $p p$ hadron colliders, the main production process
which has been studied for neutralinos and charginos is the associated
production $q \bar{q}' \rightarrow \tilde{\chi}^{\pm} \tilde{\chi}^0$.
In $R_p$-conserving theories like the MSSM or mSUGRA, measurable rates
are expected mainly in the case of $\tilde{\chi}_1^{\pm} \tilde{\chi}_2^0$
and only for certain regions of the parameter space.
In presence of $\Rp$ interactions of course, the process 
$q \bar{q}' \rightarrow \tilde{\chi}_1^{\pm} \tilde{\chi}_1^0$
involving the lowest $\tilde{\chi}^0$ mass eigenstate could also become
observable. In addition, $\Rp$ could allow for pair production of neutralinos 
and charginos in $q \bar{q} \rightarrow \gamma, Z$ annihilation processes 
to become observable. The production cross-sections would then depend on the
gaugino and higgsino components as discussed above for $l^+l^-$ 
annihilation.


The final states resulting from the decay of pair-produced neutralinos or
charginos are listed in Table~\ref{gaug_final} for the three different couplings 
\LLE, \LQD\ and \UDD.
 \begin{table}[htb]
 \begin{center}
 \begin{tabular}{|c|c|c|c|c|} \hline  \hline    
 &  &  &  &  \\
  gauginos     &  decay mode & \LLE              & \LQD                    & \UDD  \\
 \hline
  \XOI \ \XOI  &  direct     & 4$\ell$ + \Emiss  & 1$\ell$ + 4$j$ + \Emiss &  6$j$ \\
               &             &                   & 2$\ell$ + 4$j$          &       \\
               &             &                   & 4$j$ + \Emiss           &       \\
 \hline
  \XPI \ \XMI  &  direct     & 2$\ell$ + \Emiss  & 1$\ell$ + 4$j$ + \Emiss &  6$j$ \\
               &             & 4$\ell$ + \Emiss  & 2$\ell$ + 4$j$          &       \\
               &             & 6$\ell$           & 4$j$ + \Emiss           &       \\
 \hline
 \hline
  \XOII \ \XOI & indirect    & 4$\ell$ + \Emiss  & 1$\ell$ + 4$j$ + \Emiss
                                                                  & 8$j$           \\
               &             & 4$\ell$ + 2$j$ + \Emiss  & 1$\ell$ + 6$j$ + \Emiss
                                                               & 6$j$ + 2$\ell$    \\
               &             & 6$\ell$ + \Emiss         & 2$\ell$ + 4$j$ + \Emiss
                                                               & 6$j$ + \Emiss     \\
               &             &                          & 2$\ell$ + 6$j$   &       \\
               &             &                          & 3$\ell$ + 4$j$ +\Emiss & \\
               &             &                          & 4$\ell$ + 4$j$   &       \\
               &             &                          & 6$\ell$ +\Emiss  &       \\
 \hline
  \XPI \ \XMI  & indirect    & 4$\ell$ + 4$j$ + \Emiss  & 1$\ell$ + 6$j$ + \Emiss 
                                                                           & 10$j$ \\
               &             & 5$\ell$ + 2$j$ + \Emiss  & 1$\ell$ + 8$j$ + \Emiss 
	                                                & 8$j$ + 1$\ell$ + \Emiss  \\
               &             & 6$\ell$ + \Emiss         & 2$\ell$ + 4$j$ + \Emiss 
	                                                & 6$j$ + 2$\ell$ +\Emiss   \\     
               &             &                         & 2$\ell$ + 6$j$ + \Emiss & \\
               &             &                          & 2$\ell$ + 8$j$         & \\
               &             &                         & 3$\ell$ + 4$j$ + \Emiss & \\ 
               &             &                         & 3$\ell$ + 6$j$ + \Emiss & \\
               &             &                         & 4$\ell$ + 4$j$ + \Emiss & \\
               &             &                         & 8$j$ + \Emiss           & \\
               &             &                         &                         & \\
 \hline \hline
 \end{tabular}
\caption{{\it Neutralino and chargino pair production final states in 
              case of \Rp~decays. The notations $l$, \Emiss and $j$ correspond 
	      respectively to charged lepton, missing energy from at least one 
	      neutrino and jet final states.}}
 \label{gaug_final}
\end{center}
\end{table}

At first approximation, with $\lambda \ne 0$ the final states are 
characterized by multi-lepton (charged leptons and escaping neutrinos)
event topologies. In contrast, with $\lambda' \ne 0$, the final states are 
likely to contain multi-jets and several more or less isolated leptons. 
One exception concerns here slepton pair production which could lead to 
four jet final states. 
Finally $\lambda'' \ne 0$, leads to final states with very high jet 
multiplicities.
Thus, the existence of either a non-vanishing 
$\lambda$ ( \LLE\ ), $\lambda'$ ( \LQD\ ), 
or $\lambda''$ ( \UDD\ ) can indeed be readily distinguished.

Of course such a simple picture applies essentially in the framework of 
MSSM of mSUGRA models where the \XOI\ is likely to be the lightest 
supersymmetric particle. It moreover has to be modulated in the presence of 
indirect (cascade) decays. For instance indirect gaugino decays when involving
an intermediate Z$^*$ or W$^*$ might lead to final states containing
jets for \LLE\ interactions or, symmetrically, containing leptons and 
neutrinos for \UDD\ interactions.

\noindent \addtocounter{sss}{1}
{\bf \thesss Searches for Gaugino Pair Production at {\boldmath{$e^+e^-$}} Colliders}   
 \addcontentsline{toc}{subsection}{\hspace*{1.2cm} \alph{sss}) 
             Searches for Gaugino Pair Production at $e^+e^-$ Colliders}

It is interesting to review what has been learned from 
studies by the experiments at the LEP \index{LEP} collider.
The analyses have been performed assuming "short lived" sparticles such
that the \Rp\ decays occur close enough to the production vertex and are
not observable. In practice this implies a LSP flight path of less than
$ {\cal{O}}(1) \cm$.
Considering the upper limits on the \Lijk\ derived from
low energy measurements (chapter~\ref{chap:indirect}), and according
to Eq.~\ref{gaulife}, the analyses 
are thus insensitive to a 
light $\widetilde{\chi}$ of mass $M_{\chi_{LSP}} \leq 10 \GeVcc$ 
(due first to the term  ${m_{\widetilde{\chi}}}^{-5}$ and second to the
term $(\beta \gamma)$ which becomes important).
When studying  $\widetilde{\chi}$ decays, for typical masses 
considered, the LEP analyses have a lower limit in sensitivity on the
$\lambda$ coupling  of the order of $10^{-4}$ to $10^{-5}$.  

In most of the \Rp\ analyses, the main background contributions come from 
the four-fermion processes and \Zg\ events. A discussion of such SM 
background contributions can be found in Ref.~\cite{yellow}.
The \Zg\ cross-section decreases with the increase of the centre-of-mass 
energy; on the other hand, the cross-sections of the four-fermion processes 
increase; in particular beyond the W$^+$W$^-$ and the \ZZ\ thresholds, 
these processes contribute significantly to the background, and lead
to final states very similar to several \Rp\ signal event
topologies.

The \LLE\ searches at LEP \index{LEP} were mainly multi-lepton analyses, with 
the missing energy in the final states most often coming from neutrinos. 
However, the indirect decay topologies may also contain hadronic 
jets (indirect decay of gauginos, see Table~\ref{gaug_final}).
The number of charged leptons in the final state varied between 4 and 6 
except for direct decays of charginos with two neutrinos which lead 
to 2 charged leptons in the final state. 
Therefore, the crucial step in the selection of the signal events 
was the electron, muon or tau identification. Electrons and 
muons are typically identified by well isolated charged tracks in 
combination with either dE/dx measurements and deposits in the 
electromagnetic calorimeters (e) or information from hadron calorimeters 
and muon chambers ($\mu$, $\tau$) of the experiments. 
Tau decays may also be identified through isolated thin hadronic jets.
In case of {\mbox{${\mathrm W}$}} or tau jets, signal selection has been
performed with the additional discrimination provided by topological
variables like the $y_{cut}$ jet resolution variable.
The two-lepton final states are difficult to separate from the SM background mainly 
coming from \Zg ~and \GG ~events and have been considered in the analyses. 
The other multi-lepton final states on the other hand, provided almost 
background free analyses with efficiencies typically between 20 and 60 \%. 
The decays producing taus in the final states were found to have lower 
efficiencies and rejection power. 
The analyses designed for signals produced with a dominant $\lambda_{133}$ 
can be applied to signals produced with other $\lambda_{ijk}$, 
and the efficiencies are either of the same order or higher.
Therefore, the weakest limits which have been derived are those resulting
from analyses performed assuming a dominant $\lambda_{133}$ 
coupling~\cite{delphi}.

For the \LQD\ searches at LEP, \index{LEP} the analyses of gaugino decays all 
included hadronic activity in the final state (at least 4 hadronic jets, 
see Table~\ref{gaug_final}) such that topological variables like jet 
resolution and thrust were used to select events in combination with missing 
energy and/or one or two identified leptons. For the topologies with missing 
energy the polar angle of the missing momentum has also been used to 
select candidate events. The analysis of gaugino decays via the couplings 
$\lambda^{\prime}_{i3k}$ and $\lambda^{\prime}_{ij3}$ also benefits from the 
presence of b-quarks, and thereby from possible background reduction via
b-tagging.
The topologies of the indirect chargino decays depend heavily on the mass 
difference between chargino and neutralino, which is a free parameter in the 
model. Sensitivity to a large range of topologies is hence needed in order to 
completely cover all possible \LQD\ scenarios. Several of the gaugino final 
state topologies closely resemble those of \Rp~sfermion decays and the same 
event selection may therefore be used to cover these channels too, e.g. 
sneutrino and slepton decays. The SM background consists mainly of 
four-fermion events decaying either to hadronic or semileptonic final states. 
Signal efficiencies typically range from over 50\% down to a few \%, 
depending on topology and selection criteria. The worst efficiency is 
generally obtained for decays into taus and light quarks. The couplings 
generating these final states (e.g. $\lambda^{\prime}_{311}$) are therefore 
used to evaluate conservative constraints on the production cross-sections. 
The excluded \LQD\ gaugino cross-sections at a 95\% confidence level are 
typically of the order of $0.5 \picob$~\cite{aleph}. 

For the LEP \index{LEP} searches in the case of a single dominant \UDD\ 
coupling, the gauginos decay into mainly hadronic final states, however, the 
indirect gaugino decays may also include leptons and missing energy, depending 
on the decay mode of the {\mbox{${\mathrm W}$}}. 
As seen from Table~\ref{gaug_final}, the number of jets expected in
the final states varies between six and ten. The selection of candidate 
events typically depends on topological variables like jet resolution, 
thrust and jet angles, thereby rejecting the major part of 
the SM~ \Zg~, \WW~ and \ZZ~ background events. 
The absence of neutrinos and missing energy in the fully hadronic final 
states also enables direct reconstruction of the gaugino masses. 
The mass reconstruction consists of assigning each reconstructed jet 
to its parent gaugino and thereafter applying a kinematic fitting algorithm. 
These algorithms are also used to reconstruct the mass of 
the {\mbox{${\mathrm W}$}}~ bosons produced at LEP~2 \index{LEP} and impose 
constraints on conserved energy and momenta in combination 
with equal masses of the pair produced gauginos.
The indirect chargino decays, again, strongly depend on the mass difference 
between chargino and neutralino, thereby making it difficult to use the same 
event selection to cover all possible scenarios. Decays into light quarks
($\lambda''_{112}$ and $\lambda''_{122}$ couplings) 
generally have lower efficiencies and are therefore used to derive 
conservative cross-section limits at a 95\% confidence 
level (CL)~\cite{delphi}.


Since none of the \Rp~gaugino searches at LEP~2 \index{LEP} show any excess of 
data above the Standard Model expectations, the results are interpreted in terms 
of exclusions of the MSSM parameters.
The gaugino pair production cross-sections are, as previously discussed, 
mainly determined by the MSSM parameters $\mu$, M$_{2}$, m$_0$ and \tanb.
The excluded gaugino cross-section at a 95\% CL for 
each experimental search channel is therefore compared to the production 
cross-section provided by the MSSM for each set of these four parameters. 
Hereby an exclusion of experimentally disproved combinations of the 
parameters is obtained for each of the performed search channels. This 
exclusion is then typically presented as contours in the $\mu$, M$_2$ plane 
for different fixed values of \tanb~ and m$_0$ \cite{aleph,delphi,l3,opal}.
The LEP~1 \index{LEP} excluded region of the ($\mu$,~M$_{2}$) contours is 
obtained from the $Z$ line-shape measurement.  
Examples of such ($\mu$,~M$_{2}$) exclusion contours are shown in 
Figures~\ref{gaug_lle}, \ref{gaug_lqd} and \ref{gaug_udd}. 
\begin{figure}[ht]
\begin{center}
\epsfig{file=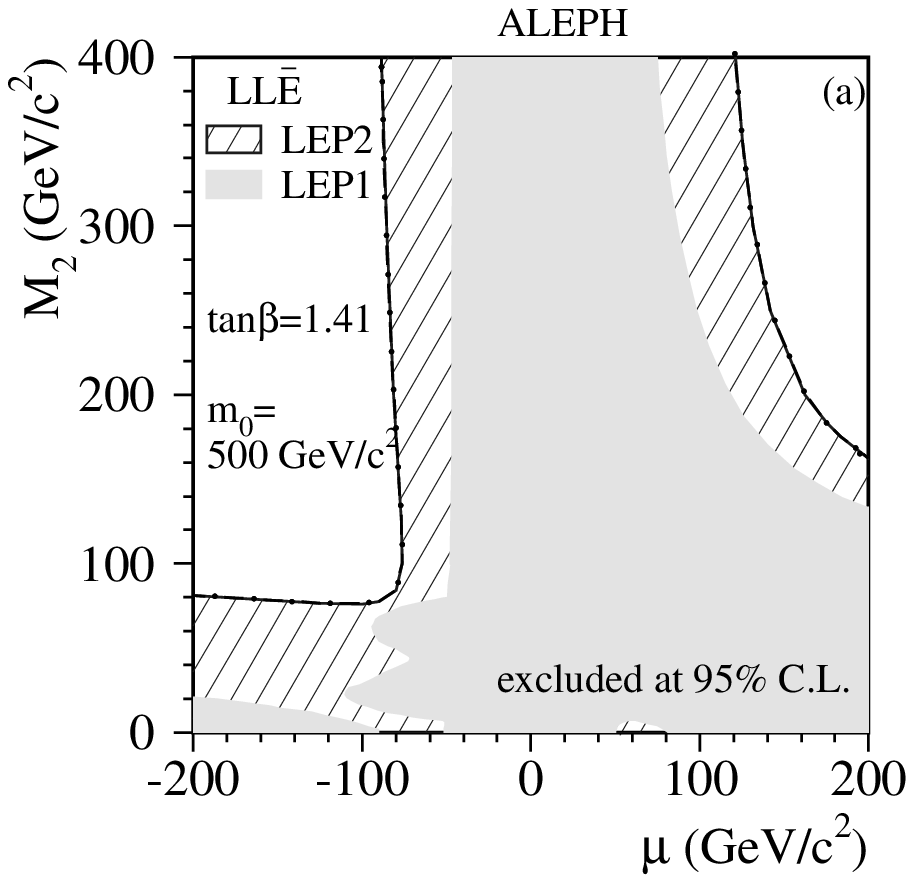,width=8cm}
\caption{{\it Regions in the ($\mu$, M$_2$) plane excluded at $95 \%$ CL 
at \tanb$ = 1.41$ and \mbox{m$_0 = 500 \GeV$} for \LLE\ coupling~\cite{aleph}. 
The dotted line is the kinematic limit for pair production 
of the lightest chargino.}} 
\label{gaug_lle}
\end{center}
\begin{center}
\epsfig{file=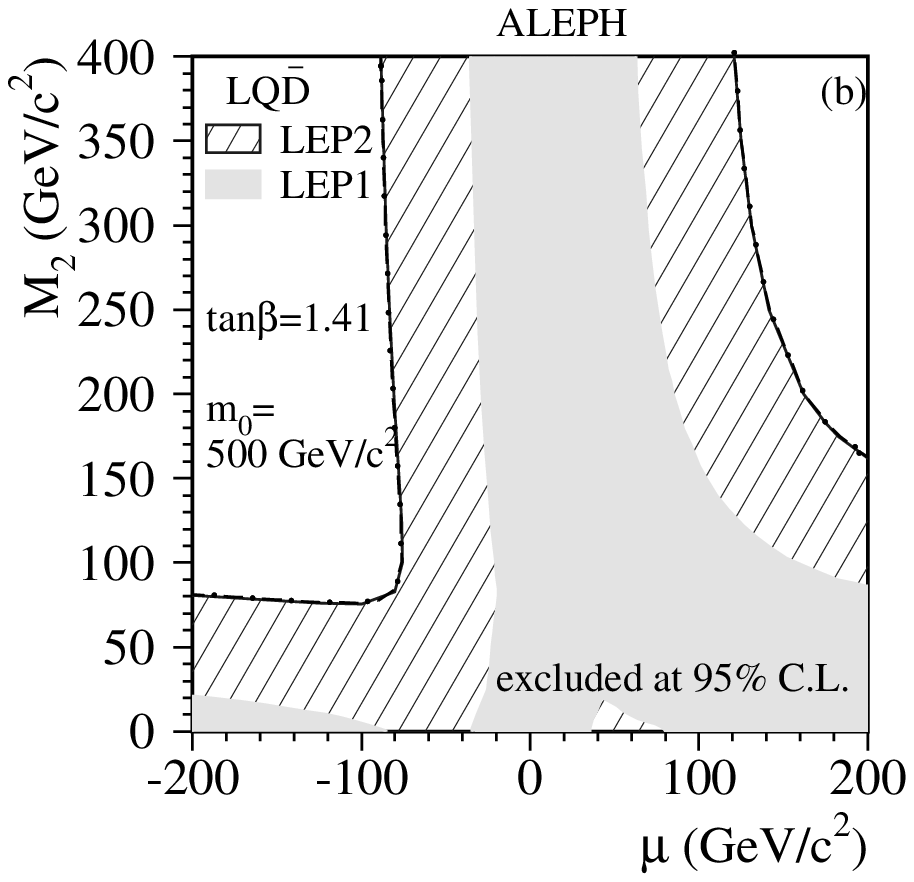,width=8cm}
\caption{{\it Regions in the ($\mu$, M$_2$) plane excluded at $95 \%$ CL 
         for \tanb$ = 1.41$ and \mbox{m$_0 = 500 \GeV$}
         in the case of the dominant 
         \LQD\ coupling~\cite{aleph}.
         The black line is the kinematic limit for pair production 
         of the lightest chargino.}} 
\label{gaug_lqd}
\end{center}
\end{figure}
\begin{figure}[htb]
\begin{center}
\epsfig{file=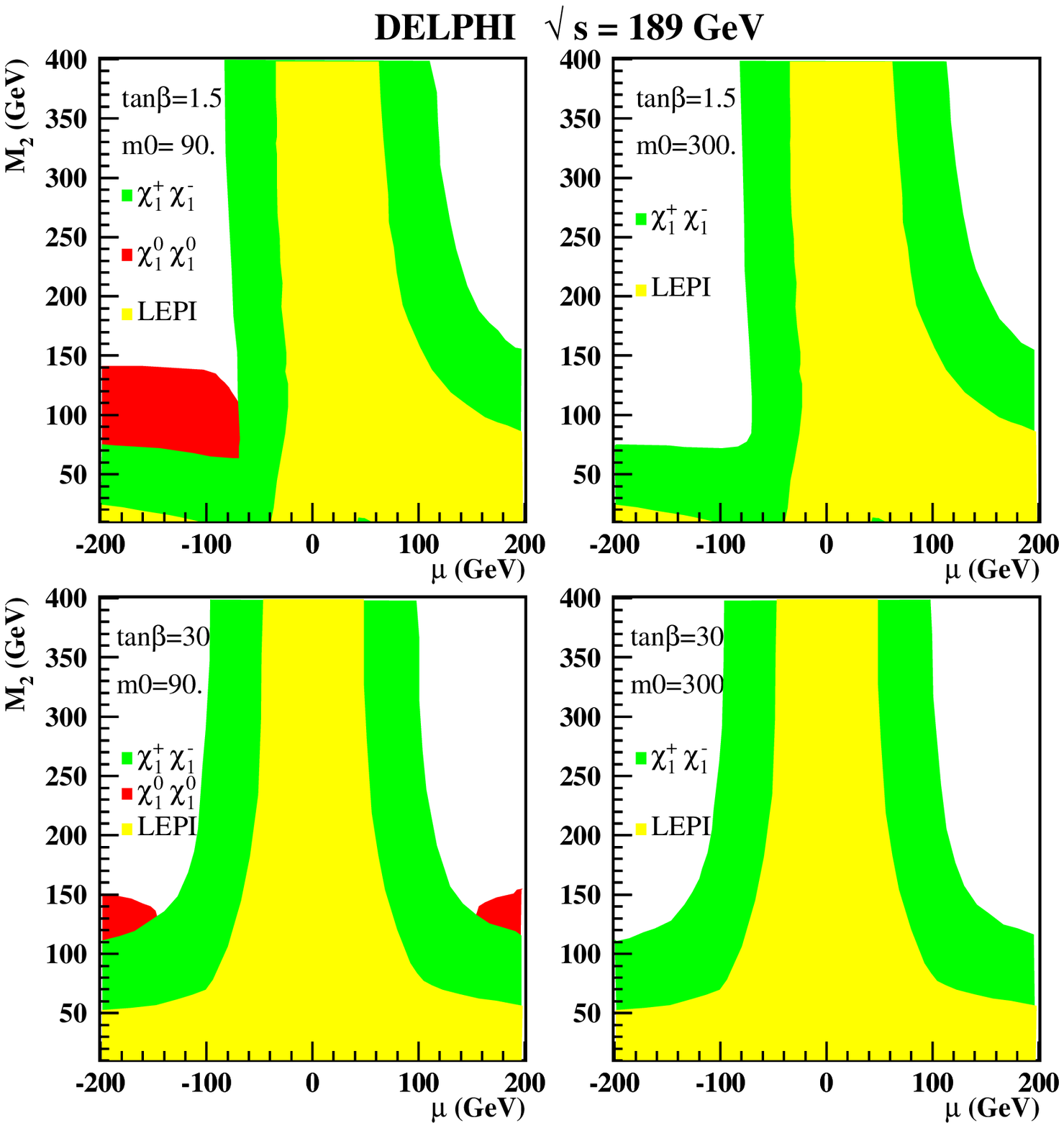,width=10cm}
\caption{{\it Regions in the ($\mu$, M$_2$) plane excluded at $95 \%$ CL 
for \tanb$ = 1.5,30$ and \mbox{m$_0 = 90,300 \GeV$}
in the case of a dominant \UDD\   
coupling \cite{delphi}.}}
\label{gaug_udd}
\end{center}
\end{figure}
The dominant contribution to the exclusion contours comes 
from the chargino pair production analyses in any \Rp~couplings.
The neutralino pair production analysis becomes relevant in case of 
low $\tan \beta$, low m$_0$,
small M$_2$ and negative $\mu$ values (Fig.~\ref{gaug_udd}) 
which means when 
the chargino pair production cross-section is suppressed by 
destructive interferences between $s$- and $t$-channels.


From the exclusion plots in the ($\mu$,~M$_{2}$) plane  
the  extraction of the minimum gaugino masses which is not excluded for the investigated 
range of parameters within the MSSM is performed.
\begin{figure}[hb]
\begin{center}
\epsfig{file=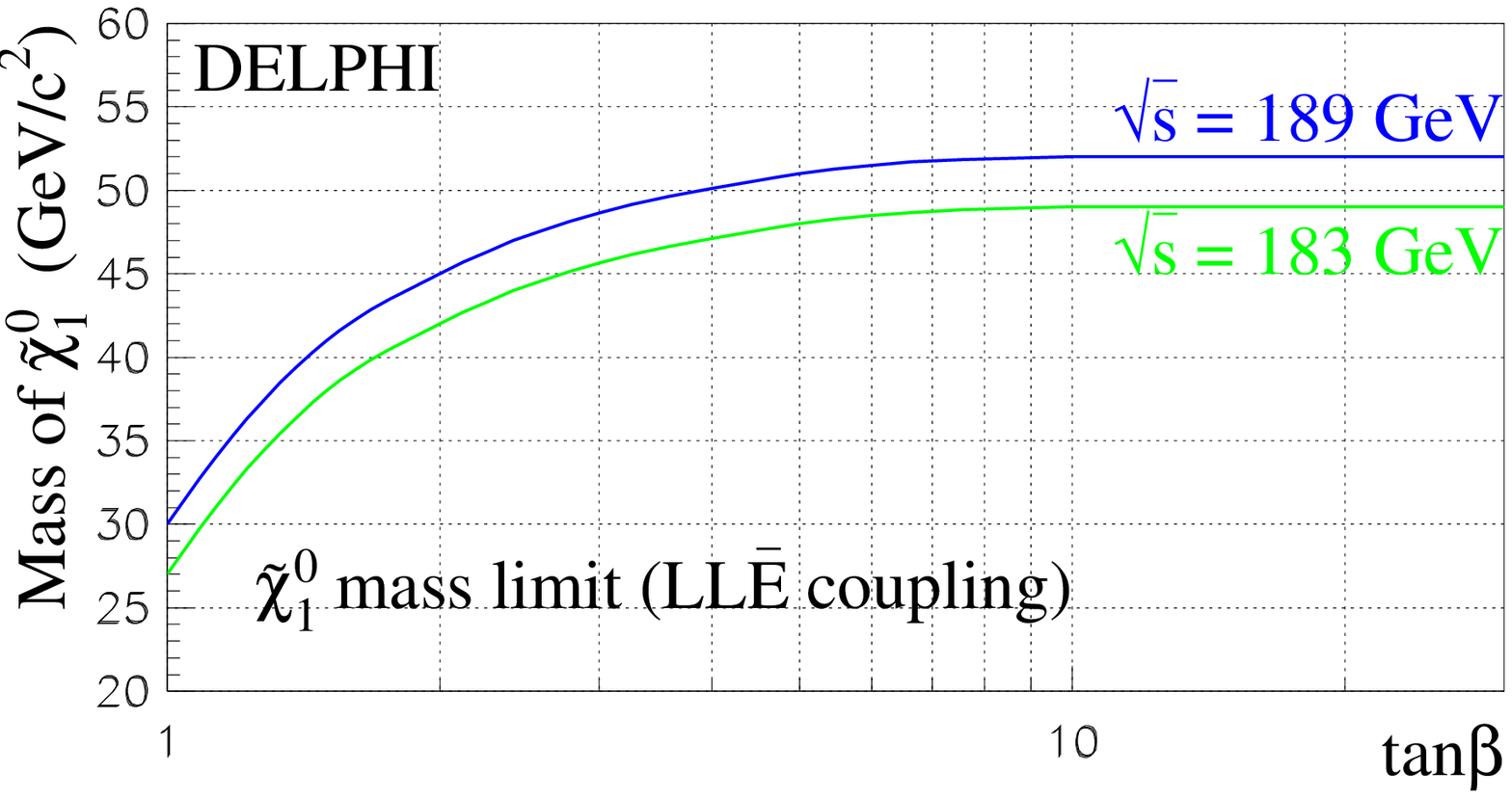,width=10cm}
\caption{{\it The lightest non-excluded neutralino mass as a function of
\tanb~ at $95 \%$~CL.
This limit is independent of the choice of m$_0$ in the 
explored range and of the generation indices $i,j,k$ of the $\lambda_{ijk}$ 
coupling \cite{delphi} and it assumes that the \XOIb in the detector.}}
\label{chilim_lle}
\end{center}
\end{figure}
These limits on the lightest chargino and
neutralino are obtained for high m$_0$ value which corresponds 
to the disappearance of neutralino  pair production  contribution.  
The mass of the lightest non-excluded gaugino is naturally shifted when \tanb~ is 
changed. In Fig.~\ref{chilim_lle}, the lightest non-excluded 
neutralino mass as a 
function of \tanb~ for the \LLE\ searches in DELPHI is shown. 
It has been checked that this result is independent of m$_0$ values.  

In this context, one of the most important results in the searches
for supersymmetry obtained with 
data taken up to a centre-of-mass energy of $208 \GeV$ at LEP~2 \index{LEP}
by the four LEP experiments is that the lightest chargino mass 
is excluded at $95 \%$~CL up to $103 \GeVcc$ and the lightest neutralino mass 
is excluded at $95 \%$~CL up to $39 \GeVcc$ 
in the framework of the MSSM with \Rp\ assuming that the \XOIb\ decays in the detector. 
These results are formally only valid in the scanned MSSM parameter space, 
i.e. for $1 \leq \rm{tan}\beta \leq 35$, m$_0 \leq 500 \GeV$, 
         $|\mu| \leq 200 \GeV$, M$_2 \leq 500 \GeV$, 
and for any coupling value from 10$^{-4}$ up to the existing limits.


We hinted above of situations where indirect (cascade) decays could
play a significant role. This could be the case for instance at future 
lepton colliders where centre-of-mass energies far beyond the current 
lower mass limits on the LSP are being contemplated. In view of the 
constraints established at LEP~2, the possibility of opening up 
large production of heavier neutralinos and charginos at a 
future 500~\GeV\ \ee\ linear collider (LC) has been studied 
in Ref.~\cite{Godbole} assuming that the lightest neutralino is the 
LSP and in presence of \Rp.
The study showed in this case (for a representative but finite number 
of points in the constrained MSSM parameter space)
that only the production modes involving the \XOI, \XOII, ${\tilde {\chi}}_1^{\pm}$
need to be considered. The \XOIII, \XOIV, and ${\tilde {\chi}}_2^{\pm}$ being almost
always beyond the reach of a $500 \GeV$ machine. 
As a consequence, the analysis would remain relatively simple, with
a limited amount of cascade decays to take into account.
Moreover, for a large part of the parameter space, 
the \XOII\ is nearly degenerate with the lighter chargino, 
and then, the number of decay chains to be considered is futher 
reduced.
The signals produced by \LLE, \LQD, \UDD\ operators have been studied
in Ref.~\cite{Godbole} and retain the basic characteristics listed
above. For \LLE\ the dominating signal remains an excess of multi-lepton
final states, with possibly substantial missing energy. 
For \LQD\ the final states contain again leptons and jets. In this case,
an algorithm to reconstruct the LSP and higher neutralino masses enables 
to identify the signal as due to supersymmetry with specific \Rp\ 
operators.
The existence of like-sign dilepton final states, originating from the
Majorana nature of neutralinos, appears to be a very promising signal
practically free of true \SM\ background sources.
For \UDD\, the final states consist again of multiple jets which are more
difficult to disentangle from a large number of \SM\ background sources.
Mass reconstruction nevertheless appears promising here also to allow to 
identify the signal as due to supersymmetry with specific \Rp\ operators.

\noindent \addtocounter{sss}{1}
{\bf \thesss Searches for Gaugino Pair Production at Hadron Colliders}  
 \addcontentsline{toc}{subsection}{\hspace*{1.2cm} \alph{sss}) 
            Searches for Gaugino Pair Production at Hadron Colliders}

Cascade decays involving trilinear \Rp\ couplings could also play a major r\^ole 
at future hadron colliders. 

Outstanding multi-lepton event signatures are expected in the presence of a 
$\lambda $ Yukawa coupling. The case of gaugino pair production with a trilepton 
signature has been investigated for the $D\emptyset$ experiment at the
Tevatron collider in Ref.~\cite{d0nagy}, in the framework of mSUGRA.

$D\emptyset$~\cite{NEWD0RPV2} also considered the dimuon and four-jets channel
occuring after \XOI decay via the $\lambda^{\prime}_{2jk}$ coupling ($j=1,2; k=1,2,3$)
where the \XOI can be produced either directly in pair or through cascade decays from
squarks or gluinos. Gluinos masses below $224 \GeVcc$ (for all squark masses and for
$tan \beta = 2$) are excluded. For equal masses of squarks and gluinos the mass limit
is $265 \GeVcc$.

The $\lambda''$ coupling implies multi-jet final states for sparticle
decays which severely challenges the sensitivity at $p\bar{p}$ and 
$p p$ colliders.

However the 1-lepton and various dileptons and 
multi-lepton event topologies that result from simultaneous  
production of all sparticles at the Tevatron assuming the LSP decays via baryon number violating operators
have been studied in~\cite{baer95} giving reaches on the gluino mass from 150~GeV up 
to 360~GeV depending on the specific topology. 

The study of the decay chain 
${\tilde q}_L \rightarrow {\tilde {\chi}}^0_2 q \rightarrow {\tilde l}_R lq 
  \rightarrow  {\tilde {\chi}}^0_1 ll q $
followed by the decay of the ${\tilde {\chi}}^0_1$ into three quarks assuming a non-zero
$\lambda''_{212}$ at the LHC  has been performed in~\cite{Allanach01}.
This study shows that even in the choice of a non-zero $\lambda''_{212}$, which is considered
as the hardest choice, the ${\tilde {\chi}}^0_2$, ${\tilde {\chi}}^0_2$ and ${\tilde q}_L$
can be detected and their masses measured and that the mass of the ${\tilde l}_R$ can be
obtained in much of the parameter space.

\index{Sparticle production!Gaugino-Higgsino pair|)}

\subsection{Sfermion Pair Production}
\label{sec:pairsf}
\index{Sparticle production!Sfermion pair|(}
 \noindent \addtocounter{sss}{1}
 {\bf \thesss Production and final states}   
  \addcontentsline{toc}{subsection}{\hspace*{1.2cm} \alph{sss}) 
              Production and final states}

As for the gaugino-higgsino production discussed above, we shall concentrate
in this section on the phenomenology associated with the presence of 
\Rp\ interactions. But we shall first briefly review the key ingredients 
for sfermion pair production via standard gauge couplings at colliders.

The sfermion mass eigenstates, \sfea\ and
\sfeb\ (f:~$q$ or $\ell$, \sfea\ lighter than \sfeb), are obtained from
the two supersymmetric scalar partners \sfeL\ and 
\sfeR\ of the corresponding left and right-handed 
\mbox{fermion~\cite{ellis,bartl}:}
\begin{center}
\begin{tabular}{lcl}
\sfea\ &=& ~\sfeL\ cos\thmixf\ + \sfeR\ sin\thmixf \\
\sfeb\ &=& --\sfeL\ sin\thmixf\ + \sfeR\ cos\thmixf 
\end{tabular}
\end{center}
where \thmixf\ is the mixing angle with 0~$\leq$~\thmixf~$\leq \pi$.
According to the equations which give the sfermion masses
(see for example in~\cite{mssm2}), the left-handed
sfermions are most often heavier than their  right-handed
counterparts.  The \sfeL--\sfeR\ mixing is related to the off-diagonal
terms of  the scalar squared-mass matrix. It is
proportional to the fermion mass, and is small compared to the
diagonal terms, with the possible exception of the third family
sfermion~\cite{drees}.  
The lightest
squark is then probably the lighter stop $\tilde{t}_1$.
This is due not only to the mixing effect but also to  
the effect of the large top Yukawa coupling; both tend to decrease the mass of
$\tilde{t}_1$~\cite{dreesmartin}.  
The lightest charged slepton is probably
the $\tilde\tau_1$. For small values of tan$\beta$, $\tilde\tau_1$ is
predominantly a $\tilde\tau_R$, and it is not so much lighter than
$\tilde{e}_R$ and $\tilde\mu_R$.


Sleptons and squarks can be pair produced in $e^+e^-$
collisions via the ordinary gauge couplings of supersymmetry with conserved 
$R$-parity provided that their masses are kinematically accessible.
They can be produced via $s$-channel $\mathrm{Z}$ or $\gamma$
exchange with a cross-section depending on the sfermion mass.
The \snue\ (\sel) can also be produced via the exchange of a chargino
(neutralino) in the $t$-channel, provided that the gaugino component 
of the chargino (neutralino) is the dominant one.
The $t$-channel contributes if the chargino (neutralino) mass is
sufficiently low, and then the cross-section depends also on the \XPM\ (\XO) 
mass and field composition and thereby on the relevant parameters of the
supersymmetric model. 
The coupling between the squarks and the $Z$ boson depends on the mixing 
angle $\theta_{\tilde{\mathrm q}}$, and it is minimal for a particular 
angle value. 
For example in the case of the stop, the decoupling between the
$\theta_{\tilde{\mathrm t}}$ and the $Z$ is maximal for
\mbox{$\theta_{\tilde{\mathrm t}} = 0.98$~rad} such that the
stop pair production cross-section is minimal. 


In general, both direct and indirect decays of sfermions can be studied
in sparticle pair production at colliders. 
The direct decay of a sfermion via a given \Rp\ coupling involves
specific standard fermions and can be (e.g. when involving the
top quark) kinematically closed.
The final states resulting from the decay of pair-produced sleptons or squarks
are listed in Table~\ref{tab.sec6.pair.sferd} 
and~\ref{tab.sec6.pair.sferi} for the three different 
couplings $\lambda$ ( \LLE\ ), $\lambda'$ ( \LQD\ ), or $\lambda''$ ( \UDD\ ).
\begin{table}[htb]
\begin{center}
\begin{tabular}{|c|c|c|c|} \hline
{sfermions}         & \LLE         & \LQD   & \UDD         \\ \hline
\snu\asnu          & 4 $\ell$      & 4 $j$ & not possible \\ \hline
\slep$_R^+$\slep$_R^-$ & $\ell \ell^\prime$ +\Emiss& not possible  & not possible \\ 
\slep$_L^+$\slep$_L^-$ & 2 $\ell$ +\Emiss& 4 $j$ & not possible \\ \hline
$\tilde{u}_L \bar{\tilde{u}}_L$, $\tilde{d}_R \bar{\tilde{d}}_R$
                   & not possible & 2$\ell$ + 2 $j$ & 4 $j$ \\
$\tilde{d}_R \bar{\tilde{d}}_R$
                   &              & 1$\ell$ + 2 $j$ + \Emiss & \\
$\tilde{d}_L \bar{\tilde{d}}_L$, $\tilde{d}_R \bar{\tilde{d}}_R$
                   &              &        2 $j$ + \Emiss & \\ \hline
\end{tabular} 
\caption{{\it Sfermion pair production final states in case of direct \Rp\
              decays. The notations $l$, \Emiss and $j$ correspond 
	      respectively to charged lepton, missing energy from at 
	      least one neutrino and jet final states.}} 
\label{tab.sec6.pair.sferd}
\end{center}
\end{table}
\begin{table}[htb]
\begin{center}
\begin{tabular}{|l|c|c|c|}\hline
sfermions           &{\LLE}          & {\LQD}              & {\UDD}         \\ \hline
\snu\asnu  & 4 $\ell$ + \Emiss &  $2\ell + 4j$ + \Emiss & $6j$ + \Emiss  \\
           &                   &  $2\ell + 4j$ + \Emiss &                \\
           &                   &  $        4j$ + \Emiss &                \\ 
\hline
\slep$^+$\slep$^-$ 
           & 6 $\ell$+ \Emiss  &  $4\ell + 4j$          & $2\ell + 6j$  \\
           &                   &  $3\ell + 4j$ + \Emiss &                \\
           &                   &  $2\ell + 4j$ + \Emiss &                \\ 
\hline
\SQ\SQB    & 4 $\ell + 2j $ + \Emiss  &  $2\ell + 6j$    & $8j$  \\
           &                &  $ \ell + 6j$ + \Emiss &                \\
           &                &  $     6j$ + \Emiss &                \\ \hline
\end{tabular}
\caption{{\it Sfermion pair production final states in case of indirect 
decays when the LSP is the lightest neutralino. The notations $l$, \Emiss and $j$ correspond 
	      respectively to charged lepton, missing energy from at 
	      least one neutrino and jet final states.}} 
\label{tab.sec6.pair.sferi}
\end{center}
\end{table}

When considering sfermion pair production and the decays of 
Table~\ref{tab.sec6.pair.sferd} and~\ref{tab.sec6.pair.sferi}
relevant for $e^+e^-$ colliders, it should be noticed that in general
the indirect decays into a chargino will not be considered.
This is because the chargino search itself provides a mass limit 
close to the kinematic limit. There is no phase space left for 
the production of (e.g.) two sleptons followed by decays involving 
charginos.
This explains why for instance at LEP, \index{LEP} the most general sfermion 
indirect decay studied has been the decay into the lightest neutralino 
considered as the LSP 
(\mbox{\snu \Ra $\nu$ \XOI}, \mbox{\slep \Ra $\ell$ \XOI}, 
 \mbox{\SQ \Ra $\rm q'$ \XOI}). 
The final states are then composed of two fermions plus the decay products
of the \Rp\ decay of the neutralino pair 
(see Table~\ref{tab.sec6.pair.sferi}).

\noindent \addtocounter{sss}{1}
{\bf \thesss Slepton Searches at lepton colliders} 
\addcontentsline{toc}{subsection}{\hspace*{1.2cm} \alph{sss})
              Results on Slepton Searches} 

\index{Sparticle production!Sfermion pair!at lepton colliders}

Here again one can profit from what has been learned from actual 
studies by the experiments at LEP \index{LEP} collider where 
both productions of sneutrino and charged slepton pairs have been
searched for. 

Early discussions on several $R$-parity-violating processes at
$e^+e^-$ colliders including charged slepton pairs can be found 
in~\cite{Workshop1,Roszkowski93}. 


Lets first consider the case of sneutrino pair production.
In the presence of \LLE\ interactions, searches for four 
charged lepton final states are performed. The six possible
configurations are listed in Table~\ref{tab.sec6.pair.4lep}.
\begin{table}[htb]
\begin{center}
\begin{tabular}{|l|lll|}\hline
final states & \multicolumn{3}{|c|}{processes and couplings} \\ 
             & \snue\asnue   & \snum\asnum     & \snut\asnut \\ \hline
eeee         &               & $\lambda_{121}$ & $\lambda_{131}$  \\
ee$\mu\mu$   & $\lambda_{121}$ & $\lambda_{122}$ & $\lambda_{132}, \lambda_{231}$ \\
ee$\tau\tau$ & $\lambda_{131}$ & $\lambda_{123}, \lambda_{231}$ & $\lambda_{133}$ \\
$\mu\mu\mu\mu$   & $\lambda_{122}$ &            & $\lambda_{232}$ \\
$\mu\mu\tau\tau$ & $\lambda_{123}, \lambda_{132}$ 
                 & $\lambda_{232}$ & $\lambda_{233}$ \\
$\tau\tau\tau\tau$ & $\lambda_{133}$ & $\lambda_{233}$ &  \\ \hline
\end{tabular}
\caption{{ \it Four lepton final states produced by the direct decay via a
             \LLE\ term of a sneutrino pair.}}
\label{tab.sec6.pair.4lep}
\end{center}
\end{table}
Event topologies containing muons and electrons allow for a high 
selectivity by applying conditions on two lepton invariant masses.
The highest efficiencies are obtained when there are at least two 
muons in the final states, and the lowest when there are taus. 
In the latter case, a certain amount of energy is missing in 
the final state, due to the neutrinos produced in the tau decays.
Then, most often, two extreme cases in the coupling choice are studied:
the first one considering that the \Labb\ or \Lbcb\ is dominant, leading
to the most efficient analyses, the second considering that the \Lacc\ 
or \Lbcc\ is dominant, leading to the least efficient analyses.
With these two analysis types, all the possible final  states
are probed.

In the presence of \LQD\ interactions, searches for four jets final states are 
performed. Here also, the absence of missing energy  in the
final state offers the possibility to reconstruct the sneutrino mass. 
Depending on the generation indices, 0, 2, and 4 jets can contain
a b-quark~(Table~\ref{tab.sec6.pair.4jets}). 
\begin{table}[htb]
\begin{center}
\begin{tabular}{|l|lll|}\hline
final states & \multicolumn{3}{|c|}{processes and couplings} \\ 
             & \snue\asnue   & \snum\asnum     & \snut\asnut \\ \hline
4 q (no b-quarks)& $\lambda^{\prime}_{1jk, j,k\neq 3}$ 
                 & $\lambda^{\prime}_{2jk, j,k\neq 3}$ 
                 & $\lambda^{\prime}_{3jk, j,k\neq 3}$ \\ 
2 q 2 b          & $\lambda^{\prime}_{1j3, j \neq 3}$, 
                   $\lambda^\prime_{13k, k \neq 3} $
                 & $\lambda^\prime_{2j3, j \neq 3}$, 
                   $\lambda^\prime_{23k, k \neq 3} $
                 & $\lambda^\prime_{3j3, j \neq 3}$,
                   $\lambda^\prime_{33k, k \neq 3} $ \\
4 b              & $\lambda^\prime_{133}$ 
                 &  $\lambda^\prime_{233}$ &  $\lambda^\prime_{333}$ \\ \hline
\end{tabular}
\caption{{\it Four jet final states produced by the direct decay via a
             \LQD\ term of a sneutrino pair.}}
\label{tab.sec6.pair.4jets}
\end{center}
\end{table}
With the possibility to tag the jets generated by b-quarks, the
analyses are very efficient.

The LEP \index{LEP} experiments have presented results on the lower limit on the 
sneutrino electron mass derived by searching for the direct decay of sneutrinos 
in
the data collected in 1999 and 2000 up to a centre-of-mass energy of $208 \GeV$.
With a \LLE\ (\LQD) coupling, the most conservative  lower limits are 
$98$ ($91$) $\GeVcc$
\cite{aleph, delphi, opal}.
For the derivation of limits, efficiencies are determined as function of 
the sneutrino mass.
In case of \snue, due to the possible $t$-channel contribution, 
they are considered for a specific set of MSSM parameters, generally
for \mbox{$\mu =-200 \GeV$}, \mbox{M$_2$ = $100 \GeV$}.
When the contribution of the $t$-channel becomes negligible, the 
\snue\asnue\ production cross-section is similar to the \snum\asnum\ 
or \snut\asnut\ ones.
Taking into account the number of expected events from the Standard
Model processes, the number of observed events, and the analysis efficiencies,
upper limits at 95\% of confidence level on the sneutrino cross-section 
are obtained as function of the sneutrino mass. Comparing these upper
limits to the expected MSSM cross-section, limits on the sneutrino
mass have been derived, as illustrated in Fig~\ref{fig.sec6.snu.dir}.
\begin{figure}[htb]
\begin{center}
\epsfig{file=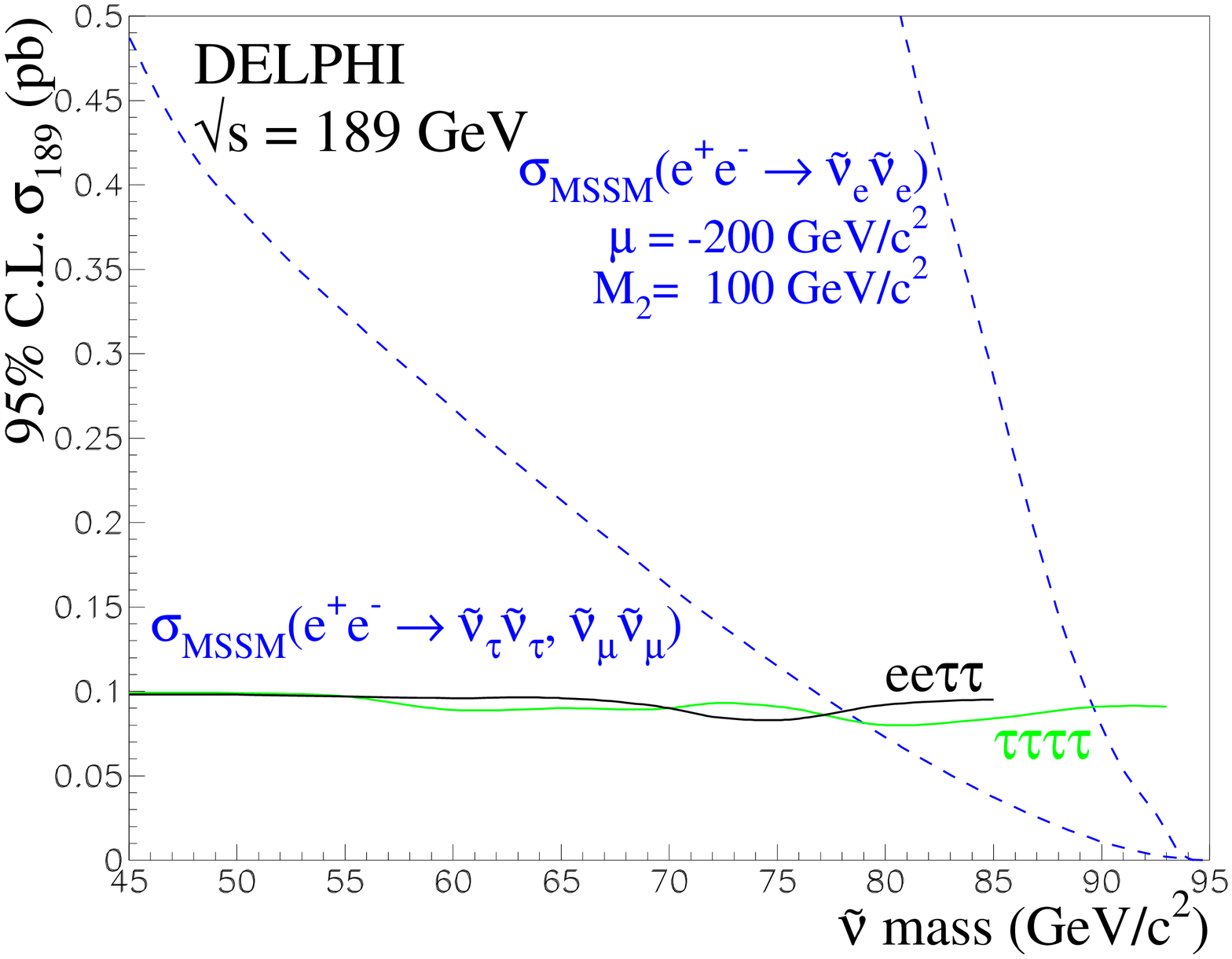,width=9.cm}
\vspace{-0.5cm}
\caption{{\it Sneutrino direct decay searches with \LLE\ coupling:
 limit on the \snu\asnu\ production cross-section as a function of the mass
for two different final states. The MSSM cross-sections 
are  reported in order
to derive a limit on the sneutrino mass in the case of direct
$\Rp$ decay. The dashed lower curve corresponds to both \snum\asnum\
and \snut\asnut\ cross-sections, which depend only on the sneutrino
mass.
The dashed upper curve on the plot is the \snue\asnue\ 
cross-section obtained for $\mu = -200 \GeV$ and M$_2 = 100 \GeV$,
the corresponding chargino mass lies between $90$ 
and $120 \GeVcc$~\cite{delphi}.}}
\label{fig.sec6.snu.dir}
\end{center}
\end{figure}
The limits obtained are much stronger that those existing in the
hypothesis of $R$-parity conservation, in which the sneutrino pair 
production  is invisible.


Right-handed charged sleptons
are mainly studied, because their production cross-section, for
a given slepton mass, is lower than 
for the left-handed ones, therefore leading to more conservative results.
The direct decay of a pair of charged sleptons lead to two charged
leptons and some missing energy. 
This low multiplicity
final state is difficult to analyse due to the high background of
low multiplicity processes.  
With a dominant $\lambda_{ijk}$ coupling constant, only the pair
produced $\tilde{\ell}_{kR}$ and $\tilde{\ell}_{iL}$ or $\tilde{\ell}_{jL}$
are allowed to directly decay. 
The decay of  $\tilde{\ell}_{kR}^+ \tilde{\ell}_{kR}^-$ gives 
$\ell_i \ell_i , \ell_i \ell_j , \ell_j \ell_j$ + \mbox{\Emiss}, 
and since $i \neq j$ two lepton flavours are mixed in
the final state (see Table~\ref{tab.sec6.pair.2lepr}). 
It is not the case in the direct decay of the supersymmetric
partners of the  
left-handed charged leptons, for which there is only one  lepton flavour in 
the $2 \ell + $\Emiss\ final state (Table~\ref{tab.sec6.pair.2lepl}).
In case of selectrons, the ${\tilde e}_L {\tilde e}_R$ production is possible
in the $t$-channel; direct decay of both selectrons is possible only
via \Laba\ (\mbox{ee, e$\mu$ +\Emiss} final state) or
via \Laca\   (\mbox{ee, e$\tau$ +\Emiss} final state).
\begin{table}[htb]
\begin{center}
\begin{tabular}{|l|lll|}\hline
final states & \multicolumn{3}{|c|}{processes and couplings} \\ 
             & \sel$_R$\sel$_R$     
             & \smu$_R$\smu$_R$ 
             & \stau$_R$\stau$_R$   \\ \hline
ee , e$\mu , \mu\mu$ + \Emiss
             & $\lambda_{121}$    
             & $\lambda_{122}$ 
             & $\lambda_{123}$ \\  
ee , e$\tau , \tau\tau$ + \Emiss 
             & $\lambda_{131}$      
             & $\lambda_{132}$ 
             & $\lambda_{133}$ \\
$\mu\mu , \mu\tau  , \tau\tau$ + \Emiss 
             & $\lambda_{231}$      
             & $\lambda_{232}$ 
             & $\lambda_{233}$ \\  \hline
\end{tabular}
\caption{{\it  Final states produced by the direct decay via a
             \LLE\ term of a pair of supersymmetric partners of the
right-handed leptons.}}
\label{tab.sec6.pair.2lepr}
\end{center}
\end{table}
\begin{table}[htb]
\begin{center}
\begin{tabular}{|l|lll|}\hline
final states & \multicolumn{3}{|c|}{processes and couplings} \\ 
             & \sell\sell   & \smul\smul     & \staul\staul \\ \hline
ee+\Emiss & $\lambda_{121,131}$ & $\lambda_{121,231}$ & $\lambda_{131,231}$  \\
$\mu\mu$+\Emiss   
        & $\lambda_{122,132}$ & $\lambda_{122,232}$ & $\lambda_{132,232}$ \\
$\tau\tau$+\Emiss 
        & $\lambda_{123,133}$ & $\lambda_{123,233}$ &
             $\lambda_{133,233}$ \\
\hline
\end{tabular}
\caption{{\it Two lepton with missing energy final states produced 
by the direct decay via a \LLE\ term of a pair of supersymmetric
partners of left-handed charged leptons.}}
\label{tab.sec6.pair.2lepl}
\end{center}
\end{table}

\begin{figure}[htb]
\begin{center}
\epsfig{file=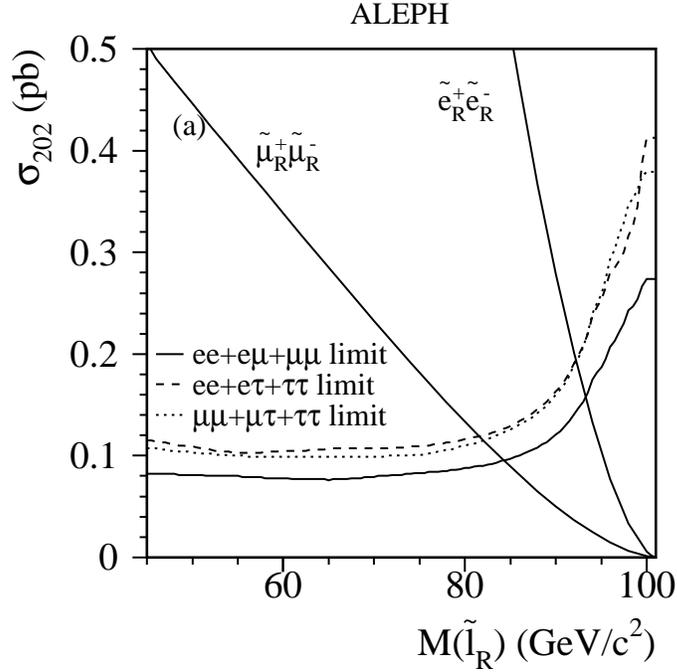,width=9.5cm}
\vspace{-.2cm}
\caption{{\it Charged slepton direct decay searches with \LLE\ couplings:
the $95 \%$ CL exclusion cross-sections for sleptons. The MSSM
cross-section for pair production of right-handed selectrons and
smuons are superimposed~\cite{aleph}.}}
\label{fig.sec6.slep.dir}
\end{center}
\end{figure}
Similarly to the sneutrino decay, the search for final states containing
mainly taus is the least efficient one. An upper limit on the cross-section 
is obtained as a function of the slepton mass.
Comparing this upper limit to the expected MSSM cross-section, limits on the 
slepton mass is deduced. In case of selectron production, the limit depends 
also on the chosen MSSM parameters, since the neutralino exchange in the 
$t$-channel may also contribute to the cross-section.  
From the data collected 
at LEP~2, \index{LEP} the ALEPH experiment derived
a lower limit of 
$96 \GeVcc$ on the \smur~\cite{aleph}, and OPAL 
obtained a limit of $74 \GeVcc$ for the same slepton mass~\cite{opal}, 
when a \LLE\ coupling is considered to be dominant. 

\begin{figure}[htb]
\begin{center}
\vspace{-0.5cm}
\epsfig{file=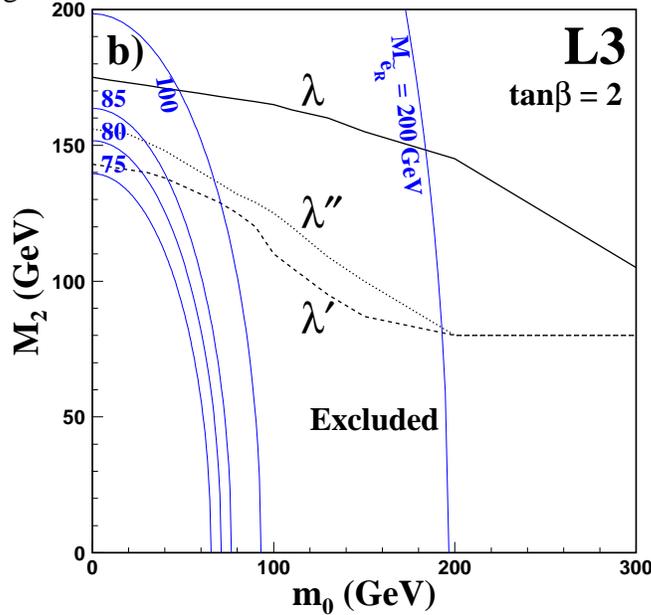,width=8.5cm}
\vspace{-0.5cm}
\caption{{\it Exclusion contours in the m$_0$--M$_2$ plane, for 
$\tan \beta = $2, at $95 \%$ CL. The lines represent the 
 isomasses of the supersymmetric partner of the right-handed electron
(labelled with the corresponding value in~$\GeVcc$). The solid and dotted
curves show the $95 \%$ CL lower limits on M$_2$ as a function of m$_0$ from
which the limits on the scalar electron mass were derived~\cite{l3}.} 
\label{fig.sec6.slepL3}}
\end{center}
\end{figure}

From exclusion contours in the $\mu$--M$_2$  plane, determined with the MSSM 
interpretation of the gaugino pair production results, after the analysis of 
the data collected up to $189 \GeV$, the L3 experiment sets indirect lower 
limit on the scalar lepton masses~\cite{l3}. Figure~\ref{fig.sec6.slepL3} 
shows how the lower limits on the mass of the supersymmetric partner of the 
right-handed electron are derived, taking into account the limits on M$_2$ 
as a function of m$_0$.


Contrary to the direct decays, the slepton indirect decays can be studied in
any choice of the dominant coupling.
As previously said, mainly the indirect decay into a neutralino (LSP)
is searched for: 
\begin{itemize}
\item indirect decay \snu \Ra $\nu$\XOI\ ,  \XOI\ $\Rp$ decay via any coupling,
\item indirect decay \slep \Ra $\ell$\XOI\ ,  \XOI\  $\Rp$ decay via
any coupling.
\end{itemize}
Then the final state depends on the slepton type (and
flavour in case of charged sleptons), and mainly on the \XOI\ LSP decay.
The efficiencies are determined 
in a m$_{\tilde{\chi}}$ versus m$_{\tilde{\nu}}$ (m$_{\tilde{\ell}}$  )  plane
as well as the upper limit on the cross-section, which
is compared to the MSSM cross-section, in order to exclude
domains in the same m$_{\tilde{\chi}}$ versus m$_{\tilde{\nu}}$
(m$_{\tilde{l}}$  ) plane (see Fig.~\ref{sec6.fig.slep.ind}).
The limit on the neutralino mass is used to set the limit on the 
sneutrino (slepton) mass in case of indirect decay.
The results obtained on data collected in 1998, 1999 and 2000~\cite{aleph,delphi,l3,opal}
are summarized in Table~\ref{sec6.tab.slep.ind.res}.
\begin{table}[htb]
\begin{center}
\begin{tabular}{ll|c|c|c|c}\hline
\multicolumn{2}{c|}{experiments} & ALEPH & DELPHI & L3        & OPAL \\ \hline
\multicolumn{2}{c|}{DATA} & 1998-2000 & 1998-2000 & 1998-2000 & 1998-2000 \\ \hline
                 &\LLE             &  89    &  85   &  78       & 81 \\
\snu$_{\mu,\tau}$& \LQD            &  78   &        &           &    \\ 
                 & \UDD            &  65   &        &  70       &     \\ \hline
           & \LLE                  &  96   &  95    &  79       & 99 \\
 \selr     & \LQD                  &  93   &        &           & 92  \\ 
           & \UDD                  &  94   &  92    &  96       &    \\ \hline
           & \LLE                  &  96   &  90    &  87       & 94 \\
 \smur     & \LQD                  &  90   &        &           & 87  \\ 
           & \UDD                  &  85   &  87    &  86       &    \\ \hline
           & \LLE                  &  95   &        &  86       & 92   \\   
 \staur    & \LQD                  &  76   &        &           &     \\ 
           & \UDD                  &  70   &        &  75       &    \\ \hline 
\end{tabular}
\caption{{ \it $95 \%$ CL lower limits (in~$\GeVcc$) on the slepton mass, considering the
slepton indirect decay in lepton and lightest neutralino only.}
\label{sec6.tab.slep.ind.res}}
\end{center}
\end{table}
\begin{figure}[p]
\begin{center}
\vspace{-0.5cm}
\epsfig{file=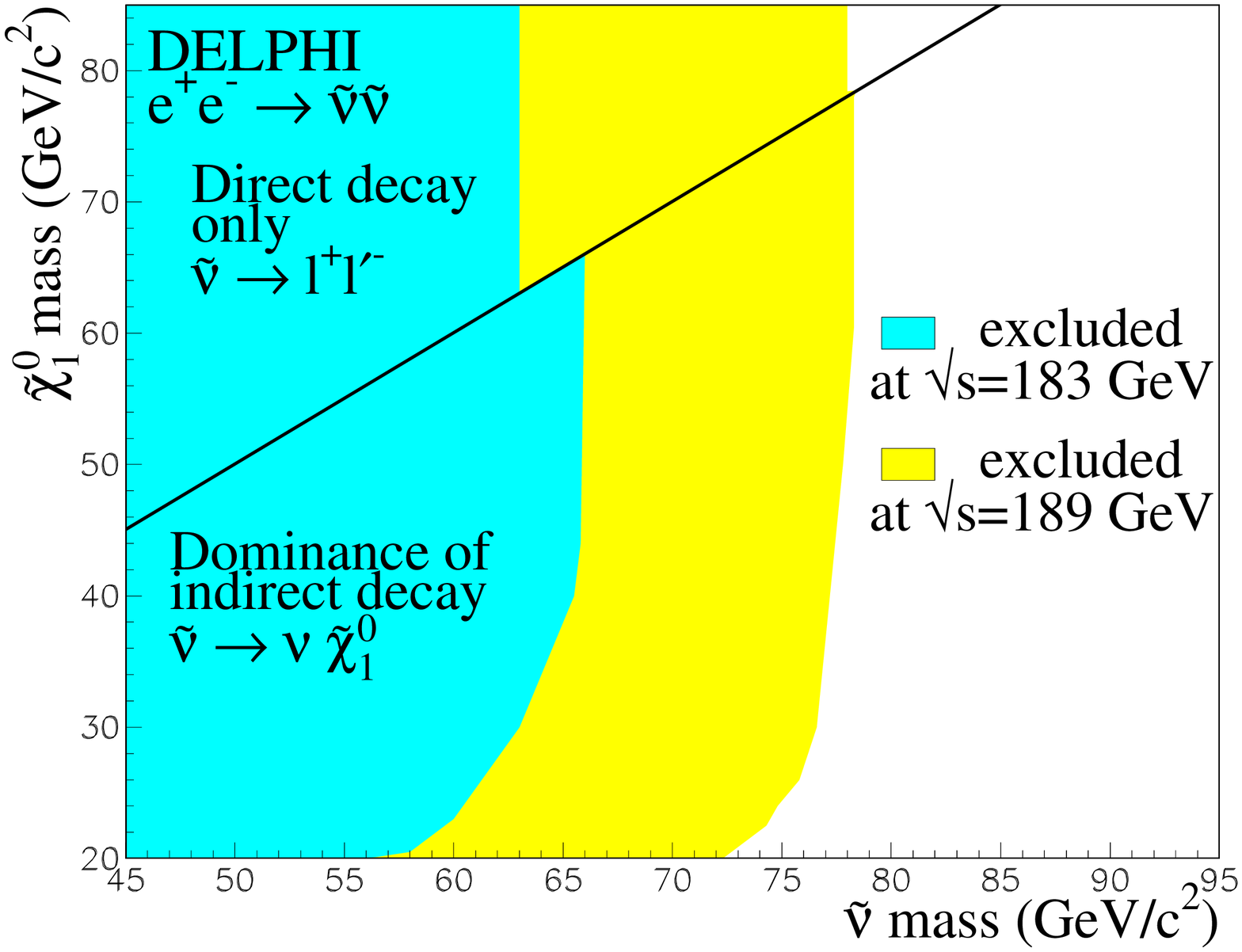,width=8.cm}
\epsfig{file=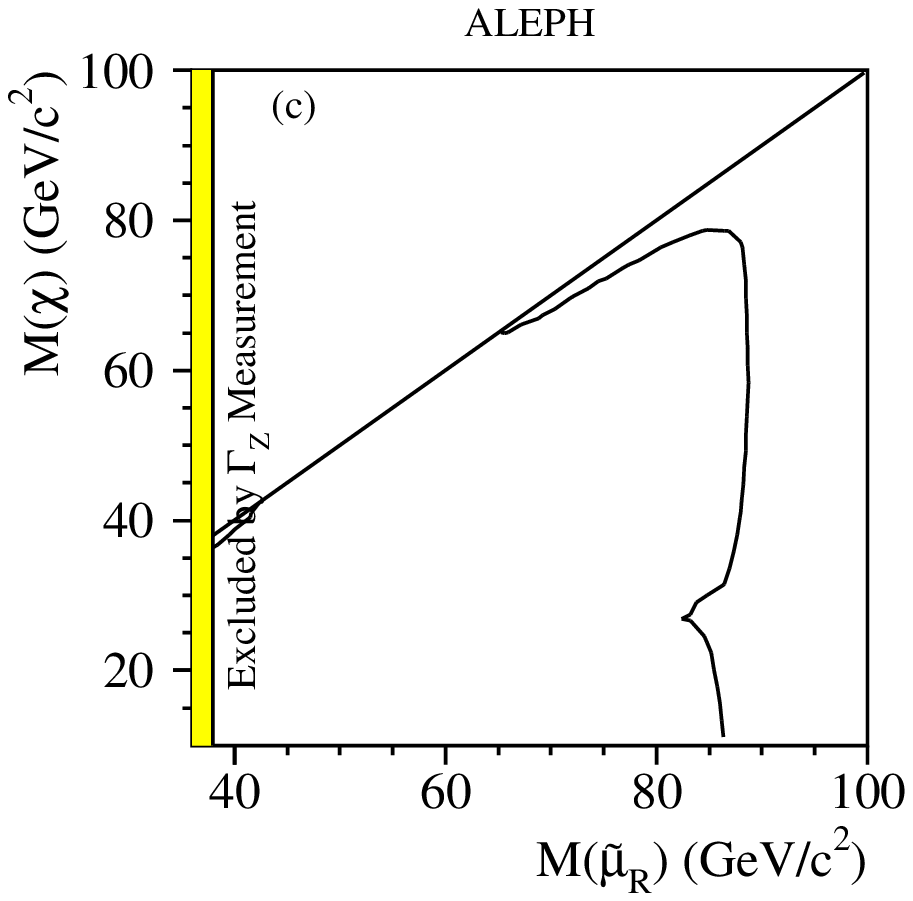,width=8.cm}
\epsfig{file=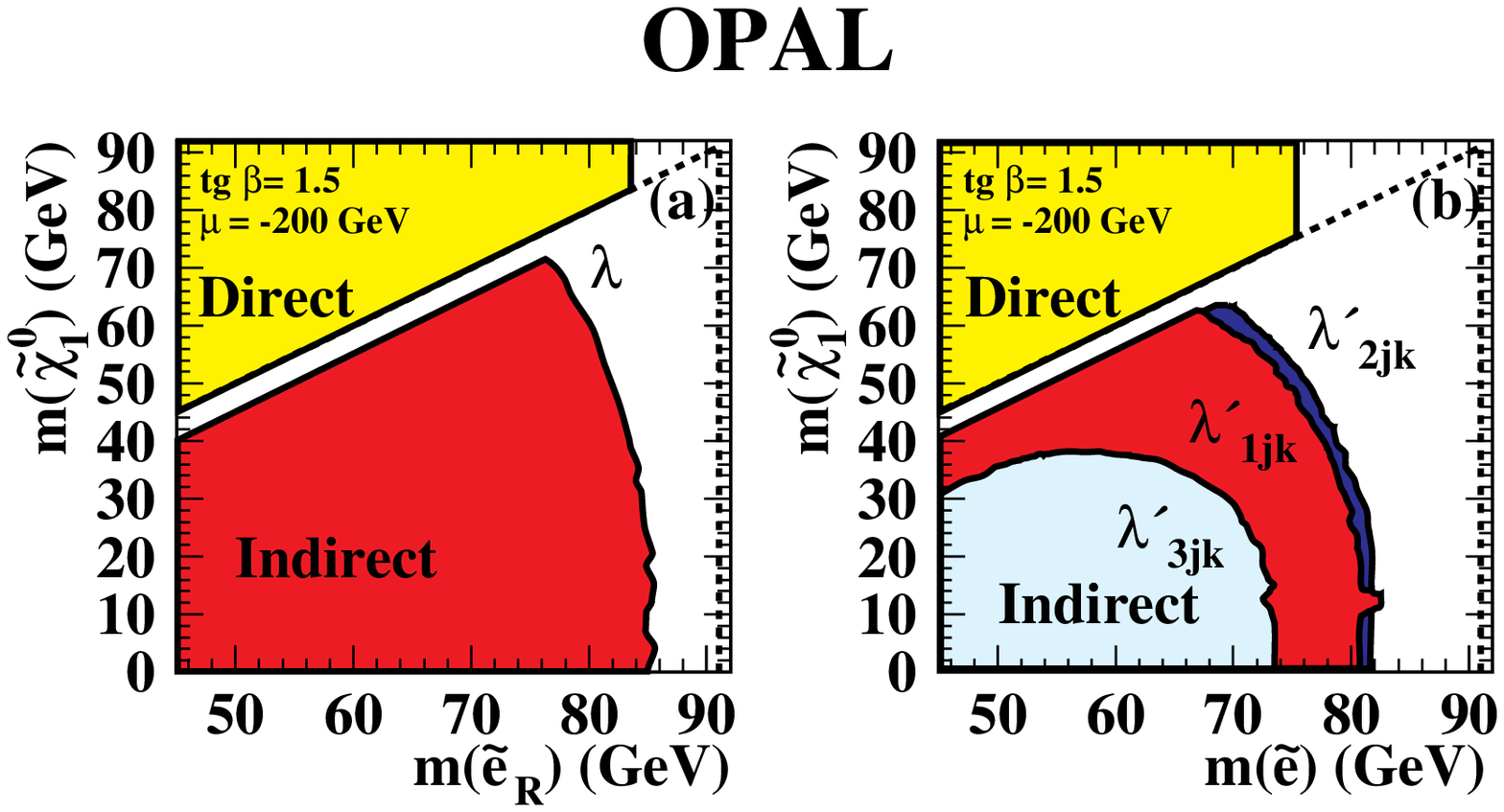,width=12.cm}
\caption{{\it Top: Search for sneutrino decaying via a dominant \LLE\
              coupling in DELPHI; excluded region at $95 \%$ CL in 
              m$_{\tilde{\nu}}$, m$_{\tilde{\chi}^0}$  plane by \snu\ pair
              production for direct and indirect decays.
              The dark grey area shows the part excluded  by the
              searches at $\sqrt s = 183 \GeV$, the light grey area the one 
	      excluded by the searches at $\sqrt s = 189 \GeV$. Middle:  
	      Search for smuon decaying via a dominant \UDD\ coupling in 
	      ALEPH;  excluded region at $95 \%$ CL in 
              m$_{\tilde{\chi}^0}$, m$_{\tilde{\mu}_R}$ plane. 
              Bottom: Search for selectron decaying indirectly via a dominant 
	      \LLE\ (left) and \LQD\ (right) coupling in OPAL; MSSM exclusion 
	      region for selectron pair production in the 
	      (m$_{\tilde{\chi}^0}$, m$_{\tilde{\ell}}$) plane at $95 \%$ CL.}} 
\label{sec6.fig.slep.ind}
\end{center}
\end{figure}


The pair production of right selectrons followed by their decay
in the presence of $R$-parity-violating couplings has been 
investigated for a 500~\GeV\  $e^- e^-$ linear collider (with possibly
highly polarized beams) in Ref.~\cite{Ghosh}.

At such a collider, a very strong suppression of the Standard Model background
is expected and this could be further reduced by exploiting specific beam 
polarizations. 
The conservation of electric charge and lepton number actually forbid
the pair production of most of supersymmetric particles at a
$e^-e^-$ collider: only selectrons can be produced via the
$t$--channel exchange of a neutralino. The pair production of
selectrons has been studied in the hypothesis of $R$-parity violation.
In case of \LLE\ operator, the decay of the pair produced right
selectrons will lead to final state consisting in 
 \mbox{$e^-e^-$ + 4$\ell^\pm$ + E$_{miss}$}, where the flavour of
 $\ell^\pm$ depend on the particular type of coupling. This kind
of final state is free from Standard Model background and permit an
easy detection at a $500 \GeV$ $e^-e^-$ collider. In case of
a dominant \LQD\ operator, the final state consists of charged
leptons, multiple jets and/or missing energy, and in addition, the
Majorana nature of the LSP gives rise to like-sign dilepton
signal, with almost no background from Standard Model.
In case of a dominant \UDD\ operator, the final state consists
of multiple jets associated to like-sign dielectrons. In both
\LQD\ and \UDD\ cases, it might be possible to give an estimate for
the LSP mass from invariant mass reconstruction.

\noindent \addtocounter{sss}{1}
{\bf \thesss Squark Searches at lepton colliders} 
\addcontentsline{toc}{subsection}{\hspace*{1.2cm} \alph{sss})
              Results on Squark Searches} 

\Rp\ decays of pair-produced squarks have been searched for in \ee 
collisions at \index{LEP} LEP~2. Special emphasis has been given to the 
$\tilde{t}$ and $\tilde{b}$ as they could possibly be the lightest squarks.

For squarks of the third generation, the \sfeL--\sfeR\ mixing cannot
be neglected. Hence a mixing angle must be taken into account for the pair 
production cross-section. This parameter will consequently enter as a 
free parameter when deriving experimental squark mass limits.

The results obtained at LEP~2 \index{LEP} for the searches of both squark direct 
and indirect decays are reviewed in the following.

For small couplings ($<O(10^{-1})$) the $\tilde{t}$ hadronises into colourless 
bound states before decaying (see section~\ref{sec:dirsfdecays}),
producing additional hadronic activity in the 
decay. Another consequence of the small width of the $\tilde{t}$ indirect 
decay ($\tilde{t} \rightarrow c \XOI$) is that, unusually, the direct decay 
dominates over the indirect decay for a large range of coupling values 
($>O(10^{-5})$). On the contrary the indirect decay of the $\tilde{b}$ 
($\tilde{b} \rightarrow  b \XOI$) dominates whenever kinematically 
possible.


As no quark superfield enters in the \LLE\ term of the \Rp\ superpotential, 
there is no direct two-body decay of squark via a $\lambda$ coupling. 
The \LQD\ and \UDD\ terms can be responsible for squark direct decays. 
In the first case, the final states consist of two jets and charged 
lepton(s) and/or missing energy: the three possibilities are listed in 
Table~\ref{tab.sec6.pair.sferi}.
In the stop pair production searches, only the channels with two charged 
leptons were considered.
Highest efficiencies are obtained for final states containing
electrons ($\lambda_{13k}^\prime$) or muons ($\lambda_{23k}^\prime$); 
final states with two taus ($\lambda_{33k}^\prime$)
are more problematic and consequently have lower efficiency 
and lead to weakest limits.
Using the efficiencies determined for different stop masses,
an upper limit on the stop pair production cross-section can be set
at $95 \%$~CL as a function of the  stop mass. 
Then, considering the cross-section for the stop pair production 
(e.g. in the framework of the MSSM) and in case of no mixing and maximal  
decoupling to the $Z$ boson, a lower limit 
on the stop mass can be derived.
In the tau channel, a stop mass lower than 
$96 \GeVcc$ has been excluded 
by OPAL~\cite{opal} for any mixing angle using the data recorded in 1999 and 2000.
At a centre-of mass energy from 
189 to $209 \GeV$, considering also the tau 
channel, but in a no mixing scenario, stop masses lower than 
$97 \GeVcc$ are excluded by ALEPH~\cite{aleph}.

Via a \UDD\ term, the stop decays directly into two down quarks 
and the sbottom into an up and a down quark.
The signature for the pair production of squarks is therefore four 
hadronic jets. 
Selections for these types of topologies rely mainly on reconstructing 
the mass of the decaying squark by forcing the event to four jets 
and forming the invariant masses between pairs of jets. 
For couplings involving a b-quark in the final state ``b-tagging'' 
algorithms based on requiring large impact parameter tracks are also 
helpful to separate signal from the large background coming from two-quark 
and four-quark Standard Model processes.  
In ``flavour blind'' searches, the cross-section limits degraded in the 
range of W mass. For direct decay via a \UDD\ interaction 
($\lambda^{\prime\prime}$ coupling), stop masses lower than 
$77 \GeVcc$ have been excluded by OPAL~\cite{opal}
for any mixing angle using the data recorded in 1999 and 2000. 

Assuming that the lightest neutralino \XOI is the LSP, the squark indirect 
decay into a quark and a  neutralino with the subsequent 
\mbox{\Rp\ decay} of the neutralino, has been studied. 
As any squark field can couple to the \XOI, 
there are no restrictions related to  the squark ``chirality''. 
Final states for each coupling consist of the 
corresponding fermions from the neutralino pair 
decay plus two jets.

In case of \LLE\ coupling, the relevant final states are two hadronic 
jets + 4 leptons + missing energy.
Six jets are expected with a \LQD\ coupling, together with two charged 
leptons and no missing energy or 0-1 lepton + missing energy. 
Pure hadronic final states consisting of eight jets are expected 
in case of \UDD\ couplings. 

Using the efficiencies determined in the 
(m$_{\tilde{t}}$, m$_{\tilde{\chi}^0_1}$) plane,
upper limit on the stop pair production cross-section can be derived
as a function of the  stop and neutralino masses (taking \mbox{$\mu = -200 \GeV$}  
and \mbox{$\tan \beta = 1.5$}). 
Considering the MSSM cross-section for the stop pair production in case of 
no mixing and in case of a maximal decoupling to the $Z$ boson exclusion 
contours were derived in the m$_{\tilde{t}}$, m$_{\tilde{\chi}^0_1}$ plane.
By combining the exclusion contours with the result on
the neutralino mass limit, a lower bound on stop mass can be derived.

\begin{figure}[htb]
\begin{center}
\vspace{-0.5cm}
\epsfig{file=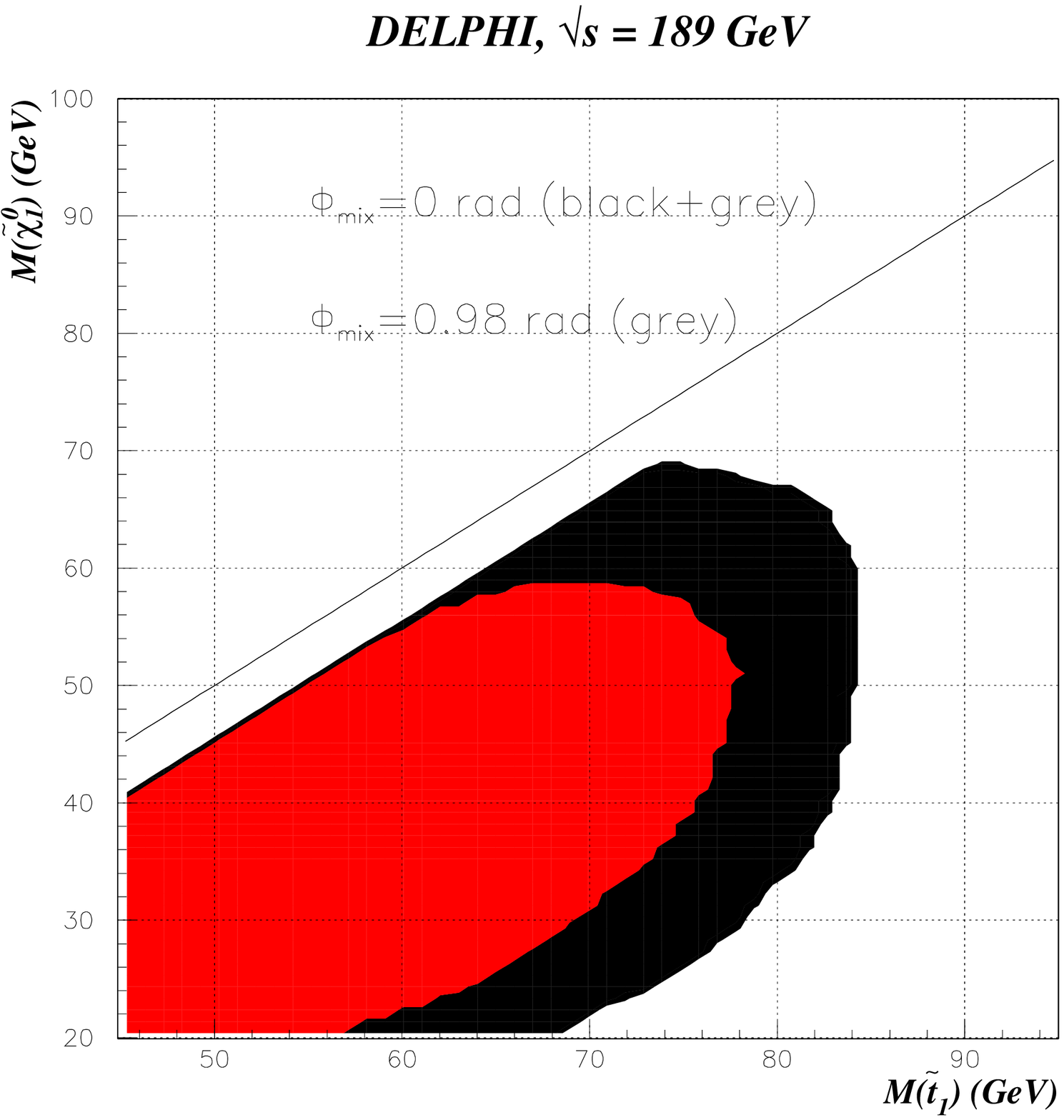,width=8.5cm}
\epsfig{file=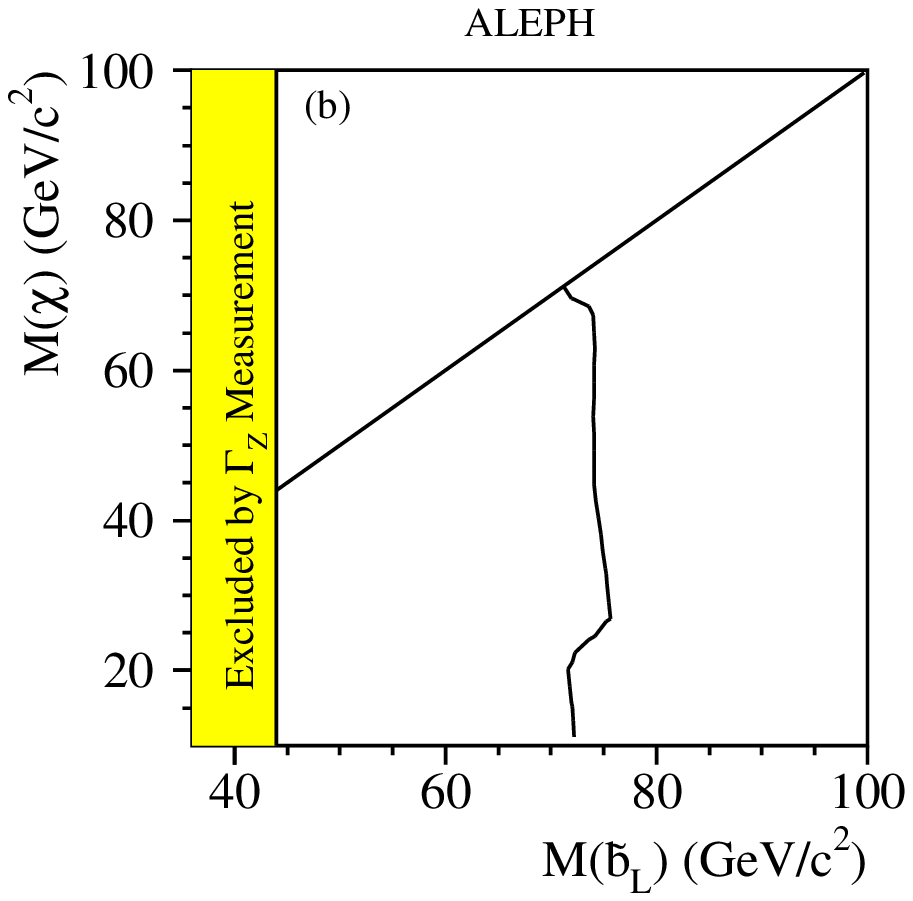,width=8.cm}
\caption{{\it Top: Search for stop decaying indirectly via a dominant \UDD\
              operator in DELPHI; excluded region at 95\% CL in 
              m$_{\tilde{\chi}^0}$ versus m$_{\tilde{\t}}$ plane; 
              the largest excluded area corresponds to the case 
              of no mixing and the smallest one  to the case with the mixing 
              angle which gives a minimum cross-section. 
              Bottom:  Search for left-handed sbottom decaying indirectly via 
	      a dominant \UDD\ operator in ALEPH;  the 95\% CL exclusion 
	      cross-sections is shown in the 
	      m$_{\tilde{\chi}^0}$ versus m$_{\tilde{b}_L}$ plane.}}
\label{sec6.fig.squ.ind}
\end{center}
\end{figure}
From the analysis of the events collected at a centre-of-mass energy
from $189 \GeV$ to $209 \GeV$, in the hypothesis of a dominant \LLE\ coupling,
a left-handed stop lighter than $91 \GeVcc$ at $95 \%$~CL
has been excluded by ALEPH~\cite{aleph}.
ALEPH, DELPHI and L3 have performed the 
search for stop and sbottom indirect decays in the
hypothesis of a dominant \UDD\ coupling.
Using the limit on the neutralino mass ($32 \GeVcc$) from the gaugino
searches, DELPHI set lower bounds on the squark masses with
m$_{\tilde{q}} -$~m$_{\tilde{\chi}^0_1} > $~$5 \GeVcc$~(Fig.~\ref{sec6.fig.squ.ind}).
The lower mass limit on the stop (sbottom) is 
$87 \GeVcc$ ($78 \GeVcc$) in case of no mixing, and 
$77 \GeVcc$ in case of maximal Z-decoupling.
The study of indirect decay of left-handed stop and sbottom
by ALEPH lead to exclude stop and sbottom lighter than $71 \GeVcc$ 
at 95\% CL~(Fig.~\ref{sec6.fig.squ.ind}).


In view of the limitations posed on centre-of-mass energies and luminosities 
by $e^+e^-$ collider technologies, the case of a high energy $\mu^+\mu^-$
collider using storage rings has been considered.
The phenomenology of supersymmetry with \Rp\ at a ${\mu}^+{\mu}^-$ collider 
resembles very much to the one at a $e^+e^-$ collider. 
$R$-parity violation can manifest itself via either a) pair-production of 
supersymmetric particles followed by \Rp\ decays or
b) resonant and non-resonant production of a single supersymmetric particle or finally
c) virtual effects in four fermions processes. The case (a) is discussed below.
The cases (b) and (c) will be discussed 
in sections~\ref{sec:sproduct} to~\ref{sec:virtuprod}.

The discussion of pair-production of supersymmetric particles followed by \Rp\ 
decays for ${\mu}^+{\mu}^-$ colliders is analogous in most of the aspects to 
the one for $e^+e^-$ colliders and can be found in~\ref{sec:sdecays}.
However, unlike $e^+e^-$ colliders, a particular feature of ${\mu}^+{\mu}^-$ 
colliders stems from the possible $s$-channel production of the $CP$-even 
($h^0$ and $H^0$) or $CP$-odd ($A^0$) Higgs bosons of e.g. the MSSM.
The Higgs boson produced would then decay into a pair of supersymmetric 
particles in processes like
$h^0 (H^0, A^0) \rightarrow {\tilde \chi}^+ {\tilde \chi}^-$,
                           ${\tilde \chi}^o {\tilde \chi}^o$,
                           ${\tilde u}^c {\tilde u}$,
			   ${\tilde d}^c {\tilde d}$,
			   ${\tilde l}^c {\tilde l}$, etc,
where ${\tilde u}$ and ${\tilde d}$ denote generically up-type down-type 
squarks respectively.
The supersymmetric particles themselves would then undergo \Rp\ decays 
according to Table~\ref{tab.sec6.rpv.sfermions.channels}
and~\ref{tab.sec6.rpv.gauginos.channels}.
The analysis of these specific pair production of supersymmetric particles
is governed by the analysis of the Higgs bosons
decay widths with respect to the parameters of the minimal supersymmetric extension
of the Standard Model into consideration \cite{mugun}.
The results of this analysis have then
to be merged with the more familiar analysis of pair production of
supersymmetric particles from ordinary
processes arising from ${\mu}^+{\mu}^-$ annihilation (either $\gamma$ and $Z$ in the
s-channel or sfermions or gauginos in the t-channel) as in the
case of $e^+e^-$ annihilation.
\par
Additionnal Higgs bosons decays may also come into consideration such as
$H^{\pm} \rightarrow W^{\pm} h^0 $, $A^0 \rightarrow Z h^0  $,
$H^0 \rightarrow h^0 h^0 $ and
$H^0 \rightarrow A^0 A^0 $ 
which may lead to the production of a pair of supersymmetric particles
in association with a gauge boson or to the production of four supersymmetric 
particles and thus to more complicated signature when \Rp\ decays are 
switched on.
We refer the reader to \cite{mugun} for the calculation of the cross-section
${\mu}^+{\mu}^- \rightarrow higgs $ as well as the Higgs bosons total width.

\noindent \addtocounter{sss}{1}
{\bf \thesss Sfermion and Gluino Pair Production at Hadron Colliders}
\addcontentsline{toc}{subsection}{\hspace*{1.2cm} \alph{sss}) 
   Sfermion and Gluino Pair Production at Hadron Colliders}
\index{Sparticle production!Sfermion pair!at hadron colliders}

Following the observation of an excess of high $Q^2$ events at HERA 
experiments~\cite{highq1,highq2},
the CDF collaboration examined two scenarii with $\lambda'_{121} \ne 0$
using $107 \picob^{-1}$ of data~\cite{cdf2}: 
\begin{equation}
p\overline{p} \rightarrow \tilde{g}\tilde{g} 
\rightarrow (c{\tilde{c}}_L)(c{\tilde{c}}_L)
\rightarrow c(e^\pm d) c(e^\pm d) 
\label{uno}
\end{equation}
and 
\begin{equation}
p\overline{p} \rightarrow \tilde{q}\tilde{\overline{q}} \rightarrow 
(q\ch0)(\overline{q}\ch0)
\rightarrow q(dce^\pm)\overline{q}(dce^\pm) ~.
\label{due}
\end{equation} 
For process (\ref{uno}) assumptions made were 
$m_{\tilde{q}} > m_{\tilde{g}} > m_{{\tilde{c}}_L} = 200 \GeVcc$, where 
degenerate mass for all up-type squarks
(except ${\tilde{c}}_L$) and all right-handed down-type squarks is denoted by 
$m_{\tilde{q}}$. The masses of all left-handed down-type squarks were obtained
by using the HERA motivated relations given in \cite{explhighq1}. 
For analysing process (\ref{due}) assumptions made were 
$m_{{\tilde{\chi}}_1^\pm} > m_{\tilde{q}} > m_{{\tilde{\chi}}_0}$ and 
$m_{{\tilde{\chi}}_1^\pm} \simeq 2 m_{{\tilde{\chi}}_0} $. Five 
degenerate squarks and the stop were treated separately. In the case 
of the process involving stop to ensure 100 \% branching ratio for the decay
${\tilde{t}}_1 \rightarrow c \ch0$ when $m_{{\tilde{t}}_1} < m_t$, an 
additional condition  that 
$m_{{\tilde{\chi}}_1^\pm} > m_{{\tilde{t}}_1} - m_b$ was imposed.

The Majorana nature of gluino and neutralino implies processes 
(\ref{uno}) and (\ref{due})
each yield like sign dielectron and opposite sign dielectron with equal 
probability. Since the like-sign dilepton signature has very little SM background,
for both processes (\ref{uno}) and (\ref{due}) CDF searched for events with 
like sign dielectrons and at least two jets. 

\begin{figure}
\begin{center}
\leavevmode
\centerline{\psfig{figure=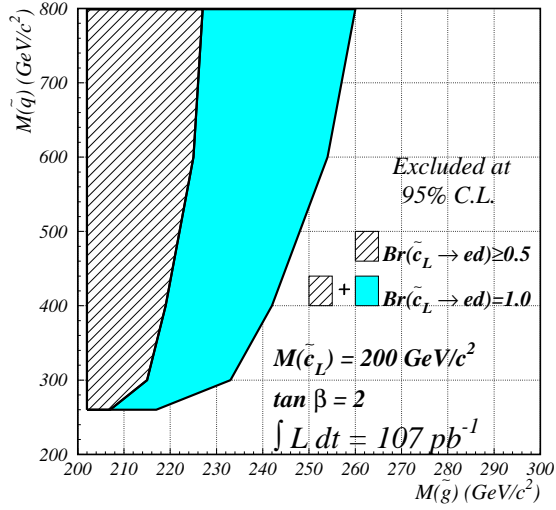,height=3in}}
\end{center}
\caption{{\it
Exclusion region in the plane $m_{\tilde{g}}$--$m_{\tilde{q}}$ for the 
charm squark analysis of the CDF collaboration. }}
\label{fig:cdf1}
\end{figure}

Analysis of $107 \picob^{-1}$ data yields no 
event with an expected contribution of $0.3\pm 0.3$ events from backgrounds.
Exclusion contours obtained for process (\ref{uno}) are shown in 
Fig.~\ref{fig:cdf1} in the $m_{\tilde{g}}$--$m_{\tilde{q}}$ plane for 
different assumptions for the $Br ({\tilde{c}}_L \rightarrow ed)$. 
The region below $m_{\tilde{q}} \leq 260 \GeVcc$
is not excluded because in this region ${\tilde{b}}_L$ becomes lighter than
${\tilde{c}}_L$, hence suppressing the decay $\tilde{g} \rightarrow 
c{\tilde{c}}_L$.  Figure~\ref{fig:cdf2} (bottom) shows 
the 95\% CL upper limit on the cross-section times branching ratios 
(obtained for process (\ref{due}) ) along with 
the NLO prediction~\cite{nlo} for the cross-section, as a 
function of squark masses. 95\% CL lower limits are given for two different
masses of the lightest neutralino. 
\begin{figure}
\begin{center}
\leavevmode
\centerline{\psfig{figure=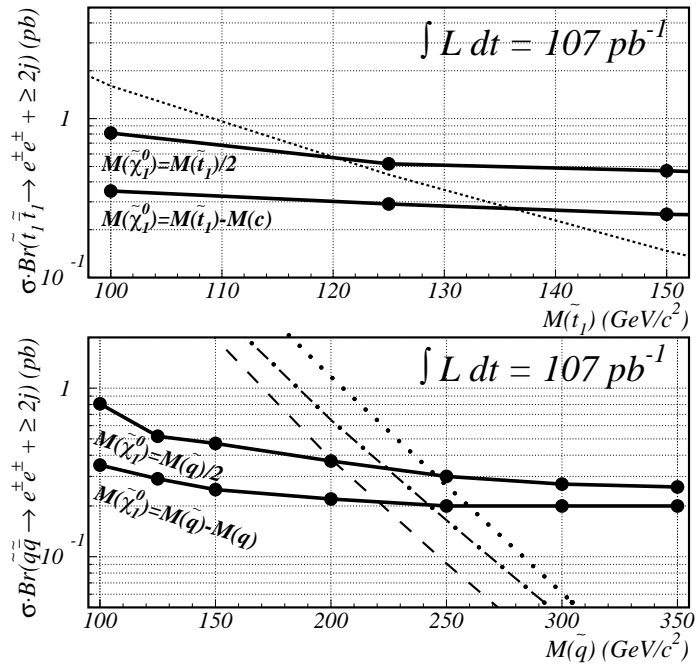,height=4in}}
\end{center}
\caption{{\it
Bottom) upper limits on the cross-section times branching ratio
for the production of 5 degenerate squark flavours decaying to 
electrons and jets via neutralinos (solid lines).
Also shown is the theoretical prediction for $\sigma \cdot Br$
for three values of the gluino mass: $200 \GeVcc$~(dotted line), 
$500 \GeVcc$~(dot-dashed line), and $1 \TeVcc$~(dashed line).
Top) upper limits on the cross-section times branching ratio
for stop pair production decaying to electrons and jets
via neutralinos (solid lines). The dashed curve is the theoretical 
prediction. }}
\label{fig:cdf2}
\end{figure}

A lower limit on the degenerate squark mass was found to be in the range of 
$200$--$260 \GeVcc$ depending on the mass of the lightest neutralino and 
gluino (range of gluino mass considered was $200$--$1000 \GeVcc$). 
Figure~\ref{fig:cdf2} (top) shows also a similar plot in the case of the stop. 
The mass of the stop was excluded below $135 (120) \GeVcc$ for a heavy 
(light)  neutralino. 
The analysis for process (\ref{due}) has been performed for the Tevatron 
Run II in~\cite{DREINER99}: it shows that squark masses up to 
$380 \GeVcc$ should be tested.
Finally, one point to note here is that, although the analysis for process 
(\ref{due}) assumed only one \Rp\ coupling $\lambda'_{121}$ to be non-zero, 
as the analysis does not depend on the quark flavours, the results are 
equally valid for any $\lambda'_{1jk}$ ($j$=1, 2 and $k$ = 1, 2, 3) couplings.

$D\emptyset$~\cite{NEWD0RPV} considered squark pair production 
leading in $\Rp$-SUGRA to like-sign dielectron events accompanied by jets, 
and has ruled out $M_{\tilde{q}} < 243 \GeVcc$ (95 \% CL) when assuming 
five degenerate squark flavours.
The $D\emptyset$ analysis covers all $\lambda'_{1jk}$ couplings.
From a similar analysis by CDF~\cite{CDFRPV} but restricted to 
$\lambda'_{121} \ne 0$, one can infer that a cross-section five times 
smaller would lead to a $M_{\tilde{q}}$ limit of $\simeq 150 \GeVcc$ 
depending on the gluino and $\tilde{\chi}_0$ masses. 

CDF also considered separately~\cite{CDFRPV,cdf2} the pair production 
of a light stop $\tilde{t}_1$ assuming a decay into $c$\XOIb and 
excluded $M_{\tilde{t}} < 135 \GeVcc$. 
To translate this constraint in one relevant for $\lambda'_{13k} \ne 0$, 
it should be noted that in this latter case, $\Rp$-decays of the 
$\tilde{t}$ would dominate over loop decays into $c$\XOI. 
Moreover, $\Rp$-decays would themselves be negligible compared to 
$\tilde{t} \rightarrow b \tilde{\chi}^+_1$ decays as soon as this becomes allowed, 
i.e. if $M(\tilde{t}_1) > M(\tilde{\chi}^+_1)$ and if the $\tilde{t}_1$ eigenstate 
possesses a sizeable admixture of $\tilde{t}_L$. 
The subsequent decays of the $\tilde{\chi}^+_1$  would then lead to final states 
similar to those studied by CDF for $\tilde{t}_1 \rightarrow c \tilde{\chi}^0_1$.
Thus, $130 - 150 \GeVcc$ appears to be reasonable rough estimate of the
Tevatron sensitivity to a light $\tilde{t}$ for $\lambda'_{13k} \ne 0$. 
In summary, Tevatron and HERA sensitivities are competitive in $\Rp$-SUSY
models with five degenerate squarks, but models predicting a light
$\tilde{t}$ are better constrained at HERA provided that $\lambda'_{13j}$ 
is not too small. 

In the above mentioned search by the $D\emptyset$ experiment~\cite{NEWD0RPV2} 
in the dimuon plus four-jets channel (occuring after \XOI decay via the 
$\lambda^{\prime}$ coupling, see section~\ref{sec:pairgg}), it was seen that
squark masses below $240 \GeVcc$ (for all gluino masses and for
$tan \beta = 2$) are excluded. For equal masses of squarks and gluinos 
the mass limit is $265 \GeVcc$

\begin{figure}
\begin{center}
\leavevmode
\centerline{\psfig{figure=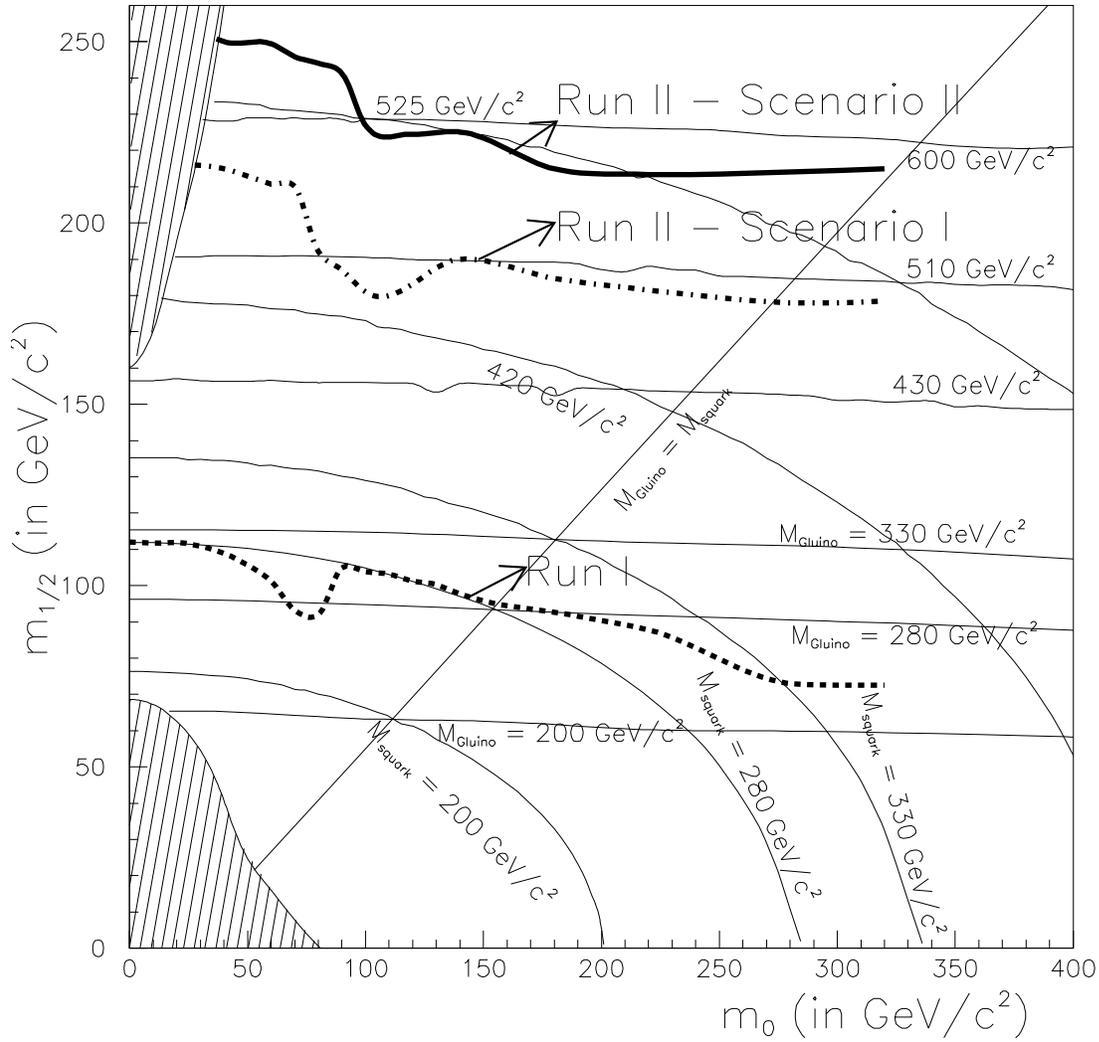,width=16cm}}
\end{center}
\caption{{\it Estimated exclusion contour for Tevatron Run I and II, 
within the mSUGRA framework, in the ($m_0$, $m_{1/2}$) plane
for $\tan \beta$ =2, $A_0=0$ and $\mu <0$, from the dielectron and four jets 
channel. Scenario I corresponds to a background of $36\pm 4 \pm 6$ events 
(direct scaling from Run I) while scenario II uses the background of 
$15\pm 1.5 \pm 1.5$ events (scaling, but with improvements in the detector
taken into account).}}
\label{fig:rpvrun2el}
\end{figure}
In contrast with the above studies of the $\lambda'$ coupling
constants which were based on the analysis of a given 
superpartner pair production, the $D\emptyset$ Collaboration has
performed a study of the $\lambda'_{1jk}$ and $\lambda'_{2jk}$ 
($j$ =1, 2 and $k$ = 1, 2, 3) couplings 
based on the Monte Carlo simulation of all the superpartner 
pair productions \cite{DREINER99,runii2}. In this work, it was assumed 
that the LSP is the lightest neutralino. The exclusion plot
obtained in \cite{DREINER99,runii2} 
in the case of a single dominant \Rp\ coupling of type 
$\lambda'_{1jk}$ is presented in Fig.\ref{fig:rpvrun2el}. 
In this case, the studied final state is composed of 2 $e^{\pm}$ 
and at least 4 jets. We observe in Fig.\ref{fig:rpvrun2el} 
that the $D\emptyset$ Collaboration is expected to search 
for squarks of mass up 
to $575 \GeVcc$ and gluinos of mass up to $520 \GeVcc$. In the case
of a single dominant \Rp\ coupling of type $\lambda'_{2jk}$, it was
shown in \cite{DREINER99,runii2} that the analysis
of the dimuon plus four jets signature leads to the expectation that squarks 
of mass up to  $640 \GeVcc$ and gluinos of mass up to $560 \GeVcc$ will 
be tested by the $D\emptyset$ Collaboration during the Tevatron Run II.
\index{Sparticle production!Sfermion pair|)}

\subsection{Effects of Bilinear {\boldmath{\Rp}} Interactions}
\label{sec:bilineff}

Search for spontaneous \Rp\ violation  has been 
performed at LEP~2 \index{LEP} by the DELPHI experiment~\cite{delphi_spont}. 

This search
is based on the model described in Refs.~\cite{masiero90,romao92},
in which \Rp\ breaking is  parametrized 
by effective bilinear terms $\mu_i L_i H_u$
(see section~\ref{sec:spontaneous} for a theoretical review of this model).
The most important phenomenological implication of the model 
is the existence of a massless Majoron\index{Majoron} $J$ which is the LSP. 
Therefore the Majoron\index{Majoron} enters in the  chargino and neutralino  decays
and the branching ratios of these new processes depend on an effective
bilinear term parameter denoted by  $\epsilon$ 
in Refs.~\cite{masiero90,romao92,delphi_spont}. 
In the case where the $\epsilon$ parameter is sufficiently high, roughly for 
$\epsilon > 10\ $GeV 
the chargino decay is fully dominated by the Majoron\index{Majoron} chanel:  
$\widetilde \chi^{\pm} \rightarrow \tau^{\pm} J$.
Therefore, the experimental signature of the chargino pair production
in this scenario is two acoplanar taus and missing momentum 
from the undetected Majorons\index{Majoron}.
One should note, however, that such values of the $\epsilon$ parameter are
incompatible with the cosmological bound on stable neutrino masses, and can 
arise only  in the context of exotic scenarios with a heavy decaying 
neutrino ($m_{\nu} \sim 1 \MeV$).

The search for spontaneous \Rp\ violation in DELPHI was performed 
in the MSSM framework~\cite{delphi_spont}.
An upper limit at $95 \%$ CL on the chargino production cross-section of 
0.3 pb and on the chargino mass of $94 \GeVcc$ (close to the kinematic limit 
for 1998 data) has been obtained.


Before closing the discussion on the possible effects of bilinear
\Rp\ interactions at lepton colliders, it is interesting to come back to 
the case of resonant higgs production at ${\mu}^+{\mu}^-$ colliders discussed
above in the presence of trilinear \Rp\ interactions.
The effects of bilinear terms from spontaneously broken $R$-parity would lead
to invisible signature when $ h^0 \rightarrow JJ$ where $J$ stands for the 
Majoron\index{Majoron}.
Furthermore, possible signatures with missing energies, when for example
${\mu}^+{\mu}^- \rightarrow H^0  \rightarrow h^0 h^0$, followed by
$h^0 \rightarrow JJ $ for one of the $h^0$
and by $h^0 \rightarrow  {\tilde \chi}^+_1 {\tilde \chi}^-_1$
for the  other $h^0$,
with the subsequent decay ${\tilde \chi}^+_1 \rightarrow  {\tau}^+ J$,
deserve further studies.


Effects of bilinear \Rp\ interactions could be observed already at existing
hadron colliders.
Current data from Tevatron Run I are already sensitive to these decays and 
effectively limits the total branching ratio of top decays in different channels 
than $t \rightarrow W^{+} b$ to approximately 25\%. 
The Tevatron Run II data will enhance the sensitivity for alternative 
top decays to branching ratios of $10^{-3}-10^{-2}$ depending on the decay 
mode. If the stop is lighter than the lightest chargino it may decay 
dominantly through $\tilde{t}_{1} \rightarrow \tau^{+} + b$. 
By interpreting the stop as a third generation leptoquark, the exclusion 
obtained from leptoquark searches can in this case be 
applied \cite{leptoq, leptoq2}. 
This leads to an exclusion of scalar-stop masses below $80-100 \GeVcc$ from the 
Run I data. Here too, the Run II data will improve the sensitivity to a wider 
region of the SUSY parameter space.

\section{Single Sparticle Production}
\label{sec:sproduct}
\index{Sparticle production!Single|(}

New $\Rp$ trilinear interactions enter directly in sparticle production
and decays via basic tree diagrams as was illustrated in 
Fig.~\ref{fig:triplel}.
The corresponding complete interaction Lagrangian is discussed in detail 
in section~\ref{sec:rpvgeneral} and appendix~\ref{chap:appendixB}.
The most striking feature of \Rp\ is to allow for single production
of supersymmetric particles. For given centre-of-mass collider 
energies, this extends the discovery mass reach for supersymmetric 
matter beyond that of superpartner pair production in $R$-parity conserving 
models.

A list of the $s$-channel processes allowed at lowest order in 
$e^+e^-$, $ep$ and $p\bar{p}$ collisions is given in Table~\ref{tab:rpvres}.
%
\begin{table*}[htb]
  \renewcommand{\doublerulesep}{0.4pt}
  \renewcommand{\arraystretch}{1.2}
  \begin{center}
  \begin{tabular}{||c||c|c|lr||}
  \hline \hline
  \multicolumn{5}{||c||}{Resonant Production of Sfermions at Colliders} \\
  \multicolumn{5}{||c||}{(lowest-order processes)} \\
   \hline
   Collider  &  Coupling     &  Sfermion           & \multicolumn{2}{|c||}{Elementary Process} \\
             &               &  Type               & \multicolumn{2}{|c||}{} \\
   \hline
    $e^+e^-$ &$\lambda_{1j1}$
                             & $\tilde{\nu}_{\mu}, \tilde{\nu}_{\tau}$ & 
               $l^+_i l^-_k \rightarrow \tilde{\nu}_j$ & $i=k=1 \, , \, j=2,3$ \\
   \hline
  $pp$, $p \bar{p}$ &$\lambda'_{ijk}$
                    & $\tilde{\nu}_{e}, \tilde{\nu}_{\mu}, \tilde{\nu}_{\tau}$
		    & $d_k \bar{d}_j \rightarrow \ \tilde{\nu}_i$ & $i,j,k = 1,\ldots,3$ \\
              &     & $\tilde{e}, \tilde{\mu}, \tilde{\tau}$
	 & $u_j \bar{d}_k \rightarrow \ \tilde{l}_{iL}$ & $i,k = 1,\ldots,3 \, , \, j = 1,\ldots,2$ \\
              & $\lambda''_{ijk}$
                    & $\tilde{d}, \tilde{s}, \tilde{b}$
	 & $\bar{u}_i \bar{d}_j \rightarrow \ \tilde{d}_{k}$ & $i,j,k = 1,\ldots,3 \, , \, j \neq k$\\
              &     & $\tilde{u}, \tilde{c}, \tilde{t}$
	 & $\bar{d}_j \bar{d}_k \rightarrow \ \tilde{u}_{i}$ & $i,j,k = 1,\ldots,3 \, , \, j \neq k$\\
   \hline
  $e p$ & $\lambda'_{1jk}$
                    & $\tilde{d}_R, \tilde{s}_R, \tilde{b}_R$
	     & $l^-_1 u_j \rightarrow \tilde{d}_{kR}$ & $j=1,2$ \\
        & $\lambda'_{1jk}$
                    & $\tilde{u}_L, \tilde{c}_L, \tilde{t}_L$
		    & $l^+_1 d_k \rightarrow \tilde{u}_{jL}$ & $i,j,k = 1,\ldots,3$\\
   \hline \hline
  \end{tabular}
  \caption {{\it
         Sfermions $s$-channel resonant production at colliders. Charge conjugate 
	 processes (not listed here) are also possible. Real $\tilde{\nu}$ production 
	 at an $e^+e^-$ collider can only proceed via $\lambda_{121}$ or $\lambda_{131}$.
	 In $e \gamma$ collisions all $\lambda_{ijk}$ couplings where $i$, $j$ or $k$ is equal
         to one can be probed. The $e \gamma$ collision mode opens new possibilities, such as
	 the single production of the $\tilde{\nu}_e$ via $\lambda_{i11}$ or the single slepton production.
	 Single squark production is also possible via a $\lambda'_{1jk}$ coupling.
         Real $\tilde{q}$ production
	 at an $ep$ collider is possible via any of the nine $\lambda'_{1jk}$ couplings.}}
  \label{tab:rpvres}	 
\end{center}
\end{table*}
%
The $L$-violating (\LV) term \LLE\ couples sleptons 
and leptons~(Fig.~\ref{fig:triplel}a). It allows for resonant production of
$\tilde{\nu}$ at $l^+l^-$ colliders and for the direct $\Rp$-decay of
sleptons $\tilde{l}^{\pm} \rightarrow {l'}^{\pm} \nu$ and
$\tilde{\nu} \rightarrow l^+ {l'}^-$.
The \LV\ term \LQD\ couples squarks to lepton-quark pairs 
(Fig.~\ref{fig:triplel}b) and sleptons to quark pairs. 
It allows for resonant production of $\tilde{q}$ at $ep$ colliders and 
$\tilde{\nu}$ or $\tilde{l}^{\pm}$ at $pp$ colliders.
Direct $\Rp$-decay of squarks via $\tilde{q} \rightarrow l q'; \nu q''$ or
sleptons via $\tilde{l}^{\pm}$ or $\tilde{\nu} \rightarrow q q'$ are made
possible. The $B$-violating (\BV) term \UDD\ couples squarks to 
quark pairs (Fig.~\ref{fig:triplel}c).
It allows for resonant production of $\tilde{q}$ at $pp$ colliders and
direct $\Rp$-decay of squarks via $\tilde{q} \rightarrow q' q''$.

Moreover, as seen in sections~\ref{sec:dirsfdecays} and~\ref{sec:dirghdecays}, 
there is a gap between the constraints obtained via the detection of the 
displaced vertex and the low-energy experimental constraints on the 
\Rp\ couplings.
This domain can be tested through the study of the single
production of supersymmetric particles.
Indeed, the cross-sections of such reactions are directly proportional
to a power of the relevant \Rp\ coupling constant(s), which allows
to determine the values of the \Rp\ couplings.
Therefore, there exists a complementarity between the displaced vertex 
analysis and the study of singly produced sparticles, since these
two methods allow to investigate different ranges of values 
of the \Rp\ coupling constants.

For values beyond ${\cal{O}}(10^{-4})$, single sparticle production 
will allow in favorable cases to determine or constrain specific 
$\lambda$, $\lambda'$ or $\lambda''$ couplings as will be discussed 
in more details in sections~\ref{sec:singlell} and~\ref{sec:singlelp}.
Otherwise, the presence of such a large \Rp\ coupling will become 
trivially manifest through the decay of pair produced sparticles.

\subsection{Single Sparticle Production at Leptonic Colliders} 
\label{sec:singlell}
\index{Sparticle production!Single!at leptonic colliders}

At leptonic colliders resonant (as well as non-resonant) production of
single supersymmetric particles involve the $\lambda_{ijk}$ couplings.

Early discussions on several $R_p$-violating processes at
$e^+e^-$ colliders including single supersymmetric particles production can be found 
in~\cite{Workshop1,Roszkowski93}. 

At a $e^+ e^-$ leptonic colliders, the sneutrinos $\tilde \nu_{\mu}$ and 
$\tilde \nu_{\tau}$ can be produced at resonance through 
the couplings $\l_{211}$ and $\l_{311}$, respectively. 
The sneutrino may then decay either via an \Rp\ 
interaction~\cite{dimopoulos88}, for example through $\l_{ijk}$ as, 
$\tilde \nu^i \to \bar l_j l_k$, or via gauge interaction as, 
$\tilde \nu^i_L \to \tilde \chi^+_a l^i$, or, 
$\tilde \nu^i_L \to \tilde \chi^0_a \nu^i_L$. 
The sneutrino partial width is given in Eq.~(\ref{larg1})
for the leptonic decay channel and is given in the following equation 
for the gauge decay channel~\cite{barger89}:

\begin{eqnarray}
  \Gamma(\tilde \nu^i_L \to \tilde \chi^+_a l^i, \ \tilde \chi^0_a \nu^i_L) = 
         {C g^2 \over 16 \pi} m_{\tilde \nu^i_L} 
         (1-{m_{\tilde \chi^+_a}^2 \over m_{\tilde \nu^i_L}^2})^2,
 \label{widthformula}
\end{eqnarray}  
where $C=\vert V_{a1}\vert ^2$ for the decay into chargino and $C=\vert
N_{a2}\vert ^2$, for the neutralino case, with $V_{a1}$ and $N_{a2}$ 
the mixing matrix elements written in the notations of \cite{haber85}. 
For reasonable values of $\l_{ijk}$ ($\leq 0.1$) and most of the region 
of the \susyq\ parameter space, the decay modes of Eq.~(\ref{widthformula}) 
are dominant, if kinematically accessible~\cite{lola,barger89}. 
In the SUGRA parameter space, if $m_{\tilde \nu^i_L} > 80 \GeVcc$, with 
$M_2 = 80 \GeV$, $\mu = 150 \GeV$ and $\tan \beta = 2$, the total 
sneutrino width is higher than $100 \MeV$ which is comparable to
or greater than the typical expected experimental resolutions. 
The cross-section formula, for the sneutrino production in the $s$-channel, 
is the following~\cite{barger89},
\begin{eqnarray}
  \s(e^+ e^- \to \tilde \nu^i_L \to X)= { 4 \pi s \over m_{\tilde
  \nu^i_L}^2} 
  {\Gamma(\tilde \nu^i_L \to e^+ e^-)  \Gamma(\tilde \nu^i_L \to X) \over
  (s-m_{\tilde \nu^i_L}^2)^2  + m_{\tilde \nu^i_L}^2 \Gamma_{\tilde
  \nu^i_L}^2},
 \label{Xsectf1}
\end{eqnarray}
where $\Gamma(\tilde \nu^i_L \to X)$ generally denotes the partial width for the sneutrino
decay into the final state $X$. At sneutrino resonance, Eq.~(\ref{Xsectf1}) 
takes the form,
\begin{eqnarray}
  \s(e^+ e^- \to \tilde \nu^i_L \to X)= { 4 \pi \over m_{\tilde
  \nu^i_L}^2} 
  B(\tilde \nu^i_L \to e^+ e^-)  B(\tilde \nu^i_L \to X),
 \label{Xsectf2}
\end{eqnarray}
where $B(\tilde \nu^i_L \to X)$ generally denotes the branching
ratio for sneutrino decay into a final state $X$. 

Diagrams for the single sparticle production at leptonic colliders
are shown in Fig.~\ref{fig:singlee}.
\begin{figure}
\begin{center}
\hspace*{-0.5cm} \epsfig{file=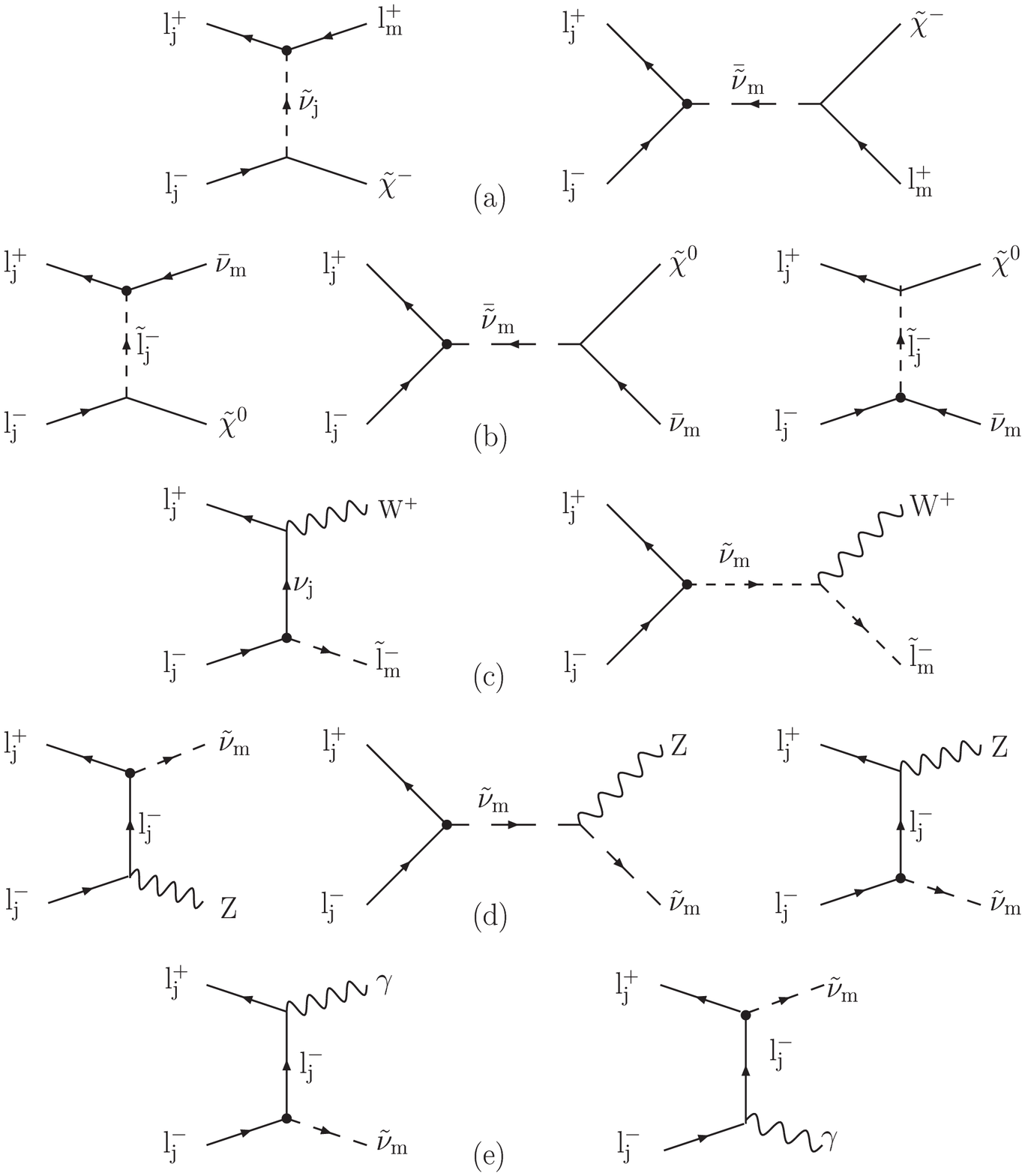,width=1.00\textwidth}
\end{center}
\caption{{\it Diagrams for the single production processes at 
              leptonic colliders, namely,
              $l_j^+l_j^- \to \tchi^- l_m^+$ (a), 
              $l_j^+l_j^- \to \tchi^0 \bar \nu_m$ (b), 
              $l_j^+l_j^- \to \tilde l^-_{mL} W^+$ (c),
              $l_j^+l_j^- \to \tilde \nu_{mL} Z $ (d) and 
              $l_j^+l_j^- \to \tilde \nu_{mL} \ \g $ (e). 
              The circled vertex corresponds to the \Rp\ interaction, with 
              the coupling constant $\l_{mjj}$, and the
	      arrows indicate the flow of the lepton number.}}
\label{fig:singlee}
\end{figure}

The case of resonant production of single supersymmetric particles at 
${\mu}^+{\mu}^-$ colliders resembles very much the one
of $e^+e^-$ colliders described above.
The relevant decay widths and cross-section formulae are obtained 
respectively from Eq.~\ref{widthformula},~\ref{Xsectf1} and~\ref{Xsectf2} 
by replacing $e^+$ and $e^-$ by ${\mu}^+$ and ${\mu}^-$.

The gauge decays of a resonantly produced sneutrino at leptonic colliders lead to
single chargino or neutralino production, or to the production of a lighter slepton
in association
with an electroweak gauge boson when this is kinematically allowed.
Away from the sneutrino resonance, other diagrams contribute to the
single production of sparticles.
The $t$-channel exchange of a slepton can lead to single chargino or
neutralino production, and the $t$- or $u$-channel exchange of a lepton
allows for single slepton production.

Single chargino and neutralino productions both receive contributions from the 
resonant sneutrino production at $e^+ e^-$ colliders 
(see Fig.~\ref{fig:singlee}a and b).
The single production of a chargino, $e^+ e^- \to \tilde \chi^{\pm}_a
l^{\mp}_j$ (via $\l_{1j1}$), receives a contribution from the $s$-channel 
exchange of a $\tilde \nu_{jL}$ sneutrino and another one from the exchange 
of a $\tilde \nu_{eL}$ sneutrino in the $t$-channel 
(see Fig.~\ref{fig:singlee}a). 
The single neutralino production, $e^+ e^- \to \tilde \chi^0_a \nu_j$ 
(via $\l_{1j1}$), occurs through the $s$-channel $\tilde \nu_{jL}$ sneutrino 
exchange and also via the exchange of a $\tilde e_L$ slepton in the 
$t$-channel or a $\tilde e_R $ slepton in the $u$-channel 
(see Fig.~\ref{fig:singlee}b).  
The single $\tilde \chi^{\pm}_1$ ($\tilde \chi^0_1$) production rate
is reduced in the higgsino dominated region $\vert \mu \vert \ll M_1,M_2$
where the $\tilde \chi^{\pm}_1$ ($\tilde \chi^0_1$) is dominated by its
higgsino component, compared to the wino dominated domain
$\vert \mu \vert \gg M_1,M_2$ in which the $\tilde \chi^{\pm}_1$
($\tilde \chi^0_1$) is mainly composed by the higgsino \cite{Chemsin}.
In addition, the single $\tilde \chi^{\pm}_1$ ($\tilde \chi^0_1$) production
cross-section depends weakly on the sign of the $\mu$ parameter at large
values of $\tan \beta$.
However, as $\tan \beta$ decreases the rates increase (decrease)
for $sign(\mu)>0$~($<0$).
This evolution of the rates with the $\tan \beta$ and $sign(\mu)$ parameters
is explained by the evolution of the $\tilde \chi^{\pm}_1$ and
$\tilde \chi^0_1$ masses in the supersymmetric parameter space~\cite{Chemsin}.

For $\l_{1j1}=0.05$, $50 \GeV < m_0 < 150 \GeV$ and $50 \GeV < M_2 < 200 \GeV$ 
in a SUGRA parameter space, the off-pole values of the cross-sections are 
typically~\cite{Chemsin} of the order of $100 \femtob$ ($10 \femtob$) for the single 
chargino production and of $10 \femtob$ ($1\femtob$) for the single neutralino 
production at~$\sqrt s=200 \GeV$ ($500 \GeV$) 
(see Fig.~\ref{fig:singlechi}).

\begin{figure}[htb]
  \begin{center}                                                                
  \vspace*{-0.2cm}

  \hspace*{-0.5cm} \epsfig{file=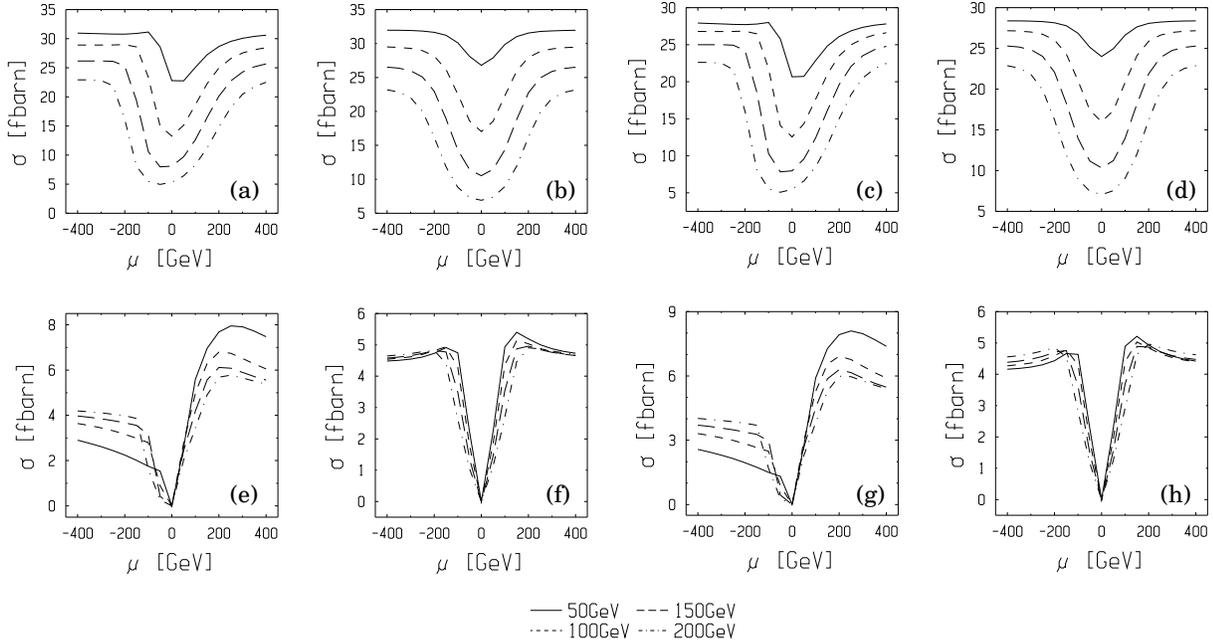,width=1.00\textwidth}

  \end{center}
  \vspace*{-0.5cm}

 \caption{{\it The integrated cross-sections \cite{Chemsin} for the processes 
  (a,b,c and d)
  $e^{+} e^{-} \rightarrow {\tilde {\chi}}^{-}_{1} l^{+}_{j}$
  and
  (e,f,g and h)
  $e^{+} e^{-} \rightarrow {\tilde {\chi}}^{0}_{1} {\bar {\nu}}_{j}$,   
  at a centre-of-mass energy of $500 \GeV$, are shown as
  a function of $\mu$ for discrete choices of the remaining parameters:
  (a,e) $\tan \beta = 2$, $m_0 = 50 \GeVcc$, 
  (b,f) $\tan \beta = 50$, $m_0 = 50 \GeVcc$,
  (c,g) $\tan \beta = 2$, $m_0 = 150 \GeVcc$,
  and (d,h) $\tan \beta = 50$, $m_0 = 50 \GeVcc$, with $\lambda_{1j1} = $ 0.05.
  The windows conventions are such that $\tan \beta = 2,50$ 
  horizontally and  $m_0 = 50,150 \GeVcc$ vertically. The different curves
  refer to the value of $M_2$ of $50 \GeVcc$ (continuous line),
  $100 \GeVcc$ (dot-dashed line), $150 \GeVcc$ (dotted line), as indicated at the bottom
  of the figure.}}
 \label{fig:singlechi}
\end{figure}
\begin{figure}[htb]
\begin{center}
\epsfig{file=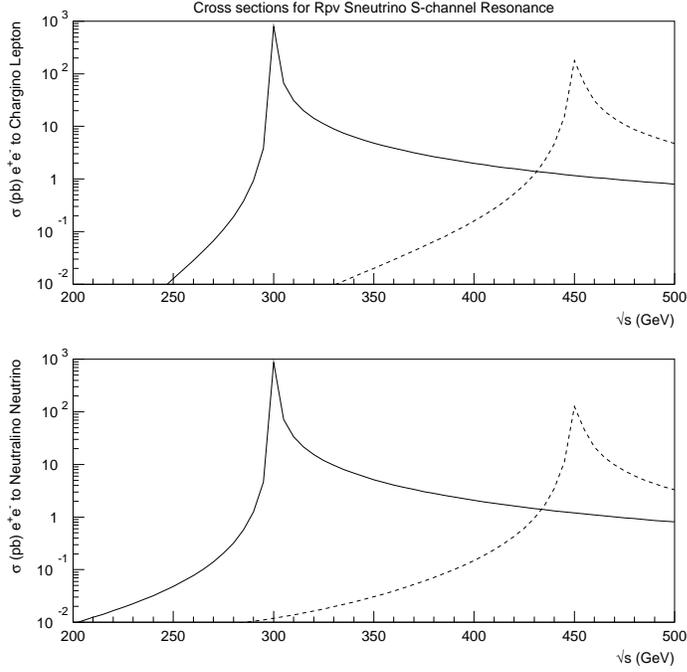,width=10cm}  
\vspace{-0.4cm}

\caption{{\it Cross sections of the single charginos and neutralinos 
  productions as a function of the centre-of-mass energy for 
  $m_{\tilde \nu} = 300 \GeVcc$ (full curves) and 
  $m_{\tilde \nu} = 450 \GeVcc$ (dashed curves) with
  $m_{\tilde e} = 1 \TeVcc$, $M_2 = 250 \GeV$, $\mu = -200 \GeV$, 
  $\tan \beta=2$ and $\l_{1j1}=0.1$.
  The rates values are calculated by including the ISR effect and by summing 
  over the productions of the different $\tilde \chi^{\pm}_i$ and 
  $\tilde \chi^0_j$ eigenstates which can all be produced for this MSSM point.}} 
 \label{fig:ISRe}
\end{center}
\end{figure}
At the sneutrino resonance, the cross-sections of the single gaugino 
production are important: using Eq.~(\ref{Xsectf2}), the rate for the 
neutralino production in association with a neutrino is of the order of $3 \ 10^3$ 
in units of the QED point cross-section, $R=\s_{pt}=4 \pi \alpha^2 / 3 s$, 
for $M_2 = 200 \GeV$, $\mu = 80 \GeV$, $\tan \beta=2$ and $\l_{1j1}=0.1$ at 
$\sqrt s=m_{\tilde \nu^j_L} = 120 \GeV$ \cite{barger89}.
The cross-section for the single chargino production reaches 
$2 \ 10^{-1} \picob$ at $\sqrt s=m_{\tilde \nu^j_L}=500 \GeV$,
for $\l_{1j1}=0.01$ and $m_{\tilde \chi^{\pm}} = 490
\GeVcc$~\cite{Workshop1,Workshop2}.
The Initial State Radiation (ISR) lowers the single gaugino production 
cross-section at the $\tilde \nu$ pole but increases greatly the single 
gaugino production rate in the domain $m_{gaugino}<m_{\tilde \nu}<\sqrt s$. 
This ISR effect can be observed in Fig.~\ref{fig:ISRe} which shows the 
single charginos and neutralinos productions cross-sections as a function 
of the centre-of-mass energy for a given MSSM point~\cite{lola}.

The experimental searches of the single chargino and neutralino productions 
have been performed at the LEP \index{LEP} collider at various centre-of-mass
energies~\cite{Arnoud,Fab1,alepha}.
The off-pole effects of the single gaugino productions rates are 
at the limit of observability at the LEP \index{LEP} collider even with the 
integrated luminosity of LEP~2.
Therefore, the experimental analyses of the single gaugino 
productions have excluded values of the $\l_{1j1}$ couplings smaller 
than the low-energy bounds only at the sneutrino resonance point 
$\sqrt s = m_{\tilde \nu}$ and, due to the ISR effect, in a range of 
typically $\Delta m_{\tilde \nu} \sim 50 \GeVcc$ around the 
$\tilde \nu$ pole.
Finally, for the various sneutrino resonances, the sensitivities on 
the $\l_{1j1}$ couplings which have been derived from the LEP \index{LEP} data 
reach values of order $10^{-3}$.
The experimental analyses of the single chargino and neutralino productions 
will be continued at the future linear collider. Using its polarisation capability
as well as the specific kinematics of the single chargino production 
allows to reduce the expected background from pair productions 
of supersymmetric particles.
As an example, this background reduction allows to improve the sensitivity 
to the $\l_{121}$ coupling obtained 
from the $\tilde \chi^{\pm}_1 \mu^{\mp}$ production study at linear 
colliders~\cite{TESLAworks} for $\sqrt s=500 \GeV$ and ${\cal L}=500$~fb$^{-1}$
which then amounts to values of the order of $10^{-4}$ at the sneutrino resonance
and can improve the 
low-energy constraint over a range of 
$\Delta m_{\tilde \nu} \approx \sim 500 \GeVcc$ around the $\tilde \nu$ pole~\cite{Tesgm}.
Due to the high luminosities reached at linear colliders, the off-resonance 
contributions to the cross-section play an important role in the single 
$\tilde \chi^{\pm}_1$ production analysis.

The slepton and the sneutrino can also be singly produced via the coupling 
$\l_{1j1}$ in the (non-resonant) reactions 
$e^+ e^- \to \tilde l^{\mp}_{jL} W^{\pm}$,
$e^+ e^- \to \tilde \nu^j_L Z^0$ and $e^+ e^- \to \tilde \nu^j_L \gamma$.
Those reactions receive contributions from the exchange of a charged or neutral
lepton of the first generation in the $t$- or $u$-channel 
(see Fig.~\ref{fig:singlee}).
The single productions of a sneutrino accompanied by a $Z$ or a $W$
boson also occur through the exchange in the $s$-channel of a $\tilde \nu^j_L$ 
sneutrino which can not be produced on-shell.
When kinematically allowed, these processes have some rates of order 
$100 \femtob$ at $\sqrt s = 200 \GeV$ and $10 \femtob$ at $\sqrt s = 500 \GeV$, 
for $\l_{1j1}=0.05$ and various masses of the scalar \susyq\ 
particles~\cite{Chemsin}.

The production of single gaugino and the non-resonant production of single 
slepton (either charged or neutral) similar to those in Fig~\ref{fig:ISRe}
are relevant at ${\mu}^+{\mu}^-$ colliders.
The only difference stems from the initial states particles which, being
muons instead of electrons, allows to test different $\lambda$ couplings as 
given in Table~\ref{tab:mucolsingledecay}.\\
\begin{table}[htb]
\begin{center}
\begin{tabular}{||c|c|c|c||c|c|c|c||} \hline
 \multicolumn{4}{||c||}{$e^+e^-$ colliders}
 & \multicolumn{4}{c||}{${\mu}^+{\mu}^-$ colliders}         \\ \hline \hline
coupling & final state & exchange & channel & coupling & final state & exchange & channel \\ \hline
 ${\lambda}_{121}$ & ${\tilde \chi}^{\pm}_a$ $\mu^\mp$  & ${\tilde {\nu}}_{e}$     & t   &
 ${\lambda}_{212}$ & ${\tilde \chi}^{\pm}_a$ $e^\mp$    & ${\tilde {\nu}}_{\mu}$   & t   \\
                   & ${\tilde \chi}^{\pm}_a$ $\mu^\mp$  & ${\tilde {\nu}}_{\mu}$   & s   &
                   & ${\tilde \chi}^{\pm}_a$ $e^\mp$    & ${\tilde {\nu}}_{\e}$    & s   \\
                   & ${\tilde \chi}^{0}_b$ $\nu_{\mu}$  & ${\tilde {e}}$           & t+u   &
                   & ${\tilde \chi}^{0}_b$ $\nu_e$      & ${\tilde {\mu}}$         & t+u   \\
                   & ${\tilde \chi}^{0}_b$ $\nu_{\mu}$  & ${\tilde {e}}$           & s     &
                   & ${\tilde \chi}^{0}_b$ $\nu_e$      & ${\tilde {\mu}}$         & s     \\ 
                   & ${\tilde \mu}^{\pm} W^{\mp}$       & ${\nu_{e}}$              & t     &
                   & ${\tilde e}^{\pm} W^{\mp}$         & ${\nu_{\mu}}$            & t     \\
                   & ${\tilde \mu}^{\pm} W^{\mp}$       & ${\tilde {\nu}}_{\mu}$   & s     &
                   & ${\tilde e}^{\pm} W^{\mp}$         & ${\tilde {\nu}}_{e}$     & s     \\
                   & ${\tilde {\nu}}_{\mu} Z$           & $e$                      & t+u   &
                   & ${\tilde {\nu}}_{e} Z$             & $\mu$                    & t+u   \\
                   & ${\tilde {\nu}}_{\mu} Z$           & ${\tilde {\nu}}_{\mu}$   & s     &
                   & ${\tilde {\nu}}_{e} Z$             & ${\tilde {\nu}}_{e}$     & s     \\
                   & ${\tilde {\nu}}_{\mu} \gamma$      & $e$                      & t+u   &
                   & ${\tilde {\nu}}_{e} \gamma$        & $\mu$                    & t+u   \\  \hline
 ${\lambda}_{131}$ & ${\tilde \chi}^{\pm}_a$ $\tau^\mp$ & ${\tilde {\nu}}_{e}$     & t     &
 ${\lambda}_{232}$ & ${\tilde \chi}^{\pm}_a$ $\tau^\mp$ & ${\tilde {\nu}}_{\mu}$   & t     \\
                   & ${\tilde \chi}^{\pm}_a$ $\tau^\mp$ & ${\tilde {\nu}}_{\tau}$  & s     &
                   & ${\tilde \chi}^{\pm}_a$ $\tau^\mp$ & ${\tilde {\nu}}_{\tau}$  & s     \\
                   & ${\tilde \chi}^{0}_b$ $\nu_{\tau}$ & ${\tilde {e}}$           & t+u   &
                   & ${\tilde \chi}^{0}_b$ $\nu_{\tau}$ & ${\tilde {\mu}}$         & t+u   \\
                   & ${\tilde \chi}^{0}_b$ $\nu_{\tau}$ & ${\tilde {\tau}}$        & s     &
                   & ${\tilde \chi}^{0}_b$ $\nu_{\tau}$ & ${\tilde {\tau}}$        & s     \\
                   & ${\tilde \tau}^{\pm} W^{\mp}$      & ${\nu_{e}}$              & t     &
                   & ${\tilde \tau}^{\pm} W^{\mp}$      & ${\nu_{\mu}}$            & t     \\
                   & ${\tilde \tau}^{\pm} W^{\mp}$      & ${\tilde {\nu}}_{\tau}$  & s     &
                   & ${\tilde \tau}^{\pm} W^{\mp}$      & ${\tilde {\nu}}_{\tau}$  & s     \\
                   & ${\tilde {\nu}}_{\tau} Z$          & $e$                      & t+u   &
                   & ${\tilde {\nu}}_{\tau} Z$          & $\mu$                    & t+u   \\
                   & ${\tilde {\nu}}_{\tau} Z$          & ${\tilde {\nu}}_{\tau}$  & s     &
                   & ${\tilde {\nu}}_{\tau} Z$          & ${\tilde {\nu}}_{\tau}$  & s     \\
                   & ${\tilde {\nu}}_{\tau} \gamma$     & $e$                      & t+u   &
                   & ${\tilde {\nu}}_{\tau} \gamma$     & $\mu$                    & t+u   \\  \hline \hline
\end{tabular}
\caption{{\it {Resonant and non-resonant single production of gauginos and sleptons at
$e^+e^-$ colliders and
${\mu}^+{\mu}^-$ colliders.
The indices $a$ and $b$ run as follow $a=1,2$ and $b=1,4$.
The ${\tilde \chi}^{\pm}_a$ and the ${\tilde \chi}^{0}_b$ can further cascade decay through
ordinary gauge decays till either the ${\tilde \chi}^{\pm}_1$ or the ${\tilde \chi}^{0}_1$ is reached.
The ${\tilde \chi}^{\pm}_1$ can either decay into $W^{\pm} {\tilde \chi}^{\pm}_1$ or via virtual
sfermion exchange and then with the $\lambda$ coupling involved in the single production. 
The ${\tilde \chi}^{0}_1$
can also further decay
with the $\lambda$ coupling involved in the single production.
The sleptons can also decay either directly via the $\lambda$ coupling involved in their single production
or into leptons and gauginos followed by the gauginos decay via the same $\lambda$ coupling.
The multilepton final state can then be deduced using 
Table~\ref{tab.sec6.rpv.sfermions.channels} and
Table~\ref{tab.sec6.rpv.gauginos.channels}.}}}
\label{tab:mucolsingledecay}
\end{center}
\end{table}

Leptonic colliders allows also for lepton-photon collisions in which sleptons 
and squarks can also be singly produced thus opening additional
perspectives for \Rp\ coupling studies.
For example~\cite{Allanach98} the processes 
$e^{\pm} \gamma \to l^{\pm} \tilde \nu, \tilde l^{\pm} \nu$,
involving an on-shell photon radiated from one of the colliding leptons, allow to probe 
the $\l_{122},\l_{123},\l_{132},\l_{133}$ and $\l_{231}$ \Rp\ couplings which are otherwise
not involved in the single sparticle productions from $e^+ e^-$ reactions.

The slepton or sneutrino production occurs either via the exchange of a charged 
lepton in the $s$-channel or the exchange of a charged slepton or lepton  
in the $t$-channel.
Since the $t$-channel is dominant and $m_{\tilde l}>>m_l$, the 
slepton production is about two order of magnitude less than the 
sneutrino production which is
$\s(e^+ e^- \to \tilde \nu_j e \tau)=300\femtob\ $ at $\sqrt s= 500 \GeV\ $.

In lepton-photon collisions, single squark production occurs via  
$\l'$ couplings as shown for example in Fig.~\ref{fig:singleph} for $e \gamma$
collisions.
\begin{figure}
\begin{center}
    \begin{tabular}{p{0.40\textwidth}p{0.55\textwidth}}
      \vspace{-4.0cm}
      \caption{{\it \label{fig:singleph}
                    Example diagram for single squark production in 
		    electron-photon collisions.}} &    
       \mbox{\epsfxsize=0.5\textwidth \epsffile{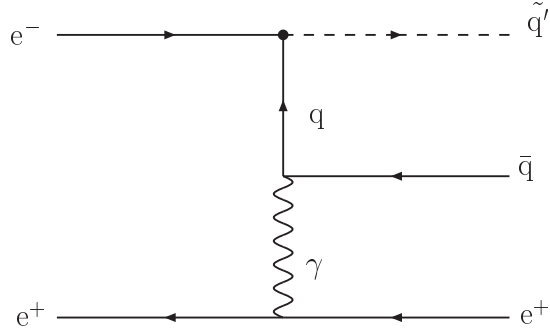}}
    \end{tabular}      
\end{center}
\end{figure}
When the produced squark directly decays via
$\l'$ into a lepton and a quark the final state  
consists of one hard mono-jet with one well isolated energetic electron, and 
eventually a soft jet in the forward region of the detector 
in the case where the initial electron which scatters the quasi real photon 
escapes detection. 

\subsection{Single Sparticle Production at Lepton-Hadron Colliders} 
\label{sec:singlelp}
\index{Sparticle production!Single!at lepton-hadron colliders}

An $lp$ collider provides both leptonic and baryonic quantum numbers in the 
initial state and is thus ideally suited for \Rp\ SUSY searches involving 
$\lambda'_{ijk}$. 
Among the twenty-seven possible $\lambda'_{ijk}$ couplings, each of the nine 
couplings with $i=1$ can lead to direct squark resonant production through 
$e$-$q$ fusion at an $ep$ collider such as HERA.
The phenomenology of such processes was first investigated theoretically 
in Refs.~\cite{Butterworth91,Kon91a,Kon91b, Butterworth93}.
Search strategies taking into account in general the indirect \Rp\
squark decay modes were discussed in Refs.~\cite{Butterworth93, Perez95, Perez96}.

The production processes are listed in Table~\ref{tab:sqprod} in the case
of an incident $e^+$ beam. 
%
%
\begin{table*}[htb]
  \renewcommand{\doublerulesep}{0.4pt}
  \renewcommand{\arraystretch}{1.2}
 \begin{center}
 \begin{tabular}{p{0.40\textwidth}p{0.60\textwidth}}
    \vspace{-1.0cm}
    \caption{{\it \label{tab:sqprod}
         Direct resonant production of squarks at an $ep$ collider
	 via a \Rp\ $\lambda'_{1jk}$ coupling. The processes are listed
	 for an incident $e^+$ beam. Charge conjugate processes are
	 accessible for an incident $e^-$ beam.}} &
   \begin{tabular}{||c||c|c||}
   \hline \hline
   $\lambda'_{1jk}$ & \multicolumn{2}{c||}{production process} \\
   \hline
   111 & $e^+ +\bar{u} \rightarrow \bar{\tilde{d}}_R$
       &$e^+ +d \rightarrow \tilde{u}_L $\\
   112 & $e^+ +\bar{u} \rightarrow \bar{\tilde{s}}_R$
       &$e^+ +s \rightarrow \tilde{u}_L $\\
   113 & $e^+ +\bar{u} \rightarrow \bar{\tilde{b}}_R$
       &$e^+ +b \rightarrow \tilde{u}_L $\\
   121 & $e^+ +\bar{c} \rightarrow \bar{\tilde{d}}_R$
       &$e^+ +d \rightarrow \tilde{c}_L $\\
   122 & $e^+ +\bar{c} \rightarrow \bar{\tilde{s}}_R$
       &$e^+ +s \rightarrow \tilde{c}_L $\\
   123 & $e^+ +\bar{c} \rightarrow \bar{\tilde{b}}_R$
       &$e^+ +b \rightarrow \tilde{c}_L $\\
   131 & $e^+ +\bar{t} \rightarrow \bar{\tilde{d}}_R$
       &$e^+ +d \rightarrow \tilde{t}_L $\\
   132 & $e^+ +\bar{t} \rightarrow \bar{\tilde{s}}_R$
       &$e^+ +s \rightarrow \tilde{t}_L $\\
   133 & $e^+ +\bar{t} \rightarrow \bar{\tilde{b}}_R$
       &$e^+ +b \rightarrow \tilde{t}_L $\\
   \hline \hline
  \end{tabular}
  \end{tabular}
\end{center}
\end{table*}

In $e^+p$ collisions, the production of $\tilde{u}_L^j$ squarks of the $j^{th}$
generation via $\lambda'_{1j1}$ is especially interesting as it involves a valence 
$d$ quark of the incident proton.
In contrast, for $e^-p$ collisions where charge conjugate processes are accessible, 
the $\lambda'_{11k}$ couplings become of special interest as they allow for the
production, involving a valence $u$ quark, of $\tilde{d}_R^k$ squarks of the 
$k^{th}$ generation.
As an illustration, the Fig.~\ref{fig:epsqxsect} shows the production cross-sections 
in $e^{\pm}p$ collisions for the ``up''-like squarks $\tilde{u}^j_L$ via $\lambda'_{1j1}$ 
($j=1 \ldots 3$) compared to that for the ``down''-like squarks $\bar{\tilde{d}}^k_R$
via $\lambda'_{11k}$ ($k=1 \ldots 3$)). 
\begin{figure}[t]
\vspace{-0.3cm}

  \begin{center}
    \begin{tabular}{p{0.40\textwidth}p{0.60\textwidth}}
      \vspace{-4.0cm}
      \caption[]{{\it  \label{fig:epsqxsect}
                Squark production cross-sections in $e^{\pm}p$ collisions 
		calculated~\cite{Perez96} for a coupling $\lambda'=0.1$ and 
		a collider centre-of-mass energy of
		$\sqrt{s_{ep}} = 300 \GeV$ . }} &
      \mbox{\epsfxsize=0.5\textwidth \epsffile{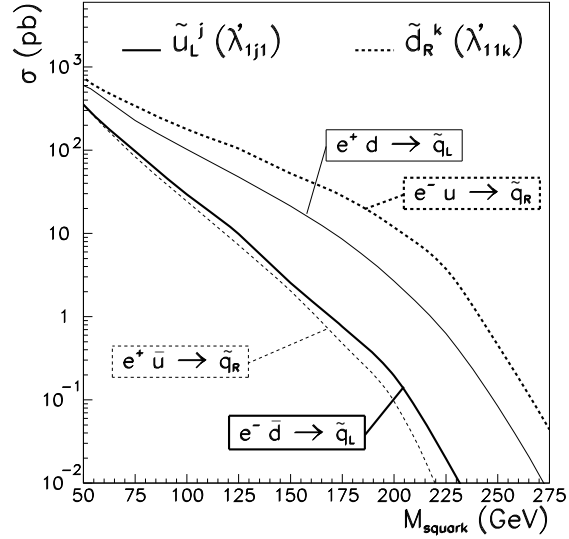}}
    \end{tabular}
  \end{center}
\vspace{-0.5cm}
\end{figure}
The cross-sections are calculated~\cite{Perez96} for coupling values of $\lambda'=0.1$
and for an available centre-of-mass energy of $\sqrt{s_{ep}}$ of $300 \GeV$ 
characteristic of the HERA collider.
By gauge symmetry, only $\tilde{u}_L$-like or $\tilde{d}_R$-like squarks
(or their charge conjugates) can be produced in $ep$ collisions.
Since superpartners of left- and right-handed fermions may have different
allowed or dominant decay modes, the dichotomy between the resonant production 
of $\tilde{u}_L$-like squarks in $e^+p$ collisions and that of 
$\tilde{d}_R$-like squarks in $e^-p$ collisions implies that the 
detailed analysis will differ between $e^-$ and $e^+$ incident beams.

In the case of a direct $\Rp$ decay through a $\lambda'$ coupling, the 
squarks which have been resonantly produced in $ep$ collisions behave
as leptoquarks~\cite{Buchmuller87}. The $\tilde{u}_L$ may couple to an 
$e^++d$ pair via a Yukawa coupling $\lambda'_{111}$ in a way similar to 
the coupling of the first generation $\tilde{S}_{1/2,L}$ 
leptoquark of charge $|{\rm Q}_{em}| = 2/3$.
Via the same coupling, the $\tilde{d}_R$ couples to $e^-+u$ or $\nu_e+d$ 
pairs, thus behaving like the first-generation $S_{0,L}$ leptoquark
of charge $|{\rm Q}_{em}| = 1/3$.
As a general consequence, it is possible to translate constraints
on the $\lambda$ couplings of leptoquarks into constraints on the 
$\lambda'_{1jk}$ couplings of squarks in $R_p$-violating supersymmetry. 
For real squark production, this translation is limited
to coupling values $\lambda' \gsim \sqrt{4 \pi \alpha}$. For smaller
values, the branching ratio into leptoquark-like final states rapidly 
drops as squarks will prefer indirect $\Rp$ decays.
Such a re-interpretation of leptoquark constraints from early HERA data 
has been performed in Refs.~\cite{H1lqsq1994,H1lqsq1999,ZEUSlqsq2000}.
For virtual squark exchange in the case where 
\mbox{$M_{\tilde{q}} \gg \sqrt{s_{ep}}$}, constraints can be established 
via four-fermion leptoquark-like contact interaction analysis as will be 
discussed in section~\ref{sec:fourf}.

In the case of indirect $\Rp$ decays, the squarks in a first stage decay
through gauge couplings into a quark and a gaugino-higgsino 
($\tilde{\chi}^0$, $\tilde{\chi}^+$) or, if $M_{\tilde{g}} \ll M_{\tilde{q}}$,
into a quark and a gluino. 
Such squark decays involving $\Rp$ couplings were discussed in detail in
section~\ref{sec:sdecays}. 
Here again, at an $ep$ collider, the dichotomy between the production
of $\tilde{u}_L$ and $\tilde{d}_R$ will have important phenomenological
consequences. 
While the $\tilde{u}_L$ might decay via 
$\tilde{u}_L \rightarrow u \tilde{\chi}_l^0$ or $\tilde{\chi}_m^+$, 
the $\tilde{d}_R$ mainly decays via 
$\tilde{d}_R \rightarrow d \tilde{\chi}_l^0$, the $\tilde{b}_R$ decay into a chargino
being also possible via the higgsino component of the latter.

Depending on the mixing parameters in the neutralino and chargino sectors,
the dominating event topologies to be expected might depend on whether
the collider is running in $e^+p$ or in $e^-p$ mode.
Decays of $\tilde{\chi}^0$ and $\tilde{\chi}^+$ mass eigenstates were
discussed in detail in section~\ref{sec:sdecays}.

Detailed dicussion on the event topologies expected at an $ep$ collider
for single squark production in the presence of $\Rp$ couplings can be found 
in Refs.~\cite{Butterworth93,Perez95,Perez96}.

%
%

Real or virtual squark exchange in the $s$-channel contributes to the 
single production of neutralinos or charginos. These can also be
singly produced via $\Rp$ interactions in lowest order processes 
involving sleptons or sneutrinos.
The $\tilde{\chi}^0$ can be produced via $t$-channel slepton exchange 
or via $u$-channel squark exchange.
The $\tilde{\chi}^+$ can be produced via $t$-channel sneutrino exchange.

The Fig.~\ref{fig:epchixsect} shows the neutralino production cross-sections 
in $e^{\pm}p$ collisions expected for a coupling value $\lambda'_{11k} = 0.5$
and for an available centre-of-mass energy of $\sqrt{s_{ep}}$ of $300 \GeV$ 
characteristic of the HERA collider. 
\begin{figure}[t]
\vspace{-0.3cm}

  \begin{center}
    \begin{tabular}{p{0.40\textwidth}p{0.60\textwidth}}
      \vspace{-4.0cm}
      \caption[]{{\it  \label{fig:epchixsect}
                Cross-sections for $\tilde \chi^0$ production in $e^{\pm}p$ 
		collisions from $t$-channel selectron exchange 
		calculated~\cite{Perez99} for a coupling $\lambda'_{1j1} =0.5$ 
		and for collider centre-of-mass energy of 
		$\sqrt{s_{ep}} = 300 \GeV$ . }} &
      \mbox{\epsfxsize=0.5\textwidth \epsffile{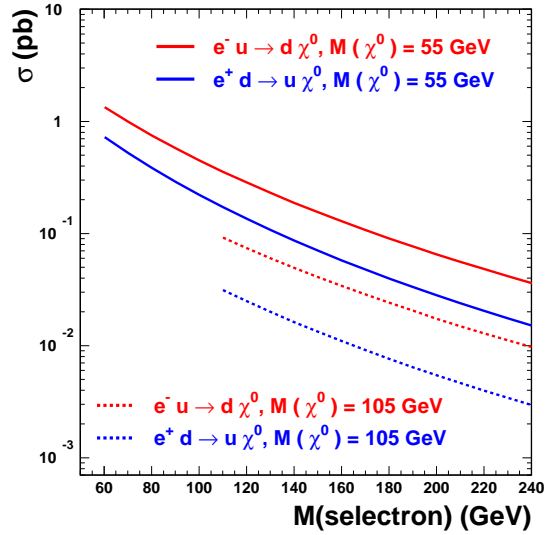}}
    \end{tabular}
  \end{center}
\vspace{-0.5cm}
\end{figure}
The cross-sections are calculated~\cite{Perez99} for selectron exchange only
in the framework of the MSSM augmented by a single non-vanishing $\lambda'$ 
coupling, for two values of $M_{\tilde{\chi}^o}$ and for $\tan \beta = 1$.
Such cross-sections could be expected in case 
$M_{\tilde{e}} \ll M_{\tilde{q}}$. When both $s$-channel $\tilde{q}$ 
exchange and $t$-channel $\tilde{e}$ contribute,
the interference between these cannot be neglected. For example at HERA~II,
constructive interference between squark and selectron exchange processes could
contribute~\cite{Perez99} up to $20 \%$ of the total $\tilde \chi^o$ production 

%
%

Searches for single squark production have been performed at HERA$_I$ under
the hypothesis of a single dominant $\lambda'_{1jk}$ coupling.
The constraints obtained~\cite{H1rpv1996,H1rpv2001,H1rpv2004} by the H1 experiment 
are shown in Fig.~\ref{fig:lpvsm}. Similar results 
were obtained~\cite{ZEUSrpv2000} by the ZEUS experiment.
All possible event topologies (multijets and lepton and/or missing energy)
resulting from the direct or indirect sparticle decays involving such coupling 
have been considered in the analysis. 
The HERA$_I$ results are compared to the best existing indirect 
bounds~\cite{reviews2} from low-energy experiments. 
\begin{figure}[htb]
  \begin{center}                                                                
  \begin{tabular}{cc}
    
  \hspace*{-0.2cm} \epsfig{file=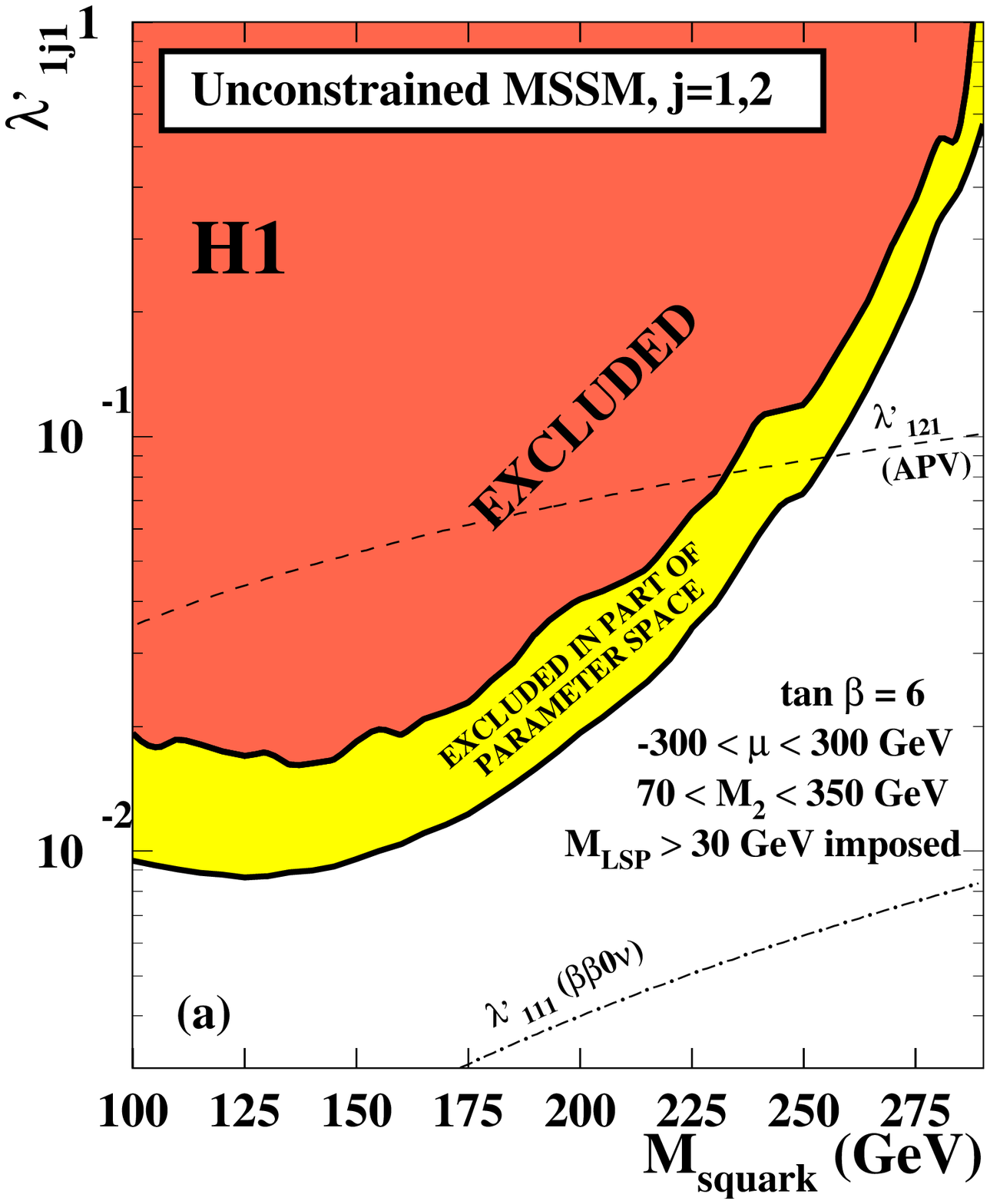,width=0.45\textwidth}
&
  \hspace*{-0.5cm} \epsfig{file=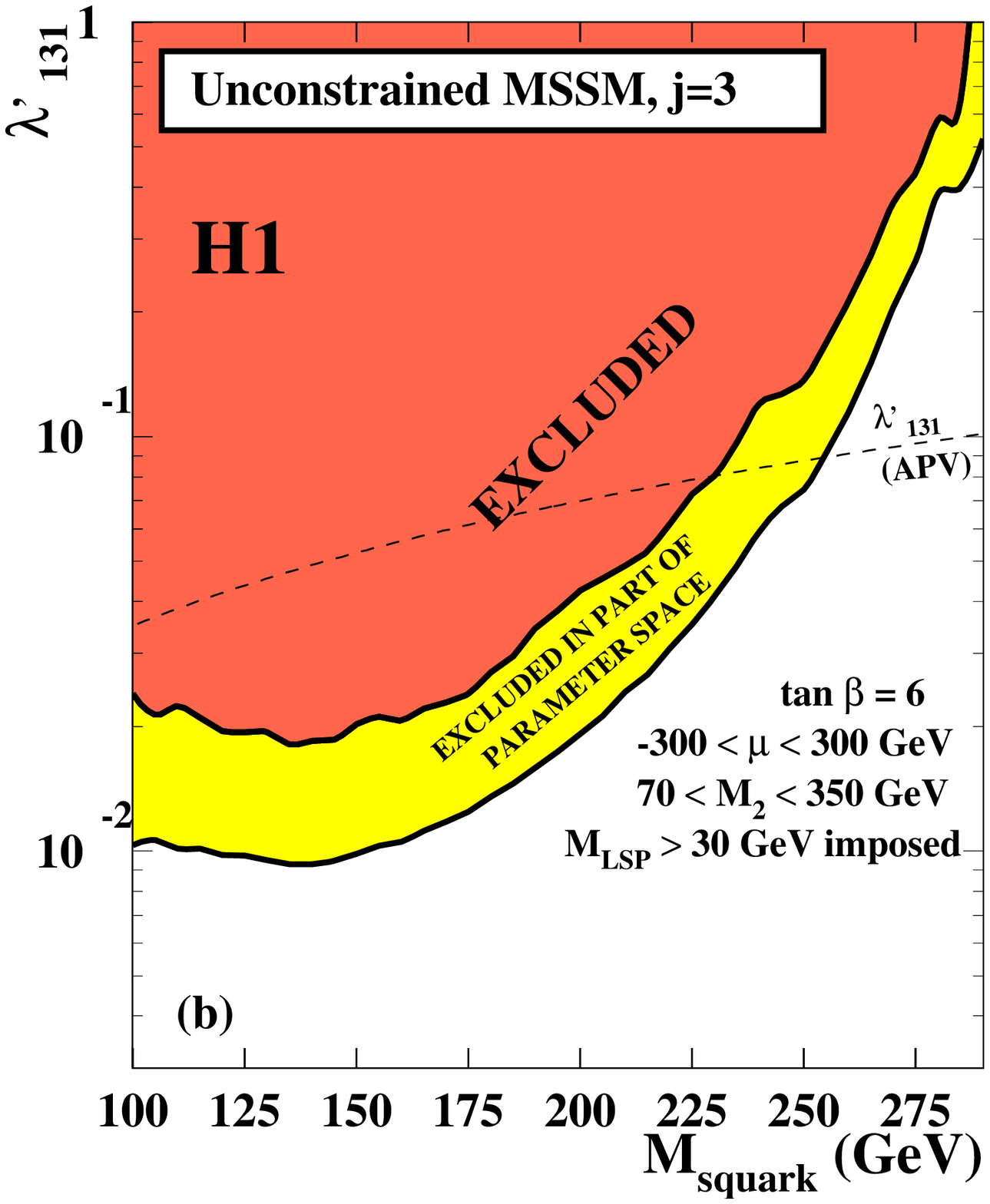,width=0.45\textwidth}

  \end{tabular}
  \end{center}
 \vspace*{-1.0cm}
 
 \hspace*{5.0cm} (a) \hspace*{7.5cm} (b) \\
 
 \vspace*{-0.5cm}

  
 \caption[]{ \label{fig:lpvsm}
            Upper Limits (95\% CL) on a) the coupling $\lambda'_{1j1}$ with
	    $j=1,2$ and b) $\lambda'_{131}$ as a function of the squark mass
	    for $\tan{\beta} = 6$ in the unconstrained MSSM.
	    The limits are obtained from a scan of the $\mu$ and $M_2$
	    parameters within $-300 < \mu < 300 \GeV$ and 
	    $70 < M_2 < 350 \GeV$ and imposing that the lightest sparticle
	    (LSP) has a mass $M_{LSP}$ above $30 \GeV$.
	    The dark shaded area is excluded for any parameter values.
	    The light shaded area is excluded for some parameters values.
	    The dashed-dotted curve is the indirect upper bound~\cite{reviews2} on 
	    $\lambda'_{111}$ derived from constraints on neutrinoless 
	    double-beta decays~\cite{Hirsch95,Balysh95}.
	    The dashed curves are the indirect upper 
	    bounds~\cite{reviews2} on $\lambda'_{1j1}$ derived from 
	    constraints on atomic-parity violation~\cite{Wood97}.}
\end{figure}
The $\lambda'_{111}$ coupling is seen to be very severely constrained by 
the non-observation of neutrinoless double-beta decay. 
The most stringent low-energy constraints on $\lambda'_{121}$ and 
$\lambda'_{131}$ come from atomic-parity violation measurements.
From these HERA~I results, it can be infered that HERA~II 
could offer a sensitivity reach beyond the domain excluded by 
indirect constraints for 2$^{nd}$ and 3$^{rd}$ generation squarks.

The HERA results analysed in the framework of $\Rp$ mSUGRA are shown in 
Fig.~\ref{fig:stophera} and compared to complementary $\Rp$ SUSY searches 
made at LEP~2 \index{LEP} and Tevatron Run~I colliders.
The searches were performed here also under the hypothesis of a single 
dominant  $\lambda'_{1jk}$ coupling. 
The results are presented as excluded domains in the parameter space of the model.
\begin{figure}[htb]

  \begin{center}
  \begin{tabular}{cc}

  \hspace*{-0.2cm} \epsfig{file=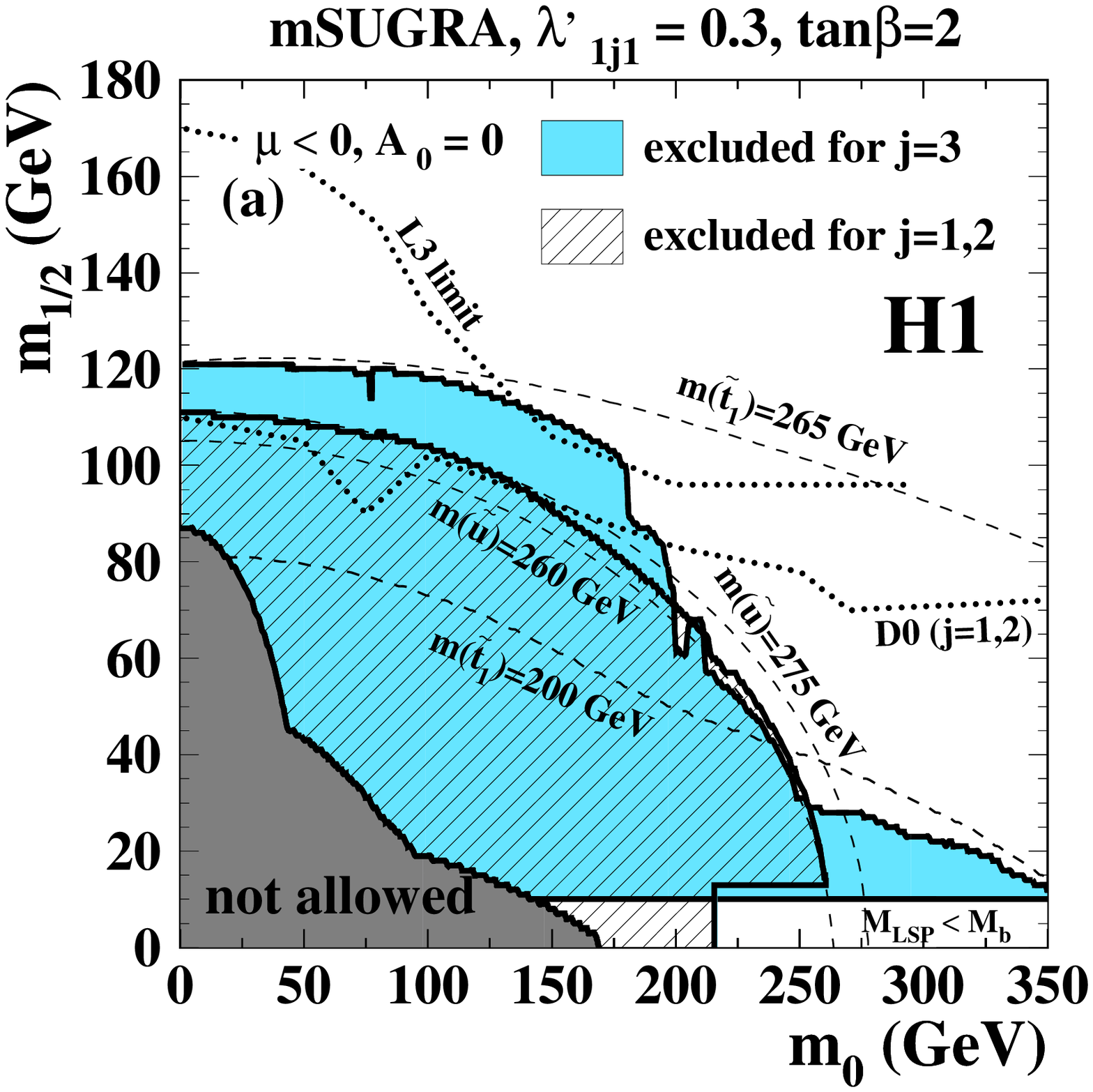,width=0.45\textwidth}
&
  \hspace*{-0.5cm} \epsfig{file=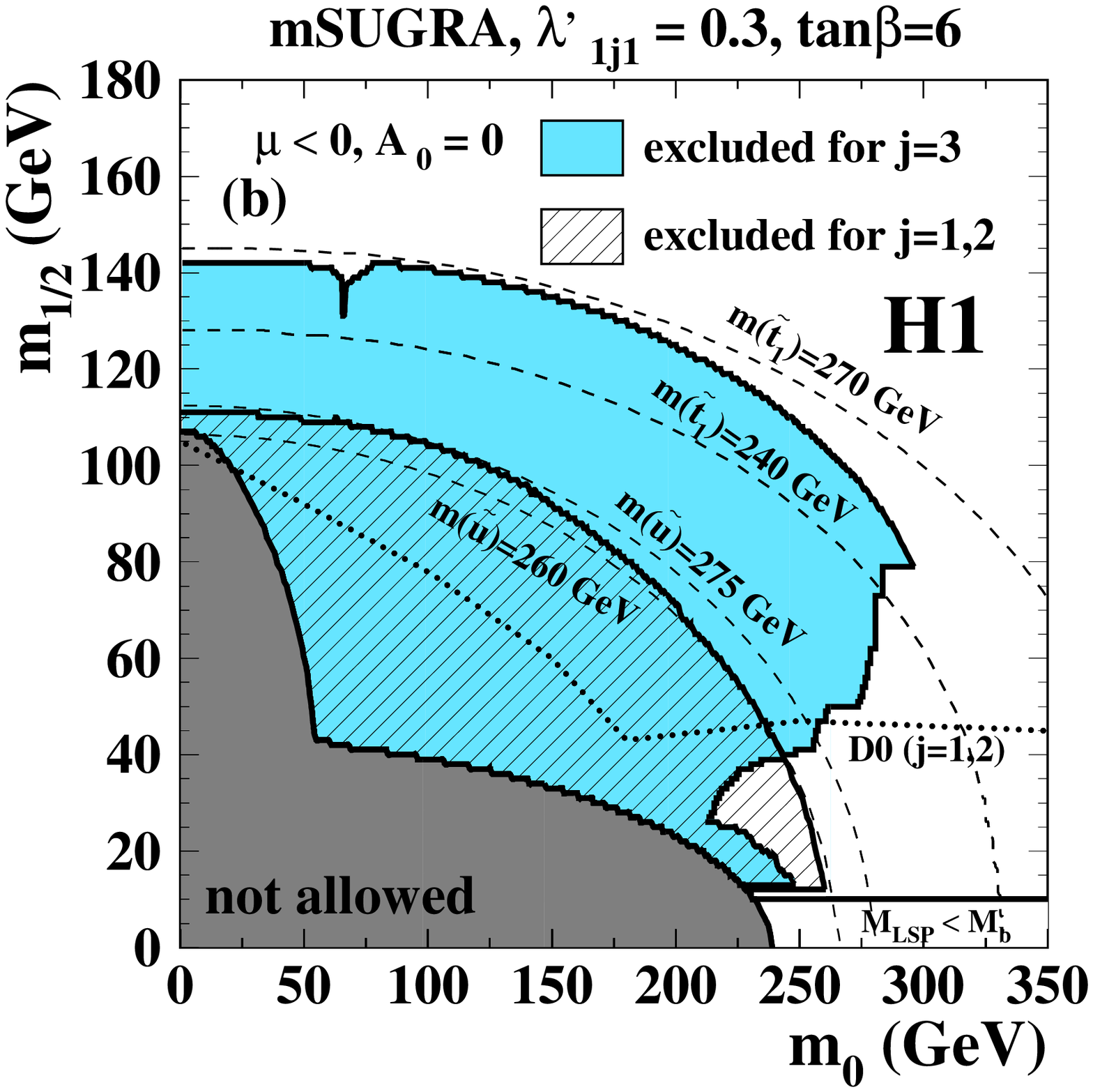,width=0.45\textwidth}

  \end{tabular}
  \end{center}

 
 
  
  \caption[]{ \label{fig:stophera}
              {\it Constraints on squark production via $\lambda '_{1j1}$ in
	      $R_p$-violating SUSY in the parameter space of Minimal 
	      Supergravity. Excluded domains obtained by the 
	      H1~\cite{H1rpv2004} (shaded area) and D$\emptyset$ (dotted curves) 
	      experiments are shown for 
	      a) $\tan{\beta} = 2$ and 
	      b) $\tan{\beta} = 6$.
	      In a) the limit obtained by the L3 experiment at LEP~2 
              is also shown as the upper dotted curve.
	      Contours of constant values for the light stop mass are drawn
	      as dashed curves. 
              The shaded region marked ``not allowed'' corresponds to points 
	      in the parameter space where the radiative electroweak symmetry 
	      breaking does not occur (or which lead to unphysical Higgs or 
	      sfermion masses). Also marked as ``not allowed'' in this
	      particular analysis are cases where the LSP is the sneutrino.}} 
\end{figure}
The constraints from the D$\emptyset$~\cite{NEWD0RPV} experiment at the 
Tevatron were obtained from a search for $\tilde{q}$ pair production through
gauge couplings. 
The D$\emptyset$ analysis profits in this framework from an approximate mass 
degeneracy implicitly extended to five $\tilde{q}$ flavours 
($\tilde{d}$,$\tilde{u}$,$\tilde{s}$,$\tilde{c}$,$\tilde{b}$) and both 
(partners) chiralities ($\tilde{q}_L$,$\tilde{q}_R$).
The $\Rp$ couplings are assumed to be significantly smaller than the gauge 
couplings, so that direct $\Rp$ decays are suppressed and each squark rather
decays back into a quark and the LSP through gauge couplings.
The only effect of the $\Rp$ couplings is then to make the LSP unstable. 
The D$\emptyset$ analysis is further restricted to $\Rp$ coupling 
values $\gsim 10^{-3}$ to guarantee a negligible decay length of the LSP.
In the domains considered, the LSP is almost always the lightest 
neutralino $\XOI$. The $\XOI$ decays via $\lambda'_{1jk}$ into a
first-generation lepton ($e$ or $\nu_e$) and two quarks. The analysis is restricted
to $j=1,2$ and $k=1,2,3$ and, in practice, the D$\emptyset$ selection of 
event candidates requires like-sign di-electrons accompanied by multiple jets.
The constraints from the L3 experiment at LEP \index{LEP} were obtained from a 
search for pair production through gauge couplings of neutralinos 
($e^+e^- \rightarrow \tilde{\chi}_m^0 
\tilde{\chi}_n^0$ with $m=1,2$ and $n=1, \ldots ,4$), charginos ($e^+e^- 
\rightarrow \tilde{\chi}_1^+ \tilde{\chi}_1^-$) and scalar leptons ($e^+e^- 
\rightarrow \tilde{l}_R^+ \tilde{l}_R^-$, $\tilde{\nu} \tilde{\nu}$).
The $\Rp$ couplings contribute here again in opening new decay modes for 
the sparticles. A negligible decay length of the sparticles through these
decay modes is ensured by restricting the analysis to coupling values 
$\gsim 10^{-5}$.
All possible event topologies (multijets and lepton and/or missing energy)
resulting from the direct or indirect sparticle decays involving
the $\lambda'_{ijk}$ couplings have been considered in the L3 analysis. 

For the set of mSUGRA parameters with $\tan \beta = 2$, the Tevatron
experiment excludes squarks with masses $M_{\tilde{q}} < 243 \GeV$ 
(95 \% CL) for any value of $M_{\tilde{g}}$ and a finite value 
($\gsim 10^{-3}$) of $\lambda'_{1jk}$ with $j=1,2$ and $k=1,2,3$. 
The sensitivity decreases for the parameter set with a larger value of 
$\tan \beta$ due in part to a decrease of the photino component of the 
LSP, which implies a decrease of the branching fraction of the LSP into 
electrons, and in part to a softening of the final-state particles for 
lighter charginos and neutralinos.
The best sensitivity at $\tan \beta = 2$ is offered by LEP \index{LEP} for any 
of the $\lambda'_{ijk}$ couplings.
HERA offers a best complementary sensitivity to the coupling 
$\lambda'_{131}$ which allows for resonant stop production via 
positron-quark fusion  $e^+ d \rightarrow \tilde{t}_1$.
The HERA constraints (shown here for a coupling
of electromagnetic strength, i.e. \mbox{$\lambda'_{131}=0.3$}) 
extend beyond LEP \index{LEP} and Tevatron constraints towards larger $\tan 
\beta$.
\subsection{Single Sparticle Production at Hadron-Hadron Colliders} 
\label{sec:singlepp}
\index{Sparticle production!Single!at hadron colliders}

The SUSY particles can be produced as resonances at hadron colliders
through the \Rp\ interactions. This is particularly attractive as
hadron colliders allow to probe for resonances over a
wide mass range given the continuous energy distribution of the colliding 
partons.
If a single $\Rp$ coupling is dominant, the resonant
SUSY particle may decay through the same coupling involved in its production,
giving a two quark final state at the partonic level. However, it is 
also possible that the decay of the resonant SUSY particle is mainly due 
to gauge interactions, giving rise to a cascade decay. A review focusing
on Tevatron Run-II can be found in~\cite{DREINER99}.

$\bullet$ {\bf Single sparticle production via {\boldmath{$\l'$}} }

First, a resonant sneutrino can be produced in $d \bar d$ annihilations 
through the constant $\l'_{ijk}$. The associated formula can be written
as follows \cite{Qui}:
%
%
\begin{eqnarray}
%
%
\s (d_k \bar d_j \to  \tilde \nu^i \to X_1 X_2)= \frac{4}{9} 
\frac{\hat{s}}{m_{\tilde \nu^i}^2}
{\frac 
{\pi \Gamma_{d_k \bar d_j} \Gamma_f}{(\hat{s}-m_{\tilde \nu^i}^2)^2 + 
m_{\tilde \nu^i}^2 \Gamma_{\tilde \nu^i}^2}}\;,
\label{snwi1}
\end{eqnarray} 
where $\Gamma_{d_k \bar d_j}$, and $\Gamma_f$ are the partial width of
the channels, $\tilde \nu^i \to d_k \bar d_j$,
and, $\tilde \nu^i \to X_1 X_2$, respectively, $\Gamma_{\tilde \nu^i}$ is
the total width of the sneutrino, 
$m_{\tilde \nu^i}$ is the sneutrino mass and $\hat{s}$ is the square of
the parton centre-of-mass energy. The factor
${1/9}$ in front is from matching the initial colours, and
$\Gamma_{d_k \bar d_j}$ is given by,
\begin{eqnarray}
\Gamma_{d_k \bar d_j} = \frac{3}{4} \alpha_{\l'_{ijk}} m_{\tilde \nu^i},
\label{snwidth}
\end{eqnarray}
where $ \alpha_{{\l'}_{ijk}}= {\l'}^2_{ijk} / 4 \pi $. To compute the rate
at a $p \bar p$ collider,
the usual formalism of the parton model of hadrons can be used
\cite{Eich}: 
\begin{eqnarray}
\s (p \bar p \to \tilde \nu^i \to X_1 X_2) = \sum_{j,k} \int_{\tau_0}^1
{\frac {d \tau}{\tau}} ({\frac {1}{s}}{\frac {d L_{jk}}{d \tau}}) \ 
\hat{s} \ \s (d_k \bar d_j \to  \tilde \nu^i \to X_1 X_2),
\label{hsec}
\end{eqnarray}
where $s$ is the centre-of-mass energy squared, $\tau_0$ is given by
$\tau_0=(M_{X_1}+M_{X_2})^2/s$ and $\tau$ is defined by
$\tau=\hat{s} / s=x_1 x_2$, $x_1,x_2$ denoting the longitudinal momentum
fractions of the initial partons $j$ and $k$,
respectively. The quantity ${dL_{jk}/ d \tau}$ is the parton
luminosity defined by,
\begin{eqnarray}
{\frac {dL_{jk}}{d \tau}}= \int_{\tau}^1 {\frac {d x_1}{x_1}} [f_j^{\bar
p}(x_1) f_k^p(\tau / x_1)+
f_j^p(x_1)f_k^{\bar p}(\tau / x_1)],
\label{plum}
\end{eqnarray}
where the parton distribution $f_j^h(x_1)$ denotes the probability of
finding a parton $j$ with momentum fraction $x_1$ inside a hadron h,
and generally depends on the Bjorken variable, $Q^2$, the square of the
characteristic energy scale of the process under consideration. 
In order to see the effects of the parton distributions on the resonant sneutrino
production, some values of the rates are given in the following \cite{Dim2}:
For instance, with an initial state $d \bar d$ for the hard process,
the cross-section value is $\s(p \bar p \to \tilde \nu^i)=8.5 \nanob$ 
for a sneutrino mass of $100 \GeVcc$ and a coupling, $\l'_{i11}=1$ at
$\sqrt s =2 \TeV$.
For identical values of the parameters and of the centre-of-mass energy,
the cross-section is $\s(p \bar p \to \tilde \nu^i)=4 \nanob$ with
an initial state, $d \bar s$, and 
$\s(p \bar p \to \tilde \nu^i)=0.8 \nanob$ with an initial state,
$d \bar b$.
\begin{figure}[htb]
\begin{center}
\epsfig{file=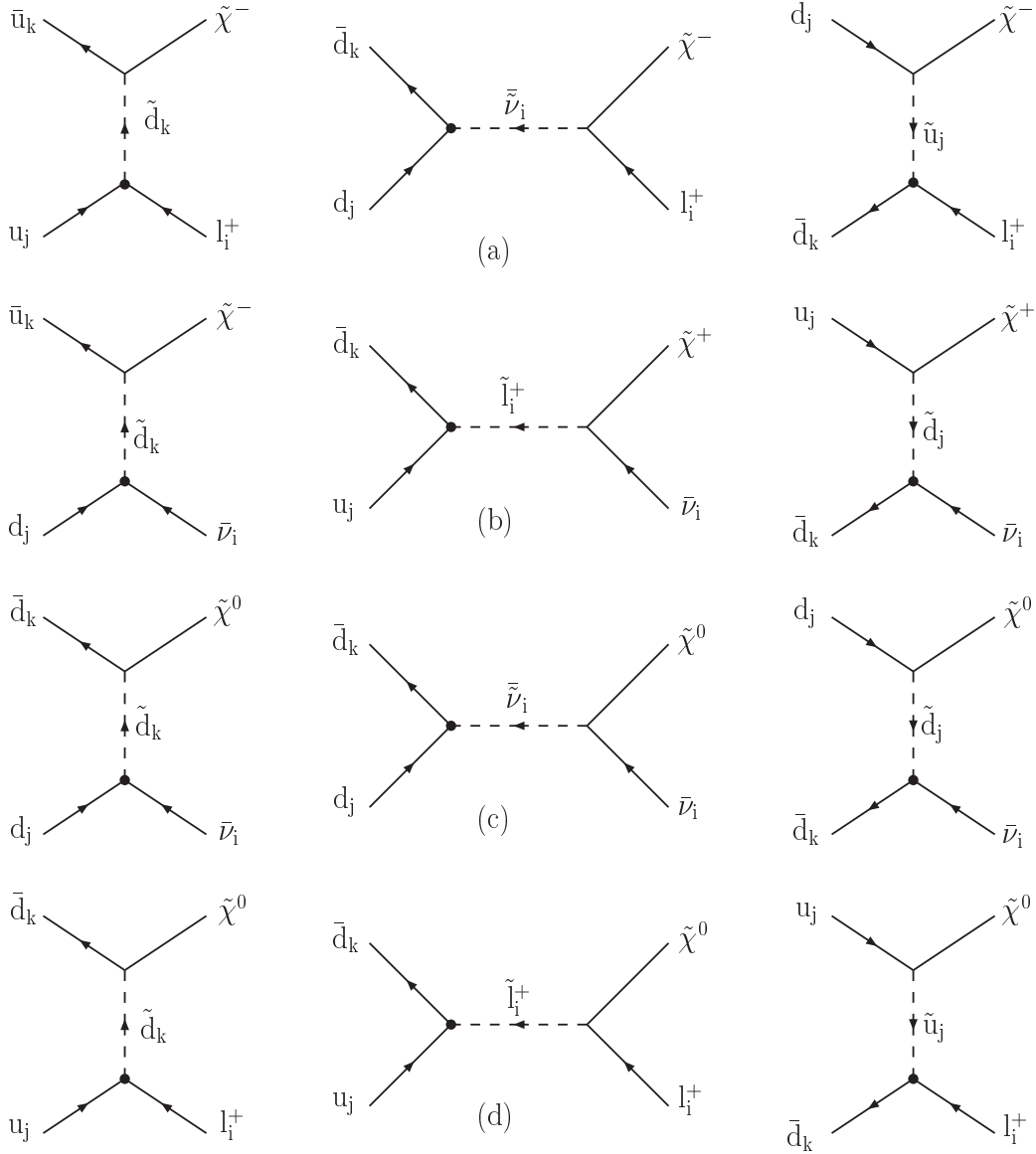,width=0.90\textwidth}  

\caption{{\it Diagrams for the four single superpartner production
              reactions involving $\l'_{ijk}$ at hadron colliders which
              receive a contribution from a resonant \susyq\ particle 
	      production.
              The $\l'_{ijk}$ coupling constant is symbolised by a small 
	      circle and the arrows indicate the flow of the lepton 
	      or baryon number.}}
\label{fig:singllp}
\end{center}
\end{figure}
The charged slepton can also be produced as a
resonance at hadron colliders from an initial state
$u_j \bar d_k$ and via the constant $\l'_{ijk}$.
The cross-section value is $\s(p \bar p \to \tilde l^i_L)=2 \nanob$
for $m_{\tilde l^i_L} = 100 \GeVcc$, $\sqrt s=2 \TeV$ and
$\l'_{i11}=1$ (\cite{Dim2,Kal}).

The single production of SUSY particles via $\l'$ occurring through 
{\it two-to-two}-body processes, offers the opportunity to study the 
parameter space of the \Rp\ models with a quite high sensitivity at hadron
colliders.

In Fig.~\ref{fig:singllp}, all the single superpartner productions which 
occur via $\l'_{ijk}$ through {\it two-to-two}-body processes at 
hadron colliders and receive a contribution from a resonant SUSY particle 
production are presented \cite{More}. 
The spin summed amplitudes of those reactions including the higgsino 
contributions have been calculated in \cite{More}. 
In a SUGRA model, the rates of the reactions presented in 
Fig.~\ref{fig:singllp} depend mainly on the $m_0$ 
and $M_2$ parameters.

\begin{figure}[htb]
\begin{center}
\epsfig{file=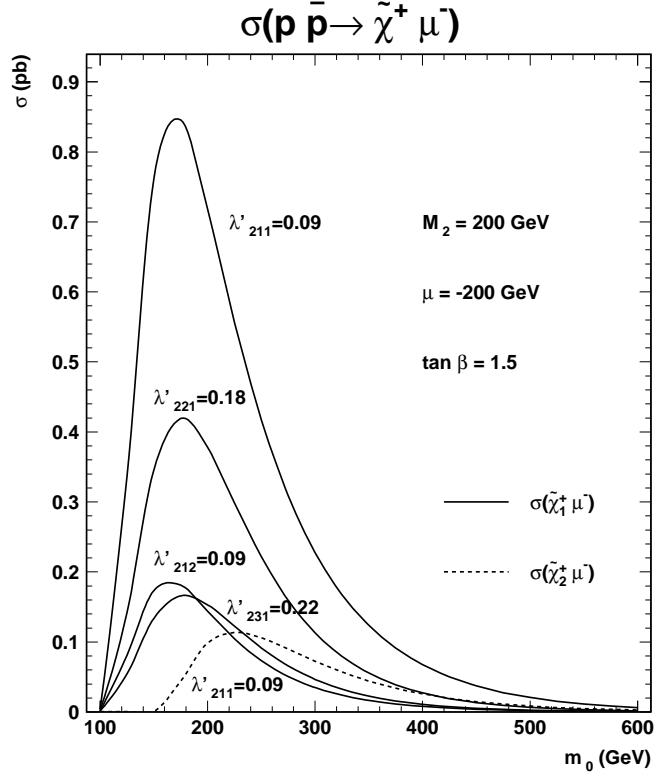,width=10cm}   
\vspace{-0.4cm}

\caption{{\it Cross-sections (in $pb$) of the single chargino productions  
              $p \bar p \to \tilde \chi^+_{1,2} \mu^-$ as a function of the 
              $m_0$ parameter (in $\GeVcc$).
              The centre-of-mass energy is taken at $\sqrt s=2$~TeV and the 
              considered set of parameters is: $\l'_{211}=0.09$,
              $M_2=200 \GeVcc$, $\tan \beta=1.5$ and $\mu=-200 \GeVcc$.
              The rates for the $\tilde \chi^+_1$ production via the
              \Rp\ couplings $\l'_{212}=0.09$, $\l'_{221}=0.18$ and 
	      $\l'_{231}=0.22$ are also given. 
              The chosen values of the \Rp\ couplings correspond to the 
              low-energy limits for a squark mass of 
	      $100 \GeVcc$~\cite{reviews}.}}
\label{fig:XScl}
\end{center}
\end{figure}

In Fig.~\ref{fig:XScl}, the variations of the
$\sigma(p \bar p \to \tilde \chi^+_{1,2} \mu^-)$ cross-sections with $m_0$
for fixed values of $M_2$, $\mu$ and $\tan \beta$
and various \Rp\ couplings of the type $\l'_{2jk}$
at Tevatron Run~II in a SUGRA model are shown \cite{More}. 
The \Rp\ couplings giving the highest cross-sections have been considered.
The $\sigma(p \bar p \to \tilde \chi^+_{1,2} \mu^-)$ rates decrease when
$m_0$ increases since then the sneutrino becomes heavier and more energetic
initial partons are required in order to produce the resonant sneutrino.
A decrease of the cross-sections also occurs at small values of $m_0$, the
reason being that when $m_0$ approaches $M_2$ the $\tilde \nu$ mass is getting
closer to the $\tilde \chi^{\pm}$ masses so that the phase space factors
associated to the decays $\tilde \nu_{\mu} \to \chi^{\pm}_{1,2} \mu^{\mp}$ 
decrease.
The differences between the $\tilde \chi^+_1 \mu^-$
production rates occurring via the various $\l'_{2jk}$ couplings
are explained by the different partonic luminosities. 
Indeed, as shown in Fig.~\ref{fig:singllp} the hard process associated 
to the $\tilde \chi^+_1 \mu^-$ production occurring through the 
$\l'_{2jk}$ coupling constant has a partonic initial state 
$\bar q_j q_k$. 
The $\tilde \chi^+_1 \mu^-$ production via the $\l'_{211}$ coupling has first 
generation quarks in the initial state which provide the maximum partonic
luminosity.

\begin{figure}[htb]
\begin{center}
\epsfig{file=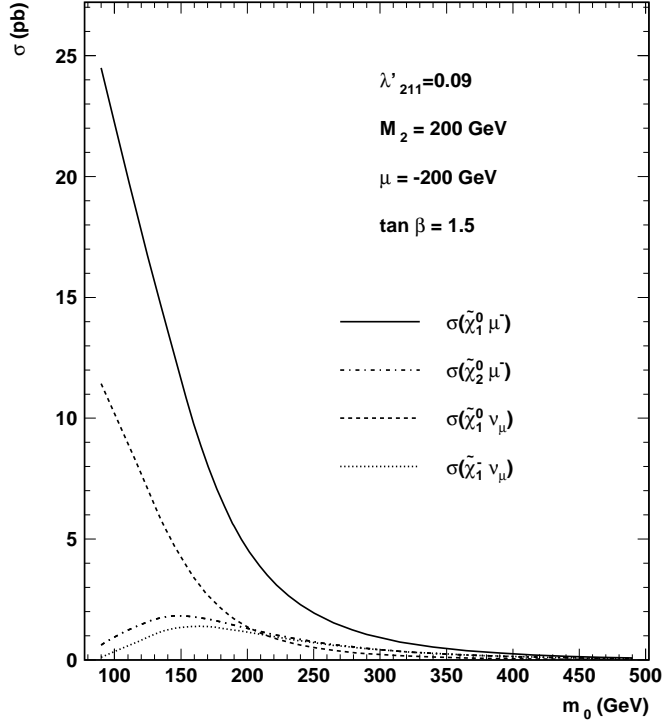,width=10cm}  
\vspace{-0.4cm}

\caption{{\it Cross-sections (in $pb$) of the $\tilde \chi^-_1 \nu$, 
              $\tilde \chi^0_{1,2} \mu^-$ and $\tilde \chi^0_1 \nu$  
	      productions at Tevatron Run~II as a function of the $m_0$ 
              parameter (in $\GeVcc$).
              The centre-of-mass energy is taken at $\sqrt s=2$~TeV 
              and the considered set of parameters is: $\l'_{211}=0.09$,
              $M_2=200 \GeVcc$, $\tan \beta=1.5$ and $\mu=-200 \GeVcc$.}}
\label{fig:allXS}
\end{center}
\end{figure}

In Fig.~\ref{fig:allXS}, the variations of the rates of the reactions 
$p \bar p \to \tilde \chi^-_1 \nu$, $p \bar p \to \tilde \chi^0_{1,2} \mu^-$ 
and $p \bar p \to \tilde \chi^0_1 \nu$ with the $m_0$ parameter in a 
SUGRA model are shown \cite{More}.
From this figure one can see that the single neutralino productions 
do not decrease at small $m_0$ values in contrast with the 
single chargino productions (see also Fig.~\ref{fig:XScl}). 
This is due to the fact that in SUGRA scenarios the $\tilde \chi^0_1$ and 
$\tilde l_L$ ($\tilde l_L=\tilde l^{\pm}_L,\tilde \nu_L$) 
masses are never close enough to induce a significant decrease of the  
phase space factor associated to the decay
$\tilde l_L \to \tilde \chi^0_1 l$ ($l=l^{\pm},\nu$). 
By analysing Fig.~\ref{fig:XScl} and Fig.~\ref{fig:allXS}, 
one can also see that the $\tilde \chi^- \nu$ 
($\tilde \chi^0 \mu^-$) production rate is larger than the
$\tilde \chi^+ \mu^-$ ($\tilde \chi^0 \nu$) production rate. 
The explanation is that in $p \bar p$ collisions
the initial states of the resonant charged slepton production
$u_j \bar d_k, \bar u_j d_k$ have higher partonic luminosities than the 
initial states of the resonant sneutrino production
$d_j \bar d_k, \bar d_j d_k$.

The neutralino production in association with a charged lepton via
$\lambda'$ (see Fig.~\ref{fig:singllp}d) 
is an interesting case at Tevatron \cite{Dim2}.
The topology of the events consists of an isolated lepton
in one hemisphere balanced by a lepton
plus two jets in the other hemisphere, coming from the neutralino decay
via $\l'$. The \SM\ background arising
from the production of two jets plus a $Z^0$, decaying into two leptons,
has a cross-section of order $10^{-3} \nanob$
\cite{Eich}, and can be greatly reduced by excluding lepton pairs with
an invariant mass equal to the $Z^0$ mass. 
The other source of \SM\ background, which is the Drell-Yan mechanism
into $2$ leptons accompanied by 2 jets, is
suppressed by a factor, $10^{-6}/ \alpha_{\l}$. Moreover, the signal can
be enhanced by looking at the invariant mass 
of the 2 jets and the lepton in the same hemisphere, which should peak
around the neutralino mass.

The single production via $\lambda'$ of the neutralino together with a 
charged lepton can also generate clean signatures free from large 
\SM\ background, containing two like-sign charged leptons 
\cite{dreiner1,DREINER99,Rich1,Rich2,houches,More,Roy,Drein7228}. 
As a matter of fact, the neutralino has a decay channel into a lepton 
and two jets through the coupling $\l'_{ijk}$ and due to its 
Majorana nature, the neutralino decays to the charge conjugate
final states with equal probability: $\Gamma (\tilde \chi^0 \to l_i u_j
\bar d_k)=\Gamma(\tilde \chi^0 \to \bar l_i \bar u_j d_k)$. Therefore,
the lepton coming from the production can 
have the same sign than the one coming from the neutralino decay. 
Since $\l'_{111}$ has a strong indirect bound, it is
interesting to consider the \cc\ $\l'_{211}$,
which corresponds to the dimuons production with an initial state $u
\bar d$ or $\bar u d$ (see Fig.~\ref{fig:singllp}d) composed of first 
generation quarks.
The analysis of the like sign di-taus signature generated by the 
$\tilde \chi^0 \tau^{\pm}$ production through the $\l'_{311}$ coupling
(see Fig.~\ref{fig:singllp}d) suffers from a reduction of the 
selection efficiency due to the tau-lepton decay.
Besides, the study of the $\tilde \chi^0_1 \mu^{\pm}$ production via 
$\l'_{211}$ in a scenario where the $\tilde \chi^0_1$ is the LSP is 
particularly attractive since then the $\tilde \chi^0_1$ can only 
undergo \Rp\ decays.
It was found that in a SUGRA model, such a study can probe values of the 
$\l'_{211}$ coupling at the $5 \sigma$ discovery level down to 
$2 \ 10^{-3}$ ($10^{-2}$) for a muon-slepton mass of
$m_{\tilde \mu_L}=100 \GeVcc$ ($m_{\tilde \mu_L}=300 \GeVcc$) 
with $M_2=100 \GeVcc$, $2 <\tan \beta <10$ 
and $\vert \mu \vert < 10^3 \GeVcc$ at Tevatron Run~II 
assuming a luminosity of ${\cal L}=2 fb^{-1}$ \cite{DREINER99,Rich1},
and down to $2 \ 10^{-3}$ ($10^{-2}$) for 
$m_{\tilde \mu_L}=223 \GeVcc$ ($m_{\tilde \mu_L}=540 \GeVcc$) 
with $m_{1/2}=300 \GeVcc$, $A=300 \GeVcc$, $\tan \beta=2$
and $sign(\mu)>0$ at the LHC assuming a luminosity of
${\cal L}=10 fb^{-1}$ \cite{Rich2,houches}.
It was also shown in \cite{More}, by using a detector response simulation, 
that the study of the single LSP production at Tevatron Run~II
$p \bar p \to \tilde \chi^0_1 \mu^{\pm}$
would allow to probe $m_{1/2}$ values up to $\sim 850 \GeVcc$ and $m_0$ values 
up to $\sim 550 \GeVcc$ at the $5 \sigma$ discovery level, in a SUGRA scenario
where $sign(\mu)<0$, $A=0$, $\tan \beta=1.5$ 
$\l'_{211}=0.05$ and assuming a luminosity of ${\cal L}=2 fb^{-1}$.
In the case where one considers the Standard Model background combined with
the background generated by the superpartner pair production \cite{Drein7228},
the single $\tilde \chi^0_1$ production study based on the like sign
dilepton signature
analysis still allows to test large ranges of the SUGRA parameter space
at Tevatron Run~II or LHC, for $\lambda'_{211}$ values of the same order of
its present limit.

Besides, the like sign dilepton signature analysis based on the 
$\tilde \chi^0_1 \mu^{\pm}$ production (see Fig.~\ref{fig:singllp}d) allows
the $\tilde \chi^0_1$ and $\tilde \mu^{\pm}_L$ mass reconstructions since 
the decay chain $\tilde \mu^{\pm}_L \to \tilde \chi^0_1 \mu^{\pm}, 
\ \tilde \chi^0_1 \to \mu^{\pm} u d$ can be fully  
reconstructed~\cite{More,Drein7228}.
Based on the like sign dilepton signature analysis,
the $\tilde \chi^0_1$ ($\tilde \mu^{\pm}_L$) mass can be measured with a
statistical error of $\sim 11 (20) \GeVcc$ at the Tevatron Run~II~\cite{More}.

The single $\tilde \chi^{\pm}_1$ production in association with a 
charged lepton (see Fig.~\ref{fig:singllp}a) is another interesting 
reaction at hadron colliders. 
In a scenario where $\tilde \chi^0_1$ is the LSP and
\mbox{$m_{\tilde \nu},
       m_{\tilde l},m_{\tilde q}>m_{\tilde \chi^{\pm}_1}$},  
this single production receives a contribution from the 
resonant sneutrino production and the singly produced
chargino decays into quarks and leptons with branching 
ratios respectively of 
\mbox{$B(\tilde \chi^{\pm}_1 \to \tilde \chi^0_1 d_p u_{p'}) \approx 70\%$} 
($p=1,2,3;p'=1,2$) and 
\mbox{$B(\tilde \chi^{\pm}_1 \to \tilde \chi^0_1 l^{\pm}_p \nu_p) 
\approx 30\%$} due to the colour factor.
The neutralino decays via $\l'_{ijk}$ either into a lepton as, 
\mbox{$\tilde \chi_1^0 \to l_i u_j \bar d_k,\bar l_i \bar u_j d_k$},
or into a neutrino as, 
\mbox{$\tilde \chi_1^0 \to \nu_i d_j \bar d_k, \bar \nu_i \bar d_j d_k$}.
Hence, if both the $\tilde \chi^{\pm}_1$ and 
$\tilde \chi^0_1$ decay into charged leptons,
the $\tilde \chi^{\pm}_1 l_i^{\mp}$ 
production can lead to the three charged leptons
signature which has a small Standard Model background
at hadron colliders \cite{pol00,houches,More,Roy,gia1}.
The study of the three leptons signature generated by the 
$\tilde \chi^{\pm}_1 \mu^{\mp}$ production via the
$\l'_{211}$ coupling constant is particularly 
interesting for the same reasons as above.
The sensitivity to the $\l'_{211}$ coupling obtained
from this study at Tevatron Run~II would reach a 
maximum value of $\sim 0.04$ for \mbox{$m_0 \approx 200 \GeVcc$}
in a SUGRA model with $M_2=200 \GeVcc$, $sign(\mu)<0$, $A=0$ 
and $\tan \beta=1.5$, assuming
a luminosity of ${\cal L}=2 fb^{-1}$ \cite{More}.
\begin{table}
\begin{center}
\begin{tabular}{|c|c|c|c|c|c|c|c|c|}
\hline
$\l'_{211}$ & $\l'_{212}$ & $\l'_{213}$ & $\l'_{221}$ & $\l'_{222}$ 
& $\l'_{223}$ & $\l'_{231}$ & $\l'_{232}$ & $\l'_{233}$  \\
\hline
0.01 & 0.02 & 0.02 & 0.02 & 0.03 & 0.05 & 0.03 & 0.06 & 0.09 \\

\hline
\end{tabular}
\caption{{\it
Sensitivities on the $\l'_{2jk}$ coupling constants 
for $\tan\beta$=1.5, $M_1 = 100 \GeV$, $M_2 = 200 \GeV$, $\mu = -500 \GeV$,
$m_{\tilde q}=m_{\tilde l}=300 \GeVcc$ and
$m_{\tilde \nu}=400 \GeVcc$,
assuming an integrated luminosity of ${\cal L}=30 fb^{-1}$.}}
\label{tabrev}
\end{center}
\end{table}
The sensitivities on the $\l'_{2jk}$ couplings that can be obtained from 
the trilepton analysis based on the $\tilde \chi^{\pm}_1 \mu^{\mp}$ production 
at the LHC for a given set of MSSM parameters are shown in 
Table~\ref{tabrev} \cite{pol00}.
For each of the $\l'_{2jk}$ couplings the sensitivity has been obtained 
assuming that the considered coupling was the single dominant one.
The difference between the various results presented in this table 
is due to the fact that each $\l'_{2jk}$ coupling involves a specific 
initial state (see Fig.~\ref{fig:singllp}a) with its own parton density. 
Besides, all the sensitivities shown in Table~\ref{tabrev} improve 
greatly the present low-energy constraints.
The trilepton analysis based on the
$\tilde \chi^{\pm}_1 e^{\mp}$ ($\tilde \chi^{\pm}_1 \tau^{\mp}$) 
production would allow to test the $\l'_{1jk}$ ($\l'_{3jk}$) 
couplings constants. While the sensitivities obtained on the 
$\l'_{1jk}$ couplings are expected to be of the same order of 
those presented in Table~\ref{tabrev}, the sensitivities on the
$\l'_{3jk}$ couplings should be weaker due to
the tau-lepton decay.
The results presented in Table~\ref{tabrev} illustrate
the fact that even if some studies on the single superpartner 
production via $\l'$ at hadron colliders 
(see Fig.~\ref{fig:singllp}) only concern the 
$\l'_{211}$ coupling constant, the analysis of a given single 
superpartner production at Tevatron or LHC allows to probe 
many $\l'_{ijk}$ coupling constants down to values
smaller than the corresponding limits from low-energy data.

Besides, the three leptons final state study based on the
$\tilde \chi^{\pm}_1 \mu^{\mp}$ production 
(see Fig.~\ref{fig:singllp}a) allows to reconstruct
the $\tilde \chi^0_1$, $\tilde \chi^{\pm}_1$ and 
$\tilde \nu$ masses \cite{pol00,houches,More,Roy,gia1}.
Indeed, the decay chain $\tilde \nu_i \to \tilde \chi^{\pm}_1 l^{\mp}_i$,
$\tilde \chi^{\pm}_1 \to \tilde \chi^0_1 l^{\pm}_p \nu_p$,
$\tilde \chi^0_1 \to l^{\pm}_i u_j d_k$ can be fully 
reconstructed since the produced charged
leptons can be identified thanks to their flavours and signs.
Based on the trilepton signature analysis,
the $\tilde \chi^0_1$ mass can be measured with a
statistical error of $\sim 9 \GeVcc$ at the Tevatron Run~II 
\cite{Roy,More} 
and of $\sim 100 \MeVcc$ at the LHC \cite{gia1,houches,pol00}.
Furthermore,
the width of the Gaussian shape of the invariant mass
distribution associated to the $\tilde \chi^{\pm}_1$  
($\tilde \nu$) mass is of $\sim 6 \GeVcc$ ($\sim 10 \GeVcc$) 
at the LHC for the MSSM point defined by $M_1 = 75 \GeV$, 
$M_2 = 150 \GeV$, $\mu = -200 \GeV$, $m_{\tilde f} = 300 \GeVcc$ 
and $A=0$ \cite{gia1,houches,pol00}.\\
Let us make a general remark concerning the
superpartner mass reconstructions based on
the single superpartner production studies at hadron colliders.
The combinatorial background associated to these mass
reconstructions is smaller than in the mass reconstructions
analyses based on the supersymmetric particle pair production 
since in the single superpartner production studies only 
one cascade decay must be reconstructed.

At hadron colliders, some supersymmetric particles can also be singly
produced through {\it two-to-two}-body processes which generally do not 
receive contribution from resonant superpartner production \cite{More}.
Some single productions of squark
(slepton) in association with a gauge boson 
can occur through the exchange of a quark in the $t$-channel or a squark
(slepton) in the $s$-channel via $\l''$ 
($\l'$). From an initial state $g \ q$, a squark (slepton) can also be
singly produced together with a quark
(lepton) with a \cc\ $\l''$ ($\l'$) via the exchange of a quark or a
squark in the $t$-channel, and of a quark in
the $s$-channel. Finally, a gluino can be produced in association with
a lepton (quark) through a \cc\ $\l'$ ($\l''$) via the exchange of a
squark in the $t$-channel (and in the $s$-channel).

Let us enumerate the single scalar particle and gluino productions occurring 
via the {\it two-to-two}-body processes which involve the $\l'_{ijk}$ coupling 
constants~\cite{More} (one must also add the charge conjugate 
processes):
\begin{itemize}
\item The gluino production $\bar u_j d_k \to \tilde g l_i$ via the exchange
of a $\tilde u_{jL}$ ($\tilde d_{kR}$) squark in the $t$-($u$-) channel.
\item The squark production $\bar d_j g \to \tilde d_{kR}^* \nu_i$ via the
exchange of a $\tilde d_{kR}$ squark ($d_j$ quark) in the $t$-(~$s$-~) channel.
\item The squark production $\bar u_j g \to \tilde d_{kR}^* l_i$ via the
exchange of a $\tilde d_{kR}$ squark ($u_j$ quark) in the $t$-($s$-) channel.
\item The squark production $d_k g \to \tilde d_{jL} \nu_i$ via the exchange
of a $\tilde d_{jL}$ squark ($d_k$ quark) in the $t$-($s$-) channel.
\item The squark production $d_k g \to \tilde u_{jL} l_i$ via the exchange
of a $\tilde u_{jL}$ squark ($d_k$ quark) in the $t$-($s$-) channel.
\item The sneutrino production $\bar d_j d_k \to Z \tilde \nu_{iL}$ via the
exchange
of a $d_k$ or $d_j$ quark ($\tilde \nu_{iL}$ sneutrino) in the $t$-($s$-)
channel.
\item The charged slepton production $\bar u_j d_k \to Z \tilde l_{iL}$ via
the exchange
of a $d_k$ or $u_j$ quark ($\tilde l_{iL}$ slepton) in the $t$-($s$-) channel.
\item The sneutrino production $\bar u_j d_k \to W^- \tilde \nu_{iL}$
via the exchange of a $d_j$ quark ($\tilde l_{iL}$ slepton) in the
$t$-($s$-) channel.
\item The charged slepton production $\bar d_j d_k \to W^+ \tilde l_{iL}$
via the exchange of a $u_j$ quark ($\tilde \nu_{iL}$ sneutrino) in the 
$t$-($s$-) channel.
\end{itemize}
One must also add to this list the $g d_k \to t \tilde l_i$ reaction which
occurs via the $\lambda'_{i3k}$ coupling through the exchange of a $d_k$ 
quark in the $s$-channel and a top quark in the $t$-channel \cite{Kneur5443}.

Among these single productions only the $\bar u_j d_k \to W^- \tilde 
\nu_{iL}$ and  
$\bar d_j d_k \to W^+ \tilde l_{iL}$ reactions can receive
a contribution from a resonant sparticle production.
However, in most of the SUSY models,
as for example the supergravity or the gauge
mediated models, the mass difference
between the so called left-handed charged slepton and the
left-handed sneutrino
is due to the D-terms so that it is fixed by the relation
$m^2_{\tilde l^{\pm}_L}-m^2_{\tilde \nu_L}=\cos 2 \beta M_W^2$ \cite{Iban}
and thus it does not exceed the $W$ boson mass.
In scenarios with large $\tan \beta$ values, a scalar particle of the 
third generation produced as a resonance can generally decay into the 
$W$ boson due to the large mixing in the third family sfermions sector.
For instance, in the SUGRA model with
a large $\tan \beta$ a tau-sneutrino produced as a resonance 
in $d_k \bar d_j \to \tilde \nu_{\tau}$ through $\lambda'_{3jk}$
can decay as $\tilde \nu_{\tau} \to W^{\pm} \tilde \tau^{\mp}_1$,
$\tilde \tau^{\mp}_1$ being the lightest stau.

Similarly, the single scalar particle and gluino productions occurring via 
the {\it two-to-two}-body processes which involve the $\l''_{ijk}$ coupling 
constants cannot receive a contribution from a resonant scalar particle 
production for low $\tan \beta$.
Indeed, the only reactions among these {\it two-to-two}-body processes 
which can receive such a contribution are of the type
$q q \to \tilde q \to \tilde q W$.
In this type of reaction, the squark produced in the $s$-channel, is produced
via $\l''_{ijk}$ and is thus either a Right squark $\tilde q_R$, which does
not couple to the $W$ boson, or the squarks $\tilde t_{1,2}$, $\tilde b_{1,2}$.
However, the single gluino productions occuring via the {\it two-to-two}-body
processes which involve the $\lambda''_{ijk}$ coupling constants can receive 
a contribution from a resonant scalar particle production.

Therefore, the single scalar particle and gluino productions occurring via 
the {\it two-to-two}-body processes are generally non resonant single 
superpartner productions, as already mentioned at the beginning of this
section.
These non resonant single superpartner productions have typically smaller
cross-sections than the reactions receiving a contribution from a 
resonant superpartner production.
For instance, with $m_{\tilde q}=250 \GeVcc$,
the cross-section $\sigma(p \bar p \to \tilde u_L \mu)$
is of order $\sim 10^{-3}pb$ at a centre-of-mass
energy of $\sqrt{s}=2 \TeV$, assuming an \Rp\ coupling of
$\l'_{211}=0.09$ \cite{More}.
However, the non resonant single productions
can lead to interesting signatures. For instance, the production, 
$q \bar q \to \tilde f W$ leads to the final
state $2l+2j+W$ for a non vanishing \Rp\ \cc\ $\l'$ and to 
the signature $4 j +W$ for a $\l''$ \cite{dreiner1}. 
Furthermore, the non resonant single productions 
are interesting as their cross-section
involves only few SUSY parameters, namely one or two scalar
superpartner(s) mass(es) and one \Rp\ coupling constant. 

The $D\emptyset$ collaboration searched for single slepton production
through the $\lambda^{\prime}_{211}$ coupling in the two muons and two
hadron jets channel~\cite{NEWD0RPV3}. In the absence of any evidence
for an excess of events with respect to expectation from
Standard Model processes, bounds on \SUGRA\ parameters $m_{1/2}, m_{o}$ 
have been set. 
Sneutrinos and smuons masses up to $280 \GeVcc$ have been excluded.

$\bullet$ {\bf Single sparticle production via {\boldmath{$\l''$}} }

The $B$-violating couplings $\lambda''_{ijk}$ allows 
for resonant production of squarks at hadron colliders.
Either a squark $\tilde u_i$ or $\tilde d_k$ can be produced at the 
resonance from an initial state, $\bar d_j \bar d_k$ or 
$\bar u_i \bar d_j$, respectively. For $m_{\tilde d^k_R} = 100 \GeVcc$, 
$\sqrt s =2 \TeV$ and $\l''_{11k}=1$, the rate of the down squark 
production at the Tevatron is 
$\s(p \bar p \to \tilde d^k_R)=25 \nanob$~\cite{Dim2}.
%
\begin{figure}[htb]
\begin{center}
\vspace*{1.0cm}

\epsfig{file=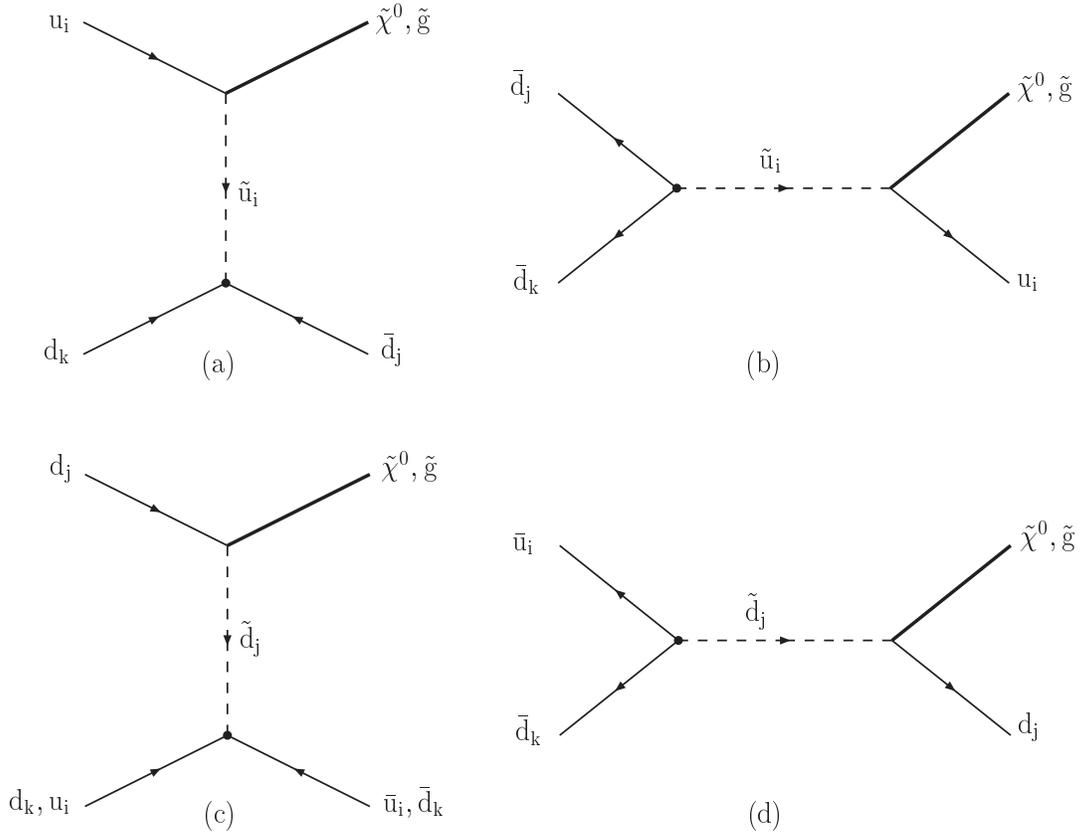,width=0.90\textwidth}


 \caption{{\it Diagrams for the single neutralino production
               reactions involving $\l''_{ijk}$ at hadron colliders.
               The $\l''_{ijk}$ coupling constant is symbolised by a small 
 	       circle and the arrows indicate the flow of the baryon number.}}
 \label{fig:singllpp1}
 \end{center}
 \end{figure}
\begin{figure}[htb]
\begin{center}
\vspace*{1.0cm}

\epsfig{file=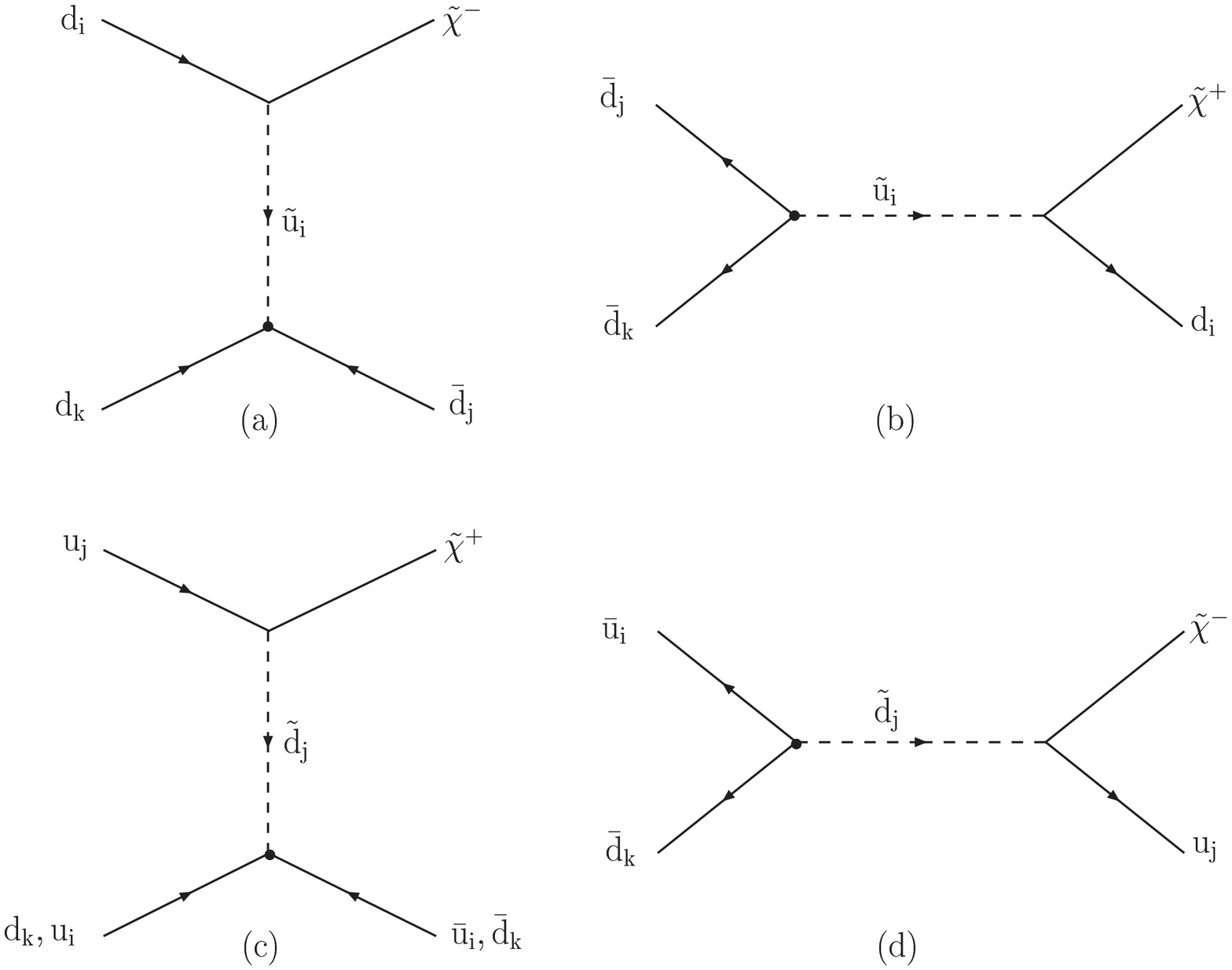,width=0.90\textwidth}

 \caption{{\it Diagrams for the single chargino production
               reactions involving $\l''_{ijk}$ at hadron colliders.
               The $\l''_{ijk}$ coupling constant is symbolised by a small 
 	       circle and the arrows indicate the flow of the baryon number.}}
 \label{fig:singllpp3}
 \end{center}
 \end{figure}
\begin{figure}[htb]
\begin{center}
\vspace*{1.0cm}

\epsfig{file=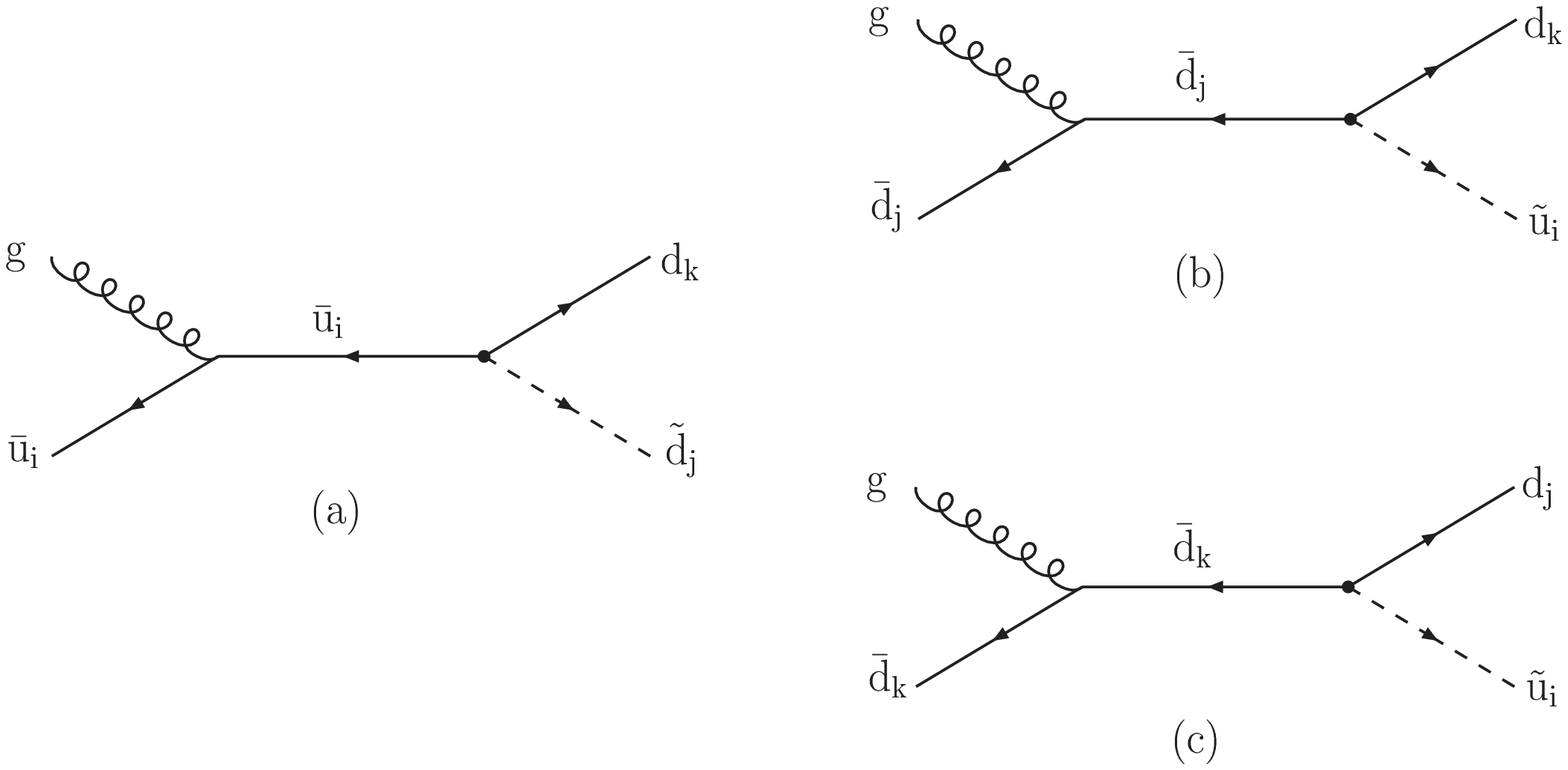,width=0.90\textwidth}

 \caption{{\it Diagrams for the resonant production of squarks
               involving $\l''_{ijk}$ at hadron colliders.
               The $\l''_{ijk}$ coupling constant is symbolised by a small 
 	       circle and the arrows indicate the flow of the baryon number.}}
 \label{fig:singllpp4}
 \end{center}
 \end{figure}
\begin{figure}[htb]
\begin{center}
\vspace*{1.0cm}

\epsfig{file=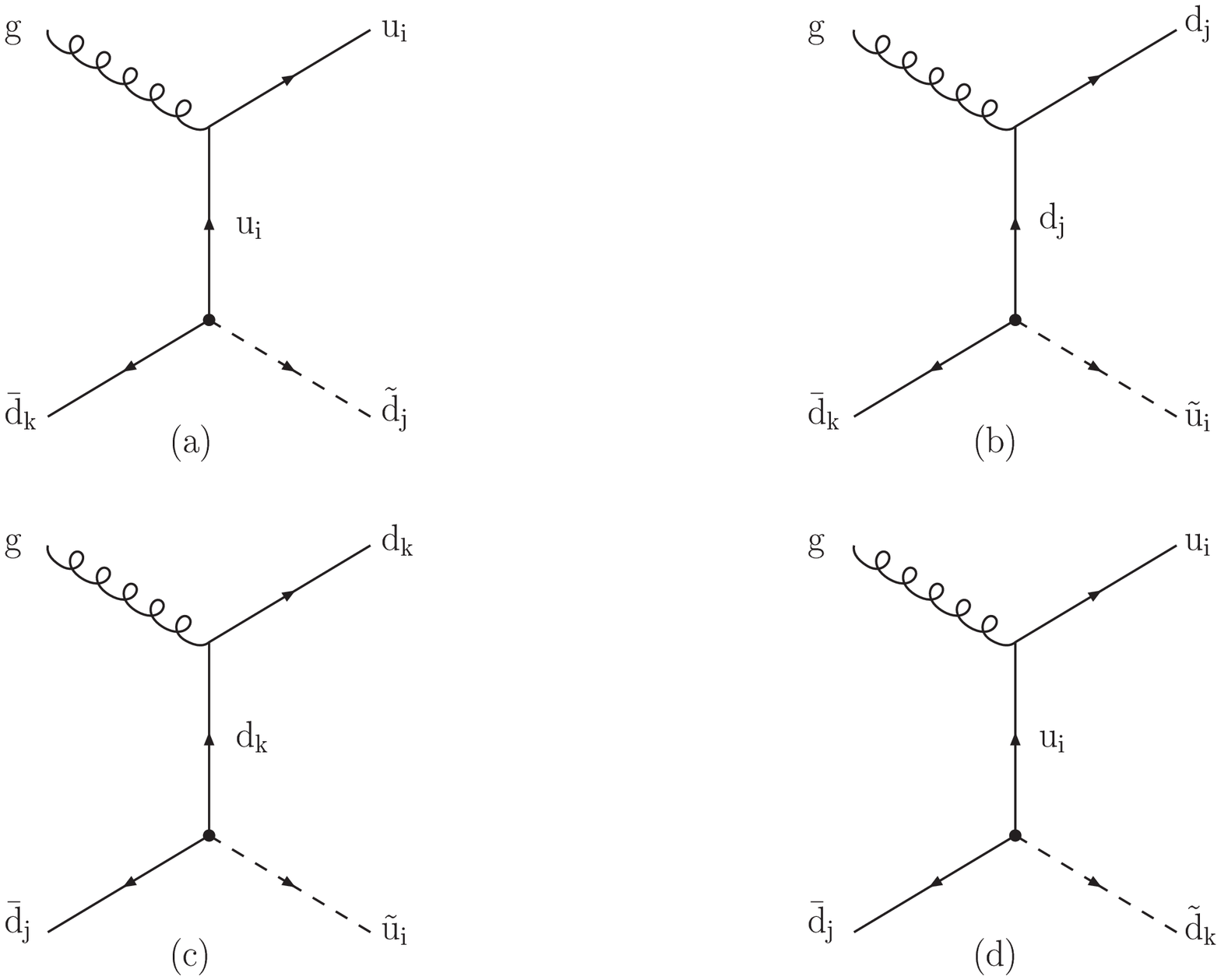,width=0.90\textwidth}

 \caption{{\it Diagrams for the non-resonant production of squarks
               involving $\l''_{ijk}$ at hadron colliders.
               The $\l''_{ijk}$ coupling constant is symbolised by a small 
 	       circle and the arrows ndicate the flow of the baryon number.}}
 \label{fig:singllpp5}
 \end{center}
 \end{figure}
\begin{figure}[htb]
\begin{center}
\vspace*{1.0cm}

\epsfig{file=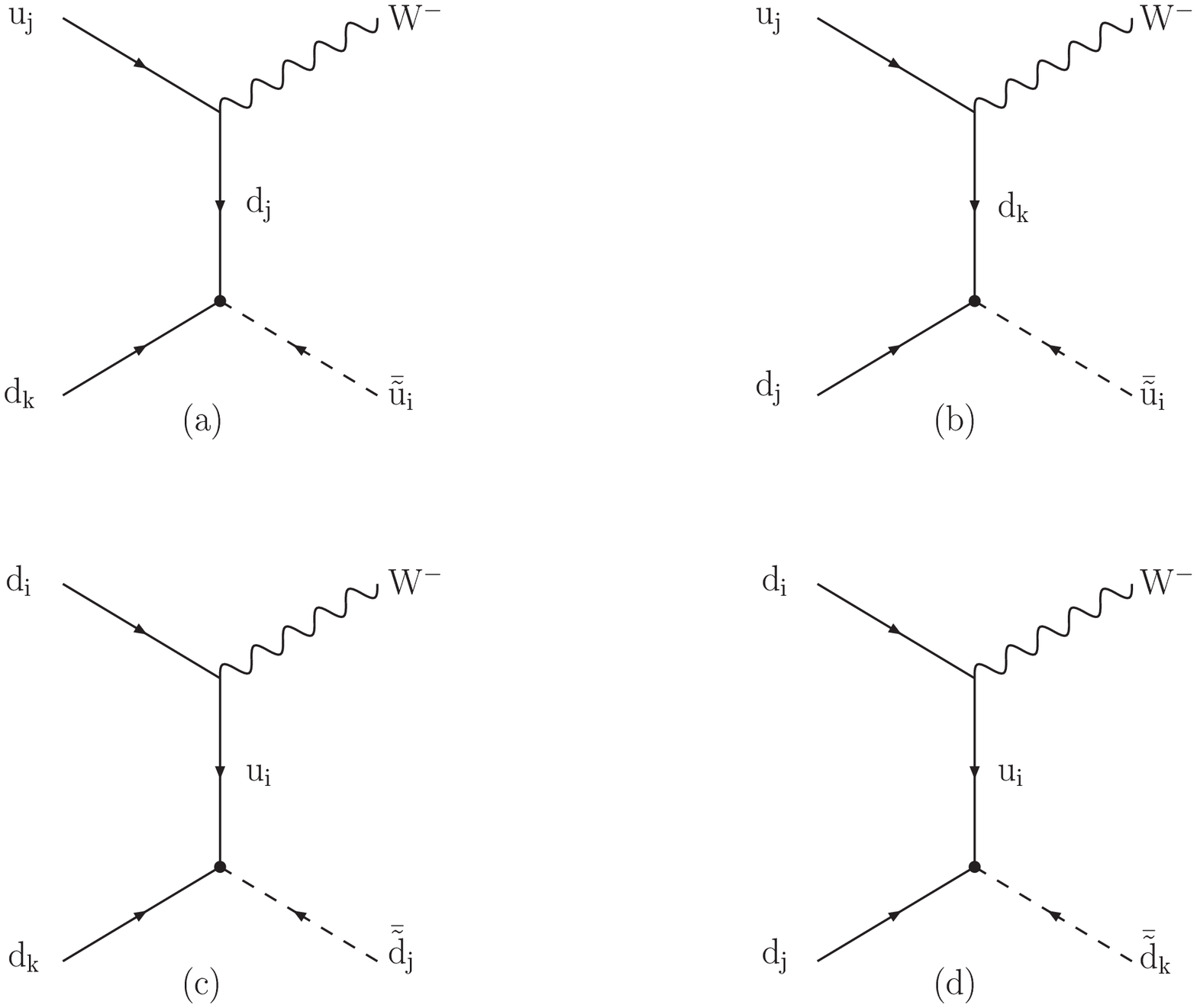,width=0.90\textwidth}

 \caption{{\it Diagrams for the associated $\tilde{q}-W$ production 
               involving $\l''_{ijk}$ at hadron colliders.
               The $\l''_{ijk}$ coupling constant is symbolised by a small 
 	       circle and the arrows indicate the flow of the baryon number.}}
 \label{fig:singllpp6}
 \end{center}
 \end{figure}
%
%
\begin{figure}[htb]
\begin{center}
\vspace*{1.0cm}

\epsfig{file=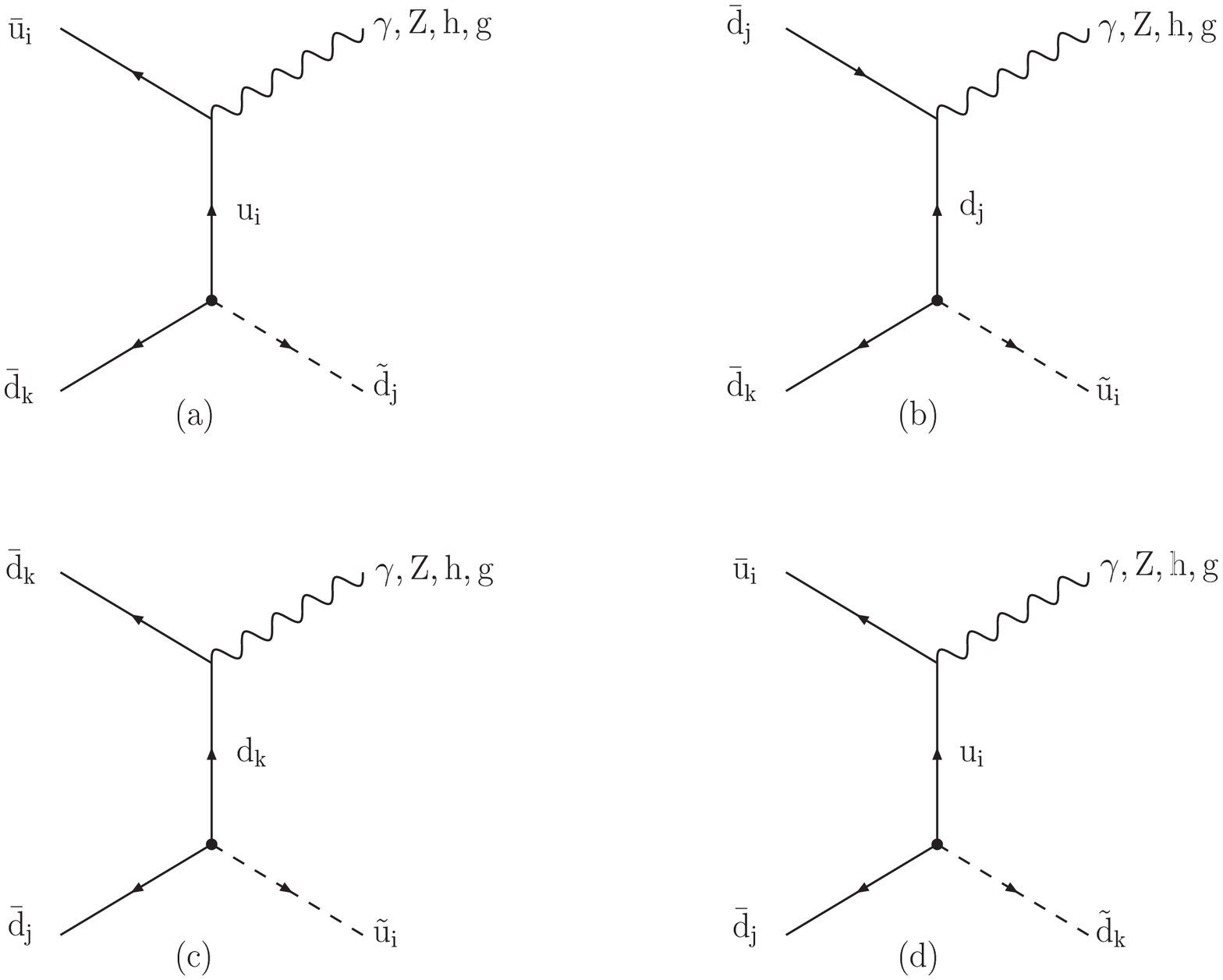,width=0.90\textwidth}

 \caption{{\it Diagrams for the associated $\tilde{q}-\gamma$, 
               ($-Z$, $-h$ and $-g$) production involving $\l''_{ijk}$ at hadron
	       colliders.
               The $\l''_{ijk}$ coupling constant is symbolised by a small 
 	       circle and the arrows indicate the flow of the baryon number.}}
 \label{fig:singllpp7}
 \end{center}
 \end{figure}
%

For $m_{\tilde t_1}=600 \GeVcc$, $\sqrt s =2 \TeV$
and $\l''_{323}=0.1$, the
rate of the resonant stop production is 
$\s(p \bar p \to \tilde t_1)=10^{-3}$ picobarns \cite{Berg}. Note
that this rate is higher than the stop pair
production rate at the same centre-of-mass energy and for the same stop mass, which 
is of order $\s(p \bar p \to \tilde t_1 \tilde t_1)=10^{-6}$ picobarns.

The single superpartner production can also occur as a {\it two-to-two}-body process,
through an \Rp\ coupling $\l''$ and  an ordinary gauge interaction vertex:
In baryon-number-violating models, any gaugino (including gluino) can be 
produced in association with a quark, in quark-quark scattering, by the
exchange of a squark in the $s$-, $t$- or $u$-channel. 

For example, let us consider the photino and gluino production
\cite{Dim2}: The rate values in the 
$t$- and $u$-channel are, 
$\s (p \bar p \to \tilde{\gamma} q)=2 \ 10^{-2} \nanob$, and,
$\s (p \bar p \to \tilde g q)=3 \ 10^{-1} \nanob$, for, $m_{\tilde
q}=m_{\tilde g}=m_{\tilde{\gamma} }=100 \GeVcc$,
$\sqrt s =2 \TeV$ and $\l''_{111}=1$. The photino or gluino which is
produced will then decay into three 
jets via the $\l''$ coupling, resulting in a four jets final state. The
corresponding QCD background is strong:
It is estimated to be about $10 \nanob$ for $\sqrt s =2 \TeV$ \cite{Arg}. Of
course, the ratio signal over background can
be enhanced considerably by looking at the mass distribution of the
jets: the QCD 4 jets are produced relatively 
uncorrelated, while the trijet mass distribution of the signal should
peak around the gaugino mass. However, 
one of the three jets may be too soft to be measured or
jet coalescence may occur,
especially for small values of the gaugino mass. The study of this
example bring us to the conclusion that, due to high QCD background, 
the analysis of the single production via $\l''$ remains difficult.

Nevertheless, there are some specific cases where the final state can be 
clear and free from a large background. For instance, 
a $\tilde \chi^+_1$ chargino can be produced via $\l''_{3jk}$ through the 
resonant production of a top squark as $\bar d_j \bar d_k \to \tilde t_1 
\to b \tilde \chi^+_1$, $\tilde t_1$ being the lightest top squark, and then 
decay into the lightest neutralino plus leptons
as $\tilde \chi^+_1 \to \bar l_i \nu_i \tilde \chi^0_1$ \cite{Berg,DREINER99}.
Due to the stop resonance, this reaction can reach high rate values.
The cascade decay demands the mass hierarchy, $m_{\tilde t_1} > 
m_{\tilde \chi^+_1} > m_{\tilde \chi^0_1}$, to be respected,
and by consequence is not allowed in all regions of the \SUGRA\ parameter
space. Assuming $\l''_{3jk}$ to be the single dominant \Rp\ coupling constant
and the $\tilde \chi^0_1=LSP$ to be lighter than the top quark, 
the $\tilde \chi^0_1$ should then be treated as a stable particle. 
Then, the signal for our process would be
very clear since it would consist of a tagged b-quark jet, a lepton and 
missing transverse energy. The \SM\ background for such a signature comes
from the single top quark production, via $W \ g$ fusion, and the production 
of a $W$ gauge boson in association with $b \bar b$, $c \bar c$ or a 
jet faking a b-quark jet. 
Experimental studies lead to the conclusion that values of $\l'' > 0.03-0.2$ 
and $\l''>0.01-0.03$ can be excluded at the $95 \%$ confidence level for, 
$180 \GeVcc < m_{\tilde t_1} < 285 \GeVcc$, at the Tevatron Run~I 
($\sqrt s=1.8 \TeV$ and $\int {\cal L} dt = 110 \picob^{-1}$) and for, 
$180 \GeVcc < m_{\tilde t_1} < 325 \GeVcc$, at the Run II of the
Tevatron ($\sqrt s=2 \TeV$ and $\int {\cal L} dt = 2 \femtob^{-1}$), 
respectively.
This result is based on the leading-order
CTEQ-4L parton distribution functions \cite{CTEQ4L} and holds for the
normalisation,
$\l''=\l''_{312}=\l''_{313}=\l''_{323}$, and for the point of a minimal
\SUGRA\ model, $m_{1/2}=150 \GeV, \ A_0=-300 \GeV$ and $\tan \beta =4$. 
The constraints obtained on $\l''$ are stronger than the present low energy
bounds.

Another particularly interesting reaction has been studied in \cite{Yu7220}:
the single gluino production $\bar d_j \bar d_k \to t \tilde g$ which
can receive a contribution from the resonant stop production via the
$\lambda''_{3jk}$ coupling. In certain regions of the mSUGRA parameter
space, this single gluino production can reach rates at LHC of order
$10^{2}fb$ (for $\lambda''_{3jk}=10^{-1}$) thanks to the contribution
coming from the resonant $\tilde t_2$ production, $\tilde t_2$ being
the heavier top squark. The interesting point is that in these mSUGRA
domains the branching ratios of the decays $\tilde g \to t b \tilde
\chi^{\pm}_1$
and $\tilde g \to t \bar t \tilde \chi^0_1$ reach also significant values
thanks to the exchange of the virtual $\tilde t_1$ (the lighter top squark)
which is the lighter squark and has a mixing angle near $\pi / 2$.
By consequence, the process $pp \to t \tilde g$ ($\bar t \tilde g$) can
simultaneously have large cross section values at LHC and produce in a
significant
way a clear signature containing 3 $b$ quarks, at least 2 charged leptons
and some missing energy (due to the top quark decay $t \to b l \nu$).
Since the background associated to this final state can be greatly reduced
thanks to the large $b$-tagging efficiency available at the LHC ($\sim 50
\%$), the study of the reaction $pp \to t \tilde g$ ($\bar t \tilde g$) 
should provide an effective test of the $\lambda''_{3jk}$ coupling constant.
\index{Sparticle production!Single|)}
%

\section{Virtual Effects involving {\boldmath{$\Rp$}} Couplings} 
\label{sec:virtuprod}

In a scenario where none of the \susyq\ particles can be directly produced 
at colliders with a significant cross-section, because of very high 
masses or unfavorable couplings with the \SM\ particles, the effects 
induced by \Rp\ could turn out to be felt only in indirect processes 
involving virtual sparticle exchange.

In contrast to single sparticle production for which a \Rp\ coupling
only enter at one vertex when calculating total production rates,
\Rp\ contributions (via additional sparticle exchange) to \SM\ processes 
are suppressed in proportion to the square of the Yukawa coupling.
These processes generally imply high statistics inclusive measurements
as in the case of fermion pair production and effective four-fermion
contact interactions discussed in section~\ref{sec:fourf}.

\subsection{Fermion Pair Production} 
\label{sec:fourf}

For sparticle masses far above the kinematical reach of a given collider,
\Rp\ interactions could manifest themselves through effective four-fermion
contact interactions interfering with \SM\ fermion pair production
processes.  


%
%

At leptonic colliders dilepton production can occur in the presence of a unique
(or largely dominant) \Rp\ coupling.
The resonant sneutrino $\tilde \nu_{\mu}$ or $\tilde \nu_{\tau}$ production 
via $\l_{121}$ or $\l_{131}$ respectively followed by a decay through the 
same coupling constant (i.e. $\tilde \nu^i \to \bar l_j l_k$ via $\l_{ijk}$)
would lead to a spectacular signature such as an excess of events 
Bhabha scattering events~\cite{dimopoulos88}.
For example the cross-section of Bhabha scattering
including the $\tilde \nu_{(\mu,\tau)} $ sneutrino $s$-channel exchange and 
the interference terms reaches $3 \picob$ at 
$\sqrt s=m_{\tilde \nu_{(\mu,\tau)}}=200 \GeV$~\cite{Kalept,Kalseul,KalZ} 
for $\Gamma_{\tilde \nu_{(\mu,\tau)}}=1 \GeV$ and 
$\l_{1(2,3)1}=0.1$.

Table~\ref{tab:mucoltwoandfourfermion1} shows the accessible $\l$ couplings at  
$e^+e^-$ and ${\mu}^+{\mu}^-$ colliders 
and fermion pair production to which a single dominant $\l$ coupling can contribute.
Except few exceptions, $e^+e^-$ and ${\mu}^+{\mu}^-$ colliders allow to access the
same $\l$ couplings. The difference in center-of mass energies and luminosities between these two
types of leptonic
colliders will determine the explorable domain of these couplings.

\begin{table}[htb]
\begin{center}
\begin{tabular}{|c||c|c|c||c|c|c||} \hline
 & \multicolumn{3}{c||}{$e^+e^-$ colliders}
 & \multicolumn{3}{c||}{${\mu}^+{\mu}^-$ colliders}         \\ \hline \hline
coupling & final state & exchange & channel & final state & exchange & channel \\ \hline
$\lambda_{121}$ &  $e^+e^-$        & ${\tilde {\nu}}_{\mu}$    & s   & $e^+e^-$        & ${\tilde {\nu}}_{e}$      & t   \\
                &  $\mu^+ \mu^-$   & ${\tilde {\nu}}_{e}$      & t   &    -            &      -                    &  -  \\ \hline
$\lambda_{122}$ &  $\mu^+ \mu^-$   & ${\tilde {\nu}}_{\mu}$    & t   & $e^+e^-$        & ${\tilde {\nu}}_{\mu}$    & t   \\
                &       -          &          -                &  -  & $ \mu^+ \mu^-$  & ${\tilde {\nu}}_{e}$      & s+t \\ \hline
$\lambda_{123}$ &  $\tau^+ \tau^-$ & ${\tilde {\nu}}_{\mu}$    & t   & $\tau^+ \tau^-$ & ${\tilde {\nu}}_{e}$      & t   \\ \hline
$\lambda_{131}$ &  $\tau^+ \tau^-$ & ${\tilde {\nu}}_{e}$      & t   &       -         &            -              &   - \\ 
                &  $e^+e^-$        & ${\tilde {\nu}}_{\tau}$   & s   &       -         &            -              &   - \\ \hline
$\lambda_{132}$ &  $\mu^+ \mu^-$   & ${\tilde {\nu}}_{\tau}$   & t   & $e^+e^-$        & ${\tilde {\nu}}_{\tau}$   & t   \\
                &       -          &          -                & -   & $\tau^+ \tau^-$ & ${\tilde {\nu}}_{e}$      & t   \\ \hline
$\lambda_{133}$ &  $\tau^+ \tau^-$ & ${\tilde {\nu}}_{\tau}$   & t   &     -           &          -                & -   \\ \hline
$\lambda_{231}$ &  $\tau^+ \tau^-$ & ${\tilde {\nu}}_{\mu}$    & t   & $e^+e^-$        & ${\tilde {\nu}}_{\tau}$   & t   \\
                &  $\mu^+ \mu^-$   & ${\tilde {\nu}}_{\tau}$   & t   &      -          &          -                &   - \\ \hline
$\lambda_{232}$ &   -              & -                         & -   & $\mu^+ \mu^-$   & ${\tilde {\nu}}_{\tau}$   & s+t \\
                &         -        &             -             & -   & $\tau^+ \tau^-$ & ${\tilde {\nu}}_{\mu}$    & t   \\ \hline
$\lambda_{233}$ &  -               & -                         & -   & $\tau^+ \tau^-$ & ${\tilde {\nu}}_{\tau}$   & t   \\ \hline \hline
\end{tabular}
\caption{{\it {Accessible $\l$ couplings at  $e^+e^-$ and ${\mu}^+{\mu}^-$ colliders 
and fermion pair production to which a single dominant coupling can contribute.}}}
\label{tab:mucoltwoandfourfermion1}
\end{center}
\end{table}

The observation of an excess of high $Q^2$ events at HERA 
experiments~\cite{highq1,highq2} and its interpretation in terms
of \Rp\ interactions has been followed by 
numerous discussions on dilepton production at \index{LEP}  
LEP~\cite{Kalept,Kalseul,KalZ,Chou,DELPHI1,OPALsm} which are beyond the scope of 
this review.

Di-jets production can also occur at leptonic colliders in the presence of
a unique $\l'$ coupling through the 
exchange of a squark in the $t$-channel. The $b \bar b$ and $ c \bar c$
production via $\l'_{1k3}$ and $\l'_{12k}$ respectively are of particular
interest due to the possibility of tagging bottom, charm or light quarks (u,d,s)
at the experiment level~\cite{Richa}.
Table~\ref{tab:mucoltwoandfourfermion2} shows the accessible $\l'$ couplings at
$e^+e^-$ and ${\mu}^+{\mu}^-$ colliders 
and fermion pair production to which a single dominant $\l'$ coupling can contribute.
In contrast to the case of $\l$ couplings, $e^+e^-$ and ${\mu}^+{\mu}^-$ colliders
access completely different sets of $\l'$ couplings.
More specifically, in the case of 
$\l'$ couplings, fermion pair production 
allows to explore only $\lambda^{\prime}_{1jk}$ at $e^+e^-$ colliders and 
$\lambda^{\prime}_{2jk}$ at ${\mu}^+{\mu}^-$ colliders.
\par
Preliminary studies have been performed in~\cite{feng98} focusing on the study of 
${\mu}^+{\mu}^- \rightarrow {\mu}^+{\mu}^-$ via ${\tilde {\nu}}_{\tau}$ involving
the $\lambda_{232}$ coupling and ${\mu}^+{\mu}^- \rightarrow b \bar b$
via ${\tilde {\nu}}_{\tau}$
involving the product $\lambda_{232} \lambda^{\prime}_{333}$.
In this case it has been found that once the mass of the ${\tilde {\nu}}_{\tau}$ is
known from earlier stage of a $e^+e^-$ collider or the ${\mu}^+{\mu}^-$ collider and
once fixing the center-of-mass energy at ${\tilde {\nu}}_{\tau}$ resonance or
around the resonance with the ${\mu}^+{\mu}^-$ collider, 
one can explore $\lambda_{232}$ down to $10^{-4}$ with an integrated luminosity
of $3$~fb$^{-1}$ and a beam energy resolution of $0.1$~\%.
\par
Further preliminary studies have been performed in~\cite{choudmuon}.

\begin{table}[htb]
\begin{center}
\begin{tabular}{|c||c|c|c||c|c|c||} \hline
 & \multicolumn{3}{c||}{$e^+e^-$ colliders}
 & \multicolumn{3}{c||}{${\mu}^+{\mu}^-$ colliders}         \\ \hline \hline
coupling & final state & exchange & channel & final state & exchange & channel \\ \hline
$\lambda^{\prime}_{111}$  &$d\bar d$ & ${\tilde u}_L$ & t & -         & -              & -   \\
$\lambda^{\prime}_{111}$  &$u\bar u$ & ${\tilde d}_R$ & t & -         & -              & -   \\ \hline
$\lambda^{\prime}_{112}$  &$s\bar s$ & ${\tilde u}_L$ & t & -         & -              & -   \\
$\lambda^{\prime}_{112}$  &$u\bar u$ & ${\tilde s}_R$ & t & -         & -              & -   \\ \hline
$\lambda^{\prime}_{113}$  &$b\bar b$ & ${\tilde u}_L$ & t & -         & -              & -   \\
$\lambda^{\prime}_{113}$  &$u\bar y$ & ${\tilde b}_R$ & t & -         & -              & -   \\ \hline
$\lambda^{\prime}_{121}$  &$d\bar d$ & ${\tilde c}_L$ & t & -         & -              & -   \\
$\lambda^{\prime}_{121}$  &$c\bar c$ & ${\tilde d}_R$ & t & -         & -              & -   \\ \hline
$\lambda^{\prime}_{122}$  &$s\bar s$ & ${\tilde c}_L$ & t & -         & -              & -   \\
$\lambda^{\prime}_{122}$  &$c\bar c$ & ${\tilde s}_R$ & t & -         & -              & -   \\ \hline
$\lambda^{\prime}_{123}$  &$b\bar b$ & ${\tilde c}_L$ & t & -         & -              & -   \\
$\lambda^{\prime}_{123}$  &$c\bar c$ & ${\tilde b}_R$ & t & -         & -              & -   \\ \hline
$\lambda^{\prime}_{131}$  &$d\bar d$ & ${\tilde t}_L$ & t & -         & -              & -   \\
$\lambda^{\prime}_{131}$  &$t\bar t$ & ${\tilde d}_R$ & t & -         & -              & -   \\ \hline
$\lambda^{\prime}_{132}$  &$s\bar s$ & ${\tilde t}_L$ & t & -         & -              & -   \\
$\lambda^{\prime}_{132}$  &$t\bar t$ & ${\tilde s}_R$ & t & -         & -              & -   \\ \hline
$\lambda^{\prime}_{133}$  &$b\bar b$ & ${\tilde t}_L$ & t & -         & -              & -   \\
$\lambda^{\prime}_{133}$  &$t\bar t$ & ${\tilde b}_R$ & t & -         & -              & -   \\ \hline \hline
$\lambda^{\prime}_{211}$  & -        & -              & - & $d\bar d$ & ${\tilde u}_L$ & t   \\
$\lambda^{\prime}_{211}$  & -        & -              & - & $u\bar u$ & ${\tilde d}_R$ & t   \\ \hline
$\lambda^{\prime}_{212}$  & -        & -              & - & $s\bar s$ & ${\tilde u}_L$ & t   \\
$\lambda^{\prime}_{212}$  & -        & -              & - & $u\bar u$ & ${\tilde s}_R$ & t   \\ \hline
$\lambda^{\prime}_{213}$  & -        & -              & - & $b\bar b$ & ${\tilde u}_L$ & t   \\
$\lambda^{\prime}_{213}$  & -        & -              & - & $u\bar y$ & ${\tilde b}_R$ & t   \\ \hline
$\lambda^{\prime}_{221}$  & -        & -              & - & $d\bar d$ & ${\tilde c}_L$ & t   \\
$\lambda^{\prime}_{221}$  & -        & -              & - & $c\bar c$ & ${\tilde d}_R$ & t   \\ \hline
$\lambda^{\prime}_{222}$  & -        & -              & - & $s\bar s$ & ${\tilde c}_L$ & t   \\
$\lambda^{\prime}_{222}$  & -        & -              & - & $c\bar c$ & ${\tilde s}_R$ & t   \\ \hline
$\lambda^{\prime}_{223}$  & -        & -              & - & $b\bar b$ & ${\tilde c}_L$ & t   \\
$\lambda^{\prime}_{223}$  & -        & -              & - & $c\bar c$ & ${\tilde b}_R$ & t   \\ \hline
$\lambda^{\prime}_{231}$  & -        & -              & - & $d\bar d$ & ${\tilde t}_L$ & t   \\
$\lambda^{\prime}_{231}$  & -        & -              & - & $t\bar t$ & ${\tilde d}_R$ & t   \\ \hline
$\lambda^{\prime}_{232}$  & -        & -              & - & $s\bar s$ & ${\tilde t}_L$ & t   \\
$\lambda^{\prime}_{232}$  & -        & -              & - & $t\bar t$ & ${\tilde s}_R$ & t   \\ \hline
$\lambda^{\prime}_{233}$  & -        & -              & - & $b\bar b$ & ${\tilde t}_L$ & t   \\
$\lambda^{\prime}_{233}$  & -        & -              & - & $t\bar t$ & ${\tilde b}_R$ & t   \\ \hline \hline
$\lambda^{\prime}_{3jk}$  &-         & -              & - & -&-&-\\ \hline \hline
\end{tabular}
\caption{{\it {Accessible $\l'$ couplings at $e^+e^-$ and ${\mu}^+{\mu}^-$ colliders 
and fermion pair production processes to which a single dominant coupling can contribute.}}}
\label{tab:mucoltwoandfourfermion2}
\end{center}
\end{table}

%
At hadron colliders \Rp\ reactions can induce contributions to 
Standard Model di-jets or di-leptons production processes. 
First, the jets pair production receives contributions from reactions 
involving either $\l'$ or $\l''$ coupling constants. 
As a matter of fact, a pair of quarks can be produced through the $\l''$ 
couplings with an initial state $u d$ or $\bar u \bar d$ 
($d d$ or $\bar d \bar d$) by the exchange of a $\tilde d$ ($\tilde u$) 
squark in the $s$-channel, and also with an initial state
$u \bar u$ or $d \bar d$ ($u \bar d$ or $\bar u d$) by the exchange of a 
$\tilde u$ or $\tilde d$ ($\tilde d$) squark in the $t$-channel. 
If the $s$-channel exchanged particle is produced on
shell the resonant diagram is of course dominant with respect 
to the $t$-channel diagram. 
The dijet channel can also be generated via the $\l'$ couplings from an 
initial state $u \bar d, \bar u d$ or $d \bar d$ through the exchange
of a $\tilde l$ or $\tilde \nu$ slepton (respectively) in the $s$-channel. 

If the dominant mechanism for either the slepton or the squark decay
leads to two jets, the resonant production of such a scalar particle 
would result in a bump in the two-jet invariant mass
distribution~\cite{dimopoulos88,Dim2} which would be a very clean 
signature. However the dijet production through \Rp\ \ccs\ will be hard to 
study at LHC unless the narrow resonances are copiously produced given
the severe expected QCD background~\cite{dreiner1,Bin}. This was discussed
in more details above in section~\ref{sec:sproduct}.

Top quark pair production appears to be a particular case of fermion pair production at hadron
colliders because if kinematically allowed new decay channels such as $t \to d \tilde d_R$ and 
$t \to d \tilde l_L$
can open up.
The amplitudes for top quark pair production involve diagrams with an initial state
$d_k \bar d_k$ with either a
$\tilde l^i_L$ slepton exchange (via $\l'_{i3k}$) or a $\tilde d^i_R$ squark exchange 
The \susyq\ parameter space region allowed at a $ 95 \% $ confidence level by
the D0 and CDF data \cite{data1} on $ t \bar t $ production cross-section have 
also been obtained in \cite{Sri} and are shown in Fig.~\ref{topair:fig1} in the
plane $ \l'_{i31}/m_{\tilde l^i_L} $ and in Fig.~\ref{topair:fig2} in the plane
$  \l''_{31i}/m_{\tilde d^i_R} $.
Furthermore, \Rp\ interactions being chiral, one expects the two top quarks
to be polarized thus providing an additional handle to probe the details
of \Rp\ couplings~\cite{hikasa99} since the polarization of the top quark pairs
is very small in the Standard Model.

\begin{figure}
\begin{center}
\leavevmode
\centerline{\psfig{figure=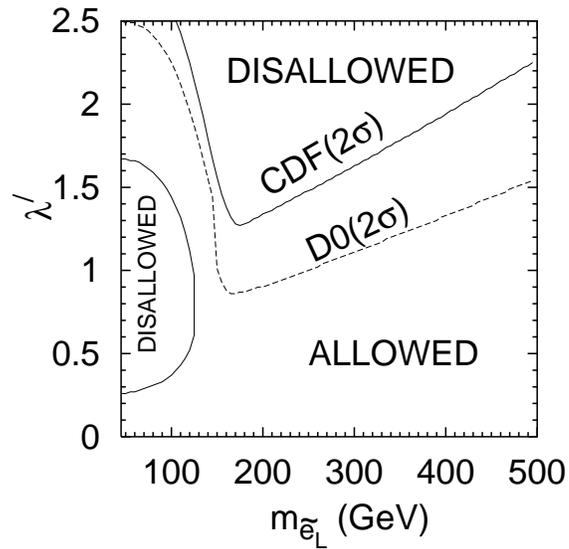, height=3in}}
\end{center}
\caption{{\it
Allowed regions in the plane of $\lambda'_{i3k}$ and the mass of the
left slepton in a lepton-number-violating scenario. Solid 
(dashed) lines correspond to the 2-$\sigma$ bounds from the 
CDF (D0) collaborations.}}
\label{topair:fig1}
\end{figure}

\begin{figure}
\begin{center}
\leavevmode
\centerline{\psfig{figure=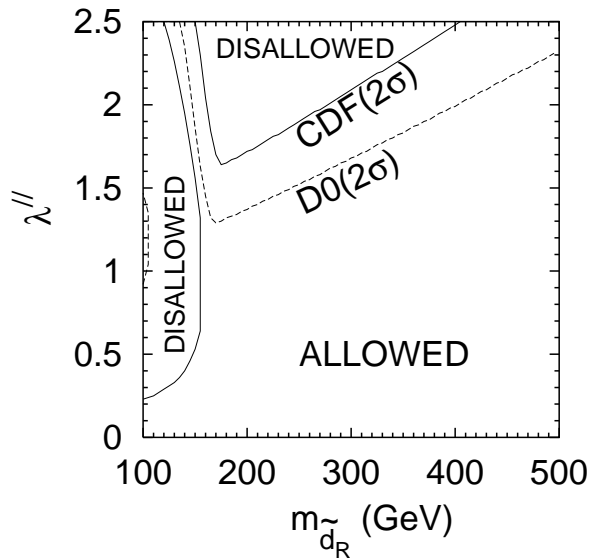, height=3in}}
\end{center}
\caption{{\it
Allowed regions in the plane of $\lambda''_{3ki}$ and the mass
of the right $d$-squark in a baryon-number-violating scenario. Solid 
(dashed) lines correspond to the 2-$\sigma$ bounds from the 
CDF (D0) collaborations.}}
\label{topair:fig2}
\end{figure}

More complicated decay chains of the top quark such as 
the double cascade decays $ t \to
\tilde l^+_i d_k ,\ \tilde l^+_i \to \tchi^0 +e_i ,\
\tchi^0 \to \nu _ i b \bar d_k + \bar \nu _ i \bar b d_k $ where the
top quark and neutralino \Rp\ decay  processes are both controlled by the coupling
constants $\l '_{i3k}$  can lead to two  potentially observable effects
in the leptonic events namely a deviation from lepton universality and
(for $ k=3$) an excess of $ b$ quark hadron events
A study
based on the comparison of the ratio of branching fractions for single
$e$ to single $\mu $ events
$ B(t\bar t \to (e +jets) / B(t\bar t \to \mu +jets) $ to the experimental 
ratio of events $ N(e+jets)/ N(\mu +jets) = 1 {+a\choose -b} $ 
from the one charged lepton and two b-quark jets final state of the CDF top quark sample
of Tevatron Run I gives the
bound~\cite{agashe} $\l'_{13n}< 0.41, \ [n=1,2] $.  

Another method of
analysis based on an identification of this ratio with the ratio of
the experimental to theoretical  total production cross sections yields~$\l'_{13n}<0.48 $.
An analysis of the hadron $b$ quarks events
yields~\cite{agashe}~$\l'_{133} < 0.41$.

Alternatively~\cite{fp2} the top quark \Rp\ decay channel 
$t\to b \tilde \tau ^+$ initiated by the $\l '_{333}$ coupling 
leads to signature which can not be confused
with the Standard Model decay channel and can compete with it. This
induces a reduction of the observed Standard Model  
$t\bar t$ event rates. The correction factor reads:
\begin{eqnarray}
 R_B \simeq  1.12 \ \l ^{'2} _{333} (1-{m_{\tilde
\tau _L} ^2 \over m_t ^2 } )^{-2} 
\label{eq:barger}
\end{eqnarray}
Similarly the hadron two-body decay channels $ t \to \bar d_j +\bar
{\tilde d} _{kR}$ with $\l ''$ couplings have an
impact on the $ t\bar t$ events through a modification in the fraction
of hadron top quark decays.
Performing a similar analysis to the one above for the \Rp\ decay modes
$ t\to \bar b \tilde {\bar s} $ initiated by the $\l ''_{323}$
coupling where the $ t \bar t$ pairs cascade  down 
to a 5 jets final state leads to an induced reduction
factor on the multiple jet signal of~$ (1 +0.16 \l ^{''2}_{323}).$
Aside from ruling out the associated \Rp\ coupling constants, one can evade
a conflict  with  the experimental  observations by closing the relevant
decay channels by assuming stau or squark masses larger
than  $150 \GeVcc$.

Before closing this subsection on fermion pair production one has to keep in mind
that allowing for more than one dominant \Rp\ coupling leads to further possibilities
for fermion pair production at both leptonic and hadron colliders. 

At leptonic colliders, dilepton production involving two dominant 
$\l_{ijk}$ couplings such as for example \mbox{$e^+ e^- \to \mu^+ \mu^-$} involving
$\l_{131}$ and $\l_{232}$ with $s$-channel $\tilde \nu_{\tau}$ exchange
or \mbox{$e^+ e^- \to \tau^+ \tau^-$}
involving $\l_{131}$ and $\l_{232}$ with a $s$-channel $\tilde \nu_{\mu}$ exchange
have been considered~\cite{Kalept,DELPHI1}. Dijet production can also occur 
in processes involving  $\l_{ijk}$ and $\l'_{ijk}$ couplings with
$s$-channel $\tilde \nu$ exchange. For example \mbox{$e^+ e^- \to b \bar b$} involving
$\l_{131}$ and $\l'_{333}$  or \mbox{$e^+ e^- \to d \bar d$}
involving $\l_{131}$ and $\l'_{311}$  both with $s$-channel $\tilde \nu_{\tau}$ exchange
have been discussed in~\cite{Arnoud,Kal}. Since the angular distribution of the $d$ and $\bar d$
jets is nearly isotropic on the sneutrino resonance, the strong
forward-backward asymmetry in the \SM\ continuum, $A_{FB}(b) \approx
0.65$ at $\sqrt s= 200 \GeV$,
is reduced to $\approx 0.03$ on top of the sneutrino resonance
\cite{Kal}. 

Studies involving products of $\l$ coupling, products of $\l'$ and products
of $\l$ with $\l'$ couplings at ${\mu}^+{\mu}^-$ colliders 
have been performed in~\cite{choudmuon}.

At hadron colliders the third generation slepton resonant production
i.e. $\tilde \nu_{\tau}$ tau-sneutrino (neutral current)
and $\tilde \tau$ stau (charged current) involving weakly constrained 
$\l'_{311} \l_{3jk}$ \ccs\ thus leading to lepton pair production, have 
been considered in~\cite{Rizz} for both the Tevatron and LHC colliders.
The reach in terms of the slepton mass ranges from 800 $\GeVcc\ $ at the Tevatron Run II
to 4 $\TeVcc\ $ at the LHC for sizeable values of $X=\l'_{311} \l_{3jk} B_{l}$, $B_{l}$
being the leptonic branching ratio, from $X \approx 10^{-3}$ down to
$X = 10^{-(5-8)}$ the latter for small slepton masses of the order of hundred $\GeVcc\ $.
In the particular case of  
$e^+ e^- $ 
production, existing Tevatron data~\cite{data2} 
from the CDF detector on the $e^+ e^- $ production 
have been exploited in \cite{Kal} to derive the following bounds on the product
$\l'_{311} \l_{311}$ (with some theoretical uncertainties coming from
the knowledge of K factor for slepton production):
\begin{eqnarray}
(\l'_{311} \l_{311})^{1/2}<0.08 \ \Gamma_{\tilde \nu_{\tau}}^{1/4}
\label{bound1}
\end{eqnarray}
for sneutrino masses in the range $120-250 \GeVcc$
where $\Gamma_{\tilde \nu_{\tau}}$ denotes the sneutrino width in units
of $\GeV $. 
The particular cases of $\mu^+ \mu^-$ and $\tau^+ \tau^-$ productions have
been considered in~\cite{shalom2} based on the total cross-section studies
above a given threshold on the dilepton invariant mass in order to get rid
of the background from the $s$-channel $Z$ resonance contribution.

Futhermore, the distinction between a scalar or a new gauge boson resonance
can be performed~\cite{Rizz} by testing the lepton universality and by
measuring the forward-backward asymmetry which is expected to be
zero in the case of a resonant scalar production and non zero in the case
of a new gauge boson resonance as well as the leptonic charge
asymmetry defined as:
\begin{eqnarray}
A(\eta )={{ {{dN_+ } \over { d \eta }} - {{dN_- } \over {d \eta }} } 
\over {         {{dN_+ } \over {d \eta }} + {{dN_- } \over {d \eta }} }}
\label{chas}
\end{eqnarray}  
where $N_{\pm}$ are the number of positively/negatively charged leptons
of a given rapidity $\eta$. 
The presence of the slepton tends to drive the leptonic charge
asymmetry to smaller absolute values while 
a new $W'$ gauge boson substantially increases
the magnitude of this asymmetry. At the Tevatron Run II, the minimum value 
of the product $\l \l'$ for which the asymmetry differs
significantly from the \SM\ expectation
is $0.1$, for a luminosity of $2 \femtob^{-1}$,
assuming $m_{\tilde l} = 750 \GeVcc$ and $\Gamma_{\tilde l} /
m_{\tilde l} = 0.004$.

\subsection{{\boldmath{$\Rp$}} Contributions to FCNC}
\label{fcncrp}
\index{FCNC and $\Rp$|(}

In the Standard Model \fcnc\ effects arise through loop diagrams. They
are strongly suppressed~\cite{Axel,Clem,Gana}
because of the CKM matrix unitarity and the
quark mass degeneracy (except the top quark) relative to the $Z$ boson mass.
In several supersymmetric extensions of the Standard Model like the MSSM 
the large \fcnc\ effects are expected to be reduced by assuming 
either a degeneracy of the soft supersymmetry breaking scalars masses 
or an alignment of the fermion and scalar superpartners 
mass matrices~\cite{Binetruy98supp}.
In addition \fc\ decay rates such as
$Z \to q_{J} \bar q_{J'}$ through triangle diagrams involving squarks and 
gluinos have been found to be small with respect the Standard 
Model predictions~\cite{Dun,Mukho}. 
\par
The \Rp\ interaction, because of its non-trivial flavour structure, 
opens up 
the possibility of observable \fc\ effects at the tree level.
\par
The \Rp\ interactions contributions to the $Z$ boson flavour off-diagonal 
decays branching ratios were discussed in Section \ref{sec:fcnczpole}.
\par
At colliders, these \fc\ \Rp\ processes occur through the exchange of a 
\susyq\ scalar particle in the $s$- or $t$-channel and lead to fermion pair
productions $f_J f_{J^{\prime}}$ with $J \neq J'$.

Furthermore, in minimal supersymmetric extension of the Standard Model 
without degeneracies for sleptons masses, \fc\ effects can be induced in 
the \susyq\ particle pair production involving \Rp\ interactions.


%
At leptonic colliders, with centre-of-mass energies above the $Z$ boson pole,
single top quark production such as $l_i^+ l_i^- \to t \bar c, \ \bar t c$
occuring via the exchange of a $\tilde d_{kR}$ squark in the $t$-channel
through the \Rp\ couplings $\l'_{i2k}$ and $\l'_{i3k}$ offers a clean
opportunity to observe one of these tree level \fcnc\ 
effects~\cite{Mahanta,JYi,chela,Chemfer,JYigg}. Indeed
single top quark production occuring at the one loop level 
in the Standard Model~\cite{Axel,Clem,Gana} is suppressed with respect
to $b \bar s$ production since it does not receive large
contributions from heavy fermions in the loop. Moreover the MSSM contribution 
has been shown to be small compared to the
\SM\ one~\cite{Dun,Mukho}. 
The cross-section of $e^+ e^- \to t \bar c  + \bar t c$ is shown
in
Fig.~\ref{fig:eetc} from~\cite{JYi} 
for 
$\l'_{12k} \l'_{13k} = 0.01 $ which is the order of 
magnitude of the low-energy 
constraint on this product of \Rp\ couplings for 
$m_{\tilde f} = 100 \GeVcc$.
\begin{figure}[htb]
\begin{center}
\epsfig{file=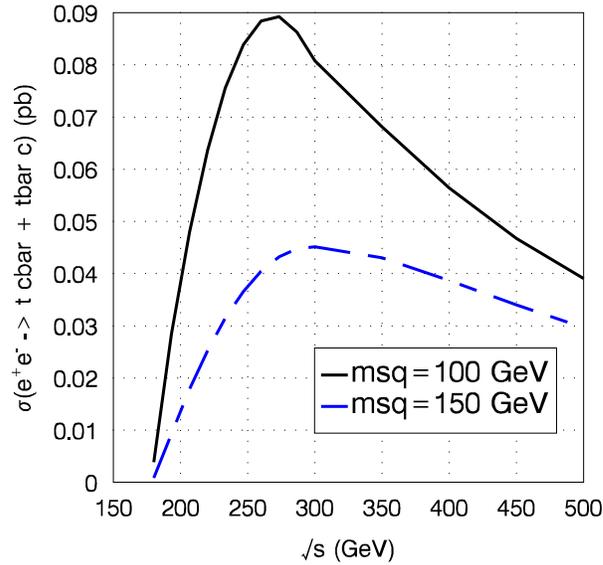,width=10cm}  
\caption{{\it Cross section of the reaction $e^+ e^- \to 
t \bar c + \bar t c$ as a function of the centre-of-mass energy  
for $\l'_{12k} \l'_{13k}=0.01$. The solid line corresponds to
$m_{\tilde d_{kR}} = 100 \GeVcc$ and the dashed line to 
$m_{\tilde d_{kR}} = 150 \GeVcc$.}} 
\label{fig:eetc}
\end{center}
\end{figure}

The reaction $e^+ e^- \to t \bar c + \bar t c$ receives also
contributions at one loop level from the $\l''$ interactions 
\cite{Chemfer,JYi} in which a $\tilde d_R$ squark is involved
with the $\l''_{2jk}$ and $\l''_{3jk}$ coupling constants. 
In particular
the combination $\l''_{223}$ $\l''_{323}$ with a low energy constraint of 
$\l''_{223}$ $\l''_{323} < 0.625 $
which is less stringent than
the constraints of the other $\l''_{2jk}$ $\l''_{3jk}$ combinations
can lead to cross-sections as big as 1~fb for $m_{\tilde d_{kR}} = 100 \GeVcc$.
\par
The $t \bar c / \bar t c$ production can also occur at one loop level 
via photon-photon reactions $e^+ e^- \to 
\gamma \gamma \to t \bar c + \bar t c$ which involve the products of \Rp\ couplings 
\mbox{$\l'_{i2k} \l'_{i3k}$} when $\tilde l_{iL}$ 
sleptons or $\tilde d_{kR}$ squarks are exchanged in the loop
and the products \mbox{$\l''_{2jk} \l''_{3jk}$}
when $\tilde d_R$ squarks run in the loop. 
Again the combinations 
\mbox{$\l'_{323} \l'_{333}$}
(\mbox{$\l''_{223} \l''_{323}$}) which have less stringent 
low-energy constraints than the other \mbox{$\l'_{i2k} \l'_{i3k}$}
(\mbox{$\l''_{2jk} \l''_{3jk}$}) combinations lead to cross-sections
which are about an order of magnitude below the cross-sections
of Fig.~\ref{fig:eetc} from tree level diagrams involving
\mbox{$\l'_{12k}$ $\l'_{13k}$}.
A combination of the results from the $e^+ e^-$ and 
$\gamma \gamma$ collisions would allow to distinguish between
the $\l'$ and $\l''$ effects on the $t \bar c / \bar t c$ 
production.
\par
On the experimental side the $t \bar c$ or $ \bar t c$ production
can lead to $b c l \nu$ final state. The background 
from Standard Model processes such as $e^+ e^- \to 
W^+ W^- \to b c l \nu$ can then be significantly reduced by observing that
the c-quark 
has a fixed energy given by~\cite{chela}: 
\begin{eqnarray}
E(c)=(s+m_t^2-m_c^2) / 2 \sqrt s.
\label{eq:ec}
\end{eqnarray}  
Searches for $t \bar c$ or $ \bar t c$ production have been
performed at LEP$_{II}$ \index{LEP} along these lines.
However they have not yet allowed to 
put a more stringent constraint on $\l'_{12k} \l'_{13k}$ 
than those coming from low energy.
Searches for $t \bar c$ or $ \bar t c$ production will be performed
at the futur linear collider. 
The study of the final state $b c l \nu$ would allow to probe 
values of the product $\l'_{12k} \l'_{13k}$ down to $\sim 0.1$ 
for $m_{\tilde d_{kR}} = 1 \TeVcc$ at a linear collider with a centre-of-mass
energy of $\sqrt s = 500 \GeV$ and a luminosity of 
${\cal L}=100 fb^{-1}$~\cite{chela}.
\par
The $t \bar c$ or $ \bar t c$ production can occur at $\mu^+ \mu^-$ colliders as well.
The cross-section for such a production is shown
in Fig.~\ref{fig:mmtc} from~\cite{JYi} for 
\mbox{$\l'_{223} \l'_{233} = 0.065$} which is equal to its low-energy limit for 
$m_{\tilde f} = 100 \GeVcc$. An additionnal motivation for this choice
of \Rp\ couplings is provided by the observation that
among the possible $\l'_{22k} \l'_{23k}$ combinations
the $\l'_{223} \l'_{233}$ one has the less stringent low-energy constraint.
\begin{figure}[htb]
\begin{center}
\epsfig{file=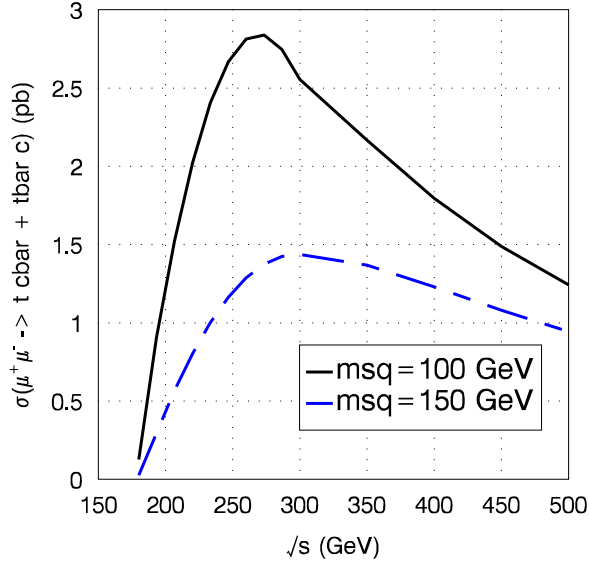,width=10cm}  
\caption{{\it Cross section of the reaction $\mu^+ \mu^- \to 
t \bar c + \bar t c$ as a function of the centre-of-mass energy  
for $\l'_{223} \l'_{233}=0.065$. The solid line corresponds to
$m_{\tilde d_{kR}} = 100 \GeVcc$ and the dashed line to 
$m_{\tilde d_{kR}} = 150 \GeVcc$.}} 
\label{fig:mmtc}
\end{center}
\end{figure}

Finally \fc\ effects in sfermion pair production can be
investigated in high precision measurements planned to be performed for example at 
future leptonic linear colliders~\cite{NLC}. 
The \Rp\ interactions can
generate such effects through the exchange 
of a neutrino in the $t$-channel in slepton pair production
$e^+ e^-\to \tilde l_J \tilde l^*_{J'}$ ($J \neq J'$).
The flavour non-diagonal rates vary in the range 
$\s_{JJ'} \approx ({\L \over 0.1})^4 (2-20) \ \femtob$~\cite{Chemsca} 
with $\L=\l, \l'$ for sleptons masses $m_{\tilde l} < 400 \GeVcc$
as one covers centre-of-mass energy regions from the $Z$ boson pole up 
to the~$\TeV region$. 
Due to the strong dependence on the \Rp\ couplings, the flavour non-diagonal 
rates reach smaller values than the rates obtained in the flavour oscillations
approach~\cite{Arkani} which range between $250 (100)$ and $0.1 (0.01) 
\femtob$ for $\sqrt s=190 (500) \GeV$.


At hadron colliders \fc\ lepton pair productions $l_j l_{j^{\prime}}$ as well as  
quark pair productions $q_j q_{j^{\prime}}$
($j \neq j'$) are both expected to be challenging to search for 
since the environment in terms of background is not as clean as the environment at 
leptonic colliders. 

For example \fc\ lepton pair productions occur from an initial state
$d_j \bar d_k$ ($d_k \bar d_k$) through the exchange
of a $\tilde \nu^i_L$ sneutrino ( $\tilde u^j_L$ squark) in the 
$s$-channel ($t$-channel) via the couplings product $\l'_{ijk}\l_{iJJ'}$
($\l'_{Jjk}\l'_{J'jk}$), or, from an initial state $u_j \bar u_j$ 
through the exchange of a $\tilde d^k_R$ squark in the 
$t$-channel via the couplings product $\l'_{Jjk}\l'_{J'jk}$.
More specific studies on \fc\ lepton pair productions remain to be done.

More striking signatures of \Rp\-induced \fcnc\ effects could be
observed in rare decays of the top quark as discussed in
section~\ref{sec:hfccons}.

Finally the possibility of single top quark production via squark 
and slepton exchanges to probe several combinations of \Rp\
couplings at hadron colliders has been studied in \cite{Ste,Dat,Oak,Chiap}.
Initial state partons such as $u \bar d $ are particularly
relevant for $p \bar p$ colliders such as the Tevatron while the $u d$ initial
state system
is more relevant for $p p$ colliders such as the LHC.
\par
The single top quark production $u_i \bar d_j \to t \bar b$
can occur via the exchange of a $\tilde d_R^k$ squark in
the $t$-channel, through the product of couplings $\l''_{i3k} \l''_{3jk}$.
The choice of the initial state of the reaction $u_i \bar d_j \to t \bar b$
fixes the flavour indices of the \ccs\ product $\l''_{i3k} \l''_{3jk}$ because
of the antisymmetry of the generation indices of the coupling constants $\l''$. Furthermore, because of the
low energy constraints and the low parton luminosities, the only
product of interest is $\l''_{132} \l''_{312}$.
Assuming the observability criteria
$\Delta \s / \s_0 > 20\% $ where $\Delta \s$ is the
\Rp\ cross-section and $\s_0$ is the \SM\ cross-section,
Table~\ref{tab:tevsingtop} from~\cite{Dat} shows the sensitivities
on $\l''_{132}$ $\l''_{312}$
at the upgraded Tevatron for various~$m_{\tilde s_R}$.
\begin{table}
\begin{center}
\begin{tabular}{|c|c|c|c|c|c|c|c|c|}
\hline
$m_{\tilde s_R}$ in $\GeVcc$ & 100 & 200 & 300 & 400 & 500 & 600 & 700 & 800 \\ \hline
$\l''_{132}$ $\l''_{312}$ & 0.01& 0.02& 0.03&0.04 & 0.06 & 0.08 & 0.1 & 0.13 \\ \hline
\hline
\end{tabular}
\caption{{\it
Sensitivities on the product $\l''_{132}$ $\l''_{312}$ for various $m_{\tilde s_R}$
at the upgraded Tevatron from the process 
$u_1 \bar d_1 \to {\tilde s_R} \to t \bar b$ from~\cite{Dat}.}}
\label{tab:tevsingtop}
\end{center}
\end{table}
\par
The single top quark production $u_j \bar d_k
\to t \bar b$ can also occur through the exchange of a $\tilde l_L^i$ slepton in
the $s$-channel via the couplings product $\l'_{ijk}\l'_{i33}$.
The dominant process \mbox{$ u \bar d \to \tilde l_L^i \to t \bar b$}
which involves the sum of couplings
\mbox{$\l'_{111}\l'_{133}+\l'_{211}\l'_{233}+\l'_{311}\l'_{333}$} has been
considered in \cite{Oak}. 
According to~\cite{Oak}
values of $\l'$ couplings below the low energy bounds can be probed 
if the slepton mass lies
in the range 
$200 \GeVcc < m_{\tilde l_L^i} < 340 \GeVcc$ for the upgraded Tevatron
and 
in the range $200 \GeVcc <m_{\tilde l_L^i}< 400 \GeVcc$ for the LHC.
Although larger parton momenta are allowed at the LHC the result is 
not really improved at LHC because of the relative suppression of the 
$\bar d$ quark structure function compared to the $d$ quark one.
\par

Turning to the case of $u_i d_j$ initial state partons, 
the single top quark production can also occur 
through the exchange of a $\tilde d^k_R$ squark in the $s$-channel via the
couplings product $\l''_{ijk} \l''_{33k}$. Table~\ref{tavlhc1} gives an example
of the cross-section obtained from different initial parton states at the LHC. 
\begin{table}[htb] 
\begin{center} \vspace*{-4mm}
\begin{tabular}{|c|c|c|c|c|c|}
\hline
Initial partons   & $cd$    &  $cs$    &  $ub$   &  \multicolumn{2}{ c|}{$cb$} \\ 
\hline 
Exchanged particle   
&$\tilde s$ & $\tilde d$ &$ \tilde s$ & $ \tilde d$ & $ \tilde s$ 
\\ \hline 
Couplings    
& $\lambda''_{212}\lambda''_{332}$   & $\lambda''_{212}\lambda''_{331}$   
& $\lambda''_{132}\lambda''_{332}$   & $\lambda''_{231}\lambda''_{331}$   
& $\lambda''_{232}\lambda''_{332}$   \\ \hline 
Cross-section in pb    
& 3.98  & 1.45   & 5.01  &  \multicolumn{2}{ c|} {0.659}\\
\hline
\end{tabular} 
\caption{{\it Cross section in $pb$ of
the reaction $u_i d_j \to \tilde d^k_R \to t b$ at LHC 
for a squark of mass of $600 \GeVcc$ assuming 
and $\l''_{ijk}=0.1$ and $\Gamma_{R_p} ({\tilde q} ) = 0.5 \GeV$
where $\Gamma_{R_p} ({\tilde q} ) $ is
the width of the exchanged squark due to $R$-parity conserving decay.}} 
\label{tavlhc1} 
\end{center}
\end{table} 
%
Sensitivities on the coupling product $\l''_{212} \l''_{332}$ 
at the upgraded Tevatron and at the LHC have been obtained in~\cite{Oak}.  
A more detailed simulation
has been performed in~\cite{Chiap} and the sensitivities 
on the coupling product $\l''_{212} \l''_{332}$ are shown
in Fig.~\ref{fig:deandrea}.
The reaction $u_j d_k \to t b$ receives also a contribution from the
exchange
of a $\tilde l^{\pm}_{iL}$ slepton in the $u$-channel
via the $\l'_{ij3}$ and $\l'_{i3k}$ couplings \cite{Chiap}.
\par
Supersymmetric particle masses reconstruction have been also performed
within the framework of single top production in~\cite{Chiap}.
\par
To summarize, the studies of single top quark production at hadron
colliders~\cite{Ste,Dat,Oak,Chiap} tend to indicate that the  
LHC is better at probing the $B$-violating
couplings $\lambda^{\prime \prime}$ whereas the Tevatron and the LHC have a
similar sensitivity to $\lambda'$ couplings. Furthermore, this is the only framework
in which the constraints on $\l''$ from physics at colliders are 
comparable or better than the low energy bounds on the $\l''$ coupling
constants.

\begin{figure}[htb]
\begin{center}
\epsfig{file=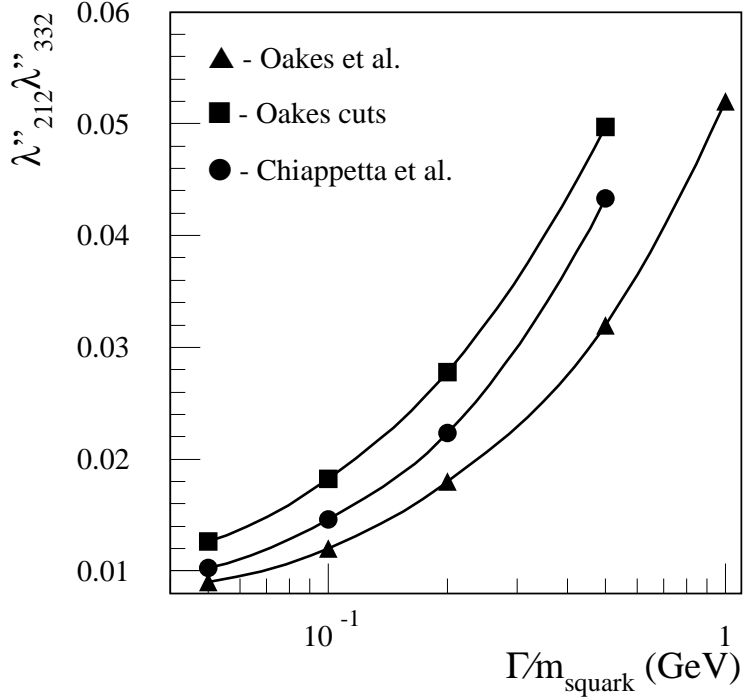,width=10cm}  
\caption{{\it Sensitivity limits on the $\lambda''_{212}\lambda''_{332}$ 
Yukawa couplings obtained from the analysis of the reaction 
$cd \to \tilde s^* \to t b$ at the LHC 
after 1 year with low luminosity for $m_{\tilde s}= 300 \GeVcc$,
found in \cite{Chiap} (circles) and in \cite{Oak} (triangles). 
The squares indicate the results obtained in \cite{Chiap} by applying 
the simplified cuts used in \cite{Oak}.}}
\label{fig:deandrea}
\end{center}
\end{figure}

\index{FCNC and $\Rp$|)}

\subsection{{\boldmath{$\Rp$}} Contributions to {\boldmath{$CP$}} Violation}
\label{indlept}
\index{Violation of $C P$!at colliders|(}

As already discussed in section~\ref{secxxx2c}
the \Rp\ \ccs\ can have a complex phase and hence be by
themselves an independent source of $CP$ violation motivating
many studies on low energy \Rp\ physics.
These can still lead to new tests of $CP$ 
violation in combination with the other possible source of 
complex phase in supersymmetric extensions 
of the Standard Model such as the MSSM even if one assumes that the \Rp\ interactions
are $CP$ conserving.
For instance the \Rp\ couplings can bring a dependence on
the CKM matrix elements due to the fermion mass matrix transformation 
from current basis to mass basis.
\par
A study of $CP$ violation effects in association with sneutrino
flavour oscillations has been carried out 
in section~\ref{sec:sneutrinos}. 
The $CP$ violation effects from \Rp\ couplings in
the $K^o {\bar K}^o$ system and in 
hadrons decays asymmetries
has been discussed
in section~\ref{secxxx2c}.
The $CP$ asymmetries at the $Z$ boson pole 
has been discussed in section~\ref{sec:Aszpole}.
\par
At colliders, $CP$ violation effects from \Rp\ couplings can also be 
further studied from fermion pair productions either \fc\ or non \fc.
These effects are either controlled 
by interference terms between tree and loop amplitudes
in the case of $CP$ asymmetries
or directly considered from tree level processes.
\par
Furthermore, if both non-degeneracies and mixing angles between all slepton
flavours and if the $CP$ odd phase do not vanish, $CP$ violation 
asymmetries can also be observable in \susyq\ particles pair production. 
The $R$-parity odd interactions can provide an alternative mechanism for
explaining $CP$ violation asymmetries in such productions through
possible $\psi$ $CP$ odd phase incorporated in the relevant dimensionless 
coupling constant.

At leptonic colliders the effects of \Rp\ interactions on the $CP$
asymmetries in the processes 
$l^+ l^- \to f_J \bar f_{J'}$, with $J \neq J'$, were calculated in 
\cite{Chemfer}. The \Rp\ contributions to these $CP$ asymmetries are controlled by 
interference terms between tree and loop level amplitudes.
The discussion of loop amplitudes was restricted to the photon and 
$Z$ boson vertex corrections.
The off $Z$ boson pole asymmetries is given by:
\begin{eqnarray}
{\cal A}_{JJ'}={ \sigma_{JJ'}-\sigma_{J'J} \over \sigma_{JJ'}+\sigma_{J'J} }, 
\label{CPasym}
\end{eqnarray}
where $\sigma_{JJ'} = \sigma(l^+ l^- \to f_J \bar f_{J'})$. 
Defining $\psi$ 
as the $CP$ odd phase, these asymmetries lie at
${\cal A}_{JJ'} \approx (10^{-2}-10^{-3}) \sin \psi $  
for leptons and
quarks irrespective of whether one deals with light or heavy flavours.

The $CP$ asymmetries ${\cal A}_{JJ'}$ 
depend on a ratio of different \Rp\ coupling constants and are
therefore less sensitive to the absolute magnitude of these
couplings than the flavour changing rates $\sigma_{JJ'}$ which involve 
higher power of the \Rp\ constants.
The particular dependence of the $CP$ asymmetries on the
couplings is of the form $Im(\l \l^* \l \l^*)/\l^4$ and may thus 
lead to strong enhancement or suppression factors depending on the
largely unknown flavour hierarchical structure 
of the involved Yukawa couplings.
For example the study of single top production $l^+ l^- \to t \bar c$ with
$t \to b W \to b l \nu$ allows to 
learn about $CP$ violation in the quark sector.
In this reaction 
the $CP$ violation can be probed through the asymmetry
defined in Eq.~(\ref{CPasym}) or via the following flavour
off-diagonal $CP$ asymmetry \cite{chela}:
\begin{eqnarray}
   {\cal A}_{+-}   ={ {d \sigma^+ \over d E_l}-{d \sigma^- \over d E_l} 
 \over    {d \sigma^+ \over d E_l}+{d \sigma^- \over d E_l} }, 
\label{CPtop}
\end{eqnarray}
where 
\mbox{$\sigma^+=\sigma(l^+ l^- \to t \bar c \to b \bar c \bar l \nu)$}, 
\mbox{$\sigma^-=\sigma(l^+ l^- \to \bar t c \to \bar b c l \bar \nu)$}
and $E_l$ is the energy of the produced charged lepton.
The values of this $CP$ asymmetry ${\cal A}_{+-}$ range typically
in ${\cal A}_{+-} \approx (10^{-2}-10^{-3}) \sin \psi$ for
$E_l<300 \GeVcc$ \cite{chela}. These 
${\cal A}_{JJ'}$ and ${\cal A}_{+-}$ $CP$ asymmetries
can be enhanced up to 
$\sim 10^{-1} \sin \psi$ if the \Rp\ 
coupling constants exhibit
large hierarchies with respect to the generations.
\par
Turning now briefly to $CP$ violation 
asymmetries in \susyq\ particles pair production,
as in the case of \fc\ fermion pair production, the \Rp\ contributions to these $CP$
asymmetries in scalar particles pair production are controlled by 
interference terms between tree and loop level amplitudes.
For example the flavour non-diagonal $CP$ asymmetries ${\cal A}_{JJ'}$ 
for the slepton pair production, 
$e^+ e^-\to \tilde l_J \tilde l^*_{J'}$ ($J \neq J'$) are predicted to be  
${\cal A}_{JJ'}\approx (10^{-2}-10^{-3}) \sin \psi$~\cite{Chemsca}. 

\par
Finally, the \Rp\ interactions can give rise to $CP$ violation effects at 
tree level in the non~\fc\ reaction $e^+ e ^- \to \tau^+ \tau^-$ via the observation 
of the double spin correlations of the produced tau-leptons pair.
\par
This possibility studied in~\cite{shalom} stands out as an 
very interesting issue by itself since previous studies of
$CP$-violating effects in the process $e^+ e ^- \to \tau^+ \tau^-$ 
which can happen for instance in models with multi-Higgs doublet or in
leptoquark, Majorana $\tilde \nu$ or 
supersymmetry models, all occur at one loop level. 
\par
Here the $CP$ asymmetries are generated from the exchange of a
resonant $\tilde \nu_{\mu}$ sneutrino in the $s$-channel via the real 
coupling $\l_{121}$ and the complex coupling $\l_{323}$ if a 
$\tilde \nu_{\mu} - \tilde {\bar \nu}_{\mu}$ mixing exists. 
This sneutrino mixing \index{Sneutrino!mixing} can generate both $CP$-even and 
$CP$-odd spin asymmetries which are forbidden in the Standard Model and that can 
be measured for $\tau$ leptons at leptonic colliders.
The observation of such asymmetries would provide explicit information
about three different aspects of new physics: 
$\tilde \nu_{\mu} - \tilde {\bar \nu}_{\mu}$ mixing, $CP$ violation and 
\Rp\ .
The sneutrino-antisneutrino mixing phenomena 
which have been gaining some interest
recently~\cite{hirschosc,haber1,Mur} is interesting since it is closely 
related to the generation of neutrino masses \cite{hirschosc,haber1}.
The polarisation asymmetries from double spin correlations of the produced
tau-leptons pair provide a feasible alternative for establishing the mass
splitting between 
the $CP$ even $\tilde \nu^{\mu}_+$ and $CP$ odd $\tilde \nu^{\mu}_-$
muon-sneutrino mass eigenstates~\cite{shalom}.
These polarisation asymmetries depend on the relative values of the real 
part, $a$, and the imaginary part, $b$, of the complex \cc\ $\l_{323}$.
At the next linear collider with $\sqrt s=500 \GeV$~\cite{shalom},
with the simultaneous measurement of the $CP$ conserving and 
$CP$-violating asymmetries,
the whole range $0 \leq {b \over a+b} \leq 1$ can be
probed to at least $3 \s$ in the $m_{\tilde \nu_{\mu}}$ mass range
of $20 \GeVcc$ around resonance 
i.e.~${\sqrt s} - 10 \GeV$~$ <m_{\tilde \nu_{\mu}}< $~${\sqrt s} + 10 \GeV$
even for a small mass splitting of~1~$\GeVcc$.


At hadron colliders, in analogy to the case of the leptonic colliders, the 
resonant production of a sneutrino gives also rise to the possibility 
of having $CP$ violation effects at tree level~\cite{shalom2}.
\par
If the $\tau$ spins can be measured, $CP$ violation effects in the polarisation
asymmetries of the hard process 
$d_j \bar d_k \to \tilde \nu_{\mu} \to \tau^+ \tau^-$ can be observed at the Tevatron.
The \Rp\ \cc\ $\l'_{2jk}$ which enters this subprocess is chosen real,
while $\l_{233}$ is taken complex in order to generate $CP$ asymmetries.
However, at hadron colliders, spin asymmetries deserves a careful discussion.
The spin asymmetries change sign around 
$\sqrt s \approx m_{\tilde \nu^{\mu}_{\pm}}$ so that
one has to integrate over ${\sqrt {\hat {s}}}$ of the initial parton system.
In consequence the spin asymmetries seem too small to be measurable.
Nevertheless a two-step measurement helps in overcoming this problem. In a first
step one has to determine the mass of the resonant sneutrino by measuring
the $\tau \tau$ invariant mass distribution and in a second step  
one has to integrate the absolute values of the polarization 
asymmetriesl~\cite{shalom2}.
\par
At the Tevatron Run IIA (IIB) with ${\cal L}=2 \femtob^{-1}$ ($30 \femtob^{-1}$),
taking their low energy bounds as the values of 
$\l'_{2jk}$ and 
$\vert \l_{233} \vert $ and including all ${j,k}$ combinations 
in $d_j \bar d_k$ fusion,
the $CP$ conserving and $CP$-violating asymmetries may be detected with a 
sensitivity above $3 \s$ over the mass range 
$155 \GeVcc <m_{\tilde \nu^{\mu}_{\pm}} < 400 \GeVcc$ 
($155 \GeVcc <m_{\tilde \nu^{\mu}_{\pm}}<300 \GeVcc$) if 
$\Delta m_{\tilde \nu_{\mu}}=\Gamma_{\tilde \nu_{\mu}}$
($\Delta m_{\tilde \nu_{\mu}}=\Gamma_{\tilde \nu_{\mu}}/10$) where
$\Gamma_{\tilde \nu_{\mu}}$ is the 
sneutrino width.
\par
Moreover the entire range $0 \leq {b \over a+b} \leq 1$
can be practically covered for $m_{\tilde \nu^{\mu}_-}=200 \GeVcc$ at the 
Tevatron with at least $3 \s$ standard deviations 
for $\Delta m_{\tilde \nu_{\mu}}=\Gamma_{\tilde \nu_{\mu}}$
($\Delta m_{\tilde \nu_{\mu}}=\Gamma_{\tilde \nu_{\mu}}/4$).

\par
These results show that in contrast to the case of leptonic colliders~\cite{shalom}, 
the $CP$ odd and $CP$ even spin asymmetries can be observed over 
a wide $\tilde \nu_{\mu}$ sneutrino mass range of about $300 \GeVcc$.
\index{Violation of $C P$!at colliders|)}

\cleardoublepage
\chapter{CONCLUSIONS AND PROSPECTS}
\label{chap:conclusions}

After the great successes of spontaneously broken gauge theories and of the 
Standard Model, supersymmetric theories of particles and interactions
constitute one of the best motivated frameworks for the discussion of new 
physics beyond the Standard Model. 
The reasons are profound and fundamental -- although none is definitely 
conclusive, especially in view of the fact that the new $R$-odd
superpartners have escaped, for a long time now, all experimental efforts 
to disclose their existence.

Among the reasons to consider supersymmetry is our desire to see bosons 
and fermions play similar r\^oles, although this is against all immediate 
evidence, known bosons and fermions having very different properties\,!
Indeed the supersymmetry algebra did not allow us to relate directly 
known bosons with known fermions, and we had to invent, instead, 
a whole new zoo of ``supersymmetric particles'', squarks and sleptons, 
gluinos, charginos and neutralinos, etc., so as to allow us to view the
world as possibly supersymmetric. 
These objects are, precisely, the new $R$-odd particles.
Not only do we have to ``double everything'' -- which was once 
considered as evidence against supersymmetry --\, but in the usual 
framework of spontaneously broken gauge theories additional Higgses 
should also be introduced, with their associated higgsinos\,!
And the whole construction assumes the existence of new 
self-conjugate Majorana fermions, often considered as ugly beasts, 
only Dirac fermions being known in Nature\,!
\,Is all this too high a price to pay\,? Only the future \,-- and
experiments --\, will tell.

But what can such supersymmetric theories do for us\,? 
Plenty of things, many of them well-known, according to different arguments 
all based on the nice and attractive features of supersymmetric theories.
There are also, unfortunately, less nice features, as the reader
who went through detailed discussions of the many possible
supersymmetry-breaking terms in \hbox{$R$-parity} conserving 
and $R$-parity-violating theories will certainly have noticed.

Among the attractive features is the fact that, in supersymmetric theories
\,-- which are closely related with gravitation --\, the Higgs potential 
is largely determined by the supersymmetry.
The quartic Higgs boson self-coupling ($\,\lambda\,$ in the Standard Model), 
or rather self-couplings (for two Higgs doublets), instead of being arbitrary, 
are now fixed by $\,g^2$ and $\,g^2+g^{\prime\, 2}$, 
\,a fact at the origin of many relations involving massive gauge bosons 
and Higgs bosons. The new particles introduced also allow for
an appropriate high-energy convergence of the three $SU(3)$, $SU(2)$ and $U(1)$ 
gauge couplings, whose values get unified, as it would be the case
in a grand-unified theory.
Supersymmetric theories also have improved convergence properties 
at the quantum level, leading to hopes of solving or alleviating 
the hierarchy problems associated with the extreme smallness of the 
cosmological constant $\Lambda$, or the smallness of $m_W$ and $m_Z$ compared 
to the GUT or Planck scales (although these hints towards solutions would 
still have to survive supersymmetry-breaking). 
Supersymmetry usually also appears as a necessary ingredient in the 
construction of consistent string (and brane) theories \,-- 
and, even without having to consider strings and branes at all,
shows us the way towards new spacetime dimensions...

The fundamental motivations for supersymmetric theories are and remain strong, 
even if we still don't know which particular model, within the general class
of supersymmetric theories of weak, electromagnetic and strong interactions,
should effectively be chosen. 
While the allowed parameter space of the popular Minimal version 
of the Supersymmetric Standard Model has now very seriously shrinked,
we have known from the beginning that other ingredients 
(such as an extra singlet superfield coupled to the two Higgs doublets 
$\,H_d\,$ and $\,H_u$) could naturally be present, with no special 
reason to stick to the ``MSSM''.
\,In addition, we still have very little insight on how
supersymmetry should be broken. 
In the absence of a really satisfactory, consistent and predicting mechanism, 
one generally chooses the option of parametrizing the various possible 
supersymmetry-breaking terms. 
Even soft terms are numerous, and this led to the introduction of a large 
number of arbitrary parameters in supersymmetric theories, 
as the price to pay for our ignorance.

Yes, but what about $R$-parity, the subject of this review\,?
As discussed in chapter~\ref{chap:intro}, one of the initial difficulty with 
supersymmetry was the absence of Majorana fermions in Nature, all known 
fermions appearing as Dirac particles carrying additive quantum numbers, 
baryon number $B$ and lepton number $L$, both very well conserved. 
When trying to implement supersymmetry we had to cope with the fact that 
these conserved $B$ and $L$ appear as carried by fundamental 
fermions only \,-- quarks and leptons --\, not by bosons\,!
Still an additive $R$-quantum number might tentatively 
have been interpreted as a lepton number, 
if we could have used supersymmetry to relate the photon with a neutrino.
However, once supersymmetry transforms the photon into a ``photino'', 
the gluons with gluinos, quarks with squarks, etc., this $R$-number,
if it survives at all (under the form of a discrete $R$-parity character), 
must be given a different interpretation.
While the ordinary particles of the Standard Model are $R$-even,
their superpartners, including the various squarks and sleptons, are $R$-odd
\,-- with $R_p = (-1)^R$.
\,But we may have introduced the wolf inside the sheephold since 
$B$ and $L\,$ get now carried, not only by fundamental fermions
(which would make it easy to understand their conservation),
but also by fundamental bosons, the new ($R$-odd) squarks and sleptons\,!
If these are not well behaved we shall certainly face severe problems with 
$B$ and $L$ non-conservation.

Good behavior is, as we saw, closely connected with $R$-parity, even 
if $R$-parity may ultimately turn out not to be exactly conserved.
Indeed $R$-parity conservation, or possibly non-conservation, is 
related with $B$ and $L$ conservation laws, as 
easily seen by reexpressing $R$-parity as $\,(-1)^{2\,S}\,(-1)^{3B+L}$. 
\,A conserved $R$-parity would prevent unwanted direct exchanges 
of \hbox{spin-0} squarks and sleptons between ordinary particles. 
It would, also, prevent neutrinos from mixing with the photino or, more 
generally, the various neutralinos, etc..

With no $R$-parity at all, i.e. if $R$-parity is not even an 
approximate symmetry of the superpotential and of the supersymmetry-breaking terms 
(or in the absence of
analogous symmetries that would play a similar r\^ole 
in excluding unwanted interactions),        
supersymmetric theories are not phenomenologically viable, since they would
lead, in general, to much too large $B$- and/or $L$-violating processes \,-- 
e.g. a much too fast proton decay, or too large neutrino masses.

$R$-parity, on the other hand, naturally excludes unwanted 
$B$- and/or $L$-violating terms
from the superpotential, and from the supersymmetry-breaking terms.
It leads to the famous ``missing-energy'' signature 
of supersymmetric theories at colliders, and to the stability of the 
Lightest Supersymmetric Particle, the LSP, generally thought to be the lightest neutralino.
It then provides us, for free, with a stable weakly-interacting non-baryonic Dark Matter 
candidate.
Quite remarkably, such a candidate is naturally present
for structural reasons, without being introduced ``by hand'' 
for the sole purpose of obtaining Dark Matter.

$R$-parity may well be viewed as having a very fundamental origin, 
in relation with the reflection symmetry 
$\,\theta \to -\,\theta\,$ in superspace, 
or with the existence of extra dimensions which may be responsible 
for supersymmetry breaking by dimensional reduction.
It is, on the other hand, often criticized by tenants of an opposite attitude, 
explaining that they don't see anything fundamental in this symmetry.
And that all possible terms compatible with gauge symmetries
should therefore be included in the superpotential; 
and also added in the Lagrangian density, as supersymmetry-breaking terms.

This certainly leads, in general, to a complete disaster, which necessitates 
the reintroduction of $R$-parity or $R$-parity-like 
symmetries, at least for parts of the Lagrangian density or as approximate 
symmetries. 
Actually some other symmetries (coming e.g. from higher energy ...)
could mimic the effects of an $R$-symmetry or $R$-parity 
in excluding a certain number of terms from the superpotential 
and the supersymmetry-breaking terms, while still allowing others, 
possibly with small or even extremely small coefficients. 
This makes it worthwhile to study possible violations 
of $R$-parity, within supersymmetric theories.
And to discuss in a systematic way the constraints existing on the possible 
\,\Rp\, terms, taking into account all data,
originating from astrophysics and cosmology 
as well as from accelerator experiments.

It is clear that $R$-parity violations are certainly allowed, but only
provided they are sufficiently well hidden and therefore not too large\,!
From the cosmological point of view the most drastic 
\,-- and in general regretted --\,
consequence of $R$-parity violation is that the LSP should normally be 
unstable, and must then in general be abandoned as a favorite Dark Matter 
candidate (unless of course its lifetime were extremely long, 
at least of the order of the age of the universe).
If the LSP is really unstable, however, one has to make sure that 
the \,\Rp\, interactions responsible for its decay are sizeable 
enough for this decay to occur before nucleosynthesis.
In this case, the LSP is no longer constrained to be 
electrically neutral and uncolored.
\,\Rp\, interactions, which would also violate the $B$ and/or $L$
symmetries, may allow for new baryogenesis scenarios.
Conversely, one has to make sure that these new \,\Rp\, interactions are 
sufficiently small so as not to erase the baryon asymmetry needed 
to understand the origin of matter in our universe.

The most flagrant penalties for too much \,\Rp\, are, as we have discussed,
too large $B$- and/or $L$-violating processes, leading for example to a too 
short lifetime for the proton, or too large masses for the neutrinos.
But neutrinos are now known to have small masses anyway, and it is tempting 
to speculate that these very small neutrino masses may have something to do 
with a very small mixing between the neutrino and neutralino sectors, 
that would be induced in \,\Rp\, theories, in the presence of $L$-violating
interactions. 

Small neutrino masses, as well as neutrino oscillations from one flavour 
to another, could then be viewed as originating from the effect of large
neutralino masses, transmitted to the neutrino sector through (sufficiently 
small) \,\Rp\, interactions.
This certainly constitutes an appealing alternative to the familiar 
see-saw mechanism, as a framework in which to discuss the properties 
of neutrinos, masses and oscillations, as well as possible magnetic moments.
It may be in fact closely related to the general question of the origin of 
the mixing between the three quark and lepton families.
The question of $R$-parity conservation, or non-conservation 
(or of how it might turn out to be slightly 
violated), may then simply appear as one of the aspects 
of a much more general ``flavour problem''.
This is indeed quite crucial, but also not easy to solve!

Ongoing experiments such as MiniBoone at FERMILAB, possible future 
experimentation close to a nuclear reactor \`a la CHOOZ and future
long baseline projects as the US NUMI, CERN to Gran Sasso
in Europe and T2K in Japan are expected to give highly
valuable informations on neutrino oscillations.
Exploiting the $\beta$-decay of tritium as in the futur
KATRIN spectrometer in Germany and using the search for $0\nu\beta\beta$
decay as planned by the NEMO3 and the GENIUS experiments will also bring
fundamental informations for the understanding of neutrino mass spectrum.
On the longer term, projects involving very high intensity neutrino beams
like beta-beams and megaton Cerenkov detectors are expected to further
help determine the parameters of the neutrino sector and hopefully bring
some information on $CP$ violation in this sector.
These data will be exploited in order to solve
at least a part of this ``flavour problem''.

Supersymmetric particles have been searched for intensively in a large variety 
of accelerator experiments, most notably at $\ e^+e^-\,$, $ \ e^+p\,$ 
and $\ p\,\bar p\,$ colliders \,-- 
both under the assumptions of a conserved, or violated, $R$-parity.
No direct experimental sign of supersymmetry has been found yet, 
and it is known, from \index{LEP} LEP, HERA and Tevatron experiments, 
that superpartners should be heavier than about 100 GeV at least,
excepted may be for some of the neutral ones, 
both in $R_p$-conserving and $R_p$-violating theories.

Furthermore, sets of bounds for the parameters of 
\,\Rp\, interactions have been discussed, 
both from the indirect searches for such interactions, 
and from the direct production of the new sparticles \,-- either isolated 
or in pairs --\, in a situation of $R$-parity violation.
The most characteristic signature of supersymmetry is then no longer the 
missing energy-momentum carried away by the two unobserved LSP's.
These bounds have been obtained and discussed, either as bounds on every 
single \,\Rp\, coupling constant considered isolately, or as bounds on 
products of two such \,\Rp\, coupling constants.

A large number of the many \,\Rp\, coupling constants and parameters 
still remain unconstrained. Imaginative efforts to find new processes that 
might fill in the remaining gaps in the information will require a 
concerted effort between theory and experiment. One needs to identify 
processes, allowing for significant contributions from the \Rp\ interactions,
where a high experimental sensitivity, also taking into account the 
uncertainties in the Standard Model predictions, is attainable.
Several measurements aiming at detecting rare processes are expected to be
performed soon and should further extend the search for \,\Rp\, effects.
Just to cite a few, the searches for $\mu \rightarrow e$ conversion
either with the $\mu \rightarrow e \gamma$ decay as chased by the MEG
experiment at PSI or with $\mu N \rightarrow e N$ conversion processes as
considered by the MECO project at BNL are expected to gain 2 to 3 orders of
magnitude in sensitivity with respect to their predecessors 
(the MEGA and SINDRUM2 experiments at PSI).
Other promising examples are offered by $B$ meson and $\tau $ lepton rare
decay processes.
If some coupling constants happen to be of the order of $ 10^{-1}$, 
\,this could be enough to lead to observable effects 
at high energy colliders. 

The prospects on the long term are encouraging. 
Thanks to the ongoing experiments such as BABAR at SLAC and 
BELLE at KEK, both aiming at very high $B$ meson production statistics
corresponding to several hundreds fb$^{-1}$ of integrated luminosity, 
experimental measurements of rare (``forbidden'') decay
processes are expected to gain several orders of magnitude in sensitivity.
This kind of gain is also expected for planned projects such as CKM at 
FERMILAB, CLEO upgrades at Cornell, KOPIO at BNL and JHF at Tokai for high 
intensity kaon beams, as well as detectors such as LHCB at CERN and BTEV 
at FERMILAB, dedicated to $B$ physics.
In parallel the CDF and $D\emptyset$ experiments at Tevatron Run II at FERMILAB
are expected to gain 2 to 3 orders of magnitude in sensitivity for B physics
with respect to Run I thus providing further tests for \Rp\ interactions.
As for more direct searches both CDF and $D\emptyset$ are expected to 
extend their searches for supersymmetry with 
\Rp\ effects in both the single supersymmetric particle production mode
and the more conventional pair production mode followed by \Rp\ decays.
Factors of $\,10 \,- \,100\, $ improvements in accuracy
are also anticipated for high precision measurements of magnetic or
electric dipole moments as, for example, the $10^{-28} e.$cm region for
the electric dipole moment of the neutron (to be explored with the spallation
ultra-cold neutron source (SUNS) at PSI). 
Some progress is expected thanks to the high energy leptonic
colliders for high precision $Z$ boson physics observables, especially with the
possibility of the high luminosity option of a future linear electron 
collider running at the $Z$ boson resonance.
Our theoretical understanding of supersymmetry and of physics beyond the
Standard Model is likely to deepen in the meantime. 
On a different front it is likely, also,
that we shall learn more about the properties 
of the Dark Matter and possibly its nature.

Ultimately if supersymmetry is indeed a symmetry of Nature along
the lines presented here, there is no substitute for a direct observation 
of the superpartners.
The best hope is that superpartners could show up 
directly, in a few years from now, at the LHC $\,p\,p\,$ collider at CERN, 
revealing directly the presence of supersymmetry 
as a fundamental symmetry of the world of particles and interactions.
One would then expect a wealth of new results,
on the mass spectrum of the new particles as well as on their production and 
decay properties, which should all be more precisely measured at a future 
linear electron collider.
These data should be crucial to help us understand the actual mechanism
which breaks supersymmetry, and to discover whether $R$-parity is conserved 
or not. And, in the last case, 
how and how much it turns out to be violated.
In particular the unstability of the LSP 
associated with \,\Rp\, could be observable, especially if \,\Rp\, 
interactions were effectively responsible for neutrino masses.
Beyond the possible discovery of supersymmetry, the knowledge about
the conservation or possible violations of $R$-parity is expected to be
essential for the understanding
of several fundamental problems in particle physics, and cosmology.


\cleardoublepage                                                         %
\section*{Acknowledgements}                                              %

This review emerged from common efforts initiated in the framework 
of the $R$-Parity Working Group of the French {\it Groupement de 
Recherche en Supersym\'etrie} (GDR). We wish to thank P.~Bin\'etruy
who headed the GDR for his continuous support. We are grateful to all
members of the $R$-Parity Working Group for useful discussions.
We wish to acknowledge in particular the contributions at early stages 
of this review work of F.~Brochu, P.~Coyle, D.~Fouchez, P.~Jonsson,
F.~Ledroit-Guillon, A.~Mirea, E.~Nagy, R.~Nicolaidou, N.~Parua 
and G.~Sajot.

We also whish to thank the {\it Institut National de Physique Nucl\'eaire 
et de Physique des Particules} (IN2P3), the {\it Centre National de la
Recherche Scientifique} (CNRS) and the {\it Commissariat \`a  l'Energie
Atomique} (CEA) for their support.

\appendix 

\cleardoublepage                                                       %
\chapter{Notations and Conventions}                                    %
\label{chap:appendixA}                                                 %


In the following, the notations and conventions used throughout this 
review are presented.

The three $SU(3)_C \times SU(2)_L \times U(1)_Y$ gauge couplings of 
the \SM\ 
are denoted by $g_3$,  $g$ and $g'$ respectively and the electroweak
mixing angle by $\theta_W$. We use the metric $(+, -, -, -)$.

The superpartners of matter, Higgs and gauge fields in the \SSM\ are
denoted as follows: 

$\bullet$  Scalar partners of left-handed quark fields {\footnotesize$\left(\!
\begin{array}{c} u_{iL} \cr d_{iL} \end{array}\! \right)$} (squarks) by 
${\tilde Q}_i =$ {\footnotesize$\left(\! \begin{array}{c}
{\tilde u}_{iL} \cr {\tilde d}_{iL} \end{array}\! \right)$}, and scalar
partners of right-handed quark fields $u_{iR}$, $d_{iR}$ by ${\tilde u}_{iR}$,
${\tilde d}_{iR}$ ($i=1,2,3$ is a family index).
Similarly, the superpartners of the left-handed lepton fields
{\footnotesize$\left(\! \begin{array}{c} \nu_{iL} \cr l_{iL} \end{array}\!
\right)$} (sleptons) are denoted by ${\tilde L}_i =$ {\footnotesize$\left(\!
\begin{array}{c} {\tilde \nu}_{iL} \cr {\tilde l}_{iL} \end{array}\! \right)$},
and those of the right-handed leptons $l_{iR}$ by ${\tilde l}_{iR}$. The
corresponding superfields are denoted with capital letters $Q_i=$
{\footnotesize$\left(\! \begin{array}{c} U_i \cr D_i \end{array}\! \right)$},
$L_i=$ {\footnotesize$\left(\! \begin{array}{c} N_i \cr E_i \end{array}\!
\right)$}, $U_i^c$, $D_i^c$, $E_i^c$.
Since we use left-handed chiral superfields only, right-handed fermion fields
and their scalar partners are described by 
the corresponding $CP$ conjugate fields (for example the scalar and fermion
components of $U_i^c$ are ${\tilde u}_i^c \equiv (\widetilde{u_{iR}})^\star$ 
and $u_i^c \equiv C (\overline{u_{iR}})^T$, respectively).

$\bullet$ The two Higgs doublets of the \SSM\ are denoted by $h_d =$
{\footnotesize$\left(\! \begin{array}{c} h^0_d \cr h^-_d \end{array}\!
\right)$} and $h_u =$ {\footnotesize$\left(\! \begin{array}{c}
h^+_u \cr h^0_u \end{array}\! \right)$}, the corresponding Weyl fermions
(higgsinos) by ${\tilde h}_d$ and ${\tilde h}_u$, and the corresponding
superfields by $H_d$ and $H_u$. The Higgs \VEVs\ are $<h^0_d> = v_d /\!\!$
{\footnotesize$\sqrt{2}$} and $<h^0_u> = v_u /\!\!$ {\footnotesize$\sqrt{2}$}
(we adopt the normalization $\phi = (a + ib) /\!\!$ {\footnotesize$\sqrt{2}$} 
for complex scalar fields), and the ratio
of \VEVs\ is $\tan \beta = v_u/v_d$. The five physical Higgs states of the
\MSSM, in which no other superfield than the ones mentioned here are
introduced, include two neutral scalars ($CP$-even) denoted by $h$ (for the
lightest one) and $H$, a charged Higgs boson $H^{\pm}$ and a pseudoscalar
($CP$-odd) Higgs boson $A$.

$\bullet$ The Majorana fermion partners of the gluons (gluinos) are denoted by 
${\tilde g^a}$ ($a=1 \cdots 8$); similarly, the superpartners of the
$SU(2)_L \times U(1)_Y$ gauge bosons are gaugino fields denoted by
${\tilde W^i}$ ($i=1,2,3$) and ${\tilde B}$.
Alternatively, one can define the fermionic partners of the photon,
$Z$ and $W^{\pm}$ gauge fields: two Majorana spinors ${\tilde \gamma}
\equiv \sin \theta_W {\tilde W^3} + \cos \theta_W {\tilde B}$ and
${\tilde Z} \equiv \cos \theta_W {\tilde W^3} - \sin \theta_W {\tilde B}$,
and a Dirac spinor ${\tilde W}^{\pm} \equiv ({\tilde W^1}
\mp i {\tilde W^2}) / \sqrt{2}$.
The mass eigenstates of the higgsino-gaugino system, the neutralinos and
the charginos, are denoted by ${\tilde \chi}_{l}^0$ ($l=1 \cdots 4$) and
${\tilde \chi}_{l'}^{\pm}$ ($l'=1,2$), respectively. 

The chiral superfields are normalized so that the lowest term in the $\theta$,
$\bar \theta$ expansion of the left-handed chiral superfield $\Phi$ associated
with the complex scalar field
$\phi = (a + ib) /\!\!$ \nolinebreak {\footnotesize$\sqrt{2}$} is
$\Phi|_{\theta = \bar \theta = 0} = (a + ib) /\!\!$ {\footnotesize$\sqrt{2}$}.
We adopt the following convention for the contraction of two $SU(2)_L$
doublets $\Phi$ and $\Psi$: $\Phi \Psi \equiv \epsilon_{ab} \Phi^a \Psi^b
= \Phi^1 \Psi^2 - \Phi^2 \Psi^1$, where $a,b = 1,2$ are $SU(2)_L$ indices,
$\epsilon_{ab} = - \epsilon_{ba}$ is the totally antisymmetric tensor
(with $\epsilon_{12} = +1$), and $\Phi^1$ (resp. $\Phi^2$) denotes
the $T_3 = + \frac{1}{2}$ (resp. $T_3 = - \frac{1}{2}$) component of $\Phi$.

The discussion of $R$-parity violation does not depend, in general, 
of the particular version of the \SSM\ considered.
In the following, we nevertheless specialize for clarity on the 
minimal \susyq\ extension of the \SM\ (MSSM).
With the above notations, the renormalizable
superpotential of the MSSM reads
\begin{equation}
  W_{MSSM}\ =\ \mu\, H_u H_d\ +\ \lambda_{ij}^e\, H_d L_i E_j^c\ +\
  \lambda_{ij}^d\, H_d Q_i D_j^c\ -\ \lambda_{ij}^u\, H_u Q_i U_j^c\ ,
\label{a1}
\end{equation}
where $\mu$ is the supersymmetric Higgs mass parameter, and
$\lambda_{ij}^{u,d,e}$ denote the quark and charged lepton Yukawa coupling
matrices.
In Eq. (\ref{a1}), like in most equations of this review, a summation
over the generation indices $i,j=1,2,3$, and over gauge indices is
understood. In the absence of $R$-parity, the following \Rp\ terms may also be
added to the superpotential (\ref{a1})\ :
\begin{equation}
  W_{\Rp}\ =\ \mu_i\, H_u L_i\
  +\ \frac{1}{2}\, \lambda_{ijk}\, L_i L_j E^c_k\
  +\ \lambda'_{ijk}\, L_i Q_j D^c_k\
  +\ \frac{1}{2}\, \lambda''_{ijk}\, U^c_i D^c_j D^c_k\ .
\label{a2}
\end{equation}
The supersymmetric mass parameters $\mu_i$ as well as the trilinear couplings
$\lambda_{ijk}$ and $\lambda'_{ijk}$ violate lepton-number conservation law, 
while the couplings $\lambda''_{ijk}$ violate baryon-number conservation law. 
Gauge invariance enforces antisymmetry of the $\lambda_{ijk}$ 
($\lambda''_{ijk}$) couplings in their first (last) two 
indices: $\lambda_{ijk} = - \lambda_{jik}$
($\lambda''_{ijk} = - \lambda''_{ikj}$). To avoid unwanted  factors of 2
in scattering amplitudes, a factor $1/2$ has been introduced in the definition
of the $\lambda_{ijk}$ and $\lambda''_{ijk}$ couplings in Eq. (\ref{a2}).
It should be noted that some authors omit these factors but restrict
the sum over generation indices to $i<j$ (resp. $j<k$) in the
$\lambda_{ijk}\, L_i L_j E^c_k$ (resp. $\lambda''_{ijk}\, U^c_i D^c_j D^c_k$)
terms; this alternative writing is equivalent to our definition (\ref{a2}).

The \SSM\ makes use of a large number of parameters describing our ignorance
about the mechanism which breaks \SUSY. As is customary, we consider the
most general terms that break \SUSY\ in a soft way, i.e. without
reintroducing quadratic divergences. In the MSSM, these ``soft
\SUSY-breaking parameters'' consist of the following:


$\bullet$ The mass parameters $M_1$, $M_2$ and $M_3$ for the $U(1)_Y$, $SU(2)_L$
and $SU(3)_C$ gauginos.
 
$\bullet$ $3 \times 3$ hermitian mass matrices for
each type of squarks and sleptons, both left- and right-handed: 
$m^2_{\tilde Q}$, $m^2_{\tilde u^c}$, $m^2_{\tilde d^c}$, $m^2_{\tilde L}$,
$m^2_{\tilde l^c}$. We shall sometimes use the alternative notation
${\tilde m}_{ij}$ for $(m^2_{\tilde L})_{ij}$. When $R$-parity is broken,
there may also be Higgs-slepton mixing soft masses $\tilde m_{di}^2$.
 
$\bullet$ The "analytic" (i.e. involving only the scalar components of chiral
superfields, and not their complex conjugates) trilinear scalar couplings $A$,
with the same structure as the Yukawa couplings $\lambda$. For example,
the up-quark-type Yukawa couplings $\lambda_{ij}^u H_u Q_i U_j^c$ have
associated trilinear soft terms $A^u_{ij}\, h_u {\tilde Q}_i {\tilde u^c}_j$.
When $R$-parity is explicitly broken, there are also trilinear couplings
$A_{ijk}$, $A'_{ijk}$ and $A^{''}_{ijk}$ corresponding to the \Rp\
superpotential couplings $\lambda_{ijk}$, $\lambda'_{ijk}$ and
$\lambda^{''}_{ijk}$, with the same symmetry properties. The $A$
parameters have mass dimension 1.

$\bullet$ The soft mass parameters ${\tilde m}_d^2$ and ${\tilde m}_u^2$ for
the two Higgs doublets $h_d$ and $h_u$, and a bilinear analytic mass term 
$B h_u h_d$, corresponding to the \susyq\ Higgs mass term $\mu H_u H_d$ in 
the superpotential. There are also \Rp\ bilinear soft terms
$B_i h_u {\tilde L}_i$ corresponding to the \Rp\ mass terms
$\mu_i H_u L_i$ in the superpotential. The $B$ parameters have
mass dimension 2.

The soft \SUSY-breaking terms in the Lagrangian density of the MSSM are
then given by
\begin{eqnarray}   
  -\, {\cal L}^{soft}_{R_p} & = &
     (m^2_{\tilde Q})_{ij}\, {\tilde Q}^{\dagger}_i {\tilde Q}_j\
  +\ (m^2_{\tilde u^c})_{ij}\, {\tilde u}^{c \dagger}_i {\tilde u}^c_j\
  +\ (m^2_{\tilde d^c})_{ij}\, {\tilde d}^{c \dagger}_i {\tilde d}^c_j\
  +\ (m^2_{\tilde L})_{ij}\, {\tilde L}^{\dagger}_i {\tilde L}_j\
  +\ (m^2_{\tilde l^c})_{ij}\, {\tilde l}^{c \dagger}_i {\tilde l}^c_j\
  \nonumber  \\  & &
  + \left( A^e_{ij}\, h_d {\tilde L}_i {\tilde l}^c_j\
  +\ A^d_{ij}\, h_d {\tilde Q}_i {\tilde d^c}_j\
  -\ A^u_{ij}\, h_u {\tilde Q}_i {\tilde u^c}_j\
  +\ \mbox{h.c.} \right)  \nonumber  \\  & &
  +\ {\tilde m}_d^2\,  h_d^{\dagger} h_d\
  +\ {\tilde m}_u^2\,  h_u^{\dagger} h_u\
  + \left( B h_u h_d\ +\ \mbox{h.c.} \right)
  \nonumber  \\  & &
  +\ \frac{1}{2} M_1\, \bar{\tilde B} {\tilde B}\
  +\ \frac{1}{2} M_2\, \bar{\tilde W}^3 {\tilde W}^3\
  +\ M_2\, \bar{\tilde W}^+ {\tilde W}^+\
  +\ \frac{1}{2} M_3\, \bar{\tilde g}^a {\tilde g}^a\ ,
\label{a3}
\end{eqnarray}
where we have written the gaugino soft mass terms in a four-component notation,
with Majorana spinors $\tilde B$, $\tilde W^3$, $\tilde g^a$ and a charged 
Dirac spinor $\tilde W^+$.
In the absence of $R$-parity, additional soft supersymmetry-breaking terms
may also be introduced in the Lagrangian density, as given by:
\begin{eqnarray}   
  -\, {\cal L}^{soft}_{R\!\!\!\slash_p}\ \ =\ \ V^{soft}_{R\!\!\!\slash_p} & = &
  \frac{1}{2} A_{ijk}\, {\tilde L}_i {\tilde L}_j {\tilde l^c}_k\
  +\ A'_{ijk}\, {\tilde L}_i {\tilde Q}_j {\tilde d^c}_k 
  +\ \frac{1}{2} A''_{ijk}\, {\tilde u^c}_i {\tilde d^c}_j {\tilde d^c}_k
     \nonumber \\
  & & +\ B_i\, h_u {\tilde L}_i\
  +\ \widetilde{m}^2_{di}\, h_d^{\dagger}\, {\tilde L}_i\ +\ \mbox{h.c.}\ . 
\label{a4}
\end{eqnarray}

\cleardoublepage                                                       %
\chapter{Yukawa-like {\boldmath{\Rp}} Interactions                                 %
Associated with the Trilinear {\boldmath{\Rp}} Superpotential}                     %
\label{chap:appendixB}                                                 %


In the following, the Yukawa-like (fermion-fermion-scalar) \Rp\ interactions
associated with the trilinear \Rp\ superpotential  couplings of Eq. (\ref{a2})
are derived.. 
The latter  also give rise to $R_p$ conserving scalar interactions
that are quartic in the squark and slepton fields. However these have no
significant phenomenological effects on the low-energy physics for heavy 
superpartners,
so we do not discuss them here (see section \ref{subsec:Lagrangian}).

Let us first derive explicitly the couplings trilinear in the fields
generated by the part of the superpotential (\ref{a2}) 
given by
\begin{eqnarray}
W_{L_{i}L_{j}E^c_{k}}= \frac{1}{2} \lambda_{ijk} L_i L_j E^c_k. 
\label{eq:wdera}
\end{eqnarray}
The trilinear couplings coming from $W_{L_{i}L_{j}E^c_{k}}$ are obtained by
differentiating $W_{L_{i}L_{j}E^c_{k}}$, expressed in term of the scalar 
components $z$ of the superfields, over all the scalar fields:
\begin{eqnarray} 
{\cal L}_{L_{i}L_{j}E^c_{k}}=
 -  \frac{1}{2} \sum_{\alpha,\beta}  \frac{\partial^2 W_{L_{i}L_{j}E^c_{k}}(z)}
{\partial z_\alpha \ \partial z_\beta} \ \psi_\alpha \psi_\beta 
 -  \frac{1}{2} \sum_{\alpha,\beta}  \frac{\partial^2
 W_{L_{i}L_{j}E^c_{k}}^*(z)}
{\partial z^*_\alpha \ \partial z^*_\beta } 
\ \bar \psi_\alpha \bar \psi_\beta, 
\label{eq:wderb}
\end{eqnarray}
where the two-component spinors $\psi$ are the superpartners
of the scalar fields $z$. 
The two-component spinors $\psi$ and $\bar \psi$ belong respectively
to the $(1/2,0)$ and $(0,1/2)$ representations of the Lorentz group.
Eq.(\ref{eq:wdera}) and Eq.(\ref{eq:wderb}) lead together to,
\begin{eqnarray}
{\cal L}_{L_{i}L_{j}E^c_{k}}= 
& - & \frac{1}{2} \sum_{\alpha,\beta} \frac{\partial^2 \left[ 
\frac{1}{2} \lambda_{ijk} \left( \tilde \nu_{iL} \tilde l_{jL}
-\tilde l_{iL} \tilde \nu_{jL} \right) 
\tilde l^c_{kR} \right] } {\partial z_\alpha \ \partial z_\beta } 
\ \psi_\alpha \psi_\beta \cr
& - & \frac{1}{2} \sum_{\alpha,\beta}  \frac{ \partial^2 
\left[ \frac{1}{2} \lambda^*_{ijk} 
\left( \tilde \nu^*_{iL} \tilde l^*_{jL}
-\tilde l^*_{iL} \tilde \nu^*_{jL} \right) 
\tilde l^{c*}_{kR} \right]} {\partial z^*_\alpha \ \partial z^*_\beta } 
\ \bar \psi_\alpha \bar \psi_\beta, 
\label{eq:wderc}
\end{eqnarray} 
where $\tilde \nu$ and $\tilde l$ denote the sneutrinos and charged 
sleptons, respectively, the superscripts $^c$ denote the charge 
conjugate fields
and the superscripts $^*$ the complex conjugate fields. 
The `R' and `L' chirality indices for the scalar fields distinguish 
independent 
fields corresponding to superpartners of right- and left-handed fermions, 
respectively. 
The Lagrangian density (\ref{eq:wderc}) is equivalent to
\begin{eqnarray}
{\cal L}_{L_{i}L_{j}E^c_{k}}= 
& - & \frac{1}{2} \lambda_{ijk} 
\bigg ( \chi_{\nu_i} \chi_{l_j}  \tilde l^c_{kR}
+ \chi_{\nu_i} \eta_{l_k}  \tilde l_{jL}
+ \chi_{l_j}   \eta_{l_k}  \tilde \nu_{iL} 
- (i \leftrightarrow j) \bigg ) \cr
& - & \frac{1}{2} \lambda^*_{ijk} 
\bigg ( \bar \chi_{\nu_i} \bar \chi_{l_j}  \tilde l^{c*}_{kR}
+ \bar \chi_{\nu_i} \bar \eta_{l_k}  \tilde l^*_{jL}
+ \bar \chi_{l_j}   \bar \eta_{l_k}  \tilde \nu^*_{iL} 
- (i \leftrightarrow j) \bigg ).
\label{eq:wderd}
\end{eqnarray} 
In our notations, the two-component spinors $\chi_l$ ($\chi_{\nu}$) and
$\eta_l$ ($\eta_{\nu}$) associated with the charged lepton (neutrino) are 
related to the four-component Dirac spinors describing the charged leptons 
$l$ (neutrinos $\nu$) and antileptons $l^c$ (antineutrinos $\nu^c$) by
\begin{equation}
l=
\left (
\begin{array}{c}
\chi_l \\ \bar \eta_l
\end{array}
\right ) ,
\ \ \ 
l^c=
\left (
\begin{array}{c}
\eta_l \\ \bar \chi_l
\end{array}
\right ) ,
\ \ \
\nu=
\left (
\begin{array}{c}
\chi_{\nu} \\ \bar \eta_{\nu}
\end{array}
\right ) ,
\ \ \
\nu^c=
\left (
\begin{array}{c}
\eta_{\nu} \\ \bar \chi_{\nu}
\end{array}
\right ) .
\label{eq:wdere}
\end{equation} 
The products of two-component spinors $\psi$ and $\bar \psi$
and the products of four-component Dirac spinors $\Psi$ and 
$\bar \Psi=\Psi^\dagger \gamma_0$ are related through the equations,
\begin{eqnarray}
\bar \Psi_1 P_L \Psi_2 =\eta_1 \chi_2, \ \ \ \
\bar \Psi_2 P_R \Psi_1 =\bar \eta_1 \bar \chi_2,
\label{eq:wderf}
\end{eqnarray} 
where $P_L$ and $P_R$ are respectively the left and right chirality
projectors. By applying the relations (\ref{eq:wderf}), one can express
the \index{Lagrangian!\Rp\ trilinear interactions}Lagrangian 
density (\ref{eq:wderd}) in terms of the four-component Dirac spinors:
\begin{eqnarray}
{\cal L}_{L_{i}L_{j}E^c_{k}}= - \frac{1}{2} \lambda_{ijk}
\bigg ( \tilde  \nu_{iL}\bar l_{kR}l_{jL} +
\tilde l_{jL}\bar l_{kR}\nu_{iL} + \tilde l^\star _{kR}\bar \nu^c_{iR}
l_{jL}-(i \leftrightarrow j) \bigg ) + \ \mbox{h.c.},
\label{eq:laglambda_app}
\end{eqnarray} 
where for instance $\bar \nu^c_{iR}=\overline{(\nu^c_i)_R}$.

Similarly, the couplings trilinear in the fields
generated by the superpotential terms 
$W_{L_i Q_j D^c_k}=\lambda'_{ijk} L_i Q_j D^c_k$ 
and $W_{U_i^c D_j^c D_k^c}=\frac{1}{2} \lambda''_{ijk} U_i^c D_j^c D_k^c$ are 
found to be
\begin{eqnarray}
{\cal L}_{L_i Q_j D^c_k}= 
- \lambda '_{ijk} && \bigg ( \tilde  \nu_{iL}\bar d_{kR}d_{jL} +
\tilde d_{jL}\bar d_{kR}\nu_{iL} + \tilde d^\star _{kR}\bar \nu^c_{iR}
d_{jL} \cr && -\tilde  l_{iL}\bar d_{kR}u_{jL} -
\tilde u_{jL}\bar d_{kR}l_{iL} - \tilde d^\star _{kR}\bar l^c_{iR} u_{jL}
\bigg ) + \ \mbox{h.c.},
\label{eq:laglambdap_app}
\end{eqnarray} 
and
\begin{eqnarray}
{\cal L}_{U_i^c D_j^c D_k^c}= - \frac{1}{2} {\lambda ''}_{ijk} 
\bigg (\tilde  u^\star _{i R}\bar d_{j R}d^c_{k L} +
\tilde  d^\star _{j R}\bar u_{i R}d^c_{k L} +
\tilde  d^\star _{k R}\bar u_{i R}d^c_{j L} \bigg ) 
+ \ \mbox{h.c.} \ ,
\label{eq:laglambdapp_app}
\end{eqnarray} 
respectively.

\cleardoublepage                                                       %
\chapter{Production and Decay Formulae}                                %
\label{chap:appendixC}                                                 %

In the following, some useful formulas relevant for $R$-parity
violation searches at colliders are listed. 
This section is based to a large extent on appendix B of \cite{dreinparton}.
The formulas for the decays are organized here by particle families.

\subsubsection{Mixing}
The mixing for the first two generations of sleptons and squarks is
expected to be small to a good accuracy due to the small fermion masses
in the off-diagonal elements of the mass matrices.
On the other hand, a large mixing between the left and right
handed stops is expected because of the large top-quark mass.

For the current eigenstates $\tilde{q}^i_{L,R}$ and the mass eigenstates
$\tilde{q}^i_{1,2}$ the 
\index{Mixing!squark eigenstates}mixing matrix is 
\begin{equation}
\left( \begin{array}{c} \tilde{q}^i_L \\ \tilde{q}^i_R \end{array}\right) 
= \left( \begin{array}{cc} \cos\theta^i_q & \sin\theta^i_q \\
-\sin\theta^i_q & \sin\theta^i_q\end{array}\right)
\left( \begin{array}{c} \tilde{q}^i_1 \\ \tilde{q}^i_2 \end{array}\right).
\end{equation}
It will be denoted as $Q^i_{jk}$ where $i=u,\,d,\,s,\,
c,\,b,\,t$ is the quark flavour index. The slepton mixing matrix is similar and 
will be denoted as $L^i_{jk}$, where $i={e^-},\, {\nu_e},\,
{\mu^-},\,{\nu_\mu},\,{\tau^-},\,{\nu_\tau}$ is the 
lepton flavour index. Sfermion mixing between generations will be 
neglected. 

\subsubsection{Two body decays}
The two-body decay rate corresponding to an averaged matrix element 
$\overline{M}(A \rightarrow B+C)$ with no angular dependence is:
\begin{equation}
\Gamma(A \rightarrow B+C) = \frac{|\overline{M}(A \rightarrow B+C)|^2}  
{16 \pi m_A^3} \sqrt{\left[m_A^2-(m_B+m_C)^2\right]
\left[m_A^2-(m_B-m_C)^2\right] }~. 
\label{2bdr}
\end{equation}

\subsubsection{Three body decays}
The partial width is given by
\begin{equation}
 \Gamma(A\to 1+2+3) = \frac{1}{(2\pi)^3}\frac{1}{32 m^3_A}
 \int^{\left(m^2_{12}\right)_{max}}_{\left(m^2_{12}\right)_{min}} d\; m^2_{12}
 \int^{\left(m^2_{23}\right)_{max}}_{\left(m^2_{23}\right)_{min}} d\;
m^2_{23}|\overline{M}|^2, 
\label{3bodyform}
\end{equation}
where $m_{12}^2 \equiv (p_1+p_2)^2=m_A^2+m_3^2-2 m_A E_3$ and $p_1$, 
$p_2$ are the
4-momenta of particles 1 and 2 respectively, while $E_3$ is the energy of 
particle 3 in the rest frame of particle $A$. $m_{23}$ is defined 
in a similar way. Therefore we obtain $\left(m^2_{12}\right)_{max}=(m_A-m_3)^2$, 
$\left(m^2_{12}\right)_{min}=(m_1+m_2)^2$,
\begin{eqnarray}
\left(m^2_{23}\right)_{max}&=&(E_2+E_3)^2
-\left(\sqrt{{E_2}^2-m^2_2}-\sqrt{{E_3}^2-m^2_3}\right)^2 \; ,
\nonumber \\
\left(m^2_{23}\right)_{min}&=&(E_2+E_3)^2
-\left(\sqrt{{E_2}^2-m^2_2}+\sqrt{{E_3}^2-m^2_3}\right)^2 \; . 
\nonumber
\end{eqnarray}
$E_2=\left(m^2_{12}-m^2_1+m^2_2\right)/2m_{12}$ and
$E_3=\left(m^2_A-m^2_{12}-m^2_3\right)/2m_{12}$
are now the energies of particles 2 and 3 in the rest frame of the 
reduced variable $m_{12}$.

\section{Sfermions}
Two-body $R$-parity-violating decays\index{Decays ! sfermions direct}
of sfermions are given by (\ref{2bdr})
with the following matrix elements (averaged over spin and colour). $\alpha$ is
the mass eigenstate of the sfermion if there is mixing. $i, \, j, \, k$ are the
generation indices. 

For sneutrinos we have:
\begin{eqnarray}
|\overline{M}({\tilde \nu}_j \rightarrow \ell^+_i \ell^-_k)|^2 &=&   
|\lam|^2 (m_{\tilde{\nu}}^2 -  m_{\ell_i}^2-  m_{\ell_k}^2) \; , 
\nonumber \\
|\overline{M}({\tilde \nu}_i \rightarrow \bar{d}_j d_k)|^2 &=&   
N_c|\lam'|^2
(m_{\tilde{\nu}}^2- m_{d_j}^2- m_{d_k}^2) \; , 
\end{eqnarray}
for sleptons:
\begin{eqnarray}
|\overline{M}(\tilde{e}_{j\alpha}^- \rightarrow \bar{\nu}_i \ell^-_k)|^2 &=&   
|\lam|^2 |L^{2j-1}_{1\alpha}|^2(m_{\tilde{e}}^2-m_{\ell_k}^2)\; , 
\nonumber \\
|\overline{M}(\tilde{e}_{k\alpha}^- \rightarrow \nu_i \ell^-_j)|^2 &=&   
|\lam|^2 |L^{2k-1}_{2\alpha}|^2(m_{\tilde{e}}^2-m_{\ell_j}^2)\; , 
\nonumber \\
|\overline{M}(\tilde{e}^-_{i\alpha} \rightarrow \bar{u}_j d_k)|^2 &=&   
N_c|\lam'|^2 |L^{2i-1}_{1\alpha}|^2
(m_{\tilde{e}}^2- m_{u_j}^2- m_{d_k}^2)\; , 
\end{eqnarray}
for squarks:
\begin{eqnarray}
|\overline{M}({\tilde u}_{j\alpha} \rightarrow e_i^+ d_k)|^2 &=&   
|\lam'|^2 |Q^{2j}_{1\alpha}|^2
(m_{\tilde{u}}^2- m_{e_i}^2- m_{d_k}^2)\; , 
\nonumber \\
|\overline{M}({\tilde u}_{i\alpha} \rightarrow \bar{d}_j \bar{d}_k)|^2 
&=& (N_c-1)! |\lam''|^2 |Q^{2i}_{2\alpha}|^2
(m_{\tilde{u}}^2- m_{d_j}^2- m_{d_k}^2)\; , 
\nonumber \\
|\overline{M}({\tilde d}_{j\alpha} \rightarrow \bar{\nu}_i d_k)|^2 &=&   
|\lam'|^2 |Q^{2j-1}_{1\alpha}|^2
(m_{\tilde{d}}^2- m_{d_k}^2)\; , 
\nonumber \\
|\overline{M}({\tilde d}_{k\alpha} \rightarrow \nu_i d_j)|^2 &=&   
|\lam'|^2 |Q^{2k-1}_{2\alpha}|^2
(m_{\tilde{d}}^2- m_{d_j}^2)\; , 
\nonumber \\
|\overline{M}({\tilde d}_{k\alpha} \rightarrow e_i^- u_j)|^2 &=&   
|\lam'|^2 |Q^{2k-1}_{2\alpha}|^2
(m_{\tilde{d}}^2- m_{e_i}^2- m_{u_j}^2)\; , 
\nonumber \\
|\overline{M}({\tilde d}_{k\alpha} \rightarrow 
\bar{u}_i \bar{d}_j)|^2 &=&   
(N_c-1)! |\lam''|^2 |Q^{2k-1}_{2\alpha}|^2
(m_{\tilde{d}}^2- m_{u_i}^2- m_{d_j}^2) \; , 
\end{eqnarray}
where $N_c$ is the number of colours.

\subsubsection{Stops}
A large mixing between the left and right
handed stops is expected because of the large top--quark mass.
We show here explicitly the 
\index{Mixing!squark eigenstates}effect of the mixing 
given in general terms above. 
The mass eigenstates are~\cite{Kon94}
\begin{equation}
\left({{\sqt_1}\atop{\sqt_2}}\right)=
\left(
{\sqt_{L}\,\cos\theta_t-\sqt_{R}\,\sin\theta_t}
\atop
{\sqt_{L}\,\sin\theta_t+\sqt_{R}\,\cos\theta_t}
\right),
\end{equation}
where $\theta_t$ denotes the mixing angle of the stops:
\begin{eqnarray}
\sin 2\theta_t=
{\frac{2a_{t}\,m_{t}}
{\sqrt{(m^{2}_{\sqt_{L}}-m^{2}_{\sqt_{R}})^{2}
+4a^{2}_{t}\,m^{2}_{t}}}}, \\
\cos 2\theta_t=
{\frac{m^{2}_{\sqt_{L}}-m^{2}_{\sqt_{R}}}
{\sqrt{(m^{2}_{\sqt_{L}}-m^{2}_{\sqt_{R}})^{2}
+4a^{2}_{t}\,m^{2}_{t}}}}.
\end{eqnarray}
The $m_{\sfe_{L, R}}$ and $a_{f}$ are the SUSY mass parameters
and $m_{t}$ is the top-quark mass.
The mass eigenvalues are given by:
\begin{eqnarray}
m^{2}_{\sqt_1}&=&{\frac{1}{2}}\left\{ m^{2}_{\sqt_{L}}+m^{2}_{\sqt_{R}}
-\left[ (m^{2}_{\sqt_{L}}-m^{2}_{\sqt_{R}})^{2}
+(2a_{t}m_{t})^{2}\right]^{1/2}\right\} \; , \nonumber \\
m^{2}_{\sqt_2}&=&{\frac{1}{2}}\left\{ m^{2}_{\sqt_{L}}+m^{2}_{\sqt_{R}}
+\left[ (m^{2}_{\sqt_{L}}-m^{2}_{\sqt_{R}})^{2}
+(2a_{t}m_{t})^{2}\right]^{1/2}\right\} \; .
\end{eqnarray}

The partial width of the lightest stop $\sqt_1$ for the corresponding
$R_p$-violating decay is:
\begin{equation} 
\Gamma ( \sqt_1 \rightarrow \ell^+_i d_k) = \frac{1}{16 \pi} \lamp ^2
\cos^2 (\theta_t) \, m_{\sqt_1} \, ,
\end{equation}
if the masses in the final state are neglected.
Depending on the mass of the stop this decay mode may be competitive
with respect to the $R$-parity conserving ones 
$\sqt_1 \rightarrow$ t \XO and $\sqt_1 \rightarrow$ b \XP .

The sneutrinos may also decay via gauge interactions as 
$\tilde \nu^i_L \to \tilde \chi^+_a l^i$ or 
$\tilde \nu^i_L \to \tilde \chi^0_a \nu^i_L$.
The partial width is~\cite{barger89}:
\begin{eqnarray}
\Gamma(\tilde \nu^i_L \to \tilde \chi^+_a l^i, \ \tilde \chi^0_a \nu^i_L)= 
\frac{C g^2}{16 \pi} 
m_{\tilde \nu^i_L} (1-\frac{m_{\tilde \chi^+_a}^2}{m_{\tilde \nu^i_L}^2})^2,
\label{widthform2}
\end{eqnarray}  
where $C=\vert V_{a1}\vert ^2$ for the decay into chargino and $C=\vert
N_{a2}\vert ^2$, for the neutralino case, with $V_{a1}$ and $N_{a2}$ 
the mixing matrix elements.

The cross section for the sneutrino production in the $s$-channel at $e^+ e^-$
colliders, is
\begin{eqnarray}
\sigma (e^+ e^- \to \tilde \nu^i_L \to X)= \frac{4 \pi s}{m_{\tilde
\nu^i_L}^2} 
\frac{\Gamma(\tilde \nu^i_L \to e^+ e^-)  \Gamma(\tilde \nu^i_L \to X)}{
(s-m_{\tilde \nu^i_L}^2)^2  + m_{\tilde \nu^i_L}^2 \Gamma_{\tilde
\nu^i_L}^2},
\label{ffXsectf1}
\end{eqnarray}
where $\Gamma(X)$ generally denotes the partial width for the sneutrino
decay into the final state $X$. At sneutrino resonance, Eq.(\ref{ffXsectf1}) 
takes the form,
\begin{eqnarray}
\sigma (e^+ e^- \to \tilde \nu^i_L \to X)= \frac{4 \pi}{m_{\tilde
\nu^i_L}^2} B(\tilde \nu^i_L \to e^+ e^-)  B(\tilde \nu^i_L \to X),
\end{eqnarray}
where $B(X)$ denotes the partial width for sneutrino decay
into a final state $X$. 

The $\tilde{\nu}$ production $d \bar d$ annihilations through 
$\lambda'_{ijk}$ is~\cite{Qui}: 
\begin{eqnarray}
\sigma (d_k \bar d_j \to  \tilde \nu^i \to X_1 X_2)= \frac{4}{9} {\frac 
{\pi \Gamma_{d_k \bar d_j} \Gamma_f}{(\hat{s}-m_{\tilde \nu^i}^2)^2 + 
m_{\tilde \nu^i}^2 \Gamma_{\tilde \nu^i}^2}}\;,
\end{eqnarray} 
where $\Gamma_{d_k \bar d_j}$ and $\Gamma_f$ are, respectively, the partial 
widths of the channels $\tilde \nu^i \to d_k \bar d_j$ and 
$\tilde \nu^i \to X_1 X_2$, $\Gamma_{\tilde \nu^i}$ is the total width of the 
sneutrino, $m_{\tilde \nu^i}$ is the sneutrino mass and $\hat{s}$ is the
square of the parton energy in the centre of mass reference frame. 
A factor ${1/3}$ originates from the matching of the color indices in the
initial state. $\Gamma_{d_k \bar d_j}$ is given by
\begin{eqnarray}
\Gamma_{d_k \bar d_j} = \frac{3}{4} \alpha_{\lambda'_{ijk}} m_{\tilde \nu^i},
\label{snwi2}
\end{eqnarray}
where $ \alpha_{{\lambda'}_{ijk}}= {\lambda'}^2_{ijk} / 4 \pi $.

\section{Neutralinos}
\label{secneutralino}
The three-body partial width of the neutralino 
\index{Decays ! neutralino}
can be calculated using the matrix elements given in the following
together with the width formula (\ref{3bodyform}). 
The spin and colour averaged matrix 
elements are given in terms of the following functions:
\begin{eqnarray}
R(\tilde{a},m_{bc}^2) &\equiv&
\frac{1}{(m_{bc}^2-M_{\tilde{a}}^2)^2 +\Gamma_{\tilde{a}}^2 
M_{\tilde{a}}^2},\nonumber \\
S(\tilde{a},\tilde{b},m_{cd}^2,m_{ef}^2) &\equiv& R(\tilde{a},m_{cd}^2)
R(\tilde{b},m_{ef}^2) \nonumber \\
& & \left[(m_{cd}^2-M_{\tilde{a}}^2)(m_{ef}^2-M_{\tilde{b}}^2) 
+ \Gamma_{\tilde{a}} \Gamma_{\tilde{b}} M_{\tilde{a}}  
M_{\tilde{b}}\right] ,
\label{ricor1}
\end{eqnarray}
where $m_{bc}^2 = (p_b+p_c)^2$, and $M_{\tilde a}$, $\Gamma_{\tilde a}$
are the mass and the decay width of the sfermion ${\tilde a}$. 
\begin{eqnarray} 
\Psi(\tilde{a},1,2,3) &\equiv &
R(\tilde{a},m^2_{12}) \left(m^2_{12}-m^2_1-m^2_2\right) \nonumber \\
&& \left[\left(a^2(\tilde{a})+b^2(\tilde{a})\right)
\left(m^2_A+m^2_3-m^2_{12}\right)+4a(\tilde{a})b(\tilde{a})m_3 m_A
\right], \nonumber \\ 
\Upsilon(\tilde{a},1,2,3) &\equiv&
S(\tilde{a}_1,\tilde{a}_2,m_{12}^2,m_{12}^2)\left(m^2_{12}-m^2_1-m^2_2\right)
\nonumber \\ &&
\left[\left(a(\tilde{a}_1)a(\tilde{a}_2)+b(\tilde{a}_1)b(\tilde{a}_2)\right)
\left(m^2_A+m^2_3-m^2_{12}\right)\right.\nonumber\\
&&\left. +2\left(a(\tilde{a}_1)b(\tilde{a}_2)
+a(\tilde{a}_2)b(\tilde{a}_1)\right) m_3 m_A \right], \nonumber\\
\Phi(\tilde{a},\tilde{b},1,2,3)& \equiv &
S(\tilde{a},\tilde{b},m_{12}^2,m_{23}^2) \left[
m_1m_3a(\tilde{a})a(\tilde{b})
\left(m^2_{12}+m^2_{23}-m^2_1-m^2_3\right) \right.\nonumber \\ &&
+m_1 m_A b(\tilde{a})a(\tilde{b})\left(m^2_{23}-m^2_2-m^2_3\right)
\nonumber \\ &&
+m_3 m_A a(\tilde{a})b(\tilde{b})\left(m^2_{12}-m^2_1-m^2_2\right)
\nonumber \\ && \left.  +b(\tilde{a})b(\tilde{b})
\left(m^2_{12}m^2_{23}-m^2_1m^2_3-m^2_A m^2_2\right) \right] ,
\label{ricor2}
\end{eqnarray} 
where $m_A$ is the mass of the neutralino,
$\tilde{a}_1$ and $\tilde{a}_2$ are the mass eigenstates of the
SUSY particle. The functions $a$ and $b$ are
gaugino-sfermion-fermion coupling constants. For the neutralino decay
the couplings $a$ and $b$ are as follows:
\begin{equation}
a({\tilde \nu}_i)=0 \; , \;\;\;\;\;\;
b({\tilde \nu}_i)=\frac{g{N'}_{l2}}{2 \cos \theta_W}\; ,
\end{equation}
\begin{eqnarray}
a({\tilde \ell}_{i\alpha})&=&
m_{\ell_i}\frac{g N_{l3}}{2M_W \cos \beta}L^{2i-1}_{1\alpha}
+L^{2i-1}_{2\alpha}\left( e  {N'}_{l1} 
-\frac{g \sin^2 \theta_W {N'}_{l2}}{\cos }\right) ,
\nonumber \\
b({\tilde \ell}_{i\alpha})&=&
m_{\ell_i}\frac{g N_{l3}}{2M_W \cos \beta}L^{2i-1}_{2\alpha}
-L^{2i-1}_{1\alpha}\left( e{N'}_{l1} 
+\frac{ g{N'}_{l2} \left(\frac{1}{2}- \sin^2 \theta_W \right)}
{\cos \theta_W} \right) ,
\end{eqnarray}
\begin{eqnarray}
a({\tilde d}_{i\alpha})&=&
m_{d_i}\frac{g N_{l3}}{2M_W \cos \beta}Q^{2i-1}_{1\alpha}
-Q^{2i-1}_{2\alpha}\left( e e_d {N'}_{l1} 
-\frac{g e_d \sin^2 \theta_W {N'}_{l2}}{\cos \theta_W}\right) ,
\nonumber \\
b({\tilde d}_{i\alpha})&=&
m_{d_i}\frac{g N_{l3}}{2M_W \cos \beta}Q^{2i-1}_{2\alpha}
+Q^{2i-1}_{1\alpha}\left( e e_d {N'}_{l1} 
-\frac{g {N'}_{l2}\left(\frac{1}{2}+ e_d \sin^2 \theta_W\right)}
{\cos \theta_W} \right) ,
\end{eqnarray}
\begin{eqnarray}
a({\tilde u}_{i\alpha})&=&
m_{u_j}\frac{g N_{l4}}{2M_W \sin \beta}Q^{2j}_{1\alpha}
-Q^{2j}_{2\alpha} \left( e e_u {N'}_{l1} 
-\frac{g e_u \sin^2 \theta_W {N'}_{l2}}{\cos \theta_W}\right) ,
\nonumber \\
b({\tilde u}_{i\alpha})&=&
m_{u_i}\frac{g N_{l4}}{2M_W \sin \beta}Q^{2i}_{2\alpha}
+Q^{2i}_{1\alpha}\left( e e_u {N'}_{l1} 
+\frac{g {N'}_{l2}\left(\frac{1}{2}- e_u \sin^2 \theta_W\right)}
{\cos \theta_W} \right) .
\end{eqnarray}
In terms of these functions and couplings, the averaged matrix 
elements for three-body neutralino decays can be written:
\begin{eqnarray}
\lefteqn{|\overline{M}(\widetilde{\chi}_l^0 
\rightarrow \bar{\nu}_i \ell_j^
+\ell_k^-)|^2 =}  & \nonumber \\
&& {\lam}^2 \left[
\Psi({\tilde \nu}_i,\ell_j,\ell_k,\nu_i)
+\sum_{\alpha=1,2} |L^{2j-1}_{1\alpha}|^2
\Psi({\tilde \ell}_{j\alpha},\nu_i,\ell_k,\ell_j) 
\right. \nonumber \\ 
&&+\sum_{\alpha=1,2}|L^{2k-1}_{2\alpha}|^2
\Psi({\tilde \ell}^*_{k\alpha},\nu_i,\ell_j,\ell_k) 
\nonumber \\
&& +2 L^{2j-1}_{11}L^{2j-1}_{12}
\Upsilon({\tilde \ell}_j,\nu_i,\ell_k,\ell_j) 
+2 L^{2k-1}_{21}L^{2k-1}_{22}\Upsilon({\tilde \ell}^*_k,\nu_i,\ell_j,\ell_k)
\nonumber \\
&& -\sum_{\alpha=1,2} 2L^{2j-1}_{1\alpha}
\Phi({\tilde \ell}_{j\alpha},{\tilde \nu}_i,\nu_i,\ell_k,\ell_j)
- \sum_{\alpha=1,2} 2L^{2k-1}_{2\alpha}
\Phi ({\tilde \ell}^*_{k\alpha},{\tilde \nu}_i,\nu_i,\ell_j,\ell_k)  
\nonumber \\
&& \left. - \sum_{\alpha , \beta =1,2} 
2L^{2j-1}_{1\alpha}L^{2k-1}_{2\beta}
\Phi({\tilde \ell}^*_{k\beta},{\tilde \ell}_{j\alpha},\ell_j,\nu_i,\ell_k) 
\right] , 
\end{eqnarray}

\begin{eqnarray}
\lefteqn{|\overline{M}({\widetilde{\chi}}_l^0 
\rightarrow \bar{\nu}_i \bar{d}_j d_k)|^2 =} & \nonumber \\
&& {\lam'}^2 N_c  \left[ \Psi({\tilde \nu}_i,d_j,d_k,\nu_i)
+\sum_{\alpha=1,2} |Q^{2j-1}_{1\alpha}|^2 \Psi({\tilde d}_{j\alpha},
\nu_i,d_k,d_j) \right. \nonumber \\
&& +\sum_{\alpha=1,2} |Q^{2k-1}_{2\alpha}|^2 \Psi({\tilde d}^*_{k\alpha},
\nu_i,d_j,d_k) \nonumber \\
&& +2Q^{2j-1}_{11}Q^{2j-1}_{12}\Upsilon({\tilde d}_j,\nu_i,d_k,d_j) 
+2Q^{2k-1}_{21}Q^{2k-1}_{22}\Upsilon({\tilde d}^*_k,\nu_i,d_j,d_k)
\nonumber \\
&& - \sum_{\alpha=1,2} 2Q^{2j-1}_{1\alpha} \Phi({\tilde d}_{j\alpha},
{\tilde \nu}_i,\nu_i,d_k,d_j)- \sum_{\alpha=1,2} 2Q^{2k-1}_{2\alpha}
\Phi ({\tilde d}^*_{k\alpha},{\tilde \nu}_i,\nu_i,d_j,d_k)\nonumber \\
&& \left.- \sum_{\alpha , \beta=1,2} 2Q^{2j-1}_{1\alpha}Q^{2k-1}_{2\beta}
\Phi({\tilde d}^*_{k\beta},{\tilde d}_{j\alpha},d_j,\nu_i,d_k) \right] ,
\end{eqnarray}
\begin{eqnarray}
\lefteqn{|\overline{M}(\widetilde{\chi}_l^0 
\rightarrow \ell^+_i \bar{u}_j d_k)|^2 =} & \nonumber \\
&& {\lam'}^2 N_c  \left[ \sum_{\alpha=1,2} |L^{2i-1}_{1\alpha}|^2 
\Psi({\tilde \ell}_{i\alpha},u_j,d_k,\ell_i)
+\sum_{\alpha=1,2} |Q^{2j}_{1\alpha}|^2 \Psi({\tilde u}_{j\alpha},
\ell_i,d_k,u_j) \right. \nonumber \\
&& +\sum_{\alpha=1,2} |Q^{2k-1}_{2\alpha}|^2 
\Psi({\tilde d}^*_{k\alpha},\ell_i,u_j,d_k) 
+2 L^{2i-1}_{11}L^{2i-1}_{12} \Upsilon({\tilde \ell}_i,u_j,d_k,\ell_i) 
\nonumber \\
&& +2 Q^{2j}_{11}Q^{2j}_{12} \Upsilon({\tilde u}_j,\ell_i,d_k,u_j) 
+2 Q^{2k-1}_{21}Q^{2k-1}_{22}\Upsilon({\tilde d}^*_k,\ell_i,u_j,d_k) 
\nonumber \\
&& -\sum_{\alpha , \beta =1,2} 2L^{2i-1}_{1\alpha}Q^{2j}_{1\beta}
\Phi({\tilde u}_{j\beta},{\tilde \ell}_{i\alpha},\ell_i,d_k,u_j) 
\nonumber \\
&& - \sum_{\alpha , \beta =1,2} 2L^{2i-1}_{1\alpha}Q^{2k-1}_{2\beta}
\Phi({\tilde d}^*_{k\beta},{\tilde \ell}_{i\alpha},\ell_i,u_j,d_k)
\nonumber \\
&& \left.- \sum_{\alpha , \beta =1,2} 2Q^{2j}_{1\alpha}Q^{2k-1}_{2\beta}
\Phi({\tilde d}^*_{k\beta},{\tilde u}_{j\alpha},u_j,\ell_i,d_k) \right] ,
\end{eqnarray}

\begin{eqnarray}
\lefteqn{|\overline{M}
(\widetilde{\chi}_l^0\rightarrow \bar{u}_i \bar{d}_j \bar{d}_k)|^2 =}
 & \nonumber \\
 && {\lam''}^2 N_c!  \left[  
  \sum_{\alpha=1,2}|Q^{2i}_{2\alpha}|^2
\Psi({\tilde u}^*_{i\alpha},d_j,d_k,u_i)
 +\sum_{\alpha=1,2}|Q^{2j-1}_{2\alpha}|^2\Psi({\tilde
d}^*_{j\alpha},u_i,d_k,d_j) \right. \nonumber \\  &&
+\sum_{\alpha=1,2}|Q^{2k-1}_{2\alpha}|^2\Psi({\tilde
d}^*_{k\alpha},u_i,d_j,d_k)  + 2
Q^{2i}_{21}Q^{2i}_{22}\Upsilon({\tilde u}^*_i,d_j,d_k,u_i) \nonumber \\   
&& +2 Q^{2j-1}_{21}Q^{2j-1}_{22}  \Upsilon({\tilde d}^*_j,u_i,d_k,d_j)     +2
Q^{2k-1}_{21}Q^{2k-1}_{22} \Upsilon({\tilde d}^*_k,u_i,d_j,d_k) \nonumber \\ 
&& - \sum_{\alpha , \beta =1,2} 2Q^{2i-1}_{2\alpha}Q^{2j-1}_{2\beta}
\Phi({\tilde d}^*_{j\beta},{\tilde u}^*_{i\alpha},u_i,d_k,d_j)\nonumber \\
 &&- \sum_{\alpha , \beta =1,2} 2Q^{2i-1}_{2\alpha}Q^{2k-1}_{2\beta}
\Phi({\tilde d}^*_{k\beta},{\tilde u}^*_{i\alpha},u_i,d_j,d_k)\nonumber \\ 
 &&  \left.- \sum_{\alpha , \beta =1,2} 2Q^{2j-1}_{2\alpha}Q^{2k-1}_{2\beta}
\Phi({\tilde d}^*_{k\beta},{\tilde d}^*_{j\alpha},d_j,u_i,d_k) \right] . 
\end{eqnarray}

\section{Charginos}
\label{secchargino}
Three-body 
\index{Decays ! chargino}decays of the chargino are obtained in 
terms  of the same functions (\ref{ricor1},\ref{ricor2}) given in the previous
section (\ref{secneutralino}), but the coefficients $a$ and $b$ for the 
couplings are as follows: 
\begin{eqnarray}
a({\tilde \ell}_{i\alpha})&=&0 \;\;\;\;\;\;\;\;\;  \;\;\;\;\;\;\;\;\;
\;\;\;\;\;\;\;\;\;\;\;\;
b({\tilde \ell}_{i\alpha})=L^{2i-1}_{1\alpha} U_{l1}
- \frac{U_{l2} L^{2i-1}_{2\alpha} m_{e_i}} {\sqrt{2}M_W\cos\beta}
\nonumber \\
a({\tilde \nu}_i)&=&\frac{U_{l2} m_{e_i}}{\sqrt{2}M_W\cos\beta}
\;\;\;\;\;\;\;\;\; \;\;\; b({\tilde \nu}_i)=V^*_{l1} \nonumber \\
a({\tilde u}_{i\alpha})&=&-\frac{m_{d_i} U_{l2} Q^{2i}_{1\alpha}}
{\sqrt{2}M_W\cos\beta} \;\;\;\;\;\;\;\;\; b({\tilde u}_{i\alpha})=
V^*_{l1}Q^{2i}_{1\alpha}-\frac{m_{u_i} V^*_{l2} Q^{2i}_{2\alpha}}
{\sqrt{2}M_W\sin\beta} \nonumber \\
a({\tilde d}_{i\alpha})&=&-\frac{m_{u_i} V^*_{l2} Q^{2i-1}_{1\alpha}}
{\sqrt{2}M_W\sin\beta} \;\;\;\;\;\;\;\;\; b({\tilde d}_{i\alpha})=
Q^{2i-1}_{1\alpha} U_{l1} -\frac{U_{l2} Q^{2i-1}_{2\alpha} m_{d_i}}
{\sqrt{2}M_W\cos\beta} \; .
\end{eqnarray}
The averaged matrix elements are:
\begin{eqnarray}
\lefteqn{|\overline{M}(\widetilde{\chi}_l^+ \rightarrow \bar{\nu}_i
\ell^+_j\nu_k)|^2 = } & \nonumber \\
&& \frac{g^2 \lam^2}{2} \left[
\sum_{\alpha = 1,2} |L^{2k-1}_{2\alpha}|^2 \Psi({\tilde \ell}^*_{k\alpha},
\nu_i,\ell_j,\nu_k) +2L^{2k-1}_{21} L^{2k-1}_{22} 
\Upsilon({\tilde \ell}^*_k,\nu_i,\ell_j,\nu_k) \right] , 
\end{eqnarray}
\begin{eqnarray}
\lefteqn{|\overline{M}(\widetilde{\chi}_l^+ \rightarrow \nu_i \nu_j 
\ell^+_k)|^2=} & \nonumber \\
&& \frac{g^2 \lam^2}{2} \left[ \sum_{\alpha = 1,2} |L^{2i-1}_{1\alpha}|^2 
\Psi({\tilde \ell}_{i\alpha},\nu_j,\ell_k,\nu_i)
+\sum_{\alpha = 1,2} |L^{2j-1}_{1\alpha}|^2
\Psi({\tilde \ell}_{j\alpha},\nu_i,\ell_k,\nu_j)\right. \nonumber \\
&& 2L^{2i-1}_{11} L^{2i-1}_{12}\Upsilon({\tilde \ell}_i,\nu_j,\ell_k,\nu_i)
+2L^{2j-1}_{11} L^{2j-1}_{12}\Upsilon({\tilde \ell}_j,\nu_i,\ell_k,\nu_j) 
\nonumber \\
&& \left.+\sum_{\alpha , \beta = 1,2}  L^{2i-1}_{1\alpha}L^{2j-1}_{1\beta} 
\Phi({\tilde \ell}_{j\beta},{\tilde \ell}_{i\alpha},\nu_i,\ell_k,\nu_j) 
\right] ,
\end{eqnarray}
\begin{eqnarray}
|\overline{M}(\widetilde{\chi}_l^+ \rightarrow \ell^+_i \ell^+_j \ell^-_k)|^2 
= &\frac{g^2 \lam^2}{2} \left[\Psi({\tilde \nu}_i,\ell_j,\ell_k,\ell_i)
+ \Psi({\tilde \nu}_j,\ell_i,\ell_k,\ell_j)\right. \nonumber \\
&\left. +2 \Phi({\tilde \nu}_j,{\tilde \nu}_i,\ell_i,\ell_k,\ell_j) \right] ,
\end{eqnarray}
\begin{eqnarray}
\lefteqn{|\overline{M}(\widetilde{\chi}_l^+ \rightarrow \bar{\nu}_i 
\bar{d}_j u_k)|^2 =} & \nonumber \\
&& \frac{g^2 {\lam'}^2 N_c}{2} \left[ \sum_{\alpha = 1,2} 
|Q^{2k-1}_{2\alpha}|^2 \Psi({\tilde d}^*_{k\alpha},\nu_i,d_j,u_k)
+2 Q^{2k-1}_{21}Q^{2k-1}_{22} \Upsilon({\tilde d}^*_k,\nu_i,d_j,u_k) 
\right] ,
\end{eqnarray}
\begin{eqnarray}
\lefteqn{|\overline{M}(\widetilde{\chi}_l^+ \rightarrow \ell^+_i 
\bar{u}_j u_k)|^2 = } & \nonumber \\
&&\frac{g^2{\lam'}^2 N_c }{2} \left[ \sum_{\alpha = 1,2} 
|Q^{2k-1}_{2\alpha}|^2\Psi({\tilde d}^*_{k\alpha},\ell_i,u_j,u_k)
+2 Q^{2k-1}_{21}Q^{2k-1}_{22} \Upsilon({\tilde d}^*_k,\ell_i,u_j,u_k) 
\right] ,
\end{eqnarray}
\begin{eqnarray}
\lefteqn{|\overline{M}(\widetilde{\chi}_l^+ \rightarrow \ell^+_i 
\bar{d}_j d_k)|^2 =} & \nonumber \\
&& \frac{g^2 {\lam'}^2 N_c}{2} \left[ \Psi({\tilde \nu}_i,d_j,d_k,\ell_i)
+\sum_{\alpha = 1,2} |Q^{2j}_{1\alpha}|^2 \Psi({\tilde u}_{j\alpha},
\ell_i,d_k,d_j)\right. \nonumber \\
&& \left. +2 Q^{2j}_{11} Q^{2j}_{12} \Upsilon({\tilde u}_j,\ell_i,d_k,d_j) 
+2\sum_{\alpha = 1,2} Q^{2j}_{1\alpha} \Phi({\tilde u}_{j\alpha},
{\tilde \nu}_i,\ell_i,d_k,d_j) \right] ,
\end{eqnarray}
\begin{eqnarray}
\lefteqn{|\overline{M}(\widetilde{\chi}_l^+ \rightarrow \nu_i u_j 
\bar{d}_k)|^2 =} & \nonumber \\
&& \frac{g^2 {\lam'}^2 N_c}{2} \left[ \sum_{\alpha = 1,2} 
|L^{2i-1}_{1\alpha}|^2 \Psi({\tilde \ell}_{i\alpha},u_j,d_k,\nu_i)
+\sum_{\alpha = 1,2} |Q^{2j-1}_{1\alpha}|^2 \Psi({\tilde d}_{j\alpha},
\nu_i,d_k,u_j)\right. \nonumber \\
&& +2 L^{2i-1}_{11}L^{2i-1}_{12}\Upsilon({\tilde \ell}_i,u_j,d_k,\nu_i) 
+2 Q^{2j-1}_{11} Q^{2j-1}_{12} \Upsilon({\tilde d}_j,\nu_i,d_k,u_j) 
\nonumber \\
&& \left. +2\sum_{\alpha , \beta = 1,2} L^{2i-1}_{1\alpha}Q^{2j-1}_{1\beta}
\Phi({\tilde d}_{j\beta},{\tilde \ell}_{i\alpha},\nu_i,d_k,u_j) \right] , 
\end{eqnarray}
\begin{eqnarray}
\lefteqn{|\overline{M}(\widetilde{\chi}_l^+ \rightarrow 
u_i u_j d_k)|^2 = } & \nonumber \\
&&\frac{g^2 N_c!}{2(1+\delta_{ij})} \left[ {\lambda''}_{jik}^2 
\sum_{\alpha = 1,2} |Q^{2i-1}_{2\alpha}|^2\Psi({\tilde d}^*_{i\alpha},
u_j,d_k,u_i) +{\lam''}^2\sum_{\alpha = 1,2} |Q^{2j-1}_{2\alpha}|^2
\Psi({\tilde d}^*_{j\alpha},u_i,d_k,u_j) \right.\nonumber \\
&&+2{\lambda''}_{jik}^2Q^{2i-1}_{21}Q^{2i-1}_{22}
\Upsilon({\tilde d}^*_i,u_j,d_k,u_i) +2{\lam''}^2Q^{2j-1}_{21}
Q^{2j-1}_{22}\Upsilon({\tilde d}^*_j,u_i,d_k,u_j) \nonumber \\
&& \left. + 2{\lam''} {\lambda''}_{jik}\sum_{\alpha , \beta = 1,2} 
Q^{2i-1}_{2\alpha} Q^{2j-1}_{2\beta} \Phi({\tilde d}^*_{j\beta},
{\tilde d}^*_{i\alpha},u_i,d_k,u_j) \right] ,
\end{eqnarray}
\begin{eqnarray}
\lefteqn{|\overline{M}(\widetilde{\chi}_l^+ \rightarrow \bar{d}_i \bar{d}_j 
\bar{d}_k)|^2 =}& \nonumber \\
&& \frac{g^2 N_c!}{2(1+\delta_{ij}+\delta_{jk}+\delta_{ik})} \left[ 
{\lam''}^2 \sum_{\alpha = 1,2} |Q^{2i}_{2\alpha}|^2 
\Psi({\tilde u}^*_{i\alpha},d_j,d_k,d_i) \right.\nonumber \\
&& +{\lambda''}_{jki}^2\sum_{\alpha = 1,2} |Q^{2j}_{2\alpha}|^2
\Psi({\tilde u}^*_{j\alpha},d_i,d_k,d_j) \nonumber \\
&&+{\lambda''}_{kij}^2\sum_{\alpha = 1,2} |Q^{2k}_{2\alpha}|^2
\Psi({\tilde u}^*_{k\alpha},d_i,d_j,d_k) +2 {\lam''}^2Q^{2i}_{21}
Q^{2i}_{22}\Upsilon({\tilde u}^*_i,d_j,d_k,d_i) \nonumber \\
&&+2 {\lambda''}_{jki}^2Q^{2j}_{21}Q^{2j}_{22}
\Upsilon({\tilde u}^*_j,d_i,d_k,d_j) +2{\lambda''}_{kij}^2Q^{2k}_{21}
Q^{2k}_{22}\Upsilon({\tilde u}^*_{k\alpha},d_i,d_j,d_k) \nonumber \\
&& -2{\lam''}{\lambda''}_{jki} \sum_{\alpha , \beta = 1,2} 
Q^{2i}_{2\alpha}Q^{2j}_{2\beta}\Phi({\tilde u}^*_{j\beta},
{\tilde u}^*_{i\alpha},d_i,d_k,d_j) \nonumber \\
&& - 2{\lam''}{\lambda''}_{kij}\sum_{\alpha , \beta = 1,2} 
Q^{2i}_{2\alpha}Q^{2k}_{2\beta}\Phi({\tilde u}^*_{k\beta},
{\tilde u}^*_{i\alpha},d_i,d_j,d_k) \nonumber \\
&& \left. - 2{\lambda''}_{jki}{\lambda''}_{kij}\sum_{\alpha ,\beta = 1,2} 
Q^{2j}_{2\alpha}Q^{2k}_{2\beta}\Phi({\tilde u}^*_{k\beta},
{\tilde u}^*_{j\alpha},d_j,d_i,d_k) \right] . 
\end{eqnarray}

\vfill 
\clearpage 
{\Large\normalsize}

\vfill 
\clearpage 
 
\printindex 
 
\end{document}